\documentclass[a4paper,11pt]{book}
\usepackage[utf8]{inputenc}
\usepackage[T1]{fontenc} 
\usepackage{lmodern} 
\usepackage[margin=28mm,includeheadfoot,bindingoffset=5mm]{geometry}[2010/03/13]

\usepackage{graphicx}
\usepackage{amsmath}
\usepackage{bbold}
\usepackage{amssymb} 
\usepackage{textcomp} 
\usepackage[most]{tcolorbox} 
\usepackage{enumitem} 
\usepackage{xcolor}
\usepackage{float}
\usepackage{lscape}
\usepackage{physics} 
\usepackage{stmaryrd}
\usepackage{wasysym} 
\usepackage{tikz}
\usepackage{cite}
\usepackage{hyperref}

\usepackage{stmaryrd}
\usepackage{trimclip}

\makeatletter
\DeclareRobustCommand{\shortto}{%
  \mathrel{\mathpalette\short@to\relax}%
}

\newcommand{\short@to}[2]{%
  \mkern2mu
  \clipbox{{.5\width} 0 0 0}{$\m@th#1\vphantom{+}{\shortrightarrow}$}%
  }
\makeatother

\usepackage{thcover}

\tcbset{enhanced,colback=red!5!white, colframe=red!75!black,fonttitle=\bfseries}
\graphicspath{{figures/}} 
\usepackage[explicit, clearempty]{titlesec}
\titleformat{\chapter}[display]{\bfseries\filright}{\huge\chaptername~\thechapter}{20pt}{\Huge#1}
\titleformat{name=\chapter, numberless}[display]{\bfseries\filright}{}{0pt}{\Huge#1}[\addcontentsline{toc}{chapter}{#1}]

\begin{document}
\frontcover
\setcounter{tocdepth}{-1}
\tableofcontents
\addtocontents{toc}{\protect\setcounter{tocdepth}{2}}%

\mainmatter






\chapter*{Introduction}

\section*{General context}

\noindent Photons, the quanta of the electromagnetic field, do$\,$not$\,$interact$\,$with$\,$each$\,$other$\,$in$\,$vacuum. In dielectric materials however, the situation$\,$is$\,$different,$\,$as$\,$photons$\,$are$\,$likely$\,$to$\,$couple$\,$with the electric charges inside the medium and, thus, to polarize it locally. As polarized$\,$matter radiates light, a polarization field appears in response to the$\,$incoming$\,$electromagnetic$\,$one. This response is nonlinear in most materials in the sense that their polarizabilities depend, in a nonlinear manner, on the strength of the incoming light field. In practice nevertheless, only$\,$lasers$\,$are$\,$sufficiently$\,$intense$\,$to$\,$modify$\,$in$\,$such$\,$a$\,$way$\,$the$\,$optical$\,$properties$\,$of$\,$a$\,$material. That is why the beginning of the research in nonlinear optics is often considered to$\,$be$\,$the discovery$\,$of$\,$second-harmonic$\,$generation$\,$by$\,$Franken$\,$and$\,$co-workers~\cite{1:Franken}$\,$(1961),$\;$shortly$\,$after the first experimental demonstration of working laser by Maiman in 1960~\cite{2:Maiman}. Since$\,$then, higher optical intensities are usable and a vast array of nonlinear$\,$effects$\,$have$\,$been$\,$explored, with wide-ranging applications from optical frequency conversion~\cite{3:Jain} to light storage~\cite{4:Fleischhauer} and quantum information processing~\cite{5:Lukin}. 
\vspace{6pt}
\newline
\noindent Under specific conditions, the$\,$effective$\,$photon-photon$\,$interaction$\,$arising$\,$from$\,$the$\,$nonlinear polarizability of certain materials (such as photorefractive crystals, thermo-optic liquids or warm alkaline vapors for instance) makes light behave as a fluid. Indeed, the paraxial propagation of a laser field in a Kerr medium $-$ whose refractive index depends$\,$on$\,$the$\,$input light intensity $-$ is governed by the nonlinear Schr\"{o}dinger equation. The$\,$later$\,$shares$\,$strong similarities with the Gross-Pitaevskii equation, describing the dynamics of quantum$\,$fluids such as atomic Bose-Einstein condensates (BECs), and can, thus, be$\,$reformulated$\,$into$\,$a$\,$set of$\,$hydrodynamic equations. In$\,$this$\,$hydrodynamic$\,$analogy,$\,$the$\,$laser$\,$field$\,$can$\,$be$\,$regarded$\,$as fluid of light whose density and velocity are respectively defined by the$\,$laser$\,$intensity$\,$and its phase gradient. The first branding of a coherent light$\,$field$\,$as$\,$a$\,$photon$\,$fluid$\,$dates$\,$back$\,$to the early nineties, where the time-evolution of the electromagnetic field in$\,$a$\,$laser$\,$cavity$\,$was reformulated into a Ginzburg-Landau equation \cite{6:Brambilla,7:Staliunas}. In$\,$that$\,$latter$\,$case,$\,$the$\,$photon-photon interaction was mediated through the nonlinear refractive index of the cavity gain$\,$medium. In the following decade, a seminal attempt to observe superfluid-like behaviours in these cavity systems~\cite{8:Vaupel} has been followed by a series of theoretical articles by Chiao \textit{et al.}~\cite{9:Chia0,10:Chia0}. Surprisingly, no other experiments were reported thereafter, possibly because large non-linearities and high-Q factor cavities were hardly available at the time.  
\newpage

\noindent Interestingly, modern research on quantum fluids of light has shifted towards$\,$the$\,$study$\,$of exciton-polaritons in micro-cavities, thanks to progress in semi-conductor nano-structures manufacturing. In these systems, the photon entering the mico-cavity strongly mixes with electron/hole pairs (the excitons) in the bulk material. This leads to the creation$\,$of$\,$massive interacting bosonic quasi-particles, known as exciton-polaritons. The optical nonlinearity arises here from the Coulomb interaction between the excitons. The spatial confinement, provided$\,$by$\,$the$\,$cavity,$\,$moreover$\,$assigns$\,$an$\,$non-zero effective$\,$mass$\,$to$\,$photons.$\,$Remarkably, the space-time evolution of the polaritons wave-function follows a similar dynamics to that of interacting atomic BECs, but includes additional detrimental non-equilibrium$\,$features arising$\,$from$\,$its$\,$intrinsic$\,$dissipative$\,$nature.$\;$After$\,$the$\,$observation$\,$of$\,$polariton$\,$BECs\cite{11:Kasprzak,12:Baumberg,13:Balili} the hydrodynamical aspects of polariton fluids gained a lot of interest because of the easy experimental$\,$access$\,$they$\,$were$\,$offering$\,$to$\,$many-body$\,$physics.$\;$For$\,$instance,$\;$a$\,$series$\,$of$\,$works on the superfluid aspects of these photon fluids has led to the experimental$\,$observation$\,$of dissipationless flows around$\,$a$\,$defect\cite{14:Amo,15:Amo}.$\;$The$\,$nucleation$\,$of$\,$quantized$\,$vortex/anti-vortex pairs at the breakdown of superfluidity\cite{16:Sanvitto}, as well as the generation of dark solitons past an obstacle in the supersonic regime\cite{17:Amo}, have also been reported. All the previous cited articles have$\,$been$\,$published$\,$by$\,$our$\,$group$\,$at$\,$LKB$\,$and$\,$co-workers$\,$in$\,$between$\,$2009$\,$and$\,$2011.
\vspace{6pt}
\newline
\noindent However,$\;$exciton-polariton$\,$systems$\,$suffer$\,$from$\,$fundamental$\,$limitations.$\;$First,$\;$they$\,$require continuous pumping due$\,$to$\,$dissipation$\,$and$\,$losses$\,$(light$\,$fatally$\,$leaks$\,$out$\,$of$\,$the$\,$micro-cavity) which$\,$thus$\,$makes$\,$the$\,$dynamics$\,$of$\,$polariton$\,$fluids$\,$intrinsically$\,$driven-dissipative.$\;$Moreover, the$\,$effective$\,$photon-photon$\,$interaction$\,$strength$\,$is$\,$fully$\,$governed$\,$by$\,$the$\,$exciton-exciton$\,$one, which is$\,$only$\,$marginally$\,$tunable.$\;$In$\,$order$\,$to$\,$overcome$\,$both$\,$these$\,$limitations,$\;$an$\,$alternative approach based on cavityless systems can be used. As mentioned previously, the paraxial propagation of a laser field inside$\,$a$\,$bulk$\,$nonlinear$\,$material$\,$can$\,$be$\,$described$\,$as$\,$a$\,$fluid$\,$of$\,$light evolving along the propagation direction as a normal fluid will do over time. These$\,$systems are commonly referred to as \textit{propagating} or \textit{paraxial photon fluids}. After the$\,$observation$\,$by Swartzlander \textit{et al.} of quantized vortex$\,$pairs$\,$created$\,$by$\,$the$\,$instability$\,$of$\,$dark$\,$soliton$\,$stripes in nonlinear liquids~\cite{19:Swartzlander}, the hydrodynamics features of paraxial fluids of light have been extensively investigated theoretically. An exhaustive list of the works published$\,$in$\,$this$\,$field can$\,$be$\,$find$\,$in$\,$the$\,$review$\,$by$\,$Carusotto$\,$and$\,$Ciuti~\cite{18:Carusotto}.$\;$Especially$\,$strong$\,$attention$\,$was$\,$revived toward$\,$solitonic$\,$structures~\cite{20:Josserand,21:Firth,22:Michinel,23:Paz}. Surprisingly, few experimental studies have specifically investigated the hydrodynamic features of light$\,$propagating$\,$in$\,$bulk$\,$nonlinear$\,$mediums$\,$and most of them have been performed in the Fleischer's group from 2007$\,$to$\,$2012.$\,$For$\,$instance, the generation, the propagation as well as the interaction of optical dispersive shock-waves has$\,$first$\,$been$\,$studied$\,$in$\,$photorefractive$\,$crystals$\,$by$\,$Jia$\,$\textit{et$\,$al.}~\cite{24:Jia} and$\,$Wan$\,$\textit{et$\,$al.}~\cite{25:Wan} in$\,$2007. Wan and co-workers have also reported the spontaneous nucleation of quantized vortices in the wake of an obstacle in a 2-dimensional geometry~\cite{27:Wan} and$\,$the$\,$tunneling$\,$of$\,$density$\,$waves through a small potential barrier in the 1-dimensional case~\cite{26:Wan}. Their studies have$\,$laid$\,$the groundwork$\,$for$\,$studying$\,$superfluid$\,$effects$\,$in$\,$paraxial$\,$photon$\,$fluids.$\,$Recently,$\,$there$\,$has$\,$been a renewed interest in this field, with the observation of quantized vortices past an obstacle in thermo-optics liquids~\cite{2-23Vocke} and the optomechanical demonstration of frictionless flows$\,$of light in photorefractive crystals by Michel \textit{et al.}~\cite{3-3Michel}.  

\newpage

\section*{Motivations}

\noindent This thesis aims at studying the hydrodynamical properties of paraxial photon$\,$fluids$\,$in$\,$hot rubidium vapors. In this system, the nonlinear interaction between photons is provided by the nonlinear polarization of the atomic ensemble when the light frequency$\,$is$\,$tuned$\,$close$\,$to an$\,$atomic$\,$resonance. Hot rubidium vapors offer$\,$certain$\,$compelling$\,$advantages$\,$compared$\,$to other$\,$platforms$\,$used$\,$so$\,$far$\,$to$\,$generate$\,$paraxial$\,$fluid$\,$of$\,$light$\,$(that is,$\,$photorefractive$\,$crystals and thermo-optics liquids principally). First, the nonlinear interaction between photon$\,$is easily tunable over several orders of magnitude in this system$\,$by$\,$changing$\,$the$\,$vapor$\,$density (which$\,$increases$\,$exponentially$\,$with$\,$the$\,$temperature)$\,$and/or$\,$the$\,$laser$\,$frequency.$\;$Moreover, the vapor optical response is only weakly$\,$nonlocal,$\,$in$\,$contrast$\,$to$\,$thermo-optic$\,$liquids$\,$where strong nonlocalities, arising from heat diffusion inside$\,$the$\,$material,$\,$have$\,$been$\,$reported~\cite{2-23Vocke}. Such nonlocalities make the observation of quantum phenomena difficult if$\,$not$\,$impossible. We thus believe that alkaline vapors could be the first platform to lay the$\,$groundwork$\,$for studying quantum effects in propagating photon fluids.
\vspace{6pt}
\newline
\noindent Despite the fact that many nonlinear optics experiments have been performed using warm rubidium vapors (and,$\,$more$\,$generally,$\,$alkaline$\,$vapors),$\,$only$\,$few$\,$attention$\,$has$\,$been$\,$devoted to studying the hydrodynamical aspect of the nonlinear propagation of laser beams inside these systems. In addition, most of the experimental works on propagating photon fluids reported so far focus mainly on studying strongly nonlinear effects, such as spatial solitons or dispersive shock-waves. The amplitude of such nonlinear phenomena is large enough$\,$to locally change the properties of the photon fluid. Conversely, only few$\,$studies$\,$consider$\,$the linear aspect of the dynamics, that is, how small amplitude density waves travel onto the fluid of light. Throughout this manuscript,$\,$I$\,$will$\,$explore$\,$phenomena$\,$lying$\,$on$\,$this$\,$borderline between nonlinear hydrodynamics and standard nonlinear optics. The$\,$main$\,$purpose$\,$of$\,$this thesis is thus to further bridge the gap between this$\,$two$\,$disciplines,$\,$by$\,$drawing$\,$new$\,$parallels between nonlinear optics and Bogoliubov formalism principally.  

\section*{Thesis Summary}

\noindent This$\,$thesis$\,$manuscript$\,$consists$\,$of$\,$three$\,$parts.$\;$The$\,$first$\,$one$\,$(chapters$\,$2$\,$and$\,$3)$\,$is$\,$a$\,$theoretical introduction to hot rubidium$\,$vapors$\,$and$\,$paraxial$\,$photon$\,$fluids.$\;$The$\,$second$\,$part$\,$(chapter$\,$4) presents the tools$\,$and$\,$methods$\,$used$\,$to$\,$create$\,$and$\,$characterize$\,$photon$\,$fluids$\,$in$\,$our$\,$platform. The third part (chapters 4, 5 and 6) is mainly dedicating to presenting the experimental results I obtained during my thesis. 
\vspace{3pt}
\newline
\noindent \textbf{Chapter 1} $-$ The optical response of a warm rubidium vapor under a near resonance laser-excitation is described using a two level model first and $-$ at a later stage $-$ a more sophisticated 3-level scheme. Effects arising from the atomic motion (Doppler$\,$broadening, transport induced nonlocality) are taken into account in this second description.
\vspace{3pt}
\newline
\noindent \textbf{Chapter 2} $-$ The theoretical ground work for studying propagating fluids of$\,$light$\,$is$\,$laid. The analogy between the propagation of a laser beam in a Kerr medium and the evolution of a fluid of light is discussed in detail. A particular intention is devoted$\;$to$\;$studying$\;$the dynamics of small amplitude density modulations travelling$\,$onto$\,$the$\,$paraxial$\,$photon$\,$fluid. In this perspective, the dispersion relation of these density waves is derived. The effects$\,$of absorption and nonlocality on this dispersion are investigated. 

\newpage

\noindent \textbf{Chapter 3} $-$ The tools$\,$and$\,$methods$\,$used$\,$to$\,$generate$\,$and$\,$characterise$\,$propagating$\,$photon fluids in hot rubidium vapors are introduced. An$\,$important$\,$part$\,$of$\,$this$\,$chapter$\,$is$\,$dedicated to presenting the method used to measure$\,$the$\,$vapor$\,$nonlinear$\,$refractive$\,$index.$\;$The$\,$latter$\,$is based on measuring the self-phase accumulated by a Gaussian$\,$beam$\,$during$\,$its$\,$propagation inside the vapor cell. We show that counting the ring appearing in its far-field intensity distribution is enough to access the nonlinear refractive index.
\vspace{3pt}
\newline
\noindent \textbf{Chapter 4} $-$ A measurement of the dispersion relation of small amplitude density waves is reported. This is achieved by measuring the group velocity of a small amplitude wave-packet travelling onto the$\,$photon$\,$fluid.$\;$The$\,$dispersion$\,$relation$\,$exhibits$\,$a$\,$linear$\,$trend$\,$at$\,$low excitation$\,$wave-vector,$\,$whose$\,$slope$\,$defines$\,$the$\,$velocity$\,$of$\,$sound$\,$in$\,$the$\,$paraxial$\,$photon$\,$fluid. The way the sound velocity depends on the fluid density is studied and perfectly matches theory without any fitting parameter. According to the Landau criterion for superfluidity, demonstrating the existence of such a sonic regime in the dispersion relation is a key step toward the observation of superfluid flow of light.  
\vspace{3pt}
\newline
\noindent \textbf{Chapter 5} $-$ In order to probe the superfluidity of light, the way the fluid flows around a localized obstacle (\textit{ie}, a local change of refractive$\,$index)$\,$should$\,$be$\,$investigated.$\;$Chapter$\,$5$\,$is dedicated to studying how such an obstacle can be optically generated in rubidium$\,$vapors. In the first section, the 2-laser$\,$optical$\,$scheme$\,$designed$\,$to$\,$that$\,$end$\,$is$\,$theoretically$\,$described, using the dressed-state formalism. In the second section, we show that Bessel beams are suitable$\,$to$\,$generate$\,$spatially$\,$localized,$\,$collimated$\,$obstacles.$\,$Using$\,$a$\,$spatial$\,$light$\,$modulator, we demonstrate that the on-axis intensity of such beam$\,$can$\,$be$\,$tailored$\,$so$\,$as$\,$to$\,$compensate absorption locally. 
\vspace{3pt}
\newline
\noindent \textbf{Chapter 6} $-$ Preliminary results obtained by$\,$bringing$\,$all$\,$the$\,$previous$\,$ingredients$\,$together are presented. Images of the near- and far-field$\,$scattering$\,$patterns$\,$generated$\,$by$\,$making$\,$the fluid flow toward the all-optical obstacle are shown. The amount$\,$of$\,$light$\,$scattered$\,$by$\,$the defect is measured in Fourier space as function of the photon fluid density and velocity.







\chapter{Atomic vapor}

\noindent Photon-photon interactions in paraxial photon fluids arise from the light-matter$\,$coupling in nonlinear optical materials. The strength of the nonlinear interaction between photon, as well as the thickness of the nonlinear material, are key parameters in$\,$experiments$\,$which need to be made as large as possible. Hot alkaline vapors constitute a$\,$sound$\,$choice$\,$because they$\,$provide$\,$strong$\,$optical$\,$nonlinearities$\,$under$\,$a$\,$near-resonance$\,$laser$\,$excitation.$\;$Moreover, spectroscopic cells filled with high quality alkaline vapors are easy to handle and also to design according to our specific needs (isotopic concentration, cell length and shape, etc). We$\,$chose$\,$more$\,$specifically to work with rubidium vapors as$\,$an$\,$expertise$\,$in$\,$dealing$\,$with$\,$this atom has already been developed in the team for several years. The$\,$purpose$\,$of$\,$this$\,$chapter is to$\,$introduce$\,$the$\,$theoretical$\,$framework$\,$describing$\,$the$\,$optical$\,$response$\,$of$\,$a$\,$rubidium$\,$vapor under$\,$a$\,$near$\,$resonance$\,$laser$\,$excitation.$\,$To$\,$this$\,$end,$\,$I$\,$first$\,$model$\,$the$\,$rubidium$\,$level$\,$structure with$\,$a$\,$two-level$\,$description, that$\,$mainly$\,$helps$\,$introducing$\,$the$\,$basic$\,$concepts$\,$and$\,$notations. This simplistic model is improved afterward, by taking into account the optical pumping between the rubidium ground states. The effects of the high vapor temperatures on the optical response are also investigated.

\newpage

\section{Atomic structure}

Rubidium belongs to the alkali metal group of the periodic table and possesses therefore one valence electron on its outermost shell. Two rubidium isotopes  are present$\;$on$\;$Earth: rubidium 85 ($^{_{85}}$Rb, $72\%$) and rubidium 87 ($^{_{87}}$Rb, $28\%$), which is slightly radioactive, with a half-life$\;$time of 49 billion years (safe enough). Natural-abundance mixtures and isotopically pure vapors were used in the$\,$lab$\,$(more$\,$than$\,99\%\,$purity).$\,$In$\,$the$\,$experiments$\,$presented in this manuscript, we optically addressed the rubidium $D$-lines exclusively, either or both the $D_{1}$ ($5^{2}S_{1/2} \rightarrow 5^{2}P_{1/2}$)$\,$and$\,$the $D_{2}$ ($5^{2}S_{1/2} \rightarrow 5^{2}P_{3/2}$)$\,$component.$\;$The$\,$optical$\,$properties of the rubidium $D$-lines have been gathered in review by D.A.$\,$Steck$\,$\cite{2-1Steck,2-2Steck}$\,$for$\,$rubidium$\,$85 and rubidium 87. In this first section, I will introduce the reader to the$\,$fine$\,$and$\,$hyperfine structures of rubidium $D_{1}$ and $D_{2}$ lines, that basically determine the optical$\;$response of the atomic vapor under a near-resonance laser excitation.

    \subsection{Fine structure $-$ LS coupling}
    
The doublet structure of the rubidium D-line arises from the coupling between the angular orbital momentum $\mathbf{L}$ of the valence electron and its spin angular momentum $\mathbf{S}$. The total electron angular momentum $\mathbf{J}$ is then defined by: $\mathbf{J} = \mathbf{L}+\mathbf{S}$; the associated quantum number $J$ lies in the range $\vert L-S \vert \! \leq \! J \! \leq \! L+S $. In the ground state of rubidium, $L = 0$ and $S = 1/2$ so $J = 1/2$; in the first excited state, $L = 1$ and $S = 1/2$ so $J = 1/2$ or $3/2$. The fine coupling leads thus to the splitting of the first rubidium excited$\;$state$\;$into two fine states ($5^{2}P_{1/2}$ and $5^{2}P_{3/2}$) and gives, accordingly, a doublet structure to the D-line. 

\begin{figure}[h]
\center
\includegraphics[width=\columnwidth]{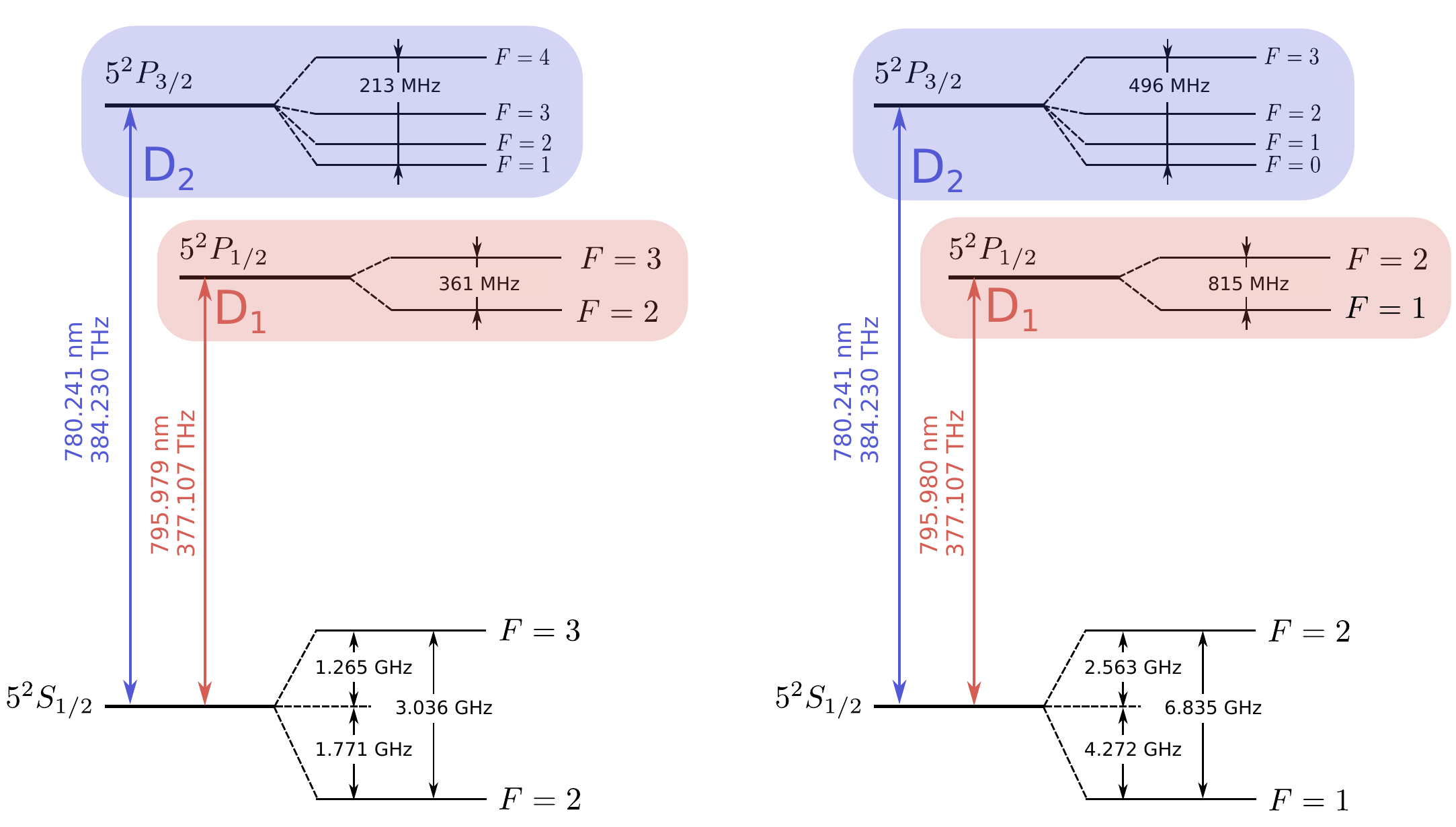} 
\caption{$^{85}$Rb (left) and $^{87}$Rb (right) D-line hyperfine structure.}
\label{fig:RbLevelSchematic}
\end{figure}

    \subsection{Hyperfine structure $-$ IJ coupling}
We can go one step further by introducing the atomic hyperfine structure, which arises from the coupling between the total electron angular momentum, $\mathbf{J}$, and the nuclear$\;$spin angular momentum, $\mathbf{I}$. As before, the quantum number $F$, associated to$\,$the$\,$atomic$\,$angular momentum $\mathbf{F} = \mathbf{J} + \mathbf{I}$, lies in between $\vert J-I \vert$ and $J+I$. The nuclear spin differs from one isotope$\,$to$\,$the$\,$other:$\,I = 5/2\,$for$\,^{_{85}}$Rb$\,$and$\,3/2\,$for$\,^{_{87}}$Rb.$\,$In$\,$the$\,$ground$\,$state,$\,$the$\,$total$\,$electron angular momentum is $1/2$ and $F$ can therefore takes two different values: $F = 2$ or $F = 3$ for $^{_{85}}$Rb and $F = 1$ or $F = 2$ for $^{_{87}}$Rb. The ground state of both isotopes separates thus into two hyperfine states. Moreover, it is straightforward to show that each of the excited states $5^{2}P_{1/2}$ and $5^{2}P_{3/2}$ splits into two and$\;$four$\;$hyperfine$\;$states respectively. Let's for instance consider the case of $^{_{85}}$Rb:
\begin{itemize}
    \item [$\bullet$] For $5^{2}P_{1/2}$ ($D_{1}$ line), $F$ can take any integer values between $I-J = 2$ and $I+J = 3$. The fine state $5^{2}P_{1/2}$ splits thus into $2$ hyperfine states labeled by $F = 2$ and $F = 3$.    
    \item [$\bullet$] For $5^{2}P_{3/2}$ ($D_{2}$ line), $F$ can take any integer values from $I-J = 1$ to $I+J = 4\;$and $5^{2}P_{3/2}$ splits therefore into $4$ hyperfine states labeled by $F = 1,2,3,4$ respectively.
\end{itemize}
When no external magnetic field is applied, each of the hyperfine ground and excited states are degenerate, as they contain $2F + 1$ magnetic sublevels each,$\,$labelled$\,$by$\,$the$\,$total$\,$angular momentum projection $m_{F}$. The schematic of the $D$-line hyperfine level structure for both $^{_{85}}$Rb and $^{_{87}}$Rb is depicted on figure~\ref{fig:RbLevelSchematic}. The transition frequencies$\;$are$\;$reported,$\;$as$\;$well$\;$as the frequency difference between (i) the hyperfine ground states and (ii) the furthest apart excited states for in each of the $D$-lines. 

\begin{figure}[h]
\center
\includegraphics[scale=0.4]{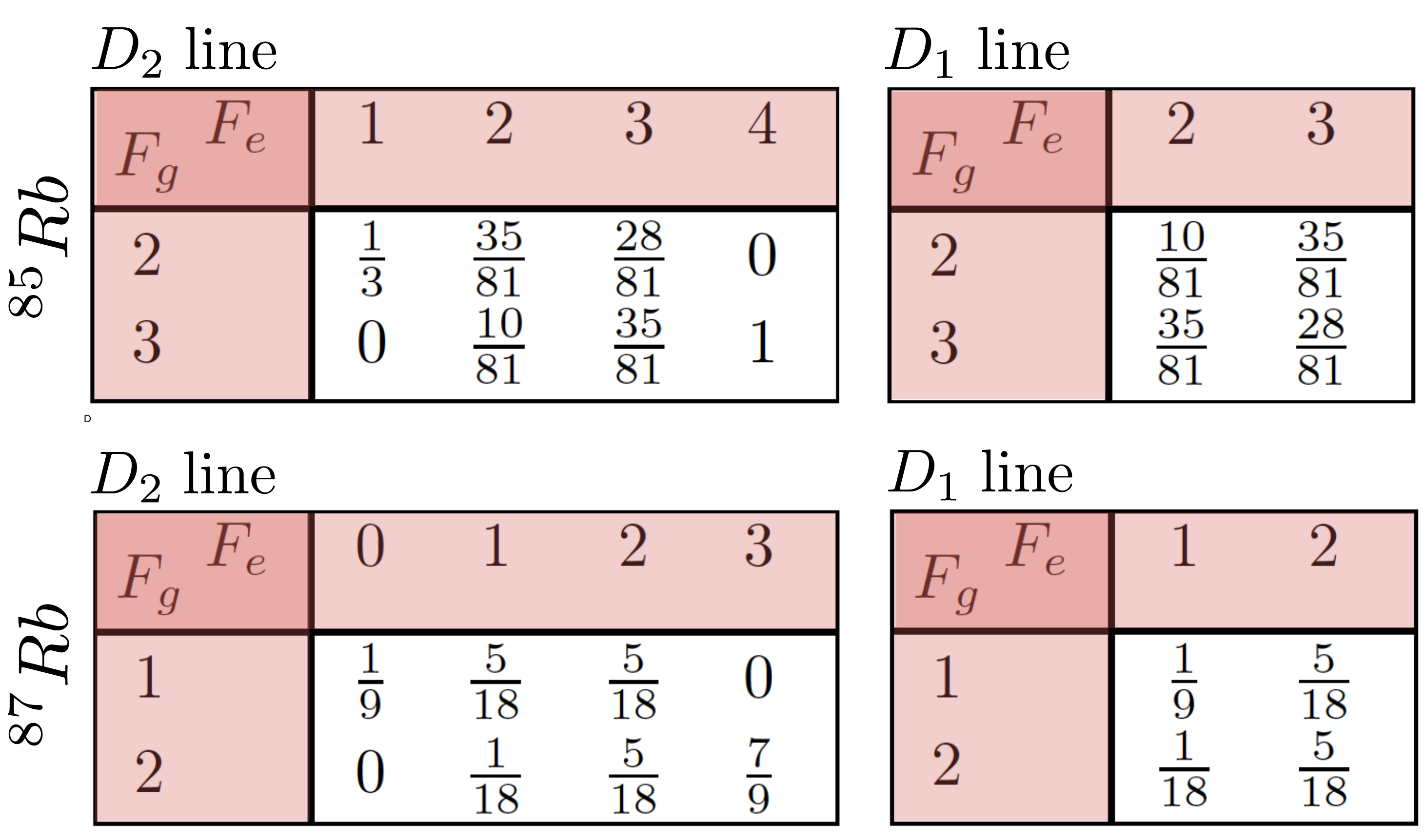} 
\caption{Line strengths $C_{_{F}}^2$ of the $D$-lines for $^{_{85}}$Rb and $^{_{87}}$Rb~\cite{2-9Siddons}}
\label{tab:TabStrength}
\end{figure}

\newpage 

    \subsection{Line strength: reduction of the dipole operator}

The optical response of the vapor under a monochromatic laser excitation is a key$\;$feature we need to characterize in order to accurately describe our experiments.  This response$\;$is linked to the polarizability (or the dielectric$\,$susceptibility) of the material, which strongly depends on the dipole strength of the transition we optically address. In order to$\;$evaluate the dipole strength associated to the transition between the ground and excited states $\ket{F_{g}, m_{F_{g}}}$ and $\ket{F_{e}, m_{F_{e}}}$, we must calculate the matrix element:
\begin{equation}
    \bra{F_{g}, m_{F_{g}}} \boldsymbol{\hat{d}} \ket{F_{e}, m_{F_{e}}} = \bra{F_{g}, m_{F_{g}}} e \, \boldsymbol{\hat{r}} \ket{F_{e}, m_{F_{e}}},  
\end{equation}

\noindent where $\boldsymbol{\hat{d}}$ and $\boldsymbol{\hat{r}}$ are the dipole and position operators. In doing so, we should first factor$\;$out the angular dependence and write the matrix element as a product of Wigner $3\!-\!j$ and $6\!-\!j$ symbols times a reduced matrix element. This procedure is called "reduction of the dipole operator" \cite{2-3Edmonds} and reads as follows:
\begin{align}
\nonumber
\bra{F_{g}, m_{F_{g}}} e \, \hat{r}_{q} \ket{F_{e}, m_{F_{e}}} =& \; (-1)^{F_{e}-1+m_{F_{g}}} \, (-1)^{F_{e}+J_{g}+1+I} \, (-1)^{J_{e}+L_{g}+1+S} \, \langle L_{g} || e \, \boldsymbol{\hat{r}} || L_{e} \rangle \\
\nonumber
& \times \sqrt{\vphantom{(}2F_{g}+1 \vphantom{)}} \, \sqrt{(2 F_{e}+1)(2 J_{g}+1)} \, \sqrt{(2 J_{e}+1)(2 L_{g}+1)} \\ 
& \times 
    \begin{pmatrix}
        F_{e} & 1 & F_{g} \\ 
        m_{F_{e}} & q & -m_{F_{g}} 
    \end{pmatrix} \;
    \begin{Bmatrix}
        J_{g} & J_{e} & 1 \\ 
        F_{e} & F_{g} & I 
    \end{Bmatrix} \;
    \begin{Bmatrix}
        L_{g} & L_{e} & 1 \\ 
        J_{e} & J_{g} & S  
    \end{Bmatrix},
    \label{ReductionDipole}
\end{align}

\noindent where $q$ labels the components of $\boldsymbol{\hat{r}}$ in the spherical basis; according to its usual definition, $q$ quantifies the integer change in the angular momentum projection during the transition. The reduced dipole matrix element $d = \langle L_{g} || \, \boldsymbol{\hat{r}}|| L_{e} \rangle$ involves only the quantum number $L$ and is therefore identical for both $D$ lines. The round and curly brackets matrices in~\eqref{ReductionDipole} stand for the Wigner $3\!-\!j$ and $6\!-\!j$ symbols respectively. For linearly polarized light, $q = 0$ and the $3\!-\!j$ symbol is non-zero only when $m_{F_{e}} = m_{F_{g}}$. We can for instance express it as a function of the transition wavelength $\lambda$ and decay rate $\Gamma$ of the $D_{1}$ line. The decay$\;$rate associated to the fine-structure transition $J_{g} \rightarrow J_{e}$ reads as follows: 
\begin{equation}
\label{Gamma}
    \Gamma = \frac{\omega^{3}}{3 \pi \epsilon_{0} \hbar c^{3}} \frac{2 J_{g}+1}{2 J_{e}+1} \left| \langle J_{g} || e \, \boldsymbol{\hat{r}} || J_{e} \rangle \right|^{2},
\end{equation}

\noindent where the reduced matrix element $\langle J_{g} || e \, \boldsymbol{\hat{r}} || J_{e} \rangle$ is related to the reduced dipole factor $d\;$by: 
\begin{equation}
\langle J_{g} || e \, \boldsymbol{\hat{r}} || J_{e} \rangle = (-1)^{J_{e}+L_{g}+1+S} \, \sqrt{(2 J_{e}+1)(2 L_{g}+1)}  
    \begin{Bmatrix}
        L_{g} & L_{e} & 1 \\ 
        J_{e} & J_{g} & S  
    \end{Bmatrix} \times d.
\end{equation}

\noindent For the $D_{1}$ line (where $J_{g} = J_{e} = 1/2$) the relation above reduces to: $\langle J_{g} || e \, \boldsymbol{\hat{r}} || J_{e} \rangle = d / \sqrt{3}$. Using~\eqref{Gamma} and rearranging:
\begin{equation}
d = \langle L_{g} || e \, \boldsymbol{\hat{r}} || L_{e} \rangle = \sqrt{\frac{9 \, \epsilon_{0} \, \hbar \, \Gamma \, \lambda^3}{8 \pi^2}}.
\end{equation} 

\noindent A similar analysis leads to the same result for the $D_{2}$ line. Using the tabulated values$\,$\cite{2-1Steck} for the $D$-lines wavelengths and decay rates yields $d = 5.182 \, e \, a_{0}$ and $d = 5.177 \, e \, a_{0}$ for the $D_{1}$ and $D_{2}$ lines respectively, $a_{0}$ being the Bohr radius. 

\newpage

\noindent The dipole matrix element in equation~\eqref{ReductionDipole} reads finally: $\bra{F_{g}, m_{F_{g}}} e \hat{r}_{q} \ket{F_{e}, m_{F_{e}}} = c_{m_{F}} \, d$, where $c_{m_{F}}$ is a geometrical factor that depends on the the initial and final hyperfine states of the transition. Since no magnetic field is applied, every hyperfine state is$\;$degenerate $2F_{g}+1$ times. The total strength $f_{_{F_{g}}}^{_{F_{e}}}$ of the $F_{g}\rightarrow F_{e}$ hyperfine transition is thus obtained by averaging over the strength of all the Zeeman transitions $m_{F_{g}} \rightarrow m_{F_{e}}$ in the hyperfine manifold. We finally obtained that: 
\begin{equation}
\label{Stength}
    f_{_{F_{g}}}^{_{F_{e}}} = \frac{\sum_{m_{F}} c_{m_{F}}^{2} \, d^{2}}{2F_{g}+1} = \frac{ C_{_{F}}^{2} \, d^{2}}{2F_{g}+1},
\end{equation}
\noindent where $C_{_{F}}^{2}= \sum_{m_{F}} c_{m_{F}}^{2}$.
The $C_{_{F}}^{2}$ coefficients have been calculated for linearly polarized$\;$light and are reported in the table of figure~\ref{tab:TabStrength} for both $^{_{85}}$Rb and $^{_{87}}$Rb.
\vspace{6pt}
\newline
\noindent It is worth mentioning however that the formula \eqref{Stength} is generally an approximation as it amounts$\,$to$\,$neglecting$\,$optical$\,$pumping$\,$effects$\,$between$\,$Zeeman$\,$sublevels,$\,$such$\,$as$\,$population aligning and population trapping. In other words, we assume here that every Zeeman transition $m_{F_{g}} \rightarrow m_{F_{e}}$ is cycling, which is obviously wrong as the excited state $\ket{F_{e}, m_{F_{e}}}$ may also decay toward $\ket{F_{g}, m_{F_{e}}\!\pm\!1}$ according to the selection rules. 
\begin{itemize}
    \item [$\bullet$] In the case of a cycling hyperfine transition for which $F_{e} < F_{g}$ (the $F_{g} = 1  \rightarrow  F_{e} = 0$ transition of $^{87}$Rb for instance), the atoms get pumped into one of the dark states $\ket{F_{g}, m_{F_{g}}\!=F_{g}}$ or $\ket{F_{g}, m_{F_{g}}\!=-F_{g}}\,$(population$\,$trapping,$\,$fig.$\,$1.3(a))$\,$and$\,$remain$\,$there.
    \item[$\bullet$] When $F_{e}$ and $F_{g}$ are integers and fulfilled the condition $F_{e} = F_{g}$, the Zeeman$\;$sublevel on the excited state right edge $\ket{F_{e}, m_{F_{e}}\!=F_{e}}$ (resp. left edge $\ket{F_{e}, m_{F_{e}}\!=-F_{e}}$)$\,$may only decay toward the sublevels $\ket{F_{g}, m_{F_{g}}\!=F_{g}}$ or $\ket{F_{g}, m_{F_{g}}\!=F_{g}-1}$ (resp.$\,$toward $\ket{F_{g}, m_{F_{g}}\!=-{F_{g}}}\,$or$\,\ket{F_{g}, m_{F_{g}}\!=-F_{g}+1}$).$\;$This$\,$asymmetry$\,$in$\,$the$\,$decay$\,$process$\,$forces the atoms to accumulate in the sublevel $m_{F_{g}}=0$ (population aligning, fig.$\,$1.3(b)).
\end{itemize}


\begin{figure}[h]
\center
\includegraphics[width=\columnwidth]{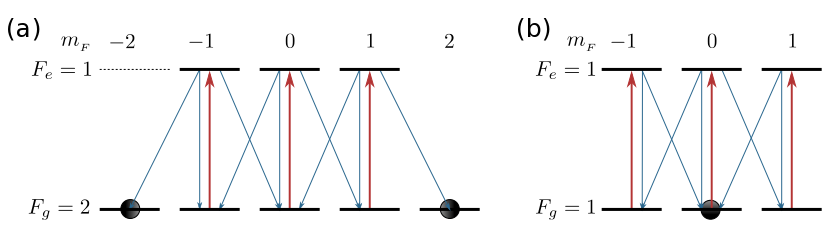} 
\caption{Optical pumping between Zeeman sublevels. The excitation laser is linearly polarized and couples ground and excited sublevels sharing the same $m_{F}$ (red arrows). The blue arrows represent all possible decay channels from $F_{e}$ to $F_{g}$. (a) Sketch of the population trapping process: atoms end up in the $m_{F_{g}} = \pm F_{g}$ dark states which are not coupled by the excitation laser with any of the excited Zeeman sublevels. (b) Sketch$\;$of$\;$the population aligning process: atoms are more likely to decay toward the $m_{F_{g}} = 0$.}
\label{fig:ZeemanPumping}
\end{figure}

\newpage

\noindent The Zeeman degeneracy can affect the atomic population in the steady-state and$\;$change, consequently, the optical response of the rubidium vapor~\cite{2-4Sagle}. However, taking the whole hyperfine structure plus the Zeeman degeneracy into account to describe the polarisation of the vapor under a near-resonance laser excitation is far to be easy and not necessary$\;$in a first instance to get a good insight about the physical process at play. Let's then continue with equation \eqref{Stength}, bearing in mind the discussion above.  

\section{Two-level atoms}
\label{sec:2Level}

\newsavebox{\auteurbm}
\newenvironment{Bonmot}[1]%
  {\small\slshape%
  \savebox{\auteurbm}{\upshape\sffamily#1}%
  \begin{flushright}}
  {\\[4pt]\usebox{\auteurbm}
  \end{flushright}\normalsize\upshape}
\begin{Bonmot}{William D. Phillips}
There is no two-level atom and rubidium is not one of them.
\end{Bonmot} 

\noindent As Bill Phillips reminds us, the level structure of the rubidium $D$-lines is everything but simple and it seems somewhat oversimplified to model it with a two-level system. However, even-though the two-level atom model remains simplistic compared to the real$\;$complexity of alkali atomic structures, I still want to introduce it in this section for mainly two reasons. Firstly, this model can effectively describe the rubidium vapor under certain conditions (I will come back to this point in detail in the next section) and, secondly, it enables to present one by one, in a simplified context, all the theoretical tools we will need to cover more complicated situations. I will in particular introduce the dielectric susceptibility $\chi$ of the medium and study how it depends on the laser intensity, which is one of the key feature for photon fluid experiments. 

\subsection{Is the two-level model relevant ?}

\noindent Before getting into the details of the calculations, let's try to assess the relevance of$\;$the two-level model in describing rubidium atoms. In order to do so, we have first$\;$and$\;$foremost to understand which atomic levels among those forming the fine and hyperfine structures of the $D$-lines are involved in this two-level description. 

\subsubsection{Hyperfine splitting}

As you may have noticed$\;$looking at figure~\ref{fig:RbLevelSchematic}, there is more than $15$ nm difference$\;$between the wavelengths of the $D_{1}$ and $D_{2}$ transition lines. Therefore, when we optically address one of the $D$-line, the contribution of the other one on the optical response of the rubidium vapor is completely negligible. On figure~\ref{fig:RbLevelSchematic}, you can also see that the hyperfine ground state splitting $\delta_{0}$ is almost 10 times larger than the characteristic excited state hyperfine splitting $\delta_{\mathrm{HF}}$. Let's focus for example on the $D_{1}\;$line of $^{_{85}}$Rb for which $\frac{\delta_{0}}{2\pi} \simeq 3.036 \; \mathrm{GHz}$, $\frac{\delta_{\mathrm{HF}}}{2\pi} \simeq 213 \; \mathrm{MHz}$ and $\delta_{0}/\delta_{\mathrm{HF}} \simeq 14.3 \gg 1$. If the laser frequency $\omega$ is red-detuned with respect to the $F_{g} = 3 \rightarrow F_{e}$ transition frequency $\omega_{D_{1}}$, the laser detuning $\Delta$, defined by:
\begin{equation}
    \Delta = \omega - \omega_{D_{1}} ,
\end{equation}

\noindent is negative. In addition, if $\delta_{0} > |\Delta| \gg \delta_{\mathrm{HF}}$, the laser couples much more efficiently $F_{g} = 3$ than $F_{g} = 2$ to the hyperfine excited states $F_{e} = 2$ and $F_{e} = 3$ of the $D_{1}$ line. Moreover, as$\;|\Delta| \gg \delta_{\mathrm{HF}}$, the laser field drives the $D_{1}$ line off-resonantly: light/matter interaction processes are dominated by Rayleigh scattering events involving the fine state $5^{2}P_{1/2}$ as a whole, since the laser detuning is too large to probe its hyperfine level structure in detail.

\newpage

\noindent As things stand at the moment, if the conditions $\Delta < 0$ and $\delta_{0} > |\Delta| \gg \delta_{\mathrm{HF}}$ are fulfilled, it$\;$seems actually reasonable to model the $D_{1}$ line of $^{_{85}}$Rb by a two-level atomic system where the ground state $\ket{g}$ is the hyperfine level $F_{g} = 3$ of $5^{2}S_{1/2}$ and the excited state$\;\ket{e}$ the fine level $5^{2}P_{1/2}$. In practice, the previous conditions on the$\;$laser$\;$detuning$\;$are$\;$satisfied.

\begin{itemize}
    \item [$\bullet$] In fluid of light experiments, the nonlinear interaction between photons has to be repulsive, which imposes the laser frequency $\omega$ to be red-detuned with respect$\,$to$\,$the resonance frequency (\textit{ie} $\Delta$ to be negative). As of now, you must take my word for$\;$it. I will come back to this point in paragraph 1.2.3 i. 
    \item [$\bullet$] If the laser comes closer to resonance, absorption increases and transmission through the$\,$vapor$\,$highly$\,$decreases,$\,$according$\,$to$\,$the$\,$Beer-Lambert$\,$law.$\;$This$\,$effect$\,$is$\,$enhanced if Doppler broadening is taken into account, as we will see in the$\;$next$\;$paragraph. In experiments, we always keep a transmission above $70\%$ (see paragraph 1.2.3 iii). 
\end{itemize}

\subsubsection{Doppler broadening}

\noindent In hot vapors, atoms are constantly moving and do not contribute equally to the medium optical response. If thermal motion causes an atom to move toward the incoming$\;$photons, or in other words, if the scalar product $\mathbf{k} \cdot \mathbf{v}$ is negative ($\mathbf{k}$ and $\mathbf{v}$ being respectively the laser wave-vector and the atom velocity), the laser frequency $\omega_{\mathrm{D}}$ in the moving frame of this atom will be blue-shifted with respect to $\omega$ because of Doppler effect:
\begin{equation}
    \omega_{D} = \omega - \mathbf{k} \cdot \mathbf{v} = \omega - k_{z} \, v_{z}. 
    \label{DopplerEffect}
\end{equation}
\noindent The z-axis defines here the laser optical axis and $v_{z}$ stands for the component of $\mathbf{v}$ along$\;z$. Different atom velocities result therefore in different Doppler shifts, the cumulative effect of which is the line broadening. This resulting line profile is known as a Doppler$\;$profile. The susceptibility $\chi$, that characterizes the vapor optical response, must then be averaged over the atomic velocity distribution along $z$, namely, the 1D Maxwell distribution: 
\begin{equation}
    \mathcal{P}_{_{\!\mathrm{1D}}}(v) = \sqrt{ \frac{m}{2 \pi k_{B} T}} \, \exp \left( - \frac{m \, v_{z}^{2}}{2 \, k_{B} \, T} \right),
    \label{1DMaxBolt}
\end{equation}
\noindent where $m$ is the rubidium mass, $k_{B}$ the Boltzmann constant and $T$ the vapor temperature. Using the relation~\eqref{DopplerEffect}, $\mathcal{P}$ can be expressed a function of $\omega$ and $\omega_{D}$ as follows:
\begin{equation}
    \mathcal{P}_{_{\!\mathrm{1D}}}(\omega_{D}) = \sqrt{ \frac{m}{2 \pi k_{B} T}} \, \exp \left[ -\left(\frac{\omega-\omega_{D}}{\Delta \omega_{D}}\right)^{2} \right].
\end{equation}
\noindent The width $\Delta \omega_{D}$ of this Gaussian distribution is equal to $k \,\sqrt{2 k_{B} T / m}$. When $T = 400 \; \mathrm{K}$, the Doppler linewidth $\Gamma_{D} = \Delta \omega_{D} / 2 \pi$ is about $350$ MHz for the $D_{1}$ line of $^{_{85}}$Rb. This$\;$is much larger than the natural linewidth $\Gamma \simeq 2\pi\!\times\!5.7$ MHz of the $D_{1}$ line. Even if the laser is not at resonance with zero velocity atoms, fast moving ones can thus still absorb light, because of Doppler effect. In order to minimize absorption, the laser detuning should be large compared to $\Delta \omega_{D}$. Moreover, you may have noticed that $\Gamma_{D}$ is of the same order of magnitude as $\delta_{\mathrm{HF}}$. Doppler broadening will thus smooth the hyperfine structure of the excited state in the absorption profile and, more generally, in the vapor$\,$dielectric$\,$response. The two-level description of the rubidium $D$-lines seems then to be even more appropriate to model the optical response of the atomic vapor.

\newpage

\subsubsection{Conclusion}

\noindent Let's summarize. As long as the laser detuning is negative and satisfied $\delta_{0} > |\Delta| \gg \delta_{\mathrm{HF}}$ (which also implies that $|\Delta| \gg \Delta \omega_{D}$ since $\delta_{\mathrm{HF}}$ and $\Delta \omega_{D}$ are of the$\,$same$\,$order$\,$of$\,$magnitude at $T = 400$ K by the way) the two-level model describes$\,$each$\,$of$\,$the$\,D$-lines$\,$quite$\,$accurately. However, it is worth mentioning that effect such as ground state optical pumping$\,$are$\,$not taken into account in this simple description of the rubidium level structure. If it is true that the laser couples more efficiently one ground state than the other to the excited$\;$level, the latter can still decay with equal probability toward both ground states,$\,$as$\,$the$\,$transition is not cycling. This will lead to a$\,$population$\,$transfer$\,$between$\,$ground$\,$states$\,$that$\,$can$\,$greatly impact the internal dynamics of rubidium atoms (see section 1.3). But for$\,$now,$\,$let's$\,$further investigate the features of the two-level model by deriving the optical Bloch equations.    

\subsection{Maxwell-Bloch equation for a closed two-level atom}
\label{subsec:BlochEquation2Level}

Let us consider the interaction of a monochromatic electric field $\boldsymbol{E}(t)$ with a system of $N$ two-level atoms per unit volume. In what follows, $\ket{g}$ and $\ket{e}$ stand respectively for the ground and the excited state in this two-level description, while $\omega_{eg}$ denotes the resonant transition frequency. We assume that the upper level $\ket{e}$ decays because of spontaneous emission toward the ground state $\ket{g}$ at a rate $\Gamma$, and therefore that the lifetime of the$\;\ket{e}$ is given by $\tau = 1/\Gamma$. This system is called closed since any population that leaves the upper level necessarily enters the lower one.
\vspace{6pt}
\newline
\noindent The interaction between the laser field and the two-level atom is described by a Lindblad master equation, commonly known as the \textbf{optical Bloch equation}~\cite{2-5Arecchi,2-6Brasil}: 
\begin{equation}
    \frac{\mathrm{d} \hat{\rho}}{\mathrm{d}t} = -\frac{i}{\hbar} \left[ \hat{H}, \hat{\rho} \right] + \sum_{\nu \ne 0} \left( L_{\nu} \hat{\rho} L_{\nu}^{\dagger} - \frac{1}{2} \{ L_{\nu} L_{\nu}^{\dagger}, \hat{\rho} \} \right), 
    \label{LindbladEq1}
\end{equation}
\noindent where $\hat{\rho}$ and $\hat{H}$ are respectively the density matrix operator and the system Hamiltonian. Both are Hermitian. The Lindblad operators $L_{\nu}$ are quantum jump operators,$\,$describing$\,$a random evolution of the system which suddenly changes under the environment influence (at the time scale of the evolution). 
\vspace{6pt}
\newline
\noindent The Hamiltonian $\hat{H}$ splits into a non-perturbative diagonal part $\hat{H}_{0}$ and a perturbative$\;$one $\hat{W}$ that describes the light/matter interaction. The unperturbed Hamiltonian is given by: 
\begin{equation}
    \hat{H}_{0} = \sum_{i} E_{i} \ket{i}\bra{i} = E_{g}\ket{g}\bra{g} + E_{e}\ket{e}\bra{e},    
\end{equation}
\noindent where $E_{i}$ is the energy associated to the $i$-th level. As energy is defined up to a constant, we choose to set $E_{g}$ to zero. Hence:
\begin{equation}
    \hat{H}_{0} = 0 \!\times\! \ket{g} \bra{g} + \hbar \, \omega_{eg} \ket{e} \bra{e} =  -\hbar \, \Delta \ket{e} \bra{e} + \Tilde{H}_{0}, 
    \label{AtomicHamiltonian}
\end{equation}
\noindent where the laser detuning is still defined by $\Delta = \omega - \omega_{eg}$ and $\Tilde{H}_{0} = 0 \!\times\! \ket{g} \bra{g} + \hbar \, \omega \ket{e} \bra{e}$. The role plays by $\Tilde{H}_{0}$ is explained in the next paragraph. For now, just remember how $\hat{H}_{0}$ has been rewritten in order to make explicitly appear $\Delta$ in the left hand side of~\eqref{AtomicHamiltonian}.  
Let's now focus on the off-diagonal part $\hat{W}$ of the Hamiltonian $\hat{H}$.

\newpage

\subsubsection{Dipole and rotating wave approximation}

\noindent Under the so-called \textbf{dipole approximation}, the spatial variation of the electric field at the atomic scale can be neglected. We can thus think of the atom as an electric dipole$\;$that interacts with the laser electric field through $\hat{W} = -\boldsymbol{\hat{d}}\cdot \boldsymbol{E}(t)$. This interaction Hamiltonian describes how the quantum dipole operator $\boldsymbol{\hat{d}}$ tends to align on the classical field $\boldsymbol{E}(t)$ to minimize the atom plus field overall energy. The dipole operator reads as follows:
\begin{equation}
    \boldsymbol{\hat{d}} = \mathbf{d}^{*} \ket{e} \bra{g} + \mathbf{d} \ket{g} \bra{e}.  
\end{equation}
\noindent Therefore, as $\boldsymbol{E}(t) = \mathcal{E}_{0} \, \cos(\omega t) \, \boldsymbol{\xi}$ ($\boldsymbol{\xi}$ being the laser field polarization unit vector):
\begin{align}
    \nonumber
    \hat{W}(t) =& \; -\frac{1}{2} \left[\left(\mathbf{d}^{*} \ket{e} \bra{g} + \mathbf{d} \ket{g} \bra{e} \vphantom{\mathcal{E}_{0} \left(e^{i \omega t}+e^{-i \omega t} \right)} \right)\right] \cdot \left[\mathcal{E}_{0} \left(e^{i \omega t}+e^{-i \omega t} \right) \boldsymbol{\xi} \right] \\
    =& \; -\frac{\hbar}{2} \left(\Omega^{*} \ket{e} \bra{g} + \Omega \ket{g} \bra{e} \vphantom{e^{i \omega t}+e^{-i \omega t}} \right) \times \left(e^{i \omega t}+e^{-i \omega t} \right). 
    \label{InteractionH}
\end{align}
\noindent The Rabi frequency $\Omega = \mathcal{E}_{0} \, \mu_{ge}/\hbar $ measures the strength of the light-matter interaction. For linearly polarized light, the dipole moment $\mu_{ge} = \bra{g} \boldsymbol{\hat{d}} \cdot \boldsymbol{\xi} \ket{e}$ is given by: $\mu_{ge} = \sum_{F_{e}} f_{_{F_{g}}}^{_{F_{e}}}$ ($F_{e}$ runs over the angular momentum of all the hyperfine levels composing the state $\ket{e}$). The coefficients $f_{_{F_{g}}}^{_{F_{e}}}$ are defined in~\eqref{Stength}. Embarrassingly, the interaction Hamiltonian in equation~\eqref{InteractionH} contains a explicit time dependence. The usual way for getting rid of it is to place ourselves in the interaction representation with respect to $\Tilde{H}_{0}$, defined in~\eqref{AtomicHamiltonian}. Let $\Tilde{U}$ stand for the interaction picture unitary evolution operator: 
\begin{equation}
    \Tilde{U}(t) = \mathrm{exp} \left(-i \Tilde{H}_{0}/\hbar\right) =  \ket{g} \bra{g} + \ket{e} \bra{e} e^{-i\omega t}.    
\end{equation}
\noindent The new Hamiltonian is: $\hat{H} = \Tilde{U}^{\dagger} \, \hat{H} \, \Tilde{U} = \hat{H}_0 + \Tilde{U}^{\dagger} \, \hat{W}(t) \, \Tilde{U}$ (because $\Tilde{U}$ commutes with$\;\hat{H}_{0}$). Applying this unitary transformation is like switching from a static frame to a rotating$\;$one at $\omega$ in the complex plane (in which $e^{-i \omega t}$ "rotates" around the origin). The rotating frame "unwinds" part of the evolution of the quantum state, which has at the end of the day a simpler time dependence. The coupling operator in the interaction picture reads: 
\begin{align}
    \hat{W}_{I}(t) = \Tilde{U}^{\dagger} \, \hat{W}(t) \, \Tilde{U} =& \; -\frac{\hbar}{2} \left(\Omega^{*} \, e^{i\omega t} \ket{e} \bra{g} + \Omega \, e^{-i\omega t} \ket{g} \bra{e}\right) \times \left(e^{i \omega t}+e^{-i \omega t} \right) \\
    \simeq & \; -\frac{\hbar}{2} \left(\Omega^{*} \ket{e} \bra{g} + \Omega \ket{g} \bra{e} \vphantom{e^{i\omega t}} \right).
    \label{RotatingWave}
\end{align}
\noindent In equation \eqref{RotatingWave}, we get rid of the fast oscillating terms at $\pm 2 \omega$, that intuitively$\;$average to zero in the Bloch equation \eqref{LindbladEq}. This approximation, the so-called$\;$\textbf{Rotating$\;$Wave Approximation} (RWA), is used in a wide variety of contexts, from quantum optics~\cite{2-7Mandel} to atomic physics~\cite{2-8Agarwal}. In the interaction picture and using the RWA, the time dependence of $\hat{W}$ cancels. By introducing $\hat{\rho}_{I} = \Tilde{U}^{\dagger} \, \hat{\rho} \, \Tilde{U}$, equation~\eqref{LindbladEq1} becomes: 
\begin{equation}
    \frac{\mathrm{d} \hat{\rho}_{I}}{\mathrm{d}t} = -\frac{i}{\hbar} \left[ \hat{H}_{I}, \hat{\rho}_{I} \right] + \sum_{\nu \ne 0} \left( \Tilde{L}_{\nu} \, \hat{\rho}_{I} \, \Tilde{L}_{\nu}^{\dagger} - \frac{1}{2} \{ \Tilde{L}_{\nu} \, \Tilde{L}_{\nu}^{\dagger}, \hat{\rho}_{I} \} \right), 
    \label{LindbladEq}
\end{equation}
\noindent with $\hat{H}_{I} =  0 \times \ket{g} \bra{g} - \hbar \, \Delta \ket{e} \bra{e} + \hat{W}_{I}$. A straightforwards calculation leads finally to the following set of Bloch equations for the slowly varying density matrix elements$ \, \rho_{ij}(t)$ (I will drop out the $I$ index in the following):


\newpage

\begin{align}
    \frac{\mathrm{d}\rho_{gg}}{\mathrm{d} t} =& \; \Gamma \, \rho_{ee} + \frac{i}{2} \left( \Omega^{*} \, \rho_{eg} - \Omega \, \rho_{ge} \right) \nonumber \\
    \frac{\mathrm{d}\rho_{ee}}{\mathrm{d} t} =& \; - \Gamma \, \rho_{ee} - \frac{i}{2} \left( \Omega^{*} \, \rho_{eg} - \Omega \, \rho_{ge} \right) \label{OBE2Level} \\
    \frac{\mathrm{d}\rho_{eg}}{\mathrm{d} t} =& \; -\Tilde{\gamma} \, \rho_{eg} + \frac{i \Omega}{2} \left( \rho_{gg} - \rho_{ee} \right) \nonumber
\end{align}
\noindent where $\Tilde{\gamma} = \gamma-i \Delta$. The real part of $\Tilde{\gamma}$ is the dephasing rate of the atomic dipole$\;$moment: the coherence $\rho_{eg}$ is naturally damped over time, at a rate $\gamma$. This is usually caused$\;$by fluctuations or inhomogeneities in the energy splitting between the two levels $\ket{g}$ and$\;\ket{e}$, due to random fluctuations of the external electromagnetic fields or to Rb-Rb collisions. Generally, one only takes into account collision-induced decoherence by adding a collision dephasing$\,$decay$\,$rate$\,\gamma_{\mathrm{col}}\,$to$\,\gamma\,$so$\,$that$\,\gamma = \Gamma/2 + \gamma_{\mathrm{col}}$.$\;$In equation~\eqref{OBE2Level},$\,\rho_{gg}\,$and$\,\rho_{ee}\,$represent the population of atoms in the ground and in$\;$the excited state respectively. The$\;$total atomic population is conserved and thus $\rho_{gg} + \rho_{ee} = 1$.$\,$One$\,$equation$\,$among$\,$\eqref{OBE2Level} is not required to solve the system, which can be rewritten as follows:
\begin{align}
    \Dot{\rho}_{ee}-\Dot{\rho}_{gg} =& \; -\Gamma \left(\rho_{ee} - \rho_{gg} + 1 \right) - i\left( \Omega^{*} \, \rho_{eg} - \Omega \, \rho_{ge} \right) 
    \label{OBE2LevelPopulation1} \\
    \Dot{\rho}_{eg} =& \; -\Tilde{\gamma} \, \rho_{eg} + \frac{i \Omega}{2} \left( \rho_{gg} - \rho_{ee} \right) 
    \label{OBE2LevelCoherence1}
\end{align}

\subsubsection{Steady-state solution}

\noindent The steady-state solution of the optical Bloch equations is obtained by setting the time derivatives to zero in~\eqref{OBE2LevelPopulation1} and~\eqref{OBE2LevelCoherence1}. One can then derive two time-independent coupled equations which$\;$are solved algebraically as follows:  
\begin{align}
    \rho_{ee}-\rho_{gg} =& \; -\frac{\gamma^{2}+\Delta^{2}}{\Delta^{2} + \gamma^{2} + \gamma \, \Omega^{2}/\Gamma} \\
    \rho_{eg} =& \; - \frac{\Omega}{2} \frac{\Delta - i\gamma}{\Delta^{2} + \gamma^{2} + \gamma \, \Omega^{2}/\Gamma}
    \label{OBE2LevelCoherence}
\end{align}

\subsection{Atomic polarization and susceptibility}

\begin{figure}[h]
\center
\includegraphics[width=\columnwidth]{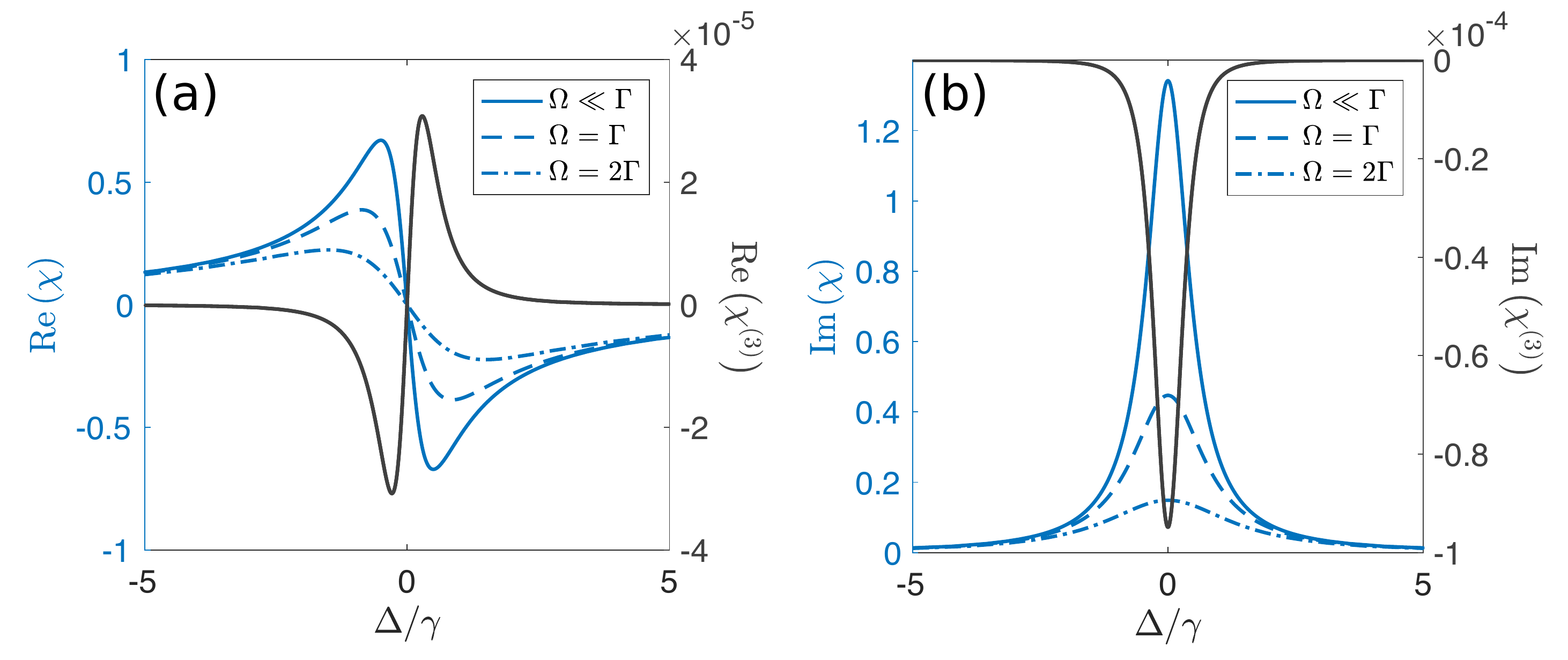} 
\caption{Real (a) and imaginary parts (b) of the total (solid, dashed and dotted blue) and third-order (solid black) dielectric susceptibilities. The total susceptibility has been plotted for different Rabi frequencies $\Omega$. Effects of power broadening and saturation$\;$on$\;$the lineshapes are clearly visible in (a) and (b). The signs of $\chi$ and $\chi^{_{(3)}}$ are opposite since$\;$the latter represents a saturation of the optical$\;$response. Plots obtained for the $D_{1}$ line of rubidium 87. Parameters: $T = 400$ K and $N = 2.5 \, 10^{13}$ atoms/cm$^{3}$.}
\label{fig:2LevelChi}
\end{figure}

\noindent The dielectric response of the atomic ensemble to the laser field excitation is described$\;$by the atomic polarization $\mathbf{P}(t)$ (\textit{ie} the dipole moment per unit volume). The applied$\;$electric field intuitively shifts electron cloud and atomic nucleus in opposite directions and thus polarizes the atom along the laser field polarisation vector $\boldsymbol{\xi}$. The atomic$\;$polarization$\;\mathbf{P}(t)$ is related to the excitation field $\mathbf{E}(t)$ through the following formula:
\begin{equation}
    \mathbf{P}(t) = \epsilon_{0} \chi\left[ \mathcal{E}_{0} \right] \mathbf{E}(t),
    \label{Polarization}
\end{equation}
\noindent where $\chi\left[ \mathcal{E}_{0} \right]$ defines the atomic susceptibility (or the atomic polarizability), which$\;$basically measures the ability of an atom to be polarized under the excitation field $\mathbf{E}(t)$. 
\vspace{6pt}
\newline
\noindent The atomic polarization can also be derived as function of the off-diagonal element $\rho_{eg}$ of the density matrix $\hat{\rho}$ (\textit{ie}, the coherence between ground and excited states)$\,$as$\,$follows~\cite{2-9Siddons}: $P(t) = 2 N \mu_{ge} \, \rho_{eg}$, where $N$ is the atomic density. By identifying the foregoing expression with equation~\eqref{Polarization}, one finally finds that: 

\newpage

\begin{equation}
    \chi = \frac{2 N}{\epsilon_{0} \mathcal{E}_{0}} \, \mu_{ge} \, \rho_{eg} = \frac{\alpha_{0}(0)}{\omega_{eg}/c} \, \frac{i-\Delta/\gamma}{1\!+\!\left( \frac{\Delta}{\gamma} \right)^{2} \!+\! \left( \frac{\mathcal{E}_{0}}{\mathcal{E}_{s}} \right)^{2}}.
    \label{DielectricSusc2Level}
\end{equation}
\noindent The linear line-center absorption coefficient: 
\begin{equation}
    \alpha_{0}(0) = \frac{\omega_{eg}}{c} \, \frac{N}{\epsilon_{0} \hbar} \, \frac{|\mu_{ge}|^{2}}{\gamma_{eg}},
    \label{AbsCoef}
\end{equation}
defines the absorption coefficient experienced by a weak enough (non-saturating) laser$\;$field propagating at resonance inside the atomic vapor. The line-center saturation field strength $\mathcal{E}_{s} = \hbar \sqrt{\gamma \, \Gamma}/\mu_{eg}$ is the value at which an on-resonance laser field makes the on-resonance absorption$\;$coefficient $\alpha_{0}(0)$ drop to half of its weak-field value~\eqref{AbsCoef}. 
\vspace{6pt}
\newline
\noindent The real and imaginary parts of the atomic polarizability $\chi$ show a standard dispersive$\;$and Lorentzian lineshape, which is not surprising as they respectively give information about the medium refractive index and absorption coefficient. They have been plotted in blue for different values of $\Omega = \mathcal{E}_{0} \, \mu_{ge}/\hbar $ in figures~\ref{fig:2LevelChi}(a) and (b). As you may have seen, both lines get broader when the Rabi frequency $\Omega$ rises up. This effect is known$\;$as$\;$power$\;$broadening. We can also notice, from~\eqref{DielectricSusc2Level} and from figures~\ref{fig:2LevelChi}(b), that the line center value of $\mathrm{Im} \left[ \chi \right]$ $-\,$and consequently, of the absorption coefficient $\alpha = \frac{\omega_{eg}}{c} \mathrm{Im} \left[ \chi \right]$ $-$ decreases with respect to its weak-field value as soon as $\Omega$ increases. This tendency of absorption to decrease when intense optical fields are applied is known as saturation.

\newpage

\subsubsection{First- and third-order susceptibilities}

\noindent As long as $\left| \mathcal{E}_{0}/\mathcal{E}_{s} \right|^{2} \ll 1\!+\! (\Delta / \gamma )^{2}$, a power series expansion in $\mathcal{E}_{0}/\mathcal{E}_{s}$ of equation~\eqref{DielectricSusc2Level} can be performed. By retaining only the zeroth and second order terms:
\begin{equation}
\chi \simeq \frac{\alpha_{0}(0)}{\omega_{eg}/c} \, \frac{i-\Delta/\gamma}{1\!+\!\left( \Delta / \gamma \right)^{2}} \left(1- \left| \frac{\mathcal{E}_{0}}{\mathcal{E}_{s}} \right|^{2} \frac{1}{1\!+\!\left( \Delta / \gamma \right)^{2}}\right).
    \label{ApproxDielectricSusc3Level}    
\end{equation}
\noindent We finally equate the foregoing expression with the standard power series expansion: $\chi = \chi^{(1)} \!+\! \frac{3}{4} \chi^{(3)} |\mathcal{E}_{0}|^2$ in order to obtain analytical expressions for the first- and third-order dielectric susceptibilities as follow:
\begin{align}
    \chi^{(1)} =& \; \frac{\alpha_{0}(0)}{\omega_{eg}/c} \, \frac{i-\Delta/\gamma}{1\!+\!\left( \Delta / \gamma \right)^{2}} 
    \label{Chi1Exp2Level} \\ 
    \chi^{(3)} =& \; -\frac{4}{3} \, \frac{1}{|\mathcal{E}_{s}|^{2}} \, \frac{\alpha_{0}(0)}{\, \omega_{eg}/c} \, \frac{i-\Delta/\gamma} {\left[ 1\!+\!\left( \Delta / \gamma \right)^{2} \right]^{2}}
    \label{Chi3Exp2Level}
\end{align}
\noindent The frequency dependence of the real and imaginary parts of $\chi^{_{(3)}}$ is illustrated in figure~\ref{fig:2LevelChi} (black lines). The signs of $\chi\;$and $\chi^{_{(3)}}$ are$\;$opposite since $\chi^{_{(3)}}$ represents a saturation$\;$of$\;$the vapor optical response. The real part of the third-order susceptibility plays a crucial role in photon fluid experiments because it controls the strength of the nonlinear interaction between photons, as we sill see later on, in the third chapter. Moreover, the sign of $\mathrm{Re}\left[\chi^{_{(3)}}\right]$ defines if this interaction is attractive or repulsive. As we can see in~\ref{fig:2LevelChi}(a),$\;$it$\;$changes$\;$sign with the laser detuning, going from negative ($\Delta < 0$) to positive values ($\Delta > 0$).$\;$By$\;$tuning the laser frequency, we are thus able to control the nature and the strength of the nonlinear photon-photon interaction easily. In practice, we always red-detuned the laser frequency form resonance however, in order to generate repulsive interaction between photons.  

\subsubsection{Intensity dependent refractive index: Kerr effect}

The dielectric susceptibility $\chi$ is related to the dielectric permitivity $\epsilon_{r}$ by:
\begin{equation}
    \epsilon_{r} = \sqrt{1 + \chi} \simeq  \sqrt{1 + \mathrm{Re}\left(\chi\right)} +\frac{i}{2} \frac{\mathrm{Im}\left(\chi\right)}{\sqrt{1+\mathrm{Re}\left(\chi\right)}}. 
\end{equation}
\noindent The expansion on the right-hand side is valid as long as the condition $| \chi | \ll 1$ is fulfilled. The real part of $\epsilon_{r}$ defines the medium refractive index: $n = \sqrt{1+\mathrm{Re}\left(\chi\right)}$. The absorption coefficient is obtained by multiplying the imaginary part of $\epsilon_{r}$ by the laser wave-vector $k_{0}$ in vacuum such that: $\alpha = k \, \mathrm{Im}\left(\chi\right)$ ($k = k_{0}/n$ is the laser wave-vector inside the medium). In nonlinear optics, the total refractive index is commonly expressed as function of the linear and nonlinear refractive indices $n_{0}$ and $n_{2}$ (using the fact that $\mathrm{Re}\left[\chi^{_{(3)}}\right] \ll \mathrm{Re}\left[\chi^{_{(1)}}\right]$):
\begin{equation}
    n = \underbrace{\vphantom{\frac{3}{8}}\sqrt{1+\mathrm{Re}\left(\chi^{(1)}\right)}}_{n_{0}}  + \underbrace{\frac{3}{8} \mathrm{Re}\left(\chi^{(3)}\right)|\mathcal{E}_{0}|^{2}}_{n_{2} \,I } = n_{0}+n_{2} \, \mathcal{I}_{0}, 
\label{RefractiveIndex}
\end{equation}
\noindent where $\mathcal{I}_{0} = \frac{1}{2} \epsilon_{0} n_{0} c |\mathcal{E}_{0}|^{2}$ and consequently $n_{2} = \frac{3}{4} \frac{\mathrm{Re}\left(\chi^{_{(3)}}\right)}{\epsilon_{0} n_{0} c}$.

\newpage

\noindent The Rubidium vapor behaves as a \textbf{third-order Kerr medium} under near-resonance$\;$laser excitation since the medium refractive index is intensity dependent. When an intense$\;$laser beam propagates in such a medium, it induces a refractive index variation which is larger at its center than at its periphery and accumulates$\,$therefore$\,$along$\,$its$\,$propagation$\,$a$\,$radially dependent nonlinear phase. This self-induced phase modulation acts on the beam as a lens would do, either by focusing (attractive interactions) or spreading (repulsive$\;$interactions) the$\;$light within the material (depending on the sign of $n_{2}$). By measuring this nonlinear phase shift as function of intensity, we can thus access the nonlinear refractive index $n_{2}$, or equivalently, the third-order susceptibility $\chi^{_{(3)}}$ (see section 2.3).

\subsubsection{Absorption versus non-linearity: how to optimize ?}

\noindent Depending on what is the intended purpose of his experiment, one should optimize either the nonlinear refractive index $n_{2}$ or the nonlinear change of refractive index $\Delta n = n_{2} \, \mathcal{I}$. At the single photon level, observing nonlinear effects requires for instance to optimize the value of $n_{2}$ in the limit $\mathcal{I} \ll \mathcal{I}_{\mathrm{sat}}$ ($\mathcal{I}_{\mathrm{sat}}$ is the saturation intensity). Reversely,$\;$in$\;$fluid$\;$of$\;$light experiments, a macroscopic number of photons is involved. In that case, we can either act on the strength of the nonlinear interaction (namely, on $n_{2}$) or on$\;$the$\;$laser$\;$intensity$\;\mathcal{I}\;$to scale up nonlinear effects.$\;$We$\,$also$\,$have$\,$to$\,$make$\,$sure$\,$that$\,$the$\,$condition $\mathcal{I}/\mathcal{I}_{\mathrm{sat}} \ll 1 +  (\Delta / \gamma )^{2}$ is fulfilled in order for the power series expansion~\eqref{ApproxDielectricSusc3Level} to be valid. 
\vspace{6pt}
\newline
\noindent Experimentally, one might think that the best configuration is to set the laser frequency close to resonance where the nonlinear refractive index $n_{2}$ varies significantly.$\;$Nevertheless, absorption increases when the laser detuning goes$\,$to$\,$zero,$\,$which$\,$results$\,$in$\,$a$\,$reduction$\,$of$\,$the average field intensity inside the vapor cell, and thus, of the nonlinear change$\,$of$\,$refractive index $\Delta n$. In practice, we allow a minimum transmission of $70 \%$ and search$\,$for$\,$the$\,$largest value of$\;n_{2}$ (in absolute value) at this fixed transmission$\;$threshold.$\;$When$\;\gamma\,$$\ll$$\Delta$,$\,$equations \eqref{Chi1Exp2Level} and~\eqref{Chi3Exp2Level} yield respectively: 
\begin{align}
    \mathrm{Im}\left[\chi^{(1)}\right] =& \; \frac{\alpha_{0}(0)}{\omega_{eg}/c}  \left(\frac{\gamma}{\Delta}\right)^{2} \propto \frac{N(T)}{\Delta^2} \\ 
    \mathrm{Re}\left[\chi^{(3)}\right] =& \; \frac{1}{|\mathcal{E}_{s}|^{2}} \, \frac{4}{3} \, \frac{\alpha_{0}(0)}{\omega_{eg}/c}  \left(\frac{\gamma}{\Delta}\right)^{3} \propto \frac{N(T)}{\Delta^3}
\end{align}
\noindent where $N(T)$ is the atomic density of the rubidium vapor at the temperature$\,T$.$\;$If$\,$we$\,$neglect the$\,$nonlinear$\,$absorption$\,$(scaling$\,$as$\,1/\Delta^{4}$),$\,$the$\,$fixed$\,$transmission$\,$condition$\,$requires$\,\mathrm{Im}[\chi^{_{(1)}}]$ and thus the ratio$\;N(T)/\Delta^{2}$ to be constant whatever$\,$the$\,$detuning;$\;\chi^{_{(3)}}\,$is$\,$then$\,$proportional to $1/\Delta$ times this constant quantity. In$\,$order$\,$to$\,$maximize$\,$the$\,$nonlinear$\,$refractive$\,$index$\,$at$\,$a given transmission, we should therefore$\,$reduce$\,$the$\,$laser$\,$detuning,$\,$which$\,$consequently$\,$leads to lower the vapor temperature (and thus the atomic density $N$) in order$\;$to$\;$keep$\;$the$\;$ratio $N(T)/\Delta^{2}$ constant. This procedure is obviously limited by our initial assumption, as the detuning should remains much larger than both the excited state$\,$hyperfine$\,$splitting $\delta_{\mathrm{HF}}$ and the Doppler broadening $\Delta \omega_{D}$.

\newpage

\section{Open three-level atomic system}

\noindent The two-level model developed in the previous section provides a good description$\;$of$\;$each rubidium $D$-line. In this model, the ground state $\ket{g}$ is defined by the upper$\;$hyperfine$\;$state of$\;$the level $5^{2}S_{1/2}$ while the excited state $\ket{e}$ is defined by the fine level $5^{2}P_{1/2}$ (resp.$\;5^{2}P_{3/2}$) of the $D_{1}$ line (resp. of the $D_{2}$ line). As mentioned previously, the two-level description is valid as long as the laser detuning $\Delta = \omega - \omega_{eg}$ satisfies the following conditions: $\Delta < 0$ and $\delta_{0} > |\Delta| \gg \delta_{\mathrm{HF}}$, where $\delta_{0}$ and $\delta_{\mathrm{HF}}$ are $-$ let us recall it $-$ the hyperfine ground and excited state splitting. Nevertheless, there is nothing preventing the fine states $5^{2}P_{1/2}\;$and $5^{2}P_{3/2}$ from decaying toward the lower hyperfine state of $5^{2}S_{1/2}$. In$\;$reality,$\;$the$\;$excited$\;$state $\ket{e}$ of the two-level description is therefore coupled to both the upper and lower hyperfine states of $5^{2}S_{1/2}$. As the laser detuning needs to be negative, we mostly couple the upper hyperfine ground state to the excited state and thus pump preferentially the atom in the lower hyperfine ground state. This$\;$population$\;$transfer$\;$between$\;$ground$\;$states,$\;$known$\;$as optical pumping, is not taken into account in the two-level description$\;$but$\;$will definitively affect the optical response of the atomic vapor. We have therefore to extend the previous description to a 3-level model in which both hyperfine ground states are involved.

    \subsection{Extension to three levels}
    
\begin{figure}[h]
\center
\includegraphics[width=\columnwidth]{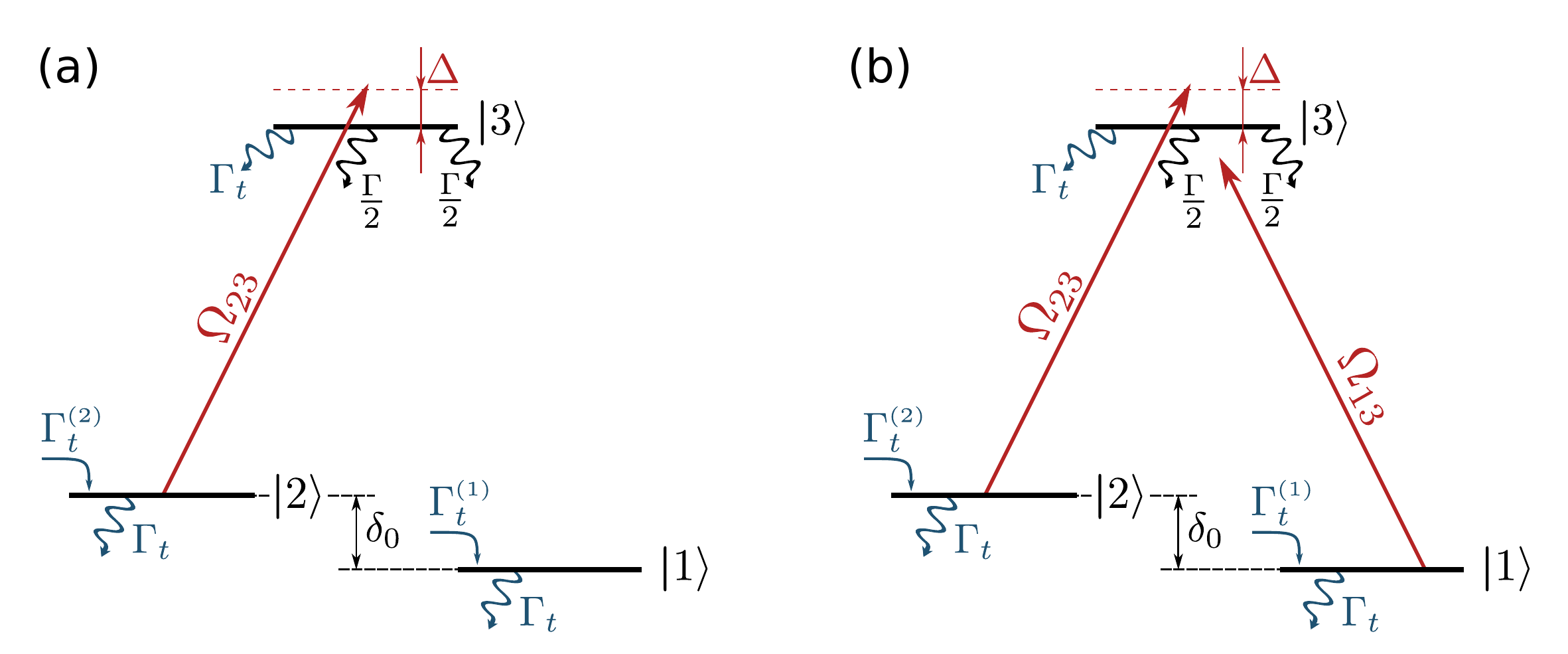} 
\caption{Three-level system with one (a) and two (b) coupling beam(s).}
\label{fig:3Level}
\end{figure}

\noindent The 3-level atomic system$\,$describing$\,$more$\,$precisely$\,$the$\,$rubidium$\,D$-lines$\,$has$\,$been$\,$sketched on figures~\ref{fig:3Level} (a) and (b). The ground states $\ket{1}$ and $\ket{2}$ stand respectively for the lower and upper hyperfine levels of $5^{2}S_{1/2}$. The laser detuning $\Delta$ is defined with respect to$\;$the $\ket{2} \rightarrow \ket{3}$ transition: $\Delta = \omega - \omega_{32}$. If the laser detuning is much larger than the$\;$hyperfine excited states splitting $\delta_{\mathrm{HF}}$ and the Doppler width $\Delta \omega_{D}$, we can safely$\;$forget about$\;$the hyperfine structure of the excited levels $5^{2}P_{1/2}$ and $\,$hyperfine$\,$transitions allowed by the selection rules between $\ket{2}$ and $\ket{3}$ is addressed by the$\,$laser$\,$field$\,$because of the large laser detuning and Doppler broadening. As in the$\,$two$\,$level$\,$description,$\,$the$\,$state$\,\ket{3}\,$in$\,$the$\,$3-level model is thus a meta excited state hiding the hyperfine complexity of  $5^{2}P_{1/2}$ and $5^{2}P_{3/2}$.

\newpage

\noindent In figure~\ref{fig:3Level}, two different situations are depicted:
\begin{itemize}
    \item [(a)] We assume the laser field couples mainly the states $\ket{2}$ and $\ket{3}$. As for the two-level description, this assumption is correct if the laser detuning $\Delta$ is negative and fulfills (as usual now) the condition $\delta_{0} > |\Delta| \gg \delta_{\mathrm{HF}}$. 
    \item [(b)] In that case, the laser field couples at the same time the ground states $\ket{1}$ and $\ket{2}\;$to the excited state $\ket{3}$. This description is more$\,$general$\,$because$\,$it$\,$allows$\,$to$\,$compute$\,$the vapor optical response at every detuning (as long as $|\Delta|$ remains large$\,$compared$\,$to $\delta_{\mathrm{HF}}$ and $\Gamma_{D}$), by$\,$adding$\,$the$\,$susceptibilities$\,$arising from the coherences between$\;$levels $\ket{1}\;$and$\;\ket{3}$ on one hand, levels $\ket{2}$ and $\ket{3}$ on the other. This kind of lambda$\,$system is widely used to model strong resonant-driving effects, such as the$\;$Electromagnetically Induced Transparency (EIT)~\cite{2-10Boller} or the Autler-Townes splitting~\cite{2-11Autler}.  
\end{itemize}
\noindent I first focus on the situation sketched in figure~\ref{fig:3Level}(a), for which the optical Bloch equations in the steady-state are analytically and easily solvable.

    \subsection{Transit and influx rates}

\noindent If spontaneous emission is the only decay process entering the model, it is not difficult to see that after a sufficiently long amount of time, all the atoms interacting with the laser field are pumped into the uncoupled ground state $\ket{1}$. In the steady-state, the coherence between levels $\ket{2}$ and $\ket{3}$ (and thus the dielectric polarizability $\chi$) vanishes in that case. In order to resolve this issue, we must take into account the finite spatial extension of the laser beam as well as the time of flight of an atom across the beam transverse section. 
\vspace{6pt}
\newline
\noindent In hot vapors, atoms are moving and therefore travel across the laser beam during a finite amount of time, which depends on their position $\boldsymbol{r_{0}}$ and velocity $\boldsymbol{v}$ when they get inside it. They are thus constantly entering and leaving the interaction area, defined by the beam cross-section, at the transit rate $\Gamma_{t}$. Everything happens as if the 3-level system sketched in figure~\ref{fig:3Level} was connected to an external atomic reservoir with which it exchanges atoms. This is basically why this model is referred to as \textbf{"open 3-level system"}. 
\vspace{6pt}
\newline
\noindent In any case, whether or not the atomic system is closed, the total population should be conserved \textit{ie} $\sum_{i} \rho_{ii} = 1$. In other words, the rate at witch levels are filled should be equal to the rate at which atoms return to the reservoir. We assume the atoms entering$\;$the$\;$beam are either in state $\ket{1}$ or in state $\ket{2}$. The filling rates $\Gamma_{t}^{_{(1)}}$ and $\Gamma_{t}^{_{(2)}}$ of $\ket{1}$ and $\ket{2}$ are likely to be unbalanced but must, in any event, fulfill$\;$the$\;$condition:$\;\Gamma_{t}^{_{(1)}} + \Gamma_{t}^{_{(2)}} = \Gamma_{t}$.

\begin{itemize}
    \item [$\bullet$] If we consider that atoms enter the beam only once, they must be initially$\,$prepared$\,$in a statistical mixture of the states $\ket{1}$ and $\ket{2}$, described by the$\,$Boltzmann$\,$statistics. At $T = 400$ K, the ground states should be equally populated in average over the atomic ensemble, since the thermal energy (in $\hbar$ units) $k_{B} T/\hbar \simeq 2\pi \times 50$ THz$\;$is$\;$much larger than$\;\delta_{0}$. However, both ground states are Zeeman degenerated as no magnetic field is applied. Let $g_{i} = 2F_{i}+1$ be the degeneracy factor of state $\ket{i}$, $F_{i}$ being the magnitude of the total atomic angular momentum in $\ket{i}$ ($i=1 \; \mathrm{or} \; 2$). As $F_{2}\;$is$\;$always bigger than $F_{1}$ (for example, $F_{1} = 2$ and $F_{2} = 3$ in $^{_{85}}$Rb), atoms are more likely to enter the beam in state $\ket{2}$. The filling rates are therefore$\,$unbalanced$\,$and$\,$read:
\begin{equation}
    \Gamma_{t}^{_{(i)}} = \frac{g_{i}}{g_{1}+g_{2}} \, \Gamma_{t} = G_{i} \, \Gamma_{t}. \label{InfluxRateMixture}
\end{equation}

\newpage

\begin{figure}[h]
\center
\includegraphics[scale=0.35]{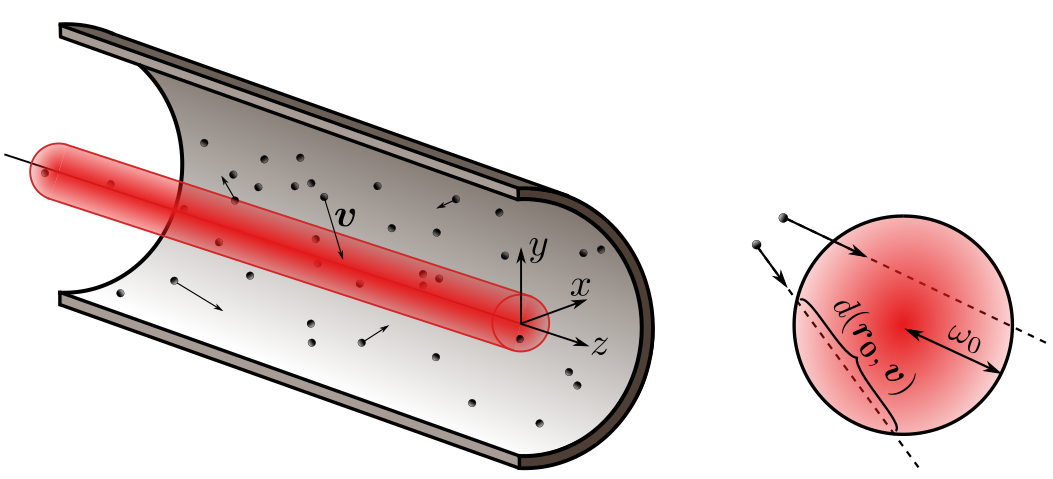} 
\caption{Left: sketch of the beam propagation inside the vapor cell. Right: beam cross$\!\,$-$\!\,$section at a given position $z$ on the optical axis. The time an atom takes to flight across$\;$the laser beam depends on its position $\mathbf{r_{0}}$ and velocity $\mathbf{v}$ when it enters the interaction region.}
\label{fig:TimeofFlight}
\end{figure}

\item [$\bullet$] We can refine this model by considering the fraction $\beta$ of atoms that return inside$\,$the beam before$\,$loosing$\,$the$\,$internal$\,$state$\,$they$\,$have$\,$been$\,$prepared$\,$in$\,$before$\,$\cite{2-12Glorieux, 2-13Gosh}.$\;$In$\,$our collision-free model, atoms$\;$eventually$\;$interact with the cell walls and may return to the interaction area in the same$\;$internal$\;$state. Paraffin-coating of the cell walls may additionally conserved the atomic coherences and polarization, since the atom/wall collisions are not phase-interrupting$\,$anymore$\,$in$\,$that$\,$case~\cite{2-14Klein}.$\;$The$\,$laser$\,$pumps$\,$atoms from state $\ket{2}$ to state $\ket{1}$; the ground state population diffusing outside the beam$\,$is therefore likely to be unbalanced. The influx rates are not constant$\,$anymore$\,$then$\,$but depend dynamically on the populations $\rho_{11}$, $\rho_{22}$ and $\rho_{33}$:
\begin{align}
    \Gamma_{t}^{_{(1)}} =& \; \left[(1-\beta) \, G_{1} + \beta \left(\rho_{11}+\frac{\rho_{33}}{2}\right) \right] \Gamma_{t},
    \label{InfluxRates1} \\
    \Gamma_{t}^{_{(2)}} =& \; \left[(1-\beta) \, G_{2} + \beta \left(\rho_{22} + \frac{\rho_{33}}{2} \right) \right] \Gamma_{t}.
    \label{InfluxRates2}    
\end{align}
\noindent The relation $\Gamma_{t}^{_{(1)}} + \Gamma_{t}^{_{(2)}} = \Gamma_{t}$ ensures that the overall atomic population is conserved, as expected. In the stationary-state, equations~\eqref{InfluxRates1} and~\eqref{InfluxRates2} can be understood as$\;$follow. When $\beta = 0$, all the atoms have lost the memory of the state$\;$they$\;$have been prepared in before returning inside the beam; \eqref{InfluxRates1} and~\eqref{InfluxRates2} reduce to~\eqref{InfluxRateMixture} in that case. If $\beta = 1$ reversely, the fraction of atoms entering in $\ket{1}\;$and$\;\ket{2}$ is given by the population diffusing outside the beam in those states plus half the population diffusing outside it in the excited state (atoms leaving the beam in $\ket{3}$ will eventually decay toward $\ket{1}$ or $\ket{2}$ with equal probability before getting inside the beam again). The weighting factor $\beta$ can be computed in practice by solving the diffusion equation that describes the atomic motion outside the beam, assuming a random distribution of the time spent by the atoms outside the interaction region~\cite{2-15Xiao}.
\end{itemize}

\noindent The influx rates defined in~\eqref{InfluxRates1} and~\eqref{InfluxRates2} will be used when the optical Bloch equations for this open 3-level system are solved numerically. For the sake of simplicity, I will keep considering that $\Gamma_{t}^{_{(1)}}$ and $\Gamma_{t}^{_{(1)}}$ are given by equation~\eqref{InfluxRateMixture} in the calculations below. 

\newpage

\noindent Let's now define a bit more precisely the transit rate $\Gamma_{t}$. An atom moving$\;$toward$\;$the$\;$beam along a certain direction will travel a different distance to get across the interaction area depending on the position $\boldsymbol{r_{0}}$ at which it enters. The mean distance $\bar{d}$ is found by averaging all possible paths through the black circle representing the laser cross section in~\ref{fig:TimeofFlight}~\cite{2-4Sagle}:
\begin{equation}
\bar{d} = \frac{2}{\omega_{0}} \int_{-\frac{\omega_{0}}{2}}^{\frac{\omega_{0}}{2}} \sqrt{\omega_{0}^{2}-x^{2}} \, \mathrm{d}x = \frac{\pi}{4}\, \omega_{0}, 
\end{equation}
\noindent where $w_{0}$ stands for the beam width at $1/e^{2}$. Of course, $\bar{d}$ does not depend on the direction along which the atoms move. The transit rate is finally obtained$\;$by$\;$averaging$\;$the$\;$quantity $v_{\perp}/\bar{d}$ over the 2D Maxwell-Boltzmann velocity distribution: 
\begin{equation}
    \mathcal{P}_{_{\!\mathrm{2D}}}(\boldsymbol{v_{\perp}}) = \frac{1}{\pi u^{2}} \exp \left[- \left(\frac{v_{\perp}}{u}\right)^{2} \right],  
\end{equation}
\noindent ($u = \sqrt{2 k_{B} T/m}$ is the most probable atomic speed in the transverse plane)$\;$which$\;$yields:
\begin{equation}
\Gamma_{t} = \int \frac{v_{\perp}}{\bar{d}} \, \mathcal{P}_{_{\!\mathrm{2D}}}(\boldsymbol{v_{\perp}}) \, \mathrm{d}\boldsymbol{v_{\perp}} = \frac{2\pi}{\bar{d}}\int_{0}^{\infty} v_{\perp}^{2} \, \mathcal{P}_{_{\!\mathrm{2D}}}(v_{\perp}) \, \mathrm{d}v_{\perp} = \frac{2}{\sqrt{\pi}} \frac{u}{\omega_{0}}
\end{equation}

\begin{figure}[h]
\center
\includegraphics[width=\columnwidth]{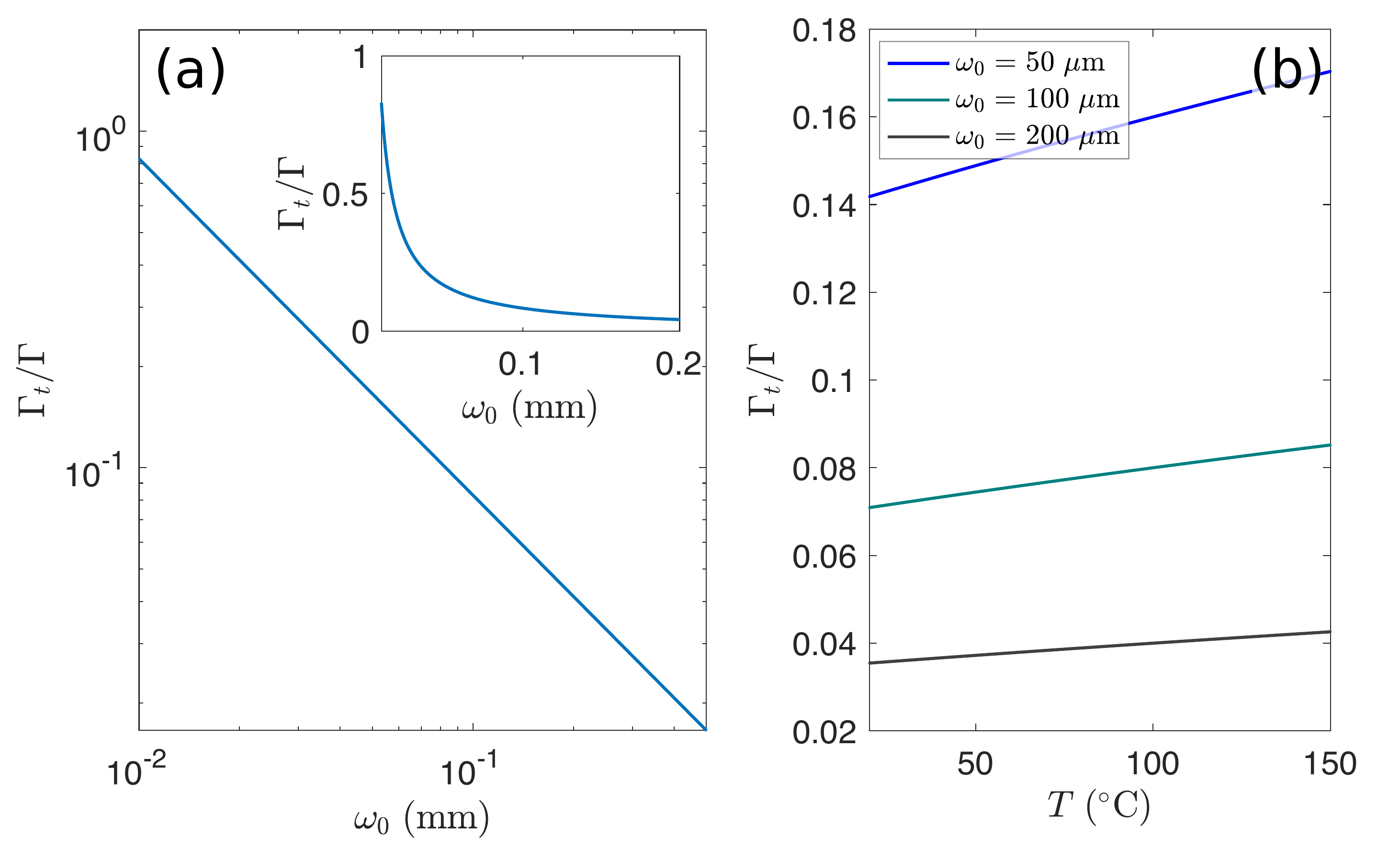} 
\caption{(a) Transit rate $\Gamma_{t}$ as function of the beam width ($T=400$ K). The transit$\;$rate has been normalized by the radiative decay rate $\Gamma$. For small beams ($\omega_{0} < 50$ $\mu$m), they are of the same order of magnitude. (b) Transit rate as function of the vapor temperature for different beam widths. At room temperature and above, $\Gamma_{t}$ only slightly varies with$\;T$. Plots obtained for the $D_{1}$ line of rubidium 87.}
\label{fig:TransitRate}
\end{figure}

\newpage

\noindent The transit rate $\Gamma_{t}$, normalized by the radiative decay rate $\Gamma$, has been plotted as function of the beam width $\omega_{0}$ and as function of the vapor temperature $T$ on figures~\ref{fig:TransitRate}$\;$(a)$\;$and$\;$(b) respectively. As you may have seen, $\Gamma_{t}$ only slightly changes with $T$ at room$\;$temperature and above. However, it strongly varies with the width of the laser beam; for $\omega_{0} = 25$ $\mu$m, $\Gamma_{t}$ is almost a third of $\Gamma$. When $\Gamma_{t}$ and $\Gamma$ are of the same magnitude, the finite$\;$transit$\;$time of the atoms across the beam starts making the spectral lines broader. This effect, which$\,$is known$\;$in the literature as "transit time broadening"~\cite{2-4Sagle, 2-16Bruvelis}, can be understood using the Heisenberg’s uncertainty principle: $\Delta \tau \, \Delta \epsilon \simeq 1$. The uncertainty $\Delta \epsilon$ on the energy$\;$of$\;$an excited state is affected not only by its spontaneous lifetime ($\Delta t \simeq 1/\Gamma$) but also$\;$by$\;$the transit$\;$time$\;$($\Delta t \simeq 1/\Gamma_{t}$).$\;$Thus,$\;$a$\;$decrease$\,$in$\,$the$\,$transit$\,$time$\,$will$\,$lead$\,$to$\,$an$\,$increase$\,$in$\,\Delta \epsilon$, which$\,$will$\,$manifest$\,$in$\,$turn$\,$as$\,$line$\,$broadening.$\;$This$\,$effect$\,$is$\,$known$\,$to$\,$limit$\,$the$\,$resolution$\,$of a variety of spectroscopic applications, such as two-photon~\cite{2-17Thomas}, and$\,$saturated$\,$absorption$\,$\cite{2-18Moon} and photon-echo spectroscopy~\cite{2-19Wang}.

\subsection{Optical Bloch equations in the interaction picture}

\begin{figure}[h]
\center
\includegraphics[width=\columnwidth]{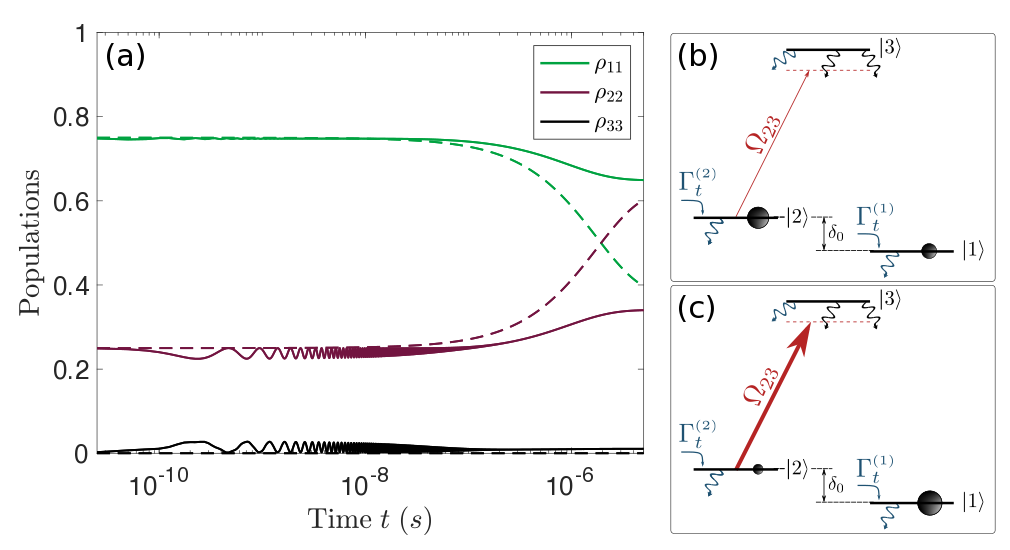} 
\caption{(a) Time evolution of the populations in states $\ket{1}$, $\ket{2}$ and $\ket{3}$ for$\;$two$\;$different laser powers: $\mathcal{P} = 1$ mW (dotted lines) and $\mathcal{P} = 500$ mW (solid$\;$lines). At$\;$low$\;$laser$\;$power, this evolution is mainly driven by the filling and transit rates. As $\Gamma_{t}^{_{(2)}} > \Gamma_{t}^{_{(1)}}$, $\rho_{22}$ increases while $\rho_{11}$ decreases. In the stationary-state, $\rho_{_{11}}^{_{\mathrm{eq}}} = G_{1}$ and $\rho_{_{22}}^{_{\mathrm{eq}}} = G_{2}$. At high$\;$laser$\;$power, optical pumping from state $\ket{2}$ to state $\ket{1}$ overcomes this filling unbalance. Rabi$\;$flopping between states $\ket{2}$ to $\ket{3}$ makes the populations $\rho_{22}$ and $\rho_{33}$ oscillate over time in that$\;$case. Both situations are depicted on figures (b) and (c). Parameters in the text.}
\label{fig:RungeKutta3Level}
\end{figure}

\noindent In order to find an expression for the dielectric susceptibility $\chi$ of the open 3-level system sketched$\,$on$\,$figure~\ref{fig:3Level},$\;$we$\,$should$\,$go$\,$through$\,$the$\,$same$\,$calculation$\,$as$\,$for$\,$the$\,$two-level$\,$system and derive the optical Bloch equations. I will not reproduce the details of this calculation here but just remind the main steps.  
\vspace{4pt}
\newline
\noindent The Hamiltonian $\hat{H}$ of the 3-level system splits into an non-perturbative diagonal part $\hat{H}_{0}$ (atomic$\,$Hamiltonian),$\,$and$\,$a$\,$perturbative$\,$off-diagonal$\,$part$\,\hat{W}\,$(atom/field$\,$coupling$\,$term). 
\vspace{-4pt}
\begin{itemize}
    \item [$\bullet$] The atomic counterpart is defined by: $\hat{H}_{0} = \sum_{i=1}^{3} E_{i} \ket{i}\bra{i}$, where $E_i$ is the energy associated to the $i$-th level. As energy is defined up to a constant, we choose to set the energy of the lower state to zero (\textit{ie} $E_{1} = 0$). Hence:
    \vspace{-9pt}
    \begin{align}
        \nonumber
        \hat{H}_{0} =& \; \hbar \omega_{21} \ket{2} \bra{2} + \hbar \omega_{31} \ket{3} \bra{3} \\
        = & \; \hbar \delta_{0} \ket{2} \bra{2} - \hbar (\Delta-\delta_{0}) \ket{3} \bra{3} + \underbrace{0 \!\times\! \ket{1} \bra{1} + 0 \!\times\! \ket{2} \bra{2} + \hbar \omega \ket{3} \bra{3}}_{\Tilde{H}_{0}}, 
        \label{DiagonalH}
    \end{align}
    \noindent Once$\,$again,$\,$we$\,$make$\,$explicitly$\,$appear$\,$the$\,$laser$\,$detuning$\,\Delta = \omega-\omega_{32}\,$by$\,$adding$\,\Tilde{H}_{0}$. 
    \vspace{-3pt}
    \item [$\bullet$] In the$\,$\textbf{dipole$\,$approximation},$\,$the$\,$atom/field$\,$interaction$\,$Hamiltonian$\,$simply$\,$reads: $\hat{W} = -\boldsymbol{\hat{d}} \cdot \boldsymbol{E}(t)$. Since only the $\ket{2} \rightarrow \ket{3}$ transition is optically addressed in~\ref{fig:3Level}$\;$(a), the dipole operator can$\,$be$\,$expressed$\,$as$\,$follows:$\,\boldsymbol{d} = \mathbf{d}_{23}^{*} \ket{3} \bra{2} + \mathbf{d}_{23} \ket{2} \bra{3}$.$\;$Therefore, as $\mathbf{E}(t) = \mathcal{E}_{0} \cos(\omega t) \boldsymbol{\xi}$ ($\xi$ being the laser polarization vector, as before): 
    \begin{equation}
    \hat{W}_{23}(t) = -\frac{\hbar}{2} \left(\Omega_{23}^{*} \ket{3} \bra{2} + \Omega_{23} \ket{2} \bra{3} \vphantom{e^{i \omega t}} \right) \times \left(e^{i \omega t}+e^{-i \omega t} \right), 
    \label{Interaction3Level}
\end{equation}
\noindent where $\Omega_{23} = \mathcal{E}_{0} \, \mu_{23} /\hbar$ is the Rabi frequency associated to the $\ket{2} \rightarrow \ket{3}$ transition. Let's$\;$remind that the dipole moment $\mu_{23} = \bra{2} \boldsymbol{\hat{d}} \cdot \boldsymbol{\xi} \ket{3}$ is given by: $\mu_{23} = \sum_{F_{3}} f_{_{F_{2}}}^{_{F_{3}}}$. In this formula, $F_{3}$ runs over the magnitude of the total atomic angular momentum in all the hyperfine levels of state $\ket{3}$. The coefficient $f_{_{F_{g}}}^{_{F_{e}}}$ are defined in~\eqref{Stength}.
\end{itemize}
\vspace{-3pt}
\noindent Within the \textbf{interaction representation} and using the \textbf{rotating wave approximation}, the explicit time dependence in~\eqref{Interaction3Level} vanishes. Indeed, if $\Tilde{U}$ stands for the interaction picture unitary evolution operator, then: 
\begin{equation}
    \Tilde{U}(t) = \exp \left( -i \Tilde{H}_{0}/\hbar \right) = \ket{1} \bra{1} + \ket{2} \bra{2} + \ket{3} \bra{3} e^{-i \omega t},
\end{equation}
\noindent and the coupling operator in the interaction picture reads:
\begin{align}
    \hat{W}_{23}^{_I} = \Tilde{U}^{\dagger} \, \hat{W}_{23}(t) \, \Tilde{U} =& \; -\frac{\hbar}{2} \left(\Omega_{23}^{*} \, e^{i\omega t} \ket{3} \bra{2} + \Omega_{23} \, e^{-i\omega t} \ket{2} \bra{3}\right) \times \left(e^{i \omega t}+e^{-i \omega t} \right) 
    \label{RotatingWave3LevelFast} \\
    \simeq & \; -\frac{\hbar}{2} \left(\Omega_{23}^{*} \ket{3} \bra{2} + \Omega_{23} \ket{2} \bra{3} \vphantom{e^{i\omega t}} \right).
    \label{RotatingWave3Level}
\end{align}
\noindent The fast oscillating terms at $\pm 2\omega$ in equation~\eqref{RotatingWave3LevelFast} are neglected as usual using$\;$the$\;$RWA. The atomic Hamiltonian $\hat{H}_{0}$ commutes with $\Tilde{U}(t)$; the total Hamiltonian in the interaction picture is thus finally time independent. In this description, the Bloch equation$\;$is$\;$given$\;$by: 
\begin{equation}
    \frac{\mathrm{d} \hat{\rho}_{I}}{\mathrm{d}t} = -\frac{i}{\hbar} \left[ \hat{H}_{I}, \hat{\rho}_{I} \right] + \sum_{\nu \ne 0} \left( \Tilde{L}_{\nu} \, \hat{\rho}_{I} \, \Tilde{L}_{\nu}^{\dagger} - \frac{1}{2} \{ \Tilde{L}_{\nu} \, \Tilde{L}_{\nu}^{\dagger}, \hat{\rho}_{I} \} \right), 
    \label{LindbladEq}
\end{equation}
\noindent where $\hat{\rho}_{I} = \Tilde{U}^{\dagger} \, \hat{\rho} \, \Tilde{U}$ and $\hat{H}_{I} =  \hbar \delta_{0} \ket{2} \bra{2} - \hbar (\Delta-\delta_{0}) \ket{3} \bra{3} + \hat{W}_{23}^{_I}$. The Lindblad operators $\bar{L}_{\nu}$ are defined in subsection 1.2.2. A straightforwards$\;$but tedious calculation leads finally to the following set of Bloch equations for the slowly varying density matrix elements$ \, \rho_{ij}(t)$ (I drop out the $I$ index in the following):
\begin{align}
        \frac{\mathrm{d}\rho_{11}}{\mathrm{d} t} =& \; -\Gamma_{t} \, \rho_{11} + \frac{\Gamma}{2} \, \rho_{33} + \Gamma_{t}^{_{(1)}} 
        \nonumber \\ 
        \frac{\mathrm{d}\rho_{22}}{\mathrm{d} t} =& \; -\Gamma_{t} \, \rho_{22} + \frac{\Gamma}{2} \, \rho_{33} + \frac{i}{2} \left( \Omega_{23}^{*} \, \rho_{32} - \Omega_{23} \, \rho_{23} \right) + \Gamma_{t}^{_{(2)}} 
        \nonumber \\
        \frac{\mathrm{d}\rho_{33}}{\mathrm{d} t} =& \; -\left(\Gamma_{t} + \Gamma \right) \rho_{33} - \frac{i}{2} \left( \Omega_{23}^{*} \, \rho_{32} - \Omega_{23} \, \rho_{23} \right) \label{Bloch3Level} \\
        \frac{\mathrm{d}\rho_{32}}{\mathrm{d} t} =& \; -\Tilde{\gamma}_{32} \, \rho_{32} + \frac{i \Omega_{23}}{2} \left( \rho_{22} - \rho_{33} \right) \nonumber \\
        \frac{\mathrm{d}\rho_{23}}{\mathrm{d} t} =& \; -\Tilde{\gamma}_{32}^{*} \, \rho_{23} - \frac{i \Omega_{23}^{*}}{2} \left( \rho_{22} - \rho_{33} \right)
        \nonumber
\label{Bloch3Level}    
\end{align}
\noindent where $\Tilde{\gamma}_{32} = \gamma_{32} - i \Delta$ and $\gamma_{32} = \Gamma/2 + \Gamma_{t}$. As$\,$you$\,$can$\,$see,$\,$the$\,$transit$\,$rate$\,\Gamma_{t}\,$enters$\,$now$\,$in the optical Bloch equations. Moreover, it appears explicitly in the coherence$\;$decay$\;$rate$\;\gamma_{32}$, which defines the $\ket{2} \rightarrow \ket{3}$ transition linewidth. The finite transit time of atoms across the laser beam induces therefore a broadening of the transition line, as mentioned$\;$before. 
The Bloch equations~\eqref{Bloch3Level} can be solved together using a 4th-order Runge-Kutta method. The numerical results are shown on figure~\ref{fig:RungeKutta3Level} (a) for the $D_{1}$ line of $^{_{87}}$Rb. The beam$\;$width at $1/e^{2}$ is $\omega_{0} = 500 \; \mu$m. The laser is $2$ GHz red-detuned with respect to the $\ket{2} \rightarrow \ket{3}$ transition and the vapor temperature is set to $415$ K. With these values for the parameters, $\Gamma_{t}/\Gamma \simeq 1.8 \%$. The degeneracy weights $G_{1}$ and $G_{2}$ are respectively equal to $3/8$ and $5/8$; therefore, $\Gamma_{t}^{_{(1)}}/\Gamma \simeq 0.7\%$ and $\Gamma_{t}^{_{(2)}}/\Gamma \simeq 1.1\%$. The computation has been performed using the influx rates defined by equation~\eqref{InfluxRateMixture} ($\beta = 0$). The populations in states $\ket{1}$, $\ket{2}$ and $\ket{3}$ have been plotted in green, purple and black respectively for two different laser$\;$power: $\mathcal{P} = 1$ mW ($\Omega/2\pi \simeq 30$ MHz, dashed lines) and $\mathcal{P} = 0.5$ W ($\Omega/2\pi \simeq 0.7$ GHz, solid$\;$lines).
\begin{itemize}
    \item [$\bullet$] At low power, the internal state dynamics is controlled by the influx and transit rates exclusively. As $\Gamma_{t}^{_{(2)}} > \Gamma_{t}^{_{(1)}}$, $\rho_{22}$ increases at the expense of $\rho_{11}$. In the steady-state, the proportion of atoms in $\ket{1}$ and $\ket{2}$ is directly given by the unbalanced$\;$degeneracy weights $G_{1}$ and $G_{2}$. During the transient regime, the population in the excited$\;$state is$\;$almost$\;$zero. The laser power is too weak to populate $\ket{3}$ as well as to pump$\;$atoms from state $\ket{2}$ to state $\ket{1}$.$\,$The$\,$low$\,$power$\,$situation$\,$has$\,$been$\,$sketched$\,$on$\,$figure~\ref{fig:RungeKutta3Level}$\,$(b). Populations$\;$in $\ket{1}$ and $\ket{2}$ has been represented by disks of different radii.
\end{itemize}

\newpage

\begin{itemize}
    \item [$\bullet$] At high power, the internal state dynamics is controlled by the laser field$\;$and$\;$by$\;$the influx and transit rates simultaneously. The laser field strongly drives the $\ket{2} \rightarrow \ket{3}$ transition even if the frequency is far red-detuned from resonance. As you can$\;$see, Rabi oscillations appear between the populations of states $\ket{2}$ and$\;\ket{3}$.$\;$In$\;$the$\;$steady-state, even if $\Gamma_{t}^{_{(2)}} > \Gamma_{t}^{_{(1)}}$, $\rho_{11}$ remains larger than $\rho_{22}$. Optical pumping from$\;\ket{2}\;$to$\;\ket{1}$ overcomes the filling rates unbalance.
    Nevertheless, $\Gamma_{t}^{_{(2)}}$ and $\Gamma_{t}^{_{(1)}}$, together with the transit rate $\Gamma_{t}$, are still playing a crucial role. If the time of flight of atoms across$\;$the laser beam was infinite, all the atoms would be pumped in $\ket{1}$ in the stationary-state. The coherence between levels $\ket{2}$ and $\ket{3}$ (and thus the susceptibility),$\;$which$\;$depends on the population difference $\rho_{33}-\rho_{22}$
    (see equation~\eqref{OBE2LevelCoherence}), would then drop to zero. The filling and transit rates are thus essential in order for this model to describe experimental observations. The high power case is sketched on~\ref{fig:RungeKutta3Level} (c).
\end{itemize}

\subsection{Steady-state solution of the Optical-Bloch equations}  

\noindent Looking at figure~\ref{fig:RungeKutta3Level}, we can observe that the 3-level system quickly converges$\;$toward a steady-state in which populations as well as coherences do not evolve over time anymore. By setting the time derivatives to zero in~\eqref{Bloch3Level} and by using the conservation$\;$of$\;$the$\;$total atomic population ($\sum_{i} \rho_{ii} = 1$), one can recast the Bloch equations into the matrix$\;$form:

\begin{equation}
    \underbrace{
    \vphantom{
    \begin{pmatrix}
        a \\ b \\ c \\ d \\ e
    \end{pmatrix}
    }
    \begin{pmatrix}
        \Gamma_{t} + \frac{\Gamma}{2}  & \frac{\Gamma}{2} & 0 & 0 \\ 
        \frac{\Gamma}{2} & \Gamma_{t} + \frac{\Gamma}{2} & -\frac{i \Omega^{*}}{2} & \frac{i \Omega}{2} \\
        \frac{i \Omega}{2} & i \Omega & -\Tilde{\gamma} & 0 \\
        -\frac{i \Omega^{*}}{2} & i \Omega^{*} & 0 & -\Tilde{\gamma}^{*}
    \end{pmatrix}}_{M} 
    \underbrace{
    \vphantom{
    \begin{pmatrix}
        a \\ b \\ c \\ d \\ e
    \end{pmatrix}
    }
    \begin{pmatrix}
        \rho_{11} \vphantom{\frac{\Gamma}{2}} \\ 
        \rho_{22} \vphantom{\frac{\Gamma}{2}} \\
        \rho_{32} \vphantom{\frac{i \Omega}{2}} \\ 
        \rho_{23} \vphantom{\frac{i \Omega}{2}} 
    \end{pmatrix}}_{\rho} =  
    \underbrace{
    \vphantom{
    \begin{pmatrix}
        a \\ b \\ c \\ d \\ e
    \end{pmatrix}
    }
    \begin{pmatrix}
        -\frac{\Gamma}{2} - G_{1} \, \Gamma_{t} \vphantom{\frac{\Gamma}{2}} \\ 
        -\frac{\Gamma}{2} - G_{2} \, \Gamma_{t} \vphantom{\frac{\Gamma}{2}} \\
        \frac{i \Omega}{2} \vphantom{\frac{i \Omega}{2}} \\
        -\frac{i \Omega^{*}}{2} \vphantom{\frac{i \Omega}{2}} 
    \end{pmatrix}}_{X}.
    \label{OBE_ST_3level}
\end{equation}
\noindent For the sake of simplicity, we define $\Tilde{\gamma} = \Tilde{\gamma}_{32}$ and $\Omega = \Omega_{32}$. The steady-state$\;$is$\;$obtained$\;$by inverting equation~\eqref{OBE_ST_3level} as follows: $\rho_{\mathrm{eq}} = M^{-1} \cdot X$, which yields the following results:
\begin{align}
    \rho_{11}^{_\mathrm{eq}} =& \; \frac{G_{1}}{\mathcal{F}(\Delta, \Omega)} \left[ 1+\left( \frac{\Delta}{\gamma} \right)^{2} + \frac{1+b}{2 \, b \, (1+a)} \left( \frac{\Omega}{\gamma} \right)^{2} + \frac{G_{2}}{G_{1}} \frac{1-b}{2 \, b \, (1+a)} \left( \frac{\Omega}{\gamma} \right)^{2} \right] 
    \label{Population11} \\
    \rho_{22}^{_\mathrm{eq}} =& \; \frac{G_{2}}{\mathcal{F}(\Delta, \Omega)} \left[ 1 + \left( \frac{\Delta}{\gamma} \right)^{2} + \frac{b}{2 \, b \, (1+a)} \left( \frac{\Omega}{\gamma} \right)^{2} \right] 
    \label{Population22} \\
    \vphantom{
    \left( \frac{\Delta}{\gamma} \right)^{2}
    } 
    \rho_{32}^{_\mathrm{eq}} =& \; \frac{G_{2}}{2} \left(\frac{\Omega}{\gamma} \right)   \frac{i-\Delta/\gamma}{\mathcal{F}(\Delta, \Omega)} \vphantom{\left[ \left(\frac{\Omega}{\gamma} \right)^{2} \right]}
    \label{Coherence23}    
\end{align}
\noindent The coefficients $a$ and $b$ basically measure the contribution of the natural and transit time broadening in the linewidth $\gamma$:  $a = \Gamma/(2 \, \gamma)$ and $b = \Gamma_{t}/\gamma$. The function $\mathcal{F}$ is defined by:
\begin{equation}
    \mathcal{F}(\Delta, \Omega) = 1 + \left( \frac{\Delta}{\gamma} \right)^{2} + \frac{1+b}{2 \, b \, (1+a)} \left( \frac{\Omega}{\gamma} \right)^{2}.   
\end{equation}

    \subsection{Dielectric susceptibility $\chi$ for the 3-level system}

\noindent In the steady-state, the dielectric susceptibility $\chi$ of the 3-level system is related to the coherence $\rho_{32}^{_\mathrm{eq}}$ between the states $\ket{2}$ and $\ket{3}$ (see equation~\eqref{DielectricSusc2Level}): 
\begin{equation}
    \chi = \frac{2 N}{\epsilon_{0} \, \mathcal{E}_{0}} \, \mu_{23}^{\vphantom{_\mathrm{eq}}} \, \rho_{32}^{_\mathrm{eq}} = \frac{\alpha_{0}(0)}{\omega_{32}/c} \, \frac{i-\Delta/\gamma}{1 + \left( \frac{\Delta}{\gamma} \right)^{2} + \left( \frac{\mathcal{E}_{0}}{\mathcal{E}_{s}} \right)^{2}}.
    \label{DielectricSusc3Level}
\end{equation}
\noindent As for the two-level system, we define the linear line-center absorption coefficient$\;\alpha_{0}(0)$ (that$\,$is, the linear absorption coefficient at resonance) by:
\begin{equation}
    \alpha_{0}(0) = G_{2} \frac{\omega_{32}}{c} \, \frac{N}{\epsilon_{0} \hbar} \, \frac{|\mu_{23}|^{2}}{\gamma},    
\end{equation}
\noindent as well as the line-center saturation field strength $\mathcal{E}_{s} = \sqrt{\frac{2 \, b \, (1+a)}{1+b}} \, \hbar \gamma/{\mu_{23}}$. Equation~\eqref{DielectricSusc3Level} is identical in all respects to equation~\eqref{DielectricSusc2Level}. This is an important result as it demonstrates that the optical response of an open 3-level system under a red-detuned laser excitation is comparable to the response of a two-level system. 
However, as mentioned several times already, the linewidth $\gamma = \Gamma/2 + \Gamma_{t}$ is broadened by the finite transit$\;$time$\;$of$\;$atoms across the beam in the open 3-level description. This effect is not describe by the over simplistic 2-level model. Moreover, the on-resonance saturation field strength $\mathcal{E}_{s}$ also depends on the transit rate $\Gamma_{t}$ in the present case. The optical response of the rubidium vapor is thus likely to change with the dimensions of the coupling beam.   

\newpage 

    \subsection{Linear and non-linear response}

\begin{figure}[h]
\center
\includegraphics[width=\columnwidth]{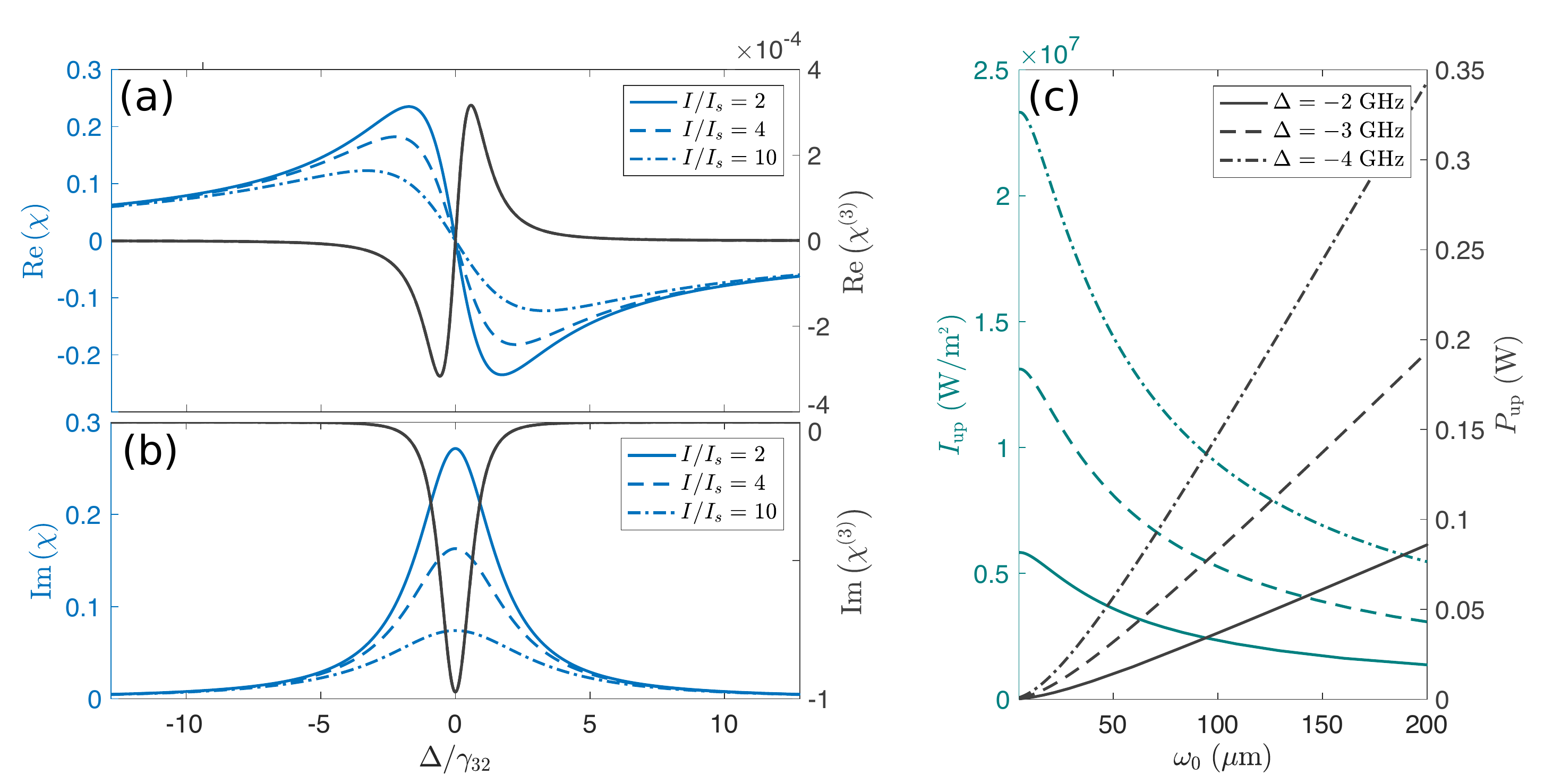} 
\caption{Real (a) and imaginary parts (b) of the total (solid,$\;$dashed$\;$and$\;$dotted$\;$blue) and third-order (solid black) dielectric susceptibilities. $\mathrm{Re}\left[\chi\right]$ and $\mathrm{Im}\left[\chi\right]$ are plotted for different values of the saturation parameter $\mathcal{I}/\mathcal{I}_{s}$. Increasing $\mathcal{I}/\mathcal{I}_{s}$ makes the line broader, as can be seen on fig.(b) (power broadening), and decreases the absorption (saturation). (c) Off-resonance intensity (cyan) and power (black) as function of the beam width $\omega_{0}\;$for different laser detunings. (a)-(b): $\omega_{0} = 500$ $\mu$m. (a)-(c):$\;T = 415$ K. Computation made for the $D_{2}$ line of an isotopically pure rubidium 87 vapor.}
\label{fig:3LevelChi}
\end{figure}

\noindent In order to derive an expression for the third-order dielectric polarizability, we should first expand equation~\eqref{DielectricSusc3Level} as a power series in $\mathcal{E}_{0}/\mathcal{E}_{s}$. This expansion is correct as$\;$long$\;$as $\left|\mathcal{E}_{0}/\mathcal{E}_{s}\right|^{2} \ll 1 +  (\Delta / \gamma )^{2}$ and read (at the second-order) as follows:
\begin{equation}
\chi \simeq \frac{\alpha_{0}(0)}{\omega_{32}/c} \, \frac{i-\Delta/\gamma}{1 + \left( \Delta / \gamma \right)^{2}} \left(1 - \left|\frac{\mathcal{E}_{0}}{\mathcal{E}_{s}}\right|^{2} \frac{1}{1 + \left( \Delta / \gamma \right)^{2}}\right).
    \label{ApproxDielectricSusc3Level}    
\end{equation}
\noindent We now equate the foregoing expression with the usual expansion: $\chi \!=\! \chi^{(1)}+ \frac{3}{4} \chi^{(3)} |\mathcal{E}_{0}|^2$ in order to obtain the first- and third-order dielectric susceptibilities :
\begin{align}
    \chi^{(1)} =& \; \frac{\alpha_{0}(0)}{\omega_{32}/c} \, \frac{i-\Delta/\gamma}{1 + \left( \Delta / \gamma \right)^{2}}, 
    \label{Chi1Exp3Level} \\ 
    \chi^{(3)} =& \; -\frac{1}{|\mathcal{E}_{s}|^{2}} \, \frac{4}{3} \, \frac{\alpha_{0}(0)}{\, \omega_{32}/c} \, \frac{i-\Delta/\gamma}{ \left[ 1 + \left( \Delta / \gamma \right)^{2} \right]^{2}}.
    \label{Chi3Exp3Level}
\end{align}
\noindent Real and imaginary parts of the total (blue solid, dashed and dotted lines) and third-order dielectric polarizabilities (black solid lines) have been plotted on figures~\ref{fig:3LevelChi} (a) and (b). As in the two-level case, power broadening and saturation strongly$\;$affect$\;$the$\;$line$\;$shape of$\;$the$\;$total$\;$susceptibility$\;$at$\;$resonance.$\;$As$\;$expected,$\;$the signs of $\chi$ and $\chi^{_{(3)}}\,$are$\;$opposite.
\vspace{6pt}
\newline
\noindent The expansion~\eqref{ApproxDielectricSusc3Level} makes sense only if the condition$\,\left|\mathcal{E}_{0}/\mathcal{E}_{s}\right|^{2} \ll 1\!+\! (\Delta / \gamma_{_{32}} )^{2}\,$is$\,$fulfilled. In experiments, the laser is typically $3$ GHz red-detuned with respect to the resonance. For a beam width of $500\;\mu$m, the transit rate$\,$is$\,2\pi \times 91\,$kHz$\,$($\Gamma_{t}/\gamma \simeq 3\%$)$\,$at$\,T =  415$ K. If$\;$we address the $D_{2}$ line of a isotopically pure $^{_{87}}$Rb vapor for example, $\Gamma = 2 \pi \times 6.07 \; \mathrm{MHz}$ and therefore, $\gamma \simeq 2\pi \times 3.13$ MHz. The dipole moment of the $\ket{2} \rightarrow \ket{3}\,$transition$\,$is$\,$given$\,$in that case by: $\mu_{23} = \sum_{F_{e}} f_{_{F_{g}}}^{_{F_{e}}} = \sqrt{1/18+5/18+7/9} \times \frac{d}{\sqrt{g_{2}}} \simeq 2.44 \times e \, a_{0}\,$(see$\,$formula~\eqref{Stength} and$\,$the$\,$table$\,$of$\,$figure~\ref{tab:TabStrength}).$\,$The$\,$saturation$\,$intensity$\,\mathcal{I}_{s} = \frac{1}{2}\, \epsilon_{0} \, c \, |\mathcal{E}_{s}|^{2}$ is then of $ 1.5 \; \mathrm{W/m}^{2}$. With the foregoing values for the parameters, we find an off-resonance saturation intensity $\mathcal{I}_{s}(\Delta) = \mathcal{I}_{s}\left[ 1 + (\Delta / \gamma )^{2} \right]$ around $1.6 \times 10^6 \; \mathrm{W/m}^{2}$. This value basically$\,$sets$\,$an$\,$upper$\,$limit on the laser intensity beyond which the expansion~\eqref{ApproxDielectricSusc3Level} starts to be invalid. It is worth mentioning that this result strongly depends on the beam width obviously. Working with smaller beams will indeed increase the off-resonance saturation$\;$intensity $\mathcal{I}_{s}(\Delta)$.$\;$It$\;$will$\;$first increase the line-center saturation intensity, scaling as $\Gamma_{t}^{4} \propto 1/\omega_{0}^{4}$ when $\omega_{0}$ goes to zero. This effect is easily understandable. When the width decreases, the rate at which fresh atoms enter the beam rises up accordingly. In other words, the average time of flight of atoms across the beam becomes shorter and shorter. The laser field should then be more intense in order to drive more efficiently the transition line over this shorter$\;$atom/field interaction time $1/\Gamma_{t}$.$\;$A$\,$second$\,$effect$\,$appear$\,$when$\,$decreasing$\,$the$\,$beam$\,$width,$\,$as$\,$it$\,$will$\,$also increase the linewidth $\gamma$ of$\;$the $\ket{2} \rightarrow \ket{3}$ transition. The atomic ensemble will thus get more resonant with the laser excitation at fixed detuning, leading to a reduction in $\mathcal{I}_{s}(\Delta)$. This effect is appreciable when the natural transition linewidth $\Gamma/2$ and the transit rate $\Gamma_{t}$ are$\;$comparable,$\;$that$\;$is, when $\omega_{0} < 4 \, u/ \sqrt{\pi} \, \Gamma \simeq 4$ $\mu$m at 400 K, which is much smaller than the beam widths we standardly use experimentally. The off-resonance saturation intensity and power $\mathcal{P}_{s}(\Delta)$ have been plotted on figure~\ref{fig:3LevelChi} (c) for different laser detunings.  


\newpage
 
    \subsection{General case: open 3-level system with two coupling fields}
    
\noindent So far, the laser field was only driving the $\ket{2} \rightarrow \ket{3}$ transition. Let's now assume it also drives the $\ket{1} \rightarrow \ket{3}$ transition, as sketched on figure~\ref{fig:3Level}(b). In that case, the steady-state solution of the optical Bloch equations is found by solving the following matrix equation:   
\begin{equation}
    \begin{pmatrix}
        -\Gamma_{t} \!-\! \frac{\Gamma}{2}  & -\frac{\Gamma}{2} & 0 & 0 & \frac{i \Omega_{13}^{*}}{2} & -\frac{i \Omega_{13}}{2} & 0 & 0 \\
        -\frac{\Gamma}{2} & \Gamma_{t} \!+\! \frac{\Gamma}{2} & 0 & 0 & 0 & 0 & \frac{i \Omega_{23}^{*}}{2} & -\frac{i \Omega_{23}}{2} \\
        0 & 0 & -\Tilde{\gamma}_{21} & 0 & \frac{i \Omega_{23}^{*}}{2} & 0 & 0 & -\frac{i \Omega_{13}}{2} \\
        0 & 0 & 0 & -\Tilde{\gamma}^{*}_{21} & 0 & -\frac{i \Omega_{23}}{2} & \frac{i \Omega_{13}^{*}}{2} & 0 \\
        i \Omega_{13} & \frac{i \Omega_{13}}{2} & \frac{i \Omega_{23}}{2} & 0 & -\Tilde{\gamma}_{31} & 0 & 0 & 0 \\
        -i \Omega_{13}^{*} & -\frac{i \Omega_{13}^{*}}{2} & 0 &-\frac{i \Omega_{23}^{*}}{2} & 0 & -\Tilde{\gamma}^{*}_{31} & 0 & 0 \\
         \frac{i \Omega_{23}}{2} & i \Omega_{23} & 0 & \frac{i \Omega_{13}}{2} & 0 & 0 & -\Tilde{\gamma}_{32} & 0 \\
         -\frac{i \Omega_{23}^{*}}{2} & -i \Omega_{23}^{*} & -\frac{i \Omega_{13}^{*}}{2} & 0 & 0 & 0 & 0 &-\Tilde{\gamma}^{*}_{32}
    \end{pmatrix} \!
    \begin{pmatrix}
        \rho_{11} \vphantom{\frac{\Gamma}{2}} \\ 
        \rho_{22} \vphantom{\frac{\Gamma}{2}} \\
        \rho_{21} \vphantom{\frac{i \Omega_{23}^{*}}{2}} \\ 
        \rho_{12} \vphantom{\frac{i \Omega_{23}^{*}}{2}} \\ 
        \rho_{31} \vphantom{\frac{i \Omega_{23}^{*}}{2}} \\ 
        \rho_{13} \vphantom{\frac{i \Omega_{23}^{*}}{2}} \\ 
        \rho_{32} \vphantom{\frac{i \Omega_{23}^{*}}{2}} \\ 
        \rho_{23} \vphantom{\frac{i \Omega_{23}^{*}}{2}}
    \end{pmatrix} \!=\!  
    \begin{pmatrix}
        -\frac{\Gamma}{2}-G_{1} \Gamma_{t}  \\ 
        -\frac{\Gamma}{2}-G_{2} \Gamma_{t}  \\
        0 \vphantom{\frac{i \Omega_{23}^{*}}{2}} \\
        0 \vphantom{\frac{i \Omega_{23}^{*}}{2}} \\
        \frac{i \Omega_{13}}{2} \vphantom{\frac{i \Omega_{23}^{*}}{2}} \\
        -\frac{i \Omega_{13}^{*}}{2} \vphantom{\frac{i \Omega_{23}^{*}}{2}} \\
        \frac{i \Omega_{23}}{2} \vphantom{\frac{i \Omega_{23}^{*}}{2}} \\
        -\frac{i \Omega_{23}^{*}}{2} \vphantom{\frac{i \Omega_{23}^{*}}{2}} 
        \label{ST_Bloch3Level_2Coupling}
    \end{pmatrix}
\end{equation}

\noindent where $\Tilde{\gamma}_{21} = \gamma_{21} + i \delta_{0}$, $\Tilde{\gamma}_{31} = \gamma_{31} - i(\Delta -\delta_{0})$ and $\Tilde{\gamma}_{32} = \gamma_{32} - i \Delta$. The decoherence rates are defined by: $\gamma_{21} = \Gamma_{t}$, $\gamma_{31} = \Gamma/2 + \Gamma_{t}$ and $\gamma_{32} = \Gamma/2 + \Gamma_{t}$. The$\,$linewidths$\,$of$\,$the$\,\ket{1} \rightarrow \ket{3}\,$and $\ket{2} \rightarrow \ket{3}$ transitions are equal since the probabilities for the excited state to decay toward the upper$\,$and$\,$lower$\,$ground$\,$state$\,$are$\,$the$\,$same$\,$in$\,$our$\,$model.$\;$This$\,$assumption$\,$is$\,$not$\,$perfectly fulfilled because some of the hyperfine transitions between ground and excited states are cycling (the $F_{g} = 3 \rightarrow F_{e} = 4$ hyperfine transition in $^{_{85}}$Rb$\;$for$\;$example).  

\newpage

\noindent The stationary-state density matrix elements are obtained by inverting equation~\eqref{ST_Bloch3Level_2Coupling}. Nevertheless, populations and coherences cannot be easily expressed analytically anymore. We cannot either use density matrix perturbation techniques, as both Rabi frequencies $\Omega_{13}$ and $\Omega_{23}$ have the same magnitude. They are actually equal in this case, as$\;\mu_{13} = \mu_{23}$. However, since $\delta_{\mathrm{HF}} \ll \delta_{0}$, we can think about breaking down this problem into pieces and see the situation sketched in fig.~\ref{fig:3Level} (b) as a composition of two open 3-level$\;$systems$\;$with one coupling field, driving either the $\ket{1} \rightarrow \ket{3}$ or the $\ket{2} \rightarrow \ket{3}$ transition. An approximate steady-state solution can then be found using the set of equations~\eqref{Population11},~\eqref{Population22} and~\eqref{Coherence23}. Within this approximation, the population in state $\ket{1}$ is given for instance by: 
\begin{equation}
    \rho_{11} \simeq \rho_{_{11}}^{_{1 \shortto 3}} + \rho_{_{11}}^{_{2 \shortto 3}} - G_{1},    
\end{equation}
\noindent where $\rho_{_{11}}^{_{i \shortto 3}}$ stands for the steady-state population of level $\ket{1}$ when the laser only addresses the $\ket{i} \rightarrow \ket{3}$ transition. Therefore, $\rho_{_{11}}^{_{2 \shortto 3}}$ is given by equation~\eqref{Population11} and $\rho_{_{11}}^{_{1 \shortto 3}}$ is obtained by (i) interchanging the indices "1" and "2" and (ii) by replacing $\Delta$ by $\Delta-\delta_{0}$ in equation~\eqref{Population22}. The degeneracy weight $G_{1}$ should be subtracted in order to take the filling of the ground state $\ket{1}$ into account only once. The steady-state population in $\ket{2}$ can be approximate in a similar way: $\rho_{22} \simeq \rho_{_{22}}^{_{1 \shortto 3}} + \rho_{_{22}}^{_{2 \shortto 3}}-G_{2}$. We can also derive an approximate expression for the coherences $\rho_{31}$ and $\rho_{32}$ using~\eqref{Coherence23} (after replacing "1" by "2" and $\Delta$ by $\Delta-\delta_{0}$ for $\rho_{31}$). The ground states coherence $\rho_{21}$ can then simply be expressed as: $\rho_{21} \!=\! \frac{i \, \Omega}{2 \gamma_{21}} \left( \rho_{31} - \rho_{32}^{*}\right)$. 
   
\begin{figure}[h]
\center
\includegraphics[width=\columnwidth]{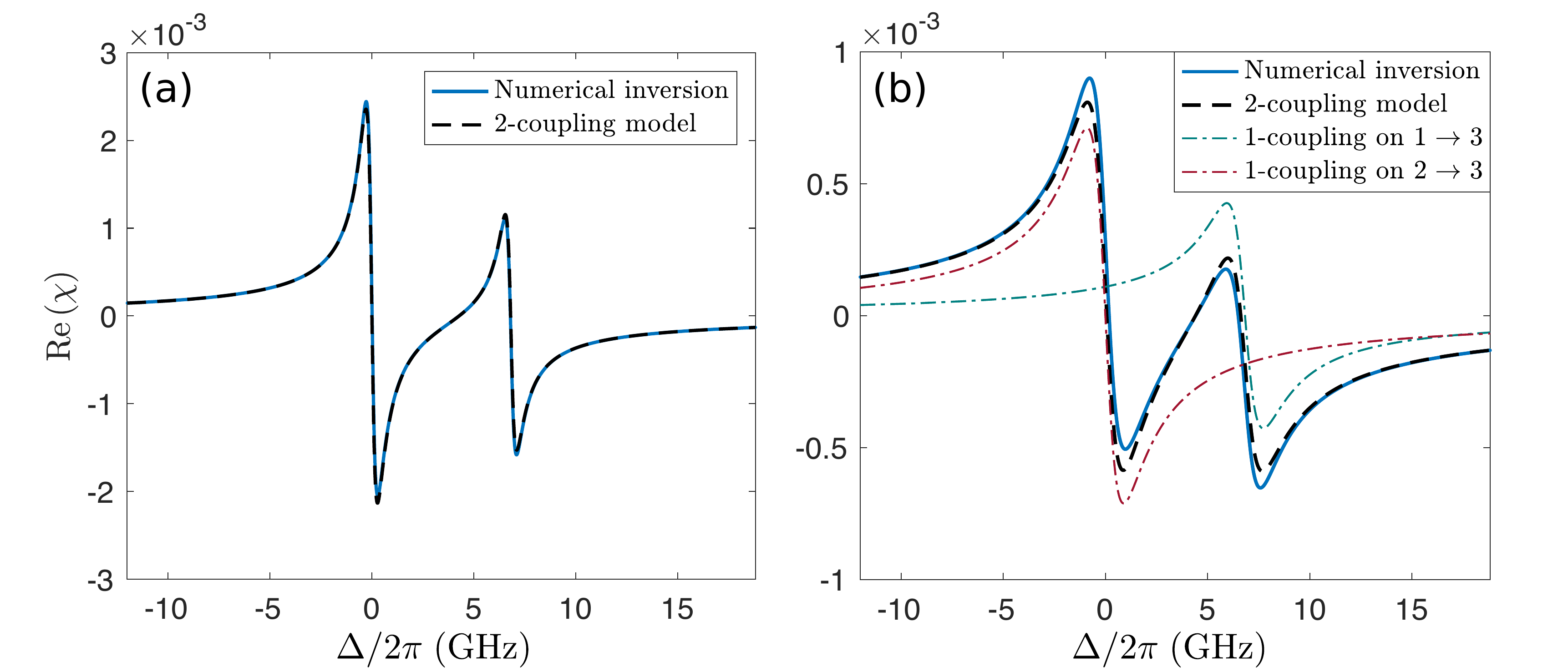} 
\caption{Comparison between the total susceptibilities computed from the 2-coupling 3-level model (black dashed line) and the numerical inversion of~\eqref{ST_Bloch3Level_2Coupling} (blue solid line). (a) Low power (10 mW). (b) High power (100 mW). The discrepancy$\;$between the black dashed and the blue solid lines starts being visible around $\Delta = 0$ and $\Delta = \delta_{0} \simeq 6.8\;$GHz. If we consider that the field only addresses the $\ket{2}\rightarrow\ket{3}$ transition,$\;$we$\;$underestimate$\;$the value of $\mathrm{Re}\left[\chi\right]$ when the laser is highly red-detuned. The contribution$\,$of$\,\ket{1}\rightarrow\ket{3}\,$to$\,\mathrm{Re}\left[\chi\right]$ is about $25\%$ when $\Delta = -2\pi\!\times\!6\;$GHz for instance. Parameters in the text.
}
\label{fig:3LevelChi2Coupling}
\end{figure}   
  
\noindent The real part of the dielectric susceptibility $\mathrm{Re}\left[\chi\right]$, obtained either by numerical inversion (blue solid line) or within the approximation above (black dashed lines), has been plotted on figure~\ref{fig:3LevelChi2Coupling} as function of $\Delta$, for $\mathcal{P} \!=\! 10 \; \mathrm{mW}$ (fig.(a), $\Omega/2\pi \simeq 0.1$ GHz)$\;$and$\;\mathcal{P} \!=\! 100 \; \mathrm{mW}$ (fig.(b), $\Omega/2\pi \simeq 0.3$ GHz). The laser drives the $D_{1}$ line of $^{_{87}}$Rb and its width is $500 \;\mu$m. 

\newpage 

\begin{itemize}
    \item [$\bullet$] As long as the coupling field is low enough to safely neglect optical pumping between ground states, the agreement between the numerical inversion of equation~\eqref{ST_Bloch3Level_2Coupling}$\;$and the theory is excellent, as can be seen on figure~\ref{fig:3LevelChi2Coupling}(a). 
    \vspace{-4pt}
    \item [$\bullet$] When optical pumping comes into play, the model fails to reproduce exactly the numerical$\,$results close to resonances (\textit{ie}, when $\Delta = 0$ and $\Delta = \delta_{0} \simeq 2\pi\!\times\! 6.8\;$GHz). However, for highly red-detuned laser frequencies (typically, when $\Delta/2\pi < - 3\;$GHz), the model describes perfectly well the vapor optical response. As you may have$\;$seen, the contribution of the $\ket{1} \rightarrow \ket{3}$ transition line on the red-detuned optical response (cyan dashed line) is not negligible; the relative error made on $\mathrm{Re}\left[ \chi \right]$ by considering only one coupling field when $\Delta/2\pi = -6\;$GHz is almost 25$\%$ for instance.
\end{itemize}

\noindent The$\;$two coupling fields model is thus able to correctly predict the dielectric$\;$susceptibility at those detunings even for strong driving, which is not the case of the one coupling field and the simplistic two-level descriptions.   

\subsection{Ballistic transport of atoms}

\noindent Contrary to what might sometimes be believed, hot alkaline vapors are dilute systems. As$\;$long as the temperature of the atomic ensemble does no exceed $150 \; ^{\circ}$C, the atomic motion is mainly ballistic and Rb-Rb collisions can be neglected as a first approximation. This ballistic transport affects the atom internal state because it forces the atom/field interaction to happen on a finite time, as described above. In this section, I will introduce other effects resulting from atomic motion $-$namely, Doppler broadening$\;$and$\;$transport-induced nonlocal dielectric response $-$ and describe how these effects impact the line$\;$shape.  

    \subsubsection{Doppler broadening}

\noindent The atomic motion along the optical axis shifts the laser frequency in the translating$\;$frame of a moving atom. This Doppler shift affects the transition linewidth (Doppler$\;$broadening) by making fast atoms be resonant with a slightly detuned laser beam. In this$\;$subsection, I will generalize equations~\eqref{Chi1Exp3Level} and \eqref{Chi3Exp3Level} by taking Doppler broadening into account. The calculation mainly follows the derivation made in~\cite{2-9Siddons}. 
\vspace{6pt}
\newline
\noindent Let's first define the line shape factors $f_{\gamma}^{_{(1)}}$ and $f_{\gamma}^{_{(3)}}$ as follow: 
\begin{equation}
f_{\gamma}^{_{(1)}} = \frac{1}{\gamma} \, \frac{i-\Delta/\gamma}{1\!+\!\left( \Delta / \gamma \right)^{2}} \hspace{0.275cm} \mathrm{and} \hspace{0.25cm} f_{\bar{\gamma}}^{_{(3)}} = \frac{1}{\gamma} \, \frac{i-\Delta/\gamma}{ \left[ 1\!+\!\left( \Delta / \gamma \right)^{2} \right]^{2}},   
\end{equation}
\noindent where $\gamma = \gamma_{31} = \gamma_{32} = \Gamma/2 + \Gamma_{t}$. The linear Doppler broadened susceptibility is obtained by convolving $\chi^{_{(1)}}(\Delta_{j})\;$with the 1D Maxwell Boltzmann velocity distribution $\mathcal{P}_{_{\mathrm{1D}}}$ \eqref{1DMaxBolt}:
\begin{equation}
    \chi_{\mathrm{D}}^{_{(1)}}(\Delta_j) = \frac{\alpha_{0}(0)}{\omega_{3j}/c} \, \gamma \underbrace{\int_{-\infty}^{\infty} f_{\gamma}^{(1)}(\Delta_{j}-k v) \, \mathcal{P}_{_{\mathrm{1D}}}(v) \, \mathrm{d}v}_{s(\Delta_{j})}, 
    \label{DopplerBroadChi3Level}
\end{equation}
\noindent where the detuning $\Delta_{j}$ is equal either to $\Delta$ or to $\Delta-\delta_{0}$ if the laser addresses the $\ket{2} \rightarrow \ket{3}$ or the $\ket{1} \rightarrow \ket{3}$ transition. 

\newpage

\noindent The magnitude of the susceptibility $\chi_{\mathrm{D}}^{_{(1)}}$ appears simply as a prefactor in equation~\eqref{DopplerBroadChi3Level}. For$\;$the$\;$sake of convenience, we define the complex function $s$ which is proportional$\;$to$\;\chi_{\mathrm{D}}^{_{(1)}}$ but does not depend on the atomic transition considered. Let's also define$\;a_{j} = \Delta_{j}/ku\;$and $b=\gamma/ku$ ($u = \sqrt{2 k_{B} T/m}$ is the most probable speed) and make the change of$\;$variable $v \rightarrow x = v/u$ in the integral $s$. By separating the real and imaginary parts of $f_{\gamma}^{_{(1)}}$ and by using the convolution theorem, the Fourier transform of the function $s$ reads finally: $S(\Tilde{a}_{j}) = \left( F_{b}^{_{\mathrm{R}}}(\Tilde{a}_{j})+ F_{b}^{_{\mathrm{I}}}(\Tilde{a}_{j})\right) P(\Tilde{a}_{j})$ where:
\begin{align}
    F_{b}^{_{\mathrm{R}}}(\Tilde{a}_{j}) =& \; -\int_{-\infty}^{\infty} \frac{1}{b} \frac{t/b}{1+(t/b)^{2}} \, e^{-i \Tilde{a}_{j} t} \, \mathrm{d}t \;=\; i \, \pi \, \mathrm{sgn}(\Tilde{a}) \, e^{-b |\Tilde{a}_{j}|}, \\ 
    F_{b}^{_{\mathrm{I}}}(\Tilde{a}_{j}) =& \; \int_{-\infty}^{\infty} \frac{1}{b} \frac{1}{1+(t/b)^{2}} \, e^{-i \Tilde{a} t} \, \mathrm{d}t \;=\; i \, \pi \, e^{-b |\Tilde{a}|}, \\
    P(\Tilde{a}_{j}) =& \; \int_{-\infty}^{\infty} g_{u}(t) \, e^{-i \Tilde{a}_{j} t} \, \mathrm{d}t \;=\; e^{- (\Tilde{a}/2)^{2}}. 
\end{align}
\noindent By taking the inverse Fourier transform of $S$ and by rearranging, we finally find that:
\begin{align}
    \chi_{\mathrm{D}}^{_{(1)}}(\Delta_{j}) =& \; \frac{\alpha_{0}(0)}{\omega_{3j}/c} \, b \left \{\mathrm{Re} \left[F\left(a_{j} + i b \right)\right] +i \, \mathrm{Im} \left[F\left(a_{j} + i b \right)\right] \right \}, \nonumber \\
    =& \; \frac{\alpha_{0}(0)}{\omega_{3j}/c} \left( \frac{\gamma}{ku} \right) \left \{\mathrm{Re} \left[F\left(\frac{\Delta_{j}}{ku} + i \frac{\gamma}{ku} \right)\right] +i \, \mathrm{Im} \left[F\left(\frac{\Delta_{j}}{ku} + i \frac{\gamma}{ku} \right)\right] \right \},  
    \label{DopplerBrodenedChi1}
\end{align}

\noindent where $F$ stands for the plasma dispersion function: $F(z) = i \, \sqrt{\pi} \, e^{-z^{2}} \, \mathrm{Erfc} (-i z)$ and$\;\mathrm{Erfc}$ for the complex complementary error function. The absorption$\;$coefficient $\alpha = k \, \mathrm{Im}(\chi_{\mathrm{D}}^{_{(1)}}) $ has a Voigt-type profile arising from the convolution of the Lorentzian absorption$\;$profile of an atom at rest and the Gaussian distribution $\mathcal{P}_{_{\mathrm{1D}}}$. By following the exact same$\;$steps, we can also derive an expression for the real part of the Doppler broadened third-order dielectric susceptibility. Introducing $z_{j} = a_{j}+ib$ and using the relation $F(-z^{*}) = - F(z)^{*}$:
\begin{align}
     \mathrm{Re} \left[\chi_{\mathrm{D}}^{(3)}(\Delta_{j})\right] =& \;  -i \, \frac{1}{|\mathcal{E}_{s}|^{2}} \, \frac{4}{3} \, \frac{\alpha_{0}(0)}{\omega_{32}/c} \, \frac{b^3}{2a} \left[ z_{j} \, F(z_{j}) + z_{j}^{*} \, F(-z_{j}^{*}) \right], \nonumber \\
     =& \; \frac{1}{|\mathcal{E}_{s}|^{2}} \, \frac{4}{3} \, \frac{\alpha_{0}(0)}{\omega_{32}/c} \left(\frac{\gamma}{ku} \right)^{2} \left(\frac{\gamma}{\Delta_{j}} \right) \mathrm{Im}\left[\left( \frac{\Delta_{j}}{ku} + i \frac{\gamma}{ku}  \right) F\left( \frac{\Delta_{j}}{ku} + i \frac{\gamma}{ku} \right)\right]\; 
     \label{DopplerBrodenedChi3}
\end{align}
\noindent For the sake of completeness, let me finally derive an expression for the total Doppler broadened dielectric susceptibility $\chi_{\mathrm{D}}^{\vphantom{(3)}}$: 
\begin{equation}
    \chi_{\mathrm{D}}^{\vphantom{(3)}}(\Delta_{j}) = \frac{\alpha_{0}(0)}{\omega_{3j}/c} \left(\frac{\gamma}{ku} \right) \left \{ \mathrm{Re} \left[ F(a_{j}+ib_{\mathcal{I}}) \right] + i \, \frac{\mathrm{Im} \left[ F(a_{j}+ib_{\mathcal{I}}) \right]}{\sqrt{1+\mathcal{I}/\mathcal{I}_{s}}} \right \},
\end{equation}
\noindent where $b_{\mathcal{I}} = b \, \sqrt{1+\mathcal{I}/\mathcal{I}_{s}}$. We$\,$recover$\,$the$\,$formula$\,$derived$\,$following$\,$the$\,$Lamb's$\,$model$\,$in$\,$\cite{2-21Close}, where the effects of gain saturation by strong driving fields in a dilute two-level atomic medium is investigated. The same equation has also been reported in~\cite{2-20Skupin}. In$\;$the$\;$latter, the authors$\,$claim$\,$that$\,$the$\,$atomic$\,$motion$\,$is$\,$not$\,$only$\,$accountable$\,$for$\,$Doppler$\,$broadening$\,$but also for a ballistic transport induced nonlocal dielectric response.$\,$The$\,$next$\,$section$\,$describes this effect in our configuration. 

    \subsubsection{Ballistic transport of excited atoms: nonlocal dielectric response}

\begin{figure}[h]
\center
\includegraphics[width=0.9\columnwidth]{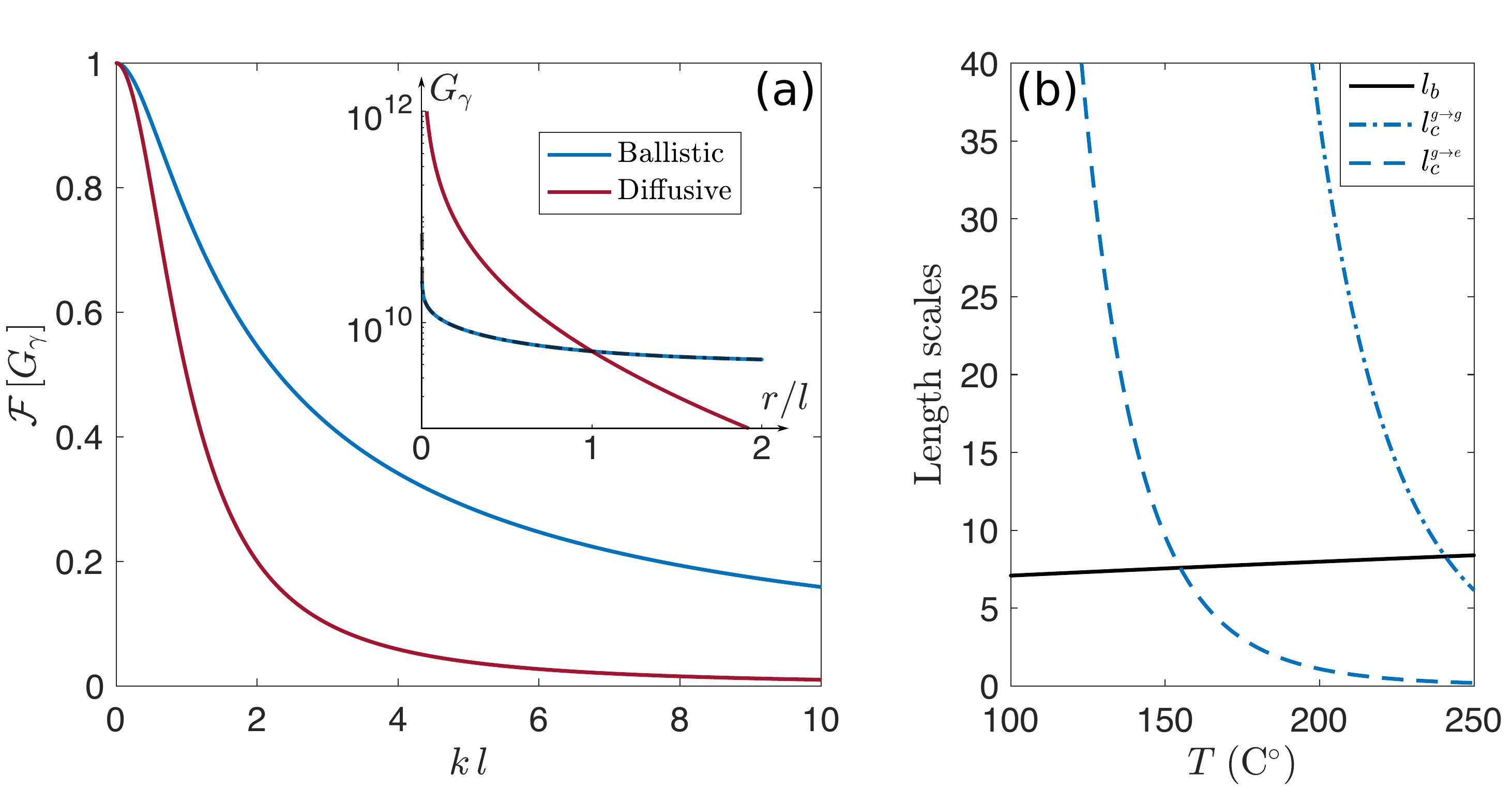} 
\caption{(a) Ballistic (blue) and diffusive (red) response functions in k-space. Inset: real space response functions, rescaled to be equal at $r=l$. (b) Transport length scales as function of the vapor temperature $T$. The ballistic transport length scale$\;l_{b}$ (black$\;$line) remains lower than the free mean path $l_{c}^{g \shortto e}$ for $T < 155\;$C$^{\circ}$. The transport$\;$of$\;$excited$\;$atoms is therefore mainly ballistic at $T=130\;$C$^{\circ}$. Figure reproduced from~\cite{2-20Skupin}.}
\label{fig:GreenFunctions}
\end{figure}

\noindent So far, I have implicitly assumed that the dielectric response of the atomic vapor$\;$was$\;$local, or, in other words, that it only depends on the field strength or intensity at a given point in space. This assumption is fulfilled as long as no intensity redistribution process makes the susceptibility at $\mathbf{r_{0}}$ depend on the surrounding field strength at $\mathbf{r_{0}} + \boldsymbol{\delta} \mathbf{r}$. In$\;$hot$\;$vapors, intensity redistribution do occur through ballistic transport of excited$\;$atoms. Even if the laser field is far detuned from resonance, fast moving atoms can still absorb photons because of Doppler effect. Those photons, after being absorbed at $\mathbf{r_{0}}$, can therefore be re-emitted at an other location in space, at $\mathbf{r_{0}} + \mathbf{v} \, \tau$, where $\tau = 1/\Gamma$ is the lifetime of the excited state. This picture is very simplistic but still gives an insight into the physics at play in the transport-induced nonlocal dielectric response of hot vapors. 
\vspace{6pt}
\newline
\noindent As mentioned in~\cite{2-21Close}, the degree of nonlocality and the nature of the redistribution process (ballistic, diffusive, ...) both depend on the characteristic length scales associated with the transport of excited atoms. The first length scale is the mean free path \textit{ie} the average distance travelled by an atom before a Rb-Rb collision: $l_{c} = 1/(\sqrt{2} \, N \, \sigma_{i \shortto j})$, where $\sigma_{i \shortto j}$ is the scattering cross section between two atoms, one in state $\ket{i}$ and the other in state$\;\ket{j}$. The collisional cross-section between atoms in the ground state is $\sigma_{g \shortto g} = 2.5 \times 10^{-17}\;$m$^{2}$. For$\;$collisions between an atom in the ground state and an other one in the excited state, the cross-section is much larger since the collision process occurs via a long range dipole-dipole interaction~\cite{2-22Lewis} and: $\sqrt{T} \sigma_{g \shortto e} = 1.8 \times 10^{-14} $ K$^{\frac{1}{2}}$m$^{2}$. At $T = 400$ K, $l_{c}^{g \shortto g} \simeq 11$ mm$\;$and $l_{c}^{g \shortto e} \simeq 32$ $\mu$m; both are much larger than the ballistic transport length scale, defined by $l_{b} = u \, \tau \simeq 7.6$ $\mu$m at 400 K. The transport of excited atoms is mainly ballistic at $400$ K and the nonlocal response of the material should then depend on $u$ and $\tau$.
\vspace{6pt}
\newline
\noindent The rate equation for the excited state population is given by the third equation in~\eqref{Bloch3Level}. One can rewrite this equation as function of the total absorption coefficient $\alpha$ as follow:
\begin{align}
    \frac{\mathrm{d} \rho_{33}}{\mathrm{d}t} =& \; -\left(\Gamma + \Gamma_{t} \right) \rho_{33} + \mathrm{Im} \left(\Omega^{*}_{23} \, \rho_{32} \right), \\
    =& \; -\left(\Gamma + \Gamma_{t} \right) \rho_{33} + \frac{1}{2} \frac{\epsilon_{0}}{\hbar} \, |\mathcal{E}|^{2} \, \mathrm{Im} \left(\chi_{32} \right), \nonumber \\ 
    =& \; -\left(\Gamma + \Gamma_{t} \right) \rho_{33} +  \frac{\mathcal{I} \, \alpha (\mathcal{I})}{\hbar \omega},
    \label{ExcitedStateDensity}
\end{align}
\noindent using the relations: $\chi_{32} \!=\! \frac{2 \mu_{23}^{\vphantom{_\mathrm{eq}}}}{\epsilon_{0} \mathcal{E}}\, \rho_{32}$ and $\alpha \!=\! \frac{\omega}{n_{0} c} \; \mathrm{Im} \, (\chi_{32})$. Equation~\eqref{ExcitedStateDensity} has been derived using the 3-level description sketched in figure~\ref{fig:3Level} (a) but is actually much more general since it only involves the total decay rate $\bar{\gamma} = \Gamma + \Gamma_{t}$ of the excited state and the local rate of excitation $\mathcal{I} \alpha (\mathcal{I}) / \hbar \omega$. The same$\,$equation$\,$describes$\,$for$\,$instance$\,$the$\,$evolution$\,$of$\,$the$\,$excited state population when both the $\ket{1} \rightarrow \ket{3}$ and the $\ket{2} \rightarrow \ket{3}$ transitions are simultaneously driven by the laser field. In that case, $\alpha = k \, \mathrm{Im} \left( \chi_{31} + \chi_{32} \right)$. Working$\;$within the \textbf{paraxial approximation}, I thus suppose that $\rho_{33}$ only substantially varies in the transverse plane, where $\mathbf{r} = x \boldsymbol{\hat{x}} + y \boldsymbol{\hat{y}}$ and $\mathbf{v} = v_{x} \boldsymbol{\hat{x}} + v_{y} \boldsymbol{\hat{y}}$. One can then rewrite eq.\eqref{ExcitedStateDensity} using the differential formula $\frac{\mathrm{d}}{\mathrm{d} t} = \frac{\partial}{\partial t} + \mathbf{v} \cdot \boldsymbol{\nabla}_{\perp}$ (where $\boldsymbol{\nabla}_{\perp}$ is the gradient operator in the $(\boldsymbol{\hat{x}},\boldsymbol{\hat{y}})$ plane):
\begin{equation}
    \frac{\partial \rho_{33}}{\partial t} + \mathbf{v} \cdot \boldsymbol{\nabla}_{\perp} \rho_{33} + \underbrace{\left(\Gamma + \Gamma_{t} \right)}_{\bar{\gamma}} \rho_{33} = \frac{\mathcal{I} \, \alpha_{\mathrm{D}}(\mathcal{I})}{\hbar \omega}.
    \label{ExcitedStateDensityFullEq}
\end{equation}
\noindent This equation can be solved with$\,$the$\,$Green$\,$function$\,$formalism.$\;$Let's$\,G_{\bar{\gamma}}^{b}\,$be$\,$the$\,$solution$\,$of:
\begin{equation}
    \frac{\partial \rho_{33}}{\partial t} + \mathbf{v} \cdot \boldsymbol{\nabla}_{\perp} \rho_{33} + \bar{\gamma} \, \rho_{33} = \delta(t)\, \delta(\mathbf{r}) \, \mathcal{P}_{_{\mathrm{2D}}}(v),
\end{equation}
\noindent where $\mathcal{P}_{_{\mathrm{2D}}}$ stands for the 2D Maxwell-Boltzmann velocity distribution, as usual.$\;$Averaging the Green function $G_{\bar{\gamma}}^{b}$ over $\mathcal{P}_{_{\mathrm{2D}}}$ yields: 
\begin{equation}
    \bar{G}_{\bar{\gamma}}^{b}(\mathbf{r}, t) = \frac{1}{\pi u^{2}} \, e^{-\gamma \, t} \left[\int_{\mathbb{R}^{2}} \delta\left(\mathbf{r} - \mathbf{v} \, t \right) \, e^{-(v/u)^{2}} \, \mathrm{d}\mathbf{v}\right] = \; \frac{1}{\pi u^{2}} \frac{e^{-\bar{\gamma} \, t}}{t^{2}} e^{- \left(\frac{|\mathbf{r}|}{u t}\right)^{2}}.
\end{equation}
\noindent One can then obtain the spatial distribution of the excited state population by convolving the averaged Green function $\bar{G}_{\bar{\gamma}}^{b}$ with the local rate of excitation $\mathcal{I}  \alpha_{\mathrm{D}}(\mathcal{I} )/\hbar \omega$ of~\eqref{ExcitedStateDensity}: 
\begin{equation}
    \rho_{33}(\mathbf{r}, t) = \frac{1}{\hbar \omega} \int_{-\infty}^{t} \mathrm{d}t' \int_{\mathbb{R}^{2}} \mathrm{d}\mathbf{r'} \, \bar{G}^{b}_{\bar{\gamma}}(\mathbf{r}\!-\!\mathbf{r'}, t\!-\!t') \left \{ \mathcal{I} (\mathbf{r'}, t') \, \alpha \left[ \mathcal{I} (\mathbf{r'}, t') \right] \right \} 
\end{equation}
\noindent Let's assume the rate of excitation is constant over time and localized in space at $\mathbf{r_{0}}\;$such that $\mathcal{I} (\mathrm{r_{0}}, t) = \delta(\mathrm{r_{0}}) \, \mathcal{I} _{0}$. By making successively the$\,$substitutions$\,t\rightarrow t+t_{0}\,$and$\,t \rightarrow \xi = \frac{r}{u} \frac{1}{t}$ (where $r = | \mathbf{r} \!-\! \mathbf{r_{0}}|$), one can finally derive an expression for the spatial distribution of the excited state population in the steady-sate: 
\begin{equation}
    \rho_{33}(\mathbf{r}) = \frac{\mathcal{I}_{0} \, \alpha(\mathcal{I}_{0})}{\hbar \omega} \frac{1}{\pi \, u \, r} \int_{0}^{\infty} e^{-\bar{\gamma} r/u \xi} e^{-\xi^{2}} \mathrm{d}\xi,
\end{equation}
\noindent as well as the steady-state Green function and its spatial Fourier transform:  
\begin{align}
    \begin{cases}
        \bar{G}_{\bar{\gamma}}^{b}(\mathbf{r}\!-\!\mathbf{r_{0}}) =& \frac{1}{\pi \, u \, r} \int_{0}^{\infty} e^{-\bar{\gamma} r/u \xi} \, e^{-\xi^{2}} \mathrm{d}\xi, \\
        \mathcal{F}\!\left[ \bar{G}_{\bar{\gamma}}^{b} \right]\!(k) =& \frac{\sqrt{\pi}}{\bar{\gamma}} \frac{e^{1 / (k \, l_{b})^{2}}}{k \, l_{b}} \mathrm{Erfc \left[ 1/(k \, l_{b}) \right].}  
    \end{cases}
\end{align}
\noindent The Green function $\bar{G}_{\bar{\gamma}}^{b}$ is well normalized as $ \bar{\gamma} \int G_{\bar{\gamma}}^{b}(\mathbf{r}) \, \mathrm{d}\mathbf{r} = 1$. 
\vspace{6pt}
\newline
\noindent $\bar{G}_{\bar{\gamma}}^{b}$ and $\mathcal{F}\!\left[ \bar{G}_{\bar{\gamma}}^{b} \right]$ have been plotted on figure~\ref{fig:GreenFunctions} (a) as function of the scaled coordinates $r/l_{b}$ and $k \, l_{b}$ respectively. As mentioned in~\cite{2-20Skupin}, the ballistic response (blue curve) falls$\;$off much more rapidly than the diffusive one (red curve) in real space.$\;$Diffusion is thus$\;$much more efficient at spreading the intensity in the transverse plane$\;$than$\;$ballistic$\;$transport. 
In order to complete the description of the transport-induced nonlocality, let's$\;$derive an expression for the nonlocal dielectric susceptibility which also takes Doppler broadening into account. When the laser intensity varies spatially in the transverse plane,$\;$we$\;$can$\;$use the Green function $\bar{G}_{\bar{\gamma}}^{b}$ to write the steady-sate dielectric response as follow:
\begin{align}
    \begin{split}
        \hspace{-0.3cm} \mathrm{Re} \left[\chi_{\mathrm{D}}(\mathbf{r})\right] =& \; \chi_{0}^{\vphantom{\!_{(2)}}} \left( \mathrm{Re}  \left[ F(a + i b) \right] + \bar{\gamma} \int_{\mathbb{R}^{2}}  G_{\bar{\gamma}}^{b}(\mathbf{r}\!-\!\mathbf{r_{0}}) \times \right. \\
            & \hspace{3.6cm} \left. \left \{ \mathrm{Re} \left[ F\left(a + i b_{\mathcal{I}}(\mathbf{r_{0}})\right) \right]-\mathrm{Re} \left[ F(a + i b) \right] \right \} \mathrm{d}\mathbf{r_{0}} \vphantom{\int_{\mathbb{R}^{2}}} \right),
    \end{split}
    \label{ImNonlocalSusceptibility}
\end{align}
\begin{align}
    \begin{split}
        \hspace{-0.3cm} \mathrm{Im} \left[\chi_{\mathrm{D}}(\mathbf{r})\right] =& \; \chi_{0}^{\vphantom{\!_{(2)}}} \left( \vphantom{\left \{ \frac{\mathrm{Im} \left[ F\left(a + i b_{\mathcal{I}}(\mathbf{r_{0}})\right), \right]}{\sqrt{1+\mathcal{I}(\mathbf{r_{0}})/\mathcal{I}_{s}}}-\mathrm{Im} \left[ F(a + i b) \right] \! \right \}} \mathrm{Im}  \left[ F(a + i b) \right] + \bar{\gamma} \int_{\mathbb{R}^{2}} G_{\bar{\gamma}}^{b}(\mathbf{r}\!-\!\mathbf{r_{0}}) \times \right. \\
            & \hspace{3.2cm} \left. \left \{ \frac{\mathrm{Im} \left[ F\left(a + i b_{\mathcal{I}}(\mathbf{r_{0}})\right), \right]}{\sqrt{1+\mathcal{I}(\mathbf{r_{0}})/\mathcal{I}_{s}}}-\mathrm{Im} \left[ F(a + i b) \right] \! \right \} \mathrm{d}\mathbf{r_{0}} \right),
    \end{split}
    \label{ImNonlocalSusceptibility}
\end{align}
\noindent where $\bar{\gamma} = \Gamma/2 + \Gamma_{t}$ (which is the decoherence decay rate) and  $\chi_{0}^{\vphantom{\!_{(2)}}} = \frac{\alpha_{0}(0)}{\omega_{32}/c} \, \frac{\bar{\gamma}}{k u}$. 

\newpage

\noindent For high vapor temperatures (when $ T>450$ K), the transport of excited$\;$atoms$\;$is$\;$diffusive: the free mean path $l_{c}^{g \shortto e}$ becomes smaller than $l_{b}$, as can be seen on~\ref{fig:GreenFunctions} (b). In that case, the evolution of the excited state population is given by the following$\;$diffusion$\;$equation:
\begin{equation}
    \frac{\partial \rho_{33}}{\partial t} - \mathcal{D} \boldsymbol{\nabla}^{2}_{\perp} \rho_{33} + \bar{\gamma} \, \rho_{33} = \frac{\mathcal{I} \, \alpha_{\mathrm{D}}(\mathcal{I})}{\hbar \omega},
\label{ExcitedStateDensityFullEqDiff}    
\end{equation}
\noindent where $\mathcal{D}$ is the diffusion constant. Following the exact same steps as before, we can derive an expression for the steady-state diffusive Green function $G_{\bar{\gamma}}^{d}$ (and its Fourier transform):
\begin{align}
    \begin{cases}
        \bar{G}_{\bar{\gamma}}^{d}(\mathbf{r}\!-\!\mathbf{r_{0}}) =& \frac{1}{4 \pi \mathcal{D}} \int_{0}^{\infty} \frac{1}{t} \, e^{-\bar{\gamma} t} \, e^{-r^{2}/(4 \mathcal{D} t)} \, \mathrm{d}t = \frac{1}{2 \pi \mathcal{D}} \, \mathcal{K}_{0} \! \left( \frac{r}{\sqrt{\mathcal{D} \tau}}\right), \\
        \mathcal{F}\!\left[ \bar{G}_{\bar{\gamma}}^{d} \right]\!(k) =& 1/\left( 1 + \mathcal{D} \tau \, k^{2} \right),
    \end{cases}
\end{align}
\noindent where $\mathcal{K}_{0}$ is the zeroth-order modified Bessel function of the second kind.$\;$The$\;$same$\;$type of response function has been used to describe the nonlocal dielectric response of thermo-optic materials for example$\;$\cite{2-23Vocke}. In order to compare the response functions for ballistic and diffusive$\;$transport, $\bar{G}_{\bar{\gamma}}^{d}$ and $\mathcal{F}\!\left[ \bar{G}_{\bar{\gamma}}^{d} \right]$ have also been plotted as function of $r/l_{d}$ and $k \, l_{d}$ on figure~\ref{fig:GreenFunctions} (a) (we simply assume here that $\sqrt{\mathcal{D} \tau} = l_{c}^{g \shortto e}$).
\vspace{6pt}
\newline
\noindent 
\noindent The nonlocal dielectric response of materials under a laser excitation should be considered in order to correctly describe the dynamics of$\,$a$\,$photon$\,$fluid.$\;$In~\cite{2-24Vocke} for$\,$instance,$\,$the$\,$effects of nonlocality on the dispersion of density waves propagating on a fluid of light is reported. The nonlocality plays also an important role in stabilizing nonlinear phenomena such as transverse solitary waves~\cite{2-25Suter} or the nonlinear propagation of more complex laser fields, such as vortex (Laguerre) and dipole (Hermite) beams\cite{2-20Skupin,2-26Skupin}.






\chapter{Photon fluid in the 2D+1 propagating geometry}

\noindent In the previous chapter, I have described in detail the response of a rubidium$\,$vapor$\,$under a near-resonance laser excitation. However, we have up to now put aside$\,$the$\,$question$\,$of$\,$how such a laser field propagates through the vapor cell. This chapter is dedicated to discussing the analogy between this nonlinear propagation and the evolution of$\,$a$\,$paraxial$\,$photon$\,$fluid. I first introduce the concept of a fluid of light in the propagating geometry$\,$by$\,$deriving$\,$the nonlinear Schr\"{o}dinger equation. In this equation, the electric field of the laser beam can be regarded as a fluid flowing in the plane perpendicular$\,$to$\,$the$\,$propagation$\,$axis,$\,$which$\,$plays the role of a time axis in this hydrodynamical description$\,$of$\,$nonlinear$\,$optics.$\;$The$\,$study$\,$of the propagation of small amplitude density waves travelling onto the photon fluid$\,$provides crucial insights into this many body system.$\;$This$\,$chapter$\,$presents$\,$thus$\,$also$\,$the$\,$theoretical framework required to describe the photon fluids elementary excitations, by introducing, in particular, the Bogoliubov transform$\,$and$\,$the$\,$so-called$\,$Bogoliubov$\,$dispersion$\,$relation.

\newpage

\section{Nonlinear Schr\"{o}dinger equation in optics}

\noindent A fluid of light refers to a weakly interacting gas of photons which is formed$\,$by$\,$a$\,$laser$\,$beam propagating through an optical nonlinear medium. The mean-field dynamics of this many body Bose gas follows a Nonlinear Shr\"{o}dinger Equation (NLSE)~\cite{3-1Boyd,3-2Carusotto}. In this section, I$\,$derive the NLSE using both the slowly varying envelope and the paraxial$\;$approximations. The laser field $\boldsymbol{E}(\boldsymbol{r}, t)$ is monochromatic and linearly polarized and propagates along$\;z\;$in a$\;$Kerr nonlinear medium. Those$\,$materials$\,$exhibit$\,$an$\,$intensity-dependent refractive index, which comes from a non-zero third-order dielectric susceptibility $\chi^{_{(3)}}$. In$\,$chapter$\,$1,$\,$we$\,$have seen how to obtain such an intensity-dependent optical response in hot rubidium vapors, by$\,$tuning$\,$the$\,$frequency$\,$of$\,$a$\,$laser$\,$field$\,$close$\,$to$\,$an$\,$atomic$\,$resonance.$\;$The$\,$following$\,$derivation is general however, since it only requires a non-zero $\chi^{_{(3)}}$ and remains$\,$correct$\,$for$\,$other$\,$sorts of Kerr medium, such as thermo-optic liquids~\cite{2-24Vocke} and photorefractive crystals~\cite{3-3Michel}. 

\subsection{Propagation equation in a non-linear medium}

\noindent Starting from the Maxwell equations, we can show that the electric field $\boldsymbol{E}(\boldsymbol{r}, t)$ evolves according to the well-known nonlinear wave equation:
\begin{equation}
    \boldsymbol{\nabla}^{2} \boldsymbol{E} - \frac{1}{c^{2}} \frac{\partial^{2} \boldsymbol{E}}{\partial t^{2}} = \frac{1}{\epsilon_{0} c^{2}} \frac{\partial^{2} \boldsymbol{P}}{\partial t^{2}},
    \label{NonlinearWaveEquation}
\end{equation}
\noindent where $c$ and $\epsilon_{0}$ are the speed of light and the dielectric permitivity in vacuum,$\,$respectively. The polarization $\boldsymbol{P}$ describes the dielectric response of the material to the field$\;$excitation. This response is not necessarily linear and tends generally to saturate with$\,$the$\,$applied$\,$field. In nonlinear$\;$optics, the polarization is therefore usually expressed as a power series in the field strength (when $\boldsymbol{E}$ is sufficiently weak): 
\begin{align}
    \boldsymbol{P}(\boldsymbol{r}, t) =& \; \epsilon_{0} \, \boldsymbol{\chi}\left[ \boldsymbol{E}(\boldsymbol{r}, t) \right] \boldsymbol{E}(\boldsymbol{r}, t) \\
    =& \; \underbrace{\epsilon_{0} \, \boldsymbol{\chi}^{_{(1)}} \!\cdot\! \boldsymbol{E}(\boldsymbol{r}, t)}_{\boldsymbol{P}^{_{(1)}}(\boldsymbol{r}, t)} + \underbrace{\epsilon_{0} \, \boldsymbol{\chi}^{_{(2)}} \!:\! \boldsymbol{E}(\boldsymbol{r}, t)^2}_{\boldsymbol{P}^{_{(2)}}(\boldsymbol{r}, t)} + \underbrace{\epsilon_{0} \, \boldsymbol{\chi}^{_{(3)}} \!:\! \boldsymbol{E}(\boldsymbol{r}, t)^3}_{\boldsymbol{P}^{_{(3)}}(\boldsymbol{r}, t)} + \dots 
    \label{PolarizationVector}
\end{align}
\noindent The $(n+1)$th-rank tensor $\boldsymbol{\chi}^{_{(n)}}$ describes the $n$th-order susceptibility of the optical medium. The dielectric response of the material is fully characterized by the set $\left\{\boldsymbol{\chi}^{_{(1)}}, \boldsymbol{\chi}^{_{(2)}}, \boldsymbol{\chi}^{_{(3)}}, ... \right\}$ of all the susceptibilities. The $n$th-order polarization $\boldsymbol{P}^{_{(n)}}$ is$\;$linked to the$\;$electric field $\boldsymbol{E}$ by the following tensor product:
\begin{align}
    \begin{split}
        \hspace{-0.2cm}
        P_{j}^{_{(n)}}(\boldsymbol{r}, t) = \epsilon_{0} \sum_{i_{1} \dots i_{n}} \int_{-\infty}^{\infty} \chi_{j \, i_{1} \dots i_{n}}^{_{(n)}}(\boldsymbol{r}\!-\!\boldsymbol{r}_{1}, \dots, \boldsymbol{r}\!-\!\boldsymbol{r}_{n}; t\!-\!t_{1}, \dots, t\!-\!t_{n}) \times \\
            & \hspace{-5.5cm} E_{i_{1}}(\boldsymbol{r}_{1}, t_{1}) \dots E_{i_{n}}(\boldsymbol{r}_{m}, t_{n}) \, \mathrm{d}\mathbf{r}_{1} \dots \mathrm{d}\mathbf{r}_{1} \, \mathrm{d}t_{1} \dots \mathrm{d}t_{n}
    \end{split}
    \label{TensorProduct1}
\end{align}
\noindent The indices $i_{1} \dots i_{n}$ run over the three Cartesian components of the electric field $\boldsymbol{E}$.
\vspace{6pt}
\newline
\noindent In practice, equation~\eqref{TensorProduct1} can be highly simplified by considering:
\begin{itemize}
    \vspace{-6pt}
    \item [$\bullet$] the physical properties of the optical medium (such as the material symmetries);
    \vspace{-6pt}
    \item [$\bullet$] the assumptions made regarding the laser field (polarization, monochromaticity).
\end{itemize}

\newpage

\noindent Rubidium vapors (and more broadly alkaline vapors) are for$\;$instance \textbf{centro-symmetric} optical mediums. All the even-order nonlinear$\;$susceptibilities ($\boldsymbol{\chi^{_{(2)}}}$, $\boldsymbol{\chi^{_{(4)}}}$, $\boldsymbol{\chi^{_{(6)}}}$, ...)$\;$must therefore vanish. Rubidium vapors are also \textbf{isotropic} materials for which the polarization is aligned with the applied field$\;\boldsymbol{E}$.$\,$As the laser field is linearly polarized (let's say,$\;$along$\,x$), the polarization $\boldsymbol{P}$ must only have one non-zero component (along$\;x$). The tensor nature of the nonlinear interaction can thus be left out. Moreover, if the dielectric response of the material is \textbf{local}, we get rid of the integration over spatial coordinates. As we have$\;$seen in the first$\;$chapter$\;$however, ballistic transport of atoms makes the dielectric response of hot rubidium vapors intrinsically nonlocal. For the sake of simplicity, I will keep assuming the medium is local$\;$here; the effects of nonlocality $-$ on the dynamics of the photon fluid elementary$\;$excitations, for$\;$instance $-$ are discussed in section 2.4.2. The expression of the $n$th-order polarization vector is drastically$\,$simplified$\,$as$\,$a$\,$result$\,$of$\,$the $\,$above$\,$discussion: 
\begin{equation}
        \hspace{-0.2cm}
        P^{_{(n)}}(\boldsymbol{r}, t) = \epsilon_{0} \int_{-\infty}^{\infty} \chi^{_{(n)}}(t\!-\!t_{1}, \dots, t\!-\!t_{n}) \times E(\boldsymbol{r}, t_{1}) \dots E(\boldsymbol{r}, t_{n}) \, \mathrm{d}t_{1} \dots \mathrm{d}t_{n}.
    \label{TensorProduct2}
\end{equation}
\noindent Rubidium vapors are \textbf{dispersive} optical mediums; the nonlinear response will therefore depend on the laser frequency $\omega$ and the integration over temporal coordinates remains. By neglecting polarization degrees of freedom, the monochromatic laser field simply reads: $E(\mathbf{r}, t) = \frac{1}{2} \left[\mathcal{E}(\mathbf{r}) \, e^{i \omega t} + \mathcal{E}^{*} (\mathbf{r}) \, e^{-i \omega t} \right]$, where $\mathcal{E}$ is the complex envelope of the electric$\;$field. Replacing $E(\mathbf{r}, t)$ in~\eqref{TensorProduct2} by the foregoing expression and using equations~\eqref{NonlinearWaveEquation} and~\eqref{PolarizationVector} yield the following stationary equation for $\mathcal{E}$: 
\begin{equation}
    \boldsymbol{\nabla}^{2} \mathcal{E}(\mathbf{r}) + \frac{\omega^{2}}{c^{2}} \left[1 + \chi^{_{(1)}}(\omega) \right] \mathcal{E}(\mathbf{r}) = - \frac{3}{4} \frac{\omega^{2}}{c^{2}} \,\chi^{_{(3)}}(\omega) \left|\mathcal{E}(\mathbf{r})\right|^{2} \, \mathcal{E}(\mathbf{r}).
    \label{StationaryWaveEquation}
\end{equation}
In equation~\eqref{StationaryWaveEquation}, only the first- and third-order polarizations are taken into account$\;$while higher order contributions ($P^{_{(5)}}$, $P^{_{(7)}}$,...) are neglected. Moreover, we only keep in$\;P^{_{(3)}}$ the terms oscillating at $+\omega$, since we only concern about nonlinear wave-mixing$\;$processes conserving$\,$the$\,$input$\,$laser$\,$frequency.$\;$That$\,$is$\,$why$\,$the$\,$factor$\,3\,$appears$\,$on$\,$the$\,$left$\,$hand$\,$side$\,$of equation~\eqref{StationaryWaveEquation}, as $(^{3}_{2}) = 3$ is the number of four-wave mixing mechanisms producing$\,$a$\,$3rd- order polarization oscillating at $+\omega$~\cite{3-1Boyd}. So$\,$as$\,$to$\,$rewrite$\,$equation$\,$\eqref{StationaryWaveEquation}$\,$more$\,$aesthetically, one usually defines the linear dielectric permitivity of the material $\epsilon_{r}(\omega) = 1 + \chi^{_{(1)}}(\omega)$:
\begin{equation}
    \boldsymbol{\nabla}^{2} \mathcal{E}(\mathbf{r}) + k_{0}^{2} \, \epsilon_{r}(\omega) \, \mathcal{E}(\mathbf{r}) = - \frac{3}{4} \frac{\omega^{2}}{c^{2}} \,\chi^{_{(3)}}(\omega) \left|\mathcal{E}(\mathbf{r})\right|^{2} \, \mathcal{E}(\mathbf{r}).
    \label{StationaryWaveEquation2}
\end{equation}
where $k_{0} = \omega/c$ stands for the laser wave-vector in vacuum. Let's also introduce the linear refractive index $n_{0}(\omega) = \sqrt{\mathrm{Re}\left[\epsilon_{r}(\omega) \right]}$ and absorption coefficient $\alpha(\omega) = k_{0} \, \mathrm{Im} \left[\epsilon_{r}(\omega) \right]/n_{0}$.

\subsection{Paraxial approximation for the slowly-varying field envelope}

\noindent Equation~\eqref{StationaryWaveEquation2} describes the evolution of the envelope of a monochromatic and linearly polarized laser field inside a Kerr-type medium. This equation is general as no assumption has been made on the envelope so far. In practice however, the wave propagation is often limited to within a small angle from the optical axis, defined by the $z$-direction let's say, along which the beam propagates. In that case, the field amplitude $\mathcal{E}(\mathbf{r}_{\perp}, z)$ slowly varies in the transverse$\;$plane (that is, on the plane perpendicular to the $z$-axis) and the so-called \textbf{paraxial approximation} can be performed. The field envelope reads thus as follows: 
\begin{equation}
    \mathcal{E}(\mathbf{r_{\perp}}, z) = \mathcal{E}_{0}(\mathbf{r_{\perp}}, z) \, e^{i k(\omega) z},
    \label{SlowVarying}
\end{equation}

\newpage

\noindent where $\mathcal{E}_{0}(\mathbf{r_{\perp}}, z)$ is a \textbf{slowly-varying} function of $z$, or, in other words, a function varying$\;$on a length scale much larger than the optical wavelength $\lambda$. The paraxial approximation is valid as long as $| \boldsymbol{\nabla}_{\perp}^{2} \mathcal{E}_{0} |/k^{2} \sim \left| \partial_{z} \, \mathcal{E}_{0} \right|/k \ll 1$, where $k(\omega) = n_{0} \, k_{0}$ is the laser wave-vector in the nonlinear medium. When this condition is fulfilled, the second-order derivative $\partial_{z}^{2} \mathcal{E}_{0}$ can be neglected in the equation describing the evolution of the slow-varying envelope $\mathcal{E}_{0}$, which finally takes the form of a nonlinear Schr\"{o}dinger equation (NLSE):
\begin{equation}
    i \, \partial_{z} \, \mathcal{E}_{0}(\mathbf{r_{\perp}}, z) = \left[-\frac{1}{2 k} \boldsymbol{\nabla}_{\perp}^{2} - \frac{i \alpha}{2} - \frac{3}{8} \frac{k}{n_{0}^{2}} \, \chi^{_{(3)}}(\omega) \left| \mathcal{E}_{0} \left(\mathbf{r_{\perp}}, z \right) \right|^{2} \, \right]  \mathcal{E}_{0}(\mathbf{r_{\perp}}, z),
    \label{NLSE}
\end{equation}
\noindent where $\boldsymbol{\nabla}_{\perp}$ is the gradient with respect to the transverse spatial coordinates, $\mathbf{r_{\perp}} = (x, y)$. For the sake of completeness, we can also take into account the effect of a local modulation of the linear refractive index, $\delta n$, on the propagation of the slowly-varying field envelope. Such a modulation can either act as a repulsive$\,$obstacle$\,$for$\,$the$\,$light$\,$beam$\,$(if$\,\delta n\,$is$\,$negative) or as a wave-guide (if $\delta n$ is positive). In our system, it can be$\,$optically$\,$generated$\,$by$\,$locally driving another rubidium transition, using a second laser field tuned close to resonance. This situation is extensively investigated in chapter 5. Including $\delta n$ in the equation~\ref{NLSE} is straightforward~\cite{3-2Carusotto} and finally yields: 
\begin{equation}
     i \, \partial_{z} \, \mathcal{E}_{0}(\mathbf{r_{\perp}}, z) = \left[-\frac{1}{2 k} \boldsymbol{\nabla}_{\perp}^{2} - \frac{i \alpha}{2} - k \, \frac{\delta n(\mathbf{r_{\perp}}, z)}{n_{0}} - \frac{3}{8} \frac{k}{n_{0}^{2}} \, \chi^{_{(3)}}(\omega) \left| \mathcal{E}_{0} \left(\mathbf{r_{\perp}}, z \right) \right|^{2} \, \right]  \mathcal{E}_{0}(\mathbf{r_{\perp}}, z).
    \label{FullNLSE}
\end{equation}

\subsection{Comparison with the Gross-Pitaevskii equation}

\noindent If linear absorption is negligible ($\alpha \simeq 0$), that is, if the system is conservative, the$\;$NLSE is mathematically$\,$analogous$\,$to$\,$the$\,$Gross–Pitaevskii$\,$equation$\,$(GPE).$\;$This$\,$equation$\,$describes for instance the space-time evolution of the macroscopic wave-function $\Psi(\mathbf{r}, t)$ of a dilute atomic Bose-Einstein condensate (BEC) in the Hartree-Fock approximation:
\begin{equation}
    i \hbar \, \partial_{t} \, \Psi(\mathbf{r}, t) = \left[ -\frac{\hbar^{2}}{2 \, m} \boldsymbol{\nabla}^{2} + \mathcal{V}(\mathbf{r}) + g \left| \Psi(\mathbf{r}, t) \right|^{2} \right] \Psi(\mathbf{r}, t), 
    \label{GPE}
\end{equation}
\noindent where $\hbar$ is the reduced Planck constant, $m$ the boson mass, $\mathcal{V}$ an external potential and $g = 4\pi \hbar^{2} a_{s}/m$ the coupling constant, proportional to the s-wave scattering$\;$length $a_{s}$~\cite{3-4Pitaevskii}. 

\subsubsection{Space-time mapping}

As$\;$you may have noticed, equations~\eqref{NLSE} and~\eqref{GPE} are indeed pretty similar.$\;$However, while the GPE describes the evolution of the wave-function $\Psi$ over the real$\;$time, the$\;$NLSE describes how the electric field envelope $\mathcal{E}_{0}$ propagates in space, along the optical axis. Therefore, the $z$-direction plays the role of an effective time $\tau$ in the NLSE so that$\;$every transverse plane along the optical axis can be regarded as a snapshot of the nonlinear "time$\;$evolution" of the laser beam inside the medium. This seemingly$\;$elementary$\;$space-time mapping $z \leftrightarrow \tau = z n_{0}/c $ has profound consequences when one tries to build$\,$from~\eqref{NLSE} a fully quantum field theory~\cite{3-2Carusotto, 3-5Larre}.$\;$Once the $z$-direction$\,$has$\,$been$\,$mapped$\,$into$\,$a$\,$time$\,$axis, the only difference remaining between the NLSE and the GPE lies in their$\,$dimensionality. While the GPE describes the time evolution of the condensate wave-function in the three dimensions of space (3D+1 geometry), the "time evolution" of $\mathcal{E}_{0}$ intrinsically involves$\;$only two$\;$spatial$\;$dimensions$\;$defining$\;$the$\;$transverse$\;$plane$\;$(2D+1$\;$geometry).

\subsubsection{Effective mass and coupling constant}

In order to complete the analogy between NLSE and GPE, let's derive an expression$\;$for the mass and the coupling constant which characterize the weakly interacting photon gas formed by the laser field inside the Kerr medium. We first define the normalized envelope:
\begin{equation}
    \overline{\mathcal{E}}_{0}(\mathbf{r_{\perp}}, z) = \mathcal{E}_{0}(\mathbf{r_{\perp}}, z)\bigg/\left[ \int_{\mathcal{S}} \left| \mathcal{E}_{0}(\mathbf{r_{\perp}}, z) \right|^{2} \mathrm{d} \mathbf{r}_{\perp}\right]^{_{\frac{1}{2}}},  
    \label{Normalization}
\end{equation}
\noindent where$\,\mathcal{S}\,$stands$\,$for$\,$the$\,$surface$\,$of$\,$the$\,$medium$\,$cross-section.$\;$The$\,$amplitude$\,$square$\,$of$\,\overline{\mathcal{E}}_{0}\,$is$\,$the electric field density. By definition, its integral over $\mathcal{S}$ is one.$\;$The$\,$integral$\,$in$\,$equation$\,$\eqref{Normalization} is a conserved quantity, as it is proportional to the laser field input$\;$power$\,\mathcal{P}_{0}$ whatever the position $z$ on the optical axis (as long as linear absorption is zero). Therefore, $\mathcal{E}_{0}$ can be replaced by $\overline{\mathcal{E}}_{0}$ in~\eqref{NLSE} and using the space-time mapping yields: 
\begin{equation}
    i\hbar \, \partial_{\tau} \, \overline{\mathcal{E}}_{0}(\mathbf{r_{\perp}}, \tau) = \left[-\frac{\hbar^{2}}{2 (\hbar k/c)} \, \boldsymbol{\nabla}_{\perp}^{2} - \hbar \omega \, \delta n(\mathbf{r_{\perp}}, z) - \hbar \omega \, n_{2} \mathcal{P}_{0} \left| \overline{\mathcal{E}}_{0} \left(\mathbf{r_{\perp}}, z \right) \right|^{2} \, \right]  \overline{\mathcal{E}}_{0}(\mathbf{r_{\perp}}, z).
    \label{NLSETemporal}
\end{equation}
\noindent The nonlinear refractive index $n_{2}$ is defined by: $n_{2} = 2 \, \Tilde{n}_{2} / (c \, \epsilon_{0} \, n_{0})$ with $\Tilde{n}_{2} = 3 \, \chi^{_{(3)}} / (8 \, n_{0})$. One can then readily identify the effective mass $\overline{m}$ and coupling constant $\overline{g}$ in~\eqref{NLSETemporal}:
\begin{equation}
\overline{m} = \hbar \, k /c \hspace{0.25cm} \mathrm{and} \hspace{0.25cm} \overline{g} = - \hbar \omega \, n_{2} \mathcal{P}_{0}, 
\end{equation}
where $2 \, \mathcal{P}_{0}/(c \, \epsilon_{0} \, n_{0}) \!=\! \int_{\mathcal{S}} \left| \mathcal{E}_{0}(\mathbf{r_{\perp}}, z) \right|^{2} \mathrm{d} \mathbf{r}_{\perp}$. A focusing (resp. defocusing) Kerr$\;$nonlinearity, for which $n_{2} > 0$ (resp. $n_{2} < 0$), can therefore be regarded in this analogy as an$\;$attractive (resp. repulsive) photon-photon interaction, mediated by the atomic ensemble in$\;$our$\;$case. The nonlinear refractive index $n_{2}$ plays thus a crucial role in photon fluid physics since it controls the strength and nature (either attractive or repulsive) of the effective interaction between photons. Throughout my thesis, I$\,$only$\,$dealt$\,$with$\,$defocusing$\,$nonlinearities,$\,$which make the photon gas stable against modulational$\,$instabilities~\cite{3-6Yuen,3-7Agrawal}.$\;$Let's$\,$finally$\,$mention that the index modulation $\delta n$ in~\eqref{NLSE} acts as an external potential on the paraxial photons:
\begin{equation}
    \overline{\mathcal{V}}(\mathbf{r}_{\perp}) = -\hbar\omega \, \delta n(\mathbf{r}_{\perp}, z)    
\end{equation}
which is either attractive ($\delta n > 0$) or repulsive ($\delta n < 0$) depending on the sign of $\delta n(\mathbf{r}_{\perp},z)$. We can then think about trapping the photon gas in the $(x, y)$ plane, in order to observe the optical analog of breathers in 2-dimensional BECs~\cite{3-8Saint} for example, or about studying the dynamics of the interacting photon gas in disordered potentials.
\vspace{6pt}
\newline
\noindent It might be interesting to compare the typical values of the coupling constant obtained in 2-dimensional BECs and in interacting photon gas. To that end, we define $N \Tilde{g} = 2 \, \overline{m} \, \overline{g} /\hbar^{2}$ which is an adimensional quantity introduced usually in 2D$\;$BECs to evaluate the strength of$\,$the$\,$nonlinear$\,$interactions.$\;$In$\,$that$\,$case,$\,N\,$stands$\,$for$\,$the$\,$number$\,$of$\,$bosons$\,$inside$\,$the$\,$BEC. Regarding interacting photon gas, what makes sense is$\,$not$\,$the$\,$number$\,$of$\,$photons$\,$inside$\,$the Kerr medium but rather the \textbf{flux of photons}, $\Phi = \mathcal{P}_{0}/\hbar\omega$, through the$\,$cell$\,$entrance$\,$plane ($\hbar \omega$ being the energy of a single photon).  When $n_{2} = -5 \!\times\! 10^{-11}\,$m$^{2}$/W and $\mathcal{P}_{0} = 500 \, \mathrm{mW}$  $-$ which are typical experimental values $-$ $N\Tilde{g}$ is equal to $3.2 \!\times\! 10^{3}$. In 2D atomic BECs, $N\Tilde{g}$ is of the same order of magnitude. It reaches $4 \!\times\! 10^{3}$ in~\cite{3-8Saint} for$\,$instance.$\;$Tuning$\,$the$\,$value$\,$of the product $N\Tilde{g}$ in 2D atomic BEC experiments requires to change the intensity $I$ of the laser beams confining the 2D condensate in the desired direction (since$\,\Tilde{g}\,$scales$\,$as$\,I^{_{1\!/\!4}}\,$\cite{3-8Saint}).

\newpage

\noindent In photon gas experiments, we have control over both the flux of photons $\Phi$, by$\,$tuning$\,$the laser power $\mathcal{P}_{0}$, and over the nonlinear index$\,$of$\,$refraction$\,n_{2}$,$\,$by$\,$tuning$\,$the$\,$laser$\,$frequency$\,\omega$ in rubidium vapors$\,$for$\,$instance.$\;$It$\,$seems$\,$therefore$\,$that$\,$photon$\,$gas$\,$in$\,$propagating$\,$geometry are versatile and highly tunable systems to investigate the optical counterpart of nonlinear many-body phenomena$\,$arising$\,$in$\,$2-dimensional$\,$BECs.$\;$Nevertheless,$\,$it$\,$is$\,$worth$\,$mentioning that the nonlinear "time evolution" of the photon gas in this analogy is intrinsically limited by the length $L$ of the Kerr medium. This may potentially prevent us from observing nonlinear phenomena which establish on time-scales longer than $L n_{0}/c$.

\subsubsection{Kinetic, interaction and potential energy}

\noindent It might be useful to introduce the kinetic, the interaction and the potential energies of the interacting photon gas, which are respectively defined as follows: 
\begin{align}
    E_{\mathrm{kin}}\left[\overline{\mathcal{E}}_{0}\right] =& \; \frac{\hbar^{2}}{4 \, \overline{m}} \int_{\mathcal{S}} \left|\boldsymbol{\nabla}_{\perp} \overline{\mathcal{E}}_{0} \right|^{_{2}} \mathrm{d} \mathbf{r_{\perp}} \\ 
    E_{\mathrm{int}}\left[\overline{\mathcal{E}}_{0}\right] =& \; \frac{\overline{g}}{4} \int_{\mathcal{S}} \left|\overline{\mathcal{E}}_{0}(\mathbf{r_{\perp}})\right|^{_{4}} \mathrm{d} \mathbf{r_{\perp}} \\ 
   \hspace{0.25 cm} E_{\mathrm{pot}}\left[\overline{\mathcal{E}}_{0}\right] =& \; \frac{1}{2} \int_{\mathcal{S}} \overline{V}(\mathbf{r}_{\perp}) \left| \overline{\mathcal{E}}_{0} \right|^{_{2}} \mathrm{d} \mathbf{r_{\perp}}  
\end{align}
\noindent When the system is conservative (no loss, $\alpha = 0$), the total energy $E_{\mathrm{tot}} = E_{\mathrm{kin}}+ E_{\mathrm{int}}+E_{\mathrm{pot}}$ is conserved during the evolution. One retrieves the left hand side of equation~\eqref{NLSETemporal} by computing the functional derivative $\delta E_{\mathrm{tot}}/ \delta \overline{\mathcal{E}}_{0}$, as expected. In this$\,$manuscript,$\,$I$\,$almost exclusively$\,$study$\,$situations$\,$for$\,$which$\,E_{\mathrm{int}} \gg E_{\mathrm{kin}}$.$\,$This$\,$latter$\,$condition$\,$defines$\,$the$\,$so-called \textbf{Thomas-Fermi} or \textbf{hydrodynamic regime}~\cite{3-9Dalfovo}, in which the envelope$\,$of$\,$the$\,$electric$\,$field behaves as a 2-dimensional photon fluid in the transverse plane.      


\section{Hydrodynamic analogy}

\noindent The analogy between NLSE and GPE indicates that it is somehow$\;$possible to describe a laser beam propagating in a Kerr medium as a fluid of light flowing in the transverse$\;$plane. In this section, I will explicitly transpose the NLSE into a set of hydrodynamic equations, using the Madelung transform~\cite{3-10Madelung}. In addition to$\,$being$\,$highly$\,$aesthetic,$\,$the$\,$hydrodynamic formulation of the NLSE provides an easy understanding about the optical counterparts of a broad range of classical fluid phenomena, such as, for instance, the transition from laminar to turbulent flow~\cite{3-12Turitsyna}, undular bores~\cite{3-11Fatome} or Rayleigh-Taylor instabilities~\cite{3-13Jia}.  

    \subsection{Madelung transform}
    
\noindent The \textbf{Madelung transform} enables one to express the electric field envelope $\mathcal{E}_{0}$ as$\;$function of its non-normalized density $\rho$ and phase $\Phi$ as follows:
\begin{equation}
    \mathcal{E}_{0}(\mathbf{r_{\perp}}, z) = \sqrt{\rho(\mathbf{r_{\perp}}, z)} \, e^{i \Phi(\mathbf{r_{\perp}}, z)}
    \label{MadelungTransform}
\end{equation}
\noindent The density $\rho = \left|\mathcal{E}_{0}\right|^{_{2}}$ is proportional to the laser$\;$intensity $\mathcal{I}_{0}$. Using$\,$\eqref{MadelungTransform}, equation$\,$\eqref{NLSE} yields the following set of hydrodynamic equations for $\rho\;$and$\;\Phi$: 

\newpage

\begin{align}
\label{Continuity}
    \frac{\partial \rho}{\partial \tau} + \boldsymbol{\nabla}_{\perp}\cdot \left( \rho \boldsymbol{v} \right) + \Tilde{\alpha} \rho \; = & \; 0  \\
\label{Euler}    
    \frac{c}{n_{0} k} \frac{\partial \Phi}{\partial \tau} + \frac{1}{2} v^{2} - \frac{c^{2}}{n_{0}^{2}} \!  \left(\frac{n_{2}}{n_{0}} \rho + \frac{1}{2 k^2} \frac{\boldsymbol{\nabla}_{\perp}^2 \sqrt{\rho}}{\sqrt{\rho}}\right)\! \; = & \; 0 
\end{align}
\noindent where $\Tilde{\alpha} = \alpha \, c / n_{0}$. The effective time is still defined by $\tau = z \, n_{0}/c$ and the nonlinear$\;$index of refraction by $\Tilde{n}_{2} = 3 \, \chi^{_{(3)}} / (8 \, n_{0})$. As you may have noticed, equations~\eqref{Continuity} and~\eqref{Euler} look respectively like the continuity and Euler equations, describing how an$\,$incompressible fluid of density $\rho(\mathbf{r}_{\perp}, \tau)$ locally flows in the plane $(x, y)$ at a velocity $\boldsymbol{v}(\mathbf{r}_{\perp}, \tau) = \frac{c}{n_{0} \, k} \boldsymbol{\nabla}_{\perp} \Phi$.
\begin{itemize}
    \vspace{-6pt}
    \item [$\bullet$] The continuity equation$\,$\eqref{Continuity} refers to the non-conservation$\,$of$\,$the$\,$mass$\,$so$\,$to$\,$speak, because of linear losses. The fluid density decays exponentially during propagation, as expected from the Beer-Lambert law: $\rho(\mathbf{r}_{\perp}, \tau) = \rho(\mathbf{r}_{\perp}, 0) \, \exp(-\Tilde{\alpha} \tau)$.
    \vspace{-6pt}
    \item[$\bullet$] The second term inside the bracket on$\,$the$\,$right$\,$hand-side$\,$of$\,$the$\,$Euler$\,$equation~\eqref{Euler} is the so-called quantum pressure~\cite{3-9Dalfovo} and$\,$does$\,$not$\,$have$\,$any$\,$counterpart$\,$in$\,$real$\,$fluids. It opposes any stretching or contraction of the fluid over distance smaller than the healing length $\xi$, defined in subsection 2.3.2. As long as the density is slowly varying in the transverse plane, the quantum pressure can be neglected; reversely, it starts dominating the dynamics in regions of rapidly changing density.

\end{itemize}
\noindent Seen from the hydrodynamic analogy perspective, the propagation geometry appears$\;$to be a simple and straightforward implementation of a photon fluid. The initial fluid density and its flow velocity can be easily tuned controlling the transverse intensity$\;$distribution and the spatial phase profile of the incident laser beam at the medium entrance plane (using a Spatial Light Modulator (SLM) for instance).

    \subsection{Speed of sound}

\noindent In liquids, the sound consists of compression waves whose speed $-$ the$\,$sound$\,$velocity$\,c_{s} \, -$ depends$\,$on$\,$the$\,$fluid$\,$compressibility$\,$and$\,$density.$\;$According$\,$to$\,$the Newton-Laplace$\,$formula, $\,c_{s} = \sqrt{\mathcal{K}/\rho}\,$\cite{3-14Rayleigh},$\,\mathcal{K}\,$being$\,$the$\,$bulk$\,$modulus$\,$of$\,$the$\,$fluid.$\;$This$\,$quantity$\,$measures$\,$how$\,$resistant to compression the liquid is. It is basically defined as the ratio of the infinitesimal pressure increase to the resulting relative decrease of the volume. In$\,$other$\,$words,$\,$if$\,P\,$stands for the pressure inside the fluid, the$\,$bulk$\,$modulus$\,\mathcal{K}\,$reads:$\;\mathcal{K} = \rho \, \partial P/ \partial \rho$.$\;$In$\,$paraxial$\,$photon$\,$fluids, the repulsive interactions between photons create a local bulk pressure $P = n_{2} \, c^{2} \rho^{2}/ (2 \, n_{0}^{2})$. This$\,$pressure$\,$is$\,$related$\,-\,$through$\,$the$\,$Newton-Laplace$\,$formula$\,-\,$to$\,$the$\,$\textbf{speed$\,$of$\,$sound}: 
\begin{equation}
    c_{s}^{2} = \frac{\partial P}{\partial \rho} =  \left(\frac{c}{n_{0}}\right)^{_{2}} \frac{\Delta n}{n_{0}}
    \label{SpeedSound}
\end{equation}    
\noindent $\Delta n = \Tilde{n}_{2} \, \rho = n_{2} \, \mathcal{I}_{0}$ being the nonlinear change of refractive index and $\mathcal{I}_{0}$ the laser$\;$intensity.
The fact that there is a well defined sound velocity in propagating photon fluids necessarily implies that density modulations$\,$in$\,$those$\,$system$\,$behave$\,$as$\,$sound-like$\,$collective$\,$excitations, under certain conditions. Phonons in crystals also propagate at a$\;$given$\;$speed$\;$of$\;$sound$\;c_{\mathrm{ph}}$, independently of their wavelength $\Lambda$, as long as the latter is larger than the microscopic details$\;$of the lattice. In other words, the group velocity $v_{g}(k) = \partial \omega_{\mathrm{ph}}/\partial k\;$does$\;$not$\;$depend on the phonons wave-vector $k = 2\pi/\Lambda$ at long wavelengths: $v_{g}(k) \simeq c_{\mathrm{ph}}$. This results in$\;$a linear dependence of the phonons frequency $\omega_{\mathrm{ph}}(k)$ on the wave-vector $k$: $\omega_{ph}(k) \simeq c_{\mathrm{ph}} \, k$. The same kind of linear dispersion relation at low wave-vectors is expected for density waves in propagating photon fluids. In the next section, I will precisely derive a$\;$formula for this dispersion relation $-$ the so-called \textbf{Bogoliubov dispersion relation} $-$ which$\;$indeed scales linearly at low values of the wave-vector.

\section{Bogoliubov dispersion in a lossless local medium}

\noindent In most cases, even if the Madelung formulation of the NLSE is highly aesthetic and$\;$gives a physical insight into the hydrodynamical nature of the dynamics, the coupled system of equations~\eqref{Continuity} and~\eqref{Euler} cannot be solved analytically. It is then of particular interest to study how small density modulations propagate on top of an uniform background$\;$fluid. In that case, the continuity and the Euler equations can be linearized. Using the so-called Bogoliubov transform, I will show that density waves in propagating photon fluids obey the well known Bogoliubov dispersion relation. The first clear experimental observation of this dispersion in fluids of light is reported in the next chapter.

\subsection{Derivation from the Euler's equations}
\label{subsec:derivationEuler}

\noindent In the following section, equations \eqref{Continuity} and~\eqref{Euler} will be directly expressed as function of the propagation distance $z$ (while bearing in mind the mapping $z \leftrightarrow \tau$):     
\begin{align}
\label{Continuity2}
    \frac{\partial \rho}{\partial z} + \frac{1}{k} \boldsymbol{\nabla}_{\perp} \! \cdot \! \left( \rho \boldsymbol{\nabla}_{\perp} \Phi \right) + \alpha \rho \; = & \; 0 \,  \\
\label{Euler2}    
    \frac{\partial \Phi}{\partial z} + \frac{1}{2 k} \left( \boldsymbol{\nabla}_{\perp} \Phi \right)^{2} - k \frac{n_{2}}{n_{0}} \rho - \frac{1}{2 k} \frac{\boldsymbol{\nabla}_{\perp}^2 \sqrt{\rho}}{\sqrt{\rho}} \! \; = & \; 0 \, 
\end{align}
When small amplitude density waves propagate over an uniform background fluid at$\;$rest, the density and phase of the overall system can be expressed as follows:
\begin{align}
    \rho(\mathbf{r_{\perp}}, z) =& \; \rho_{0}(z) + \delta \rho(\mathbf{r_{\perp}}, z) \\
    \Phi(\mathbf{r_{\perp}}, z) =& \; \Phi_{0}(z) \!+\! \delta \Phi(\mathbf{r_{\perp}}, z)
\end{align}
\noindent where $\delta \rho \ll \rho_{0}$ and $\delta \Phi \ll \Phi_{0}$ respectively. At the zeroth-order, equations~\eqref{Continuity2} and~\eqref{Euler2} simply reads: $\frac{d \rho_{0}}{d z} - \alpha \rho_{0}$ (i) and $\frac{d \Phi_{0}}{d z} = \, k_{0} n_{2} \rho_{0}$ (ii) ($k_{0}$ is$\,$the$\,$laser$\,$wave-vector$\,$in$\,$vacuum). The first equation (i) accounts for the exponential decay of the background density$\,$because of linear losses: $\rho_{0}(z) = \rho_{0}(0) e^{-\alpha z}$. The second equation (ii) describes$\;$the$\;$evolution$\;$of$\;$the phase accumulated by the background field envelope along propagation.$\,$In a Kerr$\;$medium, the refractive index depends on the laser intensity. We thus expect the phase accumulated by the background fluid at a distance $z$ from the entrance plane to be equal to $k_{0} \, \Delta n \, z$, where $\Delta n = n_{2} \, \rho$ is the nonlinear change of refractive index. As $\Delta n$ depends here on the propagation distance because of linear losses, the self-induced zeroth-order phase shift $\Phi_{0}$ is then finally given by: $\Phi_{0}(z) = k_{0} \langle \Delta n(z') \rangle_{z} \, z$, where $\langle \Delta n(z') \rangle_{z} = \frac{1}{z} \int_{0}^{z} \Delta n (z') \mathrm{d}z'$ is the average of $\Delta n(z)$ over $z$. You might be surprised by the fact that there is no contribution in $\Phi_{0}$ accounting for the linear phase accumulated by the laser beam along propagation. 
Equation~\eqref{SlowVarying} provides an explanation for this: by introducing the slow-varying envelope, we choose to describe the physics from the translating frame at $c/n_{0}$ so to speak, in which no linear phase is accumulated, by definition. 

\newpage

\noindent At the first perturbation order in $\delta \rho$ and $\delta \Phi$, equations~\eqref{Continuity2} and~\eqref{Euler2} read:
\begin{align}
\label{ContinuityFirstOrder}
    \frac{\partial \delta \rho}{\partial z} + \frac{\rho_{0}}{k} \boldsymbol{\nabla}_{\perp}^{2} \delta \Phi + \alpha \delta \rho \; = & \; 0 \\
\label{EulerFirstOrder}    
    \frac{\partial \delta \Phi}{\partial z}  - k \frac{n_{2}}{n_{0}} \delta \rho - \frac{1}{4 k} \frac{\boldsymbol{\nabla}_{\perp}^2 \delta \rho}{\rho_{0}} \! \; = & \; 0 
\end{align}
\noindent Drawing on the second-quantization protocol of the Bose field operator one usually carries out in dilute BECs~\cite{3-15Petrov}, we rewrite $\delta \rho$ and $\delta \Phi$ as follows:
\begin{align}
\label{FourierDensity}
    \delta \rho (\mathbf{r_{\perp}}, z) =& \, \sqrt{\rho_{0}} \int \frac{\mathrm{d} \mathbf{k_{\perp}}}{(2 \pi)^{2}} \left[ a(\mathbf{k_{\perp}}) f_{+}(\mathbf{k_{\perp}}, z) e^{-i \mathbf{k_{\perp}} \!\cdot\! \mathbf{r_{\perp}}} + \bar{a}(\mathbf{k_{\perp}}) f_{+}^{*}(\mathbf{k_{\perp}}, z) e^{i \mathbf{k_{\perp}} \!\cdot\! \mathbf{r_{\perp}}} \right] \\
\label{FourierPhase}    
    \delta \Phi (\mathbf{r_{\perp}}, z) =& \, \frac{1}{2 i \sqrt{\rho_{0}}} \int \frac{\mathrm{d} \mathbf{k_{\perp}}}{(2 \pi)^{2}} \left[ a(\mathbf{k_{\perp}}) f_{-}(\mathbf{k_{\perp}}, z) e^{-i \mathbf{k_{\perp}} \!\cdot\! \mathbf{r_{\perp}}} - \bar{a}(\mathbf{k_{\perp}}) f_{-}^{*}(\mathbf{k_{\perp}}, z) e^{i \mathbf{k_{\perp}} \!\cdot\! \mathbf{r_{\perp}}} \right]
\end{align}
\noindent By$\,$reinstating$\,$the$\,$foregoing$\,$expressions$\,$of$\,\delta \rho\,$and$\,\delta \Phi\,$in~\eqref{ContinuityFirstOrder}$\,$and~\eqref{EulerFirstOrder},$\,$one$\,$finally$\,$obtains a Bogoliubov-de Gennes matrix equation on the Fourier amplitudes $f_{+}$ and $f_{-}$:
\begin{equation}
    \label{BogEq}
    i \frac{\partial}{\partial z}
    \begin{pmatrix} 
    f_{+} \\
    f_{-}  
    \end{pmatrix}
    = -\left[ i\frac{\alpha}{2} +  \mathcal{H}_{\mathbf{k_{\perp}}} \right] \!
    \begin{pmatrix} 
    f_{+} \\
    f_{-}  
    \end{pmatrix} \!, \; \mathrm{where} \;
    \mathcal{H}_{\mathbf{k_{\perp}}} = 
    \begin{pmatrix} 
    0 & -\frac{k_{\perp}^{2}}{2 k} \\
    -\frac{k_{\perp}^{2}}{2 k} + 2 k_{0} \Delta n & 0
    \end{pmatrix}
\end{equation}

    \subsection{Dispersion relation}
    
\begin{figure}[h]
\center
\includegraphics[width=0.95\linewidth]{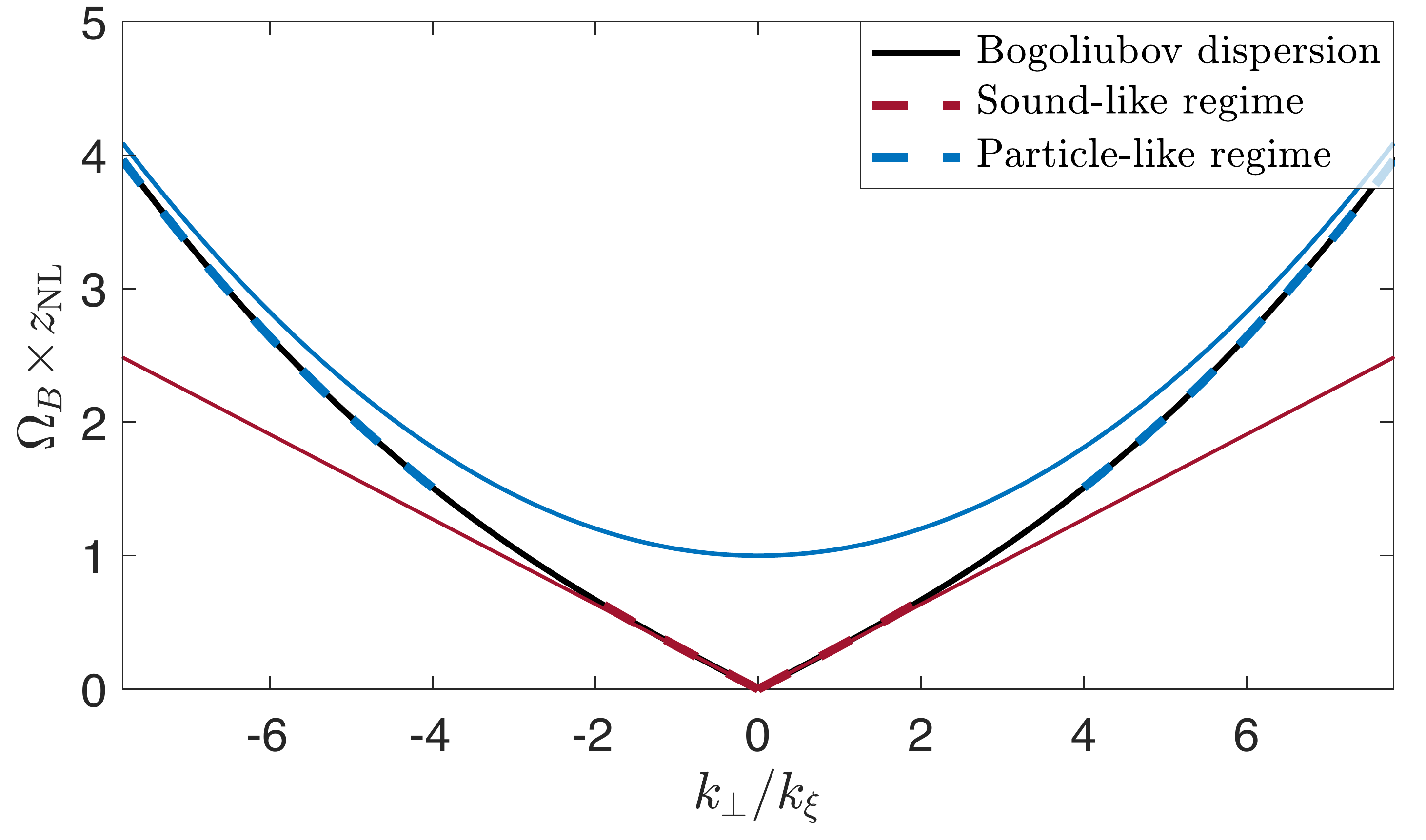} 
\caption{Bogoliubov dispersion relation (black solid line) obtained for a fluid at rest. The asymptotic sound- and particle-like regimes have been plotted in red and$\;$blue$\;$solid. The transverse wave-vector has been scaled by $k_{\xi} = 2\pi/\xi$ and $\Omega_{B}$ by $1/z_{\mathrm{NL}} = 1/ k_{0} |\Delta n$|.}
\label{fig:BogDispRegimes}
\end{figure}

\noindent We first assume that $\alpha = 0$. In this ideal lossless situation, the$\,$matrix$\,\mathcal{H}_{\mathbf{k_{\perp}}}\,$is$\,$homogeneous. The Fourier components $f_{\pm}$ of the density and phase fluctuations$\,$are$\,$therefore$\,$plane$\,$waves, whose$\,$amplitude$\,$and$\,$longitudinal$\,$wave-vector$\,$depend$\,$on$\,k_{\perp}$:$\,f_{\pm} (\mathbf{k_{\perp}}, z) = f_{\pm}(\mathbf{k_{\perp}}) e^{i \Omega_{B}(\mathbf{k_{\perp}}) z}$. Equation \eqref{BogEq} reduces finally to an eigenvalue equation whose solutions, the branches$\,$of the \textbf{Bogoliubov$\,$dispersion$\,$relation}$\,\Omega_{B}(\mathbf{k_{\perp}})$,$\,$are$\,$found$\,$by$\,$diagonalizing$\,$the$\,$matrix$\,\mathcal{H}_{\mathbf{k_{\perp}}}$: \begin{equation}
    \label{BogDisp}
    \Omega_{B}(\mathbf{k}_{\perp})-\mathbf{k_{\perp}} \!\cdot\! \boldsymbol{v} = \pm \sqrt{-\frac{n_{2}}{n_{0}} \, \rho_{0} \, k_{\perp}^{2} + \frac{k_{\perp}^4}{4 k^{2}}}.
\end{equation}
For defocusing Kerr nonlinearity ($n_{2} < 0$), $\Omega_{B}(k_{\perp})$ is a real function of $k_{\perp}$, which$\,$stands$\,$for the transverse wave-vector of the plane wave density modulation. The dispersion relation links usually the wave frequency to the wave-vector. In propagating photon fluids however, $\Omega_{B}(k_{\perp})$ is an inverse length, since the system evolves along $z$ and not over time as usual. Similarly, the sound velocity $c_{s}$, the background transverse speed $\boldsymbol{v}$ or the density wave group velocity $\boldsymbol{v_{g}}$ are measured in adimensional units, as they have the physical meaning of propagation angles with respect to the $z$-axis. Nevertheless, $\Omega_{B}(k_{\perp})$ can be regarded$\;$as a frequency through the $z \leftrightarrow \tau$ mapping. The relation \eqref{BogDisp} describes then the response frequency of the fluid of light to a small density fluctuation $\delta \rho$ whose  wave-vector is $\mathbf{k}_{\perp}$. The term ${\mathbf{k_{\perp}} \!\cdot\! \boldsymbol{v}}$ describes the shift in this response frequency because of the Doppler effect, when the background fluid is not at rest anymore.
\vspace{6pt}
\newline
\noindent When the wavelength of the modulation $\Lambda = 2 \pi/k_{\perp}$ is larger than the \textbf{healing length}:
\begin{equation}
    \xi \approx \frac{1}{k_{\xi}} = \frac{1}{k} \sqrt{\frac{n_{0}}{|\Delta n |}},
    \label{HealingLength}
\end{equation}
\noindent the Bogoliubov dispersion relation is linear and density excitations propagate as$\;$collective sound$\;$waves.$\,$This regime is entirely characterized by the speed of sound:$\;c_{s} = \sqrt{-\Delta n/n_{0}}$ which scales as the square root of the fluid density. Conversely, when $\Lambda$ is smaller than$\;\xi$, the dispersion relation becomes quadratic. Excitations have then a particle-like behaviour: they propagate in the transverse plane as "massive" free particles. Nonlinear interactions only affect in that case the effective time $\tau = L \,n/c$ ($L$ being the length of the medium) over which those particles propagate $-\;$and therefore the dynamical phase accumulated$\;-$ by modifying the refractive index $n = n_{0} + n_{2}\,\mathcal{I}_{0}$. The Bogoliubov dispersion relation has been plotted in black solid on figure~\ref{fig:BogDispRegimes} for a background fluid at rest (that is, for $\boldsymbol{v} = \textbf{0}$). The transverse wave-vector has been scaled by $k_{\xi}$ and $\Omega_{B}$ by the inverse of the so-called \textbf{nonlinear length}, defined by $z_{\mathrm{NL}} = 1/k_{0} |\Delta n|$\cite{3-16Ghofraniha}. The red lines and the blue parabola represent the asymptotic sound- and particle-like regimes respectively.

\subsection{Landau criterion for superfluidity}
\label{subsec:Landau}

\noindent The \textbf{Landau criterion for superfluidity} states that below some critical flow velocity$\;v_{c}$, the background fluid cannot transfer kinetic$\,$energy$\,$by$\,$exciting$\,$density$\,$waves$\,$anymore$\,$\cite{3-17Schmitt}. Let's look at equation~\eqref{BogDisp} to understand what it is all about. Spontaneous emission$\;$of elementary excitations (namely energy dissipation), can occur if and only if$\,$such$\,$a$\,$process is energetically favorable. Emitting a density wave at $\mathbf{k_{\perp}}$ on a background fluid flowing at$\;\boldsymbol{v}$ in the transverse plane costs an energy $\Omega_{B}|_{_{\boldsymbol{v} \ne \mathbf{0}}}(\mathbf{k_{\perp}}) = {\mathbf{k_{\perp}} \!\cdot\! \boldsymbol{v}} + \Omega_{B}|_{_{\boldsymbol{v} = \mathbf{0}}}(\mathbf{k_{\perp}})$. In order for dissipation to be energetically favorable, this energy cost $\Omega_{B}|_{_{\mathbf{v} \ne \mathbf{0}}}(\mathbf{k_{\perp}})$ should be negative:
\begin{equation}
    \label{Landau}
    \mathbf{k_{\perp} \!\cdot\! \mathbf{v}} + \Omega_{B}|_{_{\mathbf{v} = \mathbf{0}}}(\mathbf{k_{\perp}})  < 0.   
\end{equation}

\noindent This condition is fulfilled only if $\mathbf{k_{\perp} \!\cdot\! \mathbf{v}} < 0$ and if $\Omega_{B}|_{_{\mathbf{v} = \mathbf{0}}}(\mathbf{k_{\perp}}) < |\mathbf{v}| |\mathbf{k_{\perp}}|$. In other words, equation~\eqref{Landau} is satisfied when density waves are emitted upstream and the fluid$\,$velocity $|\mathbf{v}|$ exceeds the critical speed defined by:
\begin{equation}
    v_{c} = \underset{\mathbf{k_{\perp}}}{\mathrm{min}} \left \{ \frac{\Omega_{B}|_{_{\mathbf{v} = \mathbf{0}}}(\mathbf{k_{\perp}})}{|\mathbf{k_{\perp}}|} \right \}.
    \label{CriticalSpeed}
\end{equation}
\noindent For a particle-like dispersion for which $\Omega_{B}|_{_{\mathbf{v} = \mathbf{0}}}(\mathbf{k_{\perp}}) \propto k_{\perp}^{2}$, this second condition is fulfilled for arbitrary small fluid velocities and density waves are created as soon as an obstacle is dropped into the flow. Nevertheless, for a sound-like dispersion, equation \eqref{CriticalSpeed} states that excitation are emitted only if $v > v_{c} = c_{s}$. This$\;$is$\;$the Landau criterion for superfluidity: the minimum flow velocity requires to excite a wave by scattering on an obstacle is $c_{s}$; below this limit, the kinetic energy of the fluid is too low to excite any density fluctuation. In order to prove that photon fluids in the propagating geometry are likely to be superfluid, we$\,$should thus first demonstrate the existence of such a critical velocity$\,v_{c}\,$experimentally. In chapter 4, we report the first observation$\,$of$\,$a$\,$sound-like$\,$regime$\,$in$\,$the$\,$dispersion$\,$relation of density waves in paraxial photon fluids, which is enough to ensure the existence of $v_{c}$, according$\,$to$\,$the$\,$Landau criterion.$\;$It$\,$should$\,$thus$\,$be$\,$possible$\,$to$\,$observe$\,$superfluidity$\,$of$\,$light using our platform. In this perspective, preliminary results will be discussed$\,$in$\,$chapter$\,$6.

\subsection{Derivation from the NLSE}
\label{subsec:derivationNLSE}

\noindent For the sake of completeness, it has to be stressed that the Bogoliubov dispersion$\;$relation can be derived directly by linearizing the NLSE, assuming $\mathcal{E}_{0}(\mathbf{r_{\perp}}, z) = \mathcal{E}_{0}(z) +\delta \mathcal{E}(\mathbf{r_{\perp}}, z)$. One can use the Madelung transform once again to express the field envelope as: 
\begin{equation}
    \mathcal{E}_{0}(\mathbf{r_{\perp}}, z)  = \sqrt{\rho_{0}(z) + \delta \rho(\mathbf{r_{\perp}}, z)} \, e^{i \left[ \Phi_{0}(z) + \delta \Phi(\mathbf{r_{\perp}}, z) \right]},  
\end{equation}
\noindent which leads at the first-order expansion in $\delta \rho$ and $\delta \Phi$ to the following expression for $\mathcal{E}_{0}$:
\begin{equation}
    \label{FluidVSBog}
    \mathcal{E}_{0}(\mathbf{r_{\perp}}, z) = \underbrace{\vphantom{\left[\frac{1}{2} \frac{\delta \rho (\mathbf{r_{\perp}}, z)}{\sqrt{\rho_{0}(z)}}\right]} \sqrt{\rho_{0}(z)}e^{i \Phi_{0}(z)}}_{\mathcal{E}_{0}(z)} + \underbrace{\left[ \frac{1}{2} \frac{\delta \rho (\mathbf{r_{\perp}}, z)}{\sqrt{\rho_{0}(z)}} + i \sqrt{\rho_{0}(z)} \, \delta \Phi (\mathbf{r_{\perp}}, z) \right] e^{i \Phi_{0}(z)}}_{\delta \mathcal{E} (\mathbf{r_{\perp}}, z)}.
\end{equation}
\noindent By$\,$reinstating$\,$equations$\,$\eqref{FourierDensity}$\,$and$\,$\eqref{FourierPhase}$\,$in$\,$the$\,$right-hand$\,$side$\,$of$\,$equation$\,$\eqref{FluidVSBog},$\,$one$\,$gets:
\begin{equation}
    \label{BogTransform}
    \mathcal{E}_{0} (\mathbf{r_{\perp}}, z) = \mathcal{E}_{0}(z) + e^{i \Phi_{0}(z)} \! \int \frac{\mathrm{d} \mathbf{k_{\perp}}}{(2 \pi)^{2}} \left[ u(\mathbf{k_{\perp}}, z) \, b(\mathbf{k_{\perp}}) \, e^{-i \mathbf{k_{\perp}} \!\cdot\! \mathbf{r_{\perp}}} + \bar{v}(\mathbf{k_{\perp}}, z) \, \bar{b}(\mathbf{k_{\perp}}) \, e^{i \mathbf{k_{\perp}} \!\cdot\! \mathbf{r_{\perp}}} \right],
\end{equation}
\noindent where $u = \frac{1}{2}(f_{+} \!+\! f_{-})$ and $v = \frac{1}{2}(f_{+} \!-\! f_{-})$ are the so-called Bogoliubov amplitudes which obey the normalization condition $|u|^{2} - |v|^{2} = \mathrm{Re} (\bar{f}_{+} f_{-}) = 1$, in the lossless case$\;$($\alpha = 0$). In the lossy case (when $\alpha \ne 0$), $|u_{\mathbf{k}_{\perp}}|^{2} - |v_{\mathbf{k}_{\perp}}|^{2} = N(\mathbf{k}_{\perp}, z)$, where $N(\mathbf{k}_{\perp}, z)$ is a non-trivial normalization$\,$function$\,$that$\,$depends$\,$on$\,\alpha\,$\cite{3-19Larre}. The$\,$same$\,$transformation$\,$was$\,$used$\,$in$\,$1958$\,$by Nikolay Bogolyubov so as to find solutions of the BCS$\,$theory$\,$in$\,$homogeneous$\,$systems$\,$\cite{3-18Bog}. By replacing in the NLSE equation~\eqref{NLSE} $\mathcal{E}_{0}$ by its expression equation~\eqref{BogTransform}, one easily obtains a coupled system of equations on the Bogoliubov amplitudes $u$ and $v$:

\newpage

\begin{align}
    \left \{ -\frac{\partial \Phi_{0}}{\partial z} u + i \frac{\partial u}{\partial z} \right\} e^{i\Phi_{0}} =  -\left \{ \frac{k_{\perp}^{2}}{2 k_{0}} + i \frac{\alpha}{2} \right\} u \, e^{i\Phi_{0}} - k_{0} \, n_{2} \left( 2 |\mathcal{E}_{0}|^{_{2}} \, u \, e^{i\Phi_{0}} + \mathcal{E}_{0}^{_{2}} \, v \, e^{-i\Phi_{0}} \right) \\
    \left \{ \frac{\partial \Phi_{0}}{\partial z} v + i \frac{\partial v}{\partial z} \right\} e^{-i\Phi_{0}} =  \left \{ \frac{k_{\perp}^{2}}{2 k_{0}} - i \frac{\alpha}{2} \right\} v \, e^{-i\Phi_{0}} + k_{0} \, n_{2} \left( 2 |\mathcal{E}_{0}|^{_{2}} \, v \, e^{-i\Phi_{0}} + {\mathcal{E}_{0}^{*}}^{_{2}} \, u \, e^{i\Phi_{0}} \right)
\end{align}
\noindent which can be simplified using the fact that $\mathcal{E}_{0} = \sqrt{\rho_{0}} \, e^{i \Phi_{0}}$ and $\frac{\mathrm{d} \Phi_{0}}{\mathrm{d} z} = k_{0} \, n_{2} \, |\mathcal{E}_{0}|^{_{2}}$:
\begin{equation}
    \label{BogEqNLSE}
    i \frac{\partial}{\partial z}
    \begin{pmatrix} 
    u \\
    v  
    \end{pmatrix}
    = -\left[ i\frac{\alpha}{2} +  \overline{\mathcal{H}}_{\mathbf{k_{\perp}}} \right] \!
    \begin{pmatrix} 
    u \\
    v  
    \end{pmatrix} \!, \; \mathrm{where} \;
    \overline{\mathcal{H}}_{\mathbf{k_{\perp}}} = 
    \begin{pmatrix} 
    \frac{k_{\perp}^{2}}{2 k} + k_{0} \, \Delta n & k_{0} \Delta n \\
    - k_{0} \Delta n & -\frac{k_{\perp}^{2}}{2 k} - k_{0} \Delta n
    \end{pmatrix}.
\end{equation}
\noindent One finally retrieves the Bogoliubov dispersion relation~\eqref{BogDisp} by diagonalizing $\overline{\mathcal{H}}_{\mathbf{k_{\perp}}}$. 

\section{Bogoliubov dispersion in a lossy nonlocal medium}

\noindent So far, we have derived the Bogoliubov dispersion relation in the ideal lossless$\,$case,$\,$even$\,$if the derivations in$\,$subsections$\,$\ref{subsec:derivationEuler} and$\,$\ref{subsec:derivationNLSE} take$\,$linear$\,$losses$\,$into$\,$account.$\;$In$\,$experiments, linear absorption cannot be neglected usually. Moreover, as mentioned in$\,$paragraph$\,$1.3.8, photon-photon interactions in warm rubidium vapors are nonlocal because of the ballistic transport of excited atoms. In the next section, the Bogoliubov dispersion relation~\eqref{BogDisp} is therefore generalized in order to take both these effects into account.  

    \subsection{Lossy nonlinear medium ($\alpha \ne 0$)}

\noindent We first generalize the Bogoliubov dispersion relation~\eqref{BogDisp} for a lossy$\,$local$\,$Kerr$\,$medium. Suppose the absorption coefficient $\alpha$ is non-zero but sufficiently small to$\,$make$\,$the$\,$evolution of the field envelope $\mathcal{E}_{0}$ adiabatic along the $z$-axis. The eigenvectors of$\;\mathcal{H}_{\mathbf{k_{\perp}}}\!(z)$ defined in equation \eqref{BogEq} are then slowly-varying functions of $z$ that strictly follow the$\,$variations$\,$of the corresponding eigenvalues along $z$~\cite{3-19Larre}. The adiabatic solutions of equation\eqref{BogEq}$\,$may thus be written as follow: $f_{\pm}(\mathbf{k}_{\perp}, z)  =  \Tilde{f}_{\pm}(\mathbf{k}_{\perp}, z) e^{i \Omega_{\mathrm{eff}}(\mathbf{k}_{\perp}) z}$ where:
\begin{align}
    \label{EffectiveBogDispAbs}
    \Omega_{\mathrm{eff}}(\mathbf{k}_{\perp}) = & \,  \langle \Omega_{B}(\mathbf{k}_{\perp}, z')\rangle_{z}, \\
    \label{BogDispAbs}
    \Omega_{B}(\mathbf{k}_{\perp}, z) = & \, i\frac{\alpha}{2} + \sqrt{-\frac{n_{2}}{n_{0}} \, \rho_{0}(z) \, k_{\perp}^{2} + \frac{k_{\perp}^4}{4 k^{2}}}, \\
    \label{FourierComponent}
    \Tilde{f}_{\pm}(\mathbf{k}_{\perp}, z) \propto & \left(\frac{k_{\perp}^{2}/(2 k)}{\Omega_{B}(\mathbf{k}_{\perp}, z)- i \alpha/2}\right)^{\pm \frac{1}{2}},
\end{align}
\noindent for a background fluid at rest ($\boldsymbol{v} = \mathbf{0}$). The density fluctuation amplitude exponentially decays because of$\,$the$\,$linear$\,$absorption.$\;$The$\,$latter$\,$also$\,$affects$\,$the$\,$background$\,$density$\,$which decreases similarly along the $z$-axis, according to the Beer-Lambert law.$\,$The sound$\,$velocity decreases consequently from one plane to the next since it depends on the square root of the fluid density. Everything happens as if sound-waves were moving slower and slower as "time" goes$\;$by. One can then define an effective sound velocity by averaging $c_{s}(z)$ over$\;z$:  $c_{s, \mathrm{eff}} = \langle \sqrt{-\Delta n(z')/n_{0}} \rangle_{_{\!L}} = c_{s}(0) \, z_{\mathrm{eff}}(\alpha) / L $, where $z_{\mathrm{eff}}(\alpha) = 2\left[1 - \exp(-\alpha L/2 )\right]/\alpha$.
\vspace{6pt}
\newline
\noindent In our experiments, the main effect of absorption is to multiply the sound velocity $c_{s}\,$of$\,$the lossless$\,$case$\,$by$\,$the$\,$scaling$\,$factor$\,z_{\mathrm{eff}}(\alpha) / L$.$\;$It$\,$thus$\,$mainly$\,$changes$\,$the$\,$slope$\,$of$\,$the$\,$dispersion in the sonic regime without affecting its shape in-depth.$\;$In$\,$other$\,$words,$\,$if$\,$the$\,$input$\,$intensity $\mathcal{I}_{0}$ is multiplied by $L/z_{\mathrm{eff}}(\alpha)$, the curve of the effective dispersion relation $\Omega_{\mathrm{eff}}(\mathbf{k_{\perp}})$ will almost exactly translate on the curve of $\Omega_{B}|_{_{\alpha = 0}}(\mathbf{k_{\perp}})$, obtained for $\Delta n = n_{2} \, \mathcal{I}_{0}$.  

\subsection{Lossy nonlinear medium with effective nonlocal interactions}

\begin{figure}[h]
\center
\includegraphics[scale=0.55]{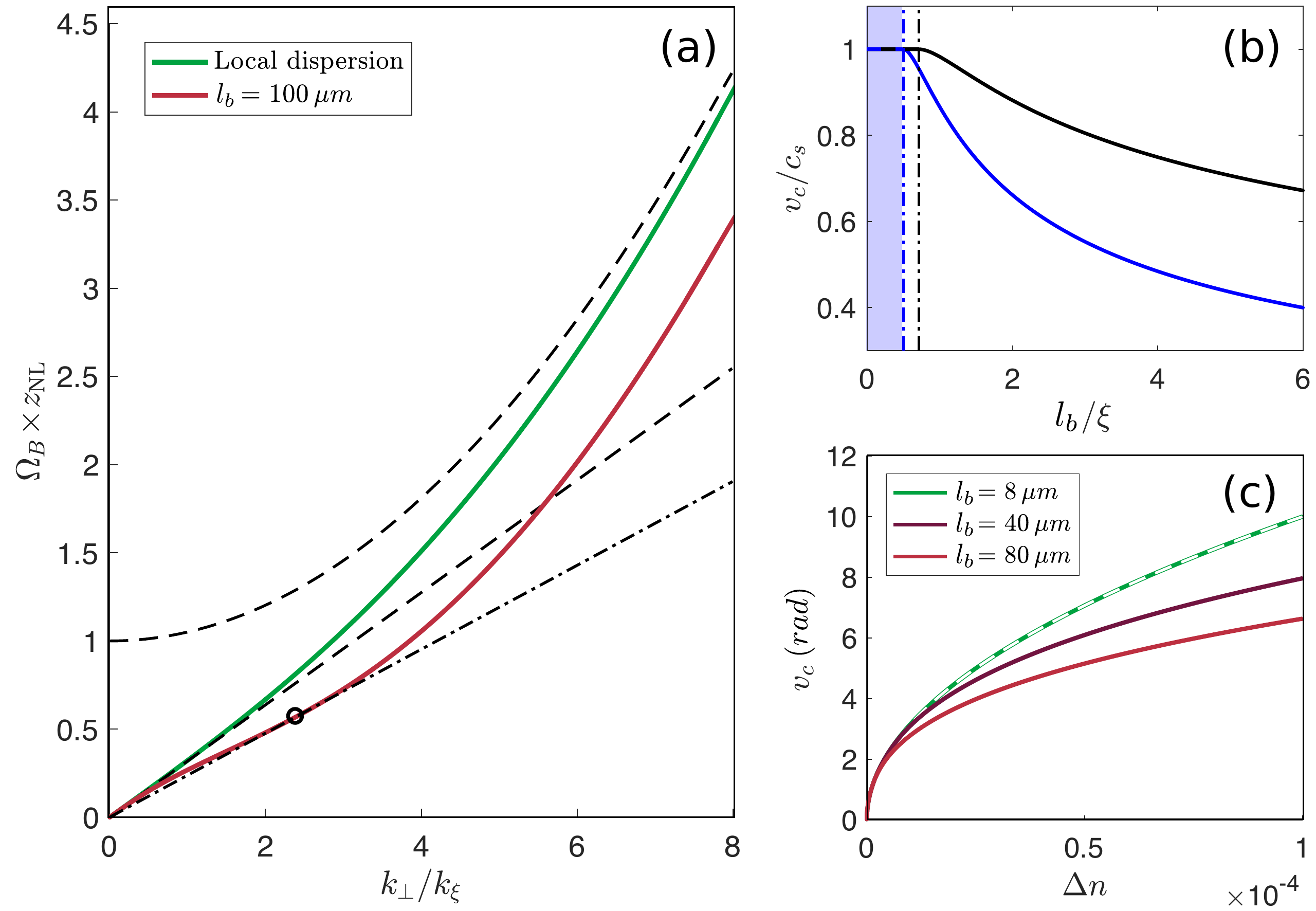} 
\caption{(a) Local (green) and nonlocal (red) Bogoliubov dispersion relations obtained in the lossless case. Asymptotic behaviours are plotted in dashed black for the local case. For high enough nonlocalities, an inflexion point appears (circle) and the critical velocity starts being lower than $c_{s}$. (b) Variation of $v_{c}/c_{s}$ as function of $l_{b}/\xi$ for ballistic$\,$(black$\,$line) and diffusive nonlocality (blue line). In the ballistic case, $v_{c} = {c_{s}}$ as long as $l_{b} < 0.71 \, \xi$, while, in the diffusive one, $v_{c} = {c_{s}}$ as long as $l_{b} < \xi/2$ precisely ~\cite{3-20Vocke}. (c) Critical velocity$\,v_{c}$ as function of $\Delta n$. The white dashed line plots the behaviour of $c_{s}$ for comparison.
}
\label{fig:BogDispersion}
\end{figure}

\noindent In the previous paragraph, the dielectric response of the medium was supposed to$\;$be$\;$local. The nonlinear change of refractive index $\Delta n$ at a given position $\boldsymbol{r}_{\perp}$ in the transverse$\;$plane was thus only depending on the laser intensity at that point and$\,$not$\,$on$\,$the$\,$intensity$\,$nearby. Let's assume from now on that the medium optical response is nonlocal, which$\,$is$\,$by$\,$the$\,$way the case in a wide variety of Kerr nonlinear mediums, such as hot alkaline vapors~\cite{2-20Skupin} but also thermo-optic liquids~\cite{2-23Vocke}. The NLSE~\eqref{NLSE} reads then as follow:
\begin{equation}
\label{NLSENonLocal}
    i \partial_{z} \, \mathcal{E}_{0}(\mathbf{r}, z) = -\left[ \frac{1}{2 k} \boldsymbol{\nabla}_{\perp}^{2} + \frac{i \alpha}{2} + k \frac{n_{2}}{n_{0}} \int_{\mathcal{S}} G(\mathbf{r}-\mathbf{r'}) \left|\mathcal{E}_{0}(\mathbf{r'}, z)\right|^{2} \mathrm{d}\mathbf{r'} \right] \mathcal{E}_{0}(\mathbf{r}, z),
\end{equation}
where $G$ is the nonlocal response function in real space. Equation~\eqref{BogEq} is still correct$\;$in the nonlocal case if the fluid density $\rho = |\mathcal{E}_{0}|^{_{2}}$ is replaced by the integral in the right-hand side of equation~\eqref{NLSENonLocal}, which is the convolution of $\rho_{0}$ with $G$. Let $\Tilde{G}^{b}_{\gamma}$ stand for$\,$the$\,$Fourier transform of the ballistic response function $\bar{G}^{b}_{\gamma}$ in Rubidium vapors:  
\begin{equation}
    \Tilde{G}_{\gamma}^{b}(k_{\perp}) = \mathcal{F}\!\left[ \bar{G}_{\gamma}^{b} \right]\!(\mathbf{k_{\perp}}) = \frac{\sqrt{\pi}}{\gamma} \, \frac{e^{1 / (k_{\perp} \, l_{b})^{2}}}{k_{\perp} \, l_{b}} \mathrm{Erfc} \left[ 1/(k_{\perp} \, l_{b}) \right].
\end{equation}
\noindent This expression$\,$has$\,$been$\,$derived$\,$in$\,$the$\,$first$\,$chapter$\,$(see$\,$paragraph$\,$1.3.8$\,$ii,$\,$equation$\,$(1.75)). The ballistic transport length scale is defined by $l_{b} = u \, \tau$, where $u = \sqrt{2 k_{B} T/m}\,$stands$\,$for the most probable speed of the atoms in the transverse plane (at the vapor temperature$\,T$) and $\Gamma = \Gamma + \Gamma_{t}$ for the total decay rate of the excited state addressed by the laser field. Using the convolution theorem, the dispersion relation~\eqref{BogDispAbs} can be rewritten as follows:
\begin{equation}
    \label{eq:BogDispTotal}
    \Omega_{B}(\mathbf{k}_{\perp}, z)-{\mathbf{k_{\perp}} \!\cdot\! \boldsymbol{v}} = \, i\frac{\alpha}{2} + \sqrt{- \, \frac{n_{2}}{n_{0}} \, \rho_{0}(z) \left[ \gamma \, \Tilde{G}_{\gamma}^{b}(k_{\perp}) \right] k_{\perp}^{2} + \frac{k_{\perp}^4}{4 k^{2}}}.
\end{equation}
\noindent As previously, the term ${\mathbf{k_{\perp}} \!\cdot\! \boldsymbol{v}}$ describes the shift of the density wave frequency because of Doppler effect when it propagates at $\mathbf{k_{\perp}}$ on top of a moving background. Equation~\eqref{eq:BogDispTotal} is the most general expression for the Bogoliubov dispersion relation, taking into account absorption$\;$and nonlocality at the same time. The dispersions$\;$obtained$\;$in$\;$local$\;$(green$\;$line) and nonlocal (red line) Kerr medium have been plotted as function of $k_{\perp}/\xi$ on figure~\ref{fig:BogDispersion}(a) in the lossless case. The dashed black lines represent the sound-like and the particle-like asymptotic regimes in the local$\;$case.$\,$The ballistic$\;$transport$\;$length$\;$scale$\;l_{b}\;$is$\;$about$\,7.5 \, \mu$m at $T = 400 \, \mathrm{K}$. For clarity, the red curve has been plotted for $l_{d} = 100$ $\mu$m in order to clearly observe the effects of nonlocality on the dispersion.  

\newpage

\noindent The Landau criterion for superfluidity (see subsection~\ref{subsec:Landau}) states that below a$\,$critical$\,$flow velocity, the photon fluid cannot dissipate energy anymore by emitting density waves and behaves thus as a superfluid in the transverse plane.$\;$In$\,$the$\,$local$\,$case,$\,$this$\,$critical$\,$velocity$\,v_{c}$ is equal to the speed of sound $c_{s}$. In the nonlocal case however,$\;$the$\;$situation$\;$is$\;$different. For strong enough nonlocality, an inflexion point (black circle) appears on$\;$the$\;$nonlocal dispersion$\;$curve. The speed of sound $c_{s}$ is$\;$still well defined as $\Omega_{B}|_{_{\mathbf{v} = \mathbf{0}}}(\mathbf{k_{\perp}}) \simeq c_{s} \, k_{\perp}\,$when $k_{\perp} \!\ll\! k_{\xi}\,$(indeed, $\Tilde{G}_{\gamma}^{b}(k_{\perp}) \simeq 1/\gamma + O(k_{\perp}^{2})$ in that case). However, it is not anymore equal to the critical speed, which is given by the slope of the tangent to the nonlocal$\,$dispersion curve at$\,$the$\,$inflexion$\,$point$\,$now$\,$(black$\,$dashed-dotted$\,$line)$\,$and$\,$therefore,$\,v_{c} < c_{s}$.$\;$In$\,$order$\,$to be superfluid, the photon fluid must flow toward any obstacle with a velocity lower than $v_{c}$; the nonlocality has thus reduced the range of velocities for which superfluid flows of light can be$\,$observed.$\;$Let's$\,$be$\,$more$\,$quantitative.$\;$According$\,$to$\,$equation$\,$\eqref{CriticalSpeed},$\,$the$\,$critical$\,$speed is obtained by calculating the minimum of the phase velocity $v_{\mathrm{ph}}(\mathbf{k_{\perp}}) = \Omega_{B}|_{_{\mathbf{v} = \mathbf{0}}}(\mathbf{k_{\perp}}) / |\mathbf{k_{\perp}}|$. 

\newpage
\noindent On figure~\ref{fig:BogDispersion}(b), the critical speed (normalized by the sound velocity $c_{s}$) has been plotted as function of $l_{b} / \xi$ in black solid. Two regimes can be identified: 
\vspace{-6pt}
\begin{itemize}
    \item[(1)] For weak enough nonlocalities, that is, when $l_{b}/\xi$ is lower than $0.71$, the minimum$\;$of $v_{\mathrm{ph}}$ is obtained for $k_{\perp} = 0$ and is still equal to the speed of sound $c_{s}$. The nonlocality slightly affects the shape of the dispersion relation but the "local" Landau criterion for superfluidity remains valid. 
    \vspace{-6pt}
    \item[(2)] For higher nonlocalities (right side of the black dashed-dotted line), the critical$\;$speed starts decreasing slowly when $l_{b} / \xi$ increases.
\end{itemize}  
\vspace{-6pt}
\noindent On figure~\ref{fig:BogDispersion}(b), the critical speed obtained by considering a diffusive nonlocality has also been plotted (blue solid line), for comparison. This is, for instance, the kind of nonlocality encountered in thermo-optic liquids, where the heat diffusion inside the material$\,$makes$\,$the optical response nonlocal~\cite{2-23Vocke}. By using a distributed loss model~\cite{3-20Vocke}, we can$\,$show$\,$that$\,$the Fourier$\,$transform$\,$of$\,$the$\,$response$\,$function$\,$is$\,$Lorentzian$\,$in$\,$that$\,$case:$\,\Tilde{G}^{d}_{\sigma}(\mathbf{k_{\perp}}) = 1/(1+\sigma^{2} k_{\perp}^{2})$ ($\sigma$ is the range of the nonlocal interaction). This diffusive response function falls off much more rapidly than the ballistic one in Fourier space (as mentioned in$\,$paragraph$\,$1.3.8$\,$ii). Consequently, it is not surprising to observe that the critical speed$\,$decreases$\,$more$\,$slowly$\,$for ballistic nonlocality than for diffusive one. Moreover, the threshold value at which$\,v_{c}\,$starts decreasing is higher in our case than for thermo-optic liquids, for which it lies$\,$at$\,\sigma/\xi = 0.5$ (blue dashed-dotted line). Much larger effects on the dispersion relation are thus expected in fluids of light propagating in these systems than in rubidium vapors.
\vspace{6pt}
\newline
\noindent Let's finally briefly comment the figure~\ref{fig:BogDispersion}(c), where the critical$\,$velocity$\,v_{c}\,$has$\,$been$\,$plotted as function of the nonlinear change of refractive index $\Delta n = n_{2} \mathcal{I}_{0}$, for$\,$a$\,$ballistic$\,$nonlocality. As you can see, it perfectly matches the speed$\,$of$\,$sound$\,$(white$\,$dashed$\,$line) when $l_{b} = 8 \, \mu$m. In this case, when $\Delta n$ ranges from 0 to $1\!\times\!10^{-4}$, the ratio $l_{b}/\xi$ varies from 0 to 0.64 and remains$\,$below 0.71. In other words, at every value$\,$took$\,$by$\,\Delta n\,$on$\,$the$\,$graph$\,$of$\,$figure$\,$\ref{fig:BogDispersion}(c), the critical speed is equal to the speed of sound when $l_{b} = 8 \, \mu$m. This is$\,$not$\,$anymore$\,$true when $l_{b} = 40 \, \mu$m of $80 \, \mu$m, as $l_{b}/\xi$ varies then up to $3.2$ and $6.4$, respectively. In$\,$both$\,$cases, there is thus a critical value of the nonlinear change of refractive index, $\Delta n_{c}$, at which an inflexion point appears in the dispersion. Thereupon, we expect the$\,$critical$\,$speed$\,$to$\,$remain lower than the speed of sound when further increasing $\Delta n$. This is indeed$\,$what$\,$we$\,$observe.








\chapter{Atomic medium characterization}

\noindent In chapter 2, I have shown how the envelope of an intense laser$\,$field,$\,$propagating$\,$through$\,$a Kerr-type$\,$medium,$\,$can$\,$be$\,$regarded$\,$as$\,$a$\,$2D$\,$photon$\,$fluid$\,$flowing$\,$into$\,$the$\,$plane$\,$perpendicular to the optical axis. The main motivations of this work is to study experimentally some of the hydrodynamical properties of those photon fluids in hot rubidium vapors. But before showing the results we obtained in that respect, I$\,$would$\,$like$\,$to$\,$introduce$\,$the$\,$experimental tools$\,$and$\,$methods used to produce and characterise such paraxial fluids of light. I first give some technical details about the glass cells containing the rubidium vapor and about$\;$the home-made heating system designed to control its temperature. I then briefly present the laser sources used to generate propagating photon fluids and, more generally,$\;$to$\;$address the rubidium $D$-lines. In a second part, the measurements performed to access the$\;$vapor temperature $T$, the atomic density $N$ and the nonlinear refractive index $n_{2}$ are presented.   

\newpage

\section{Experimental tools}

\subsection{Rubidium cells and oven design}
    \label{subsec:RbCell}
    
\noindent The glass cells we use have all been manufactured by \textit{Triad Technology} which provides high purity reference cells for spectroscopic applications.$\,$They$\;$are$\;$cylindrical$\;$(1"$\,$diameter)  
and closed on both sides by $1 \, \mathrm{mm}\;$thick anti-reflective coated windows at 780 and 795 nm. The length of the cells ranges from 1 to 7.5 cm depending on$\;$the experiment we carry$\;$out. In order to heat the rubidium vapor at high temperatures, the cells are disposed inside a copper cylinder (5 mm thick, 10 cm long) which is enveloped by Kapton flexible heaters (wattage: 5 or 10 W/in at 28 V) from \textit{Omega}. The copper cylinder has a good thermal inertia protecting the cell from fast temperature$\;$variations. The whole is placed into an aluminum enclosure with flat 1 mm thick anti-reflective coated windows on both sides$\;$to maintain optical access to the cell. The temperature is monitored by a sensor set close to the cell tip (thermocouple). Setting up a feedback loop was not required, as the heating system was efficient enough to stabilize the cell temperature at $\pm 1^{\circ}$C during experiments. Let's$\;$finally mention that about $6\%$ of laser input power is lost in$\,$reflections$\,$on$\,$the$\,$cell$\,$and enclosure$\,$windows$\,$(8$\,$interfaces),$\,$when$\,$the$\,$laser$\,$is$\,$highly$\,$detuned$\,$from$\,$the$\,$rubidium$\,D$-lines. 

\begin{figure}[h]
\center
\includegraphics[scale=0.135]{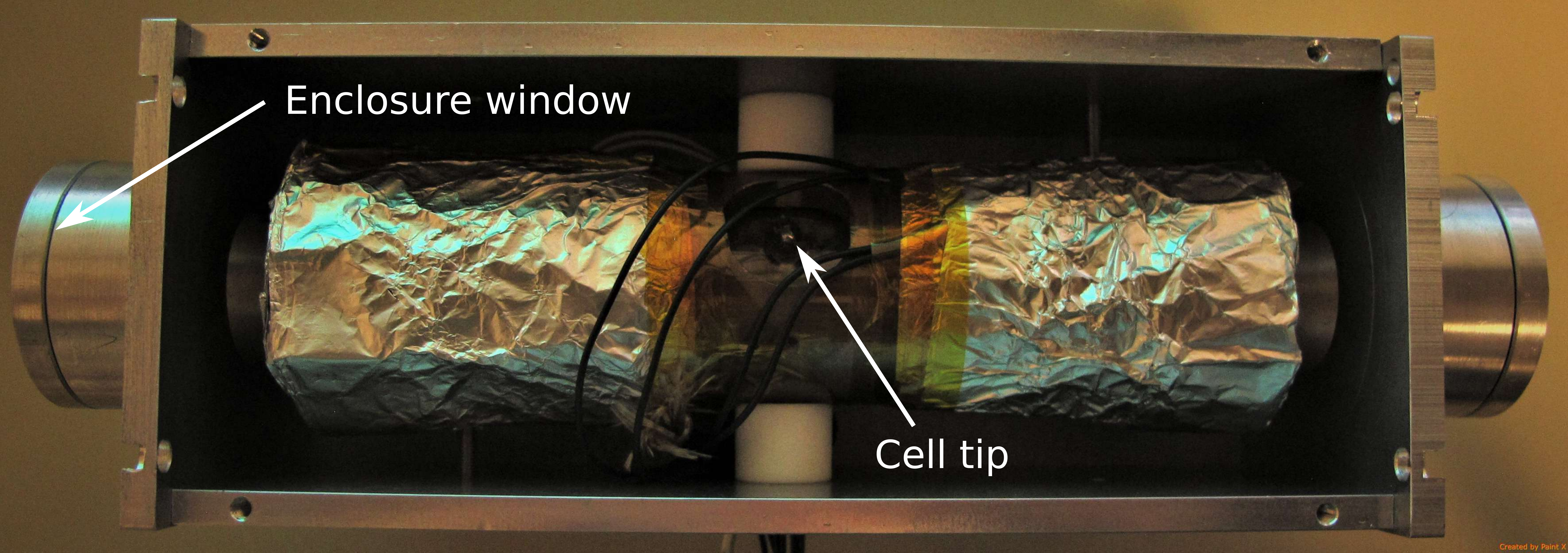} 
\caption{Heating system. A 7.5 cm long cell is placed inside $5$ mm thick$\;$copper$\;$cylinder. Flexible heaters are arranged$\;$all$\;$around and covered with several$\;$layers of aluminum foils to reduce losses by thermal radiation and provide a better heating efficiency. The cell tip (visible at the center of the picture) is about $80^{\circ}$C.}
\label{fig:Oven}
\end{figure}

\noindent The heating system is designed to prevent rubidium from condensing on the cell$\;$windows. In the cell, part of rubidium is liquid and the rest is in vapour phase. Despite$\,$the$\,$fact$\,$that condensation is very unlikely to happen at temperatures above the rubidium melting point ($T_{m} = 39.3^{\circ}$C), small droplets can still form locally at the center of the cell windows, where$\;$the$\;$glass$\;$temperature$\;$is$\;$lower. Things are getting worse then as a kind of avalanche nucleation process leads the rubidium nearby to condense a well. If condensation appear only on one window, increasing slightly the voltage applied across the heater rolled around the window at issue is often sufficient to remove it. If it appears on both$\,$windows,$\,$we$\,$have then to displace the cold point from their$\,$centers$\,$to$\,$the$\,$cell$\,$tip.$\;$As$\,$you$\,$can$\,$see$\,$on$\,$figure$\,$\ref{fig:Oven}, a$\;$1.5$\;$cm$\;$diameter$\;$hole has been drilled in the copper cylinder to$\,$let$\,$the$\,$cell$\,$tip$\,$be$\,$in$\,$contact with the air inside the enclosure, which maintains it at a lower temperature. By doing so, the rubidium will preferentially condenses inside of it and not on the windows anymore. 

\newpage

\noindent Of course, the temperature recorded with the thermocouple (silver wire on the picture~\ref{fig:Oven}) is not the absolute temperature of the rubidium vapor $T$. In order to estimate parameters such as the atomic density $N$, the atoms average speed $u$ or the transport length scales (and thus to characterize the vapor optical response), the absolute temperature$\,$is$\,$required. It can be extrapolated in practice by fitting the linear transmission$\;$spectrum, which is Doppler broadened by the atomic motion in the cell, as we will see in subsection 3.2.2. 

\subsection{Laser sources}

\noindent The$\,$transverse$\,$plane$\,$dynamics$\,$of$\,$nonlinear$\,$phenomena$\,$is$\,$controlled$\,$by$\,$the$\,$nonlinear$\,$change of refractive index $\Delta n = n_{2} \, \mathcal{I}_{0}$. In order for this dynamics to be conservative, linear$\;$losses should be low, which is possible only when the laser frequency is detuned far off-resonance. In that case, the nonlinear refractive index $n_{2}$ is small as well, because it scales as $1/\Delta^{_{3}}$ ($\Delta\,$being$\,$the$\,$laser$\,$detuning), as discussed in the paragraph 1.2.3 iii. Moreover,$\,$being$\,$in$\,$the hydrodynamical$\;$regime$\,$(defined$\,$in$\,$section$\,$2.1.3$\,$iii)$\,$requires$\,$large$\,$values$\,$of$\,\Delta n\,$and$\,$thus$\,$high laser intensities. At$\,$the$\,$end$\,$of$\,$the$\,$day,$\,$generating$\,$paraxial$\,$photon$\,$fluids$\,$in$\,$rubidium$\,$vapors requires powerful CW laser sources which can be easily tuned on a wide$\,$range$\,$of$\,$frequencies (several gigahertz) around the desired atomic resonance. Two kinds of sources satisfying these conditions are used in the lab. The first is a continuous-wave Ti-Sapphire laser and the second an amplified external-cavity diode laser. Both sources provide an output$\,$power greater than $2$ W and an easy control over the laser frequency, which can$\;$be$\;$widely$\;$tuned around the rubidium $D_{1}$ and $D_{2}$ line. In this subsection, I will give some$\,$technical$\,$details about$\,$these$\,$two$\,$sources,$\,$starting$\,$with$\,$the$\,$Ti-sapphire$\,$laser,$\,$which$\,$is$\,$the$\,$one$\,$I$\,$predominantly used throughout my thesis.    

\subsubsection{SolsTiS Ti-sapphire laser}

\noindent SolsTiS is a tunable narrow linewidth continuous-wave Ti-Sapphire laser from \textit{M Squared}. It consists of a monolithic ring cavity in which a crystal of sapphire, doped with$\;$Ti$^{_{3+}}$ ions, has been introduced. This crystal is pumped with a Verdi V10 manufactured by \textit{Coherent}, which is a $10$ W frequency-doubled Nd:YVO4 laser. Because of the large gain bandwidth of the crystal, lasing effect can be reached$\,$for$\,$a$\,$broad$\,$range$\,$of$\,$wavelengths,$\,$extending$\,$from 670 to 990 nm. In order to ensure the single-frequency operation of the SolsTiS cavity, hole burning effects in the gain medium must be removed. This is achieved in SolsTiS by using the so-called bow tie cavity geometry, together with an optical isolator, which$\;$forces the ring cavity to operate uni-directionally. This results in a traveling wave that ensures a minimum of spatial-hole burning.
\vspace{6pt}
\newline
\noindent In$\,$order$\,$to$\,$tune$\,$the$\,$SolsTiS$\,$output$\,$wavelength,$\,$a$\,$motorized$\,$birefringent$\,$filter$\,$(BRF)$\,$is$\,$used. This filter introduces a wavelength dependent loss into the cavity. The wavelength tuning is then performed by simply rotating the BRF, which provides a relatively rapid but coarse frequency adjustment however.$\,$If$\;$a$\;$finer$\;$control$\;$over$\;$the$\;$wavelength$\;$is$\;$needed,$\,$the$\;$SolsTiS intracavity Fabry–Pérot étalon can be used. The étalon introduces a spectral loss into the cavity that is a much sharper function of frequency than the BRF. Thus, by electronically adjusting the étalon spacing (that is, its tilt angle), the SolsTiS output wavelength may be finely adjusted. In order to ensure the long-term single mode$\,$operation$\,$of$\,$the$\,$ring$\,$cavity, an electronic servo locking of étalon is provided. 

\newpage

\begin{figure}[h]
\center
\includegraphics[width=\columnwidth]{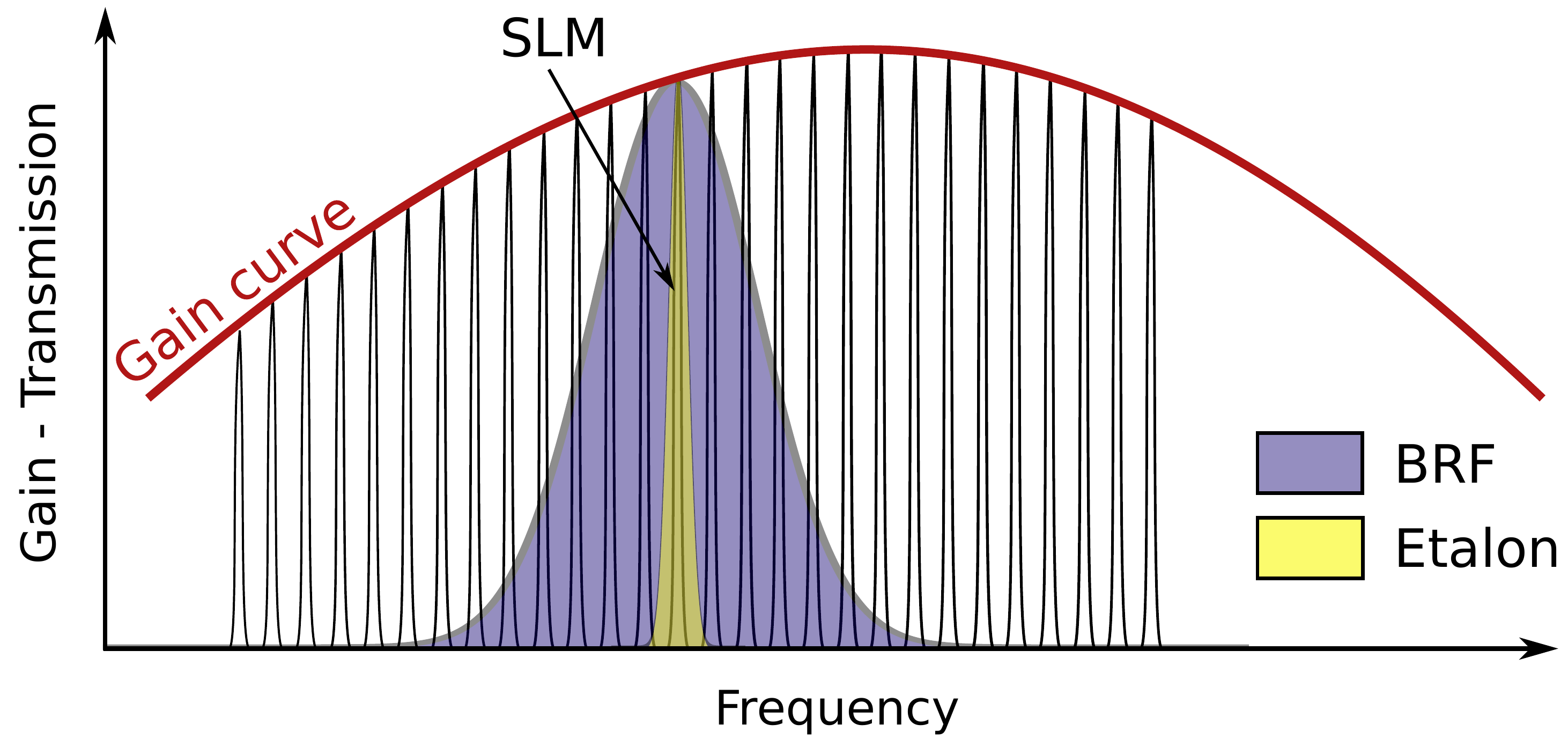} 
\caption{Principle of the single-mode operation of the Ti-Sapphire laser. The gain of the crystal is plotted in red as function of frequency. The laser frequency comb (black$\;$peaks) represents all the longitudinal modes (LM) sustained by the ring cavity. The Lyot filter (or birefringent filter, BRF) transmits only certain modes. The étalon, whose spacing is electrically stabilized, finally selects a single longitudinal mode (SLM). }
\label{fig:TiSaph}
\end{figure}

\noindent The servo locking of the étalon prevents the cavity from jumping from one longitudinal mode to the other. However, this locking does not compensate for the long-term frequency drifts of the cavity itself. In order to solve this issue, the SolsTiS cavity has been locked on a high stability, high finesse reference cavity. This locking reduces the laser linewidth to less than $50$ KHz. Slaving the SolsTiS to this internal reference is performed by:
\begin{itemize}
\vspace{-6pt}
    \item [(1)] directing a small fraction of its output power to the reference cavity;
    \vspace{-6pt}
    \item [(2)] locking its output frequency to a reference cavity fringe, mounting one of the SolsTiS cavity mirror on a fast piezo-electric transducer (PZT).
\end{itemize}
\vspace{-6pt}
\noindent With the SolsTiS cavity slaved to the reference cavity, the laser output frequency can then be scanned (or offset) by scanning (or adjusting) the reference cavity length itself
with a high degree of precision. Moreover, temperature-induced changes in the reference cavity length are compensated, further enhancing the stability of the reference cavity$\,$and, consequently, the stability of the SolsTiS output laser frequency. Nevertheless, it is worth noting that this internal reference is not locked to an absolute reference, such as an atomic absorption$\;$line.$\;$This might seem a bit surprising at first but as we always highly detuned the laser frequency far off-resonance, the remaining drift in the Solstis$\;$frequency$\,-\,$which$\;$is about $50$ MHz/hr/$^{\circ}$C$\,-\,$does not have any influence in practice.$\,$Moreover, we$\;$continuously monitor the laser frequency in experiments using a lambda-meter (either the WSB-10 from \textit{High Finesse} or the LW10 from \textit{Resolution Spectra}). This enables us to reset (by hand) the laser frequency as soon as it drifts too far from the initial desired value. Let's finally mention that the stability of the internal locking depends strongly on the injection of the Verdi inside the SolsTiS cavity. When the alignment is optimized (at a given frequency), the output power should be maximized (typically, $\mathcal{P}_{0} \simeq 3.2$ W at 780 nm) as well as the robustness of the SolsTiS internal locking system.

\subsubsection{Amplified external-cavity diode laser}

\begin{figure}[H]
\center
\includegraphics[width=0.8\linewidth]{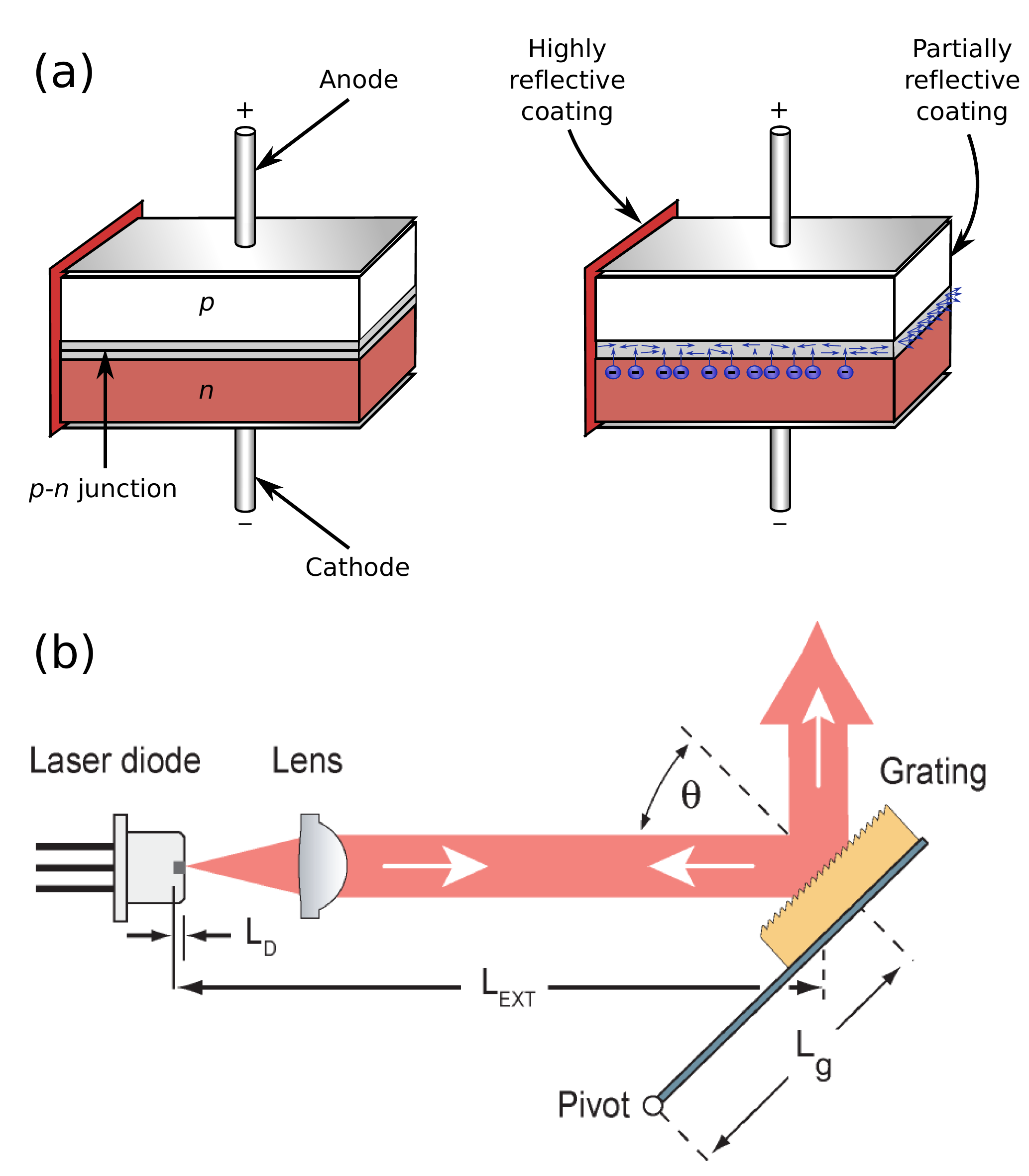} 
\caption{(a): Sketch$\,$of$\,$a$\,$double$\,$heterostructures$\,$diode$\,$laser$\,$source.$\;$A$\,$layer$\,$of$\,$a$\,$low$\,$band-gap material is sandwiched between two layers of a high band-gap material. Each$\;$of$\;$the junctions between different band-gap materials is called a heterostructure, hence$\,$the$\,$name double$\,$heterostructure$\,$(DH)$\,$laser.$\;$In$\,$that$\,$respect,$\,$the$\,$diode$\,$laser$\,$described$\,$in$\,$the$\,$text$\,$may be referred to as a homojunction laser. The active region of$\,$a$\,$DH$\,$laser,$\,$where$\,$free$\,$electrons and holes exist simultaneously, is limited to within the thin middle layer. Consequently, the amount of electron hole-pairs that contributes to amplification is much higher in DH diode lasers than in homojunction ones. (b)$\,$Sketch$\,$of$\,$the$\,$Litrow$\,$configuration$\,$(from$\,$\cite{4-1Saliba}). 
The diode of figure (a) emits light that is collimated by a lens onto a diffracting grating. The first diffraction order$\,$is$\,$sent$\,$back$\,$into$\,$the$\,$laser$\,$diode,$\,$which$\,$enables$\,$us$\,$to$\,$select$\,$a$\,$single longitudinal$\,$cavity$\,$mode.$\;$Tuning$\,$the$\,$angle$\,\theta\,$allows$\,$to$\,$coarsely$\,$change$\,$the$\,$laser$\,$frequency. 
}
\label{fig:LaserDiode}
\end{figure}

\noindent Amplified external-cavity diode$\,$lasers$\,$is$\,$a$\,$compact,$\,$low$\,$cost$\,$and$\,$easy-to-handle$\,$alternative to Dye or Ti-Sapphire laser ring systems. In this paragraph, I briefly present the different blocks$\,$composing$\,$this$\,$other$\,$laser$\,$source,$\,$that$\,$we$\,$also$\,$used$\,$to$\,$generate$\,$paraxial$\,$photon$\,$fluids.
\vspace{6pt}
\newline
\noindent \textbf{Diode$\,$laser.}$\;$Diode$\,$lasers$\,$are$\,$semiconductor$\,p$-$n\,$junction$\,$devices$\,$in$\,$which$\,$lasing$\,$conditions are created at the junction by pumping a diode with an$\,$electrical$\,$current.$\;$Forward$\,$electrical bias across the laser diode causes the two species of charge carrier $-\,$holes$\,$and$\,$electrons$\,-$ to be "injected" from opposite$\,$sides$\,$of$\,$the$\,p$-$n\,$junction$\,$into$\,$the$\,$depletion$\,$region.$\;$This$\,$region, devoid of$\,$any$\,$charge$\,$carriers,$\,$forms$\,$as$\,$a$\,$result$\,$of$\,$the$\,$difference$\,$in$\,$electrical$\,$potential$\,$between $n$- and $p$-type semiconductors. By$\,$recombining,$\,$electron/hole$\,$pairs$\,$releases$\,$a$\,$photon$\,$whose energy is defined by the semiconductor band-gap. This process is spontaneous$\,$but$\,$can$\,$also be stimulated by photons passing nearby. In order to enhance stimulated recombination$\,$of electron/hole$\,$pairs,$\,$the$\,$gain$\,$medium$\,$is$\,$surrounded$\,$by$\,$a$\,$cavity,$\,$as$\,$in$\,$every$\,$other$\,$laser$\,$system.
In the simplest diode laser design, an optical waveguide $-$ trenched in the crystal surface $-$ confines the light to a relatively narrow line. The two ends of the crystal are$\,$cleaved$\,$to$\,$form a Fabry–Pérot resonator. Light reflects back and forth inside the cavity$\,$and$\,$is$\,$amplified$\,$by stimulated$\,$emission.$\;$Finally,$\,$if$\,$there$\,$is$\,$more$\,$amplification$\,$than$\,$loss,$\,$the$\,$diode$\,$starts$\,$lasing.
\vspace{2pt}
\newline The emission frequency of a diode laser$\,$can$\,$be$\,$fine-tuned$\,$by$\,$adjusting$\,$the$\,$current$\,$across$\,$the $p$-$n$ junction and its temperature. If$\,$the$\,$laser$\,$diode$\,$is$\,$tuned$\,$by$\,$adjusting$\,$the$\,$current$\,$at$\,$fixed temperature, mode hops$\,-\,$\textit{ie}$\,$jumps$\,$over$\,$large$\,$wavelength$\,$intervals$\,-\,$will$\,$occur$\,$after$\,$a$\,$short continuous dependence of the wavelength on the current.$\;$These$\,$points$\,$of$\,$mode$\,$instability can be shifted by changing the $p$-$n$ junction temperature. Most of the time nevertheless, the desired wavelength can not be reached by adjusting only these two parameters.
\vspace{6pt}
\newline
\noindent \textbf{External-cavity diode laser (ECDL).} ECDL$\,$systems$\,$are$\,$tunable$\,$laser$\,$sources$\,$based$\,$on double heterostructures diode lasers, whose operation principal is sketched on figure$\,$\ref{fig:LaserDiode}(a). The light emitted from the front facet of the laser diode is collimated with a lens$\,$and$\,$hits$\,$a reflection grating, aligned in the "Littrow" configuration~\cite{4-1Littrow}, as illustrated on figure~\ref{fig:LaserDiode}(b). The first diffraction order is reflected and focused back into the laser diode. As$\,$this$\,$optical feedback is much higher than the reflection from the diode front facet, the extended$\,$cavity, formed by the diode rear facet (highly reflective) and the reflection grating,$\,$forces$\,$the$\,$diode into single-frequency operation. Since the length of this extended cavity resonator is larger than the diode one, ECDLs provide lower phase noise and smaller emission linewidth than free-running laser diodes. Moreover, in this external-cavity configuration, the frequency$\,$of the master oscillator (that is, of the laser diode) can be coarsely tuned over several$\,$tens$\,$of nanometers by simply rotating the reflection grating.
\vspace{6pt}
\newline
\noindent \textbf{Tapered amplifier (TA).} A TA is an optical amplifier$\,$that$\,$is$\,$usually$\,$used$\,$to$\,$increase$\,$the power of the laser field generated by the ECDL system.$\;$Its$\,$operation$\,$principle$\,$is$\,$as$\,$follows. The diode output beam is injected into a ridge waveguide, which is$\,$only$\,$a$\,$few$\,$microns$\,$wide. Thereafter, the light gets into the tapered region of the gain medium,$\,$whose$\,$width$\,$increases towards the output facet of the TA (up to hundred of micrometers). This tapered$\,$structure is required to prevent$\,$the$\,$TA$\,$from$\,$being$\,$damage$\,$by$\,$large$\,$intensities.$\;$The$\,$whole$\,$tapered$\,$area is covered with an electrode for supplying the pump$\,$current$\,$(typically$\,$a$\,$couple$\,$of$\,$amperes), which$\,$makes$\,$the$\,$device$\,$amplifying,$\,$just$\,$as$\,$in$\,$a$\,$conventional$\,$semiconductor$\,$optical$\,$amplifier. The anti-reflection coating on the TA facets prevents any laser emission without seeding and ensures the single-mode operation of the amplifier. Because of its broad gain profile, the TA operates in a wide frequency range (typically, from 775 nm to 800 nm) and delivers a laser power greater than $2.5$ W, which is sufficient for our applications.


\newpage

\section{Methods I $-$ Absolute vapor temperature measurement}

\noindent As we have seen in chapter 1, many parameters characterizing the rubidium vapor depend on the absolute temperature $T$. It controls for instance the atoms velocity which$\;$impacts the Doppler width, the filling and transit rates and the transport length scales. It also sets the atomic density. Being able to accurately measure $T$ is thus of paramount$\;$importance. I start this section by introducing the Doppler-free saturated absorption$\;$spectroscopy. This technique provides an accurate frequency reference used to measure $T$ by fitting the transmission spectrum of a low power beam, as we will see in a second step. 

    \subsection{Saturated absorption spectroscopy}
 
\subsubsection{Doppler-limited spectroscopy} 
    
\noindent Frequencies at which hyperfine transitions occur can all be seen as absolute and universal frequency references. They can be used to calibrate frequency-measuring devices such as lambdameters for instance. At millikelvin temperatures, Doppler broadening is inexistent. In that case, the transmission profile of a weak$\,$probe$\,$passing$\,$through$\,$a$\,$cold$\,$rubidium$\,$gas$\,$as function of its frequency is a succession of dips. Each$\,$dip$\,$is$\,$related$\,$to$\,$a$\,$hyperfine$\,$transition. By identifying them, one is able to:
\begin{itemize}
\vspace{-6pt}
    \item [(1)] define an absolute frequency reference, by associating the minimum of one dip to the corresponding tabulated hyperfine transition frequency.
    \vspace{-6pt}
    \item [(2)] scale the frequency axis accurately, by associating the distance between two dips to the difference in the transition frequencies to which they correspond.
\end{itemize}
\vspace{-6pt}
\noindent The precision of this calibration is limited by the dips linewidth, which is typically of the order of $\Gamma \simeq 2 \pi \!\times\! 6 \; \mathrm{MHz}$. However, the situation drastically changes at room temperature. The width$\;$of the Maxwell-Boltzmann distribution is about $2\pi \!\times\! 300$ MHz at $T \simeq 20^{\circ}$C, \textit{ie} of the order of the splittings between hyperfine excited states and much larger than $\Gamma$. Because$\,$of$\,$the$\,$Doppler$\,$broadening,$\,$the$\,$dips$\,$will$\,$therefore$\,$merge$\,$with$\,$each$\,$other,$\,$making$\,$the previous identification difficult if not impossible.$\,$Moreover, in cases where steps$\,$(1)$\,$and$\,$(2) are still doable (when the laser frequency is tuned over the $D_{1}$ line of $^{87}$Rb for instance, for$\,$which the excited state splitting $\delta_{HF} \simeq 2\pi \!\times\! 815$ MHz is larger than the Doppler$\,$width) the resolution on the frequency reference is then limited by the Doppler linewidth. 

\subsubsection{Doppler-free spectroscopy} 

\indent In order$\,$to$\,$overcome$\,$the$\,$issue$\,$of$\,$Doppler$\,$broadening$\,$without$\,$cooling$\,$down$\,$the$\,$atomic$\,$vapor to millikelvin temperatures, one usually resorts to saturated absorption spectroscopy~\cite{4-2Smith} which is based on a simple pump-probe experiment. A laser beam is sent$\,$through$\,$the$\,$vapor to saturate the atomic transitions (pump) and reflects back onto a mirror. The counter-propagating reflected beam (probe) is separated from the incoming one by a beam splitter. As pump and probe address opposite velocity classes (since they counter-propagate inside the rubidium cell), only atoms having a zero velocity projection on the optical axis are resonant simultaneously with both lasers. In that case, the absorption of the probe beam is reduced by the saturation induced by the pump. Transmission peaks appear then in the transmission spectrum when the laser frequency matches one of the transition$\;$frequencies between ground and hyperfine excited states. Those peaks are clearly visible in the$\;$spectra
\newpage 
\noindent plotted on insets (a) and (c) of figure~\ref{fig:AbsProfile}, obtained by scanning the laser frequency$\;\omega\;$over the rubidium $D_{2}$ line. Four large dips are also clearly discernible. 
\vspace{-4pt}
\begin{itemize}
    \item [$\bullet$] The $1$st and $4$th (from left to right) result from the overlap of the Doppler-broadened absorption lines associated to transitions from the ground states $F_{g} = 2$ and $F_{g} = 1$ to the hyperfine states within the fine level $5^{2}P_{3/2}$ of rubidium 87 (see figure~\ref{fig:AbsProfile}(a)). The spacing between these two dips is therefore equal to the ground state hyperfine splitting in rubidium 87 ($\delta_{0} \simeq 6.8$ GHz). 
    \vspace{-6pt}
    \item [$\bullet$] The $2$nd and $3$rd result from the overlap of the Doppler-broadened absorption lines associated to transitions from the ground states $F_{g} = 3$ and $F_{g} = 2$ to the hyperfine states within the fine level $5^{2}P_{3/2}$ of rubidium 85. The spacing between the dips is thus the ground state hyperfine splitting in rubidium 85 ($\delta_{0} \simeq 3.0 \, \mathrm{GHz}$). 
\end{itemize}
\vspace{-3pt} 
\noindent In the $D_{2}$ line, three transitions are allowed between each ground state and the hyperfine excited state manifold. Three saturated absorption peaks are thus expected in each of$\;$the large dips of the transmission profiles. As you may have noticed, the saturated$\;$absorption spectra show however a more complex structure: crossover resonance peaks appear exactly in between each pair of transition peaks. By identifying each feature in these spectra and by calculating the frequency spacing between two of them, one can then calibrate the frequency axis with a much better accuracy than using Doppler-limited spectroscopy. The resolution is now given by the width of the transmission peak, which is of the order of the decoherence rate $\Gamma/2$ (or $\Gamma$ for cycling transitions), when no other phenomenon than spontaneous emission contributes to the linewidth (such as Rb-Rb collisions for$\;$example). In figure~\ref{fig:AbsProfile}, I choose the position of the $F_{g} = 3 \rightarrow F_{e} = 4$ transition peak (2) of $^{_{85}}$Rb as origin for the $x$-axis. It is $1.13$ GHz blue-detuned with respect to the $F_{g} = 2 \rightarrow F_{e} = 3$ transition peak (1) of $^{_{87}}$Rb, used to scale the frequency axis.      

\subsection{Vapor temperature and atomic density}
 
\noindent As mentioned in~\ref{subsec:RbCell}, measuring the absorption of a low power probe beam propagating through a rubidium cell is an accurate way of accessing the vapor absolute temperature $T$. The absorption of a monochromatic laser field propagating along the $z$-direction across a uniform density atomic vapour is given by the Beer-Lambert$\;$law~\cite{2-9Siddons}:  
\begin{equation}
\mathcal{I}(\mathbf{r}, z) = \mathcal{I}_{0}(\mathbf{r}) \, \exp \left \{ -\int_{0}^{z} \alpha \left[T, \mathcal{I}(\mathbf{r}, z') \right] \, z' \, \mathrm{d}z' \right \} \simeq \mathcal{I}_{0}(\mathbf{r}) \, \exp \left[ - \alpha_{0} \!\left(T\right) z \right]
\label{BeerLambert}
\end{equation}
\noindent where $\mathcal{I}(\mathbf{r}, z)$ is the laser intensity after a propagation over $z$ in the$\,$medium,$\,\alpha_{0} \!=\! k \, \mathrm{Im}\left[\chi^{_{(1)}}\right]$ the linear$\;$absorption coefficient and $\mathcal{I}_{0}(\mathbf{r})$ the laser intensity in the cell entrance plane. Equation~\eqref{BeerLambert} holds only if the intensity of the incoming field is sufficiently weak to ensure that $\alpha$ is independent of $\mathcal{I}$. In that case, the absorption profile can$\;$be easily fitted ($T$ being the only fitting parameter). A discussion on how weak the light has to be for this assumption to be valid is given in~\cite{2-9Siddons}. In practice, the input light intensity has to be way smaller than the on-resonance saturation intensity. The medium transmission is then defined$\;$by: $\mathcal{T} = \exp \left[ -\alpha_{0}\!\left( T \right) L \right]$ ($L$ being the length of the cell). In the 3-level system described$\,$in$\,$2.3, a good approximation of the dielectric susceptibility$\,\chi^{_{(1)}}\,$is$\;$obtained$\;$by$\;$summing the$\;$linear susceptibilities associated to the $\ket{1} \rightarrow \ket{3}$ and the $\ket{2} \rightarrow \ket{3}\;$transitions:$\,\chi^{_{(1)}}_{\vphantom{2 \shortto 3}} \simeq \chi^{_{(1)}}_{1 \shortto 3} + \chi^{_{(1)}}_{2 \shortto 3}$ (because $\mathcal{I} \ll \mathcal{I}_{s}$). The expression of $\chi^{_{(1)}}_{1 \shortto 3}$ and $\chi^{_{(1)}}_{2 \shortto 3}$ taking into account Doppler$\;$broadening has$\,$been$\,$derived$\,$in$\,$section$\;$2.3.$\,$Therefore,$\,$the$\,$theoretical$\,$Doppler$\,$broadened$\,$transmission of the medium is known and can be used to fit the experimental transmission spectra.

\newpage 

\begin{figure}[h]
\center
\includegraphics[width=\columnwidth]{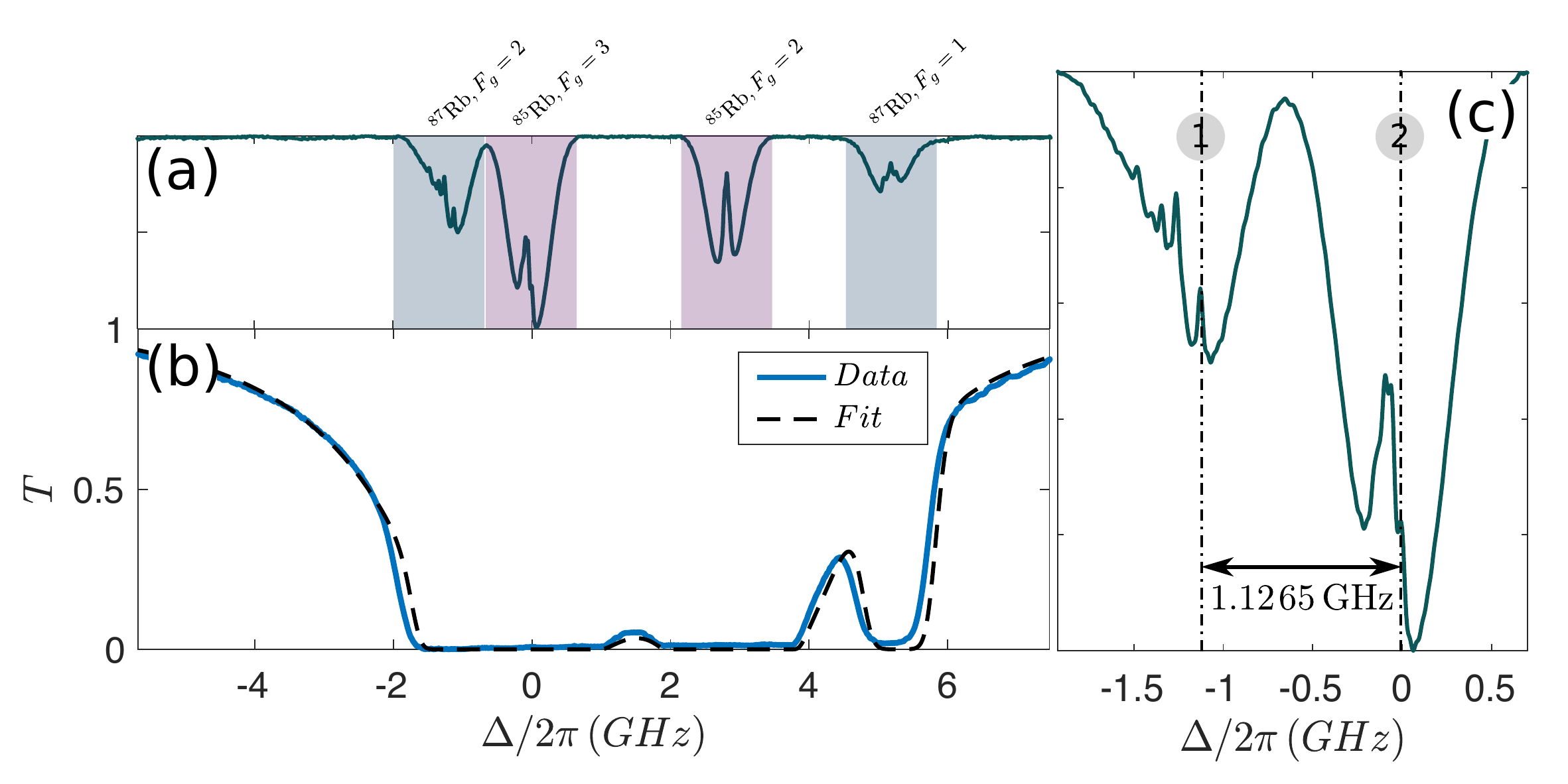} 
\caption{(a) and (c): Saturated absorption spectra$\,$obtained$\,$using$\,$an$\,$\textbf{isotopic$\,$mixture} of rubidium atoms (natural proportions), heated up at $50^{\circ}$C to increase the atomic density and the signal over noise ratio. As mentioned in the text, the $F_{g} = 3 \rightarrow F_{e} = 4$ transition peak (2) has been chosen as origin for the $x$-axis; the frequency scale has been calibrated using the frequency spacing between peaks (2) and (1), the latter corresponding to the $F_{g} = 2 \rightarrow F_{e} = 3$ transition. (b) Transmission$\,$spectrum$\,$of$\,$an$\,$isotopically$\,$pure$\,^{_{85}}$Rb$\,$vapor heated up at $134^{\circ}$C. The experimental profile (blue line) has been fitted to extract the absolute vapor temperature (dashed line). The cell is 7.5 cm long.}
\label{fig:AbsProfile}
\end{figure}

\noindent The experimental transmission profile is obtained by sending on a photo-diode a low power Gaussian beam which has propagated through the rubidium cell. The laser frequency is scanned over $15\,$GHz across the rubidium $D_{2}$ line. At high temperatures, the transmission profile looks like the blue solid curve on figure~\ref{fig:AbsProfile}(b). As you may have noticed, the$\;^{_{85}}$Rb vapor is not truly isotopically pure. The last transmission dip on the right is created$\;$by a small fraction (less than $1\%$) of $^{_{87}}$Rb atoms, at resonance with the laser field when it gets across the $F_{g} \!=\! 1 \rightarrow F_{e}$ transition of $^{_{87}}$Rb. Similarly, the transmission$\;$dip$\;$on$\;$the$\;$right gets broadened$\;$by$\;$the $F_{g} \!=\! 2 \rightarrow F_{e}$ transition of $^{_{87}}$Rb. We have therefore to extend the theoretical model to the case of an isotopic mixture of rubidium atoms in order to correctly fit the transmission spectrum. We also need$\;$an$\;$accurate$\;$frequency$\;$reference,$\;$provided$\;$by the saturated absorption spectrum of figures~\ref{fig:AbsProfile} (a)$\,$and$\,$(c). The best fit curve is shown in black dashed on inset (b). From this fit, we extract both the vapor temperature$\;$and$\;$the $^{_{87}}$Rb fraction, which is about $0.4 \, \%$ in this cell. Once $T$ is known, parameters$\;$such$\;$as the atomic density$\;N$, can be evaluated. 
\vspace{2pt}
\newline
\noindent The vapor pressure (for $298 \mathrm{K} < T < 550$ K) is given by the following formula~\cite{4-3Nesmeyanov}: 
\begin{align}
    \begin{cases}
    \label{4LevelPopulation}
        \mathrm{log}_{10} \left( \mathcal{P}_{v} \right) = 2.881 + 4.857 - \frac{4215}{T} \;\; (\mathrm{solid} \; \mathrm{phase}) \\
        \mathrm{log}_{10} \left( \mathcal{P}_{v} \right) = 2.881 + 4.312 - \frac{4040}{T} \;\; (\mathrm{liquid} \; \mathrm{phase})
    \end{cases} 
\end{align}
\noindent where $\mathcal{P}_{v}$ stands for the vapor pressure (in Torr). The atomic density $N$ is then obtained straightaway using the ideal gas law:

\begin{equation}
N = 133.323 \times \frac{\mathcal{P}_{v}(T)}{k_{B} \, T}
\end{equation}
\noindent The factor $133.323$ converts the vapour
pressure from Torr to Pascal. Since both isotopes are present in the cell, the number densities need to be calculated separately according to their abundance. The vapor pressure$\;\mathcal{P}_{v}$ and the atomic density $N$ (for an isotopically pure rubidium vapor) are plotted on the figure below.

\begin{figure}[h]
\center
\includegraphics[width=1\linewidth]{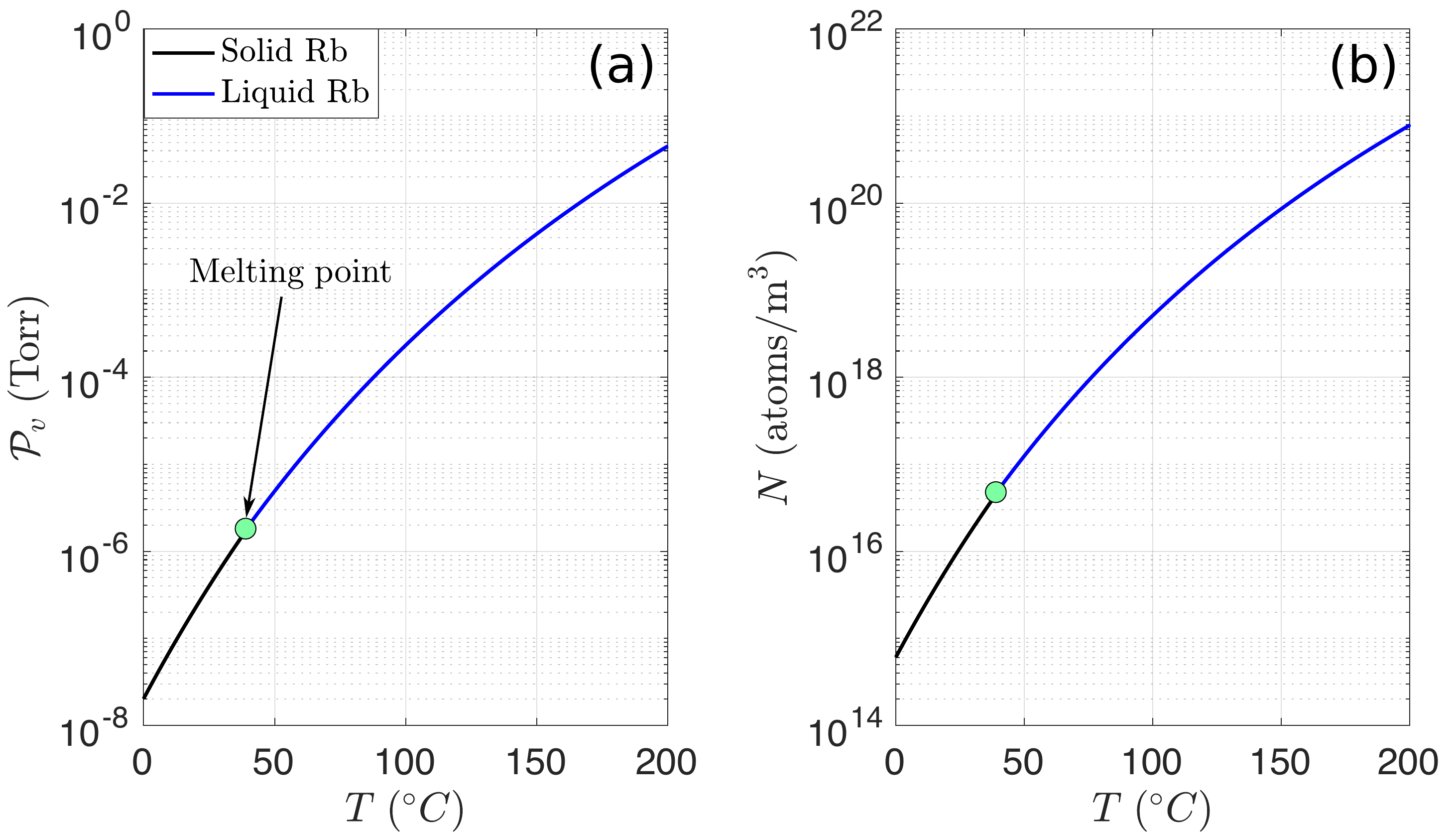} 
\caption{Vapor pressure (a) and atomic density (b) as function of the temperature $T$. The melting point of rubidium is $39.31^{\circ}$C. The density is in atoms/m$^{3}$.}
\label{fig:AtomicDensity}
\end{figure}
    
    \section{Methods II $-$ Nonlinear refractive index measurement} 
    
\noindent In propagating fluids of light, the strength of the photon-photon interaction is$\;$proportional to the third-order susceptibility $\chi^{_{(3)}}$ or, equivalently, to the nonlinear refractive index $n_{2}$. This quantity plays therefore an important role in studying the photon fluid physics and finding a convenient and precise method to measure it has been one of our main$\;$concerns during my thesis. This issue is far from being$\,$fully$\,$solved$\,$and$\,$the$\,$technique$\,$I$\,$will$\,$present$\,$is conjointly being improved$\;$by the next generation of PhD students and other teams$\;$from the fluid of light community~\cite{4-5Kaiser,4-4Boughdad}. Surprisingly, only a few methods exist to measure$\,$the nonlinear refractive index $n_{2}$ and all require the medium to be sufficiently thin in order to neglect the change in the beam shape along$\,$its$\,$propagation.$\;$The$\,$so-called$\,z$-scan$\,$technique is surely the most well-known among them~\cite{4-6Sheik, 4-7Wang}. In this subsection,$\,$I$\,$first$\,$elaborate$\,$on this thin medium approximation and the constraints$\,$it$\,$imposes$\,$on$\,$the$\,$beam$\,$width$\,$mainly. I then present in details the method we currently use in the lab to  accurately measure $n_{2}$.
   
\subsection{Thin medium approximation}  
  
\noindent When an intense Gaussian beam propagates inside a third-order Kerr medium, it induces a spatially dependent refractive index variation within the material, larger at its center than on its periphery, since $n = n_{0} + n_{2} \, \mathcal{I}(\mathbf{r},z)$. If $n_{2} < 0$ (resp. $n_{2} > 0$),$\;$the$\;$medium acts therefore as a negative (resp. positive) lens and makes the laser beam diverge (resp.$\;$focus). The$\;$\textbf{thin$\;$medium approximation} (TMA) consists in assuming self-effects do$\;$not affect the shape and width of the Gaussian intensity profile. In$\;$order to be more quantitative, let's introduce the medium effective focal length defining the propagation length above witch the TMA breaks down. Within the TMA, the intensity profile in the vicinity of the beam center reads: $\mathcal{I}(\mathbf{r},z) \simeq  \mathcal{I}_{0} \left( 1 - 2\, r^{2}/\omega_{0}^{2} \right)$. In order for this expansion to be correct, the beam should also be collimated over the length of the Kerr medium, which amounts to neglecting the z-dependence in its width. The nonlinear self-induced phase accumulated by the beam over its propagation is therefore given by: $ \Phi_{^{\mathrm{NL}}}(\mathbf{r},z)  \simeq k \, \frac{n_{2}}{n_{0}} \, \mathcal{I}_{0}  \left( 1 - 2\, r^{2}/\omega_{0}^{2} \right) z$. By identifying this expression at $z = L$ (\textit{ie}, in the$\;$cell$\;$output$\;$plane) with the phase profile of a Fresnel lens~\cite{4-8Carcole}: $\Phi_{^{\mathrm{FL}}}(\mathbf{r}) = 2 \pi \times r^{2}/2 f \lambda$ ($f$ being the focal length), one can define$\,$the medium effective focal length $f_{\mathrm{eff}}$ as follows:
\begin{equation}
    f_{\mathrm{eff}} = \frac{1}{2} \, \frac{z_{r}}{k \, (n_{2} / n_{0}) \, \mathcal{I}_{0} \, L} = \frac{1}{2} \, \frac{z_{r}}{ \Phi_{\mathrm{NL}}(\mathbf{0}, L)},
    \label{EffectiveFocalLength}
\end{equation}
\noindent where $z_{r} = \pi \omega_{0}^{2} / \lambda$ is the Rayleigh length. We assume here the on-axis$\;$intensity$\,\mathcal{I}_{0}\,$does$\;$not depend on $z$ (which is true for lossless mediums only). In lossy ones, it$\,$decays$\,$exponentially over $z$ and $\mathcal{I}_{0}$ should then be replaced by: $\langle \mathcal{I}_{0}(z') \rangle_{L} = \mathcal{I}_{0}(0) \left[ 1-\exp(-\alpha L) \right] / \alpha L$ in~\eqref{EffectiveFocalLength}. In any case, $f_{\mathrm{eff}}$ goes to infinity when $n_{2}$ tends to zero: the nonlinearity is so weak that it does not induce any lensing effect anymore. Reversely, when $n_{2}$ increases, $f_{\mathrm{eff}}$ decreases. In other words, the length scale over which self-effects$\;$start$\;$modifying$\;$the$\;$beam$\;$intensity profile decreases, as expected. The TMA finally reduces to the set of assumptions below: 
\begin{equation}
    \mathrm{(i)} \; L \ll z_{r} \hspace{0.25cm} \mathrm{and} \hspace{0.25cm} \mathrm{(ii)} \; L \ll f_{\mathrm{eff}}
\end{equation}
\noindent The first assumption (i) basically translates the fact that the beam should be collimated over a distance greater than the cell length, so as to neglect diffraction. The second$\;$one$\;$(ii) defines then an upper limit on the average self-phase accumulated at the beam center along propagation:$\,\Phi_{\mathrm{lim}} = z_{r}/(2 \, L)$,$\,$above$\,$which$\,$the$\,$TMA$\,$breaks$\,$down.$\;$For$\,$a$\,7.5 \, \mathrm{cm}\,$long$\,$cell$\,$and a beam width of $500\, \mu\mathrm{m}$ for instance, the self-phase should be much smaller than $\Phi_{\mathrm{lim}} \simeq 2 \pi$ for the$\,$TMA$\,$to$\,$apply.$\;$Because$\,\Phi_{\mathrm{lim}}\,$scales$\,$with$\,$the$\,$beam$\,$cross-section,$\,$the$\,$best$\,$is$\,$to$\,$increase $\omega_{0}$ to push it up. That is why $\omega_{0}$ is usually about $2$ mm when we measure $n_{2}$ in$\;$experiments. Now that foundations have been laid, I will present the measurement$\;$method$\;$itself.

    \subsection{Far-field measurement}
        
\noindent When a Gaussian beam passes through a Kerr medium, a concentric ring intensity pattern forms$\;$in the far field. This phenomenon has aroused wide interest in the nonlinear optics community since Callen and al.~\cite{4-9Callen} observed the far-field annular intensity$\;$distribution$\;$of a He-Ne laser beam passing through a nonlinear liquid CS$_{2}$. Similar phenomena have been observed in liquid crystals by Durbin and al.~\cite{4-10Durbin}. Recently, theoretical$\;$progress$\;$have$\;$been made in understanding the linear relationship existing between the number of rings in far-field and the on-axis nonlinear phase accumulated by the laser over its propagation inside a non-saturable Kerr medium~\cite{4-11Deng}. Such a relation has been experimentally observed in a colloidal solution of oil and gold nanoparticles~\cite{4-12Nascimento}. It provides an easy way to measure the nonlinear refractive index $n_{2}$ of the material. This technique starts being used in the fluid of light community~\cite{4-5Kaiser,4-13Fontaine} and have recently been extended to the case of anisotropic nonlinear refractive indices in~\cite{4-4Boughdad}. The following paragraph is$\,$a$\,$review$\,$on$\,$the$\,$theoretical tools$\,$developed$\,$to$\,$describe$\,$the$\,$formation$\,$of$\,$this$\,$self-induced$\,$ring-shaped$\,$pattern$\,$in$\,$far-field. A$\;$recent unpublished work by Nicolas$\;$Pavloff~\cite{4-14Pavloff} is also discussed. 

\subsubsection{Theoretical description of the self-defocusing case}
\label{subsubsec:TheoryRing}
       
\noindent Let's consider that a Gaussian beam propagates along the $z$-axis through a self-defocusing ($n_{2} < 0$) Kerr medium of thickness $L$. Let's$\;$also suppose that the beam waist lies exactly at the medium entrance plane ($z = 0$). Therefore, the radius $R(z)$ of the$\;$beam wavefront curvature is positive inside the medium (\textit{ie} for $z>0$). Both diffraction and self-defocusing make the beam diverge in that case. Conversely, when the waist is set at the output plane, they compete with each other, as the beam convergence will in some ways counterbalance self-defocusing inside the nonlinear medium.$\;$This$\,$competition$\,$can$\,$dramatically$\,$modify$\,$the far-field ring pattern~\cite{4-11Deng}. In practice however, as long$\;$as the condition $L \ll z_{r}$ is fulfilled, setting the waist at the input or the output plane does not change significantly the result. The complex envelope of the laser electric field at the medium entrance plane reads:

\begin{figure}[h]
\center
\includegraphics[width=\columnwidth]{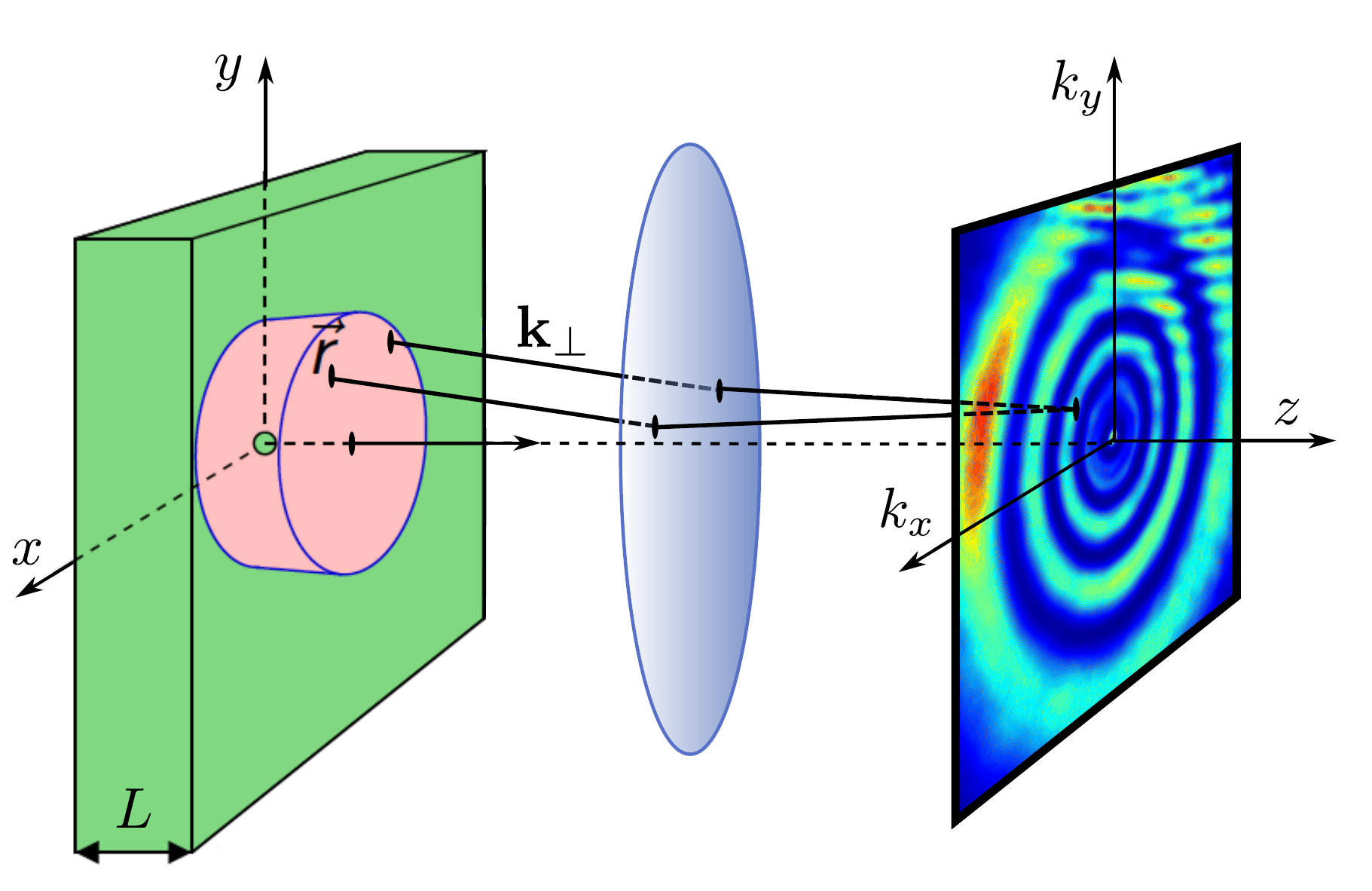} 
\caption{Sketched of the experimental setup. A Gaussian beam propagates through a Kerr medium (green slab). A lens at the exit plane focuses it. The far-field (Fourier$\,$space) is magnified and imaged on a camera. An example of the ring-shaped pattern we observe$\,$is shown on the right. Figure inspired from the N.Pavloff's talk at QFLM$\;$(2018)~\cite{4-14Pavloff}}
\label{fig:RingsExp2D}
\end{figure}  

\begin{equation}
    \mathcal{E}(\mathbf{r}, 0) = \mathcal{E}_{0} \, \exp \! \left( -\frac{r^{2}}{w_{0}^{2}}\right) 
\end{equation}
\noindent By propagating through the nonlinear medium, it accumulates an additional phase shift:
\begin{equation}
    \Delta \Phi(\mathbf{r}) = \langle \Phi_{\mathrm{NL}}(\mathbf{r}, z') \rangle_{L} = \frac{k}{n_{0}} \int_{0}^{L} \Delta n (\mathbf{r}, z') \, \mathrm{d} z',
    \label{NonlinearPhaseShift}
\end{equation}
\noindent where $\Delta n (\mathbf{r}, z) = n_{2} \, \mathcal{I}(\mathbf{r}, z)$ is the nonlinear change of refractive index. Within the TMA, the Gaussian intensity profile of the laser beam is almost not modified along propagation. The electric field envelope at the exit plane can thus be expressed as: 
\begin{equation}
    \mathcal{E}(\mathbf{r}, L) = \mathcal{E}_{0} \, \exp \! \left(-\frac{\alpha\, L}{2}\right) \exp \! \left( -\frac{r^{2}}{w_{0}^{2}}\right) \exp \! \left( - i \Phi(\mathbf{r}) \right),
    \label{RingOutputField}
\end{equation}
\noindent where $\alpha$ is, as usual, the linear absorption coefficient (nonlinear absorption is negligible). The phase shift $\Phi(\mathbf{r})$ is defined by: 
\begin{equation}
    \Phi(\mathbf{r}) = k L + \frac{k r^2}{2 R(L)} + \Delta \Phi(\mathbf{r}) \simeq k L + \langle \Phi_{\mathrm{NL}}(\mathbf{0}, z') \rangle_{L} \, \exp \! \left( -2\,r^{2} / w_{0}^{2} \right).   
\end{equation}
\noindent The Gaussian phase shift $k \, r^2/2 R(L)$, arising from the wavefront curvature, is neglected$\;$as we assume the beam is collimated along the cell ($L \ll z_{r}$). The far-field intensity pattern is obtained either by letting the beam evolve in free space (the intensity distribution can be expressed using the Fraunhofer approximation of the Fresnel-Kirchhoff$\,$diffraction$\,$formula in that case) or by imaging the focal plane of a converging lens. We$\,$prefer$\,$this$\,$second$\,$option in practice at it saves space on the optical table. The intensity distribution in k-space is then given by the following equation:
\begin{equation}
     \mathcal{I}(\boldsymbol{{k}_{\perp}}) = (2 \, \pi)^{2} \, \mathcal{I}_{0}(L) \left| \int_{0}^{\infty}   J_{0} \left( k_{\perp} r \right) \exp \left( -\frac{r^{2}}{\omega_{0}^{2}}-i \Phi(r) \right) r \, \mathrm{d}r \, \right|^{2}.
    \label{RingKSpaceIntensity}    
\end{equation}

\noindent $J_{0}$ standing for the zero-order Bessel function of the first kind and $\mathcal{I}_{0}(L) = \mathcal{I}_{0} \, \exp(-\alpha L)$.  Equation$\;$\eqref{RingKSpaceIntensity}$\,$can$\,$not$\,$be$\,$more$\,$simplified$\,$and$\,$does$\,$not$\,$provide$\,$any$\,$insight$\,$into$\,$the$\,$physical process underlying the rings pattern formation in far-field. Let's try to push further our understanding by studying the 1D case first, which is more intuitive.
\vspace{10pt}
\newline
\textbf{(a)} \textbf{1D case} $\boldsymbol{-}$ \textbf{Stationary phase approximation}
\vspace{10pt}
\newline
\noindent I now suppose that the Gaussian beam is infinitely elongated$\,$in$\,$the$\,y$-direction,$\;$as$\,$sketched on figure~\ref{fig:RingsExp1D}(a). The k-space electric field envelope is then given by:  
\begin{equation}
    \mathcal{E}_{\scriptscriptstyle{\mathrm{1D}}}(k_{x}) = \; \mathcal{E}_{0}(L) \int_{-\infty}^{\infty} \exp \left(- \frac{x^{2}}{\omega_{0}^{2}} \right) \exp \left[i(k_{x} \, x + \Phi(x)) \right] \mathrm{d}x  
    \label{RingKSpaceIntensity1D}
\end{equation}        
\noindent where $\mathcal{E}_{0}(L) = \mathcal{E}_{0} \, \exp (-\alpha L/2-ikL)$. Let's define $\Psi_{k_{x}}(x) = k_{x} \, x - \Phi(x)$. This quantity is plotted as function of $x/\omega_{0}$ on figure~\ref{fig:RingsExp1D} (b) for a given value of $k_{x}$. The$\;$slope$\;$of$\;$the$\;$black dashed line is $k_{x} \, \omega_{0}$.  As you may have noticed, $\Psi$ has two extrema $x_{1}$ and $x_{2}$, both$\;$lying in the $x>0$ half-plane when $k_{x}$ is positive. When $k_{x}$ decreases, the slope of the dashed$\;$line also do so and $x_{1}$ progressively tends to zero while $x_{2}$ goes to infinity.$\,$Using$\,$the$\,$\textbf{stationary} \textbf{phase approximation}, we can rewrite the k-space electric field envelope~\eqref{RingKSpaceIntensity1D} as$\;$follows:
 
\newpage 
 
\begin{figure}[h]
\center
\includegraphics[width=\columnwidth]{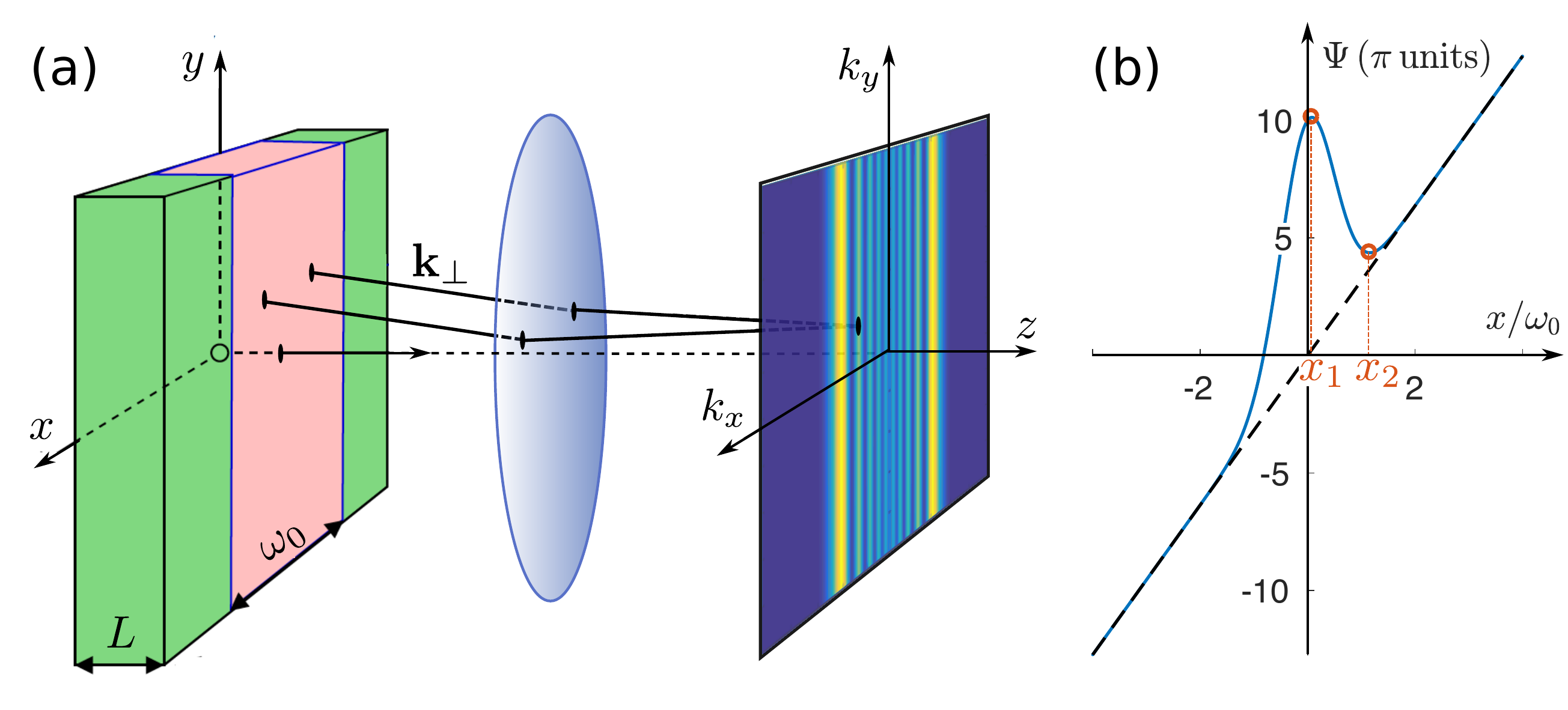} 
\caption{(a) Sketch of the 1D case. The beam is spatially elongated along the $y$-axis. A$\;$cylindrical lens in the exit plane focuses it in the $x$-direction. (b) Variation of $\Psi_{k_{x}}$ as function of $x/\omega_{0}$ for a given value of $k_{x}$. In k-space, the electric field at $k_{x}$ is obtained by summing the fields coming from the vicinity of the points having $x_{1}$ or $x_{2}$ as abscissa in the cell exit$\;$plane. Figure inspired from the Nicolas Pavloff's talk at QFLM (2018)~\cite{4-14Pavloff}}
\label{fig:RingsExp1D}
\end{figure}    
  
\begin{equation}
    \mathcal{E}_{\scriptscriptstyle{\mathrm{1D}}}(k_{x}) \simeq \sqrt{2 \pi} \, e^{-i \, k \, L} \! \sum_{\alpha = 1,2} \frac{\mathcal{E}(x_{\alpha},0)}{\sqrt{\left| \Psi''_{\!k_{x}}(x_{\alpha}) \right|}} \, \exp\left[- i \left(\Psi_{\!k_{x}}(x_{\alpha}) + \pi \, \sigma_{\alpha}/4 \right) \right],
    \label{PavloffModel}
\end{equation} 

\noindent $\Psi''_{\!k_{x}}$ being the second derivative of $\Psi_{\!k_{x}}$ and $\sigma_{\alpha}$ the sign of $\Psi''_{\!k_{x}}$ at $x_{\alpha}$. This approach is valid only if $\Psi_{k_{x}}$ has extrema, which is ensured solely when the slope of the black dashed line on \ref{fig:RingsExp1D}(b) is smaller than the maximum of $\Phi'$, that$\;$is, when: $k_{x} \, \omega_{0} < 2\left| \langle \Phi_{\mathrm{NL}}(0,z') \rangle_{L} \right| e^{-1/2}$. The intensity$\;$distribution at $k_{x}$ in Fourier space seems therefore to result from a two-wave interference process. According to the Fermat's$\,$principle,$\,$the$\,$light$\,$rays$\,$coming$\,$from$\,$a$\,$point of abscissa $x$ in the cell exit plane propagate in straight lines along a direction defined by the wave-vector $\boldsymbol{k} = k_{x} \boldsymbol{e_{x}} + k_{z} \boldsymbol{e_{z}}$. The longitudinal component of $\boldsymbol{k}$, $k_{z}$,  is given by: $k_{z} = \sqrt{k_{0}^{_{2}}-k_{x}^{_{2}}}$ ($k_{0}$ being the in-air laser wave-vector). The transverse component, $k_{x}$, is related to the derivative of $\Phi$ through: $k_{x}(x) = +\Phi'(x)$. The light rays whose transverse wave-vector are equal to $k_{x}$ come therefore from the points in the cell exit plane for which $\Psi'_{\!k_{x}}(x) = 0$. The preceding equation admits exactly two solutions,$\;$which$\;$are$\;x_{1}\;$and$\;x_{2}$. In$\,$the$\,$Fourier$\,$space, the envelope of the electric field at $k_{x}$ is$\,$thus$\,$obtained$\,$by$\,$summing$\,$the fields coming from the vicinity of the points $x_{1}$ and $x_{2}$ at the medium exit plane, which$\;$is what equation~\eqref{PavloffModel} suggests indeed. We can push a bit further assuming the amplitudes of the interfering waves are equal. By doing so, the  intensity envelope of the interference pattern will obviously be improperly described but not the oscillating trend, which only depends on the phase terms $\Psi_{\!k_{x}}(x_{1}) - \pi/4$ and $\Psi_{\!k_{x}}(x_{2}) + \pi /4$. We thus obtained:
\begin{equation}
    \mathcal{I}_{\scriptscriptstyle{\mathrm{1D}}}(k_{x}) \simeq C \left[ 1 + \sin(\Delta \Psi_{\!k_{x}}) \right],
    \label{PavloffModelSimple}
\end{equation}
where $\Delta \Psi_{\!k_{x}} = \Psi_{\!k_{x}}(x_{1})-\Psi_{\!k_{x}}(x_{2})$ and $C$ is a constant. Looking at~\ref{fig:RingsExp1D}(b),$\,$you$\,$can$\,$convince yourself that $\Psi_{\!k_{x}}(x_{2})$ goes to zero and $\Psi_{\!k_{x}}(x_{1})$ to $\left|\langle \Phi_{\mathrm{NL}}(0,z') \rangle_{L} \right|$ when $k_{x}$ tends$\,$toward$\,$zero.

\newpage

\begin{figure}[h]
\center
\includegraphics[scale=0.45]{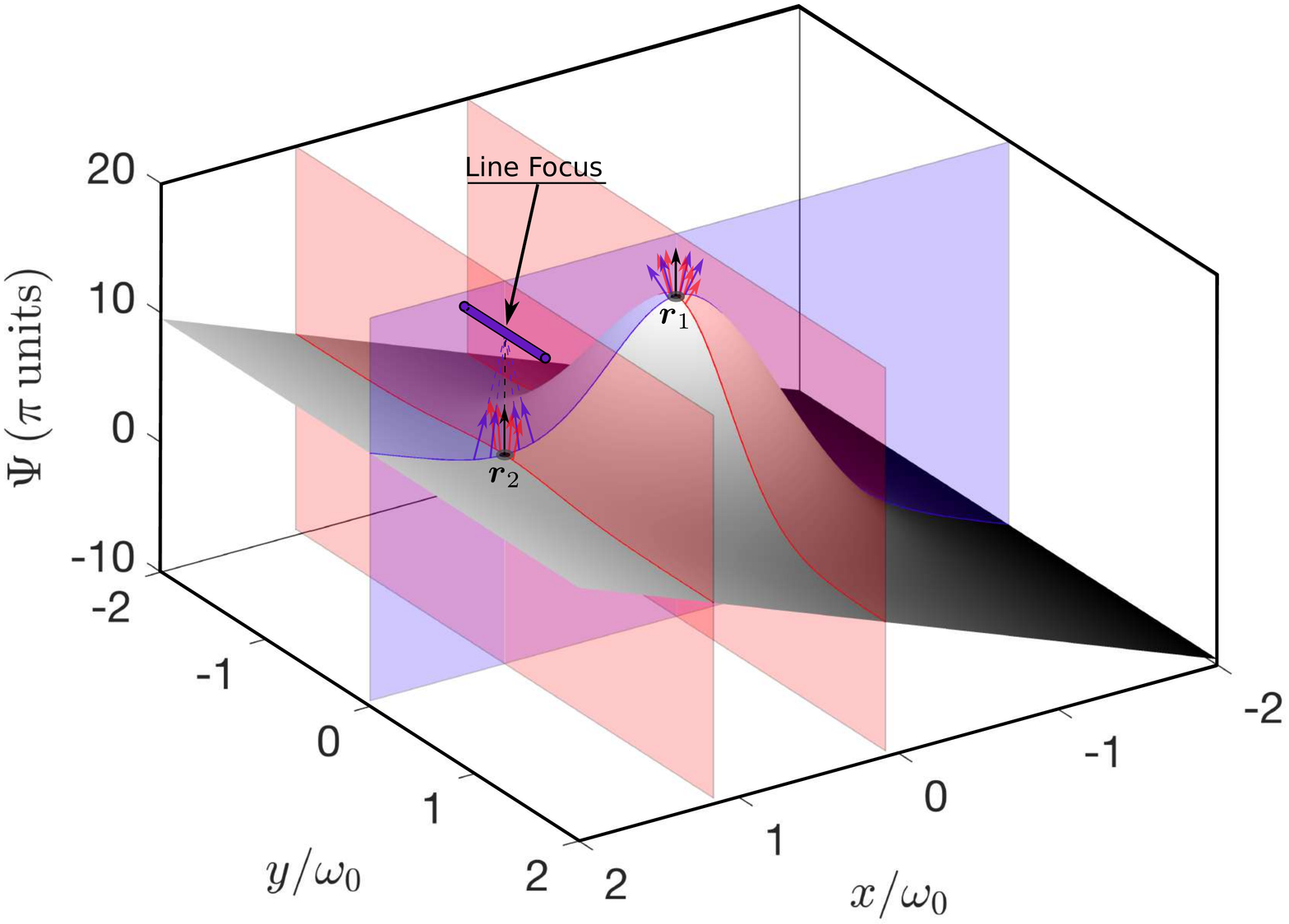} 
\caption{Variation of $\Psi_{\mathbf{k}_{\perp}}$ in the transverse plane for $\mathbf{k_{\perp}}\! = k_{x} \, \boldsymbol{{e}_{x}}$.$\,$The$\;$total$\;$electric$\;$field at $\mathbf{k_{\perp}}$ in Fourier space is obtained by summing the fields coming from the vicinity of the local maximum (at $\boldsymbol{r}_{1}$) and the saddle point (at $\boldsymbol{r}_{2}$). The light emitted at $\boldsymbol{r}_{2}$ accumulates an additional $\pi/2$ phase shift since it passes through a line focus.}
\label{fig:Phase2D}
\end{figure}

\noindent The phase difference $\Delta \Psi_{\!k_{x}}$ is therefore equal to $\left|\langle \Phi_{\mathrm{NL}}(0,z') \rangle_{L} \right|$ when $k_{x} = 0$. At low input intensity $\mathcal{I}_{0}$, the rubidium vapor behaves as a linear medium and the intensity profile in k-space$\,$is$\,$uniform$\,$according$\,$to$\,$equation~\eqref{PavloffModelSimple}$\,$(Gaussian$\,$in$\,$experiments).$\;$When$\,\mathcal{I}_{0}\,$increases, $\left|\langle \Phi_{\mathrm{NL}}(0,z') \rangle_{L} \right|$ rises$\,$accordingly,$\,$and$\,$the$\,$on-axis$\,$intensity$\,$in$\,$Fourier$\,$space,$\,\mathcal{I}_{\scriptscriptstyle{\mathrm{1D}}}(0)$,$\,$reaches$\,$a first maximum for $\left|\langle \Phi_{\mathrm{NL}}(0,z') \rangle_{L} \right| \!= \pi/2\,$and$\,$then$\,$a$\,$minimum$\,$for$\,\left|\langle \Phi_{\mathrm{NL}}(0,z') \rangle_{L} \right| \! = 3\pi/2$,$\,$etc. By extrapolating, the average on-axis nonlinear phase shift is equal to $2 n \pi + \pi/2$ when the k-space intensity distribution exhibits $2 n$ maximums plus a bright fringe at the center (that is, at $k_{x} = 0$) and to $(2 n -1)\pi + \pi/2$ when it has $2 n$ maximums plus a dark fringe$\;$at the center this time. Counting the "rings" in far-field (that is, in the 1D case, the pairs$\;$of maximums having the same transverse wave-vector in absolute value) provides therefore a way of measuring $\left|\langle \Phi_{\mathrm{NL}}(0,z') \rangle_{L} \right|$ (and also $n_{2}$, if $\mathcal{I}_{0}$ and $\alpha$ are known). Within the TMA, the ring-counting procedure in 1D can finally be formulated as follows:  
\vspace{-4pt}
\begin{itemize}
    \item [$\bullet$] when the far-field diffraction pattern is composed of $n$ rings plus a bright$\;$central$\;$spot (constructive interference at $k_{x} = 0$), the nonlinear phase shift is $2 n \pi + \pi/2$;  
    \vspace{-4pt}
    \item [$\bullet$] when the far-field diffraction pattern is composed of $n$ rings plus a dark$\;$central$\;$spot (destructive interference at $k_{x} = 0$), the nonlinear phase shift is $(2 n -1)\pi + \pi/2$;
\end{itemize}
\textbf{(a)} \textbf{General 2D case}
\vspace{10pt}
\newline
\noindent The stationary phase approximation can be generalized to the 2D case. The electric field envelope in k-space can indeed be approximated by the following formula~\cite{4-15McClure}:   
\begin{equation}
    \mathcal{E}_{\scriptscriptstyle{\mathrm{2D}}}(\mathbf{k_{\perp}}) \simeq 2 \pi \, e^{-i \, k \, L} \sum_{\alpha} \frac{\mathcal{E}(\boldsymbol{r}_{\alpha},0)}{\sqrt{\left| \mathrm{det} \left[\mathcal{H}_{\!\mathbf{k_{\perp}}\!}(\boldsymbol{r}_{\alpha})\right] \right|}} \, \exp\left[- i \left(\Psi_{\!\mathbf{k_{\perp}}\!}(\boldsymbol{r}_{\alpha}) + \pi \, \sigma \! \left[\mathcal{H}_{\!\mathbf{k_{\perp}}\!}(\boldsymbol{r}_{\alpha})\right]/4 \right) \right],
    \label{PavloffModel2D}
\end{equation} 
\noindent where $\mathbf{k_{\perp}} = k_{x} \, \boldsymbol{{e}_{x}} + k_{y} \, \boldsymbol{{e}_{y}}$ is the transverse wave-vector and $\boldsymbol{r_{\alpha}}$ the $\alpha$-th position in the medium output plane at which the gradient of $\Psi_{\!\mathbf{k_{\perp}}\!}(\boldsymbol{r}) = \mathbf{k_{\perp}} \! \cdot \! \boldsymbol{r} - \Phi(\boldsymbol{r})$ is zero. We have$\;$also introduced the Hessian matrix $\mathcal{H}_{\!\mathbf{k_{\perp}}\!}$ associated to the function $\Psi_{\!\mathbf{k_{\perp}}\!}$:
\begin{equation}
    \mathcal{H}_{\!\mathbf{k_{\perp}}\!}(\boldsymbol{r}_{\alpha}) = 
    \left. \begin{pmatrix} 
    \partial_{xx} \Psi_{\!\mathbf{k_{\perp}}}  & \partial_{xy} \Psi_{\!\mathbf{k_{\perp}}} \\
    \partial_{yx} \Psi_{\!\mathbf{k_{\perp}}} & \partial_{yy} \Psi_{\!\mathbf{k_{\perp}}} 
    \end{pmatrix} \right|_{\boldsymbol{r} = \boldsymbol{r}_{\alpha}}
\end{equation}
\noindent and the signature $\sigma$ of $\mathcal{H}_{\!\mathbf{k_{\perp}}\!}$ (which is the number of positive minus the number$\;$of negative eigenvalues of $\mathcal{H}_{\!\mathbf{k_{\perp}}}\!$). The phase $\Psi_{\!\mathbf{k_{\perp}}\!}$ has been plotted as function of the scaled coordinates $x/\omega_{0}$ and $y/\omega_{0}$ on figure~\ref{fig:Phase2D} for $\mathbf{k_{\perp}}\! = k_{x} \, \boldsymbol{{e}_{x}}$. It can be shown that whatever$\,$the$\,$initial$\,$values of $k_{x}$ and $k_{y}$, one can always define new basis vectors $\boldsymbol{{e}_{X}}$ and $\boldsymbol{{e}_{Y}}$ such$\;$that $\mathbf{k_{\perp}}\! = k_{X} \, \boldsymbol{{e}_{X}}$. Let's consider, for the sake of simplicity, that $\mathbf{k_{\perp}}\! = k_{x} \, \boldsymbol{{e}_{x}}$. This$\,$is$\,$what$\,$is$\,$done$\,$on$\,$figure~\ref{fig:Phase2D}. As long as $k_{x} \, \omega_{0} < 2\left| \langle \Phi_{\mathrm{NL}}(0,z') \rangle_{L} \right| e^{-1/2}$, $\Psi_{\!\mathbf{k_{\perp}}\!}$ exhibits only one maximum at $\boldsymbol{r}_{1}$ and one saddle point$\,$at$\,\boldsymbol{r}_{2}$,$\,$as$\,$can$\,$be$\,$seen$\,$on$\,$figure~\ref{fig:Phase2D}.$\;$As$\,$in$\,$the$\,$1D$\,$case,$\,$the$\,$intensity$\,$distribution$\,$at $\mathbf{k_{\perp}}$ in Fourier space results therefore from the interference of the waves coming from the vicinity of $\boldsymbol{r}_{1}$ and $\boldsymbol{r}_{2}$ in the cell output plane. Moreover, it is straightforward$\;$to$\;$show$\;$that $\partial_{xy} \! \left. \Psi_{\!\mathbf{k_{\perp}}} \right|_{\boldsymbol{r}_{\alpha}} \! = \partial_{yx} \! \left. \Psi_{\!\mathbf{k_{\perp}}} \right|_{\boldsymbol{r}_{\alpha}} \! = 0$ and consequently that $\sigma \! \left[\mathcal{H}_{\!\mathbf{k_{\perp}}\!}(\boldsymbol{r}_{1})\right] = -2$ while $\sigma \! \left[\mathcal{H}_{\!\mathbf{k_{\perp}}\!}(\boldsymbol{r}_{2})\right] = 0$. There is thus a constant dephasing of $\pi/2$ between the two waves interfering in far-field, exactly like in the 1D case once more. This can be intuitively understood by looking$\;$at$\;$the signs of the second-order derivatives at $\boldsymbol{r}_{1}$ and $\boldsymbol{r}_{2}$. In the vicinity of $\boldsymbol{r}_{1}$, $\partial^{2}_{x} \Psi_{\!\mathbf{k_{\perp}}}\!$ and $\partial^{2}_{y} \Psi_{\!\mathbf{k_{\perp}}}\!$ are negative which suggests that the light coming from there will tend to spread over the $(x,y)$ plane during the in-air propagation.$\,$Close$\;$to$\;$the$\;$saddle$\;$point,$\,\partial^{2}_{x} \Psi_{\!\mathbf{k_{\perp}}}$is$\,$positive$\,$while 

\newpage

\noindent $\partial^{2}_{y} \Psi_{\!\mathbf{k_{\perp}}}\!$ is negative. The light coming from the vicinity of $\boldsymbol{r}_{2}$ will thus simultaneously$\;$focus along the $x$-axis and spread over the $y$-direction during propagation. In geometrical optics, it is well known that light accumulates an additional phase $-$ the so-called Gouy$\;$phase$\;-$ when it passes through a focus. This phase shift is equal to $\pi/2$ if light propagates from $-\infty$ to $+\infty$ through a line focus and to $\pi$ when it propagates from $-\infty$ to $+\infty\;$through a point-like one. The Gouy phase shift of any focused light arises from its transverse spatial confinement which introduces, through the uncertainty principle, a spread in its transverse momenta and hence a$\;$shift in the expectation value of its longitudinal wave-vector~\cite{4-16Feng}. In$\,$the$\,$present$\,$case, the focus is a line (because $\boldsymbol{r_{2}}$ is$\,$a$\,$saddle-point).$\;$The$\,$light$\,$coming$\,$from $\boldsymbol{r}_{2}$ accumulates thus an additional $\pi/2$ phase shift with respect to the light coming from $\boldsymbol{r}_{1}$, as mentioned above. The same reasoning can be applied in 1D in$\,$order$\,$to$\,$explain$\,$the$\,\pi/2$ dephasing in far-field between the waves emitted at $x_{1}$ and $x_{2}$ in the cell exit plane.
\vspace{6pt}
\newline
\noindent We can finally, as in 1D, approximate the k-space intensity$\,$distribution$\,$in$\,$the$\,$following$\,$way: $\mathcal{I}_{\scriptscriptstyle{\mathrm{2D}}}(\mathbf{k_{\perp}}) \simeq C \left[ 1 + \sin(\Delta \Psi_{\mathbf{k_{\perp}}}) \right]$. It is easy to see that $\Delta \Psi_{\mathbf{k_{\perp}}} = \left|\langle \Phi_{\mathrm{NL}}(0,z') \rangle_{L} \right|$ when $\mathbf{k_{\perp}} = \boldsymbol{0}$. Consequently, the ring-counting procedure in 2D is the same as in 1D. Within the TMA, it can be formulated as follows:
\vspace{-4pt}
\begin{itemize}
    \item [$\bullet$] when the-far field diffraction pattern is composed of $n$ rings plus a bright$\;$central$\;$spot (constructive interference at $\mathbf{k}_{\perp} = \boldsymbol{0}$), the nonlinear phase shift is $2 n \pi + \pi/2$;  
    \vspace{-4pt}
    \item [$\bullet$] when the-far field diffraction pattern is composed of $n$ rings plus a dark$\;$central$\;$spot (destructive interference at $\mathbf{k}_{\perp} = \boldsymbol{0}$), the nonlinear phase shift is $(2 n -1)\pi + \pi/2$;
\end{itemize}

\newpage 

\noindent \textsc{remark}. It is worth mentioning that the counting procedure above is not the one usually discussed in the literature. It is commonly accepted that $\left|\langle \Phi_{\mathrm{NL}}(0,z') \rangle_{L} \right|$ is equal to $2 n \pi$ when $n$ rings plus a bright central spot are visible in far-field and to $(2 n -1)\pi$ for $n$ rings plus a dark central spot~\cite{4-5Kaiser, 4-4Boughdad}. The $\pi/2$ Gouy phase shift predicted by the theory seems to be omitted in these works where the relationship between the number of rings and the on-axis self-phase modulation has been inferred using numerical simulations. 

  
\subsubsection{Numerical simulation}
 
\begin{figure}[h]
\center
\includegraphics[scale=0.45]{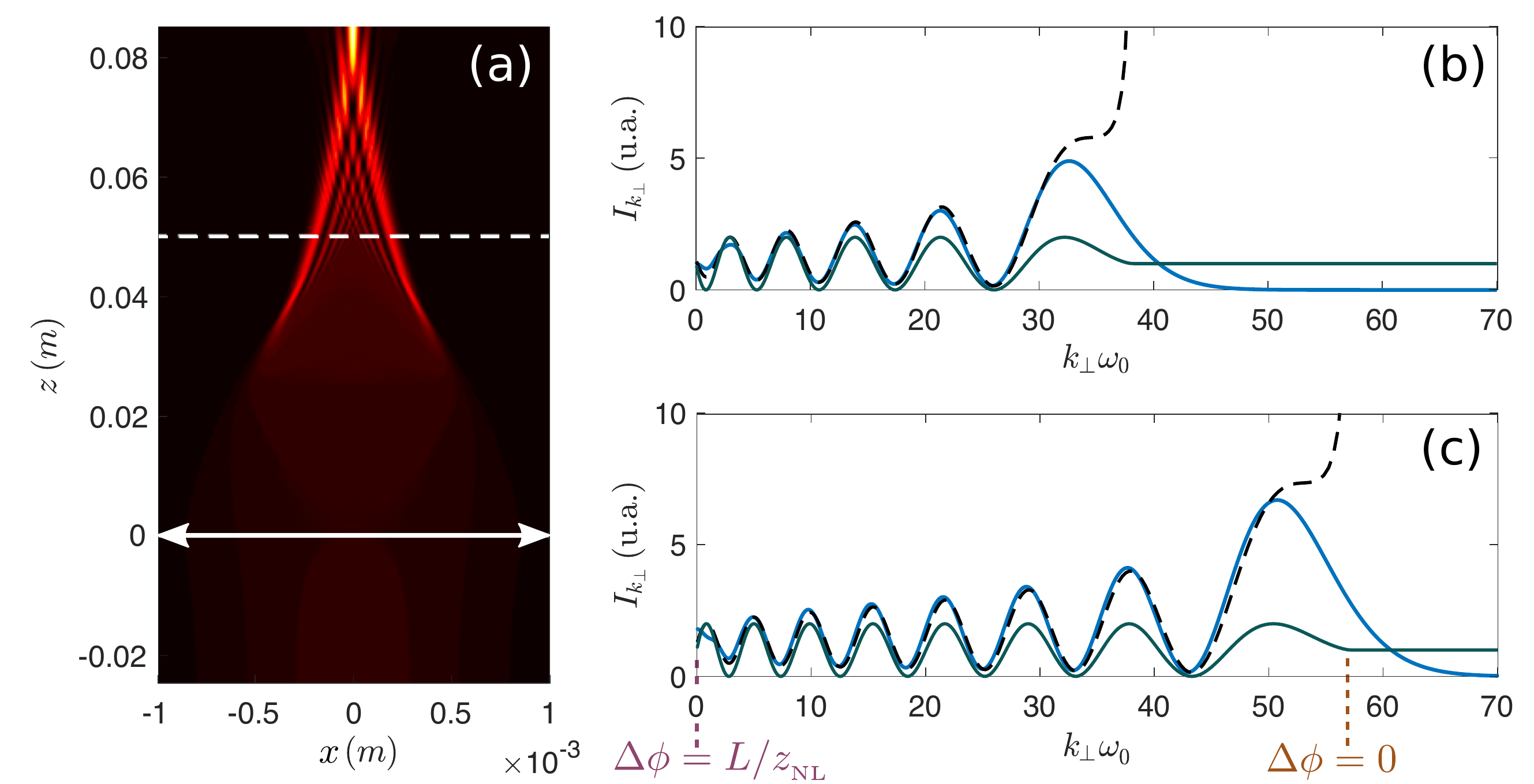} 
\caption{(a) Evolution of the intensity distribution along $z$. A lens ($f=5$ cm) focuses the beam at the nonlinear medium exit plane. Because of self-defocusing,$\,$the$\,$beam$\,$waist$\,$is not$\,$located$\,$in$\,$the$\,$focal$\,$plane$\,$(dotted$\,$line)$\,$anymore.$\;$(b-c)$\,$Comparison$\,$between$\;$simulations (blue solid lines) and the predictions of equation~\eqref{PavloffModel} (black dashed lines) and~\eqref{PavloffModelSimple} (green solid lines).$\,$Parameters: $\omega_{0} \!=\! 1\,$mm, $n_2 \!=\! 4 \, 10^{-11}\,$m$^{2}$/W, $\mathcal{I}_{0} \!=\! 10^{7}\,$W/m$^{2}\,$and$\,\alpha \!=\! 0$.}
\label{fig:Comparaison}
\end{figure}

\noindent Studying numerically the dynamics of the ring pattern formation requires to solve both the$\;$linear (LSE) and the nonlinear (NLSE) Schr\"{o}dinger equations, introduced in chapter$\;$2. In order to do so, we used the second-order split step method, which provides both good accuracy and high performance speed. I will only discuss here simulations in the 1D case (which are easier to perform), since the 2D situation is analogous.
\vspace{6pt}
\newline
\noindent We first look at the ring pattern formation in 1D by solving the $z$-evolution$\;$of the electric field envelope (i) inside the nonlinear medium with the NLSE and (ii) outside, using either the LSE to describe in detail the field evolution after the lens or a$\,$1D$\,$FFT$\,$to$\,$directly$\,$access its far-field intensity distribution (that is faster). On figure~\ref{fig:Comparaison}(a), the field$\;$intensity has been plotted as function of $z$. The nonlinear medium is 2.5 cm long here. A converging$\;$lens (white double-headed arrow, $f = 5$ cm) has been set a the medium output plane, at $z = 0$. At that point, the laser beam starts focusing. As$\,$you$\,$may$\,$have$\,$noticed,$\,$interferences$\,$do$\,$not only occur in the focal plane (white dashed line). We must thus be extremely careful when we image the Fourier space on the camera, as a small mistake on the imaged plane$\,$position can lead to misestimate the number of rings. An other possibility to get rid of this issue is to let the beam propagates freely after the cell, as mentioned before. It suffices then to image a plane far away from the medium exit$\;$plane.  

\newpage

\noindent Finally, we compare on$\,$figures$\,$\ref{fig:Comparaison}$\,$the$\,$far-field$\,$intensity$\,$profiles$\,$resulting$\,$from$\,$the$\,$numerical simulations with the theoretical predictions of equations$\,$\eqref{PavloffModel}$\,$and$\,$\eqref{PavloffModelSimple}, for on-axis nonlinear phase$\,$shifts$\,$of$\,10 \pi\,$(b)$\,$and$\,15\pi\,$(c).$\,$The$\,$intensity$\,$profile$\,$obtained$\,$with$\,$equation$\,$\eqref{PavloffModel} (black dashed line) exactly$\,$matches$\,$the$\,$results$\,$of$\,$the$\,$numerical$\,$simulations$\,$(blue$\,$solid$\,$line), when the stationary phase approach is valid, \textit{ie}, as long as $k_{x} \, \omega_{0} < 2\left| \langle \Phi_{\mathrm{NL}}(0,z') \rangle_{L} \right| e^{-1/2}$. More surprisingly, the$\;$approximate solution~\eqref{PavloffModelSimple} (green solid line)$\,$properly$\,$reproduces the oscillating trend of the far field interference pattern; to count the number$\,$of$\,$rings,$\,$it$\,$is thus enough to count the maxima of $\sin(\Delta \Psi)$ when $\Delta \Psi$ varies from zero to $\langle \Phi_{\mathrm{NL}}(0,z') \rangle_{L}$. 

\begin{figure}[h]
\center
\includegraphics[width=\columnwidth]{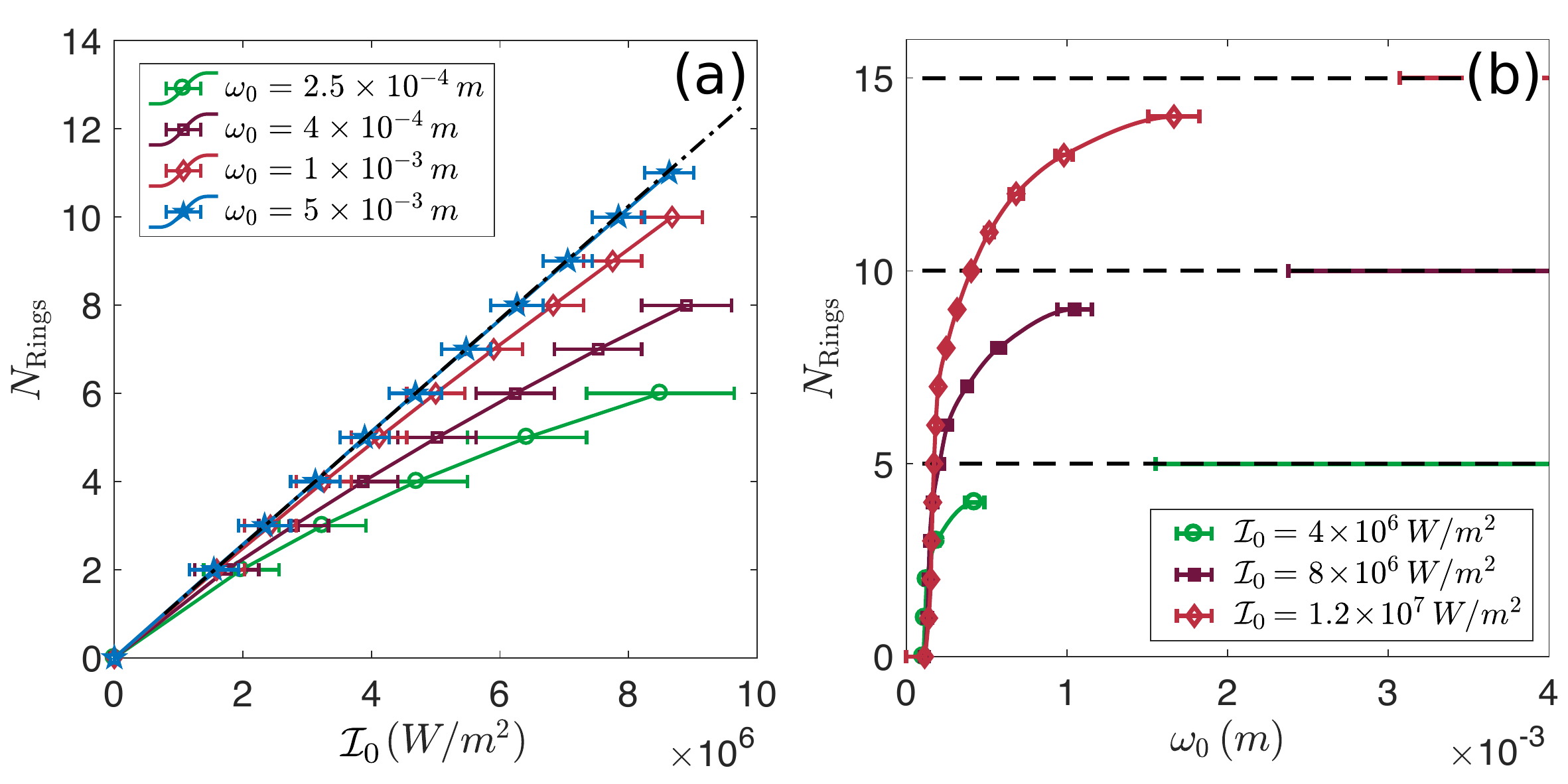} 
\caption{Number of rings in far-field as function (a) of the input on-axis intensity $\mathcal{I}_{0}$ and (b) of the beam width $\omega_{0}$. Parameters: $L = 2.5$ cm, $n_{2} = 4 \!\times\! 10^{-11}$ m$^{2}$/W and $\alpha = 0$.}
\label{fig:Ring1DSimulation}
\end{figure} 

\noindent So far, the impact of the beam width on the ring pattern formation has been neglected, as$\;$I always assumed the TMA was valid. Let's now study the effect of self-defocusing$\;$on$\;$the far-field diffraction pattern. The$\,$data$\,$points$\,$on$\,$figure~\ref{fig:Ring1DSimulation}(a)$\,$are$\,$the$\,$number$\,$of$\,$visible$\,$rings in far-field, for$\,$various$\,$beam$\,$widths$\,\omega_{0}$.$\;$The$\,$horizontal$\,$bars$\,$stand$\,$for$\,$the$\,$range$\,$of$\,$intensities over which the number of rings, $N_{\mathrm{Rings}}$, remains the same.$\;$The$\,$on-axis$\,$nonlinear$\,$phase$\,$shift $\left|\langle \Phi_{\mathrm{NL}}(0,z') \rangle_{L} \right| (= k_{0} \, n_{2} \, \mathcal{I}_{0} \, L$ in the lossless case) has also been plotted (black dashed line) in $2\pi$ units. For a 2.5 cm long cell and a beam width of $\omega_{0} = 5$ mm (blue stars), the$\;$TMA is valid as $\left|\langle \Phi_{\mathrm{NL}}(0,z') \rangle_{L} \right| \ll \Phi_{\mathrm{lim}} \simeq 320 \!\times\! 2\pi$, whatever the value of the input intensity$\,\mathcal{I}_{0}$. As you may have seen, the number of rings increases linearly with$\;\mathcal{I}_{0}\,$in$\,$that$\,$case.$\;$However, when $\omega_{0}$ decreases, the value of the input intensity at which new rings appears increases, shifting the data points rightward. This shift moreover increases with $\mathcal{I}_{0}$, as if the number of rings was saturating at high input intensities. This$\,$behaviour$\,$is$\,$expected$\,$when$\,$the$\,$TMA breaks$\,$down.$\;$In$\,$that$\,$case,$\,$we$\,$notice$\,$a$\,$spreading$\,$of$\,$the$\,$beam$\,$intensity$\,$in$\,$the$\,$transverse$\,$plane because of self-defocusing. This effect is enhanced when the beam width decreases,$\,$as$\,$the medium$\,$effective$\,$focal$\,$length$\,f_{\mathrm{eff}}\,$scales$\,$as$\,\omega_{0}^{2}$.$\;$The$\,$beam$\,$spreading$\,$results$\,$in$\,$a$\,$diminution$\,$of the on-axis intensity during propagation. The relationship between $\mathcal{I}_{0}$ and $\left|\langle \Phi_{\mathrm{NL}}(0,z') \rangle_{L} \right|$ (and therefore, between $\mathcal{I}_{0}$ and the number of rings in far-field) is thus not linear anymore. 

\newpage

\noindent In figure~\ref{fig:Ring1DSimulation}(b), the number of rings in far-field$\,$has$\,$been plotted as function of $\omega_{0}\,$this$\,$time, for different input intensities. It increases with $\omega_{0}$ until it saturates. The width at which the number of rings starts being constant indicates$\,$the minimum$\,$beam$\,$width$\,$required$\,$to fulfill$\;$the thin medium conditions and correctly$\,$measure$\,n_{2}\,$using$\,$the$\,$ring-counting$\,$method. 

\subsubsection{Experimental results}

\begin{figure}[h]
\center
\includegraphics[scale=0.45]{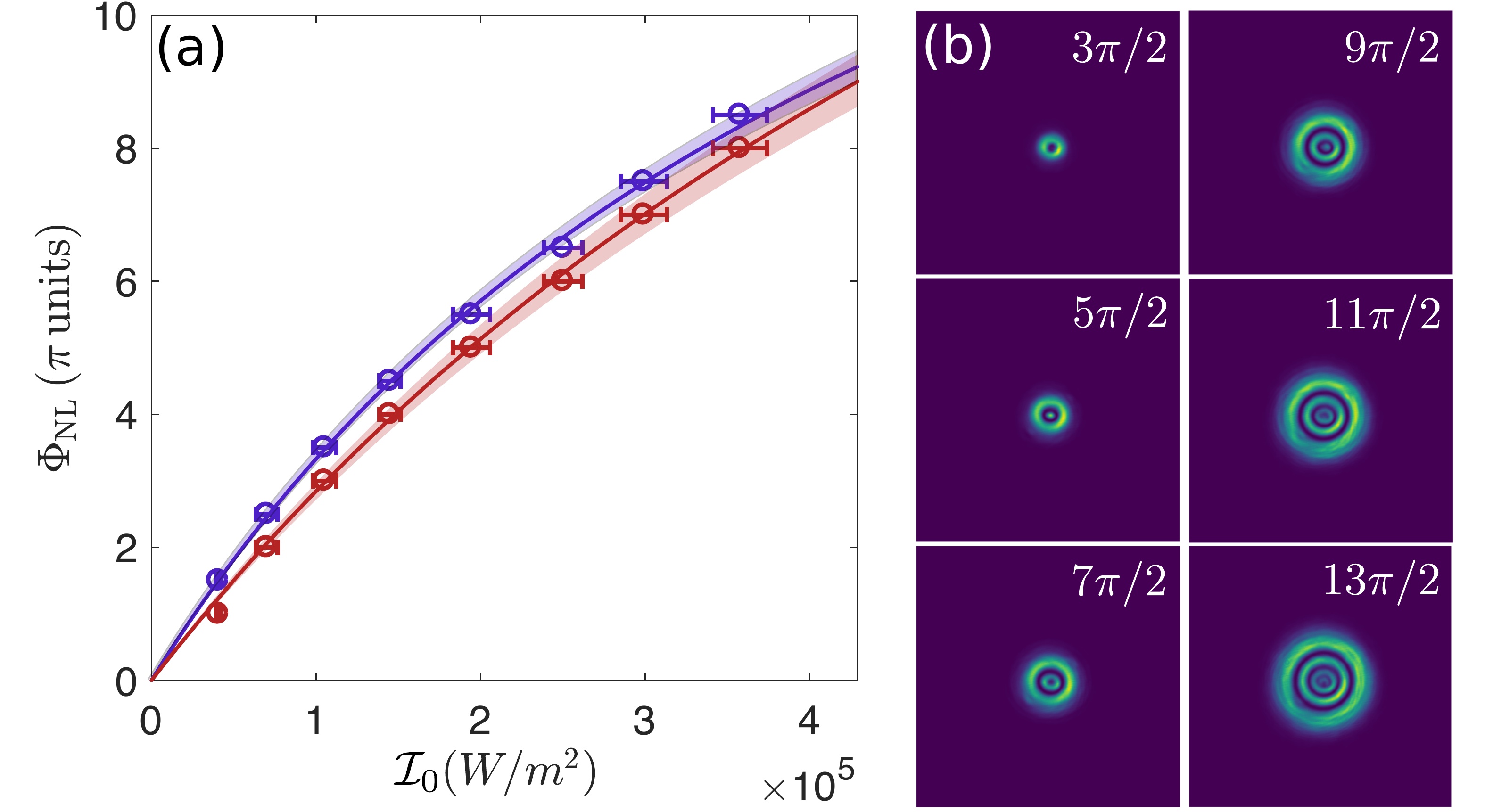} 
\caption{(a): On-axis nonlinear phase shift as function of the input intensity, measured using the counting method of paragraph 3.3.2 i(b) (blue points) and the one adopted in~\cite{4-5Kaiser, 4-4Boughdad} (red points). The data are fitted in both case using equation~\eqref{FitRing} to extract$\;n_{2}$. (b): Far-field patterns associated to the data points on the left. Parameters: $L=7.5$ cm, $T \simeq 125 \, ^{\circ}$C and $\Delta \simeq 2\pi \!\times\! 3.1$ GHz from the $F_{g} = 3 \rightarrow F'$ transition of $^{_{85}}$Rb.}
\label{fig:RingExample}
\end{figure}

\noindent In the experiments, we almost exclusively use the ring-counting technique to measure the nonlinear refractive index $n_{2}$ of the rubidium vapor. The data we got using this$\;$technique are exploited further on to support the main results of my work in the following chapters. Ring-counting curves will$\,$therefore$\,$sporadically$\,$appear$\,$in$\,$the$\,$remaining$\,$of$\,$this$\,$manuscript. I thus present here only few experimental data. On figure~\ref{fig:RingExample}(a), an example of a ring-counting curve is shown. The on-axis nonlinear phase shift $\langle \Phi_{\mathrm{NL}}(0,z') \rangle_{L}\,$has$\;$been plotted as function of the input intensity using the counting procedure of~\ref{subsubsec:TheoryRing}(b) (blue$\,$points)$\,$and$\,$the counting procedure described in~\cite{4-5Kaiser, 4-4Boughdad} (red points). Blues points are shifted up from the red ones by $\pi/2$. The uncertainty comes from our inability$\,$to$\,$accurately$\,$evaluate$\,$at$\,$which intensity the central spot is the brightest or the darkest. On$\,$figure$\,$(b),$\,$the$\,$far-field$\,$patterns associated$\,$to$\,$each$\,$points$\,$are$\,$shown.$\;$These$\,$data$\,$have$\,$been$\,$obtained$\,$with$\,$an$\,$isotopically$\,$pure $^{_{85}}$Rb vapor. The cell was 7.5 cm long and the temperature about $125\,^{\circ}$C. The beam width was set to $825$ $\mu$m. In order for the TMA to be valid, $\left|\langle \Phi_{\mathrm{NL}}(0,z') \rangle_{L} \right|$ has to be smaller than $\Phi_{\mathrm{lim}} = 8\!\times\!2\pi$. This condition is not fulfilled at high intensity. Therefore, we expect the number of rings in far-field (and thus $\left|\langle \Phi_{\mathrm{NL}}(0,z') \rangle_{L} \right|$) to saturates with $\mathcal{I}_{0}$ because of self-defocusing, as shown on figure~\ref{fig:Ring1DSimulation}(a). 
        
\newpage

\noindent \textbf{Saturation}.$\;$Rubidium$\,$vapors$\,$are$\,$saturable$\,$nonlinear$\,$mediums.$\,$At$\,$high$\,$enought$\,$intensity, higher order susceptibilities ($\chi^{_{(5)}}$, $\chi^{_{(7)}}$, ...) start contributing to the nonlinear response$\;$and make the nonlinear change of refractive index saturating. That may also explain the trend observed on figure~\ref{fig:RingExample}. In order to take the effects of self-defocusing as well$\,$as$\,$of$\,$the$\,$atomic saturation into account, we fit the data using the following model~\cite{4-5Kaiser}:
\begin{equation}
    \left|\langle \Phi_{\mathrm{NL}}(0,z') \rangle_{L} \right| = \frac{k_{0} \, n_{2} \, \mathcal{I}_{0} \, L}{1+\mathcal{I}_{0}/\mathcal{I}_{\mathrm{sat}}} 
    \label{FitRing}
\end{equation}
\noindent where $n_{2}$ and the saturation intensity $\mathcal{I}_{\mathrm{sat}}$ are fitting parameters.
\vspace{6pt}
\newline
\noindent \textbf{Comparison between the two ring-counting procedures}. By doing so, we find that $n_{2} = 2.1 \pm 0.1 \!\times\! 10^{-10}$ m$^{2}/W\,$and$\,\mathcal{I}_{\mathrm{sat}} \simeq 5.0 \!\times\! 10^{5}$ W/m$^{2}\,$using$\,$the$\,$counting$\,$procedure$\,$detailed in section 3.3.2 i(b) (blue$\,$points)$\,$and$\,n_{2} = 1.7 \pm 0.1 \!\times\! 10^{-10}$ m$^{2}/W\,$and$\,\mathcal{I}_{\mathrm{sat}} \simeq 8.3 \!\times\! 10^{5}\,$W/m$^{2}$ using the one$\,$of$\,$\cite{4-5Kaiser, 4-4Boughdad}$\,$(red$\,$points).$\;$The$\,$relative$\,$variation$\,$of$\,$the$\,$fitted$\,$value$\,$of$\,$the$\,$nonlinear refractive$\,$index$\,$obtained$\,$with$\,$the$\,$two$\,$counting$\,$methods$\,$is$\,20\%$.$\,$In$\,$this$\,$experiment,$\,$the$\,$laser was$\,3\,$GHz$\,$red-detuned$\,$from$\,$the$\,F_{g} = 3 \rightarrow F'\,$transition$\,$of$\,^{_{85}}$Rb.$\,$The$\,$theoretical$\,$value$\,$of$\,$the off-resonance$\,$saturation$\,$intensity$\,\mathcal{I}_{s}(\Delta)\,$is$\,$(with$\,$the$\,$parameters$\,$above)$\,$about$\,8\!\times\!10^{5}\,$W/m$^{2}$. As you can see, this value matches almost exactly the fitted value of the saturation$\;$intensity obtained using the counting procedure of~\cite{4-5Kaiser, 4-4Boughdad}. This is in fact quite surprising because $1/\mathcal{I}_{\mathrm{sat}} = 1/\mathcal{I}_{s}(\Delta) + 1/\mathcal{I}_{\mathrm{eff}}$, where $\mathcal{I}_{\mathrm{eff}}$ is an effective saturation$\,$intensity$\,$describing$\,$the effective contribution$\;$of self-defocusing in the saturation of $\left|\langle \Phi_{\mathrm{NL}}(0,z') \rangle_{L} \right|$. In the present$\;$case, self-defocusing plays a role for sure, as the TMA is not perfectly fulfilled. We thus$\;$expect$\;$a value of $\mathcal{I}_{\mathrm{sat}}$ lower than $\mathcal{I}_{s}(\Delta)$. In this sense, the value of $\mathcal{I}_{\mathrm{sat}}$ provided by our ring-counting method seems therefore more reasonable.
\vspace{2pt}
\newline
\noindent Moreover, if we attribute an on-axis nonlinear phase shift of $\pi$ to the far-field diffraction pattern composed of one$\,$ring$\,$plus$\,$a$\,$dark$\,$central$\,$spot,$\,$as$\,$in$\,$\cite{4-5Kaiser, 4-4Boughdad},$\,$the$\,$related$\,$point$\,$(that$\,$is, the circle of lowest intensity) does not lie on the fitted curve (red line). This$\,$observation$\,$is actually quite systematic, as if the first ring in far field was obtained for a slightly$\,$too$\,$high input intensity if we use the counting procedure described$\;$in~\cite{4-5Kaiser, 4-4Boughdad}. By shifting up the red points by $\pi/2$, this issue seems to be solved, as the point of lowest intensity lies then on the blue fitted curve. These$\;$two$\;$remarks$\;$make$\;$us$\;$think$\;$that$\;$the$\;\pi/2$ Gouy phase shift should definitely be taken into account in the ring-counting method. 

    \section{Methods III $-$ Scanning phase interferometry} 

\noindent In photon fluid experiments, measuring the phase of the fluid at$\,$the$\,$medium$\,$output$\,$plane provides useful information. It$\,$allows,$\,$for$\,$instance,$\,$to$\,$access$\,$the$\,$flow$\,$velocity$\,$distribution$\,$of the fluid after propagation, or also to highlight and study the formation and the dynamics of phase$\,$singularities (quantized vortices, solitons, etc). Such a measurement can also be used to retrieved the spatial variations of the nonlinear change of refractive$\,$index$\,$(and$\,$thus, to extract $n_{2}$)~\cite{4-17Minovich}. This possibility is further investigated by Murad Abuzarli,$\,$PhD$\,$student in the group. This short section is devoted to presenting the phase shifting interferometry, that is used in our experiments to retrieve the spatial phase distribution of the light$\,$at$\,$the medium exit plane. This phase reconstruction procedure is$\,$a$\,$generalization$\,$of$\,$the$\,$standard four-frame interferometric technique, described in~\cite{4-18Creath}.

\newpage

\noindent The experimental setup built in$\,$order$\,$to$\,$performed$\,$phase$\,$shifting$\,$interferometry$\,$is$\,$sketched on figure~\ref{fig:MachZehnder} below. A laser beam comes out from the optical fiber and is magnified$\,$to$\,$the desired$\,$size$\,$before$\,$entering$\,$a$\,$Mach-Zehnder$\,$interferometer.$\;$A$\,$polarized$\,$beam$\,$splitter$\,$(PBS) splits the incoming light beam into two parts, the reference$\,$and$\,$the$\,$fluid$\,$(which$\,$propagates through$\,$the$\,$cell).$\;$The$\,$half-wave$\,$plate$\,$($\lambda/2$)$\,$positioned$\,$in$\,$between$\,$the$\,$telescope$\,$and$\,$the$\,$PBS enables us to control the amount of light sent in each arm. Reference and fluid recombine in the $50/50$ beam splitter (BS). The length of the reference arm is modulated by setting one mirror$\,$on$\,$a$\,$piezo-actuated$\,$translation$\,$stage$\,$(referred$\,$to$\,$as$\,$PEM).$\;$The$\,$high$\,$tension$\,U(t)$ applied across the piezo-actuator is depicted in the inset on the right.$\;$A$\,$function$\,$generator produces a $0.2$ Hz triangular signal amplified by a high voltage generator. Its$\,$amplitude$\,\delta U$ is increased gradually until the reference phase is exactly scanned over $2 \pi$. In order to maximize the contrast of the interference pattern, one can tune the reference intensity by turning the half wave plate on its path. The cell exit plane is$\,$finally$\,$imaged$\,$on$\,$the camera using a second 4-$f$ telescope (depicted with a bi-convex lens for clarity).

\begin{figure}[h]
\center
\includegraphics[width=\columnwidth]{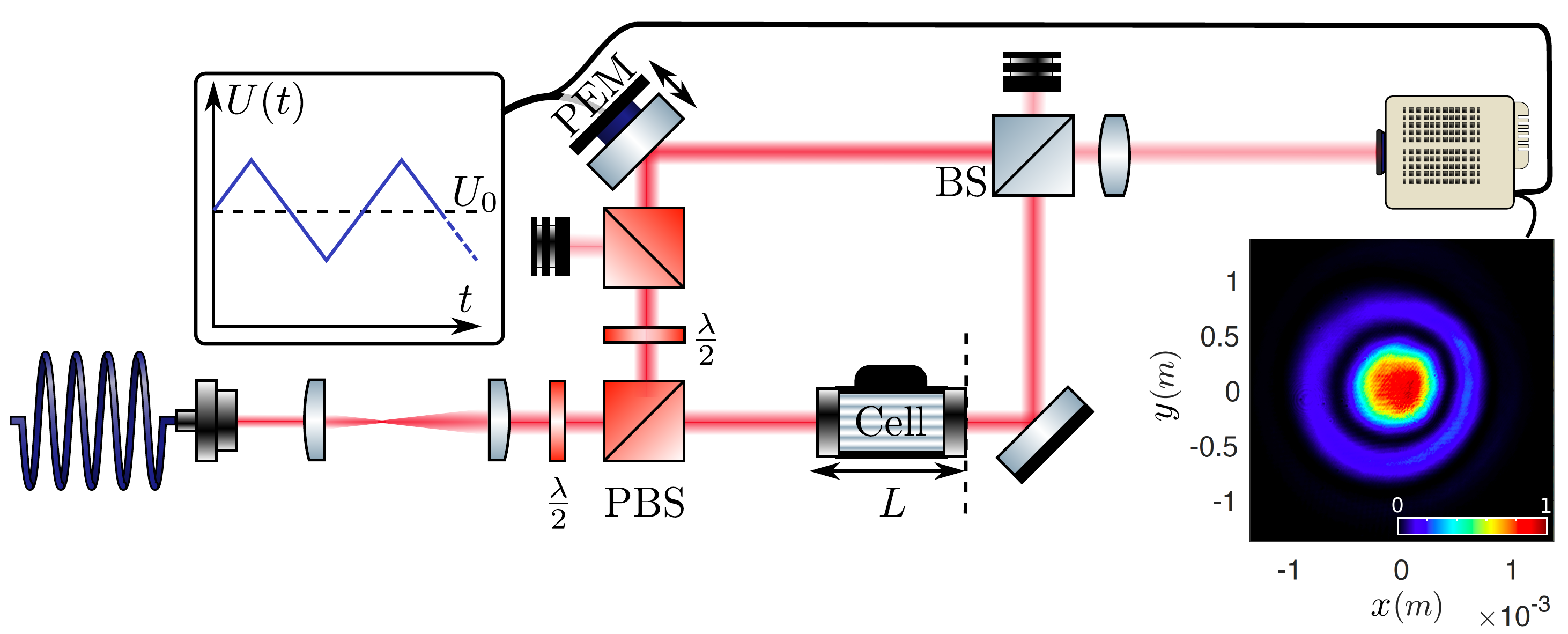} 
\caption{Experimental setup. A Mach-Zehnder interferometer is used so as to measure the beam transverse phase profile in the cell exit plane. Left inset: high voltage triangular signal making the PEM scan the phase of the reference beam. Right inset:$\,$example$\,$of$\,$an interferogram obtained when the angle between the reference and the fluid is zero.}
\label{fig:MachZehnder}
\end{figure} 

\noindent The intensity of the two-beam interference pattern at the exit plane can be described by:
\begin{equation}
    \mathcal{I}_{n}(\mathbf{r}) =  \mathcal{I}(\mathbf{r}, L) \left \{ 1 + V(\mathbf{r})\left[ \cos \left( \Phi(\mathbf{r}) - \theta_{n} \right) \right] \right \},
    \label{SetInterferograms}
\end{equation}
\noindent where $\mathcal{I}(\mathbf{r}, L)$ is the beam intensity at the medium output plane, $V(\mathbf{r})$ the fringe visibility and $\theta_{n} = \frac{2 \pi}{N} (n-1)$ ($n\;$being an integer in $\left \{ 1, ..., N \right \}$) an arbitrary reference phase which is scanned over $2 \pi$ under the PEM translation. Each of the $ \mathcal{I}_{n}(\mathbf{r})$ portrays an interferogram (see for instance the inset of fig.~\ref{fig:MachZehnder}); the complete set of $N$ interferograms is used to reconstruct $\Phi(\mathbf{r})$. By using the angle difference identity, one can rewrite~\eqref{SetInterferograms} as follows: 
\begin{equation}
    \mathcal{I}_{n}(\mathbf{r}) = I(\mathbf{r}, L) \left \{ 1+ V(\mathbf{r}) \cos \left[ \Phi(\mathbf{r}) \right] \cos(\theta_{n}) + V(\mathbf{r}) \sin \left[ \Phi(\mathbf{r}) \right] \sin(\theta_{n})\right \}
\end{equation}

\newpage

\noindent By multiplying the previous equation by (i) $\sin(\theta_{n})$ and (ii) $\cos(\theta_{n})$ before summing$\,$over$\,n$:
\begin{equation}
  \begin{split}
        \hspace{-1.0cm} \sum_{n=1}^{N}  \mathcal{I}_{n}(\mathbf{r}) \cos(\theta_{n}) =& \, \mathcal{I}(\mathbf{r}, L) \sum_{n=1}^{N} \left \{ \cos(\theta_{n}) + V(\mathbf{r}) \cos \left[ \Phi(\mathbf{r}) \right] \cos^{_{2}}(\theta_{n}) + \right. ... \\ 
        & \hspace{4.9cm} \left. V(\mathbf{r}) \sin \left[ \Phi(\mathbf{r}) \right] \sin(\theta_{n}) \cos{\theta_{n}} \right \} 
    \end{split}
    \label{cos}
\end{equation}
\begin{equation}
  \begin{split}
        \hspace{-1.0cm} \sum_{n=1}^{N}  \mathcal{I}_{n}(\mathbf{r}) \sin(\theta_{n}) =& \,  \mathcal{I}(\mathbf{r}, L) \sum_{n=1}^{N} \left \{ \sin(\theta_{n}) + V(\mathbf{r}) \sin \left[ \Phi(\mathbf{r}) \right] \sin^{_{2}}(\theta_{n}) + \right. ... \\ 
        & \hspace{4.9cm} \left. V(\mathbf{r}) \cos \left[ \Phi(\mathbf{r}) \right] \cos(\theta_{n}) \sin{\theta_{n}} \right \} 
    \end{split}
\end{equation}
\noindent The first terms and the mixed terms on the right hand side of equations~\eqref{cos} and~\eqref{sin} cancel because of the orthogonality of the trigonometric functions. We thus obtain that:
\begin{align}
    \sum_{n=1}^{N} I_{n}(\mathbf{r}) \cos(\theta_{n}) =& \, \frac{N}{2} I(\mathbf{r}, L) V(\mathbf{r}) \cos \left[ \Phi(\mathbf{r}) \right]  \\ 
    \sum_{n=1}^{N} I_{n}(\mathbf{r}) \sin(\theta_{n}) =& \, \frac{N}{2} I(\mathbf{r}, L) V(\mathbf{r}) \sin \left[ \Phi(\mathbf{r}) \right]
    \label{sin}
\end{align}
\noindent which provides an simple algorithm to process the $N$ interferograms and calculate $\Phi(\mathrm{r})$:
\begin{equation}
    \tan \left[ \Phi(\mathbf{r}) \right] = \left( \sum_{n=1}^{N} I_{n}(\mathbf{r}) \sin(\theta_{n}) \right) \bigg/ \left( \sum_{n=1}^{N} I_{n}(\mathbf{r}) \cos(\theta_{n}) \right) 
    \label{PhaseAlgorithm}
\end{equation}

\begin{figure}[h]
\center
\includegraphics[width=\columnwidth]{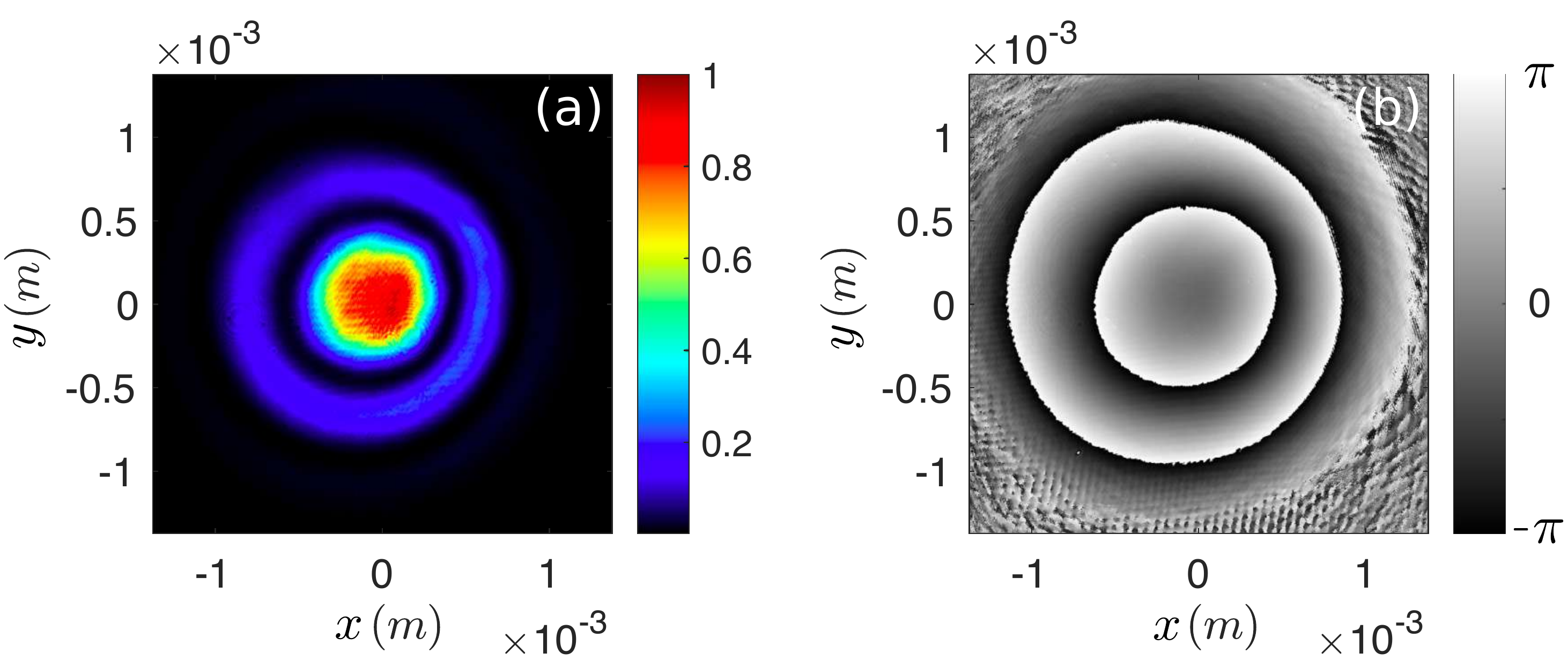} 
\caption{(a)$\,$Example$\,$of$\,$an$\,$interferogram$\,$obtained$\,$with$\,$the$\,$setup$\,$sketched$\,$in$\,$figure$\,$\ref{fig:MachZehnder}. The fluid of light is formed by a Gaussian beam whose width is $850 \, \mu$m. By propagating through a 2.5 cm long vapor cell, this beam accumulates a nonlinear phase shift $\Delta \Phi$ that is greater at its center than on its periphery. When$\,$the$\,$angle$\,$between$\,$the$\,$reference$\,$and$\,$the fluid is zero, a concentric ring-shaped pattern appears$\,$on$\,$the$\,$camera.$\;$(b)$\,$Reconstruction$\,$of the fluid phase at the cell exit plane using the phase shifting interferometry.}
\label{fig:PhaseExample}
\end{figure}

\noindent In our experiments, the trigger of the camera acquisition is set on the high voltage signal. The camera acquires a sequence of 40 images equally spaced in time on a rising ramp$\,$only (to reduce hysteresis effects). An example of an interferogram is shown on figure~\ref{fig:PhaseExample}(a). The photon fluid is formed by a Gaussian beam whose width is $850 \, \mu$m. It accumulates an additional phase shift $\Delta \Phi = \left|\langle \Phi_{\mathrm{NL}}(0,z') \rangle_{L} \right|$ by propagating through a$\,$2.5$\,$cm$\,$long $^{_{85}}$Rb$\,$cell. The angle between the reference and the fluid is zero,$\,$explaining$\,$the$\,$concentric$\,$ring-shaped pattern observed on figure~\ref{fig:PhaseExample}(a). The corresponding spatial phase modulation retrieved from equation~\eqref{PhaseAlgorithm} is shown on figure~\ref{fig:PhaseExample}(b). Reconstructing$\,$in$\,$such$\,$a$\,$way$\,$the$\,$nonlinear phase shift $\Delta \Phi$ is$\,-\,$knowing the$\,$vapor$\,$transmission$\,-\,$a$\,$way$\,$of$\,$measuring$\,n_{2}\,$accurately$\,$\cite{4-17Minovich}. It is worth mentioning that reconstructing the phase profile$\,$of$\,$the$\,$fluid$\,$at$\,$the$\,$cell$\,$exit$\,$plane using the phase shifting interferometry only requires to evaluate numerically equation~\ref{PhaseAlgorithm}, unlike$\,$phase$\,$retrieval$\,$methods$\,$using$\,$a$\,$single$\,$interferogram$\,$(that$\,$are$\,$based$\,$on$\,$FFT$\,$filtering). The scanning phase interferometry is consequently extremely robust and has no issue in reconstructing phase maps full of singularities.  

\newpage

\chapter{Dispersion of small amplitude density waves on a photon fluid}
\chaptermark{Dispersion of small amplitude density waves}

\noindent As mentioned in chapter 2, measuring how small amplitude density waves propagating$\,$onto a photon fluid disperse represents an essential step toward the observation of more$\,$striking phenomena with this system, such as superfluidity. Although this dispersion has been well characterized in atomic BEC experiments~\cite{5-1Jin,5-2Mewes,5-3Onofrio,5-4Steinhauer}, direct measurements in paraxial$\,$photon fluids was attempted~\cite{2-24Vocke} but remain elusive. I will start this fourth chapter by$\,$introducing the method used so far to measure the dispersion relation in those systems, which consists in measuring the phase velocity of density waves travelling onto the photon fluid.$\;$I$\,$will$\,$also adapt the theoretical framework developed by Pierre-\'Elie Larr\'e in~\cite{3-19Larre} to correctly$\,$analyse the experimental observations we made using this technique. In a second$\,$time,$\,$I$\,$will$\,$present the method we have developed so as to more reliably retrieve$\,$the$\,$dispersion$\,$of$\,$density$\,$waves in paraxial photon fluids. Our approach relies on the measurement of their group velocity. The results of this section have been published in two consecutive papers: "Observation of the Bogoliubov Dispersion in a Fluid of Light", Phys.Rev.Lett. 121, 183604 (2018)~\cite{4-13Fontaine} and: "Interferences between Bogoliubov excitations and their impact on the evidence of superfluidity in a paraxial fluid of light", arXiv:2005.14328 (2020)~\cite{4-13FontaineBis}.

\newpage

\section{Phase velocity measurement}

\noindent The method used by Vocke \textit{et al.}~\cite{2-24Vocke} to measure the dispersion relation of$\,$small$\,$amplitude density waves in paraxial photon fluids relies on a pump/probe experiment that has been initially proposed by Iacopo Carusotto in~\cite{3-2Carusotto}. It$\,$basically consists in measuring the phase velocity of a plane wave modulation travelling with the transverse wave-vector $\mathbf{k_{\perp}}$ onto a fluid of light. The latter is obtained by sending a wide laser beam (the pump)$\,$through$\,$the self-defocusing Kerr medium. The small amplitude plane wave modulation$\;$is created by making interfere this pump beam with a wide and weak probe field, which propagates at a small angle $\theta_{i}$ with respect to the optical axis. The resulting interference pattern can be regarded$\,$as$\,$a$\,$density$\,$wave$\,$travelling$\,$onto the photon fluid at $\mathbf{k_{\perp}}$, as$\,$soon$\,$as$\,$the$\,$beams$\,$enter the nonlinear medium. From now on, I$\,$will$\,$suppose$\,$that$\,$the$\,$interference$\,$fringes$\,$are$\,$parallel to the $y$-direction and thus that $\mathbf{k}_{\perp} = k_{0} \sin{(\theta_{i})} \, \mathbf{e}_{x}$. In what follows, I first introduce the theoretical framework to describe this experimental configuration.

\subsection{Theoretical description}

\noindent The phase velocity $v_{\mathrm{ph}}(\mathbf{k_{\perp}})$ of a plane wave propagating on top of a photon fluid$\,$is$\,$given$\,$by: 
\begin{equation}
    v_{\mathrm{ph}}(\mathbf{k_{\perp}}) =  \Omega_{B}(k_{\perp}) / k_{\perp},   
\end{equation}
\noindent where $\Omega_{B}(k_{\perp})$ is the Bogoliubov dispersion introduced in chapter 2. As$\,$the$\,$speed$\,$of$\,$sound, $c_{s}$, scales as the square root of the fluid density $\rho_{0}$ (that is, as the square root of the pump intensity $\mathcal{I}_{0}$), the phase velocity should increase with $\rho_{0}$.$\,$Therefore,$\,$the$\,$interference$\,$pattern forming the plane wave should be shifted along the $x$-direction at$\,$the$\,$medium$\,$output$\,$plane when the fluid density increases. By measuring in this plane, for different wave-vectors$\;\mathbf{k_{\perp}}$, the spatial shift $\Delta S(k_{\perp})$ between the interference fringes at low and high fluid densities, one should be able to retrieve the dispersion relation. The shift $\Delta S(k_{\perp})$ plays$\,$a$\,$crucial$\,$role in what follows. In order to further familiarize yourselves with this quantity, let me present a geometrical argument explaining qualitatively its origin.

\subsubsection{Geometric approach}

\begin{figure}[h]
\center
\includegraphics[width=\columnwidth]{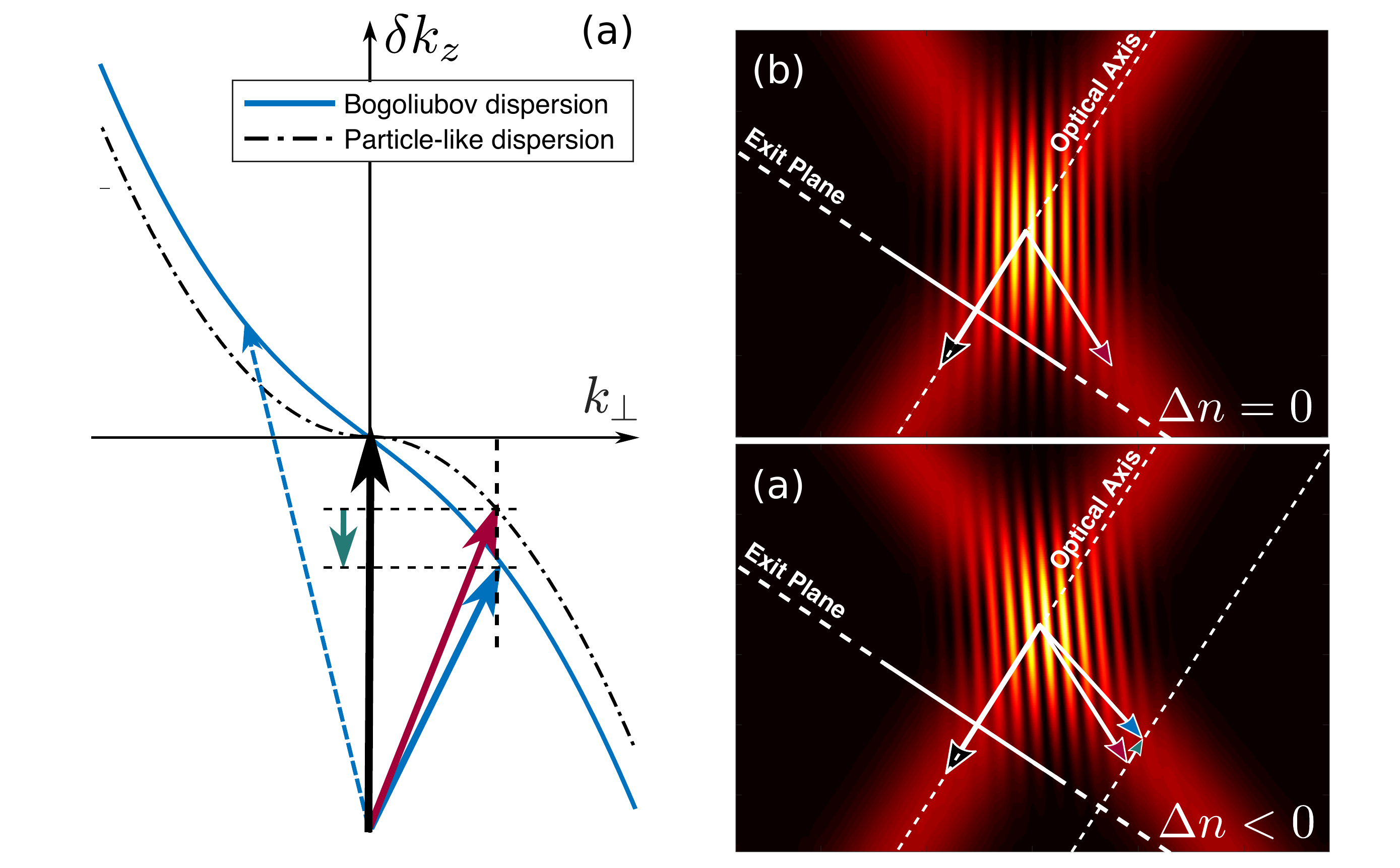} 
\caption{Qualitative explanation of the shift in the interference pattern as a result of the shortening of the probe axial wave-vector in the nonlinear scenario. Figure (a) shows the pump wave-vector (black arrow) as well as the probe one, in a linear (red arrow) and a nonlinear (blue arrow) medium. The photon fluid is a rest, which means that the pump evolves along$\,$the optical axis.$\;$Therefore,$\,$its wave-vector$\;$does$\;$not$\;$have$\,$any$\,$transverse$\;$component. The probe enters the medium with a$\,$non-zero$\,$transverse$\,$wave-vector.$\,$In$\,$the$\,$linear$\,$scenario figure (b), the magnitudes$\,$of$\,$the$\,$wave-vectors$\,$of$\,$pump$\,$and$\,$probe$\,$are$\,$equal.$\,$The$\,$two$\,$beams, propagating downward, create an interference pattern that is$\,$parallel$\,$to the vertical axis. In the nonlinear scenario sketched on figure (c), the magnitude of the probe wave-vector is smaller than the pump one, which thus tilts the interference fringes. By imaging a plane perpendicular$\,$to$\,$the$\,$optical$\,$axis$\,$(dotted$\,$line),$\,$one$\,$obverses$\,$a$\,$shift$\,$of$\,$the$\,$interference$\,$pattern when switching from the linear to the nonlinear situation. 
}
\label{fig:ShiftQualitative}
\end{figure}

\noindent In a linear medium of refractive index $n_{0}$, the$\,$longitudinal$\,$wave-vector $k_{z}$ of$\,$the$\,$probe$\,$beam is defined by $k_{z} = \sqrt{k_{\vphantom{0}}^{_{2}}-k_{\perp}^{_{2}}}$. In the paraxial approximation ($k_{\perp} \ll k_{z}$), this$\,$formula$\,$yields: $k_{z} \simeq k - k^{2}_{\perp}/2k = k + \delta k_{z}$,$\,$where$\,\delta k_{z} = - k^{2}_{\perp}/2k$.$\;$In$\,$a$\,$self-defocusing$\,$($\Delta n <0$)$\,$Kerr$\,$medium, $\delta k_{z} = - \Omega_{B}({k_{\perp}})$. Indeed, the Bogoliubov dispersion $\Omega_{B}(\mathbf{{k_{\perp}}})$ describes the variation$\;$of$\;$the longitudinal wave-vector of the probe field $\delta \mathcal{E}$ when it enters the Kerr medium supported$\,$by the pump field $\mathcal{E}$, with the transverse wave-vector $\mathbf{k_{\perp}}$.
\vspace{2pt}
\newline
\noindent Figure~\ref{fig:ShiftQualitative}(a) shows$\,$the$\,$wave-vectors$\,$of$\,$the$\,$pump$\,$and$\,$the$\,$probe.$\;$The$\,$black$\,$arrow$\,$represents the pump wave-vector, that is equal to$\,k \boldsymbol{e_{z}}$.$\;$Its$\,$head$\,$defines$\,$the$\,$origin$\,$of$\,$the$\,(k_{\perp}, \delta k_{z})\,$plane. The red and blue arrows are the probe wave-vectors in the linear and nonlinear mediums. In the linear one, $\delta k_{z} = - k^{2}_{\perp}/2k$. This is the equation of the black$\,$dotted$\,$parabola$\,$on$\,$which the head of the red arrow lies. In$\,$the$\,$nonlinear$\,$medium,$\,\delta k_{z} = - \Omega_{B}({k_{\perp}})$.$\;$The$\,$tip$\,$of$\,$the$\,$blue arrow lies thus on the negative branch of the Bogoliubov dispersion relation (blue curve). As you can see, the axial component of the probe wave-vector has been$\,$squeezed$\,$by$\,$moving from the linear$\,$to$\,$the$\,$nonlinear$\,$case.$\;$This$\,$has$\,$an$\,$impact$\,$on$\,$the$\,$interference$\,$pattern$\,$between the pump and the probe beams. 


\newpage

\noindent On$\,$figures~\ref{fig:ShiftQualitative}(b)$\,$and$\,$(c)$\,$two$\,$beams$\,$are propagating from top to bottom,$\,$inside$\,$a$\,$linear$\,$and a nonlinear self-defocusing medium respectively. For the sake of clarity, the angle between them$\,$has$\,$been$\,$exaggerated.$\;$The$\,$pump$\,$propagates$\,$along$\,$the$\,$optical$\,$axis$\,$(black$\,$dashed$\,$line). Its wave-vector has been reported on both panels$\,$(black$\,$head$\,$arrow).$\;$The$\,$probe$\,$propagates from left to right, as indicated by the red and blue$\,$head$\,$arrows,$\,$representing$\,$its$\,$wave-vector in the linear and nonlinear cases respectively. The shortening of the axial component$\,$of$\,$the probe wave-vector is shown by the small green head arrow on figure~\ref{fig:ShiftQualitative}(c). It induces$\;$a$\;$tilt of the interference fringes (that are vertical in the linear case), which in turn translates into a shift $\Delta S$ of the interference pattern in a plane perpendicular to the optical axis. 

\newpage

\noindent According to the definition of the shift, $\Delta S$ can be expressed as follows:
\begin{equation}
    \label{Shift}
    \Delta S(k_{\perp}) =  \frac{\Delta \Phi(k_{\perp})}{k_{\perp}} = \frac{\Delta \Phi_{\mathrm{nl}}(k_{\perp}) - \Delta \Phi_{\mathrm{l}}(k_{\perp})}{k_{\perp}},
\end{equation}
\noindent where $\Delta \Phi_{\mathrm{l}}$ and $\Delta \Phi_{\mathrm{nl}}$ are the differences between the phases accumulated by$\,$the$\,$probe$\,$and the pump beams, propagating inside a linear (l) and a nonlinear (nl) medium respectively. Let $\Phi_{0,\mathrm{l}}(L)$ and $\Phi_{0,\mathrm{nl}}(L)$ stand for the phase of the pump at the exit plane of a linear and a nonlinear medium of length $L$. Similarly, let $\Phi_{\mathrm{l}}(k_{\perp}, L)$ and $\Phi_{\mathrm{nl}}(k_{\perp}, L)$ be the$\,$phase$\,$of$\,$the probe in the same conditions. The phase shifts $\Delta \Phi_{\mathrm{l}}$ and $\Delta \Phi_{\mathrm{nl}}$ can then be expressed as: 
$\Delta \Phi_{\mathrm{l}}(k_{\perp}) = \Phi_{\mathrm{l}}(k_{\perp}, L) - \Phi_{0,\mathrm{l}}(L)$ and $\Delta \Phi_{\mathrm{nl}}(k_{\perp}) = \Phi_{\mathrm{nl}}(k_{\perp}, L) - \Phi_{0,\mathrm{nl}}(L)$.  
\vspace{6pt}
\newline
\noindent Some of the quantities defined previously are easy to express as function of the parameters. For instance: $\Phi_{0, \mathrm{l}}(L) = k_{0} n_{0} L$, where $n_{0}$ is the linear refractive index of the medium, and: 
$\Phi_{\mathrm{l}}(k_{\perp}, L) \simeq k_{0} n_{0} L \left[ 1 + \sin^2{(\theta_{r})}/2 \right]$, $\theta_{r}$ being the angle of refraction of the probe beam at the medium input facet. It is defined by the Snell law: $\sin(\theta_{i}) = n_{0} \sin(\theta_{r})$.$\;$It$\,$is$\,$interesting 
to notice that the transverse component $k_{\perp}$ of the probe k-vector is unchanged$\,$when$\,$the probe$\,$goes though the air/medium interface. Indeed, using the Snell law, it appears that the in-air transverse wave-vector $k_{\perp}^{_{(i)}} = k_{0} \sin(\theta_{i})$ is equal to $k_{\perp}^{_{(r)}} = k_{0} n_{0} \sin(\theta_{r})$. Therefore, 
\begin{equation}
    \Delta \Phi_{\mathrm{l}}(k_{\perp}) = \frac{1}{2} k_{0} n_{0} L \, \sin^{2}(\theta_{r}) = \frac{k_{\perp}^{2}}{2 k} L.  
    \label{LPPhaseDiff}
\end{equation}
\noindent A simple way to understand the relation~\eqref{LPPhaseDiff} is to move into the frame translating at $c/n_{0}$ along the optical axis. This is by the way what we do, in chapter 2, when we introduce the slow-varying envelope $\mathcal{E}_{0}$ of the electric field amplitude $\mathcal{E}$: $\mathcal{E}_{0}(\boldsymbol{r_{\perp}}, z) = \mathcal{E}(\boldsymbol{r_{\perp}}, z) \exp(-i k z)$.
This formula can be regarded, in some ways, as a change of observation frame. Indeed, using $\mathcal{E}_{0}$ instead of $\mathcal{E}$ is like describing the physics from the frame translating at the speed $c/n_{0}$ along $z$. In this frame, pump photons are at rest$\,$and$\,$probe$\,$photons$\,$behave$\,$as$\,$free- particles of k-vector $k_{\perp}$ and energy $\hbar \Omega=\hbar \, c \, k_{\perp}^{2}/(2 n_{0} k)$ (remembering$\,$the$\,z \leftrightarrow t\,$mapping). Those particles exist during $t = n_{0}\, L/c$ and accumulate the phase $\Omega \, t = k_{\perp}^{2} \, L /(2 k)$ during their lifetime. It seems then natural to suppose that the same applies in$\,$the$\,$nonlinear$\,$case, namely, that $\Delta \Phi_{\mathrm{nl}}(k_{\perp}) = \Omega_{B}(k_{\perp}) \, L$. This assumption has been made in~\cite{2-24Vocke} and leads to the following expression for the shift: 
\noindent  
\begin{equation}
    \label{ShiftFaccio}
    \Delta S(k_{\perp}) = \frac{k_{\perp}}{2 k} \left[ \sqrt{1 + \frac{|\Delta n|}{n_{0}} \left( \frac{2 k}{k_{\perp}} \right)^{2}} - 1 \right] L.
\end{equation}
\noindent This formula matches with the geometrical argument discussed above. By moving from the linear to the nonlinear situation, the probe wave-vector is squeezed along the optical axis by $\delta k = \Omega_{B}(k_{\perp})-k_{\perp}^{2}/2k$. The phase shift $\Delta \Phi$ is therefore given by $\Delta \Phi = \delta k \, L$ and, by using~\eqref{Shift}, we recover~\eqref{ShiftFaccio} (in absolute value). Equation~\eqref{ShiftFaccio} predicts that the shift reaches an asymptotic value, $c_{s} \, L$, when $k_{\perp} \ll k_{\xi}$, or in other words, the linear increase of $\Delta \Phi$ with $k_{\perp}$ when $k_{\perp} \ll k_{\xi}$. Conversely, $\Delta \Phi$ should tend toward $k_{0} \Delta n L$ when $k_{\perp} \gg k_{\xi}$. I remind you that $k_{\xi} = k\sqrt{\Delta n / n_{0}}$ is the inverse of the healing length $\xi$.

\newpage

\noindent This last point seems quite natural. Indeed, in the particle-like regime, plane-wave density modulations follow a quadratic dispersion as $\Omega_{B}(k_{\perp}) \approx k_{0} \, \Delta n + k_{\perp}^{2}/(2 \, k)$. The offset stems basically from the value the dispersion takes at the end of the sound-like regime, that is, when $k_{\perp} \! = \! k_{\xi}$. Because plane wave modulations propagating onto a low and$\,$a$\,$high$\,$density background fluid behave both as massive particles as long as $k_{\perp} \gg k_{\xi}$, the phase difference accumulated between them only comes from this offset and $\Delta \Phi \approx k_{0} \Delta n L$, as$\,$stated$\,$by$\,$the formula$\,$above.$\;$However,$\,$the$\,$non-zero$\,$value$\,$of$\,$the$\,$shift$\,$for$\,k_{\perp} \!\ll\! k_{\xi}\,$is$\,$not$\,$easy$\,$to understand. We have fought long and hard to observe it$\,$experimentally$\,-\,$without$\,$any$\,$success$\,-\,$before questioning the validity of equation~\eqref{ShiftFaccio}.
\vspace{6pt}
\newline
\noindent When pump and probe get inside the nonlinear medium, they spontaneously generate an idler beam, having the same frequency $\omega$, which propagates at the conjugate$\,$k-vector$\,-k_{\perp}$. This third order nonlinear wave mixing process is known as degenerate four wave mixing. Since the three beams involved in this mechanism have the same frequency, they will$\,$not fulfill phase matching conditions, except when pump and probe are$\,$copropagating,$\,$that$\,$is, when $k_{\perp} = 0$. The generation of the idler field in experiments is therefore$\,$due$\,$to$\,$the$\,$sudden change in the nonlinearity at the medium input plane.$\;$But$\,$contrarily$\,$to$\,$what$\,$is$\,$said$\,$in~\cite{2-24Vocke}, the idler beam is not suppressed shortly thereafter. Energy exchanges between probe and idler beams (by means of the pump) continuously take place all along their propagation in the nonlinear medium (probe and idler are plane waves and are thus$\,$spatially$\,$overlapping). In order to derive a reliable formula linking $\Delta S$ to $\Omega_{B}$, we need to properly describe the evolution of the probe field inside the nonlinear medium and thus to take the coupling between$\,$the probe and the idler beams into account. We can as of now state that$\,$this$\,$more exhaustive approach will not change the behaviour of the shift when $k_{\perp} \gg k_{\xi}$, since in that case degenerate for wave mixing processes are not phase$\,$matched$\,$at$\,$all$\,$and$\,$thus$\,$completely inefficient$\,$in$\,$generating$\,$the$\,$idler$\,$field.$\;$At$\,$large$\,k_{\perp}$,$\,$the$\,$impact$\,$of$\,$the$\,$idler$\,$on$\,$the$\,$propagation of the probe can therefore be neglected and the formula~\eqref{ShiftFaccio} applies, as$\,$mentioned$\,$in~\cite{5-5Ferreira}. However, we will see that this correction as drastic effect at low $k_{\perp}$.
 
\subsubsection{Full description using Bogoliubov theory} 
      
\noindent In the framework of Bogoliubov's theory, the small plane-wave modulation spontaneously excites a superposition of two Bogoliubov modes when it gets inside the nonlinear medium. Those modes are counter-propagating, in the transverse plane, at $\pm v_{\mathrm{ph}} = \pm \Omega_{B}(k_{\perp})/k_{\perp}$. In what follows, I will sometimes erroneously named these counter-propagating modes "probe" and "idler". However, I must emphasize that probe and idler fields are technically generated at the the nonlinear medium exit plane in this description. I will come back to this point in a moment. For now, let's derive, within the Bogoliubov's formalism, an exact expression for $\Delta \Phi_{\mathrm{nl}}$ and $\Delta S$, taking linear absorption into account (\textit{ie} $\alpha \ne 0$). The in-air slowly varying electric field envelope, right$\,$before$\,$the$\,$medium$\,$input$\,$plane,$\,$reads$\,$as$\,$follows:      

\begin{equation}
    \label{FieldInAir}
    \mathcal{E}_{0} (\mathbf{r_{\perp}}, z) = \mathcal{E}_{0}(z) + e^{i \Phi_{0}(z)} \int \frac{\mathrm{d} \mathbf{k_{\perp}}}{(2 \pi)^{2}} \, a(\mathbf{k_{\perp}}, z) \, e^{-i \mathbf{k_{\perp}} \!\cdot\! \mathbf{r_{\perp}}}.
\end{equation}
\noindent Let's set the phase of the pump at $z = 0^{-}$ to zero. Let's also assume that the field energy is conserved thought the air/medium interface (perfect window anti-reflection coating). The Fourier amplitudes $a(\mathbf{k_{\perp}}, 0^{-})$ and $a^{*}(-\mathbf{k_{\perp}}, 0^{-})$ are then related to the Bogoliubov operators $b(\mathbf{k_{\perp}})$ and $b^{*}(-\mathbf{k_{\perp}})$ by:
     
\newpage 

 \begin{equation}
    \label{Input}
    \begin{pmatrix} 
    a_{\mathbf{k_{\perp}}} \! (0^{-}) \\
    a^{*}_{\mathbf{-k_{\perp}}} \! (0^{-}) 
    \end{pmatrix}
    = \sqrt{n_{0}} \,
    M_{\mathbf{k_{\perp}}} \! (0)
    \begin{pmatrix} 
    b_{\mathbf{k_{\perp}}} \\
    b^{*}_{-\mathbf{k_{\perp}}} 
    \end{pmatrix} \; \mathrm{whit} \;
    M_{\mathbf{k_{\perp}}} \! (z) = 
    \begin{pmatrix} 
    u_{\mathbf{k_{\perp}}} \! (z) & v^{*}_{\mathbf{k_{\perp}}} \! (z) \\
    v_{\mathbf{k_{\perp}}} \! (z) & u^{*}_{\mathbf{k_{\perp}}} \! (z) 
    \end{pmatrix} \! ,
\end{equation}
\noindent where $u_{\mathbf{k_{\perp}}} \! (z) = \Tilde{u}_{\mathbf{k_{\perp}}} \! (z) \, e^{i \Omega_{\mathrm{eff}} z}$ and $v_{\mathbf{k_{\perp}}} \! (z) = \Tilde{v}_{\mathbf{k_{\perp}}} \! (z) \, e^{i \Omega_{\mathrm{eff}} z}$ ($\Omega_{\mathrm{eff}}$ is$\,$defined$\,$in$\,$subsection$\,$2.4.1). One can derive a relation similar to~\eqref{Input} at the second air/medium interface, when $z = L$. Combining both and setting $|u_{\mathbf{k}_{\perp}}|^{2} - |v_{\mathbf{k}_{\perp}}|^{2} = N(\mathbf{k}_{\perp}, z)$, we finally get a relation between the input and output Fourier components of the electric field envelope:
\begingroup
\allowdisplaybreaks
\begin{align}
    \nonumber
    \begin{pmatrix} 
    a_{\mathbf{k_{\perp}}} \! (L^{+}) \\
    a^{*}_{\mathbf{-k_{\perp}}} \! (L^{+}) 
    \end{pmatrix}
    =& \, \left[ \vphantom{\frac{M^{-1}_{\mathbf{k_{\perp}}} \! (0^{+})}{\sqrt{n_{0}}}} \sqrt{n_{0}} \, M_{\mathbf{k_{\perp}}} \! (L) \right] \! \cdot \! \left[\frac{M^{-1}_{\mathbf{k_{\perp}}} \! (0)}{\sqrt{n_{0}}} \right] 
    \begin{pmatrix} 
    a_{\mathbf{k_{\perp}}} \! (0^{-}) \\
    a^{*}_{\mathbf{-k_{\perp}}} \! (0^{-}) 
    \end{pmatrix} \tag{\stepcounter{equation}\theequation}  \\ 
    \label{InputOutput}
    =&  \, \frac{1}{N(\mathbf{k_{\perp}},0^{+})} \, \begin{pmatrix} 
    U_{\mathbf{k_{\perp}}} & V^{*}_{\mathbf{k_{\perp}}} \\
    V_{\mathbf{k_{\perp}}} & U^{*}_{\mathbf{k_{\perp}}} 
    \end{pmatrix} 
    \begin{pmatrix} 
    a_{\mathbf{k_{\perp}}} \! (0^{-}) \\
    a^{*}_{\mathbf{-k_{\perp}}} \! (0^{-}) 
    \end{pmatrix} \!.
\end{align}
\endgroup
\noindent The normalization constant $N(\mathbf{k_{\perp}},0^{+})$ is equal to one. Indeed, in the thin medium limit (\textit{ie} when $L \rightarrow 0$) the effect of linear absorption on the propagation of$\,$the$\,$beams$\,$is$\,$negligible and the normalization condition reduces to the one of the lossless case $|u_{\mathbf{k}_{\perp}}|^{2} - |v_{\mathbf{k}_{\perp}}|^{2} = 1$. Right before the nonlinear medium entrance plane ($z=0^{-}$), the idler beam has still not been generated. Therefore, the amplitude $a^{*}_{\mathbf{-k_{\perp}}} \! (0^{-})$ is zero, which amounts to saying that the idler is seeded by vacuum. The input-output relation \eqref{InputOutput} can easily be understood. The probe spontaneously generates two counter-propagating Bogoliubov modes at $ \pm \mathbf{k_{\perp}}$ when it enters the medium. Bogoliubov modes are eigenstates of the nonlinear dynamics; they consequently accumulate the phase $\pm \Omega_{\mathrm{eff}} L$ before reaching the medium output plane. At that point, each of the two counter-propagating Bogoliubov excitations spontaneously turns into a superposition of probe and idler fields. The idler beam is thus only generated at the medium output plane, as mentioned at the beginning of this derivation. Its k-space amplitude is given by $ a^{*}_{-\mathbf{k_{\perp}}} \! (L^{+}) = V_{\mathbf{k_{\perp}}} \, {a}_{\mathbf{k_{\perp}}} \! (0^{-})$, where the function $V_{\mathbf{k_{\perp}}}\!$ is defined by: 
\begin{equation}
    V_{\mathbf{k_{\perp}}} = \, \Tilde{u}^{*}_{\mathbf{k_{\perp}}} \! (0) \, \Tilde{v}_{\mathbf{k_{\perp}}} \! (L) e^{i\Omega_{\mathrm{eff}} L} - \Tilde{v}_{\mathbf{k_{\perp}}} \! (0) \, \Tilde{u}^{*}_{\mathbf{k_{\perp}}} \! (L) \, e^{-i\Omega^{*}_{\mathrm{eff}} L}.
    \label{VExpression}
\end{equation}
\noindent The input-output relation \eqref{InputOutput} for$\,$the$\,$probe$\,$field$\,$reduces$\,$to:$\,a_{\mathbf{k_{\perp}}} \! (L^{+}) = U_{\mathbf{k_{\perp}}} \, {a}_{\mathbf{k_{\perp}}} \! (0^{-})$,$\,$with:  
\begin{equation}
    U_{\mathbf{k_{\perp}}} = \Tilde{u}_{\mathbf{k_{\perp}}} \! (L) \, \Tilde{u}^{*}_{\mathbf{k_{\perp}}} \! (0) \, e^{i\Omega_{\mathrm{eff}} L} - \Tilde{v}^{*}_{\mathbf{k_{\perp}}} \! (L) \, \Tilde{v}_{\mathbf{k_{\perp}}} \! (0) \, e^{-i\Omega^{*}_{\mathrm{eff}} L}.  
    \label{UExpression}
\end{equation}
\noindent The phase accumulated by the probe while propagating in the medium is thus given by:
\begin{equation}
    \Phi_{\mathrm{nl}}(k_{\perp}, L^{+}) = \arg{\left(U_{\mathbf{k_{\perp}}}\right)} + \Phi_{0,
    \mathrm{nl}}(L^{+}),
    \label{NonlinearPhase}
\end{equation}
\noindent and, finally, $\Delta \Phi_{\mathrm{nl}}(k_{\perp}) = \arg{\left(U_{\mathbf{k_{\perp}}}\right)}$. When the phase matching conditions$\,$are$\,$not$\,$fulfilled, that is, when $k_{\perp} \gg k_{\xi}$, the amplitude of the "idler" beam is$\,$much$\,$lower$\,$than$\,$the$\,$"probe"$\,$one ($v_{\mathbf{k_{\perp}}} \! \ll u_{\mathbf{k_{\perp}}}$). The second term on the right hand side of~\eqref{UExpression} is negligible$\,$in$\,$that$\,$case$\,$and $\arg{(U_{\mathbf{k_{\perp}}})} \simeq \Omega_{\mathrm{eff}}(\mathbf{k_{\perp}})$.$\;$As you may have noticed, equation~\eqref{NonlinearPhase} reduces$\,$then$\,$to$\,$the$\,$relation we have intuited before in order to derive the formula \eqref{ShiftFaccio} which thus correctly describes how$\,$the$\,$shift$\,$evolves$\,$when$\,k_{\perp} \gg k_{\xi}$.$\;$But$\,$when$\,$quasi-phase$\,$matching$\,$conditions$\,$are$\,$fulfilled, the argument of $U$ depends in a complicated manner on the the amplitude and phase of the Bogoliubov amplitudes $u$ and $v$. By using the relations $\Tilde{u} = \frac{1}{2}(\Tilde{f}_{+} \!+\! \Tilde{f}_{-})$ and $\Tilde{v} = \frac{1}{2}(\Tilde{f}_{+} \!-\! \Tilde{f}_{-})$ as well as the expression for the slow-varying Fourier components $\Tilde{f}_{\pm}$, defined in 2.4.1, one can go through the tedious but straightforward computation of $\Delta \Phi_{\mathrm{nl}}$ in$\,$the$\,$general$\,$case. At the end of the day:

\newpage

\begin{figure}[h!]
\center
\includegraphics[width=\columnwidth]{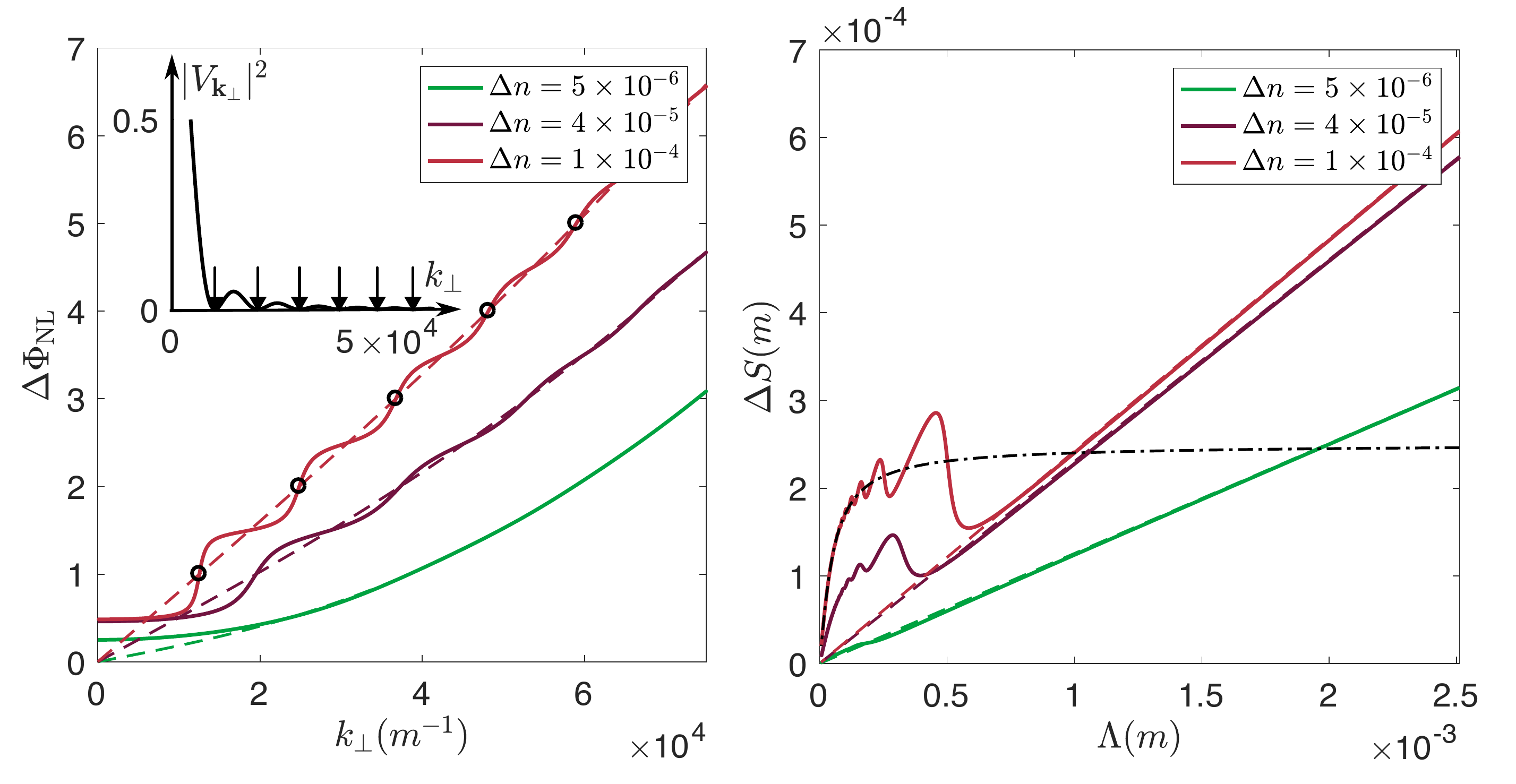} 
\caption{
Plots of the nonlinear phase shift $\Delta \Phi_{\mathrm{nl}}\,$(a) and of the shift $\Delta S\,$(b)$\,$as$\,$function of $k_{\perp}$ and $\Lambda = 2 \pi / k_{\perp}$ respectively, for various values of the nonlinear refractive index $\Delta n$. (a): The nonlinear phase shift follows globally the curve $k_{\perp} \! \shortrightarrow \! \Omega_{B}(\mathbf{k_{\perp}}) L$ (dotted lines) but exhibits a stair-like structure, which is more and more pronounced$\,$as$\,\Delta n = n_{2} \mathcal{I}_{0}\,$increases. Moreover, $\Delta \Phi_{\mathrm{nl}}$ tends toward$\,$a$\,$non-zero$\,$constant$\,$when$\,k_{\perp}\,$goes$\,$to$\,$zero$\,$(see$\,$formula~\eqref{LinearTrend}). (b): This translates into a linear increase of the shift $\Delta S$ for large modulation wavelengths. The oscillations on the right come for the stair-like structure of $\Delta \Phi_{\mathrm{nl}}$. The black dashed-dotted line shows, for comparison, the shift obtained from equation~\eqref{ShiftFaccio} when $\Delta n = 10^{-4}$. It predicts a saturation of $\Delta S$ at large $\Lambda = 2 \pi / k_{\perp}$ which is not observed experimentally.
Inset of (a): idler intensity in k-space as function of $k_{\perp}$ for $\Delta n = 10^{-4}$. The positions at which is cancels (black arrows) are reported on the plot of $\Delta \Phi_{\mathrm{nl}}$ obtained at the$\,$same$\,\Delta n$ (black$\,$circles).$\;$They$\,$correspond$\,$to$\,$the$\,$points$\,$at$\,$which$\,\Delta \Phi_{\mathrm{nl}}\,$crosses$\,$the$\,$dispersion$\,$relation. 
}
\label{fig:ThPhaseShift}
\end{figure} 

\begin{equation}
    \label{HPPhase}
    \hspace{0.1cm}\Delta \Phi_{\mathrm{nl}}(k_{\perp}) = \arctan{\left[ \frac{k_{\perp}^4 + 4 \, k^{2} \, \mathrm{Re}\left[\Omega_{B}(\mathbf{k_{\perp}},0)\right] \mathrm{Re}\left[\Omega_{B}(\mathbf{k_{\perp}},L)\right]}{2 \, k \, k_{\perp}^{2} \left \{ \mathrm{Re}\left[\Omega_{B}(\mathbf{k_{\perp}},0)\right] + \mathrm{Re}\left[\Omega_{B}(\mathbf{k_{\perp}}, L) \right] \right \}}  \! \times \! \tan \left \{ \mathrm{Re} \left[\Omega_{\mathrm{eff}}(\mathbf{k_{\perp}}, L) \right] L \right \} \right]}        
\end{equation}
\noindent which, in the lossless case ($\alpha = 0$), reads barely simpler: 
\begin{equation}
    \label{HPPhaseLossless}
    \Delta \Phi_{\mathrm{nl}}(k_{\perp}) = \arctan{\left[ \frac{k_{\perp}^4 + 4 \, k^{2} \left[ \Omega_{B}(\mathbf{k_{\perp}}) \right]^{2}}{4 \, k \, k_{\perp}^{2} \Omega_{B}(\mathbf{k_{\perp}})} \! \times \! \tan \left\{ \Omega_{B}(\mathbf{k_{\perp}}) L \right \} \right]}.
\end{equation}
\noindent The formulas~\eqref{HPPhase} and~\eqref{HPPhaseLossless} cannot be easily inverted in order to express the$\,$Bogoliubov dispersion$\,$relation$\,$as$\,$function of the nonlinear phase shift $\Delta \Phi_{\mathrm{nl}}$,$\,$or$\,$equivalently,$\,$as$\,$function of$\,$the$\,$shift$\,\Delta S\,$using$\,$the$\,$relation~\eqref{Shift}.$\;$Measuring$\,\Delta \Phi_{\mathrm{nl}}\,$to$\,$retrieve$\,$the$\,$dispersion$\,$relation$\,$of small amplitudes density waves$\,$seems$\,$therefore$\,$to$\,$be$\,$quite$\,$unsuitable.$\;$However,$\,$we$\,$can$\,$still observe some non-trivial behaviour of the nonlinear phase shift when $k_{\perp} \ll k_{\xi}$. In$\,$that$\,$case, 
$\Delta \Phi_{\mathrm{nl}}$ tends toward a non-zero value:
\begin{equation}
     \Delta \Phi_{\mathrm{nl}}(k_{\perp} \simeq 0) = \arctan \left \{ 2 \, k_{0} \, \Delta n(0) \, L \times \frac{2}{\alpha L} \frac{1 - \exp \left(-\alpha L/2 \right)}{1 + \exp \left(-\alpha L/2 \right)} \right \},  
     \label{LinearTrend}
\end{equation}
\noindent which is rather counter-intuitive from the linear optics perspective, in which we expect$\,$it$\,$to simply go to zero. Indeed, if we consider naively that the effect of nonlinearity is$\,$to$\,$shift$\,$the value of the refractive index $n_{0}$ by $\Delta n = n_{2} \mathcal{I}_{0}$, there is no reason why $\Delta \Phi_{\mathrm{nl}}\,$should$\,$not$\,$go$\,$to zero$\,$when$\,$the$\,$pump$\,$and$\,$the$\,$probe$\,$are$\,$parallel.$\,$Figure$\,$\ref{fig:ThPhaseShift}(a)$\,$shows$\,$the$\,$trend$\,$of$\,$the$\,$nonlinear phase shift as function of $k_{\perp}$ for different values of $\Delta n$ and for $\alpha = 0$. The$\,$phase$\,$shift$\,\Delta \Phi_{\mathrm{nl}}$ tends toward a non-zero value when $k_{\perp}\,$goes$\,$to$\,$zero.$\;$This$\,$asymptotic$\,$limit$\,$quickly$\,$saturates when $\Delta n$ increases, as suggested by equation~\eqref{LinearTrend}. Moreover, the nonlinear phase shift exhibits a stair-like structure, more visible at low $k_{\perp}$, which follows globally the trend of the curve $k_{\perp} \! \shortrightarrow \Omega_{B}(\mathbf{k_{\perp}}) L$ (dotted lines). It becomes$\,$more$\,$and$\,$more$\,$marked$\,$as$\,\Delta n\,$increases. The plateaus in this stair-like structure exactly lie at the transverse wave-vectors at which counter-propagating Bogoliubov modes interfere destructively in the exit plane, that is, when $k_{\perp}$ fulfills: $\Omega_{B}(k_{\perp}) L = n \pi + \pi/2$, $n$ being a positive integer. Reversely, the inflection points in between consecutive plateaus precisely mark the locations of the wave-vectors at which Bogoliubov modes constructively interfere in the output plane. These interferences occur when $\Omega_{B}(k_{\perp}) L = n \pi$ (black circles).$\;$Figure~\ref{fig:ThPhaseShift}(a)$\,$shows$\,$the$\,$evolution$\,$of$\,$the$\,$shift$\,\Delta S$ as function of the modulation wavelength $\Lambda = 2\pi/k_{\perp}$, for the$\,$same$\,$values$\,$of$\,\Delta n\,$as$\,$before. Because of the stair-like structure of $\Delta \Phi_{\mathrm{nl}}$, the$\,$shift$\,$exhibits$\,$strong$\,$oscillations$\,$when$\,\Lambda \simeq \xi$. The linear increase of $\Delta S$ with the modulation wavelength when $\Lambda \gg \xi$ results from the behaviour of $\Delta \Phi_{\mathrm{NL}}$ when $k_{\perp} \ll k_{\xi}$. The slope of this linear trend is equal$\,$to$\,$the$\,$right$\,$hand side of equation~\eqref{LinearTrend} times $1/2\pi$. For comparison, the shift obtained using the model proposed in~\cite{2-24Vocke} has been reported on the same graph (black dashed line), for$\,\Delta n = 10^{-4}$. As expected, both descriptions match in the particle-like regime (that is, for $\Lambda \ll \xi$) but instead$\,$of a linear increase at high $\Lambda$, equation~\eqref{ShiftFaccio} predicts a saturation of$\,\Delta S\,$toward$\,c_{s} L$ (which has never been observed with our setup).

\subsubsection{Quasi-particles interferences}

As we have seen in the previous paragraph, the key features in the curves$\,$of$\,\Delta \Phi_{\mathrm{nl}}\,$and$\,\Delta S$ shown on figure~\ref{fig:ThPhaseShift} can be explained in terms of interferences between counter-propagating Bogoliubov modes. A way of developing a better understanding about these quasi-particle interferences is to move from a spatial to a temporal description of the dynamics,$\,$using$\,$the $z \leftrightarrow t$ mapping once again. In doing so, we can think about the Bogoliubov modes as a quasi particle/antiparticle pair that is generated at $\tau =0\,$(where$\,\tau\,$is$\,$defined$\,$by$\,\tau = z n_{0}/c$). The particle ("idler") evolves at $-\mathbf{k_{\perp}}$ over the positive time while the antiparticle ("probe") evolves at $\mathbf{k_{\perp}}$ over the negative time (as$\,$if$\,$it$\,$was$\,$going$\,$back$\,$in$\,$the$\,$past).$\,$Indeed,$\,$in$\,$the$\,$frame translating at $c/n_{0}$ along the optical axis, probe photons seems to move backward, that$\,$is, toward the negative $z$ values, at a velocity $c \, \Omega_{B}(k_{\perp})/k$. In the lossless case, particle and antiparticle respectively$\,$accumulate$\,$a$\,$phase$\,-\Omega_{B}(\mathbf{k_{\perp}}) L\,$and$\,\Omega_{B}(\mathbf{k_{\perp}}) L\,$during$\,$their$\,$lifetime. 
As they are overlapping in the transverse plane (I assume they are plane-waves), they will totally interfere (constructively or destructively) as soon as $\mathrm{mod} \left \{ 2 \, \Omega_{B}(\mathbf{k_{\perp}}) \, L, \pi \right \} = 0$, or, in other words, when $\Omega_{B}(\mathbf{k_{\perp}}) \, L = n \pi/2$ ($n$ being a positive integer).
\vspace{6pt}
\newline
\noindent It is interesting to notice that quasi-particle interferences lead$\,$moreover$\,$to$\,$the$\,$cancellation of the idler intensity in the Fourier space. From equation \eqref{InputOutput}, one gets (when $\alpha  = 0$):

\newpage

\begin{align}
    \nonumber\frac{\mathcal{I}_{\mathbf{-k_{\perp}}}(L^{+})}{\mathcal{I}_{\mathbf{k_{\perp}}}(0^{-})} = \left| \frac{a_{\mathbf{-k_{\perp}}} \! (L^{+})}{a_{\mathbf{k_{\perp}}} \! (0^{-})} \right|^2 = \left| V_{\mathbf{k_{\perp}}} \right|^2 = & \, \left|2 \, \Tilde{u}_{\mathbf{k_{\perp}}} \Tilde{v}_{\mathbf{k_{\perp}}}\right|^2 \sin^{2}{\left[ \Omega_{B} (\mathbf{k_{\perp}}) \, L \right]} \\
    \label{IntensityRatio}
    = & \; \left( k_{0} \Delta n L \right)^{2} \mathrm{sinc}^{2} \left[ \Omega_{B} (\mathbf{k_{\perp}}) \, L \right],
\end{align}
\noindent where  $\mathrm{sinc}$ stands for the cardinal sine function. The idler intensity in$\,$Fourier$\,$space,$\,\mathcal{I}_{\mathbf{-k_{\perp}}}$, is maximum at $k_{\perp} = 0$, when phase matching$\,$conditions$\,$are$\,$perfectly$\,$fulfilled.$\;$For$\,k_{\perp} \! \ll \! k_{\xi}$, the envelope $|2 \, \Tilde{u} \; \Tilde{v}|^2$ of the k-space intensity profile scales as$\;1/k_{\perp}^4$. This is understandable, as the phase matching between pump and probe is getting worse when the probe transverse wave-vector increases. In addition, one can show that: $|U_{\mathbf{k_{\perp}}}|^{2} - |V_{\mathbf{k_{\perp}}}|^{2} = 1$ and, therefore, that probe and idler intensities in k-space are related by: $\mathcal{I}_{\mathbf{k_{\perp}}}(L^{+}) = \mathcal{I}_{\mathbf{k_{\perp}}}(0^{-}) + \mathcal{I}_{\mathbf{-k_{\perp}}}(L^{+})$.
Consequently, any variation in the idler intensity translates into a similar variation in the probe intensity, $\mathcal{I}_{\mathbf{k_{\perp}}}$. Equation$\,$\eqref{IntensityRatio} also indicates that the idler intensity $\mathcal{I}_{-\mathbf{k_{\perp}}}$ in k-space cancels at some specific values of $k_{\perp}\,$(located$\,$by$\,$the$\,$black$\,$arrows on$\,$the$\,$inset$\,$of$\,$figure$\,$\ref{fig:ThPhaseShift}(a) for instance), at which $\Omega_{B}(k_{\perp}) L = n \pi$. At those wave-vectors,$\,$we$\,$thus$\,$expect$\,$an$\,$increase in the contrast of the interference fringes at the medium exit plane, as both$\,$probe$\,$and$\,$idler are in phase. This of course depends on the$\,$length$\,$of$\,$the$\,$medium.

\begin{figure}[h!]
\center
\includegraphics[width=\columnwidth]{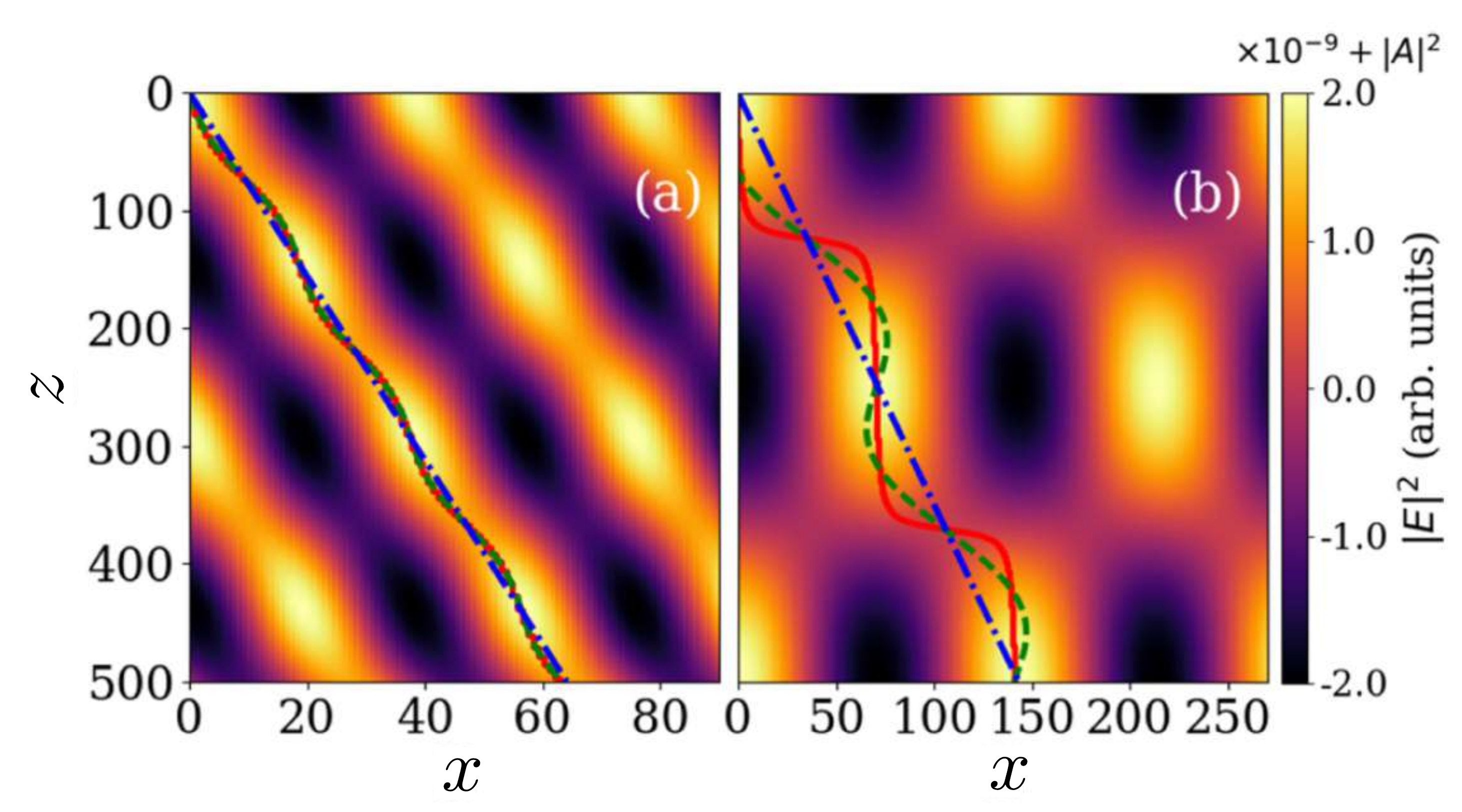} 
\caption{Figure taken from D. Ferreira \textit{et al.}~\cite{5-5Ferreira}.  The intensity of the total$\,$electric$\,$field is shown as function of the transverse$\,$position$\,x\,$and$\,$the$\,$propagation$\,$distance$\,z\,$(top$\,$view), for $k_{\perp} \gg k_{\xi}$ (a) and $k_{\perp} \ll k_{\xi}$ (b) respectively. The contrast of the interference patterns is modulated$\,$over$\,z$.$\;$We$\,$can$\,$complete$\,$the$\,$description$\,$made$\,$in~\cite{5-5Ferreira}$\,$by$\,$noticing$\,$that$\,$Bogoliubov modes$\,$destructively$\,$interfere$\,$when$\,\Omega_{B}(k_{\perp}) z = n \pi+\pi/2\,$,reducing$\,$consequently$\,$the$\,$contrast of the interference fringes. On the contrary, when $\Omega_{B}(k_{\perp}) z = n \pi$, Bogoliubov modes constructively interfere and the contrast is maximized.
}
\label{fig:Contrast}
\end{figure}

\newpage

\noindent Solving numerically the nonlinear Schr\"{o}dinger equation allows to compute the intensity$\,$of the total electric field inside the nonlinear medium in any transverse$\,$plane$\,$along$\,$the$\,z$-axis. In figure~\ref{fig:Contrast}, we$\,$present$\,$a$\,$cut$\,$along$\,$the$\,x$-axis,$\,$for$\,k_{\perp} \gg k_{\xi}\,$(a)$\,$and$\,k_{\perp} \ll k_{\xi}\,$(b)$\,$respectively. This figure is taken from~\cite{5-5Ferreira} where the feasibility of measuring the Bogoliubov dispersion $\Omega_{B}$ for$\,$propagating photon fluids in nematic liquid crystals is theoretically investigated, using the pump/probe technique described here.$\;$On$\,$figure~\ref{fig:Contrast},$\,$the$\,$background$\,$fluid$\,$density has$\,$been$\,$subtracted. The contrast of the interference fringes in (a) and (b) is modulated along the $z$-axis. When $\Omega_{B}(k_{\perp}) z = n \pi$, Bogoliubov modes constructively interfere and the contrast$\,$is$\,$increased.$\,$Reversely,$\,$when$\,\Omega_{B}(k_{\perp}) z = n \pi + \pi/2$,$\,$Bogoliubov$\,$modes$\,$destructively interfere and the contrast is reduced. Moreover, $\mathcal{I}_{\mathbf{-k_{\perp}}}\,$(and thus $\mathcal{I}_{\mathbf{k_{\perp}}}$)$\,$increases$\,$when$\,$the transverse wave-vector decreases, explaining why the variations of the contrast are bigger on figure (b) than on figure (a). 

\subsubsection{Nonlinear optics approach}

\noindent I would like to conclude this paragraph mentioning that equation~\eqref{IntensityRatio} can also directly be derived using textbook nonlinear optics.$\,$Let$\,\mathcal{E}_{pu}$,$\,\mathcal{E}_{pr}\,$and$\,\mathcal{E}_{i}\,$stand$\,$for$\,$the$\,$pump,$\,$the$\,$probe and the idler electric field envelopes. By neglecting the dependence of the various$\,$fields$\,$on the transverse spatial coordinates $x$ and $y$, the equation driving the evolution of$\,$the$\,$idler inside the nonlinear medium reads as follows (in the paraxial approximation)~\cite{3-1Boyd}:
\begin{equation}
    \label{IdlerWithoutDiffraction}
    i \frac{\partial \mathcal{E}_{i}}{\partial z} = - k \frac{n_{2}}{n_{0}} \left[ \mathcal{E}_{pu} \right]^{2} \mathcal{E}^{\,*}_{pr} \, \exp(i \, \Delta k \, z),
\end{equation}
\noindent where $k_{pu}$, $k_{pr}$ and $k_{i}$ are the pump, the probe and the idler axial wave-vectors respectively. The wave-vector mismatch $\Delta k$ is equal to $2k_{pu}-k_{pr}-k_{i}$. Since $k_{pr} = k_{i} = k_{pu}-\Omega_{B}(k_{\perp})$ (see for instance figure~\ref{fig:ShiftQualitative}(a)), $\Delta k = 2\Omega_{B}(k_{\perp})$ in the present case. When $\Delta k = 0$, that$\,$is, when $k_{\perp} = 0$ (perfect phase matching), the idler maintains a fixed phase relation with respect to the nonlinear polarization $\mathcal{P}_{i} = \frac{n_{2}}{n_{0}} \left[ \mathcal{E}_{pu} \right]^{2} \mathcal{E}^{\,*}_{pr} \, \exp(i \, \Delta k \, z)$ and is therefore$\,$able$\,$to extract energy more efficiently from the incident waves. When $\Delta k \ne 0$, the idler gets$\,$out of phase with its driving polarization $\mathcal{P}_{i} $ and part of its power can flow$\,$back$\,$into$\,$the$\,$pump.
Let $\mathcal{I}_{\mathrm{pr}}$ stand for the probe intensity.$\,$The$\,$idler$\,$intensity$\,\mathcal{I}_{\mathrm{i}}\,$at$\,$the$\,$medium$\,$exit$\,$plane$\,$is$\,$finally obtained by integrating equation~\eqref{IdlerWithoutDiffraction} from $0$ to $L$: 
\begin{equation}
\label{IdlerIntensity}
    I_{\mathrm{i}}(\Delta k) \simeq I_{\mathrm{pr}} \left(k_{0} \, \Delta n \, L\right)^{2} \left|\frac{e^{i \Delta k \, L}-1}{\Delta k \, L}\right|^{2} = I_{\mathrm{pr}} \left(k_{0} \, \Delta n \, L\right)^{2} \mathrm{sinc}^{2}\!\left[\frac{\Delta k \, L}{2} \right].
\end{equation}
\noindent Equations~\eqref{IdlerIntensity}$\,$and~\eqref{IntensityRatio}$\,$are$\,$strictly$\,$analogous.$\,$Bogoliubov's$\,$theory$\,$and$\,$nonlinear$\,$optics lead unsurprisingly to the same result but it is still interesting however to compare the understanding each of these approaches provides about the same physical phenomenon.    

\subsubsection{Discussion}

\noindent As mentioned previously, measuring the shift $\Delta S$ is not$\,$necessarily$\,$the$\,$most$\,$suitable$\,$way$\,$to retrieve the dispersion relation of density waves travelling onto propagating photon fluids. There are at least two reasons for this, that I would like$\,$to$\,$discuss$\,$before$\,$ending$\,$this$\,$section. The first is a technical limitation, that constrains the range of transverse k-vectors we can explore in practice. The second is more fundamental and concerns our ability$\,$to$\,$extract$\,$the dispersion relation of density waves from the shift, using equations~\eqref{Shift} and~\eqref{HPPhase}.  

\newpage

\begin{itemize}
    \item[$\bullet$] In order to measure the shift, pump and probe beams have to overlap all along their propagation inside the nonlinear medium. Therefore, probe and idler are also super- imposed at the medium exit plane, which is annoying as we only want$\,$to$\,$observe$\,$the interference pattern between pump and probe. Measuring the shift requires then to filter out the $-\mathbf{k_{\perp}}$ component of the field in k-space, that is, the contribution from the idler beam. In practice, the exit plane is imaged on camera using$\,$a$\,4f\,$telescope made of two lenses. The idler beam is therefore simply cut by positioning a mask$\,$in the focal plane of the first of these lenses. As the pump undergoes self-defocusing$\,$by propagating inside the nonlinear medium, its extension$\,$in$\,$k-space$\,$can$\,$be$\,$large$\,$enough to cover the range of wave-vectors over which the Bogoliubov dispersion relation is sound-like. Moreover, the mask should not cut a part of the pump beam in k-space, at$\,$the$\,$risk of distorting the real space filtered image,$\,$which$\,$drastically$\,$limits$\,$the$\,$range of usable transverse wave-vectors at the end of the day. 
    \item[$\bullet$] As outlined above, the second and more fundamental issue of the shift measurement concerns how the dispersion relation $-$ which is a priori not know $-$ is$\,$extracted$\,$from the shift using equations~\eqref{Shift} and~\eqref{HPPhase}. Once $\Delta S$ has been measured for various transverse wave-vectors $k_{\perp}$, the relation~\eqref{HPPhase} has to be inverted assuming it applies to the unknown dispersion $\Omega(k_{\perp})$. In doing so, we are not supposing directly$\,$that $\Omega = \Omega_{B}$, but however that equation~\eqref{HPPhase}, derived$\,$within$\,$the$\,$Bogoliubov$\,$framework, is still valid replacing $\Omega_{B}$ by $\Omega$. Things go consequently round and round$\,$in$\,$circles. It$\,$is$\,$absolutely essential to measure the dispersion relation of density waves using$\,$a technique that does not require any knowledge about Bogoliubov theory but shows, reversely, that this theory is appropriate to describe the dispersion relation $\Omega$.
\end{itemize}
\noindent In section 4.2, I will introduce a new$\,$experimental technique that overcomes these issues. Before that, I would like to present some results we obtained from the shift measurement. In$\,$the$\,$following$\,$subsections,$\;$I$\,$start$\,$by$\,$describing$\,$the$\,$setup$\,$and$\,$how$\,$data$\,$are$\,$post-processed, before discussing the results and putting them into$\,$perspective$\,$with$\,$numerical$\,$simulations. I conclude by briefly commenting the results obtained by Vocke \textit{et al.} in~\cite{2-24Vocke}.

\subsection{Experimental setup and data processing}

\begin{figure}[H]
\center
\includegraphics[scale=0.6]{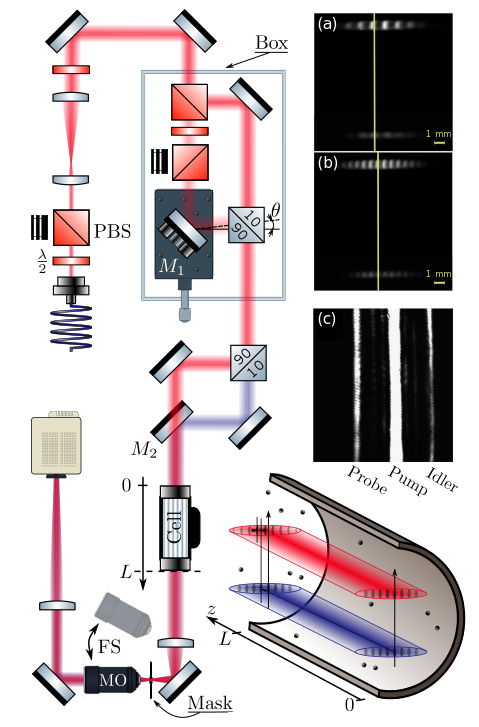} 
\caption{Experimental setup. The laser beam goes out from the fiber$\,$and$\,$is$\,$(i)$\,$magnified and (ii) elongated along the $x$-direction, before entering a Mach-Zehnder interferometer. From$\,$there,$\,$the$\,$beam$\,$splits$\,$into$\,$a$\,$high$\,$and$\,$a$\,$low$\,$power$\,$part$\,$(referred$\,$to$\,$as$\,$pump$\,$and$\,$probe). The piezo-electronically actuated mount of the mirror $M_{1}$ allows to finely tuned the angle $\theta_{i}$ between them. Pump and probe recombine and form an interference pattern whose fringes are parallel to the $y$ direction. A $90\!:\!10$ ($R\!:\!T$)$\,$beam$\,$splitter$\,$separates$\,$once$\,$again$\,$the beam in two parts after the Mach-Zehnder. The blue part is a low power reference while the red one is highly powerful. They propagate one above the$\,$other$\,$inside$\,$the$\,$vapor$\,$cell. The cell exit plane is imaged onto the camera with a $4f$ telescope.$\,$(a)$\,$and$\,$(b):$\,$output$\,$plane at low and large angle $\theta_{i}$ between pump and probe. The shift $\Delta S$ between the lower and the upper beam is clearly visible. The upper beam is broader because of self-defocusing. (c)$\,$Image$\,$of$\,$the$\,$Fourier$\,$space,$\,$obtained$\,$with$\,$a$\,$infinity$\,$corrected$\,$microscope$\,$objective$\,$(MO). Pump ($k_{\perp} = 0$), probe ($k_{\perp} = k_{0} \sin(\theta_{i})$) and idler ($k_{\perp}= -k_{0} \sin(\theta_{i})$), as well as the sidebands they generate, are visible on this image.  
} 
\label{fig:ShiftExp}
\end{figure}

\noindent The experimental setup designed to measure the shift has been sketched on figure~\ref{fig:ShiftExp}. The$\,$continuous-wave laser field at $780$ nm is$\,$provided$\,$by$\,$a$\,$Ti:Sapphire$\,$cavity,$\,$pumped$\,$with a $10$ W frequency-doubled Nd:YVO4 laser. The output laser beam is sent onto the optical table$\,$through a single-mode polarization-maintaining high$\,$power fiber.$\;$The$\,$power$\,$of$\,$the outgoing beam is controlled using a half-waveplate ($\frac{\lambda}{2}$) and$\,$a$\,$polarized$\,$beam$\,$splitter$\,$(PBS). The beam is first magnified and then highly elongated in the $x$ direction using a$\,$set$\,$of$\,$two cylindrical lenses ($f_{1} = 500$ mm and $f_{2} = 100$ mm). This cylindrical telescope is slightly misaligned in order to loosely focus the beam onto the medium input facet. In that plane, the minor axis width$\;\omega_{0,y}$ (at $1/e^{2}$) is around $500$ $\mu$m and the major$\,$axis$\,$one,$\,\omega_{0,x}$,$\,$is$\,$larger than $1$ cm. The intensity profile is thus quasi uniform along the $x$-axis.$\,$Squeezing$\,$the$\,$beam in the other direction is thus mostly a way to increase the intensity (and thus $\Delta n$) along this axis. The Rayleigh length associated to $\omega_{0,y}$ is about $1$ m which is much longer$\,$than the length of the cell$\,$($L = 7.5\,$cm).$\,$We$\,$can$\,$thus$\,$safely$\,$consider$\,$the$\,$beam$\,$as$\,$being$\,$collimated inside the nonlinear medium (as long as self-defocusing is negligible of course).

\newpage       
       
\newpage

\begin{figure}[hbt!]
\center
\includegraphics[scale=0.62]{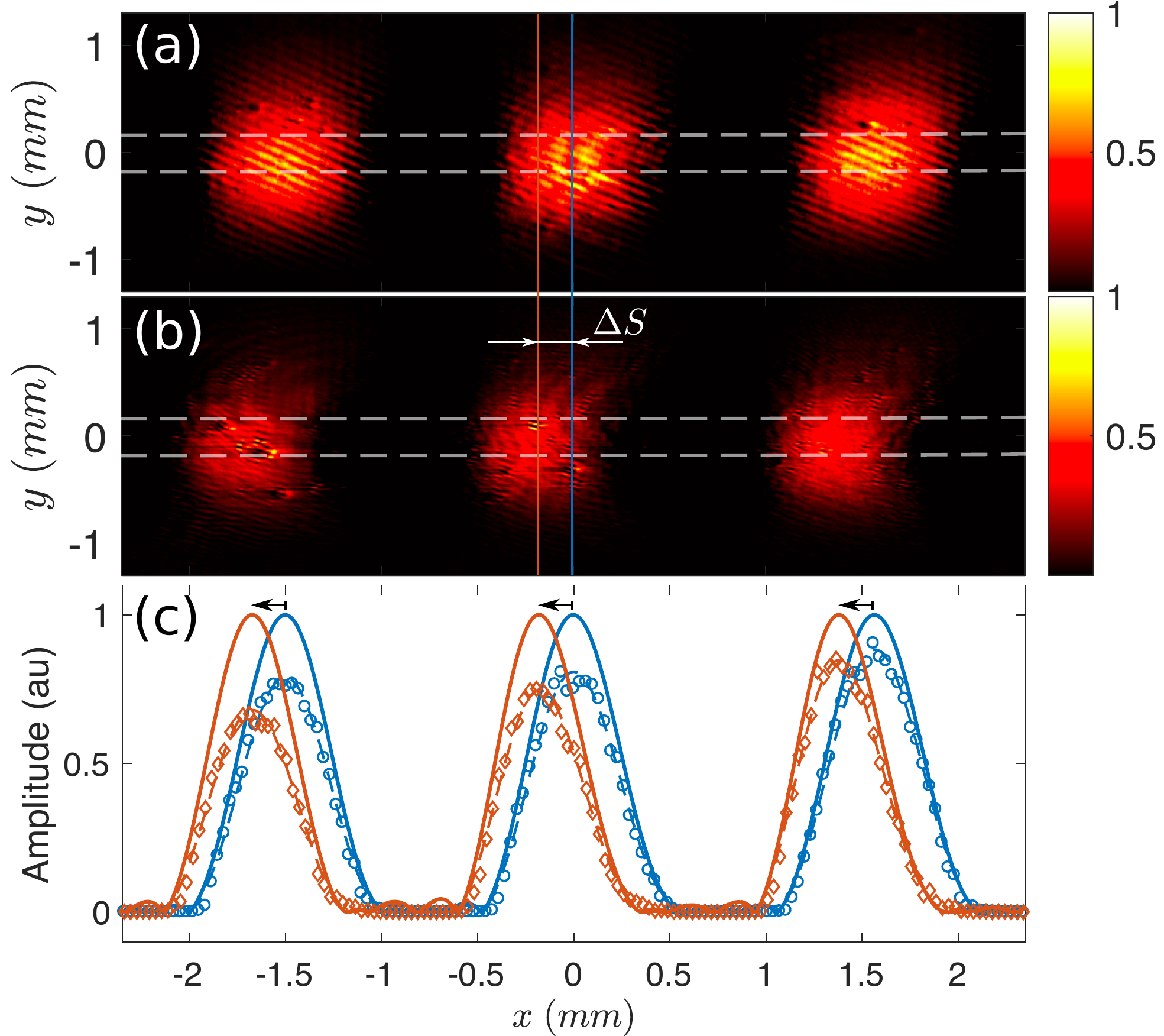} 
\caption{Data analysis.$\;$The$\,$background$\,$subtraction$\,$reveals$\,$the$\,$small$\,$amplitude$\,$density modulations that are propagating onto a low (a) and a high power background fluid (b). The blue and red data points in (c) are obtained$\,$by$\,$integrating$\,$the$\,$intensity$\,$in$\,$between$\,$the white dashed lines in (a) and (b) respectively. In order to clean$\;$the$\;$profiles,$\;$we$\;$first$\;$filter out the high frequencies noise (dashed lines) and then remove the intensity envelopes$\,$of$\,$the interference fringes (solid lines). The shift $\Delta S$ is finally computed by measuring$\,$the$\,$nearest peak-to-peak distance between the solid lines in average (black arrow).}
\label{fig:ShiftExpAnalysis}
\end{figure}

\noindent After having been shaped, the laser beam enters a Mach-Zehnder interferometer, which$\;$is protected against air turbulences by a box made of Plexiglas. At that point, the$\,$beam$\,$splits into a high power and a low power part (referred to as pump and probe$\;$in$\;$the$\;$section$\;$4.1). The probe reflects on the mirror $M_{1}$ that is hold in a piezo-actuated mount which in turn$\,$is fixed on a translation-stage. Tuning the voltage across the piezo allows to finely control the angle $\theta_{i}$ between pump and probe and thus the wavelength $\Lambda=2\pi/k_{\perp}\,$of$\,$the$\,$density$\,$waves. We make sure that the interference fringes that$\,$form$\,$when$\,$the$\,$beams$\,$recombine$\,$are$\,$parallel to the $y$-axis. On the probe arm, an other combination of half waveplate and polarized beam splitter controls the modulation depth (less than $5\%$). As soon as pump and probe exit the Mach-Zehnder interferometer, a $90\!:\!10$ ($R\!:\!T$)$\;$beam$\;$splitter$\;$splits$\;$the$\;$recombined laser beam into two parts. The blue part on figure~\ref{fig:ShiftExp} is a low power reference for which the medium response is basically linear, while the red one is highly powerful  and will thus behave as an interacting photon fluid. Blue and red beams propagates one$\,$above$\,$the$\,$other inside the vapor cell (the D-shaped mirror $M_{2}$ only reflects the reference), as sketched on the bottom right inset of figure~\ref{fig:ShiftExp}. The cell exit plane is imaged onto a CMOS camera, with a $4f$ imaging system made of two lenses$\,$of$\,$focal$\,$length$\,f_{3} = 200\,$mm$\,$and$\,f_{4} = 300\,$mm. By flipping on the beam path a microscope objective (MO), we can image the focal plane of the first lens in this $4f$ system (inset (c) of~\ref{fig:ShiftExp}) and thus precisely positioned the mask (razor blade) that filters out the idler beam.
\vspace{6pt}
\newline
\noindent In order to accurately measure the shift $\Delta S$, we need to precisely align the reference beam with respect to the high power one. The alignment procedure is as follow:
\begin{itemize}
    \item[(1)] We first make sure that both background beams (probe off) roughly propagate with the same transverse wave-vector and are correctly positioned one above the other.
    \item[(2)] We then switch the probe beam on.$\;$The$\,$next$\,$step$\,$is$\,$to$\,$align$\,$the$\,$interference$\,$fringes$\,$of the lower and upper interference patterns. We start by removing the vapor cell and make sure that bright fringes on the bottom face bright$\,$fringes$\,$on$\,$the$\,$top.$\,$Of$\,$course, by doing so, the optical axis of the lower and upper beams are not parallel anymore. We should then switch to k-space, bring back the backgrounds to the initial position ($k_{\perp} = 0$) and repeat this procedure iteratively (beam walking). We$\,$finally$\,$check$\,$that for every transverse wave-vector $k_{\perp}$ the interference fringes remain aligned before putting the cell back on the beams path.
\end{itemize}
\noindent Images obtained in the cell output plane are shown on the insets (a) and (b) of figure~\ref{fig:ShiftExp}. The contrast of the interference pattern (\textit{ie} the modulation depth) has been increased for the sake of clarity. The shift between the interference fringes is clearly visible for low (a) and high $k_{\perp}\,$(b).$\;$As you may have seen, the$\,$fringes$\,$on$\,$the$\,$upper$\,$interference$\,$pattern$\,$slightly bend toward the propagation direction of the density modulation$\,$(that$\,$is,$\,$from$\,$right$\,$to$\,$left). As the intensity profile of the upper beam along the vertical axis is Gaussian, the nonlinear phase shift the probe accumulates by propagating over it depends on$\;y$. The$\,$sound$\,$velocity is higher along the upper beam major axis than slightly above or below. Sound-waves$\,$will consequently propagate faster along this axis than along the beam edges, which as a result locally bends the fringes. In order to avoid error in the data analysis, we therefore only average the intensity profile over $30\;$pixels on both sides of the backgrounds major axis (\textit{ie}$\,$in$\,$between the white dotted line on figure~\ref{fig:ShiftExpAnalysis}).$\,$This$\,$is$\,$the$\,$maximal$\,$range$\,$for$\,$which$\,$fringe bending has negligible effects on the shift we measure. 
        
\newpage

\noindent The data analysis requires several post-processing steps that must be carefully performed. We start first by removing the background intensity distribution to keep only the density modulation on top of it, using the camera software directly. This was not a clever choice, as the background substraction function of this software is not only performing an image substraction but sets, in addition, all the negative intensities to zero, as you can see by looking$\,$at$\,$figure~\ref{fig:ShiftExpAnalysis}(c).$\;$We$\,$then$\,$integrate$\,$the$\,$intensity$\,$in$\,$between$\,$the$\,$white$\,$dashed$\,$lines on the interference patterns of figures~\ref{fig:ShiftExpAnalysis}(a) and (b), which are obtained respectively at low and high background fluid powers. The$\,$resulting$\,$profiles$\,$have$\,$been$\,$plotted on figure~\ref{fig:ShiftExpAnalysis}(c) (blue and red points). The$\,$high$\,$frequency$\,$noise$\,$is$\,$filtered$\,$out$\,$in$\,$Fourier$\,$space$\,$(dashed$\,$lines) and the interference fringes envelopes are removed to obtain the blue and red solid curves, using a Cubic spline interpolation method. The shift is finally computed$\,$by$\,$averaging$\,$the distance between the maximums of the blue and red solid lines (black arrows) over$\,$several interference fringes. The experimental$\,$results$\,$obtained$\,$with$\,$the$\,$setup$\,$sketched$\,$on$\,$figure$\,$\ref{fig:ShiftExp} and the data analysis above are presented in the next subsection.

\newpage
    
    \subsection{Experimental results and numerical simulations}

        \subsubsection{Comparison between experimental data and theory}
    
\begin{figure}[H]
\center
\includegraphics[scale=0.6]{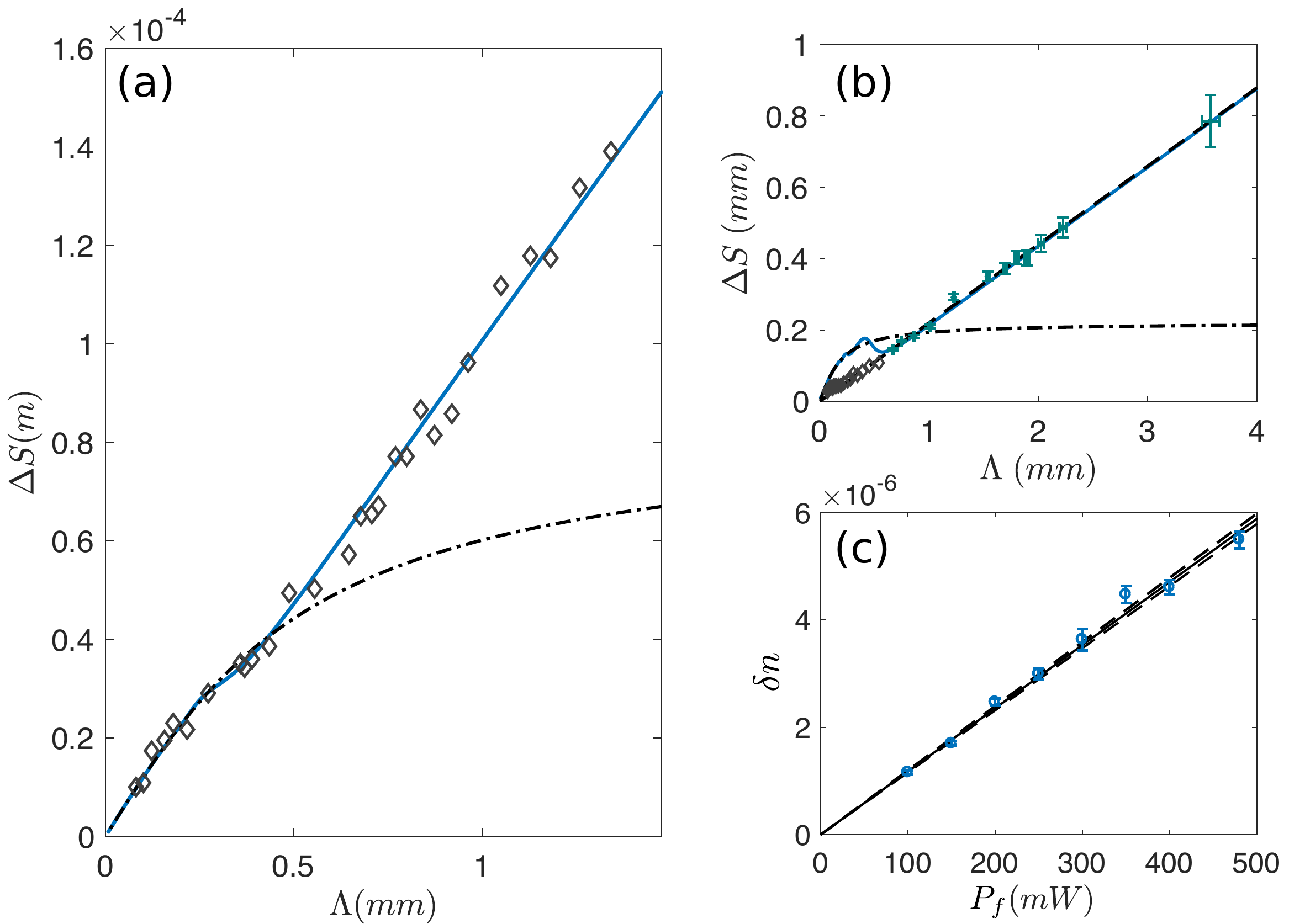} 
\caption{Shift experiment. (a): Shift $\Delta S$ as function of the modulation wavelength$\,\Lambda\,$for a fluid power of $120$ mW. The experimental data (grey diamonds) are in good agreement with the theory (blue line) for $\langle \Delta n(z') \rangle_{L} = 1.3 \!\times\! 10^{-6}$. For comparison, the shift obtained using equation~\eqref{ShiftFaccio} has also been plotted (black dashed-dotted line). The latter predicts a saturation of $\Delta S$ at large modulation wavelengths, which is not observed experimentally. (b): Same as before for a fluid power of $480$ mW this time. The theoretical curve$\;$does$\;$not match anymore with the data in the particle-like regime. (c): By fitting the shift with$\,$a$\,$line in the sound-like regime, the slope $a_{s}$ of the linear trend can be measure$\,$as$\,$function$\,$of$\,$the fluid power $\mathcal{P}_{f}$. We then check that $\delta n$, defined in equation$\,$\eqref{LinearTrend}, linearly$\,$scales$\,$with$\,\mathcal{P}_{f}$.}
\label{fig:MainShiftRes}
\end{figure}

\noindent The results we obtained with the shift experiment has been gathered together on figure~\ref{fig:MainShiftRes}. All the beams are propagating through a 7.5 cm cell filled with an isotopically$\,$pure$\,$vapor of rubidium 85, heated up at $130\,^{\circ}$C. The laser frequency is 2.6 GHz red-detuned$\,$from$\,$the $F_{g}=3 \rightarrow F'$ transition of the $^{_{85}}$Rb $D_{2}$ line. At these temperature and detuning,$\;$the$\,$vapor transmission is about $60\%$. Figure (a) shows the shift measured at low background density (grey diamonds) as function of the modulation wavelength $\Lambda$. The pump power$\,\mathcal{P}_{f}\,$is$\,$about $120$ mW and the reference one around $10$ mW. The probe power was small$\,$enough$\,$for$\,$the modulation depth to be less than $5\%$. The blue line is obtained by fitting the data using equations$\,$\eqref{HPPhase}$\,$and$\,$\eqref{Shift}.$\,$The$\,$fit$\,$provides$\,$the$\,$value$\,$of$\,$the$\,$nonlinear$\,$refractive$\,$index$\,$change: $\langle \Delta n(z') \rangle_{L} = 1.3 \!\times\! 10^{-6}$.$\,$As$\,$you$\,$can$\,$see$\,$on figure$\,$\ref{fig:MainShiftRes}(a),$\,$the$\,$agreement$\,$with$\,$theory$\,$is$\,$excellent

    
\newpage

\noindent and the shift $\Delta S$ increases linearly with the modulation wavelength, when it is$\,$larger$\,$than the healing length $\xi \simeq 300$ $\mu$m. This is an$\,$important$\,$result$\,$as$\,$it$\,$shows$\,$that$\,$the$\,$shift$\,$does$\,$not tend toward a finite value at large $\Lambda$, unlike what it is claimed in~\cite{2-24Vocke}.$\;$The$\,$shift$\,$computed with$\,$the$\,$model proposed in~\cite{2-24Vocke} $-$ equation~\eqref{ShiftFaccio} $-$ has also been plotted for comparison on the$\,$same$\,$graph$\,$(black$\,$dashed$\,$dotted$\,$line).$\;$The$\,$discrepancy$\,$between$\,$this$\,$model$\,$and$\,$the$\,$data is clearly visible in the regime where $\Lambda \gg \xi$. On figure~\ref{fig:MainShiftRes}(b), an example of$\,$data$\,$measured at higher background densities ($\mathcal{P}_{f} = 480$ mW) is shown as function of $\Lambda$. As you can$\;$see, $\Delta S$ still linearly increases$\,$with$\,$the$\,$modulation$\,$wavelength,$\,$when$\,$it$\,$is$\,$larger$\,$than$\,0.5\,$mm. 
\vspace{6pt}
\newline
\noindent The linear trend at large $\Lambda$ is theoretically described by the following equation:
\begin{equation}
    \Delta S(\Lambda) \simeq \arctan \left[ 2 \, k_{0} \, \Delta n(0) \, L \times \frac{2}{\alpha L} \frac{1 - \exp \left(-\alpha L/2 \right)}{1 + \exp \left(-\alpha L/2 \right)} \right] \frac{\Lambda}{2 \pi} \,,   
    \label{ShiftLinearTrend}
\end{equation}
\noindent derived from the formula~\eqref{LinearTrend}. Knowing$\,$the$\,$absorption$\,$coefficient$\,$($\alpha \simeq 7\,$m$^{-1}$),$\,$we$\,$can$\,$fit the linear trend of the shift at large $\Lambda$ (cyan diamonds), taking the errorbars into account. This provides the value of the nonlinear change of refractive index in$\,$the$\,$input$\,$plane$\,\Delta n (0)$. We then plot the theoretical shift obtained for this specific value$\,$of$\,\Delta n (0)\,$(blue$\,$solid$\,$line). As you can see, the experimental data a small$\,\Lambda\,$(grey$\,$diamonds)$\,$are$\,$not$\,$in$\,$good agreement with$\,$the$\,$theoretical$\,$model$\,$at$\,$high$\,$fluid$\,$densities$\,$(figure$\,$\ref{fig:MainShiftRes}(b)).$\,$I$\,$will$\,$discuss$\,$this$\,$point$\,$later. Nevertheless, we can still measure how the slope of the linear trend, $a_{s}$, evolves with $\mathcal{P}_{f}$. 

\begin{figure}[H]
\center
\includegraphics[width=\columnwidth]{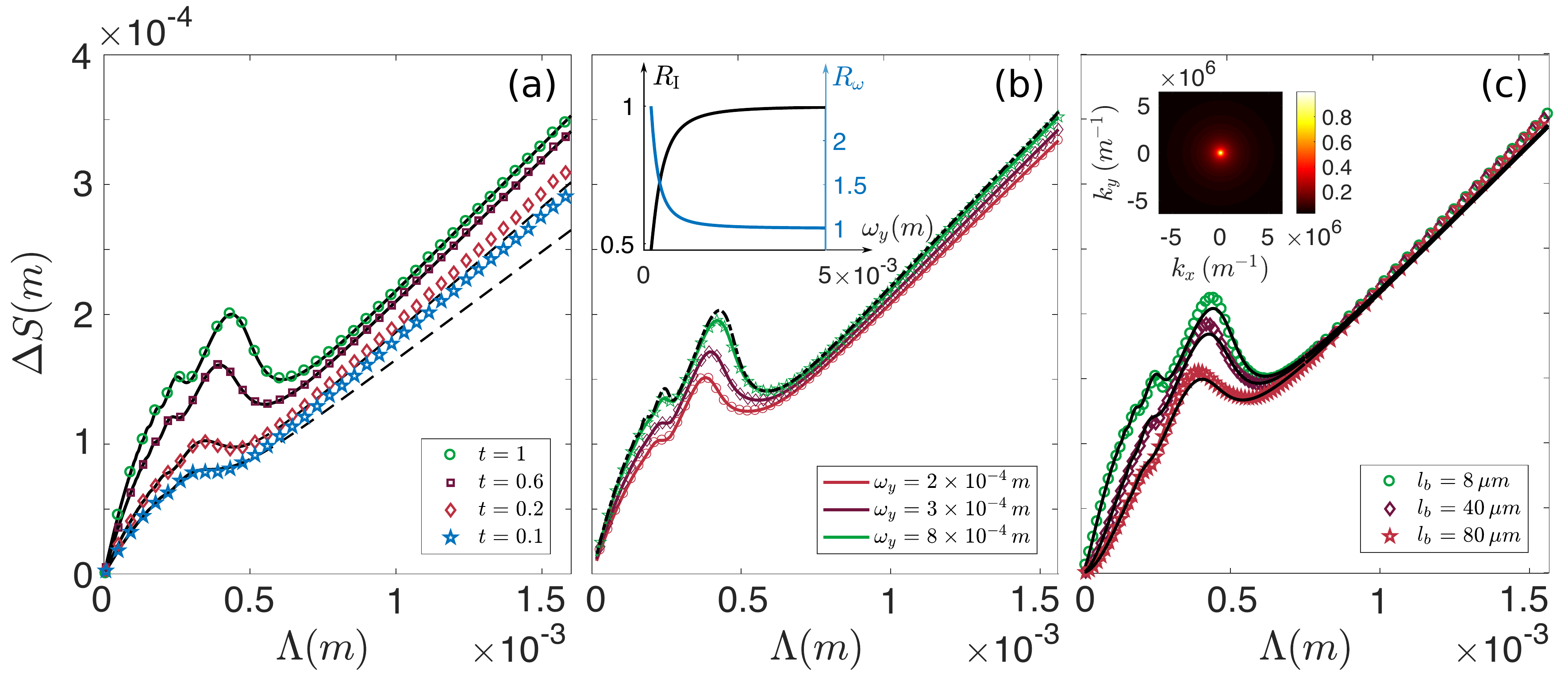} 
\caption{(a): Shift $\Delta S$ as function of the modulation wavelength $\Lambda$, for$\;$different$\;$vapor transmissions $t$. Simulations and theory match as long as the variation of the fluid density along $z$ remains adiabatic ($t > 0.5$). Absorption seems to smooth the oscillations of the shift but it only reduces the average nonlinear refractive index change $\langle \Delta n(z') \rangle_{L}$ actually. (b):$\,\Delta S\,$as$\,$function$\,$of$\,\Lambda\,$for$\,$various$\,$background$\,$widths$\,w_{y}$.$\;$The$\,$amplitude$\,$of$\,$the$\,$oscillations decreases with $w_{y}$. This behaviour is expected as$\,$self-defocusing$\,$is$\,$more$\,$likely$\,$to$\,$spread$\,$out the intensity of the background fluid$\,$in$\,$the$\,$transverse$\,$plane$\,$(and$\,$thus$\,$to$\,$reduce$\,\langle \Delta n(z') \rangle_{L}$) when $w_{y}$ decreases. (c): $\Delta S$ as function of $\Lambda$ for different nonlocal transport length scales. The oscillations are smoothed by nonlocality. However, for decent values of $l_{b}$, this effect is totally negligible. The inset of figure (c) shows the nonlocal response function in k-space. Parameters: $\Delta n(0) = 1.0 \!\times\! 10^{-5}$ and $L = 7.5$ cm.
}
\label{fig:ShiftVersus}
\end{figure} 
  
\newpage 

\noindent We already know that the nonlinear refractive index$\,$increases$\,$linearly$\,$with$\,$the$\,$fluid$\,$density and, consequently, with the pump power. We can thus invert equation~\eqref{ShiftLinearTrend} as follows: 
\begin{equation}
    \Delta n(0) = \frac{1}{4} \frac{\alpha }{k_{0}} \frac{1 + \exp \left(-\alpha L/2 \right)}{1 - \exp \left(-\alpha L/2 \right)} \, \tan \! \left[2\pi \, a_{s}(P_{f})\right], 
    \label{SlopeLinearTrend}
\end{equation}
\noindent and check that it is the case, plotting $\Delta n(0)$ as function of $P_{f}$. By doing so, we obtain$\,$the blue points on figure~\ref{fig:MainShiftRes}(c). As expected, the relationship between  $\Delta n(0)$ and $\mathcal{P}_{f}$ is linear indicating$\,$that$\,$the$\,$slope$\,$of$\,$the$\,$linear$\,$trend$\,$at$\,$large$\,\Lambda\,$correctly$\,$scales$\,$with$\,$the$\,$fluid$\,$density. 

\subsubsection{Discrepancy at low modulation wave-vectors}

\noindent As mentioned previously, the agreement between data and theory at low $\Lambda$ on figure~\ref{fig:MainShiftRes}(b) is not good. Actually, we never managed to observe experimentally the oscillations of the shift $\Delta S$ predicted by the theory at low $\Lambda$, when the fluid density$\,$is$\,$large$\,$enough.$\;$Moreover, our data does not even match the geometrical relationship between shift and modulation wavelength$\,$in$\,$the$\,$particle-like$\,$regime:$\,\Delta S(\Lambda) = k_{0} \Delta n \Lambda/2\pi$.$\;$This$\,$issue$\,$either$\,$comes$\,$from$\,$an uncontrolled systematic error in performing the experiment or from some physical$\,$process that is not taken into account in the theory yet. In order to test this second hypothesis, we numerically solve the NLSE in 2D using a second order split step method. The effect$\,$of linear$\,$absorption,$\,$self-defocusing$\,$and$\,$nonlocality$\,$on$\,$the$\,$shift$\,$are$\,$investigated.$\;$The$\,$results$\,$of the numerical simulations are grouped together on figure~\ref{fig:ShiftVersus}.$\,$They$\,$all$\,$have$\,$been$\,$performed using the same set of parameters. The background intensity, $\mathcal{I}_{0}$, is set to $2.5 \!\times\! 10^{5}$ W/m$^{2}$ and the nonlinear index of refraction, $n_{2}$, to $4 \! \times \! 10^{-11}$ m$^{2}$/W. The nonlinear change of refractive index is thus equal to $1.0 \! \times \! 10^{-5}$ in the lossless case. The background fluid is also infinitely elongated in the $x$-direction. On figure (a), $\Delta S$ has been$\,$plotted$\,$as$\,$function$\,$of$\,$the modulation wavelength for different cell transmissions $t = \exp(-\alpha L)$. The coloured points stem from numerical simulations whereas the theoretical curves are plotted in black solid. The agreement between simulations and theory is excellent as long$\,$as$\,$the$\,$variation$\,$along$\,z$ of the fluid density because of absorption remains adiabatic~\cite{3-19Larre}. Absorption seems to smooth the shift oscillations, but actually, it only reduces the average change of refractive index $\langle \Delta n(z') \rangle_{L}$. In other words,$\,$if$\,$the$\,$input$\,$intensity$\,$is$\,$multiplied$\,$by$\,\left[1-\exp(-\alpha L)\right] / \alpha L$, $\alpha$ being for instance the absorption coefficient corresponding to a cell transmission of $10 \%$, the$\,$blue$\,$stars$\,$on$\,$figure$\,$(a)$\,$will$\,$almost$\,$exactly$\,$translate$\,$onto$\,$the$\,$green$\,$circles,$\;$for$\,$which$\,t\!=\!1$. Absorption is thus not the cause of the discrepancy observed on figure~\ref{fig:ShiftExp}(b).$\;$Figure~\ref{fig:ShiftVersus}(b) shows the effect of the pump width $\omega_{0,y}$ on the shift. We clearly observe a reduction in the amplitude of the shift oscillations when $w_{0,y}$ decreases. This is due to the fact$\,$that$\,$the effective focal length $f_{\mathrm{eff}}$ of the nonlinear medium shortens in that case.$\;$Self-defocusing is then more likely to spread the background intensity in$\,$the$\,$transverse$\,$plane.$\;$This$\,$results$\,$in a diminution of the intensity along the pump major axis during propagation,$\,$which$\,$reduces $\langle \Delta n(z') \rangle_{L}$
once again. The inset of figure~\ref{fig:ShiftVersus}(b) shows the ratios $R_{\omega} = \omega_{0,y}(0)/\omega_{0,y}(L)\,$and $R_{\mathcal{I}} = \mathcal{I}_{0}(0)/\mathcal{I}_{0}(L)$ as function of the input width $\omega_{0,y}(0)$. Unsurprisingly, $R_{\omega}\;$and$\;R_{I}$ tend toward one when $z_{\mathrm{eff}} \gg L$. Reversely, when $\omega_{0,y}(0)$ decreases, $R_{\omega}$ starts decreasing as well and $R_{I}$ consequently rises up. For $\omega_{0,y}(0) \approx 500$ $\mu$m (experimental width), self-defocusing has definitely an effect but it does not smooth out the shift oscillations as on figure~\ref{fig:ShiftExp}(b).   

\newpage

\noindent Figure~\ref{fig:ShiftVersus}(c) shows the effects of nonlocality on the shift. As on figures~\ref{fig:ShiftVersus}(a) and~\ref{fig:ShiftVersus}(b), $\Delta S$ has been plotted as function of $\Lambda$, for various nonlocal ballistic length$\,$scales$\,l_{d}\,$this$\,$time. The$\,$coloured$\,$points$\,$stem$\,$for$\,$numerical$\,$simulations.$\,$The$\,$formula~\eqref{HPPhase}$\,$has$\,$been$\,$generalized using the nonlocal dispersion relation (2.46) to take into account the ballistic transport$\,$of excited$\,$atoms$\,$in$\,$our$\,$model.$\;$The$\,$theoretical$\,$predictions$\,$are$\,$plotted$\,$in$\,$black$\,$solid$\,$on$\,$figure$\,$(c) and match perfectly the simulations. As you may have noticed, nonlocality$\,$does$\,$not$\,$affect the value of $\langle \Delta n(z') \rangle_{L}$, because the way the shift increases at large $\Lambda$ remains unchanged whatever the value of $l_{d}$. The changes induced by$\,$nonlocality$\,$are$\,$therefore$\,$more$\,$structural. However, nonlocality only significantly modifies the shift$\,$for$\,$values$\,$of$\,$the$\,$ballistic$\,$transport length$\,$scale$\,$much$\,$bigger$\,$than$\,$the$\,$one$\,$expected$\,$at$\,130 \, ^{\circ}$C$\,$(which$\,$is$\,$about$\,$7.5$\,\mu$m,$\,$see$\,$1.3.8$\,$ii). Consequently, it cannot either be liable for the discrepancy observed in figure~\ref{fig:ShiftExp}(b).
\vspace{6pt}
\newline
\noindent So far, we do not have a clear understanding about why the shift is$\;$not$\;$matching$\;$the$\;$theory in the particle-like regime for some data sets. We checked experimentally if the issue was coming from the alignment procedure detailed in subsection 4.1.2 by measuring the shift in the exact same way as in experiments without the rubidium cell. In$\,$that$\,$case,$\,\Delta S\,$is$\,$zero whatever the angle between pump and probe, as expected. The shift measurement is also extremely sensitive to the phase of the pump and probe beams in the cell entrance plane. Indeed, the pump phase gradient at $z=0$ defines the initial velocity distribution of the photon fluid$\,$in$\,$the$\,$transverse$\,$plane. The$\,$latter$\,$starts$\,$thus$\,$flowing$\,$as$\,$soon$\,$as$\,$the$\,$phase$\,$of$\,$the pump beam is not uniform in the entrance plane.$\;$If the photon fluid flows$\,$along$\,$the$\,x$-axis, the phase velocity of the density waves (and therefore the shift) will certainly be modified by Doppler effect. Such a flow may come from the poor$\,$collimation$\,$of$\,$the$\,$background$\,$beam in$\,$the$\,x$-direction$\,$for$\,$instance.$\;$An$\,$other$\,$possibility$\,$is$\,$that$\,$the$\,$rubidium$\,$cell$\,$was$\,$not$\,$perfectly aligned on the optical axis, defined by the pump wave-vector. If the input window forms an angle $0< \beta \ll 1$ with the $x$-axis, the background beam progressively enters the nonlinear medium from left to right and its left$\,$end$\,$(at$\,$which$\,$the$\,$intensity$\,$falls$\,$to$\,\mathcal{I}_{0}/e^{2}$)$\,$propagates over $2 \, \omega_{0,x} \tan(\beta)$ before the right end gets inside the cell. From$\,$the$\,$medium perspective, it is as if the reference and photon fluid backgrounds had$\,$a$\,$non-zero$\,$transverse wave-vector $\boldsymbol{k_{pu}} = k_{0} \sin(\beta) \boldsymbol{e_{x'}}$, \textit{ie} a non-zero transverse speed $\boldsymbol{v} = \sin(\beta) \boldsymbol{e_{x'}}$, when they enter the$\,$cell (with $\boldsymbol{e_{x'}} = \cos(\beta)\boldsymbol{e_{x}}+\sin(\beta)\boldsymbol{e_{z}}$). The probe imprints in that case a density modulation$\,$on a moving photon fluid, which can slightly affects the shift.  

    \subsubsection{Comments on the results of Vocke \textit{et al.}}    

\noindent The shift measurement has first been performed by Vocke \text{et al.} in~\cite{2-24Vocke}, following$\,$a$\,$proposal by Carusotto~\cite{3-2Carusotto}. The setup they use is almost identical to the one sketched$\,$on$\,$figure$\,$\ref{fig:ShiftExp}.
Vocke \text{et al.} report in~\cite{2-24Vocke} a saturation of the shift $\Delta S$ at large modulation$\,$wavelengths$\,$and fit their$\,$data using equation~\eqref{ShiftFaccio}. From this fit, they retrieve the dispersion relation on one hand and measure parameters such as the nonlinear change of refractive index$\;\Delta n$ and the diffusive transport length scale $\sigma$ on the other. The latter defines the range of the nonlocal interaction between photons in the thermo-optic liquid they use as$\;$Kerr$\;$medium. In~\cite{2-24Vocke},$\,$Vocke$\,$\text{et$\,$al.}$\,$claim$\,$that$\,$interferences$\,$between$\,$counter-propagating$\,$Bogoliubov$\,$modes are suppressed by simply cutting in k-space the$\,$idler$\,$field.$\,$Moreover,$\,$they$\,$use$\,$this$\,$argument to derive equation~\eqref{ShiftFaccio} in their article, on which all their data analysis is based.$\;$It$\,$is$\,$worth mentioning however that this reasoning is wrong as it amounts to saying that the idler$\,$and the Bogoliubov mode travelling at $-\mathbf{k_{\perp}}$ are$\,$the$\,$same$\,$physical$\,$object. Indeed, by deriving carefully the input-output relation~\eqref{InputOutput}, we show that the idler$\,$field$\,$results$\,$from$\,$the$\,$mixing in the output plane of both the Bogoliubov modes moving at $+\mathbf{k_{\perp}}$ and $-\mathbf{k_{\perp}}$. Cutting$\,$the idler in k-space$\,$will$\,$thus$\,$not$\,$totally$\,$suppress$\,$the$\,$interference$\,$occurring$\,$between$\,$these$\,$modes in the cell exit plane. Consequently, as mentioned in~\cite{5-5Ferreira}, the model derived by Vocke$\,$\text{et$\,$al.} to retrieve the dispersion relation from the shift measurement is incomplete. We therefore tried to fit their data with the model developed in 4.1.1 ii following the work of Larré~\cite{3-19Larre}. Nevertheless, the saturation of the shift $\Delta S$ observed by Vocke \text{et al.} at large modulation wavelength cannot anymore be explained using this full theoretical description, even by including huge diffusive nonlocality. It should consequently come from an uncontrolled error in performing the experiments. Such a saturation can for instance be observed when the idler beam is not perfectly suppressed in k-space, which is more likely to$\,$happen$\,$in$\,$the large $\Lambda$ (\textit{ie} low $k_{\perp}$) regime. In that case, part of the idler interfere with the pump which counterbalance the shift $\Delta S$ along $+\boldsymbol{e_{x}}$ by slightly shifting the interference fringes in the opposite direction, that is, along $-\boldsymbol{e_{x}}$. This leads to an effective saturation of the shift$\,$and might explain the behaviour observed in~\cite{2-24Vocke}.

\newpage

\section{Group velocity measurement}

\noindent In the previous section, we$\,$saw$\,$that$\,$retrieving$\,$the$\,$dispersion$\,$of$\,$density$\,$waves$\,$travelling$\,$onto paraxial photon fluids from the shift measurement relies on a complex numerical inversion. This is mainly due to the fact that pump and probe have to be$\;$nearly$\;$plane-waves$\;$in$\;$this approach and, consequently, that probe and idler spatially overlap and exchange$\;$energy$\;$all along their propagation inside the Kerr medium.$\;$But$\,$what$\,$would$\,$happen$\,$if$\,$the$\,$probe$\,$beam, instead of being a plane wave, was spatially localized in the nonlinear$\,$medium$\,$input$\,$plane? In other words, how will evolve a small amplitude Gaussian$\,$wave-packet$\,$propagating$\,$at$\,$the transverse wave-vector $\mathbf{k}_{\perp}$ on top of an uniform photon fluid at rest? In$\,$what$\,$follows,$\,$I$\,$first theoretically investigate this situation and show that measuring the group velocity of such a wave-packet is a way to access the dispersion relation of density waves, without$\,$having, a$\,$priori, any knowledge about the Bogoliubov theory.$\,$I$\,$then$\,$present$\,$the$\,$results$\,$we$\,$obtained using this new experimental configuration. This section reviews and completes the paper we published in PRL: "Observation of the Bogoliubov Dispersion in a Fluid of Light"$\,$\cite{4-13Fontaine}.
In what follows, I still suppose that the probe wave-vector is parallel to the $x$-direction.

    \subsection{Introduction}
 
\subsubsection{Travelling of a wave packet onto a photon fluid} 
  
\noindent Let's imagine that the probe beam is focused in the $x$-direction but still infinitely$\,$elongated along the $y$-axis.
In that case, we can safely forget about the $y$-direction and address the problem as if it was 1-dimensional (1D+1 geometry).$\;$When$\,$it$\,$enters$\,$the$\,$nonlinear$\,$medium, the probe no longer excites$\,$plane$\,$wave$\,$density$\,$fluctuations$\,$in$\,$that$\,$case$\,$but$\,$rather$\,$a$\,$spatially localized wave-packet along the $x$-axis. The aim of this paragraph is to study$\,$how$\,$the$\,$latter travels$\,$onto$\,$some$\,$paraxial$\,$photon$\,$fluids.$\;$The$\,$wave-packet$\,$generated$\,$by$\,$the$\,$probe$\,$beam$\,$onto the background fluid can be regarded as a superposition of several plane waves, each$\,$with its$\,$own transverse wave-vector $k_{x} = k_{0} \sin(\theta_{i}) + \delta k_{x}$. The more localized the wave-packet$\,$is in real space, the larger is its extension $\delta k_{pr}$ in k-space. In dispersive mediums, it$\,$is known that a wave-packet, while moving at a constant$\,$speed$\,$(the$\,$group$\,$velocity$\,v_{g}$),$\,$spreads$\,$during its propagation. Every plane wave composing this wave-packet travels with its own$\,$velocity, which causes this spreading along the $x$-axis. This is what would happen to the probe$\;$if$\;$it propagates inside a linear medium.$\;$In$\,$that$\,$case,$\,$probe$\,$photons$\,$behave$\,$as$\,$free$\,$particles$\,$in$\,$the transverse plane and follow a quadratic particle-like dispersion $\Omega(k_{x}) = k^{2}_{x}/2k_{0}$ $-$ such as the one plotted on figure 2.1 $-$ when we adopt an effective time description of the$\,$dynamics. The phase velocity, which is defined by $v_{ph}(k_{x}) = \Omega(k_{x})/k_{x}$, depends on $k_{x}$ and$\,$varies$\,$thus in the wave-packet from one plane wave to an other. This$\,$explains$\,$why$\,$the$\,$probe$\,$undergoes diffraction in linear mediums using the fluid of light terminology.
\vspace{6pt}
\newline
\noindent In nonlinear materials, the situation is different. The quadratic dispersion $\Omega$ is replaced$\,$by the Bogoliubov dispersion relation $\Omega_{B}$.$\;$Therefore,$\;$in$\,$the$\,$sonic-regime$\,$(that$\;$is,$\,$for$\,k_{x} \!\ll\! k_{\xi}$), the phase velocity does not depend on $k_{x}$. Consequently, all the plane waves composing$\;$the Gaussian envelope of the probe beam travel at the same speed $-$ the sound velocity $c_{s}\,-$ in$\,$the transverse plane. This is true as long as $k_{\perp} + \Delta k_{x} \ll k_{\xi}$, where $k_{\perp} = k_{0} \sin(\theta_{i})\,$stands as usual for the probe transverse wave-vector. The wave-packet forming the probe beam should therefore not spread along the $x$-direction in this regime.

\newpage

\noindent More surprisingly, the group velocity $v_{g}$, which is defined by:
\begin{equation}
    v_{g}(k_{\perp}) = \frac{\mathrm{d} \Omega_{B}}{\mathrm{d} k_{\perp}},
\end{equation}
\vspace{-2pt}
\noindent is constant and equal to the sound velocity $c_{s}$ in that case.$\;$By$\,$probing$\,$the$\,$sound-like$\,$regime of the Bogoliubov dispersion relation, the probe generates a wave-packet that travels$\;$at$\;$the same speed along the $x$-axis, regardless of its angle of incidence $\theta_{i}$ in the entrance plane, as$\,$long$\,$as$\,k_{\perp} \!= k_{0} \sin(\theta_{i}) \!\ll\! k_{\xi}$.$\,$In$\,$other$\,$words,$\,$the$\,$position$\,$of$\,$the$\,$wave-packet$\,$in$\,$the$\,$medium output plane plane remains unchanged while changing $\theta_{i}$. This$\;$nonlinear$\;$refraction$\;$law, counter-intuitive$\,$from$\,$the classical optics perspective, comes from the sound-like behaviour of the Bogoliubov dispersion relation at low transverse wave-vectors.  

\subsubsection{Fundamentals of the group velocity measurement}

\begin{figure}[htbp]
\label{Groupe_Setup}
\center
\includegraphics[scale=0.34]{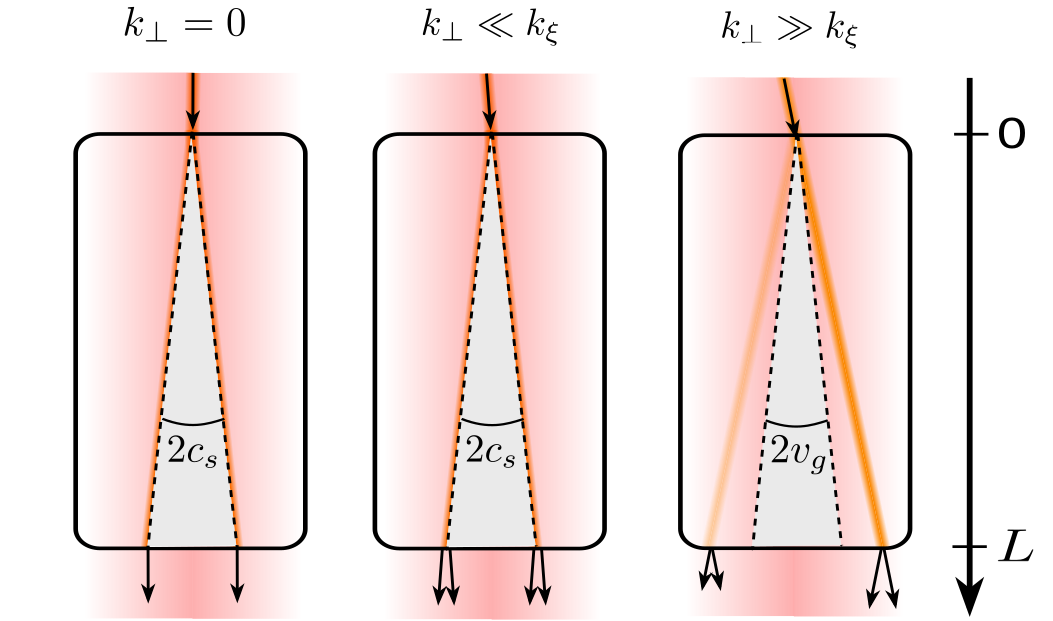} 
\caption{Sketch of the group velocity$\,$measurement.$\;$The$\,$pump$\,$(red)$\,$is$\,$a$\,$wide$\,$and$\,$intense Gaussian$\,$beam$\,$forming$\,$the$\,$photon$\,$fluid$\,$at$\,$rest$\hspace{0.05cm}$(normal$\,$incidence).$\,$The$\,$probe$\,$(orange)$\,$has$\,$a way smaller width than the pump in the $x$-direction.$\,$Both$\,$beams$\,$have$\,$the$\,$same$\,$frequency. The probe excites two counter-propagating Bogoliubov wave-packets in the input plane. As long as the the probe angle $\theta_{i}$ lies inside the cone of aperture $c_{s}$ (that is, for $k_{\perp} \ll k_{\xi}$) the excitations behave as collective phonons and move at $\pm c_{s} \, \boldsymbol{e_{x}}$ in the transverse plane. Reversely, when $\theta_{i} > c_{s}$ ($k_{\perp} \gg k_{\xi}$), the excitations behave as single particles propagating at $\pm \, v_{g}(k_{\perp}) \, \boldsymbol{e_{x}}$ along the x-axis.$\,$On$\,$the$\,$right$\,$panel,$\,$the$\,$probe$\,$beam$\,$intensity$\,$is$\,$higher$\,$than the idler one because phase match conditions are not fulfilled in that case. In situations where the conjugate beam is not visible in the exit plane, we simply measure the distance $d/2$ between the positions of the probe in the input and output planes.}
\label{Exp_Setup}
\end{figure}

\noindent In the previous paragraph, I did not mention the presence of the idler beam, which is still generated at the medium entrance plane, where the effective photon-photon interaction experiences a quench.$\,$Two$\,$counter-propagating$\,$wave-packets are therefore spontaneously created at the cell input facet (at $z=0$), when the small Gaussian perturbation forming the probe beam onto the photon fluid, that is not an eigenstate of the nonlinear dynamics, enters the medium. They$\,$then$\,$travel$\,$with$\,$the$\,$speed$\,\boldsymbol{v} = \pm v_{g}(k_{\perp}) \, \boldsymbol{{e}_{x}}\,$in$\,$the$\,$transverse$\,$plane. Measuring the distance $d(k_{\perp})$ between these two wave-packets at the$\,$propagation$\,$distance $z$ provides thus a direct access to the group velocity $v_{g}$. At$\,$the$\,$cell$\,$exit$\,$plane,$\,$this$\,$distance is given by $d(k_{\perp}) = 2 L v_{g}(k_{\perp})$. We then retrieve the dispersion relation $\Omega(k_\perp)$ by scanning the$\,$modulation$\,$wave-vector $-$ \textit{ie} by tuning the angle $\theta_{i}$ between$\,$the$\,$pump$\,$and$\,$the$\,$probe$\,-$ and integrating the group velocity from 0 to $k_{\perp}$ as follows: $\Omega(k_\perp) = \int_0^{k_\perp} v_g(q) \, \mathrm{d}q$. 
\vspace{6pt}
\newline
\noindent The fundamentals of this group velocity measurement is sketched on figure~\ref{Exp_Setup}.$\;$The$\,$pump (red) is a wide and intense Gaussian beam. It$\,$generates$\,$a$\,$quasi-uniform$\,$photon$\,$fluid$\,$at$\,$rest when it goes inside the cell. The probe (orange) is elongated in the $y$-direction in order$\;$to only sound the fluid along the $x$-axis, as previously mentioned. In this direction, the probe width $\omega_{0,x}$ is way smaller than the pump one. When it goes inside the nonlinear medium, the probe beam excites two counter-propagating Bogoliubov wave-packets. In$\,$the$\,$first$\,$case (\textit{ie} for $k_{\perp} = 0$), the two density modulations acquire a non-zero transverse speed even if the probe transverse velocity is zero initially. Because of its k-extension, the probe excites a set of plane waves, whose wave-vectors $k_{x}$ range from $-\delta k_{pr}$ to $\delta k_{pr}$, in the sonic$\,$regime$\,$of the$\,$Bogoliubov$\,$dispersion$\,$relation.$\;$Each$\,$of$\,$these$\,$plane$\,$waves$\,$generates$\,$in$\,$the$\,$cell$\,$input$\,$facet two counter-propagating Bogoliubov modes travelling at the$\,$sound$\,$velocity $c_{s}$.$\;$This$\,$process takes place all along the $x$-axis and results in the creation of two wave-packets, that move rightward and leftward at $c_{s}$. As the efficiency of such wave-mixing mechanisms$\,$depends$\,$on the phase-matching conditions, the high k-vector components of the probe envelope$\,$do$\,$not efficiently generate Bogoliubov modes, which induces an asymmetry in$\,$the$\,$wave-packets$\,$at large$\,$propagation$\,$distances.$\,$As$\,$long$\,$as$\,k_{\perp} \! \ll k_{\xi}$,$\,$or,$\,$in$\,$other$\,$words,$\,$as$\,$long$\,$as$\,$the$\,$probe$\,$angle $\theta_{i}$ remains inside the cone of aperture $c_{s}$, the probe excites in a similar way a$\,$collection$\,$of modes in the$\,$sonic$\,$part of the dispersion that are all going to move at the speed of sound in the transverse plane. In this regime, we do not expect the wave-packets$\,$to$\,$spread$\,$during their propagation and the distance $d$ between them in the$\,$exit$\,$plane$\,$should$\,$remain$\,$the$\,$same ($ d = 2 \, c_{s} \, L$). Reversely, when $k_{\perp} \gg k_{\xi}$, the excitations behave as single-particles and move along the $x$-axis$\,$at$\,\pm \, v_{g}(k_{\perp}) \, \boldsymbol{{e}_{x}}$.$\;$The$\,$distance$\,$between$\,$wave-packets$\,$centers$\,$should$\,$therefore increases linearly with the$\,$probe$\,$wave-vector$\,k_{\perp}$.$\;$Indeed,$\;$in$\,$the$\,$particle-like$\,$regime$\,$(that$\,$is, when $k_{\perp} \gg k_{\xi}$), the Bogoliubov dispersion is parabolic and $v_{g}(k_{\perp})= k_{\perp}/k_{0}$. We recover then the standard refraction law of linear optics.

\vspace{-20pt}
\subsubsection{Advantages and validity criterion}

\noindent Measuring the group velocity as function of the probe transverse wave-vector is a suitable way to access the dispersion relation $\Omega(k_{\perp})$ of density waves onto paraxial photon fluids. Indeed, this technique does not require any knowledge about the dispersion a priori and relies only on the relationship relating $\Omega$ to the group velocity $v_{g}$. Normally, $\Omega$ should be equal to the Bogoliubov dispersion relation $\Omega_{B}$. But this is not anymore$\,$a$\,$requirement$\,$as$\,$it was for the shift experiment, in the sense that the dispersion $\Omega$ can still be retrieve from the group velocity measurement otherwise. This$\,$method$\,$overcomes$\,$therefore$\,$the$\,$main$\,$issue of retrieving the dispersion from the shift $\Delta S$. Moreover, since "probe" and "idler"$\,$are$\,$spatially separated in the cell exit plane using this new pump/probe configuration, the idler beam does not need to be filtered in Fourier space anymore. The group velocity $-$ and thus the dispersion relation$\;\Omega$ $-$ can then be measured whatever the probe transverse wave-vector, even in the very low-$k_{\perp}$ regime in which pump, probe and idler overlap in k-space.

\newpage

\noindent We$\,$already$\,$know,$\;$from$\,$the$\,$shift$\,$experiment,$\;$that$\,$interactions$\,$between$\,$counter-propagating$\,$ wave-packets inside the medium can make the relationship between $d$ and $v_{g}$ more$\,$complex. If the probe width, $\omega_{0,x}$, is small enough and the nonlinear change of refractive index,$\,\Delta n$, sufficiently large, we can expect the Bogoliubov wave-packets to overlap (and thus$\,$interact) only at the beginning$\,$of$\,$their$\,$propagation$\,$inside$\,$the$\,$nonlinear$\,$medium.$\;$Once$\,$they$\,$leave$\,$this interaction zone, they propagate independently one with respect to the$\,$other$\,$(as$\,$shown$\,$on figure~\ref{fig:SimuSnellIntro}(a) for instance). In that case, we can suppose that the interaction between the wave-packets only slightly affects their positions in the cell$\,$exit$\,$plane,$\,$and$\,$thus,$\,$the$\,$distance $d$ separating them. In order for this assumption to be correct, we make$\,$sure$\,$that$\,d\,$is$\,$at$\,$least twice larger than the probe width in the output plane; the two Gaussian wave-packets$\,$are then fully separated after propagation. Moreover, we want the probe to$\,$be$\,$collimated$\,$along the nonlinear medium (which is $7.5\;$cm long in experiments). This sets thus a lower$\,$bound on the probe width $\omega_{0,x}$: $\omega_{0,x} \!>\! \sqrt{\lambda z_{r}/\pi} \approx 150$ $\mu$m (when the Rayleigh length$\,z_{r}\,$is$\,10\,\mathrm{cm}$). This constraint on the probe collimation translates directly$\,$into$\,$a$\,$constraint$\,$on$\,$its k-space extension$\,\delta k_{pr} \approx 2/\omega_{0,x}$,$\,$that$\,$should$\,$be$\,$small$\,$compared$\,$to$\,k_{\xi}$ as we want the probe$\,$to$\,$sound "locally" (in k-space) the dispersion relation. At low wave-vectors $k_{\perp}$, the group velocity$\;$is equal to the sound velocity and the constraint on the distance $d$ finally rewrites as follows: $\Delta n > n_{0} \, (\omega_{0,x}/L)^{2} \approx 4 \, 10^{-6}$. Thus, for a $7.5$ cm long vapor cell, a probe width of $150\,\mu$m and a nonlinear change of refractive index greater than $4 \, 10^{-6}$, we can safely consider that $d(k_{\perp}) = 2 v_{g}(k_{\perp}) L$ and use the group velocity measurement to access the dispersion.

\subsection{Theoretical description}

\noindent The density modulation $\delta \rho$ generated by the probe on top of the background fluid can be theoretically described using$\,$the$\,$Bogoliubov$\,$formalism.$\,$The$\,$amplitude$\,\delta \mathcal{E}\,$of$\,$the$\,$probe$\,$field, at the medium entrance plane, reads as follows: 
\begin{equation}
    \delta \mathcal{E}(\boldsymbol{r}_{\perp}, 0^{-}) = \delta \mathcal{E}_{0} \, \exp \left[ -\left( x/\omega_{0,x} \right)^{2}-\left( y/\omega_{0,y} \right)^{2} \right] \, e^{-i \mathbf{k}_{\perp} \cdot \, \boldsymbol{r}_{\perp}},
\end{equation}
\noindent where $\omega_{0,x}$ and $\omega_{0,y}$ stand respectively for the widths of the probe along the $x$- and $y$-axes.
For the sake of simplicity, we assume $\omega_{0,x}$ and $\omega_{0,y}$ do not depend on $z$, which amounts to saying that the probe is collimated in both the $x$ and $y$ directions.$\;$The$\,$in-air$\,$amplitude$\,$of the total electric field right before the medium input plane can then be$\,$expressed$\,$as$\,$follows:
\begin{equation}
    \label{FieldInAir2}
    \mathcal{E} (\mathbf{r_{\perp}}, 0^{-}) = \mathcal{E}_{0}(0^{-}) + e^{i \Phi_{0}(0^{-})} \int \frac{\mathrm{d} \mathbf{\delta k}_{\perp}}{(2 \pi)^{2}} \, \delta \Tilde{\mathcal{E}}(\mathbf{k_{\perp}} \!+ \mathbf{\delta k_{\perp}}) \, e^{-i \mathbf{\delta k_{\perp}} \cdot \, \boldsymbol{r}_{\perp}},
\end{equation}    
\noindent where $\delta \Tilde{\mathcal{E}}$ is the Fourier transform of the probe field amplitude $\delta \mathcal{E}$. The density$\,$modulation, created by the probe onto the photon fluid, is given by the formula$\;$(see$\;$subsection$\;$2.3.4): $\delta \rho(\mathbf{r_{\perp}}, z) = 2 \sqrt{\rho_{0}(z)} \, \mathrm{Re}\left[ \delta \mathcal{E}(\mathbf{r_{\perp}}, z) e^{-i \Phi_{0}(z)} \right]\!$. Using the input-output$\,$relation~\eqref{InputOutput}, we find the following expression for this density modulation $\delta \rho$:
\begin{align}
    \label{FieldInAir2}
    \delta \rho (\mathbf{r_{\perp}}, z) = & \; \frac{1}{2} \int \frac{\mathrm{d} \mathbf{\delta k}_{\perp}}{(2 \pi)^{2}} \, \left[ \left(a_{\mathbf{\delta k_{\perp}}}(L^{+}) + a^{*}_{-\mathbf{\delta k_{\perp}}}(L^{+}) \right) e^{-i \mathbf{\delta k_{\perp}} \!\cdot\! \mathbf{r_{\perp}}} + \mathrm{c.c.} \right] \nonumber \\
    \begin{split}
         = & \; \frac{1}{2} \int \frac{\mathrm{d} \mathbf{\delta k}_{\perp}}{(2 \pi)^{2}} \, \left[ \left( \vphantom{U^{*}_{\mathbf{\delta k_{\perp}}}} U_{\mathbf{\delta k_{\perp}}} + V_{\mathbf{\delta k_{\perp}}} \right) \delta \Tilde{\rho}(\mathbf{k_{\perp}} \! + \mathbf{\delta k_{\perp}}) + \dots \right. \\ 
            & \hspace{4.95cm} \left. \left(U^{*}_{\mathbf{\delta k_{\perp}}} + V^{*}_{\mathbf{\delta k_{\perp}}} \right) \delta \Tilde{\rho}(\mathbf{-k_{\perp}} \! + \mathbf{\delta k_{\perp}})   \right] e^{-i \mathbf{\delta k_{\perp}} \!\cdot\! \mathbf{r_{\perp}}},
    \end{split}
\end{align}

\newpage

\noindent where $\Tilde{\rho}(\mathbf{k_{\perp}}) = 2 \sqrt{\rho_{0}} \, \delta \Tilde{\mathcal{E}}(\mathbf{k_{\perp}})$. The foregoing expression is hard to compute. Let's then go one step further by assuming $\delta \rho$ does$\,$not$\,$depend$\,$much$\,$on$\,$the$\,y\,$coordinate.$\,$This$\,$assumption is fulfilled experimentally by sending a elliptical probe$\,$beam$\,$elongated$\,$along$\,$the$\,y$-direction such that $\omega_{0,x} \ll \omega_{0,y}$. In that case, we can address the problem as if it was 1-dimensional. Let's also neglect the linear losses ($\alpha = 0$) and consider, in a first instance, that$\,$the$\,$angle$\,$of incidence of the probe beam is zero ($k_{\perp} = 0$). The modulation$\;$density$\;$reads$\;$then: 
\begin{equation}
\label{WavePacket1Dk0}
\delta \rho_{\scriptscriptstyle{k_{\perp}\!=0}} (x, z) = \int^{\infty}_{-\infty} \frac{\mathrm{d} k_{x}}{2 \pi} \, \cos{\left( \Omega_{B}(k_{x}) \, L \right)} \, \delta \Tilde{\rho}(k_{x}, 0^{-}) \, e^{-i k_{x} x} 
\end{equation}
\noindent Equation~\eqref{WavePacket1Dk0} is derived by B.B. Baizakov \textit{et al.} in~\cite{5-6Baizakov}, where the dispersive properties of matter waves propagating onto one- and two-components Bose-Einstein condensates$\;$are theoretically investigated. In this paper, the integral$\,$on$\,$the$\,$left$\,$hand$\,$side$\,$of$\,$equation~\eqref{WavePacket1Dk0} is computed using the stationary phase approximation (for $x > c_{s} L$):  
\begin{equation}
\label{StationaryPhase}
    \delta \rho_{\scriptscriptstyle{k_{\perp}\!=0}} (x, L^{+}) \simeq \frac{\delta \Tilde{\rho}\left( \overline{k}, 0^{-} \right)}{\sqrt{2 \pi L \left| \frac{\mathrm{d^{2}} f}{\mathrm{d} k_{x}^{2}} \right|_{\overline{k}}}} \, \cos{\left[ \mathcal{F}_{\overline{k}} (x) L - \frac{\pi}{4} \right]} \, e^{i \Phi_{0}(L^{+})},
\end{equation}
\noindent where $\mathcal{F}_{k_{x}}(x) = k_{x} \, x/L - \Omega_{B}(k_{x}) $ and $\bar{k}$ stands for the transverse wave-vector$\;$at$\;$which$\;$the stationary phase condition $\mathrm{d} \mathcal{F} / \mathrm{d} k_{x} = 0$ is fulfilled. So far, we$\,$did$\,$not$\,$make$\,$any$\,$assumption regarding the k-space extension of the probe beam. If it is highly focused onto the medium input plane, it will excite$\,$counter-propagating$\,$Bogoliubov$\,$modes$\,$both$\,$in$\,$the$\,$sound-like$\,$and in the particle-like regimes of the Bogoliubov dispersion$\,$relation.$\,$As$\,$modes$\,$having$\,$different wave-vectors propagate at slightly different group-velocities (even$\,$deep$\,$in$\,$the$\,$sonic$\,$regime), they start interfering after a sufficiently long propagation distance. These interferences are described by the cosine term on the right hand side of~\eqref{StationaryPhase}. The resulting oscillations$\;$in the transverse intensity distribution make the propagation of small density wave-packets onto photon fluids look like dispersive shocks. It is worth noting that there are significant differences between these two phenomena, the first one being a purely dispersive effect while the second is highly non-perturbative and makes the nonlinearity play a$\,$crucial$\,$role. Nevertheless, when the k-extension of the probe beam is such that$\,\delta k_{pr} \ll k_{\xi}$,$\,$interferences between travelling modes start affecting the density profile for propagation distances longer than the length $L = 7.5$ cm of the$\,$medium.$\;$In$\,$that$\,$case,$\,$the$\,$cosine$\,$in$\,$the$\,$right-hand side of equation~\eqref{WavePacket1Dk0} contributes to the integral only when $k_{x}$ is small, that is, when the Bogoliubov dispersion relation is linear, so that: 
\begin{align}
    \label{WavePacket1Dk0Approx}
    \delta \rho_{\scriptscriptstyle{k_{\perp}\!=0}} (x, L^{+}) = & \, \int^{\infty}_{-\infty} \frac{\mathrm{d} k_{x}}{4 \pi}  \, \delta \Tilde{\rho}(k_{x}, 0^{-}) \, \left[ e^{-i k_{x} (x - c_{s} L)} + e^{-i k_{x} (x + c_{s} L)} \right] \nonumber \\
    \simeq & \, \frac{1}{2} \delta \rho_{0} \left[ \exp{\left(-\frac{(x-c_{s} L)^{2}}{{\omega_{0x}}^{2}}\right)} + \exp{\left(-\frac{(x+c_{s} L)^{2}}{{\omega_{0x}}^{2}}\right)} \right],
\end{align}
\noindent 
where $\delta \rho_{0} = 2 \sqrt{\rho_{0}} \, \delta E_{0}$. At the air/medium interface, the Gaussian perturbation induced by the probe beam splits into two wave-packets which propagate in opposite direction$\,$at$\,c_{s}$. This result provides therefore an easy way to measure the sound velocity and how it evolves with the photon fluid$\,$density$\,\rho_{0}$.$\,$When$\,k_{\perp}\,$is$\,$non-zero$\,$but$\,$still$\,$fulfills$\,$the$\,$condition$\,k_{\perp} \ll k_{\xi}$, 

\newpage

\noindent we expect the preceding result to hold.$\,$The$\,$centers$\,$of$\,$the$\,$counter-propagating$\,$wave-packets should still lie at $d = 2 \, c_{s} L$ one from the other because the excitation wave-vector remains inside the sound-like regime in that case, where the group velocity $v_{g}$ is not k-dependant. However, the amplitude of the "probe" envelope (that is, of the envelope of the wave-packet propagating along $+\mathbf{e}_{x}$) must then be greater$\,$than$\,$the$\,$"idler"$\,$one,$\;$because$\,$phase$\,$matching conditions are not perfectly fulfilled.
Reversely, when $k_{\perp}$ starts being much larger$\,$than$\,k_{\xi}$, equation \eqref{FieldInAir2} can be drastically simplified by noticing that $|u_{\mathbf{k}_{\perp}}\!| \! \simeq \! 1\,$and$\,|v_{\mathbf{k}_{\perp}}\!| \!\simeq\! (k_{\xi}/k_{\perp})^{2}$. Therefore, $ U_{\mathbf{k}_{\perp}} + V_{\mathbf{k}_{\perp}} \simeq \exp \left[i \Omega_B(k_{\perp}) L \right] $ and using the stationary phase method yields: 
\begin{equation}
    \label{WavePacket1DkApprox}
    \delta \rho_{\scriptscriptstyle{k_{\perp}\!\gg k_{\xi}}} (x, L^{+}) = \delta \rho_{0} \, \exp{\left( -\frac{(x-v_{g}(k_{\perp}) L)^{2}}{{\omega_{0x}}^{2}} \right)} \, \cos{\left[k_{\perp} x + \Omega_{B}(k_{\perp}) L \right]}.
\end{equation}
\noindent In that case, only the probe beam remains in the output$\,$plane. The Gaussian envelope is centered around $v_{g}(k_{\perp}) L \simeq \sin(\theta_{r}) L$, what is expected from the Snell law of linear$\;$optics. The only remaining effect of nonlinearity on the particle-like dynamics of the probe in the transverse plane is the Kerr-type phase$\,$term$\,k_{0} \, \Delta n \, L\,$it$\,$accumulates$\,$during$\,$its$\,$propagation. Indeed, this phase shift appears in the cosine function$\,$when$\,\Omega_{B}\,$is$\,$expanded$\,$for$\,k_{\perp} \!\gg k_{\xi}$. 
\vspace{6pt}
\newline
\noindent \textsc{remark} When $\alpha$ is non-zero, the results above are still correct if the group-velocity $v_{g}(k_{\perp})$ in the lossless case is replaced by its average over the propagation distance $\langle v_{g}(k_{\perp}, z) \rangle_{z}$. The sound velocity $c_{s}$ in equation \eqref{WavePacket1Dk0Approx} should then also be replaced by $c_{s,\mathrm{eff}}$, defined$\,$in subsection 2.4.1.$\,$The$\,$density$\,$modulation decays exponentially in that$\,$case$\,$along$\,$the$\,z$-axis and the right-hand side of \eqref{WavePacket1Dk0Approx} and \eqref{WavePacket1DkApprox} must therefore be multiplied by $\exp(-\alpha z)$. 

    \subsection{Numerical simulations}

\begin{figure}[H]
\label{Groupe_Setup}
\center
\includegraphics[width=\columnwidth]{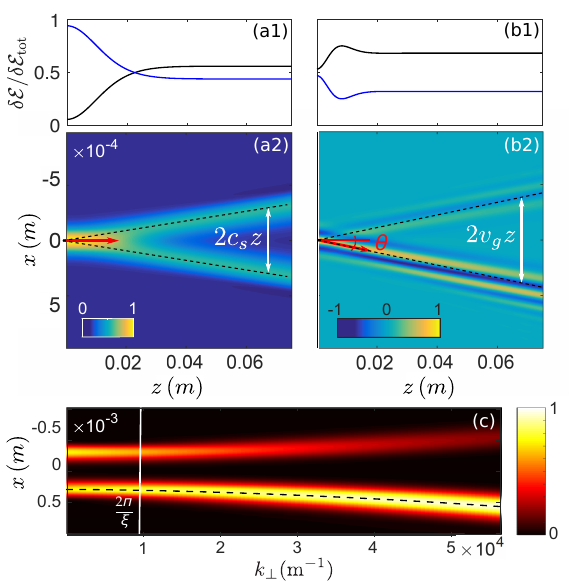} 
\caption{1D$\,$numerical$\,$simulation.$\,$(a2)$\,$and$\,$(b2):$\,$Propagation$\,$of$\,$a$\,$weak$\,$Gaussian$\,$density modulation onto an uniform background fluid (top view), in the sound-like ($\theta_{i} \approx 0$ rad) and in the particle-like regime ($\theta_{i} = 5\,$mrad). The$\,$background$\,$density$\,$has$\,$been$\,$subtracted in both cases. In (a2), the modulation created$\,$by$\,$the$\,$probe$\,$onto$\,$the$\,$photon$\,$fluid$\,$splits$\,$into two counter-propagating wave-packets (even if the probe transverse speed is zero initially). In the output plane, they are separated by $d = 2c_{s} L$.$\,$In$\,$(b2),$\,$the$\,$probe$\,$propagates$\,$as$\,$inside a linear medium. Phase matching between pump and probe gets worse when $\theta_{i}$ increases, explaining why the amplitude of the wave-packet travelling at $k_{\perp} = - k_{0} \sin(\theta_{i})$ is lower. (a1)$\,$and$\,$(b1):$\,$Interaction$\,$energy$\,\delta \mathcal{F}_{\mathrm{int}}\,$(blue$\,$line)$\,$and$\,$kinetic$\,$energy$\,\delta \mathcal{F}_{\mathrm{kin}}\,$(black$\,$line)$\,$of$\,$the density modulation as function of $z$ (normalized by its total energy $\delta \mathcal{F}_{\mathrm{tot}} = \delta \mathcal{F}_{\mathrm{int}} + \delta \mathcal{F}_{\mathrm{kin}}$). $\delta \mathcal{F}_{\mathrm{int}} $ and $\delta \mathcal{F}_{\mathrm{kin}}$ reach a stationary regime when the wave-packets on figures (a2) and (b2) are fully separated. (c) Envelope of the intensity profile in the exit plane$\,$as$\,$function$\,$of$\,k_{\perp}$. In the sound-like regime (\textit{ie} on the left side of the white line). the distance separating the wave-packets remains$\,$constant$\,$while$\,$it$\,$increases$\,$linearly$\,$with$\,k_{\perp}\,$in$\,$the$\,$particle-like$\,$regime. Parameters: $\lambda$ = 780 nm, $\Delta n = 1.3 \, 10^{-5} $, $\omega_{\mathrm{pr}}^{_{(x)}} = 150$ $\mu$m and $\alpha = 0$ (lossless case).
}
\label{fig:SimuSnellIntro}
\end{figure}

\noindent In order$\,$to$\,$illustrate$\,$more$\,$explicitly$\,$how$\,$the$\,$Bogoliubov$\,$wave-packets$\,$evolve$\,$in$\,$the$\,$medium, I$\,$numerically solved the NLSE to compute$\,$the$\,$intensity$\,$of$\,$the$\,$total$\,$electric$\,$field$\,$in$\,$any$\,$plane along the $z$-axis. To that end, I used a$\,$second-order$\,$split$\,$step$\,$algorithm$\,$for$\,$one$\,$transverse spatial dimension only (1D+1 geometry), as the probe is supposed$\,$to$\,$be$\,$infinitely$\,$elongated in the $y$-direction. The probe waist is$\;$located$\;$in$\;$the cell input plane ($z = 0$) and its width, $\omega_{\mathrm{0,x}}$, is equal to $150$ $\mu$m. On the 2D maps shown on figure~\ref{fig:SimuSnellIntro}, the uniform density $\rho_{0}$ of the background fluid is$\,$subtracted.$\;$The$\,z$-evolution$\,$of$\,$the$\,$counter-propagating$\,$Bogoliubov wave-packets is shown$\,$for$\,k_{\perp} \approx 0$ (sonic regime) and for $k_{\perp} \gg k_{\xi}$ (particle-like regime)$\,$on figures~\ref{fig:SimuSnellIntro}(a2) and (b2) respectively. In either case, the red arrow indicates the$\,$direction$\,$of incidence of$\,$the$\,$incoming$\,$probe$\,$field. Figure~\ref{fig:SimuSnellIntro}(a2) illustrates$\,$the$\,$nonlinear$\,$refraction$\,$law. Even if the probe has no transverse speed, it gives rise to a pair of counter-propagating wave-packets moving along the $x$-axis at $\pm \, c_{s}$. Of course, the total transverse momentum is conserved and equal to zero in that case, because the amplitudes of the wave-packets travelling upward and downward are the same. 
\vspace{4pt}
\newline
\noindent As you may have noticed, the wave-packets spatially overlap during a certain "time"$\,$before coming apart (at $z \approx 2$ cm here) on figure~\ref{fig:SimuSnellIntro}(a2). This can be explained looking at the energy$\,$carried by the density modulation. Because the probe has no transverse$\,$speed$\,$in$\,$the entrance plane, it only brings some additional interaction energy into the system.$\;$Part$\,$of$\,$it turns into kinetic energy, which sets the wave-packets in motion. The distance over which this energy transfer occurs turns out to be the distance at which the wave-packets$\,$separate. 

\newpage

\newpage

\noindent Let $\delta \mathcal{F}_{\mathrm{int}}$ and $\delta \mathcal{F}_{\mathrm{kin}}$ stand for the interaction and kinetic energies of$\,$the$\,$density$\,$modulation. At any position on the $z$-axis, $\delta \mathcal{F}_{\mathrm{int}}$ is obtained by subtracting the interaction energy$\,$of$\,$the background fluid $\mathcal{F}_{\mathrm{int}} \left[\mathcal{E}_{0}\right]$ from$\,$the$\,$total$\,$interaction$\,$energy$\,\mathcal{F}_{\mathrm{int}} \left[\mathcal{E}\right]$,$\,$where$\,\mathcal{F}_{\mathrm{int}}\,$is$\,$defined$\,$by: $\mathcal{F}_{\mathrm{int\!}}\left[\mathcal{E}\right] = \frac{k}{4} \, \frac{n_{2}}{n_{0}} \int_{\mathcal{S}} |\mathcal{E}(\boldsymbol{r_{\perp}})|^{_{4}} \mathrm{d} \boldsymbol{r_{\perp}}$.$\;$Similarly, $\delta \mathcal{F}_{\mathrm{kin}\!} = \mathcal{F}_{\mathrm{kin}\!}\left[\mathcal{E}\right]-\mathcal{F}_{\mathrm{kin}\!}\left[\mathcal{E}_{0}\right] = \frac{1}{4 \, k} \int_{\mathcal{S}} \left|\boldsymbol{\nabla}_{\perp} \delta\mathcal{E} \right|^{_{2}} \mathrm{d} \boldsymbol{r_{\perp}}$.
Figures~\ref{fig:SimuSnellIntro}(a1) and (b1) show the $z$-evolution of $\delta \mathcal{F}_{\mathrm{int}}$ (blue line) and $\delta \mathcal{F}_{\mathrm{kin}}$ (black line), when the probe field enters the medium with the same incidence as on figures (a2)$\,$and$\,$(b2).
Energies are normalized on both graphs by the total free energy $\delta \mathcal{F}_{\mathrm{tot}} = \delta \mathcal{F}_{\mathrm{int}} + \mathcal{F}_{\mathrm{kin}}\,$of$\,$the$\,$  density$\,$modulation.$\,$As$\,$you$\,$may$\,$have$\,$seen,$\,\delta \mathcal{F}_{\mathrm{kin}}\,$is$\,$almost$\,$zero$\,$on$\,$figure$\,$\ref{fig:SimuSnellIntro}(a1)$\,$when$\,z = 0$. The small offset results from the non-zero transverse speed of the high $k_{x}$-components of the probe field, giving a small but non-zero initial kinetic energy to the density$\,$modulation. Then, $\delta \mathcal{F}_{\mathrm{kin}}$ increases while $\delta \mathcal{F}_{\mathrm{int}}$ decreases until both of them reach a stationary regime, which occurs, indeed, when the wave-packets stop overlapping on figure~\ref{fig:SimuSnellIntro}(b1).
\vspace{4pt}
\newline
\noindent Let's$\,$now focus on figures~\ref{fig:SimuSnellIntro}(b1) and (b2), where the probe$\,$field$\,$sounds$\,$the$\,$particle-like regime of the Bogoliubov dispersion relation this time. As you can see on figure~\ref{fig:SimuSnellIntro}(b2), the wave-packet travelling upward (idler) is barely$\,$visible,$\;$as$\,$the$\,$phase$\,$matching$\,$conditions between pump and probe$\,$are$\,$not$\,$fulfilled$\,$anymore.$\;$Moreover,$\,$interference$\,$fringes$\,$appear$\,$as the fringe spacing $\Lambda = 2\pi/k_{\perp}\,$is$\,$smaller$\,$than$\,$the$\,$probe$\,$width$\,\omega_{0,x}\,$in$\,$that$\,$case.$\;$By$\,$looking$\,$at figure~\ref{fig:SimuSnellIntro}(b1), we see that the kinetic$\,$energy$\,$brought$\,$by$\,$the$\,$probe$\,$in$\,$non-zero$\,$anymore$\,$and, as in the case where $k_{\perp} \approx 0$, $\delta \mathcal{F}_{\mathrm{int}}\,$and$\,\mathcal{F}_{\mathrm{kin}}\,$stop$\,$varying along $z$ as soon as the$\,$wave-packets are spatially separated. Scanning the angle $\theta_{i}$ from$\,0\,$(a2)$\,$to$\,5\,$mrad$\,$(b2) allows to plot the envelope of the intensity profile at the medium exit plane (\textit{ie}, at $z = L$) as$\,$function$\,$of$\,k_{\perp}$. This has been$\,$done$\,$on$\,$figure$\,$\ref{fig:SimuSnellIntro}(c).$\,$As$\,$you$\,$can$\,$see,$\,$the$\,$distance$\,$between$\,$the$\,$wave-packets$\,$in the exit plane does not depend much on $k_{\perp}$ in the sound-like regime$\,$(that$\,$is,$\,$on$\,$the$\,$left$\,$side of the white solid line). Reversely, it increases when the probe starts reaching the particle-like regime of the dispersion relation. We can also notice the spreading of the wave-packets over the $x$-axis for high values of $k_{\perp}$. The theoretical distance $d/2$ between the input plane position of the probe beam ($x=0$) and the output plane position of the downward-moving wave-packet has been plotted in black dashed. As you can see, the agreement between theory$\,$and$\,$simulation$\,$is$\,$excellent.$\,$Figure~\ref{fig:SimuSnellIntro}(c)$\,$exactly$\,$illustrates$\,$what$\,$is$\,$the$\,$purpose$\,$of$\,$the experiments: measuring the distance $d$ between$\,$the$\,$wave-packets$\,$at$\,z=L\,$as$\,$function$\,$of$\,k_{\perp}$. In$\,$the$\,$following$\,$subsection,$\,$I$\,$present$\,$the$\,$experimental$\,$setup$\,$I$\,$designed$\,$and$\,$built$\,$to$\,$this$\,$end.

\subsection{Experimental setup and data processing}

\begin{figure}[H]
\center
\includegraphics[scale=0.67]{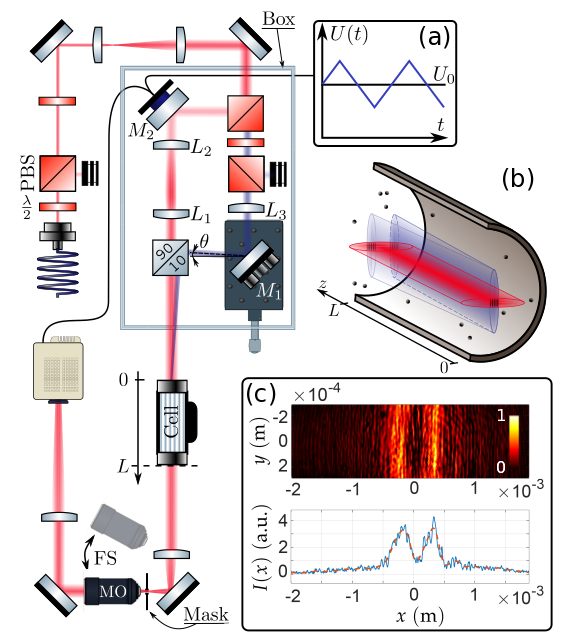} 
\caption{Experimental setup. The laser beam goes out from the fiber and is magnified before going inside a Mach-Zehnder interferometer. A PBS splits the beam into two parts, the pump (red) and the probe (blue). The pump is magnified one more time and spatially elongated along $y$-direction using a set of two cylindrical lenses. The probe is focused along the $x$-axis onto the cell entrance plane with a cylindrical lens$\;$of$\;$focal$\;$length$\,f = 1\,$m. The probe is therefore elongated in a direction perpendicular to the$\;$pump,$\;$as$\;$sketched$\;$on the inset$\,$(b). The piezo-electronically actuated mount of the mirror $M_{1}$ allows to finely tuned the angle $\theta_{i}\,$between pump$\,$and probe. The cell output plane is imaged onto the camera with a $4f$ telescope. The inset (c) shows an example of a background-subtracted image$\;$obtained for $\theta_{i} \approx 0$ rad and the related integrated  profile (blue line). On the latter, two well separated wave-packets are clearly visible. 
} 
\label{fig:SnellExp}
\end{figure}

\noindent The$\,$experimental setup$\,$is$\,$shown$\,$on$\,$figure~\ref{fig:SnellExp}. As$\,$in$\,$the$\,$shift$\,$experiment, the continuous-wave laser beam at 780 nm is produced by a Ti-Sapphire laser source (see subsection 2.1.2) and sent on the optical$\,$table$\,$using$\,$a$\,$single-mode$\,$polarization-maintaining$\,$high$\,$power$\,$fiber. The outgoing beam is magnified before entering a Mach-Zehnder interferometer, protected from air-turbulences by a box. From there, the beam splits into a high power pump (red) and a low power probe (blue). The$\,$pump$\,$is$\,$expanded$\,$twice$\,$before$\,$being$\,$loosely$\,$focused$\,$into the nonlinear medium using a set of two cylindrical lenses. The resulting beam$\;$is$\;$elongated in the $x$-direction; its widths along the $x$- and $y$-axes are $3.2$ mm and $300 \, \mu$m respectively. The pump intensity (and thus the background fluid density) is then$\,$almost$\,$uniform$\,$along$\,x$. The Rayleigh length $z_{r,y}$ associated to the pump width along $y$ is $37$ cm, which is five time longer than the length of the nonlinear medium ($L = 7.5$ cm). Consequently,$\;$we$\;$can$\;$safely suppose that the pump beam is collimated (when self-defocusing is negligible obviously).

\newpage

\newpage

\newpage
 
\noindent The probe beam is directly focused with a cylindrical lens onto$\,$the$\,$medium$\,$entrance$\,$plane. It is elliptically elongated along $y$, that is, in a direction perpendicular to the pump beam. This cross-configuration, sketched on figure~\ref{fig:SnellExp}(b), allows to only probe the fluid along the $x$-axis, where the intensity (and$\,$thus$\,\Delta n$) has been made higher by focusing the pump. At the probe waist, $\omega_{0,x} = 180 \pm 10 \, \mu$m and $\omega_{0,y} = 1.7$ mm. The value of the minor width, $\omega_{0,x}$, is chosen in order for the wave-packets to properly separate at the medium exit plane. For $\omega_{0,x} \simeq 180 \, \mu$m, the probe is also collimated along $x$ inside$\,$the$\,$medium$\,$(as$\,z_{r, x} \!\simeq\! 13\,$cm). Before recombining with the pump, the probe reflects$\,$onto$\,$the$\,$mirror$\,M_{1}$.$\,$The$\,$latter$\,$is$\,$hold in a piezo-actuated mount which allows to finely tune the angle $\theta_{i}$. The depth$\,$of$\,$the$\,$density fluctuation$\;$is$\,$adjustable$\,$too$\,$by$\,$turning$\,$the$\,$half$\,$wave-plate$\,$before$\,$the$\,$PBS$\,$on$\,$the$\,$probe$\,$arm. In experiments, it represents less than $5\%$ of the background fluid$\,$density.$\,$Pump$\,$and$\,$probe propagate then inside a 7.5$\,$cm$\,$long$\,$cell,$\,$filled$\,$with$\,$an$\,$isotopically$\,$pure$\,$vapor$\,$of$\,$rubidium$\,$85. The cell is heated up to $150\,^{\circ}$C with the homemade oven described on subsection 2.1.1. The Kerr nonlinearity$\,$is$\,$obtained$\,$by$\,$red-detuning$\,$the$\,$laser$\,$frequency$\,$from$\,$the$\,F_{g} = 3 \shortrightarrow F_{e}$ transition frequency ($\Delta = -2 \pi \!\times\!6$ GHz). The vapor transmission is then about $70\%$.
\vspace{6pt}
\newline
\noindent The cell exit plane is imaged onto a CMOS camera using a $4f$ imaging system$\,$made$\,$of$\,$two lenses of focal length $150$ mm and $500$ mm respectively. The magnification factor has been measured and is about 3.4. A microscope objective can be$\,$flipped$\,$onto$\,$the$\,$beam$\,$path$\,$so$\,$as to image the focal plane of the first lens and measure the probe transverse$\,$wave-vector$\,k_{\perp}$. Images of the pump alone (background), of the probe alone and$\,$of$\,$the$\,$k-space$\,$(without$\,$cell) are captured at every angle $\theta_{i}$. Moreover, as we only care about the envelope of the wave-packets in the exit plane, the relative phase between pump and probe is scanned$\,$over$\,2\pi$. To that end, we mount the mirror $M_{2}$ onto a piezo-actuated translation stage. By$\,$ramping up the high voltage applied across the piezo (see inset (a)), we modulate the length$\;$of$\;$the pump path in the Mach-Zehnder, and consequently the relative phase between the beams. 40 images are captured during a phase ramp.$\,$We$\,$subtract$\,$the$\,$background$\,$from$\,$each$\,$of$\,$them before integrating the resulting intensity distribution over $100$ pixels around the $x$-axis. We finally retrieve the wave-packets envelope by summing $-$ in absolute value $-\,$all$\,$the$\,$40 intensity profiles we obtained after integration. The distance $d$ between the wave-packets is measured by performing a 1D two-Gaussian fit of the intensity envelope, as long as the idler beam is visible in the exit plane. Otherwise, $d$ is directly measure from the distance between the input and output positions of the probe beam.
 
\subsection{Experimental results} 

\noindent In this subsection, I present the experimental results I obtained with the$\,$setup$\,$and$\,$the$\,$data analysis$\,$described$\,$above.$\,$I$\,$first$\,$show$\,$the$\,$dispersion$\,$relation$\,$retrieved$\,$from$\,$the$\,$group$\,$velocity measurement in$\,$a$\,$low-density$\,$photon$\,$fluid.$\,$At$\,$high$\,$densities,$\,$a$\,$discrepancy$\,$appears$\,$between the measure group velocity and the theoretical expectation. I show that$\,$nonlocality could have explained it if the values of the nonlocal ballistic transport length scale, $l_{d}$, a$\;$which notable$\,$disturbances$\,$on$\,$the$\,$group$\,$velocity$\,$arise$\,$were$\,$not$\,$so$\,$unrealistic.$\;$I$\,$finally$\,$demonstrate that this discrepancy is the experimental signature of quasi-particle interferences, that is, of interferences between the two counter-propagating Bogoliubov wave-packets.    

\newpage

\subsubsection{Dispersion relation in a low-density photon fluid}

\begin{figure}[H]
\center
\includegraphics[width=\columnwidth]{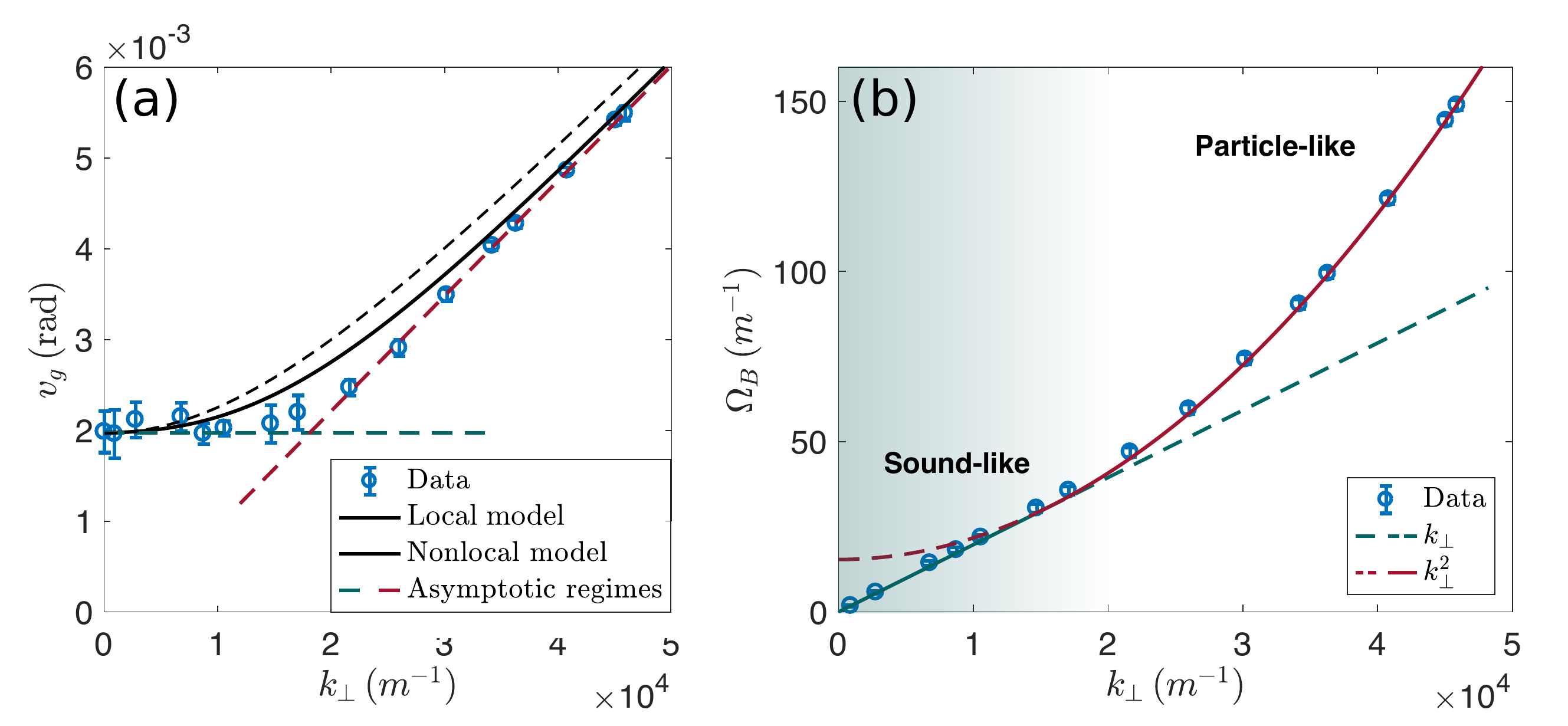} 
\caption{(a):$\,$Group$\,$velocity$\,v_{g}\,$as$\,$function$\,$of$\,$the$\,$transverse$\,$wave-vector$\,k_{\perp} =k_{0} \sin(\theta_{i})$. The blue circles represent the experimental data obtained for a fluid power $\mathcal{P}_{f} = 175$ mW. The theoretical model is plotted in black solid. It takes into account the nonlocal response of the vapor, due to the ballistic transport of fast moving excited atoms. For comparison, the group velocity$\,$predicted$\,$by$\,$the$\,$local$\,$theory$\,$is$\,$plotted$\,$in$\,$black$\,$dashed$\,$on$\,$the$\,$same$\,$graph. The dashed lines highlight the asymptotic behaviours of the group velocity, which remains constant in the sonic regime (cyan) and linearly increase in the particle-like regime (red).
(b) shows the dispersion relation obtained after integration of the group velocity in (a). It exhibits a linear increase at low $k_{\perp}$ characterized by the sound velocity $c_{s, \mathrm{eff}}$.
}
\label{fig:DispersionExp}
\end{figure}

\noindent The experimental group velocity and dispersion relation are plotted as function$\;$of$\;k_{\perp}\;$on figures~\ref{fig:DispersionExp}(a) and (b) (blue circles). The pump power is $175$ mW, which corresponds to a nonlinear change of refractive index $\Delta n(0)\,$of$\,3.9 \!\times\! 10^{-6}$ in the input plane.$\;$On$\,$figure$\,$\ref{fig:DispersionExp}(a), two regimes are clearly distinguishable. The group velocity is constant at low wave-vectors while it linearly increases at larger ones. The transition between these regimes occurs when $k_{\perp} \simeq 1.8 \!\times\! 10^{4}\,$m$^{-1}$. This value is slightly greater than $k_{\xi}$, which is about $1.6 \!\times\! 10^{4}\,$m$^{-1}$. The theoretical group velocity, obtained for $\Delta n(0) = 3.9 \!\times\! 10^{-6}$ and for $70\%$ transmission, has$\,$been$\,$plotted$\,$on$\,$black$\,$dashed.$\,$As$\,$you$\,$can$\,$see,$\,$it$\,$perfectly$\,$predicts$\,$the$\,$value$\,$toward$\,$which the measured group velocity tends when $k_{\perp}$ goes$\,$to$\,$zero.$\;$Nevertheless,$\,$the$\,$plateau$\,$observed experimentally$\,$at$\,$low$\,k_{\perp}\,$is$\,$much$\,$longer$\,$than$\,$expected.$\;$Moreover,$\,$an$\,$offset$\,$between$\,$the$\,$data and$\,$the$\,$model is clearly visible at large $k_{\perp}$. By$\,$taking$\,$nonlocality$\,$into$\,$account,$\,$which$\,$arises in hot vapors because of the ballistic transport of fast moving excited atoms,$\,$we$\,$obtain$\,$the black solid line on figure~\ref{fig:DispersionExp}(a). When $T = 150\, ^{\circ}$C, the nonlocal ballistic length scale, $l_{b}$, is about $8 \, \mu$m.  It does not sound like much but it is enough to make the nonlocal model better describe the asymptotic trend of the data at large $k_{\perp}$. However, nonlocality does not explain why the plateau at low $k_{\perp}$ is longer than theoretically expected. 



\newpage

\begin{figure}[H]
\center
\includegraphics[width=\columnwidth]{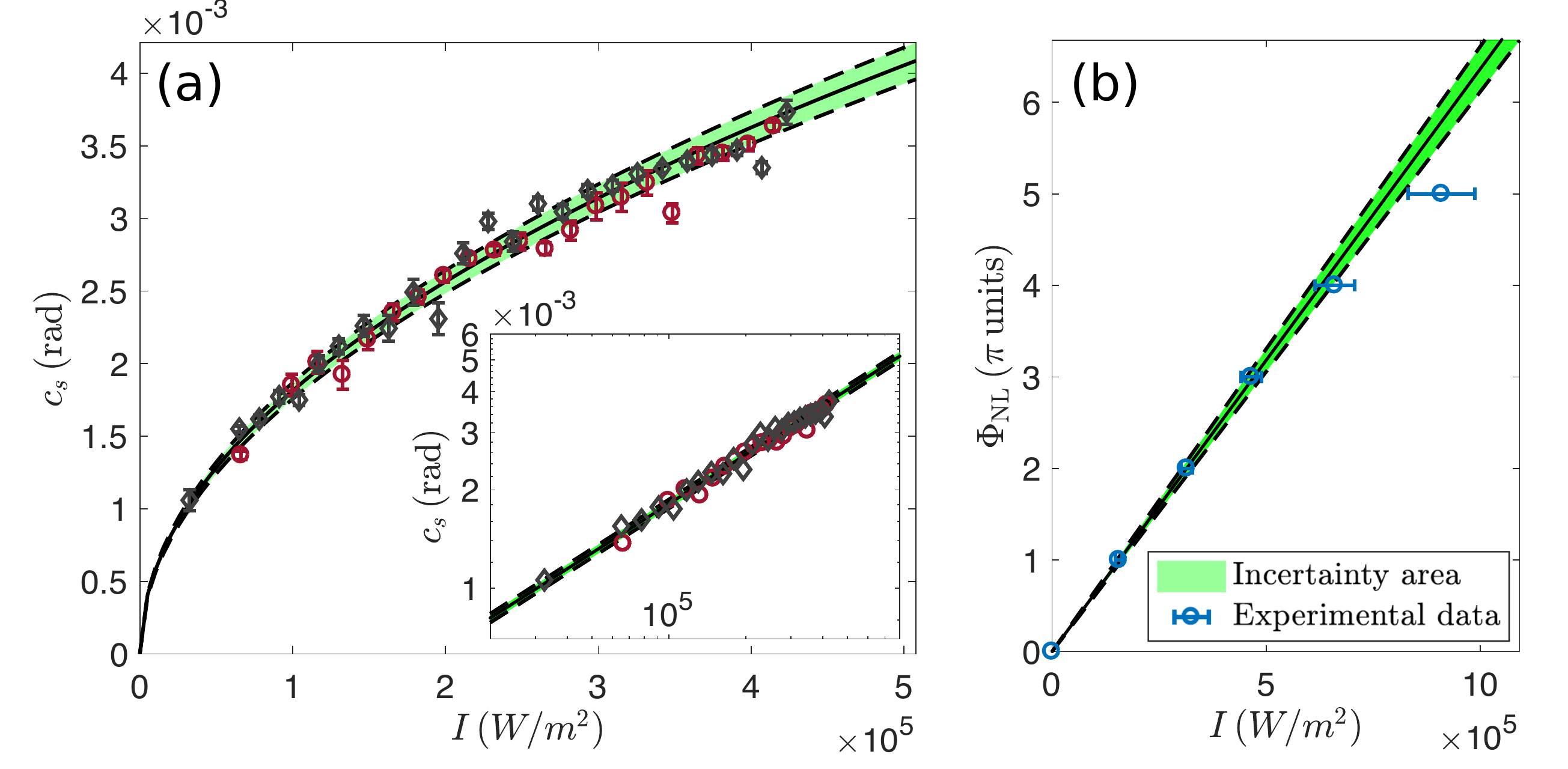} 
\caption{(a): Speed of sound $c_{s}$ as function of the pump$\,$intensity$\,\mathcal{I}_{0}$.$\,$The$\,$red$\,$circles$\,$and the grey diamonds represent two sets of data taken successively at the same laser detuning and$\,$vapor$\,$temperature.$\;$The$\,$sound$\,$velocity$\,$increases$\,$with$\,$the$\,$square$\,$root$\,$of$\,\mathcal{I}_{0}$,$\,$as$\,$expected. The theoretical prediction (black line) matches the data pretty well. No free-parameter is needed because the nonlinear refractive index $n_{2}$ has been measured independently, using$\,$the$\,$ring$\,$counting technique of subsection 2.3.2. The uncertainty on $c_{s}$ (green area) derives from the uncertainty on $n_{2}$. (b): Ring-counting measurement.$\;$The$\,$self-phase$\,\Phi_{\mathrm{NL}}$ accumulated by a Gaussian beam propagating through the cell is measured as$\,$function$\,$of the$\,$beam$\,$intensity.$\;$The$\,$nonlinear$\,$refractive$\,$index$\,$is$\,$obtained$\,$by$\,$fitting$\,$the$\,$data$\,$with$\,$a$\,$line. We find $n_{2} = 3.3 \pm 0.2 \!\times\! 10^{-11}$ m$^{2}$/W.
} 
\label{fig:SoundVelocityExp}
\end{figure}

\noindent Integrating the data on figure$\,$\ref{fig:DispersionExp}(a)$\,$yields$\,$the$\,$dispersion$\,$relation plotted$\,$on$\,$figure$\,$\ref{fig:DispersionExp}(a). As you can see, it linearly increases for $k_{\perp} < 1.8 \!\times\! 10^{4}\,$m$^{-1}$. This is an important result$\,$as it proves that small amplitude density waves behave like collective phonons for low enough excitation wave-vectors. This sound-like regime is$\,$fully$\,$characterized$\,$by$\,$the$\,$speed$\,$of$\,$sound, $c_{s,\mathrm{eff}} = \langle c_{s}(z')\rangle_{L} \!\simeq \!2.0\,$mrad.$\,$Moreover,$\,$the$\,$linear$\,$increase$\,$of$\,$the$\,$dispersion$\,$relation$\,$at$\,$low$\,k_{\perp}$ guarantees that light propagating through rubidium vapors can be superfluid, as suggested by R.Chio two decades ago~\cite{9:Chia0}. Indeed, according to the Landau criterion$\,$for$\,$superfluidity, the$\;$speed$\,$of$\,$sound$\,c_{s,\mathrm{eff}}\,$defines$\,$a$\,$critical$\,$speed$\,$below$\,$which$\,$the$\,$photon$\,$fluid$\,$cannot$\,$dissipate energy$\,$anymore by emitting sound-like$\,$excitations$\,$(see$\,$subsection$\,$2.3.3).$\;$In$\,$order$\,$to$\,$further investigate the sound-like regime of the dispersion, we set the probe wave-vector to zero. In that case, the wave-packets in the cell exit plane are separated by $2 c_{s,\mathrm{eff}} L$; we can then measure how$\,$the$\,$sound$\,$velocity$\,$increases$\,$with$\,$the$\,$background$\,$intensity.$\;$The$\,$data$\,$are$\,$shown on figure~\ref{fig:SoundVelocityExp}(a). The red circles and the grey diamonds are the results obtained from two measurements performed successively, at the same laser detuning and vapor temperature. As you may have seen, the theoretical prediction (black line) matches$\,$the$\,$data$\,$pretty$\,$well, indicating that $ c_{s,\mathrm{eff}}$ scales with the square-root of$\,$the$\,$background$\,$fluid$\,$density,$\,$as$\,$expected. It is worth mentioning that the nonlinear refractive index $n_{2}$ is measured independently (using the ring counting technique of subsection 2.3.2), which sets the only parameter$\,$in$\,$the theoretical model. The results from the ring counting measurement are$\,$shown$\,$on$\,$figure$\,$(b).

\newpage 
    
    \subsubsection{Discrepancy at high fluid densities}
    
\begin{figure}[H]
\center
\includegraphics[width=\columnwidth]{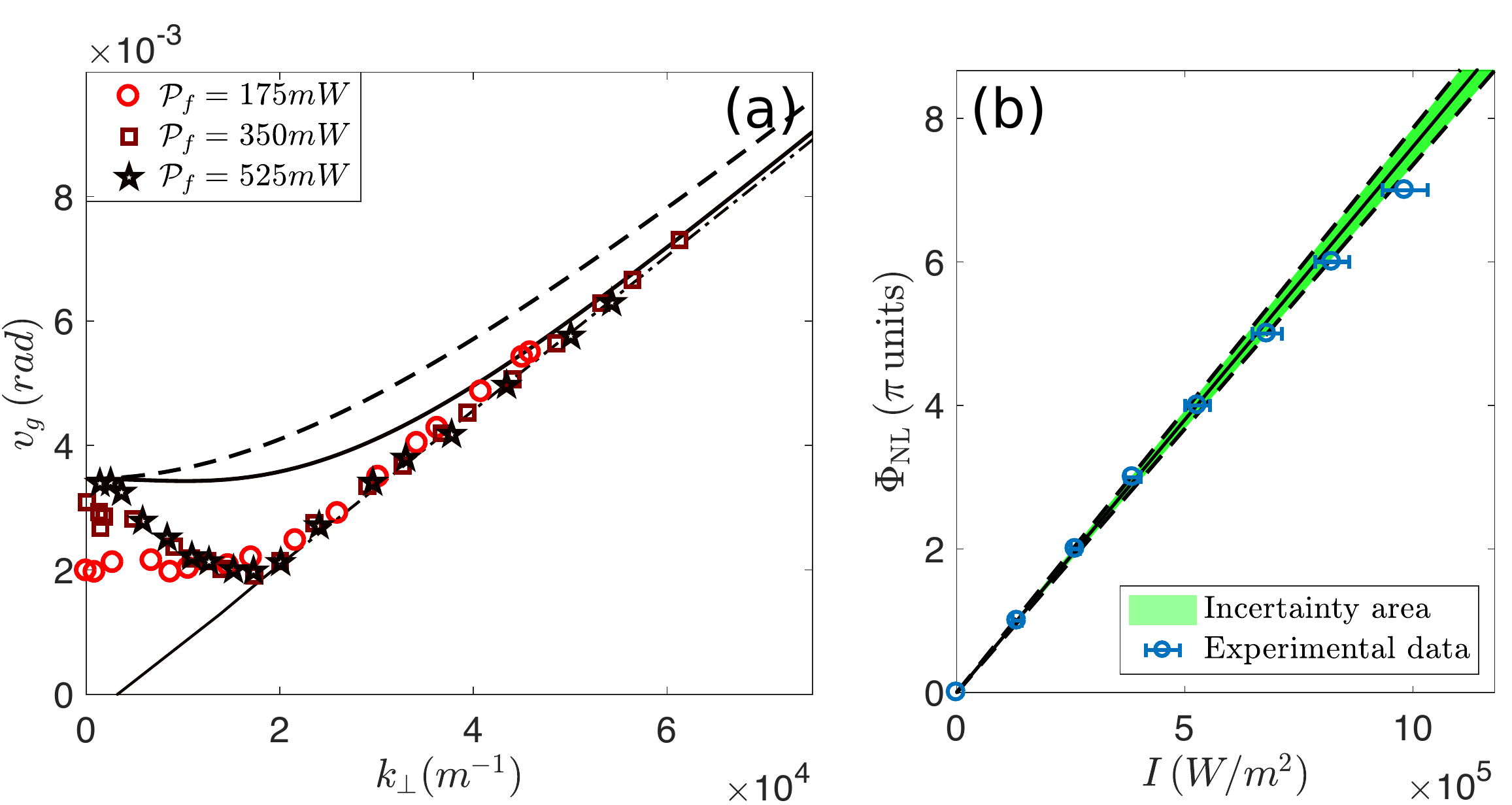} 
\caption{(a): $v_{g}$ as function of $k_{\perp}$ for different fluid powers$\,\mathcal{P}_{f}$.$\;$The$\,$experimental$\,$data shown on figure~\ref{fig:DispersionExp} have been reported on this graph (red circles). When $\mathcal{P}_{f}$ is equal$\;$to $350$ or $525$ mW, the data exhibit a dip, whose location does not seem to depend much on the fluid power (that is, on $\Delta n$). The depth of this dip increases with $\mathcal{P}_{f}$ while its bottom remains at the same height. The black dashed and solid lines$\,$are$\,$the$\,$theoretical$\,$predictions in local and nonlocal mediums respectively, for $P_{f} = 525$ mW. Taking nonlocality into account allows to better describe the asymptotic trend of the experimental$\,$data$\,$at$\,$high$\,k_{\perp}$. However, it does not explain why the dip appears. Surprisingly, the theory still match$\,$the data when $k_{\perp}$ tends to zero. (b): Ring-counting measurement. The value of $n_{2}$ used$\;$in$\;$(a) to plot the theoretical predictions is measured experimentally using the ring-counting technique of subsection 2.3.2; we find $n_{2} = 3.1 \pm 0.2 \!\times\! 10^{-11}$ m$^{2}$/W.
} 
\label{fig:SnellPower}
\end{figure}

\noindent The data shown on figures~\ref{fig:DispersionExp}(a) and (b) have been obtained at low background$\,$densities. Aside from the fact that the plateau at low $k_{\perp}$ is longer than expected on figure~\ref{fig:DispersionExp}(a), the experimental observations match well with the prediction of the Bogoliubov's theory in that case. Nevertheless, the situation is more complicated at high background densities. Figure$\,$\ref{fig:SnellPower}(a) shows$\,$the$\,$group$\,$velocities$\,$measured$\,$as$\,$function$\,$of$\,k_{\perp}\,$for$\,$various$\,$fluid$\,$powers. The$\,$data$\,$plotted on figure~\ref{fig:DispersionExp}(a) have been$\,$reported$\,$on$\,$this$\,$graph$\,$(circles).$\;$As$\,$you$\,$can$\,$see, the data points form a dip when the fluid power $\mathcal{P}_{f}$ is equal to $350\,$or$\,525\,$mW.$\,$Its$\,$location ($k_{\mathrm{dip}}\simeq 1.8\! \times \! 10^{4}$ m$^{-1}$) does not depend much on the background power and$\,$roughly$\,$match the end of the plateau formed by the data points, at low $k_{\perp}$, when $\mathcal{P}_{f} = 175\,$mW.$\;$Moreover, the depth of the dip increases with$\,\mathcal{P}_{f}\,$while$\,$its$\,$bottom$\,$remains$\,$at$\,$the$\,$same$\,$height$\,$($ 2\,$mrad). As on figure~\ref{fig:DispersionExp}(a) the theoretical group velocities in local (black dashed line) as well as in nonlocal mediums (black solid line) have been plotted as function of $k_{\perp}$ for $\mathcal{P}_{f} = 525\,$mW.
As you can see, both models correctly predicts the value of the measured group velocity when $k_{\perp}$ goes to zero. This is why the measured sound velocity still matches the theory on figure~\ref{fig:SoundVelocityExp}(a). However, neither the local nor nonlocal$\,$description$\,$is$\,$able$\,$to$\,$explain$\,$the$\,$dip.


\newpage   
        
\begin{figure}[H]
\center
\includegraphics[width=\columnwidth]{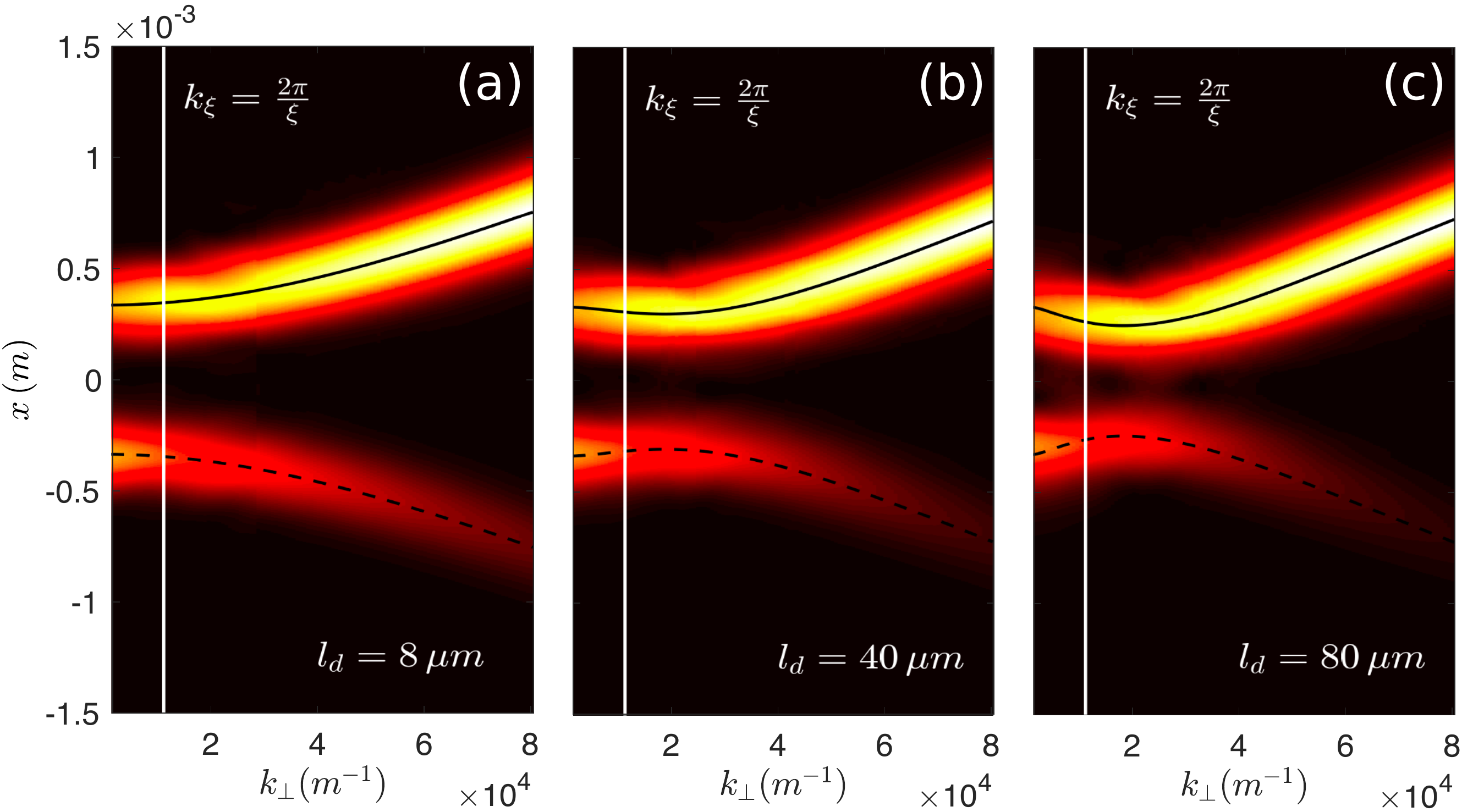} 
\caption{Envelope of the intensity profile along $x$ in the output plane as$\,$function$\,$of$\,k_{\perp}$ for different ballistic transport length scales $l_{d}$. The realistic situation is shown$\,$on$\,$the$\,$left. On figures (b) and (c), the minimum distance between the wave-packet is not anymore obtained at $k_{\perp} = 0$ but for a wave-vector slightly larger than $k_{\xi}$ (white line). This distance decreases when $l_{d}$ increases. Nonlocality could therefore have explained the dip observed on$\,$figure$\,$\ref{fig:SnellPower}.$\;$However,$\;$the$\,$values$\,$of$\,l_{d}\,$at$\,$witch$\,$this$\,$effect$\,$significantly$\,$impacts$\,$the$\,$distance between the wave-packets are very large. The theoretical predictions are plotted in black solid and black dashed. Parameters: $\Delta n = 1.0\!\times\!10^{-5}$, $L = 7.5$ cm and $\alpha = 0$.} 
\label{fig:SnellNonlocality}
\end{figure}

\noindent At first, we thought it was due to$\,$nonlocality,$\,$as$\,$it$\,$is$\,$known$\,$to$\,$modify$\,$the$\,$dispersion$\,$relation (and the group velocity) in$\,$a$\,$way$\,$that$\,$could$\,$have$\,$explained$\,$its$\,$appearance$\,$on$\,$figure$\,$\ref{fig:SnellPower}(a). The effects of diffusive nonlocality on the dispersion are investigated in~\cite{2-24Vocke} for instance. The envelope of the intensity profile along $x$ in the cell output plane has been plotted as function of $k_{\perp}$ on figure~\ref{fig:SnellNonlocality}, for different values of the ballistic transport length scale $l_{d}$.
Indeed, the distance between the wave-packets (which$\,$is$\,$proportional$\,$to$\,$the$\,$group$\,$velocity) does not continuously increase with $k_{\perp}$ on figures (b)$\,$and$\,$(c)$\,$but$\,$rather$\,$exhibits$\,$a$\,$minimum located at a wave-vector slightly bigger than $k_{\xi}$ (white line). This feature starts appearing as soon as nonlocality is strong enough to create an inflexion point in the$\,$dispersion$\,$relation (see figure 2.2) and is more and more pronounced$\,$as$\,l_{d}$ increases. However, the values$\,$of$\,l_{d}$ at which nonlocal effects induce a dip comparable to what is observed on figure~\ref{fig:SnellPower}(a)$\;$are completely unrealistic. At $T = 150 \, ^{\circ}$C, the nonlocal ballistic length scale is around $8 \,\mu$m. For such a value, nonlocality only generates a plateau at low $k_{\perp}$, as can be seen on figures \ref{fig:SnellPower}(a) (black line) and \ref{fig:SnellNonlocality}(a), and allows to better describe the asymptotic trend of the data points at high $k_{\perp}$, as already mentioned in the previous paragraph.
\vspace{6pt}
\newline
\noindent The$\,$causes$\,$beyond$\,$the$\,$emergence$\,$of$\,$the$\,$dip$\,$at$\,$high$\,$background$\,$densities$\,$on~figure$\,$\ref{fig:SnellPower}(a)$\,$lie thus elsewhere. In the next paragraph, we investigate the possibility that the dip may arise from$\,$destructive interferences between the counter-propagating Bogoliubov wave-packets.  

\newpage

$\vphantom{a}$ 
\vspace{20pt}
    
\begin{figure}[H]
\center
\includegraphics[scale=0.58]{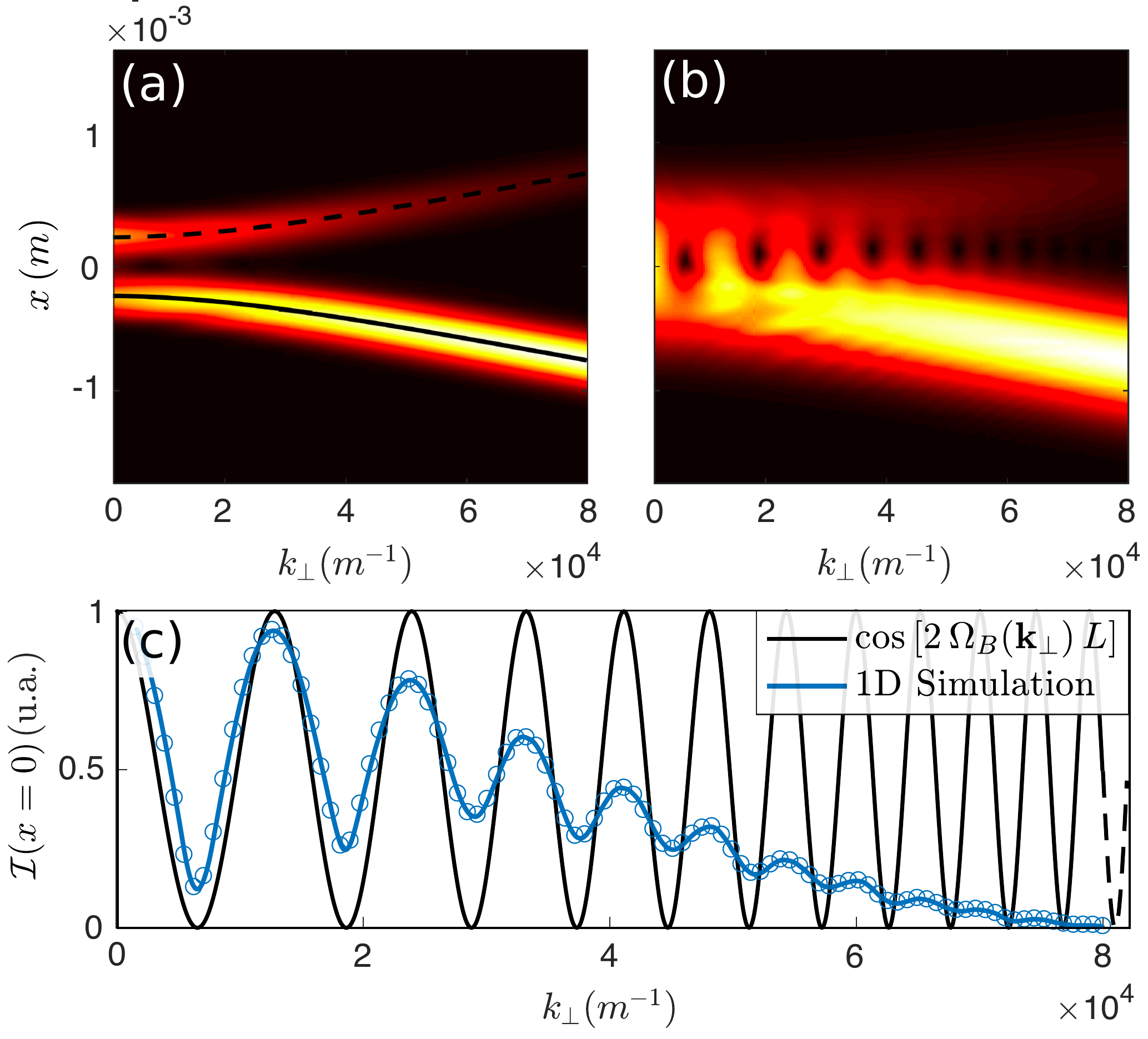}
\caption{Envelope of the intensity profile along $x$ in the output plane as$\,$function$\,$of$\,k_{\perp}$, for$\,$a$\,$probe$\,$width$\,w_{0,x}\,$of$\,150\,\mu$m$\,$(a)$\,$and$\,450\,\mu$m$\,$(b)$\,$respectively.$\;$In$\,$figure$\,$(a),$\;$the$\,$envelopes of the wave-packets are fully separated in the cell exit plane. This is not$\,$anymore$\,$the$\,$case on$\,$figure$\,$(b).$\;$Interference$\,$between$\,$the$\,$upward-$\,$and$\,$downward-moving$\,$wave-packets$\,$occurs in that case. Depending on the probe wave-vector, the envelope of the$\;$density$\;$modulation exhibits$\,$two$\,$distinct$\,$peaks$\;$(destructive$\,$interference)$\;$or$\,$just$\,$one$\;$(constructive$\,$interference). Figure (c) shows the intensity at $x=0\,$in$\,$the$\,$exit$\,$plane$\,$as$\,$function$\,$of$\,k_{\perp}\,$for$\,w_{0,x}=450\,\mu$m. It$\;$oscillates$\;$in$\;$a$\;$similar$\;$way$\;$as$\,\cos\left[ 2\Omega_{B}(k_{\perp}L\right]\,$(black$\,$solid$\,$line),$\;$that$\,$describes$\,$the$\,$beating between$\,$two$\,$counter-propagating$\,$Bogoliubov$\,$modes$\,$at$\,k_{\perp}$.$\,$The$\,$amplitude$\,$of$\,$the$\,$oscillations is damped because the overlap between the wave-packets decreases with $k_{\perp}$. Parameters:  $\Delta n = 1.0\!\times\!10^{-5}$, $L = 7.5$ cm and $\alpha = 0$.} 
\label{fig:SnellInterferenceTheory}
\end{figure}    

\newpage
        
    \subsubsection{Quasi-particle interferences: experimental evidence}
    
\begin{figure}[H]
\center
\includegraphics[width=\columnwidth]{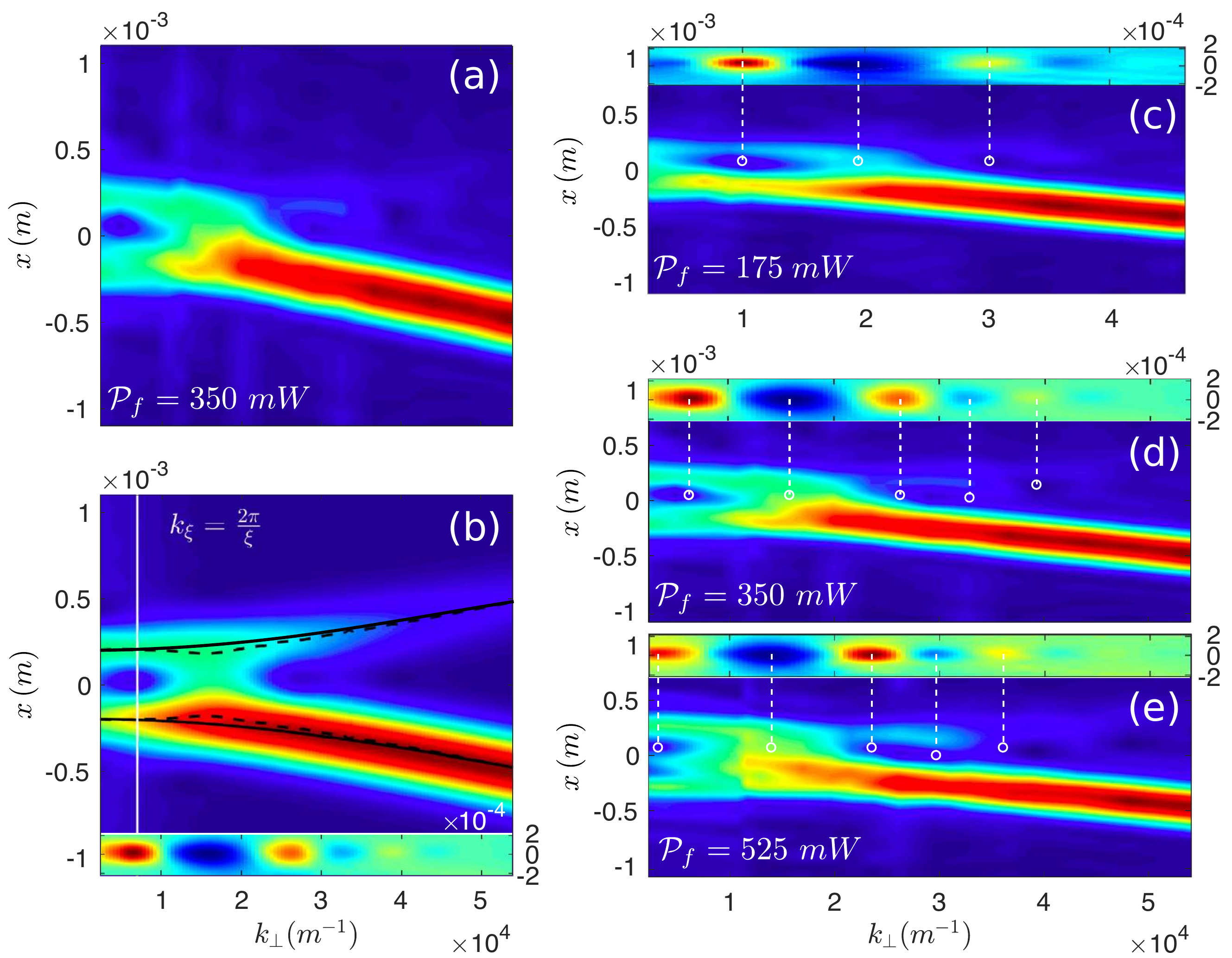} 
\caption{Quasi-particle interferences. On figure (a), the$\,$envelope$\,$of$\,$the$\,$intensity$\,$profile in the cell exit plane is shown$\,$as$\,$function$\,$of$\,k_{\perp}$,$\;$for$\,$a$\,$fluid$\,$power$\,$of$\,350\,$mW.$\;$(b)$\,$We$\,$compare these experimental data to simulation, solving the NLSE in 2D for the same beam widths and powers as in experiments. The theoretical positions of the wave-packets centers are plotted in black solid on figure (b). The black dashed lines$\,$are$\,$the$\,$positions$\,$provided$\,$by$\,$the two-Gaussian fit. The latter slightly$\,$under-estimates$\,$the$\,$spacing$\,$between$\,$the$\,$wave-packets when the$\,$destructive$\,$interference$\,$occurs.$\;$This$\,$is$\,$why$\,$a$\,$dip$\,$appears$\,$in$\,$the$\,$data$\,$of$\,$figure$\,$\ref{fig:SnellPower}. The stripe at the bottom of figure (b) shows the  Laplacian $\boldsymbol{\nabla}^{2} \, \mathcal{I}(x,k_{\perp})$ of the intensity distribution plotted above (from$\,-0.2\,$to$\,0.2\,$mm). The peaks and the dips on this stripe respectively locate the constructive and destructive$\,$interferences$\,$visible$\,$on$\,$the$\,$2D$\,$map$\,$(b). On figures (c), (d) and (e),$\,$the$\,$wave-vectors$\,$at$\,$which$\,$interferences$\,$occur$\,$in$\,$simulation$\,$and in$\,$experiments$\,$are$\,$compared$\,$for$\,\mathcal{P}_{f} = 175\,$mW$\,$(c),$\,\mathcal{P}_{f} = 350\,$mW$\,$(d)$\,$and$\,\mathcal{P}_{f} = 525$ mW$\,$(e). The 2D maps are experimental data while the strips on each figure represent the variation of $\boldsymbol{\nabla}^{2} \, \mathcal{I}(x,k_{\perp})$ from$\,-0.2\,$to$\,0.2\,$mm, obtained by simulation. 
} 
\label{fig:SnellInterference}
\end{figure}

\noindent So$\,$far,$\,$we$\,$have$\,$supposed$\,$that$\,$the$\,$wave-packets$\,$were$\,$spatially$\,$separated$\,$in$\,$the$\,$cell$\,$exit$\,$plane, or, in other words, that the criterion of paragraph 4.2.1 iii was fulfilled. If this is not the case anymore, we expect interferences between counter-propagating$\,$wave-packets$\,$to$\,$occur. This is indeed what we observe on figure~\ref{fig:SnellInterferenceTheory}(b), where the width of the probe$\,$beam $\omega_{0,x}$ at the entrance plane is $450$ $\mu$m, three times larger than on figure (a).$\,$On$\,$these$\,$two$\,$figures, the envelope of the intensity profile along $x$ in the output plane is plotted as function$\,$of$\,k_{\perp}$. While the wave-packets on figure (a) are fully separated after propagation (whatever$\,k_{\perp}$), they overlap at low wave-vectors on figure (b) and interfere therefore one with each other. Figure (c) shows a cut of the 2D map (b) along the horizontal axis ($x=0$).$\,$As$\,$you$\,$can$\,$see, the intensity $\mathcal{I}(x=0)$ (blue line) oscillates in a similar way as the function $\cos \left[ 2 \Omega_{B}(k_{\perp})L \right]$ (black line), describing the beating between two counter-propagating Bogoliubov$\,$modes$\,$at $\pm k_{\perp}$ in the transverse plane. The overlap between the two wave-packets decreases$\,$with$\,k_{\perp}$, explaining thus why the oscillations of the blue curve are damped along$\,$the$\,$horizontal$\,$axis. When interferences are destructive, the wave-packets are well separated and the 1D two-Gaussian$\,$fit,$\,$used$\,$to$\,$extract$\,$the$\,$output$\,$spacing$\,$between$\,$them,$\,$gives$\,$the$\,$expected$\,$value$\,$for$\,d$. Reversely, when interferences are constructive, the intensity in between the wave-packets strongly$\,$increases,$\,$which$\,$can$\,$induce$\,$an$\,$error$\,$in$\,$the$\,$estimation$\,$of$\,d\,$by$\,$the$\,$fitting$\,$procedure.
\vspace{6pt}
\newline
\noindent So as to check if quasi-particle interferences are observable in experiments,$\,$I$\,$have$\,$plotted$\,$on figure~\ref{fig:SnellInterference} the$\,$experimental$\,$envelope$\,$of$\,$the$\,$intensity$\,$profile$\,$(measured$\,$in$\,$the$\,$cell$\,$exit$\,$plane) as function of $k_{\perp}$. The 2D intensity distributions shown on the maps (a), (c), (d) and (e) are obtained by interpolating the experimental data, using a spline interpolation method. Let's first compare the maps (a) and (b). The latter shows the results$\;$we$\;$obtain$\;$by$\;$solving numerically$\,$the$\,$NLSE$\,$in$\,$2D$\,$for$\,$the$\,$same$\,$parameters$\,$as$\,$in$\,$(a):$\,\mathcal{P}_{f} = 350\,$mW,$\,\omega_{0,x} = 180\,\mu$m and $\omega_{0,y} = 300$ $\mu$m. We also take absorption into account in simulation. As you can$\,$see, the maps (a) and (b) look pretty similar. A destructive interference between the counter- propagating$\,$wave-packets$\,$occurs$\,$at$\,k_{\perp} \simeq 1.6 \!\times\! 10^{4}$ m$^{-1}\,$on$\,$both$\,$figures.$\,$This$\,$exactly$\,$matches the dip location on figure~\ref{fig:SnellPower}(a). Moreover, the positions of the wave-packets$\,$provided$\,$by the two-Gaussian$\,$fit$\,$on$\,$map$\,$(b)$\,$(black$\,$dashed$\,$lines)$\,$deviate$\,$from$\,$the$\,$theoretical$\,$predictions (black solid lines) when $k_{\perp}$ gets closer to $1.6 \!\times\! 10^{4}\,$m$^{-1}$. More$\,$precisely,$\,$the$\,$distance$\,$between the wave-packets$\,$is$\,$under-estimated$\,$by$\,$the$\,$fitting$\,$procedure$\,$when$\,$a$\,$destructive$\,$interference occurs at low $k_{\perp}$. This is mainly why a dip appears in the data shown on figure~\ref{fig:SnellPower}(a). 
\vspace{6pt}
\newline
\noindent We$\,$can$\,$go$\,$further by comparing the wave-vectors at which interferences$\,$occur$\,$in$\,$simulation and in experiments. The stripe at the bottom of figure (b) shows the Laplacian $\boldsymbol{\nabla
}^{2} \, \mathcal{I}(x,k_{\perp})$ of the intensity distribution plotted above it, from $-0.2$ to $0.2$ mm. The peaks lying along this stripe locate the destructive interferences while the dips locate the constructive ones. On figure (d), the 2D map (a) as well as the stripe of figure (b) have been plotted together. As you can see, the wave-vectors at which interferences occur in the simulation$\;$almost correspond to the wave-vectors at which they are observed$\,$in$\,$experiments.$\,$We$\,$can$\,$compare in a similar manner the results of the simulation with the experimental data obtained for $\mathcal{P}_{f} = 175$ mW (c) and $\mathcal{P}_{f} = 525$ mW (e) respectively. On figure (c), the simulations$\,$predict exactly the wave-vectors at which interferences$\,$occur$\,$experimentally.$\;$The$\,$agreement$\,$is$\,$less convincing on figure (e) but the simulations still predict pretty well the locations of the first destructive and the first constructive interferences.      

\newpage

\chapter{Optically induced potential in a fluid of light}

In chapter 2, the nonlinear Schr\"{o}dinger equation describing the dynamics of$\,$a$\,$photon$\,$fluid in propagating geometry has been derived. In its most general form, this equation involves an$\,$effective potential that can either trap$\,$(\textit{ie}$\,$guide)$\,$or$\,$deflect$\,$the$\,$light$\,$depending$\,$on$\,$its$\,$sign. Such a potential in photon fluids is created by a local modification of the refractive index. Inducing, in a controlled manner, local changes in the refractive index might then$\,$open$\,$up new$\,$possibilities$\,$in$\,$studying$\,$paraxial$\,$photon$\,$fluids.$\,$We$\,$can$\,$for$\,$instance$\,$think$\,$about$\,$trapping transversely the light in potentials$\;$of$\;$any$\;$shape. The photon$\;$fluid will then behave like a trapped 2-dimensional Bose-Einstein condensate, which can be interesting to investigate optical analog of many-body phenomena arising in those systems. We$\,$can$\,$also$\,$think$\,$about studying to what extent the interaction between photons affects the localization of$\,$light$\,$in random potentials~\cite{6-1Segev}. Generating a defect on the paraxial photon fluid flow (by means of a localized negative refractive index modulation) is also a way of probing superfluidity in our systems, by measuring either the cancellation of the drag force on this defect~\cite{6-2Larre, 3-3Michel}, or the amount of light it scatters at the normal/superfluid transition. This$\,$latter$\,$point$\,$is what motivates us in finding a way of controlling the refractive index felt by$\,$a$\,$fluid$\,$of$\,$light. In this chapter, I report an all-optical method to achieve$\,$this$\,$end,$\;$which$\,$requires$\,$to$\,$address at the same time the $D_{1}$ and the $D_{2}$ lines of rubidium. I thus first extend$\,$the$\,$theoretical description$\,$of$\,$chapter$\,$1 to the case of a 4-level N-type$\,$system.$\;$I$\,$therefore$\,$strongly$\,$encourage you to read chapter 1 before, as most of the notations and concepts I use here have already been introduced in the latter. In a second part, I discuss how we generate local$\,$defects$\,$in the experiments, using quasi Bessel beams; some results of this second section have been published in: "Attenuation-free non-diffracting Bessel beams", Optics Express, Vol.$\,$27, Issue 21, pp. 30067-30080 (2019)\cite{6-3Fontaine}.     

\newpage
        
\section{Modified dielectric response in a N-type atomic system}

\begin{figure}[h]
\center
\includegraphics[width=0.96\linewidth]{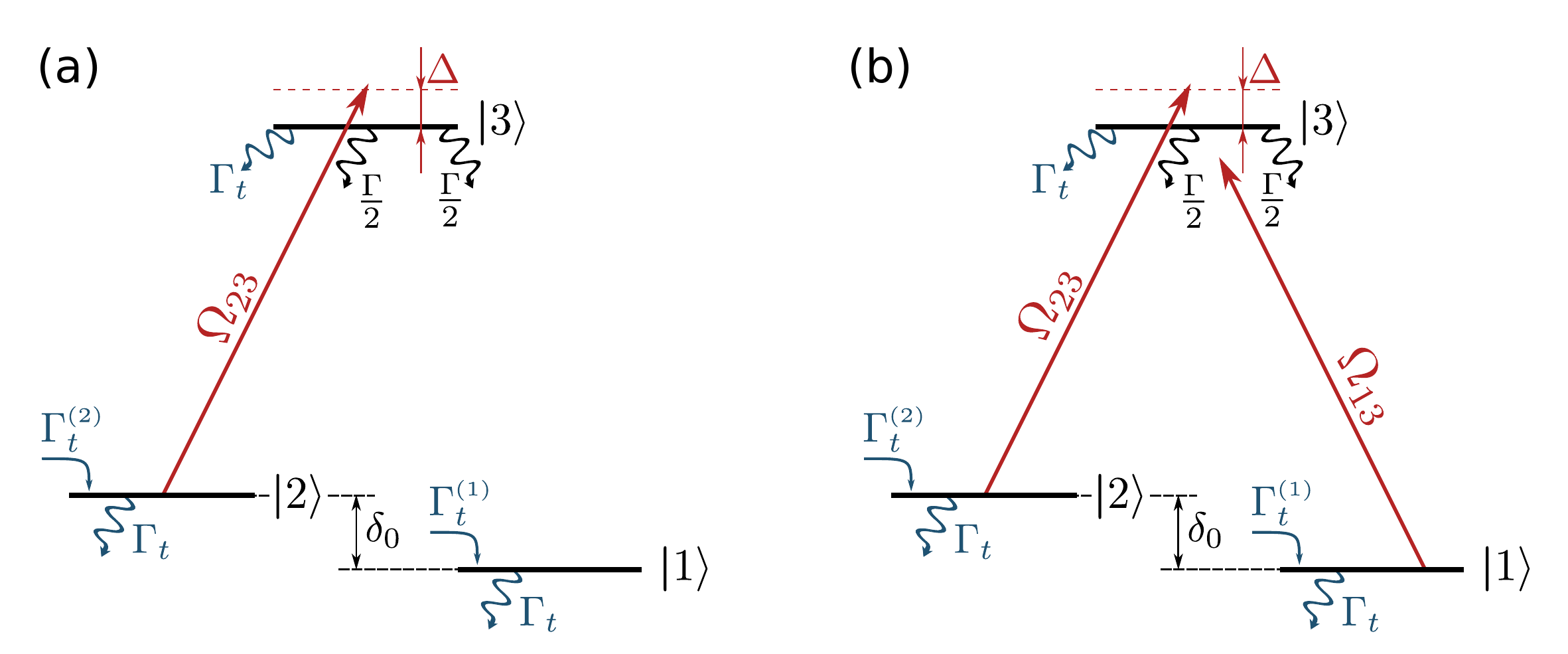} 
\caption{Sketch of the 3-level system described in$\;$chapter$\;$1.$\;$The$\,$levels$\,\ket{1}\,$and$\,\ket{2}\,$stand for the two hyperfine ground states of the $D$-lines while $\ket{3}$ stands either for the fine level $5^{2}P_{1/2}$ ($D_{1}$ line) or for $5^{2}P_{3/2}$ ($D_{2}$ line). A single laser field (red arrows) couples the ground states to the excited state. It is detuned from the$\,\ket{2} \rightarrow \ket{3}$ transition by$\,\Delta$.$\;$On$\,$figure$\,$(a), we assume it only addresses the $\ket{2} \rightarrow \ket{3}$. Reversely, this single laser field couples both ground states to the excited states on figure (b). }
\label{fig:3LevelBis}
\end{figure}

As we have seen in chapter 1, alkaline vapors are versatile platforms to study nonlinear optics phenomena since the sign and amplitude of the third-order dielectric susceptibility $\chi^{_{(3)}}$ are easily tunable changing the laser frequency. So far however, we have only slightly exploited$\;$the resources offered by the$\,$rubidium$\,$fine$\,$and$\,$hyperfine$\,$structures.$\,$On$\,$figure~\ref{fig:3LevelBis}, the sketch of the 3-level system introduced in chapter 1 has been replicated. In$\;$this$\;$model, a single coupling field addresses one of the rubidium$\;D$-lines. 

\begin{figure}[h]
\center
\includegraphics[width=\columnwidth]{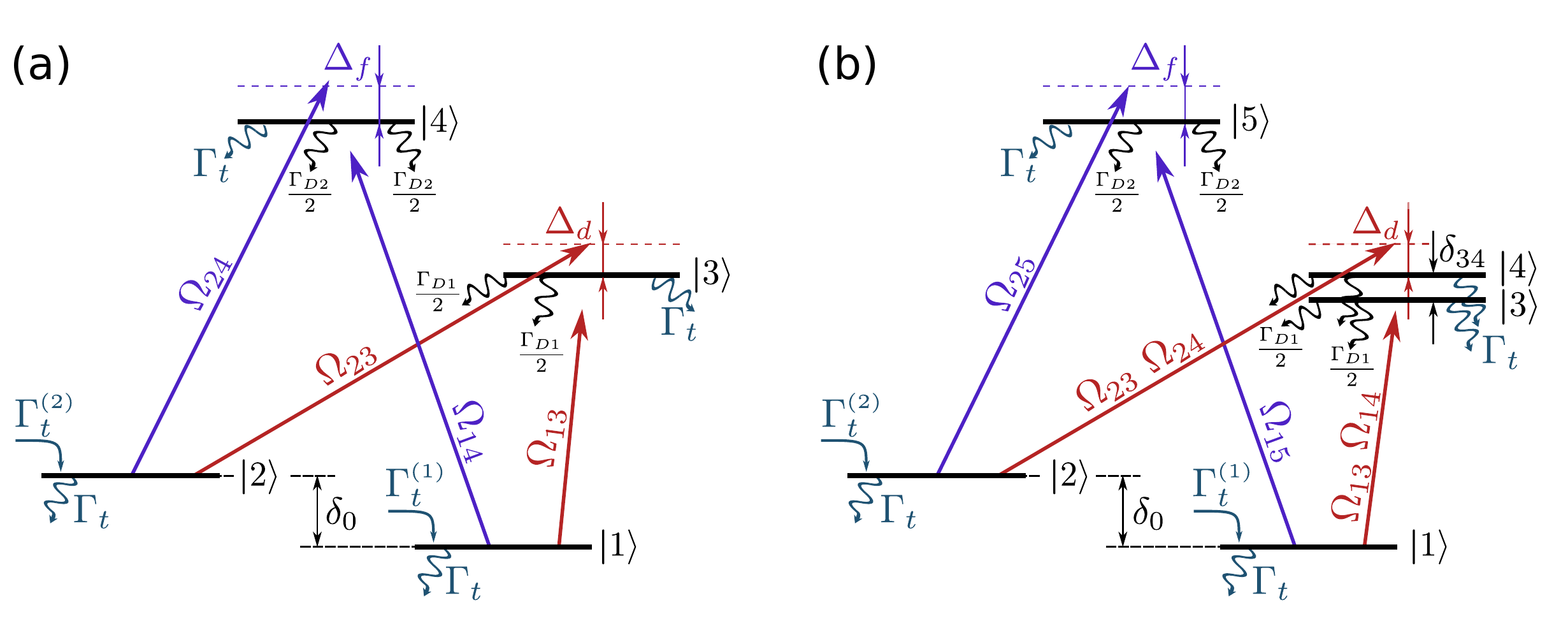} 
\caption{Four (a) and five-level system (b) in a double N-type configuration.}
\label{fig:4and5Level}
\end{figure}

\noindent If we send a second laser field on the $\ket{1} \rightarrow \ket{3}$ transition for instance, we can then locally control the atomic coherence between states $\ket{1}$ and $\ket{2}$ and therefore the strength of the nonlinear susceptibility induced by the first beam on the $\ket{2} \rightarrow \ket{3}\;$transition.$\;$This$\;$kind$\;$of "lambda" ($\Lambda$) configuration provides the basic framework for studying phenomena such$\;$as Electromagnetically Induced Transparency (EIT)~\cite{2-10Boller}. In the past years, highly$\;$enhanced self-Kerr and cross-Kerr nonlinearity~\cite{6-4Kang} have been experimentally measured by$\;$taking advantage of the atomic resources in three-level EIT systems. However, in all the schemes that involve two laser fields driving simultaneously the same $D$-line, interaction between those fields is likely to occur through four-wave-mixing or stimulated Raman processes. The strength of this interaction strongly depends on the beams wave-vectors, that$\,$defines the phase-matching condition of the wave-mixing processes. This$\,$might$\,$thus$\,$be$\,$a$\,$constraint which we do not want to concern ourselves about. In order to locally tune the linear and nonlinear dielectric susceptibilities induced by one of the two lasers inside the vapor$\;$cell, we can also think about addressing simultaneously both $D$-lines.$\;$In that case, wave-mixing is not likely to occur anymore because the beams frequencies are very different from one another (see figure 1.1). Such a configuration has been depicted on$\;$figure~\ref{fig:4and5Level}. Levels $\ket{1}$ and $\ket{2}$ still stand for the lower and upper hyperfine states of $5^{2}S_{1/2}$. 

\newpage

\begin{itemize}
    \item [$\bullet$] On figure~\ref{fig:4and5Level}(a), $\ket{3}$ and $\ket{4}$ are the fine states $5^{2}P_{1/2}$ ($D_{1}$ line) and $5^{2}P_{3/2}$ ($D_{2}\;$line). In this 4-level model, we assume that the hyperfine structure of each of these states can be neglected, which amounts to saying that the lasers$\,$detunings$\;\Delta_{d}\;$and$\;\Delta_{f}$ are large compared to the typical hyperfine splitting in $5^{2}P_{1/2}$ ($\delta_{\mathrm{D_{1}}} \simeq 213\,$MHz$\,$for$\,^{_{85}}$Rb) and in $5^{2}P_{3/2}$ ($\delta_{\mathrm{D_{2}}} \simeq 361$ MHz for $^{85}$Rb) respectively. The meaning of the indices "d" and "f" will$\;$be explained latter in this section.
    \item [$\bullet$] The splitting $\delta_{\mathrm{D_{1}}}$ between the two upper states of the $D_{1}$ line is actually$\;$significant. As you may have seen, it is almost twice the$\,$value$\,$of$\,\delta_{\mathrm{D_{2}}}$.$\;$In$\,$figure~\ref{fig:4and5Level}(b),$\,$we$\,$thus$\,$fully describe the $D_{1}$ line, $\ket{3}$ and $\ket{4}$ standing in that case for the upper$\;$and$\;$lower hyperfine states of $5^{2}P_{1/2}$ and $\delta_{34} = \delta_{\mathrm{D_{1}}}$. I still neglect the hyperfine structure$\;$of$\;5^{2}P_{3/2}$ (which is represented by the fifth level in this 5-level system).
\end{itemize}

\noindent As we will see later on, driving strongly the $D_{1}$ line with the "red" laser ($\lambda_{d} \simeq 796 \; \mathrm{nm}$) induces a modulation of the ground state populations, which in turn affects the$\;$dielectric susceptibility induced by the "blue" laser ($\lambda_{f} \simeq 780 \; \mathrm{nm}$), as $\chi$ depends on the population difference between ground and excited states. In other words, the red laser locally tunes the refractive index seen by the blue one, according to its intensity distribution in the transverse plane. This refractive index modulation $\delta n$ behaves as a repulsive ($\delta n < 0$) or an attractive ($\delta n > 0$) potential for the blue laser, depending on its sign. The$\;$areas$\;$where $\delta n < 0$ act thus as light-induced wave-guides for the blue laser. "Light$\,$guiding$\,$light"$\,$effects have been widely studied in a great variety of physical systems\cite{19:Swartzlander,6-5Morin,6-6Truscott} including$\;$hot rubidium vapors. Truscott \textit{et al.}~\cite{6-6Truscott} have demonstrated for instance the guiding of$\;$a probe beam, detuned to the red side of the $D_{2}$ line, by a powerful Laguerre-Gaussian$\;$beam addressing the $D_{1}$ line. Reversely, areas where $\delta n < 0$ act as impenetrable obstacles on which the blue light scatters. In this section, I will mainly focus on this second case and provide a theoretical description of the refractive index change felt by the blue laser when the red one strongly drives the $D_{1}$ line.$\,$The analytical calculations$\,$below$\,$are$\,$done$\,$using$\,$the 4-level model of figure~\ref{fig:4and5Level} (a) only. I will sometimes compare the results we obtain with those provided by numerical simulations of the 5-level system sketched on figure~\ref{fig:4and5Level}(b).   

\newpage

\subsection{Transit and influx rates}

\begin{figure}[h]
\center
\includegraphics[width=\columnwidth]{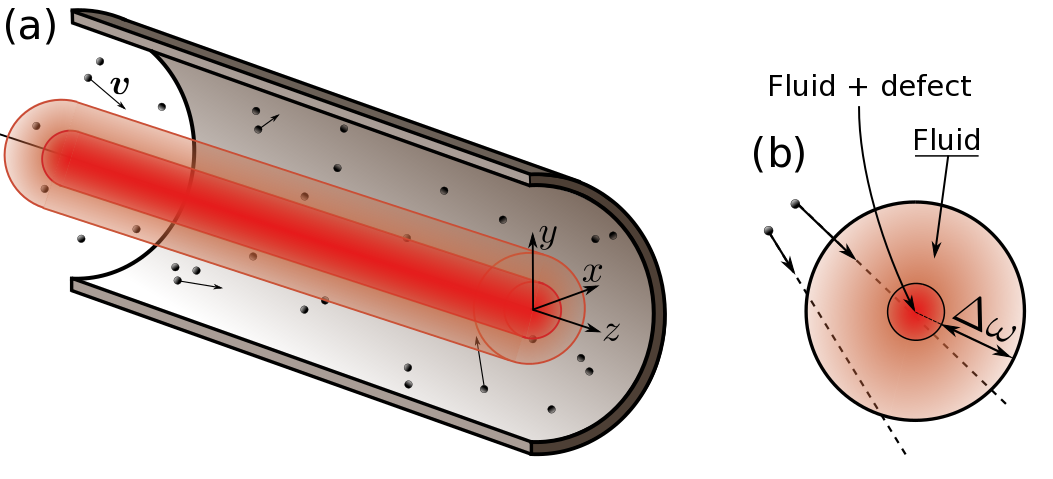} 
\caption{Beams configuration in the 4-level N-type system of figure~\ref{fig:4and5Level}. Pump (red) and probe (orange) are copropagating along the $z$ axis. In experiments, we typically set the beam widths $\omega_{0,d}$ and $\omega_{0,f}$ to $40$ $\mu$m and $500$ $\mu$m respectively.}
\label{fig:TimeofFlight2Beam}
\end{figure}

\noindent In$\,$the$\,$4-level$\,$system$\,$sketched$\,$on$\,$figure~\ref{fig:4and5Level}(a),$\,$the$\,$excited$\,$states$\,\ket{3}\,$and$\,\ket{4}\,$are$\,$likely$\,$to$\,$decay radiatively$\,$toward$\,$the$\,$ground$\,$states$\;$at$\,$the$\,$rates$\,\Gamma_{1} \simeq 2 \pi \!\times\! 5.75\,$MHz$\,$and$\,\Gamma_{2} \simeq 2 \pi \!\times\! 6.07\,$MHz respectively. As already mentioned in$\;$chapter$\;$1, atoms enter and leave the beams at the transit rate $\Gamma_{t}$; every state can thus decay toward the atomic reservoir at$\,$the$\,$same$\,$rate$\,\Gamma_{t}$. We assume that atoms enter the interacting area either in the ground state $\ket{1}$ or in $\ket{2}$.
\vspace{6pt}
\newline
\noindent 
\noindent Let us consider the situation depicted on figure~\ref{fig:TimeofFlight2Beam}. Two laser fields copropagate inside the rubidium vapor along the same optical axis. The small red disk on figure~\ref{fig:TimeofFlight2Beam}(b)$\,$stands for the transverse cross-section of the red laser in figure~\ref{fig:4and5Level}. Inside this region, both beams overlap and the atoms internal state is driven by the 4-level$\,$model$\,$sketched$\,$on$\,$figure~\ref{fig:TimeofFlight2Beam}(a). Outside this disk, only the blue beam in figure~\ref{fig:4and5Level} remains and$\;$the atoms internal state is driven by the 3-level model of figure~\ref{fig:3LevelBis}(b). Henceforward,$\,$this$\,$second$\,$beam$\,$will$\,$be$\,$referred to as the fluid ("f") or probe beam, as it will be used in experiments to create a$\,$photon$\,$fluid and probe hydrodynamical features$\,$like$\,$superfluidity.$\;$The$\,$small$\,$beam$\,$lying$\,$inside$\,$the$\,$probe will be referred to as the defect ("d") or pump beam, since it locally$\,$changes$\,$the$\,$refractive index experienced by the probe and acts, therefore, as a obstacle into the photon$\,$fluid$\,$flow. In$\,$our$\,$experiment,$\,$the$\,$probe$\,$width$\,$is$\,$much$\,$larger$\,$than$\,$the$\,$pump$\,$one:$\,$typically,$\,\omega_{0,d} = \!40 \, \mu$m and $\omega_{0,f} = 500$ $\mu$m. All the atoms reaching the defect have thus approximately travelled a distance$\,\Delta \omega = \omega_{0,f}-\omega_{0,d}\,$across$\,$the$\,$probe.$\;$The$\,$average$\,$time$\,$of$\,$flight$\,$is$\,t_{f} \simeq \Delta \omega/u \simeq 1.6\,\mu$s at $400$ K. Atoms interact then with the probe beam during $1.6$ $\mu$s only, which is, generally speaking of course, short$\;$compare$\;$to$\;$preparation$\;$time$\;\tau_{0}$. 

\newpage

\begin{figure}[h]
\center
\includegraphics[scale=0.46]{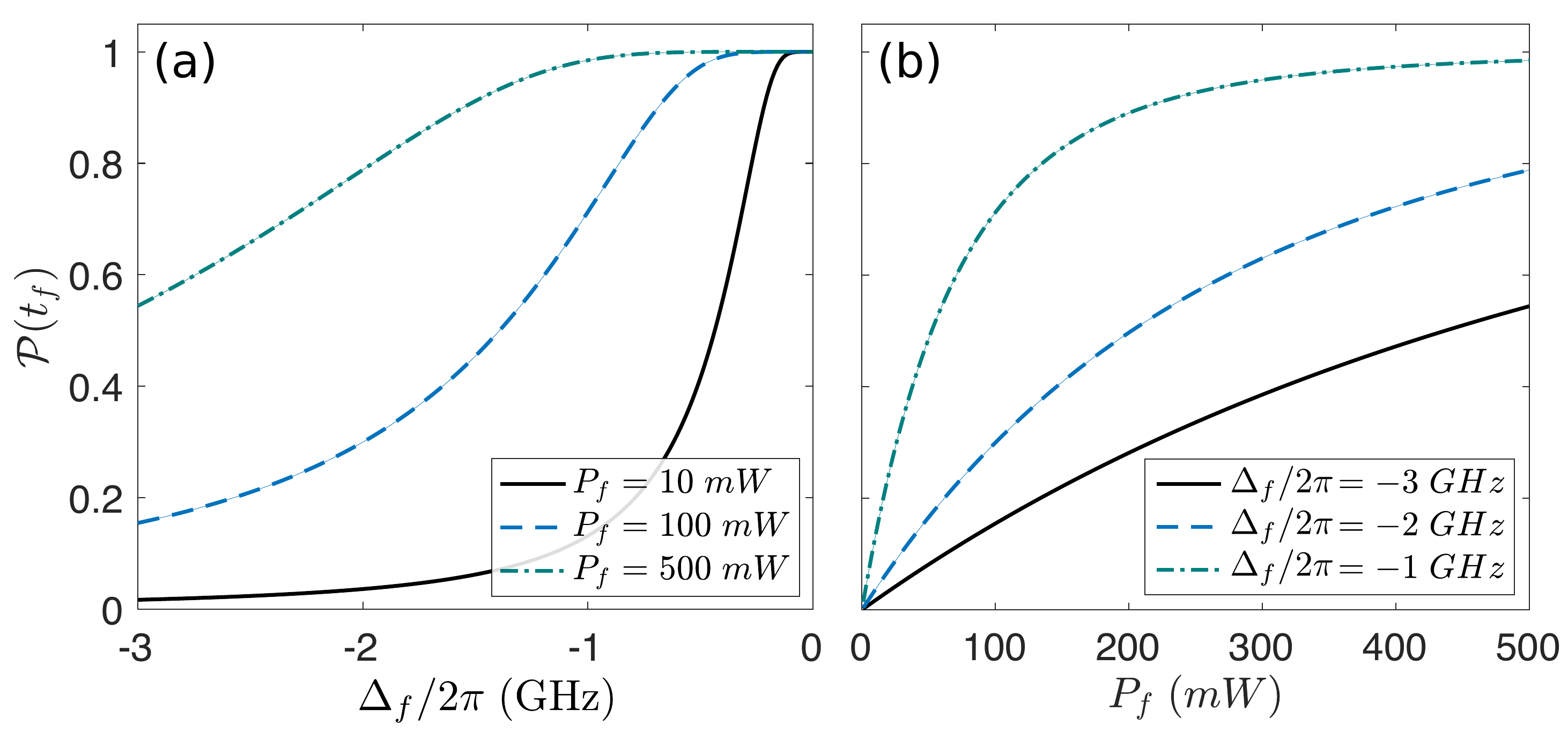} 
\caption{Fraction $\beta = \mathcal{P}(t_{f})$ of atoms prepared in the steady-state when they reach the defect beam as function of $\Delta_{f}$ (a) and $P_{f}\;$(b). At fixed fluid power, $\mathcal{P}(t_{f})$ increases$\;$by getting closer to resonance. Similarly, it increases at fixed detuning with the fluid power. Parameters: $w_{0,f} = 500$ $\mu$m, $w_{0,d} = 40$ $\mu$m and $T=400$ K.}
\label{fig:SteadyStateFraction}
\end{figure}

\noindent This parameter $\tau_{0}$ measures the time an atom needs in order to be prepared in the$\,$steady- state by the laser field. It depends$\,$on$\,$a$\,$large$\,$set$\,$of$\,$parameters,$\,$including$\,$the$\,$temperature$\,T$, the probe detuning $\Delta_{f} = \omega_{f}-\omega_{24}$ and the Rabi frequencies $\Omega_{14}$ and $\Omega_{24}$. The fraction$\;\beta$ of atoms prepared in the steady-state for such a set of parameters can be evaluated$\;$following the approach developed in~\cite{2-12Glorieux}. The probability $\mathcal{P}(t)$ to reach the equilibrium$\;$after$\;$a$\;$time$\;t$ is given by: $\mathcal{P}(t) = 1-e^{-t/\tau_{0}}$. The preparation time $\tau_{0}$ is found in practice$\;$by$\;$computing the real parts of the eigenvalues of the Bloch matrix $M$ in equation~\eqref{ST_Bloch3Level_2Coupling}; $\tau_{0}$ is finally defined as the inverse of the smallest of these real parts. In other words, $\tau_{0}$ evaluates the duration of the transient regime associated to the slowest eigenstate to reach equilibrium. A way of roughly describing the situation would therefore be to consider that a fraction $\beta = \mathcal{P}(t_{f})$ of atoms is effectively prepared in the steady-state while the remaining ones enter the defect area in their initial state, without interacting at all with the probe field. Moreover, if we assume that the atoms are $-$ before entering the probe $-$ in a mixture of states $\ket{1}$ and $\ket{2}$ described by the Boltzmann statistics, the influx rates $\Gamma_{t,d}^{_{(1)}}$ and $\Gamma_{t,d}^{_{(2)}}$ of the 4-level N-type model on figure~\ref{fig:4and5Level}(a) are given by the following formula: 
\begin{align}
    \Gamma_{t,d}^{_{(1)}} =& \; \left[(1-\beta) \, G_{1} + \beta \left(\rho_{11}^{_{\,\mathrm{in}}} + \frac{\rho_{44}^{_{\,\mathrm{in}}}}{2} \right) \right] \Gamma_{t,d},
    \label{InfluxRates4level1} \\
    \Gamma_{t,d}^{_{(2)}} =& \; \left[(1-\beta) \, G_{2} + \beta \left(\rho_{22}^{_{\,\mathrm{in}}} + \frac{\rho_{44}^{_{\,\mathrm{in}}}}{2} \right) \right] \Gamma_{t,d}.
    \label{InfluxRates4level2}    
\end{align}
\noindent In equations~\eqref{InfluxRates4level1} and~\eqref{InfluxRates4level2}, $\rho_{ii}^{_{\,\mathrm{in}}}$ stands for the proportion of atoms \textbf{prepared} by the probe laser in state $\ket{i}$ before they reach the defect area. The degeneracy weight $G_{i}\;$is$\;$still defined by: $G_{i} = g_{i}/(g_{1}+g_{2})$ ($g_{i}$ being the degeneracy factor of state $\ket{i}$). The transit$\;$rate $\Gamma_{t,d}$ is equal to $2 \, u/ \sqrt{\pi} \, \omega_{0,f}$ as usual. The fraction $\beta = \mathcal{P}(t_{f})$ has been plotted as function of the probe power and detuning on figure~\ref{fig:SteadyStateFraction} for $T = 400$ K. At this temperature and for $\omega_{0,d} = 40$ $\mu$m, the transit rate $\Gamma_{t,d}$ is about $6.4\; \mathrm{MHz}$ (\textit{ie} $\Gamma_{t,d}/\Gamma_{D_{1}} \simeq 18 \%$). 

\newpage

\begin{itemize}
    \item [$\bullet$] At high power and small detuning (\textit{ie} when $\Omega_{14},\Omega_{24} \gg \Delta_{f}$), the probe field quickly drives the atomic internal state to the steady-state.$\;$Atoms$\;$are$\;$therefore$\;$almost all prepared by the probe in that case and $\beta \simeq 1$, as$\,$can$\,$be$\,$seen$\,$on$\,$figures~\ref{fig:SteadyStateFraction}(a)$\,$and$\,$(b).
    \item [$\bullet$] Reversely, if $\Omega_{14},\Omega_{24} \ll \Delta_{f}$, there are only few atoms prepared by the probe$\;$beam. The other ones reach thus preferentially the defect in state $\ket{2}$, since $G_{1} < G_{2}$.  
\end{itemize}

\subsection{Optical-Bloch equations and steady-state solution}

\noindent Within the interaction picture and under the dipole and rotating wave approximations, one can derive the Bloch equations associated to the 4-level system sketched on~\ref{fig:4and5Level} (a), using the semiclassical approach detailed in the first chapter. At the end of the day:
\begin{align}
    \vphantom{=}&\begin{cases}
        \frac{\mathrm{d} \rho_{11}}{\mathrm{d}t} =
        -\Gamma_{t,d} \, \rho_{11} +
        \frac{\Gamma_{\!\mathrm{D1}}}{2} \, \rho_{33} + \frac{\Gamma_{\!\mathrm{D2}}}{2} \, \rho_{44} +
        \frac{i}{2} \left(\Omega_{d}^{*} \, \rho_{31} - 
        \Omega_{d} \, \rho_{13} \right) +
        \frac{i}{2} \left( \Omega_{f}^{*} \, \rho_{41} - \Omega_{f} \, \rho_{14}\right) +
        \Gamma_{t,d}^{_{(1)}} \\
        \frac{\mathrm{d}\rho_{22}}{\mathrm{d}t} =  
        -\Gamma_{t,d} \, \rho_{22} +
        \frac{\Gamma_{\!\mathrm{D1}}}{2} \, \rho_{33} + \frac{\Gamma_{\!\mathrm{D2}}}{2} \, \rho_{44} +
        \frac{i}{2} \left(\Omega_{d}^{*} \, \rho_{32} - 
        \Omega_{d} \, \rho_{23} \right) +
        \frac{i}{2} \left( \Omega_{f}^{*} \, \rho_{42} - \Omega_{f} \, \rho_{24}\right) +
        \Gamma_{t,d}^{_{(2)}} \\
        \frac{\mathrm{d}\rho_{33}}{\mathrm{d}t} =
        -\left(\Gamma_{t,d} + \Gamma_{D_{1}} \right) \rho_{33} -  
        \frac{i}{2} \left(\Omega_{d}^{*} \, \rho_{31} - 
        \Omega_{d} \, \rho_{13} \right) -
        \frac{i}{2} \left( \Omega_{d}^{*} \, \rho_{d} - \Omega_{23} \, \rho_{23}\right) \\
        \frac{\mathrm{d}\rho_{44}}{\mathrm{d}t} = -\left(\Gamma_{t,d} + \Gamma_{D_{2}} \right) \rho_{44} -  
        \frac{i}{2} \left(\Omega_{f}^{*} \, \rho_{41} - 
        \Omega_{f} \, \rho_{14} \right) -
        \frac{i}{2} \left( \Omega_{f}^{*} \, \rho_{42} - \Omega_{f} \, \rho_{24}\right) \\
        \phantom{a} \vspace{-10pt} \\
        \frac{\mathrm{d}\rho_{21}}{\mathrm{d}t} = \; -\Tilde{\gamma}_{21} \, \rho_{21} + 
        \frac{i}{2} \, \Omega_{d}^{*} \, \rho_{31} - 
        \frac{i}{2} \, \Omega_{d} \, \rho_{23} +   
        \frac{i}{2} \, \Omega_{f}^{*} \, \rho_{41} - 
        \frac{i}{2} \, \Omega_{f} \, \rho_{24} \\
        \frac{\mathrm{d}\rho_{31}}{\mathrm{d}t} = \; 
        -\Tilde{\gamma}_{31} \, \rho_{31} + 
        \frac{i}{2} \, \Omega_{d} \, \rho_{21} -
        \frac{i}{2} \, \Omega_{f} \, \rho_{34} +
        \frac{i}{2} \, \Omega_{d} \left( \rho_{11} - \rho_{33} \right) \\
        \frac{\mathrm{d}\rho_{32}}{\mathrm{d}t} = \; 
        -\Tilde{\gamma}_{32} \, \rho_{32} + 
        \frac{i}{2} \, \Omega_{d} \, \rho_{12} -
        \frac{i}{2} \, \Omega_{f} \, \rho_{34} +
        \frac{i}{2} \, \Omega_{d} \left( \rho_{22} - \rho_{33} \right) \\
        \frac{\mathrm{d}\rho_{41}}{\mathrm{d}t} = \;   -\Tilde{\gamma}_{41} \, \rho_{41} + 
        \frac{i}{2} \, \Omega_{f} \, \rho_{21} -
        \frac{i}{2} \, \Omega_{d} \, \rho_{43} +
        \frac{i}{2} \, \Omega_{f} \left( \rho_{11} - \rho_{44} \right) \\
        \frac{\mathrm{d}\rho_{42}}{\mathrm{d}t} = \;   
        -\Tilde{\gamma}_{42} \, \rho_{42} + 
        \frac{i}{2} \, \Omega_{f} \, \rho_{12} -
        \frac{i}{2} \, \Omega_{d} \, \rho_{43} +
        \frac{i}{2} \, \Omega_{f} \left( \rho_{22} - \rho_{44} \right) \\
        \frac{\mathrm{d}\rho_{43}}{\mathrm{d}t} = \;
        -\Tilde{\gamma}_{43} \, \rho_{43} + 
        \frac{i}{2} \, \Omega_{f} \, \rho_{13} +
        \frac{i}{2} \, \Omega_{f} \, \rho_{23} -
        \frac{i}{2} \, \Omega_{d}^{*} \, \rho_{41} -
        \frac{i}{2} \, \Omega_{d}^{*} \, \rho_{42}
    \end{cases} 
    \label{Bloch4Level}
\end{align}
\noindent Equations \eqref{Bloch4Level}, together with the closure relation: $\mathrm{Tr}\left( \hat{\rho} \right) = 1$, constitute a close set of Bloch equations. As the dipole moment $\mu_{13}$ and $\mu_{23}$ are equal $-\;$see$\;$for$\;$instance~\eqref{Stength}$\;-$ $\Omega_{13} \!=\! \Omega_{23} \!=\! \Omega_{d}$. Similarly, $\Omega_{14} \!=\! \Omega_{24} \!=\! \Omega_{f}$. The Rabi frequencies $\Omega_{d}$ and $\Omega_{f}$ are$\;$defined$\;$by $\Omega_{d} = E_{d} \, \mu_{d} / \hbar$ and $\Omega_{f} = E_{f} \, \mu_{f} / \hbar$, where $E_{d} = \frac{1}{2} \! \left( \mathcal{E}_{d} \, e^{i k_{d} z} +cc \right)$ and $E_{f} = \frac{1}{2} \! \left( \mathcal{E}_{f} \, e^{i k_{f} z} +cc \right)$, $\mathcal{E}_{d,f}$, $k_{d,f}$ and $\omega_{d,f}$ being respectively the complex envelopes, the wave vectors and the frequencies of the defect and probe beams. We also define the quantities $\Tilde{\gamma}_{21} = \gamma_{21} + i \delta_{0}$, $\Tilde{\gamma}_{31} = \gamma_{31} - i (\Delta_{d}-\delta_{0})$, $\Tilde{\gamma}_{32} = \gamma_{32} - i \Delta_{d}$, $\Tilde{\gamma}_{41} = \gamma_{41} - i (\Delta_{f}-\delta_{0})$, $\Tilde{\gamma}_{43} = \gamma_{43} - i \Delta_{f}$ and $\Tilde{\gamma}_{42} = \gamma_{42} - i (\Delta_{f}-\Delta_{f})$ for the sake of clarity. The coherence decay rates are given by: $\gamma_{ij} = \frac{1}{2}(\Gamma_{i}+\Gamma_{j})+\gamma_{\mathrm{col}}$, where $\Gamma_{i}$ is the decay rate of state $\ket{i}$ (for instance, $\Gamma_{3} = \Gamma_{t,d} + \Gamma_{\mathrm{D_{1}}}$) and $\gamma_{\mathrm{col}}$ the collision-induced decoherence rate. Finally, $\Delta_{d} = \omega_{d}-\omega_{23}$ and $\Delta_{f} = \omega_{f}-\omega_{24}$ are the defect and probe detunings with respect to the $\ket{2} \!\rightarrow\! \ket{3}$ and $\ket{2} \!\rightarrow\! \ket{4}\,$transitions.
\vspace{6pt}
\newline
\noindent The 4-level system reaches the steady-state after a given time of evolution. At that point, the first order time derivatives are zero in~\eqref{Bloch4Level}. We should not expect to get a compact expression for the solution of the resulting matrix equation in that$\;$case.$\;$One$\;$of$\;$the$\;$options left to us consist in solving the steady-state Bloch equations iteratively, assuming that$\;$the driving of the $D_{1}$ line by the defect is much more efficient than the driving of the $D_{2}$ line by the probe, or in other words, that $\Omega_{f} \ll \Omega_{d}$. This kind of perturbative approach has been used in~\cite{6-7Sheng} to find the steady-state solution of a N-type model in an EIT$\;$configuration: 

\newpage

\noindent two intense laser fields (namely, the coupling and switching fields) address$\;$respectively$\;$the $\ket{2} \rightarrow \ket{3}$ and $\ket{2} \rightarrow \ket{4}$ transitions, while$\,$a$\,$less$\,$powerful$\,$one$\,$probes$\,$the$\,\ket{1} \rightarrow \ket{3}\,$transition. In our case, the assumption $\Omega_{f} \ll \Omega_{d}$ is not necessarily fulfilled, as the probe laser$\,$(that$\,$is, the fluid) is far detuned from resonance, which allows$\,$to$\,$increase$\,$its$\,$intensity$\,$(and$\,$thus$\,\Omega_{f}$) while remaining below the saturation threshold $I_{s}(\Delta_{f})$. However, the linear variation$\,\delta n_{\vphantom{\mathrm{in}}}^{_{(1)}}$ of$\,$the refractive$\,$index$\,$felt$\,$by$\,$the$\,$probe$\,$beam$\,$between$\,$inside$\,$and$\,$outside$\,$the$\,$defect$\,$area$\,$does not depend on the probe strength, by definition. Therefore, the first order expansion in $\Omega_{f}$ of the stationary Bloch equations provides an exact expression for $\delta n_{\vphantom{\mathrm{in}}}^{_{(1)}}\!$ whether$\;$or$\;$not$\;$the condition $\Omega_{f} \ll \Omega_{d}$ is fulfilled. Following this comment, I thus derive in the next section the expression of $\delta n_{\vphantom{\mathrm{in}}}^{_{(1)}}\!$ and study how it varies as function of $\Delta_{d}$, $\Delta_{f}$ and $\Omega_{d}$. In parallel, I$\;$present the results obtained with the dressed-state formalism, which is not only aesthetic but provides a physical understanding about how the optical defect is generated.  

\subsection{Linear variation of the refractive index at the defect position}
\label{subsec:LinearVariation}

\subsubsection{Perturbative approach}

\begin{figure}[h]
\center
\includegraphics[width=\columnwidth]{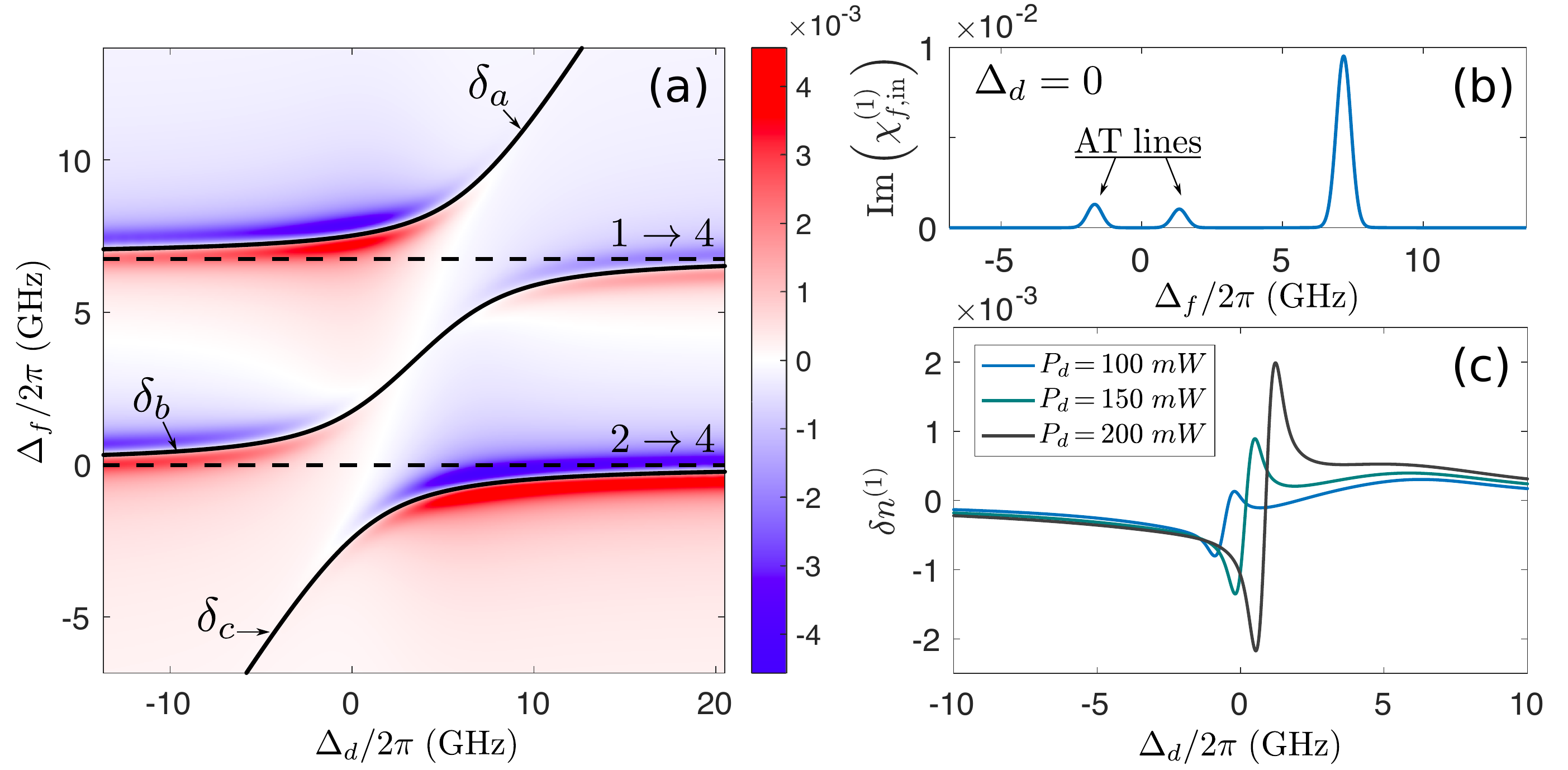} 
\caption{(a) Real part of $\chi_{_{f,\mathrm{in}}}^{_{(1)}}$. The $\ket{1} \rightarrow \ket{4}$ and $\ket{2} \rightarrow \ket{4}$ spectral lines split into$\;$two when the defect beam is at resonance with $\ket{1} \rightarrow \ket{3}$ ($\Delta_{d} = \delta_{0}$) and $\ket{2} \rightarrow \ket{3}$ ($\Delta_{d} = 0$) respectively$\;$(Autler-Town effect). The energy shifts $\hbar \delta_{a}$, $\hbar \delta_{b}$ and $\hbar \delta_{c}$ of the dressed states $\ket{a, n}$, $\ket{b, n}$ and $\ket{c, n}$ have been plotted (black lines) as function of$\;$the$\;$defect$\;$detuning. (b)$\,$Imaginary$\,$part$\,$of$\,\chi_{f,\mathrm{in}}^{_{(1)}}\,$as function of $\Delta_{f}$ (for $\Delta_{d} = 0$).$\,$The$\,$Autler-Town$\,$(AT)$\,$sidebands of the $\ket{2}\rightarrow \ket{4}$ transition are clearly visible on the left. The absorption peak on the right starts being blue-shifted from $\Delta_{f} = \delta_{0} \, (\simeq 6.8$ GHz here, since we deal with rubidium 87). (c) Linear index modulation for different defect powers $P_{d}$. Parameters: $w_{0,d} = 50$ $\mu$m, $w_{0,f} = 500$ $\mu$m and $T=415$ K. In (a-c), $\mathcal{P}_{f} = 10$ $\mu$W. In (a-b), $\mathcal{P}_{d} = 200$ mW.}
\label{fig:LinearChangeRefractive}
\end{figure}

\noindent Let$\,M\,$stands for the steady-state Bloch matrix.$\,$The steady-state matrix$\,$equation$\,$can$\,$then be decomposed as follows: $M \, \boldsymbol{\rho} = \left( M_{0} + M_{f} \right) \boldsymbol{\rho} = 0$, where $M_{0}$ and $M_{f}$ are respectively the $\Omega_{f}$-independent and -dependent parts of $M$. Since we assume $\Omega_{f} \ll \Omega_{d}$, the density matrix elements can be obtained iteratively as $\rho_{ij} = \rho_{ij}^{_{(1)}} + \rho_{ij}^{_{(2)}} + \rho_{ij}^{_{(3)}} + ...$ and the $n$th step of the expansion is given by: 
\begin{equation}
    M_{0} \, \boldsymbol{\rho}^{_{(n)}} = - M_{f} \, \boldsymbol{\rho}^{_{(n-1)}}
\end{equation}
\noindent The zeroth order equation reads: $M_{0} \, \boldsymbol{\rho}^{_{(0)}} = 0$. It basically describes the steady-state$\;$of the 4-level system sketched on figure~\ref{fig:4and5Level}(a) when the probe (blue arrows)$\,$has$\;$been$\;$switched$\,$off. In that case, the 4-level system reduces to a 3-level one and the zeroth-order equation above is the same as (1.58) when $\Omega_{13} = \Omega_{23} = \Omega_{d}$. Using the results of chapter 1, one$\;$can thus compute the zeroth order density matrix elements. The last three equations in~\eqref{Bloch4Level} can then be solved together in the steady-state. The resulting matrix equation reads:
\begin{equation}
    \begin{pmatrix}
        -\Tilde{\gamma}_{41}  & 0 \vphantom{\frac{\Gamma^{t}}{\Gamma_{t}}} & -\frac{i}{2} \, \Omega_{d} \\ 
        0 \vphantom{\frac{\Gamma^{t}}{\Gamma_{t}}} & -\Tilde{\gamma}_{42} & -\frac{i}{2} \, \Omega_{d} \\
        -\frac{i}{2} \, \Omega_{d}^{*} & -\frac{i}{2} \, \Omega_{d}^{*} & -\Tilde{\gamma}_{43} \vphantom{\frac{\Gamma^{t}}{\Gamma_{t}}}
    \end{pmatrix}
    \begin{pmatrix}
        \rho_{41}^{_{(1)}}
        \vphantom{\frac{\Gamma^{t}}{\Gamma_{t}}} \\ 
        \rho_{42}^{_{(1)}} 
        \vphantom{\frac{\Gamma^{t}}{\Gamma_{t}}} \\
        \rho_{43}^{_{\small{(1)}}} 
        \vphantom{\frac{\Gamma^{t}}{\Gamma_{t}}}
    \end{pmatrix} = -\frac{i \Omega_{f}}{2}  
    \begin{pmatrix}
        \rho_{11}^{_{(0)}} + \rho_{21}^{_{(0)}}
        \vphantom{\frac{\Gamma^{t}}{\Gamma_{t}}} \\ 
        \rho_{22}^{_{(0)}} + \rho_{12}^{_{(0)}} 
        \vphantom{\frac{\Gamma^{t}}{\Gamma_{t}}} \\
        \rho_{13}^{_{(0)}} + \rho_{23}^{_{(0)}}
        \vphantom{\frac{\Gamma^{t}}{\Gamma_{t}}}
    \end{pmatrix},
    \label{FirstOrder}
\end{equation}
\noindent and can be inverted in order to obtain $\rho_{41}^{_{(1)}}$, $\rho_{42}^{_{(1)}}$ and $\rho_{43}^{_{(1)}}$ as follows:
\begin{align}
    \rho_{41}^{_{(1)}} =& \; \Omega_{f} \,  \frac{\Tilde{\gamma}_{42} \, \Omega_{d} \left( \rho_{13}^{_{(0)}} + \rho_{23}^{_{(0)}} \right)+ i \left[ \left(\vphantom{\rho_{11}^{_{(0)}}} |\Omega_{d}|^{2} + \Tilde{\gamma}_{42} \Tilde{\gamma}_{43} \right)
    \left( \rho_{11}^{_{(0)}} + \rho_{21}^{_{(0)}} \right) - 
    |\Omega_{d}|^{2} \left( \rho_{22}^{_{(0)}} + \rho_{12}^{_{(0)}} \right) \right]}{\Tilde{\gamma}_{41} \, |\Omega_{d}|^{2} + \Tilde{\gamma}_{42} \, |\Omega_{d}|^{2} + 4 \, \Tilde{\gamma}_{41} \Tilde{\gamma}_{42} \Tilde{\gamma}_{43}}, \label{Density41} \\
    \rho_{42}^{_{(1)}} =& \; \Omega_{f} \,  \frac{\Tilde{\gamma}_{41} \, \Omega_{d} \left( \rho_{13}^{_{(0)}} + \rho_{23}^{_{(0)}} \right)+ i \left[ \left(\vphantom{\rho_{11}^{_{(0)}}} |\Omega_{d}|^{2} + \Tilde{\gamma}_{41} \Tilde{\gamma}_{43} \right)
    \left( \rho_{22}^{_{(0)}} + \rho_{12}^{_{(0)}} \right) - 
    |\Omega_{d}|^{2} \left( \rho_{11}^{_{(0)}} + \rho_{21}^{_{(0)}} \right) \right]}{\Tilde{\gamma}_{41} \, |\Omega_{d}|^{2} + \Tilde{\gamma}_{42} \, |\Omega_{d}|^{2} + 4 \, \Tilde{\gamma}_{41} \Tilde{\gamma}_{42} \Tilde{\gamma}_{43}}, \label{Density42} \\
    \rho_{43}^{_{(1)}} =& \; \Omega_{f} \, \frac{
    \Tilde{\gamma}_{41} \, \Omega_{d}^{*} \left( \rho_{22}^{_{(0)}} + \rho_{12}^{_{(0)}} \right)+
    \Tilde{\gamma}_{42} \, \Omega_{d}^{*} \left( \rho_{11}^{_{(0)}} + \rho_{21}^{_{(0)}} \right)+
    i \, \Tilde{\gamma}_{41} \Tilde{\gamma}_{41} \left( \rho_{13}^{_{(0)}} + \rho_{23}^{_{(0)}} \right) \vphantom{\left[\left( \rho_{13}^{_{(0)}} + \rho_{23}^{_{(0)}} \right) \right]}}  {\Tilde{\gamma}_{41} \, |\Omega_{d}|^{2} + \Tilde{\gamma}_{42} \, |\Omega_{d}|^{2} + 4 \, \Tilde{\gamma}_{41} \Tilde{\gamma}_{42} \Tilde{\gamma}_{43}} \label{Density43}.
\end{align}
\noindent  Inside the defect area, the linear response of the vapor to the probe field is$\;$characterized by the linear dielectric susceptibility $\chi_{f,\mathrm{in}}^{_{(1)}} = \frac{2 N}{\epsilon_{0} \mathcal{E}_{f}} \left( \mu_{14} \, \rho_{41}^{_{(1)}} + \mu_{24} \, \rho_{42}^{_{(1)}} \! \right)$. The linear variation$\;$of the refractive index felt by the probe between inside and outside the defect is thus given$\;$by:
\begin{equation}
    \delta n_{\vphantom{\mathrm{in}}}^{_{(1)}} = n_{\mathrm{in}}^{_{(1)}}-n_{\mathrm{out}}^{_{(1)}} \simeq \frac{1}{2} \, \mathrm{Re} \left[\chi_{_{f,\mathrm{in}}}^{_{(1)}} - \chi_{_{f,\mathrm{out}}}^{_{(1)}}  \right]    
\end{equation}
\noindent where $n_{\mathrm{in}}^{_{(1)}}$ and $n_{\mathrm{out}}^{_{(1)}}$ are the linear refractive indices seen by the probe inside and outside$\;$the defect area respectively. The real part of the susceptibility $\chi_{_{f,\mathrm{in}}}^{_{(1)}}$ is plotted$\;$on figure~\ref{fig:LinearChangeRefractive}(a) as function of $\Delta_{d}$ and $\Delta_{f}$, and the linear index modulation $\delta n_{\vphantom{\mathrm{in}}}^{_{(1)}}\!$ on~\ref{fig:LinearChangeRefractive}(c)$\,$as$\;$function$\;$of$\,\Delta_{d}$, at$\,$a$\,$fixed$\,$probe$\,$detuning $\Delta_{f} = -2\pi \times 2$ GHz$\,$and$\,$for$\,$different$\,$defect$\,$powers.$\,$The$\,$simulations have been performed for $w_{0,d} = 50$ $\mu$m, $w_{0,f} = 0.5$ mm and $T = 415$ K. Doppler$\;$broadening has been taken into account. 
\vspace{4pt}
\newline
\noindent As you can see, the spectrum of $\chi_{_{f,\mathrm{in}}}^{_{(1)}}$ is relatively complex. Let's analyse it step by step. When the defect beam is highly red-detuned it does not drive the $\ket{2} \rightarrow \ket{3}\,$transition$\,$very efficiently and even less the$\,\ket{1} \rightarrow \ket{3}\,$transition.$\;$In$\,$that$\,$case,$\,$cutting$\,$the$\,$2D$\,$map$\,$along$\,$some vertical line will make the resulting profile look like the plot on figure$\,$1.10(a).$\;$Nevertheless, on figure 1.10(a), the centers of the $\ket{1} \rightarrow \ket{4}$ and the $\ket{2} \rightarrow \ket{4}$ transition$\,$lines$\,$lie$\,$exactly at the transition frequencies $\omega_{14}$ and $\omega_{24}$. On figure \ref{fig:LinearChangeRefractive}(a), even$\;$when the defect beam is highly red-detuned, the lines centers are$\,$slightly$\,$blue-shifted$\,$from$\,$the$\,$transition$\,$frequencies (black dotted lines). This light-shift can be seen as the AC analog of the Stark-effect~\cite{6-8Stark}, that shifts the spectral line of$\,$atoms$\,$and$\,$molecules$\,$when$\,$submitted$\,$to$\,$constant$\,$electric$\,$field. It increases as $\Omega_{d}^{2}/4 \! \left| \Delta_{d} \right|$ at large detunings. Here, $\Omega_{d} \simeq 2\pi \!\times\!4.3\,$GHz, which explains why at tens of gigahertz detunings, the light-shift is still clearly visible.$\;$When$\,\Delta_{d} = -2\pi \!\times\! 10\,$GHz for instance, it is about $2\pi \!\times\! 470$ MHz. The defect beam is intense enough to off-resonantly drive the rubidium $D_{1}$ line. 
\vspace{2pt}
\newline
\noindent This is$\,$also$\,$reflected$\,$in$\,$the amplitude of the spectral lines. On figure 1.10(a),$\,$the$\,$amplitude of $\mathrm{Re} \left[ \chi \right]$ close to $\Delta_{f} = 0$ is larger than close to $\Delta_{f} = \delta_{0}$ (where $\delta_{0}$ is the$\,$hyperfine$\,$splitting between ground states), because atoms are more likely to get inside the beam$\,$in$\,$state$\,\ket{2}$, which$\,$is$\,$more$\,$degenerated. As the Rabi frequency$\,$of$\,$the$\,$probe$\,$is$\,$small$\,$($\Omega_{f} \simeq 2\pi \!\times\! 4.3\,$MHz), atoms will$\;$predominantly reach the defect area in state $\ket{2}$. But as soon as they get inside, they$\,$will$\,$be$\,$pumped$\,$by$\,$the defect beam, from state $\ket{2}$ to state $\ket{1}$ for instance when it is moderately red-detuned. This is why the amplitude of the real part of $\chi_{_{f,\mathrm{in}}}^{_{(1)}}$ is higher close to $\Delta_{f} = \delta_{0}$ than to $\Delta_{f} = 0$ in that case. 
\vspace{2pt}
\newline
\noindent So far, we have only discussed what happens when the defect beam is far from resonance. When it drives $\ket{1} \rightarrow \ket{3}$ (resp. the $\ket{2} \rightarrow \ket{3}$) resonantly, the $\ket{1} \rightarrow \ket{4}$ (resp. $\ket{2} \rightarrow \ket{4}$) transition line splits into two. This effect is known as the Autler-Town splitting~\cite{2-11Autler}$\,$and is qualitatively understood as follows. By driving$\;$resonantly$\;$the$\;\ket{1} \rightarrow \ket{3}$ (resp. $\ket{2} \rightarrow \ket{3}$) transition line, the defect quickly modulates the ground state population $\rho_{11}^{_{(0)}}$ (resp. $\rho_{22}^{_{(0)}}$), which oscillates at the Rabi frequency $\Omega_{d}$. The linear absorption of the$\;$probe$\;$beam$\;$on$\;$the $\ket{1} \rightarrow \ket{4}$ (resp. $\ket{2} \rightarrow \ket{4}$) transition is thus also modulated at $\Omega_{d}$, creating sidebands on the probe absorption spectrum, the Autler-Town lines, visible on the absorption profile of figure~\ref{fig:LinearChangeRefractive}(b) where the fluid frequency is scanned for $\Delta_{d} = 0$. The same features can$\,$be observed on figure~\ref{fig:LinearChangeRefractive}(a), as$\;$real$\;$and$\;$imaginary parts of the susceptibility are related to each other through the Kramers-Kronig relations. In the following paragraph, I will$\,$show how to precisely describe the Autler-Town splitting as well as the level anti-crossing visible on figure~\ref{fig:LinearChangeRefractive}(a) using the dressed-sate formalism. 

\newpage

\subsubsection{Dressed-state formalism}

\noindent The analytical expressions of the first-order density matrix elements allow us to calculate the weak probe field response for arbitrary values of the parameters and provides a detailed description of how this response is modified by the defect field.$\;$Nevertheless,$\;$the$\;$algebraic complexity of equations~\eqref{Density41},~\eqref{Density42} and~\eqref{Density43} prevents us from getting a simple physical insight into the behaviour of $\chi_{_{f,\mathrm{in}}}^{_{(1)}}$. An alternative approach is the so-called dressed-state formalism, which is particularly convenient for describing the photon-atom interaction in the strong coupling limit. Since detailed descriptions of the dressed-state theory can be found elsewhere~\cite{6-9Cohen,6-10Cohen, 6-11Cohen}, I will limit myself to applying it to the current situation.
\vspace{6pt}
\newline
\noindent The coupling between the $D_{1}$ line and the defect beam can be modeled by a 3-level$\;$system, interacting with a quasi-resonant laser field, whose Hamiltonian is: $\hat{H} = \hat{H}_{\mathrm{F}} + \hat{H}_{\mathrm{A}} + \hat{H}_{\mathrm{I}} $. $\hat{H}_{\mathrm{F}}$, $\hat{H}_{\mathrm{A}}$ and $\hat{H}_{\mathrm{I}}$ are the atomic, the field and the interaction parts of $\hat{H}$ respectively~\cite{6-8Wei}: 
\begin{align}
    \begin{cases}
    \hat{H}_{\mathrm{F}} =& \; \hbar \, \omega_{d} \, \hat{a}_{d}^{\dagger}\hat{a}_{d}^{\vphantom{\dagger}} \\
    \hat{H}_{\mathrm{A}} =& \; \hbar \, \omega_{21} \ket{2} \bra{2} + \hbar \, \omega_{31} \ket{3} \bra{3} \\ 
    \hat{H}_{\mathrm{I}} =& \; \frac{\hbar \, \Omega_{d}^{*}}{2} \, \hat{a}_{d}^{\vphantom{\dagger}} \, \ket{3} \bra{1} + \frac{\hbar \, \Omega_{d}^{\phantom{*}}}{2} \, \hat{a}_{d}^{\dagger} \, \ket{1} \bra{3} + \frac{\hbar \, \Omega_{d}^{*}}{2} \, \hat{a}_{d}^{\vphantom{\dagger}} \, \ket{3} \bra{2} + \frac{\hbar \, \Omega_{d}^{\phantom{*}}}{2} \, \hat{a}_{d}^{\dagger} \, \ket{2} \bra{3},
    \end{cases}
\end{align}
\noindent where $\hat{a}_{d}$ and $\hat{a}_{d}^{\dagger}$ are the annihilation and creation operators of a photon$\,$in$\,$the$\,$defect$\,$beam.
\newpage

\noindent We set the ground state energy$\;$to$\;$0. Let's $\ket{i, n}$ be the eigenstate of the uncoupled atom plus field Hamiltonian, where $i$ labels the atomic state while $n$ stands for the number of photons in the defect field. Under the quasi-resonance condition, the levels $\ket{1,n}$, $\ket{2,n}\,$and $\ket{3,n-1}$ become a quasi-degenerate triplet $\xi_{n}$; the energy levels$\,$of$\,$the$\,$uncoupled$\,$atom$\,$plus field system split thus into such triplets, consecutively separated by the photon energy$\;\hbar \omega_{d}$. The interaction Hamiltonian introduces couplings between states lying in$\,$the$\,$same$\,$triplet. In the basis $\left \{\ket{1,n}, \ket{2,n}, \ket{3,n-1} \right \}$, $\hat{H}$ is given by:
\begin{equation}
\hat{H}_{n} = \hat{H} - n \, \hbar \omega \, \mathbb{1}= 
    \begin{pmatrix}
        0  & 0  & \frac{\hbar \, \Omega_{d}}{2} \\ 
        0 & \hbar \delta_{0} & \frac{\hbar \, \Omega_{d}}{2}  \\
        \frac{\hbar \, \Omega_{d}^{*}}{2} & \frac{\hbar \, \Omega_{d}^{*}}{2} & \hbar (\delta_{0}\!-\!\Delta_{d})
    \end{pmatrix}, 
    \label{DressedStates}
\end{equation}
\noindent where $\mathbb{1}$ stands for the $3 \times 3$ identity matrix. The three dressed states of $\xi_{n}$ $-$ $\ket{a, n}$, $\ket{b, n}$ and $\ket{c, n}$ $-$ and their energies $-$ $\hbar \delta_{a}$, $\hbar \delta_{b}$ and $\hbar \delta_{c}$ $-$ are obtained by diagonalizing the Hamiltonian $\hat{H}_{n}$ in equation~\eqref{DressedStates}. As long as $\Omega_{d} \ne 0$, the matrix on the right-hand side of~\eqref{DressedStates} has rank 3, whatever the value of $\Delta_{d}$. The dressed states$\,$are$\,$thus$\,$always$\,$given$\,$by a linear superposition of all the unperturbed states $\ket{1,n}$, $\ket{2,n}$ and $\ket{3,n-1}$ in the triplet. The fourth level in the dressed-state representation is simply given by $\ket{4, n}$ as it is not coupled to the defect field. When the probe frequency is tuned over the whole $D_{2}\;$line, transitions from $\ket{1,n}$ or $\ket{2,n}$ to $\ket{4,n}$ occur. Moreover, the eigenstates of $\hat{H}_{n}$ contain$\;$all $\ket{1,n}$ and $\ket{2,n}$. In the dressed-state picture, there are therefore three allowed transitions between the excited state $\ket{4,n}$ and the dressed states of the $n$th multiplicity.$\;$That$\;$is$\;$why 
three peaks are visible on the absorption spectrum of figure~\ref{fig:LinearChangeRefractive}(b).$\,$The shifts$\,\hbar \delta_{i} = \hbar \delta_{0} - E_{i}$ of the dressed-state energies$\,$from$\,$the$\,$unperturbed$\,$energy$\,$of$\,$state$\,\ket{2}$,$\,\hbar \delta_{0}$,$\,$have$\,$been$\,$plotted as function of $\Delta_{d}$ (in h units) on figure~\ref{fig:LinearChangeRefractive}(a) (black lines). The dressed-state formalism correctly predicts the splitting and the level anti-crossing observed in figure~\ref{fig:LinearChangeRefractive}(a).

\subsection{Total variation of the refractive index at the defect position}

\begin{figure}[h]
\center
\includegraphics[width=\columnwidth]{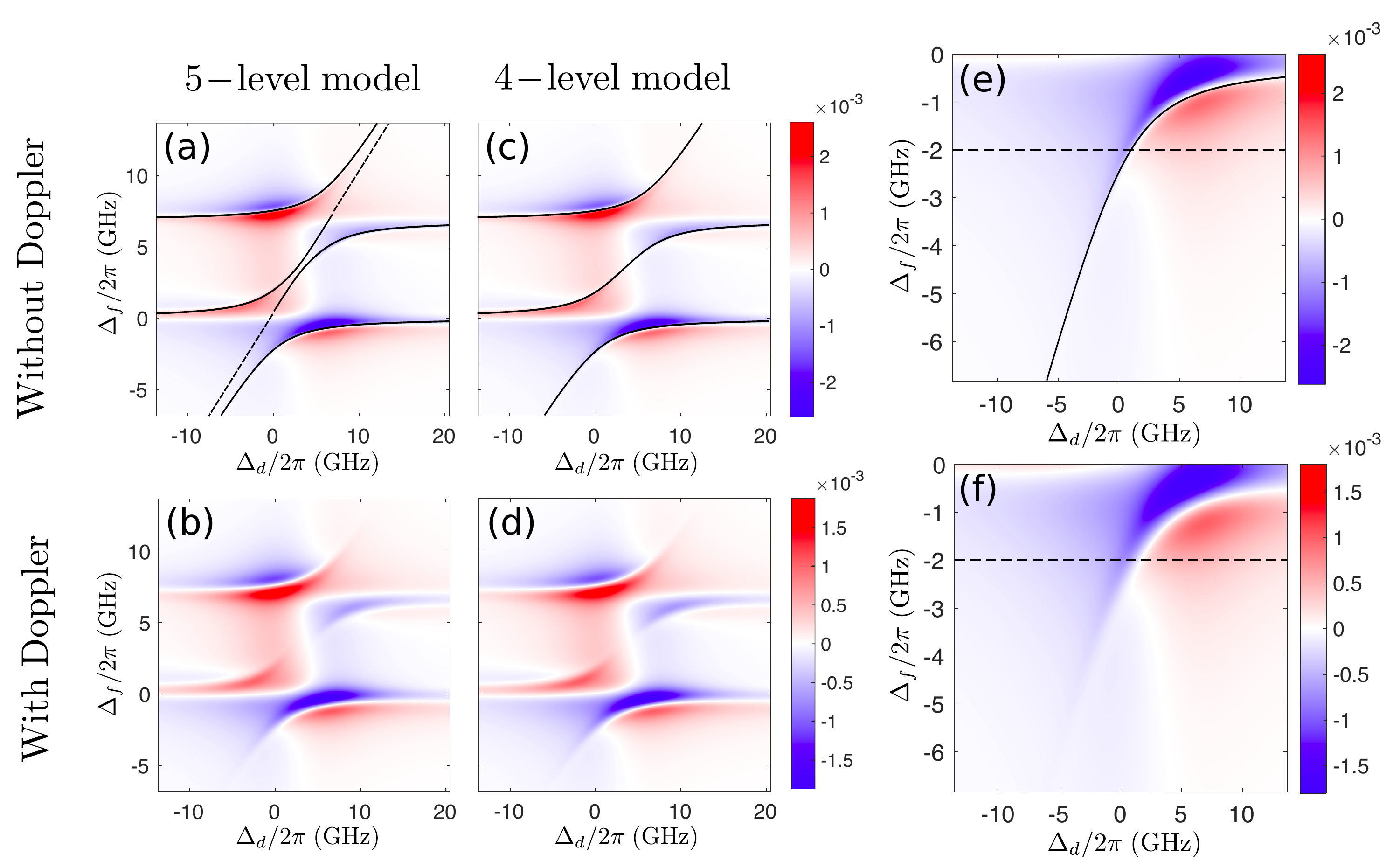} 
\caption{Total refractive index variation at the defect position. (a)-(b): Results of the calculations for the 4-level$\,$system$\,$of$\,$figure~\ref{fig:4and5Level}(a).$\;$The$\,$black$\,$lines$\,$represent$\,$the$\,$energy$\,$shift (in $h$ units) of the the three dressed states composing the triplet$\,\xi_{n}\,$(see$\,$paragraph$\,$5.1.3$\,$ii).
(c)-(d) Results of the calculations for the 4-level$\,$system$\,$of$\,$figure~\ref{fig:4and5Level}(b). The dresses-state approach introduced in the previous paragraph is extended to describe the full hyperfine structure of the $D_{1}$ line. The black lines represent the energy shift (in h units) of the four dressed states composing the quadruplet $\xi_{n}$ in that case. The central ones are separated$\,$by $\delta_{34} \simeq 815$ MHz (for rubidium 87), when $\Delta_{d} = \delta_{0}/2$. On figures (e) and (f), we$\,$zoom$\,$in$\,$the region where $\Delta_{f} < 0$ (since$\,$in$\,$our$\,$experiments,$\,\Delta_{f}\,$must$\,$be$\,$negative$\,$to$\,$ensure$\,$that$\,n_{2} < 0$). At a given probe detuning (for instance $\Delta_{f} = -2$ GHz,$\,$black$\,$dotted$\,$line),$\,$tuning$\,$the$\,$defect frequency enables us to generate either attractive (\textit{ie} guiding) or repulsive potentials into the photon fluid. Parameters: $w_{0,d} = 50$ $\mu$m, $w_{0,f} = 500\,\mu$m,$\,\mathcal{P}_{d} = 200\,$mW,$\,\mathcal{P}_{f} = 50\,$mW and $T=415$ K. Calculations performed for rubidium 87.
} 
\label{fig:ChangeRefractive2DMaps}
\end{figure}

\noindent The definition of the refractive index modulation seen by the probe beam at the defect position can be extended to take nonlinear contributions into account. The total variation of refractive index between inside and outside the defect beam cross-section is given$\;$by: 
\begin{equation}
\delta n^{\vphantom{_{(1)}}} = n_{\mathrm{in}}^{\vphantom{_{(1)}}}-n_{\mathrm{out}}^{\vphantom{_{(1)}}} \simeq \frac{1}{2} \, \mathrm{Re} \left[\chi_{_{f,\mathrm{in}}}^{\vphantom{_{(1)}}} - \chi_{_{f,\mathrm{out}}}^{\vphantom{_{(1)}}} \right].    
\end{equation}
\noindent The susceptibilities $\chi_{_{f,\mathrm{in}}}^{\vphantom{_{(1)}}}$ and $\chi_{_{f,\mathrm{out}}}^{\vphantom{_{(1)}}}$ are obtained by solving numerically the steady-state matrix$\,$equations$\,$for$\,$the$\,$4-$\,$and$\,$the$\,$3-level$\,$models$\,$respectively.$\;$The$\,$resulting$\,$refractive$\,$index modulation $\delta_{n}$ is plotted as function of $\Delta_{d}$ and $\Delta_{f}$ on figures~\ref{fig:ChangeRefractive2DMaps}(a) and (b), with (b) and without (a) Doppler broadening. These results are compared $-$ on figure~\ref{fig:ChangeRefractive2DMaps}(c)$\,$and$\,$(d)$\,-$ to the index change we get by solving the 5-level N-type model of figure~\ref{fig:4and5Level}(b) (describing more accurately the $D_{1}$line),$\;$with$\,$(d)$\,$and$\,$without$\,$(c)$\,$Doppler-broadening.$\;$As$\,$you$\,$can$\,$see, the latter smooths out the$\,$discrepancy$\,$between$\,$the$\,$4-$\,$and$\,$5-level$\,$results,$\,$indicating$\,$that$\,$the 4-level model is sufficient to describe the physics$\,$at$\,$play.$\;$As$\,$for$\,$the$\,$linear$\,$index$\,$modulation, the sign and the strength of $\delta n$ can be tuned changing the defect and the fluid frequencies. A positive$\;$(resp.$\;$negative) $\delta n$ will locally act as an attractive (resp. repulsive) potential for the photon fluid. However, in order to ensure the robustness of the fluid of light against modulational instabilities, the probe should be red-detuned from$\;$the$\;\ket{2} \rightarrow \ket{4}\;$transition

\newpage

\noindent ($n_{2}$ is negative in that case). We are therefore only interested in the region where $\Delta_{f} < 0$ on the 2D maps~\ref{fig:ChangeRefractive2DMaps}(c) and (d), magnified on figures~\ref{fig:ChangeRefractive2DMaps}(e) and (f).

\noindent The inside susceptibility $\chi_{_{f,\mathrm{in}}}^{\vphantom{_{(1)}}}$ can also be computed by using the dressed-state formalism introduced in the preceding section. This approach enables, for instance, to$\;$isolate$\;$in$\;\chi_{_{f,\mathrm{in}}}^{\vphantom{_{(1)}}}$ the contribution of each dressed-state transition $\ket{i,n} \rightarrow\ket{4,n}$, which can help us getting a deeper insight into the spectra of figure~\ref{fig:ChangeRefractive2DMaps}. In the next two paragraphs,$\;$I$\;$will$\;$therefore digress a little bit and extend the dressed-state approach of subsection~\ref{subsec:LinearVariation}.   

\newpage

\subsubsection{4-level system in the dressed-state picture}

\noindent As in subsection~\ref{subsec:LinearVariation}, let's describe the coupling between the $D_{1}$ line$\;$and$\;$the$\;$defect$\;$field using the dressed-state formalism. Let's then $\ket{a,n}$, $\ket{b,n}$ and $\ket{c,n}$ stand for the three dressed states in the triplet $\xi_{n}$. As already mentioned previously, these states can always be expressed as a linear superposition of all the unperturbed states in the multiplicity$\;\xi_{n}$: $\ket{i,n} = \alpha_{i} \ket{1,n} + \beta_{i} \ket{2,n} + \gamma_{i} \ket{3, n-1}$ ($i$ lies in $\left \{ a,b,c \right \}$), with $|\alpha_{i}|^{2} + |\beta_{i}|^{2} + |\gamma_{i}|^{2} = 1$ (normalization).$\;$The$\,$coefficients$\,\alpha_{i}$,$\,\beta_{i}\,$and$\,\gamma_{i}\,$are$\,$obtained$\,$by$\,$diagonalizing$\,\hat{H}_{n}\,$in$\,$eq.$\,$\eqref{DressedStates}. 
Using the normalization of $\ket{1,n}$ and the closure relation $\sum_{i} \ket{i,n} \bra{i,n} = \mathbb{1}$, it is straightforward to show that $\sum_{i} |\alpha_{i}|^{2} = 1$. Similarly, $\sum_{i} |\beta_{i}|^{2} = \sum_{i} |\gamma_{i}|^{2} \!=\! 1$. The$\;$decay$\;$rate$\;$from the state $\ket{4,n}$ to the dressed state $\ket{i,n}$ is given by $\Gamma_{i} = \frac{1}{2} \, (|\alpha_{i}|^{2} + |\beta_{i}|^{2}) \, \Gamma_{2}$ and $\sum_{i} \Gamma_{i} = \Gamma_{2}$. The ground states $\ket{1,n}$ and $\ket{2,n}$ both appear in the expansion of the dressed states on the unperturbed basis of $\xi_{n}$. These two states are constantly filled by fresh atoms entering the defect area. Every dressed state in the triplet is therefore associated to a filling rate $\Gamma_{t}^{_{(i)}}$ defined by: $\Gamma_{t}^{_{(i)}} = \frac{1}{2} \, |\alpha_{i}|^{2} \, \Gamma_{t,d}^{_{(1)}} +  \frac{1}{2} \, |\beta_{i}|^{2} \, \Gamma_{t,d}^{_{(2)}}$, and here again $\sum_{i} \Gamma_{t}^{_{(i)}} = \Gamma_{t,d}$. 

\begin{figure}[h]
\center
\includegraphics[width=\columnwidth]{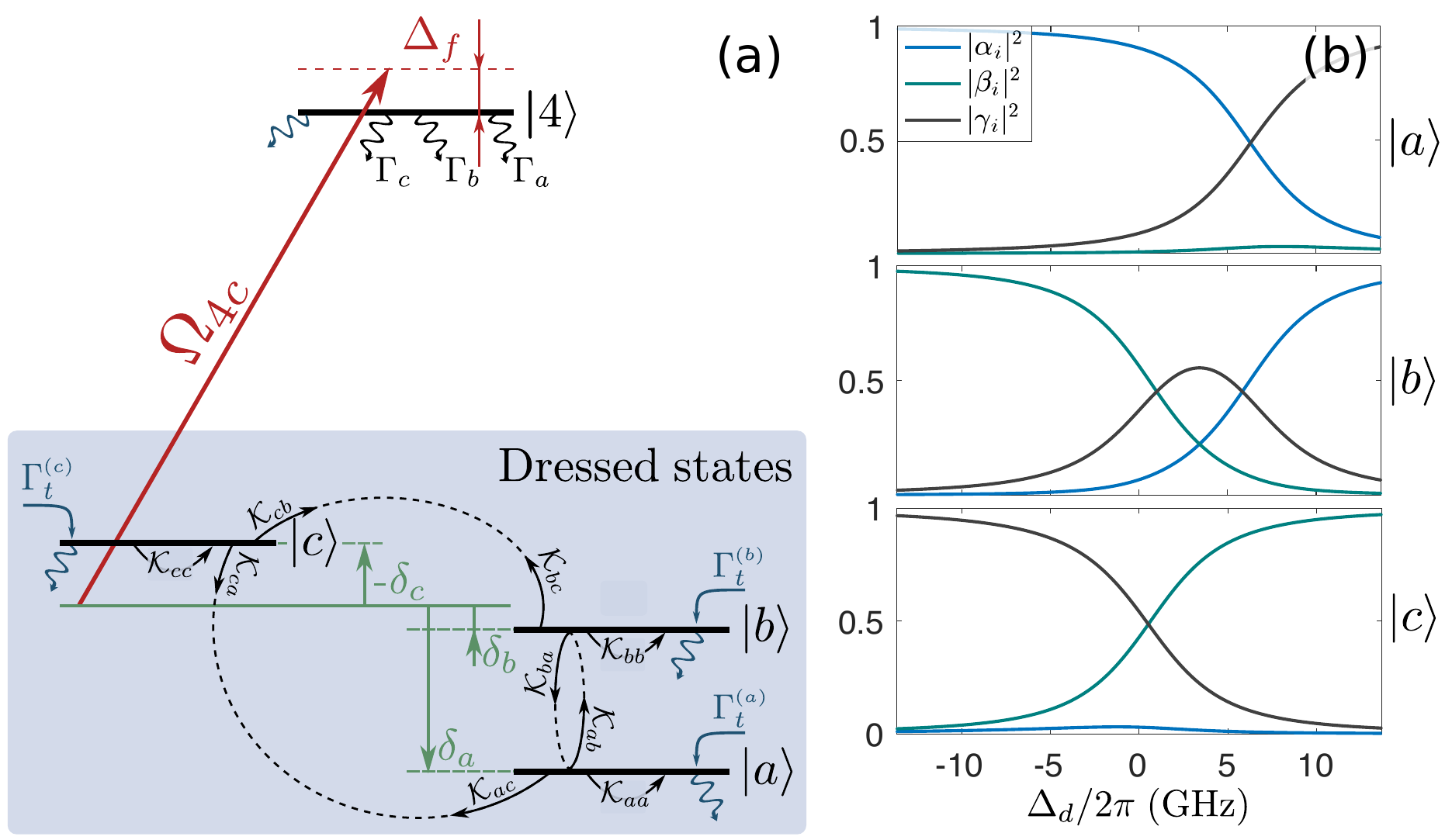} 
\caption{(a) Dressed-state model. The decay rates
$\mathcal{K}_{i,j}$ couple the dressed states $\ket{a}$, $\ket{b}$ and $\ket{c}$ together. The blue$\;$arrows stand, as usual, for the transit filling$\;$and$\;$decay$\;$rates. The energy of the dressed state $\ket{i}$ is shifted with respect$\;$to$\;$the$\;$unperturbed$\;$energy$\;$of$\;\ket{2}$ (green horizontal line) by $\hbar\delta_{i}$. I assume the probe beam couples each dressed state to the unperturbed excited state $\ket{4}$ independently (the total optical response$\;$is$\;$then$\;$the$\;$sum of the responses of all the $\ket{i} \rightarrow \ket{4}$ transition lines). The dipole moment $\mu_{4i}$ (and$\;$thus$\;\Omega_{i4}$) depends on which transition is addressed by the probe field.$\,$(b)$\,$Variation$\,$of$\,$the$\,$amplitudes $|\alpha_{i}|^{2}$, $|\beta_{i}|^{2}$ and $|\gamma_{i}|^{2}$ for $\ket{a}$ (top), $\ket{b}$ (middle) and $\ket{c}$ (bottom).}
\label{fig:DressedStateSchematic}
\end{figure}

\newpage

\noindent For the sake of correctness, we must also take the decay rate of dressed$\;$states$\;$into$\;$account. The unperturbed excited state $\ket{3,n\!-\!1}$ ends up decaying toward level $\ket{1,n}$ or level $\ket{2,n}$ and releases at that time one photon in the defect$\;$field. Since $\ket{i,n}$ contains $\ket{1,n}$, $\ket{2,n}$ and $\ket{3,n\!-\!1}$ (whatever $i$), this decay process happens$\;$not$\;$only$\;$in$\;$between$\;$dressed$\;$states but also within the states themselves. We therefore introduce $\mathcal{K}_{ij} = \frac{1}{2} |\gamma_{i}|^{2} \left( |\alpha_{j}|^{2} \!+\! |\beta_{j}|^{2}\right) \Gamma_{1}$ ($i$ and $j$ lie inside $\left \{ a,b,c \right \}$), which is the decay rate from $\ket{i,n}$ to $\ket{j,n}$; $|\gamma_{i}|^{2}$ and $|\alpha_{i}|^{2} \!+\! |\beta_{i}|^{2}$ are respectively the probability for $\ket{i,n}$ of being in the excited state $\ket{3,n\!-\!1}$ or in one of the ground states $\ket{1,n}$ or $\ket{2,n}$. The total decay rate $\mathcal{K}_{i}$ of $\ket{i,n}$ is thus$\;$finally$\;$given$\;$by: $\mathcal{K}_{i} \!=\! \sum_{j} \mathcal{K}_{i,j} \!=\! |\gamma_{i}|^{2} \, \Gamma_{1}$ and, once again, $\sum_{i} \mathcal{K}_{i} \!=\! \Gamma_{1}$. 
\vspace{6pt}
\newline
\noindent For the sake of clarity, all the notations introduced before have been gathered$\;$on$\;$figure~\ref{fig:DressedStateSchematic}. In the dressed-state picture, $\ket{a, n}$, $\ket{b, n}$ and $\ket{c, n}$ can be regarded as ground states for the excited level $\ket{4,n}$. As already mentioned, the energies of the dressed states are shifted from the unperturbed energy of level $\ket{2,n}$ (sketched by the green line on figure~\ref{fig:DressedStateSchematic}(a)) by $\hbar\delta_{a}$,$\;\hbar\delta_{b}\;$and$\;\hbar\delta_{c}$ respectively. Figure~\ref{fig:DressedStateSchematic}(b) shows the variations of $|\alpha_{i}|^{2}$, $|\beta_{i}|^{2}$ and $|\gamma_{i}|^{2}$ with$\;\Delta_{d}$ (at $\Omega_{d} = 2\pi\!\times\! 4.3$ GHz) for $\ket{a}$ (top), $\ket{b}$ (middle) and $\ket{c}$ (bottom).
\vspace{6pt}
\newline
\noindent \textsc{remark}. The decay rate between two states is proportional to the square of the dipole matrix element between these states. Being an odd operator, the dipole moment $\boldsymbol{\hat{d}}$ cannot change the number of photons in the defect field and must therefore couple $\xi_{n}$ to adjacent multiplicities $\xi_{n \pm 1}$~\cite{6-9Cohen}. Consequently, the dressed state $\ket{i,n}$ in $\xi_{n}$ cannot decay toward the other states in this multiplicity (as I have supposed above) but toward those in $\xi_{n-1}$. For a complete and rigorous treatment of spontaneous emission in a dressed-atom$\;$system, the reader may refer to~\cite{6-9Cohen, 6-11Cohen}.$\;$However,$\,$our$\;$description$\,$still$\,$provides$\,$very$\,$good$\,$results, that are in excellent agreement with those obtained by directly solving equation~\eqref{Bloch4Level} in the steady-state, as we will see later on.

\subsubsection{Optical Bloch equations in the dressed-state picture}

\noindent Let's now derive the Bloch equations associated to the 4-level dressed-state system sketched on figure~\ref{fig:DressedStateSchematic}(a).
The fluid detuning is still defined by: $\Delta_{f} = \omega_{f}-\omega_{24}$, where $\omega_{24}$ is the frequency of the unperturbed $\ket{2} \rightarrow \ket{4}$ transition. The laser detuning should then be set to $ -\delta_{i}$ in order to drive resonantly the transition from $\ket{i,n}$ to $\ket{4,n}$.$\;$The$\;$dipole$\;$moment associated to this transition is obtained by evaluating the dipole matrix element between $\ket{i, n}$ and $\ket{4,n}$: $\mu_{i4} = |\bra{i,n} \boldsymbol{\hat{d}} \ket{4,n}| = |\alpha_{i} + \beta_{i}|\, \mu_{f}$. The Rabi frequency depends thus on the transition addressed by the probe field. For$\;$the$\;$sake$\;$of simplicity, I will supposed from now on that the transitions are probed one by one by the probe beam as depicted in figure~\ref{fig:DressedStateSchematic}(a). The total optical response is thus obtained by summing$\;$the$\;$response$\;$of$\;$each individual transition. When the probe addresses the $\ket{c,n} \rightarrow \ket{4,n}$ transition for instance, the Optical Bloch equations read: 
\begin{align}
    \begin{cases}
    \label{OBEDressed}
        \frac{\mathrm{d} \rho_{aa}}{\mathrm{d}t} =&
        -\left( \Gamma_{t,d} + \mathcal{K}_{ab} + \mathcal{K}_{ac} \right) \rho_{aa} + \mathcal{K}_{ba} \, \rho_{bb} + \mathcal{K}_{ca} \, \rho_{cc} + \Gamma_{a} \, \rho_{44} + \Gamma_{t}^{_{(a)}} \\
        \frac{\mathrm{d} \rho_{bb}}{\mathrm{d}t} =& 
        \mathcal{K}_{ab} \, \rho_{aa} - \left( \Gamma_{t,d} + \mathcal{K}_{ba} + \mathcal{K}_{bc} \right) \rho_{bb} + \mathcal{K}_{cb} \, \rho_{cc} + \Gamma_{b} \, \rho_{44} + \Gamma_{t}^{_{(b)}} \\
        \frac{\mathrm{d} \rho_{cc}}{\mathrm{d}t} =& 
        \mathcal{K}_{ac} \, \rho_{aa} + \mathcal{K}_{bc} \, \rho_{bb} - \left( \Gamma_{t,d} + \mathcal{K}_{ca} + \mathcal{K}_{cb} \right) \rho_{cc} + \Gamma_{c} \, \rho_{44} + \Gamma_{t}^{_{(c)}} + \frac{i}{2} \left( \Omega_{c4}^{*} \, \rho_{4c} - \Omega_{c4} \, \rho_{c4} \right) \\
        \frac{\mathrm{d} \rho_{44}}{\mathrm{d}t} =&  - \left( \Gamma_{t,d} + \Gamma_{D_{2}} \right) \rho_{44} - \frac{i}{2} \left( \Omega_{c4}^{*} \, \rho_{4c} - \Omega_{c4} \, \rho_{c4} \right) \\
        \frac{\mathrm{d} \rho_{4c}}{\mathrm{d}t} =&  \frac{i}{2} \Omega_{c4} \left( \rho_{cc} - \rho_{44} \right) - \Tilde{\gamma}_{4c} \, \rho_{4c}
    \end{cases}
\end{align}

\begin{figure}[h]
\center
\includegraphics[width=\columnwidth]{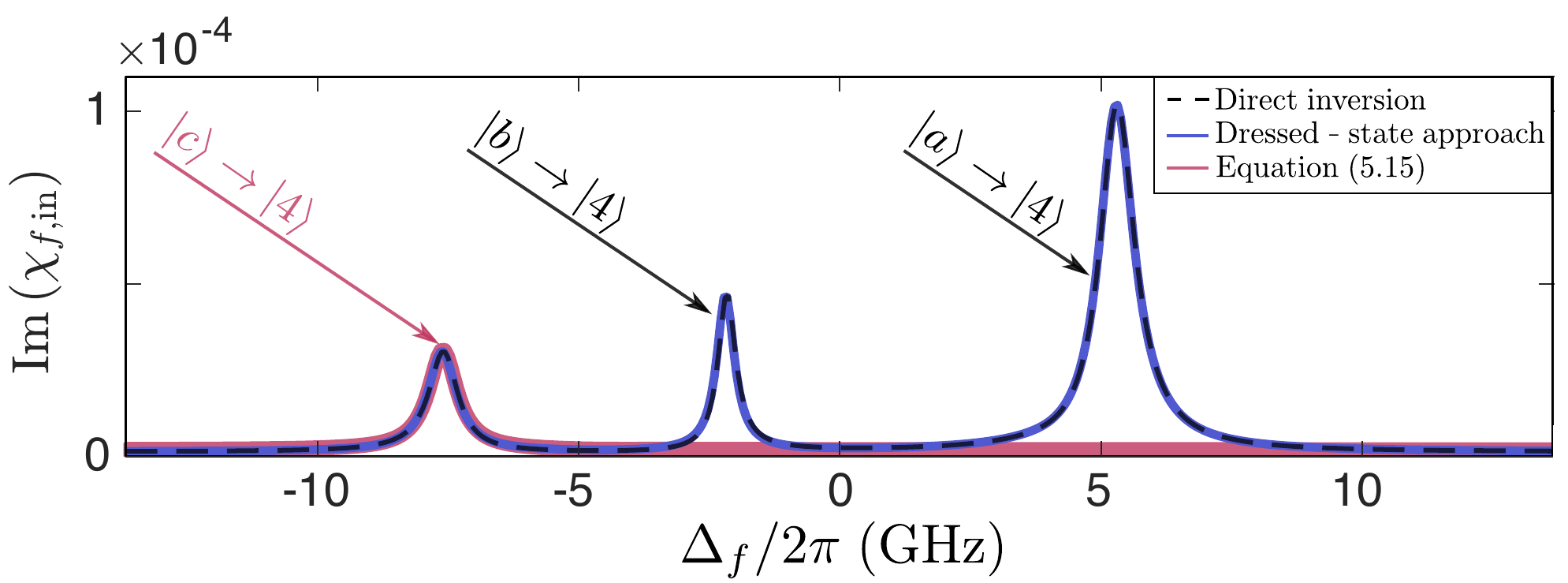} 
\caption{Comparison between the dressed-state approach and the numerical inversion of the steady-state matrix equation derived from~\ref{Bloch4Level}. Parameters are the same as for~\ref{fig:ChangeRefractive2DMaps}.}
\label{fig:ComDressed}. 
\end{figure}

\newpage

\noindent where $\Tilde{\gamma}_{4c} = (\Gamma_{2} + \mathcal{K}_{c})/2 + \Gamma_{t,d} - i(\Delta_{f} - \delta_{c})$. Since the fluid couples only $\ket{c,n}$ and $\ket{4,n}$, the dielectric susceptibility $\chi_{c4} = \frac{2 N}{\epsilon_{0} \mathcal{E}_{f}} \, \mu_{c4} \, \rho_{4c}$ is similar to the susceptibility of an atomic two-level system. We also know, by looking at the real part of $\Tilde{\gamma}_{4c}$, that the decoherence rate of the $\ket{c,n} \rightarrow \ket{4,n}$ transition (that is, the linewidth) is given by $(\Gamma_{2} + \mathcal{K}_{c})/2 + \Gamma_{t,d}$. The coherence $\rho_{4c}$ is obtained by solving the steady-state matrix equation below, derived from the Bloch equations above by setting the time derivatives to zero:

\begin{equation}
\begin{pmatrix}
        \mathcal{T}_{aa} \!-\! \left(\Gamma_{t,d} \!+\! \mathcal{K}_{a} \right) & \mathcal{T}_{ba} & \mathcal{T}_{ca}  & 0 \vphantom{\frac{i \Omega_{c4}^{*}}{2}} & 0 \vphantom{\frac{i \Omega_{c4}^{*}}{2}} \\ 
        \mathcal{T}_{ab} & \mathcal{T}_{bb} \!-\! \left(\Gamma_{t,d} \!+\! \mathcal{K}_{b} \right) & \mathcal{T}_{cb} & 0 \vphantom{\frac{i \Omega_{c4}^{*}}{2}} & 0 
        \vphantom{\frac{i \Omega_{c4}^{*}}{2}} \\
        \mathcal{T}_{ac} & \mathcal{T}_{bc} & \mathcal{T}_{cc} \!-\! \left(\Gamma_{t,d} \!+\! \mathcal{K}_{c} \right) & \frac{i \Omega_{c4}^{*}}{2} & -\frac{i \Omega_{c4}}{2} \\
        \frac{i \Omega_{c4}}{2} & \frac{i \Omega_{c4}}{2} & i \Omega_{c4} & -\Tilde{\gamma}_{4c} & 0 \vphantom{\frac{i \Omega_{c4}^{*}}{2}} \\
        \frac{i \Omega_{c4}^{*}}{2} & \frac{i \Omega_{c4}^{*}}{2} & i \Omega_{c4}^{*} & 0 & -\Tilde{\gamma}_{4c}^{*}
    \end{pmatrix} \!\!
    \begin{pmatrix}
        \rho_{aa}
        \vphantom{\frac{i \Omega_{c4}^{*}}{2}} \\ 
        \rho_{bb}
        \vphantom{\frac{i \Omega_{c4}^{*}}{2}} \\
        \rho_{cc}
        \vphantom{\frac{i \Omega_{c4}^{*}}{2}} \\
        \rho_{4c}
        \vphantom{\frac{i \Omega_{c4}^{*}}{2}} \\
        \rho_{c4}
        \vphantom{\frac{i \Omega_{c4}^{*}}{2}}
    \end{pmatrix}\! = -
    \begin{pmatrix}
        \Gamma_{t}^{_{(a)}} \!+\! \Gamma_{a} 
        \vphantom{\frac{i \Omega_{c4}^{*}}{2}} \\ 
        \Gamma_{t}^{_{(b)}} \!+\! \Gamma_{b}
        \vphantom{\frac{i \Omega_{c4}^{*}}{2}} \\
        \Gamma_{t}^{_{(c)}} \!+\! \Gamma_{c}
        \vphantom{\frac{i \Omega_{c4}^{*}}{2}} \\
        -\frac{i \Omega_{c4}}{2}
        \vphantom{\frac{i \Omega_{c4}^{*}}{2}} \\
        \frac{i \Omega_{c4}^{*}}{2}
        \vphantom{\frac{i \Omega_{c4}^{*}}{2}}    
    \end{pmatrix}
\label{DressedAtomSteady}    
\end{equation}
\noindent where $\mathcal{T}_{ij} = \mathcal{K}_{ij}-\Gamma_{j}$ (for $i$ and $j$ in $\left \{ a,b,c \right \}$). Moreover, if we suppose that $\Gamma_{1} = \Gamma_{2} = \Gamma$ (which$\,$is$\,$a$\,$reasonable$\,$assumption$\,$as $\Gamma_{1}/\Gamma_{2} \simeq  0.95$), the$\,$matrix$\,\mathcal{T}\,$becomes$\,$symmetric$\,$since: $\mathcal{T}_{ij} \simeq -\frac{1}{2} \, \left(|\alpha_{i}|^{2}+|\beta_{i}|^{2}\right) \left(|\alpha_{j}|^{2}+|\beta_{j}|^{2} \right) \Gamma$. In that case, $\chi_{c4}$ reads as follows:
\begin{equation}
    \chi_{c4} = \frac{2 N}{\epsilon_{0} \mathcal{E}_{f}} \, \mu_{c4} \, \rho_{4c} \simeq \frac{\alpha_{4c}(0)}{\omega_{4c}/c} \frac{i-2 \left( \frac{\Delta_{f}-\delta_{c}}{\overline{\gamma}_{c}+\overline{\gamma}_{4}}\right)}{1 + 4\left(\frac{\Delta_{f}-\delta_{c}}{\overline{\gamma}_{c}+\overline{\gamma}_{4}}\right)^{_{\!2}}+\left( \frac{\mathcal{E}_{f}}{\mathcal{E}_{s}} \right)^{_{\!2}}}, 
    \label{SusceptibilityC}
\end{equation}
\noindent where $\alpha_{4c}(0)$ is the linear line-center absorption coefficient and $\mathcal{E}_{s}$ the line-center$\;$saturation field strength, which is defined by: 
\begin{equation}
    \mathcal{E}_{s} = \frac{\hbar}{\mu_{c4}} \sqrt{\overline{\gamma}_{c}+\overline{\gamma}_{4}} \left[\frac{ \overline{\gamma}_{a} \overline{\gamma}_{b} \overline{\gamma}_{c} - 
    \left( \mathcal{T}_{aa} \overline{\gamma}_{b} \overline{\gamma}_{c} +
    \mathcal{T}_{bb} \overline{\gamma}_{a} \overline{\gamma}_{c} +
    \mathcal{T}_{cc} \overline{\gamma}_{a} \overline{\gamma}_{b}
    \right)
    }
    {2 \overline{\gamma}_{a} \overline{\gamma}_{b} +
    \left( \mathcal{T}_{bc} - 2 \mathcal{T}_{bb} \right)\overline{\gamma}_{a} +
    \left( \mathcal{T}_{ac} - 2 \mathcal{T}_{aa} \right)\overline{\gamma}_{b}
    }\right]^{_{\!1/2}}.
    \label{SatFieldC}
\end{equation}
\noindent In equations~\eqref{SusceptibilityC} and~\eqref{SatFieldC}, the total decay rates $\overline{\gamma}_{a}$, $\overline{\gamma}_{b}$, $\overline{\gamma}_{c}$ and $\overline{\gamma}_{4}$ of the dressed-states $\ket{a,n}$, $\ket{b,n}$, $\ket{c,n}\;$and$\;\ket{4,n}$ have been introduced. For $i$ lying in $\left \{ a,b,c \right \}$, $\overline{\gamma}_{i} = K_{i} + \Gamma_{t,d}$ whereas $\overline{\gamma}_{4} = \Gamma + \Gamma_{t,d}$. The off-resonance saturation intensity can finally be expressed$\;$as: $\mathcal{I}_{s}(\Delta_{f}) = \mathcal{I}_{s}(0) \left[1 + 4\left( \Delta_{f}-\delta_{c} \right)^{_{2}} \! / \! \left( \overline{\gamma}_{c}+\overline{\gamma}_{4} \right)^{_{2}} \right]$ where $\mathcal{I}_{s}(0) = \frac{1}{2}\epsilon_{0} c |\mathcal{E}_{s}|^{2}$. The dressed-state formalism provides therefore a value for the linewidth as well as for the saturation intensity of the $\ket{c,n} \rightarrow \ket{4,n}$ transition. At fixed defect power $P_{d}$ and detuning $\Delta_{d}$, equation~\eqref{SusceptibilityC} predicts then how the $\ket{c,n} \rightarrow \ket{4,n}$ transition line saturates and broadens increasing the fluid power, which is far from being obvious looking at the steady-sate solution of~\ref{Bloch4Level}. One can similarly derive equations for $\chi_{a4}$ and $\chi_{b4}$. The total susceptibility $\chi_{_{f,\mathrm{in}}}^{\vphantom{_{(1)}}}$ inside$\;$the defect cross-section is finally given by: $\chi_{_{f,\mathrm{in}}}^{\vphantom{_{(1)}}} = \sum_{i} \chi_{i4}$. The imaginary part of $\chi_{_{f,\mathrm{in}}}^{\vphantom{_{(1)}}}$ has$\,$been plotted on figure~\ref{fig:ComDressed} as function of $\Delta_{f}$, by using the dressed state approach (blue curve) and by directly solving the steady-sate matrix equation derived from~\ref{Bloch4Level} (dashed$\;$line). The defect detuning was set to zero and we use for the computation the same parameters as in figure~\ref{fig:ChangeRefractive2DMaps}. The$\;$agreement between both descriptions is excellent.

\newpage

\subsection{Saturation of $\delta n$}
\label{subsec:SaturationdeltaN}

\noindent In photon fluid experiments, we constantly need to change the fluid power $P_{f}$ in order to tune the nonlinear change of refractive index $\Delta n = n_{2} \, \mathcal{I}_{0}$ and access different flow regimes (superfluid, supercritical, etc). In the present situation, acting on the$\;$fluid$\;$power$\;$does$\;$not only modify the fluid properties but also the defect strength in two different ways:
\vspace{-6pt}
\begin{itemize}
    \item [$\bullet$] A change in the fluid power will first impact the preparation rates by either$\;$increasing or decreasing the number of atoms prepared in the steady-state.
    \vspace{-6pt}
    \item [$\bullet$] Increasing the fluid power can also make the refractive index modulation $\delta n$ quickly saturate and results consequently in a reduction of the defect strength.
\end{itemize}

\begin{figure}[h]
\center
\includegraphics[width=\columnwidth]{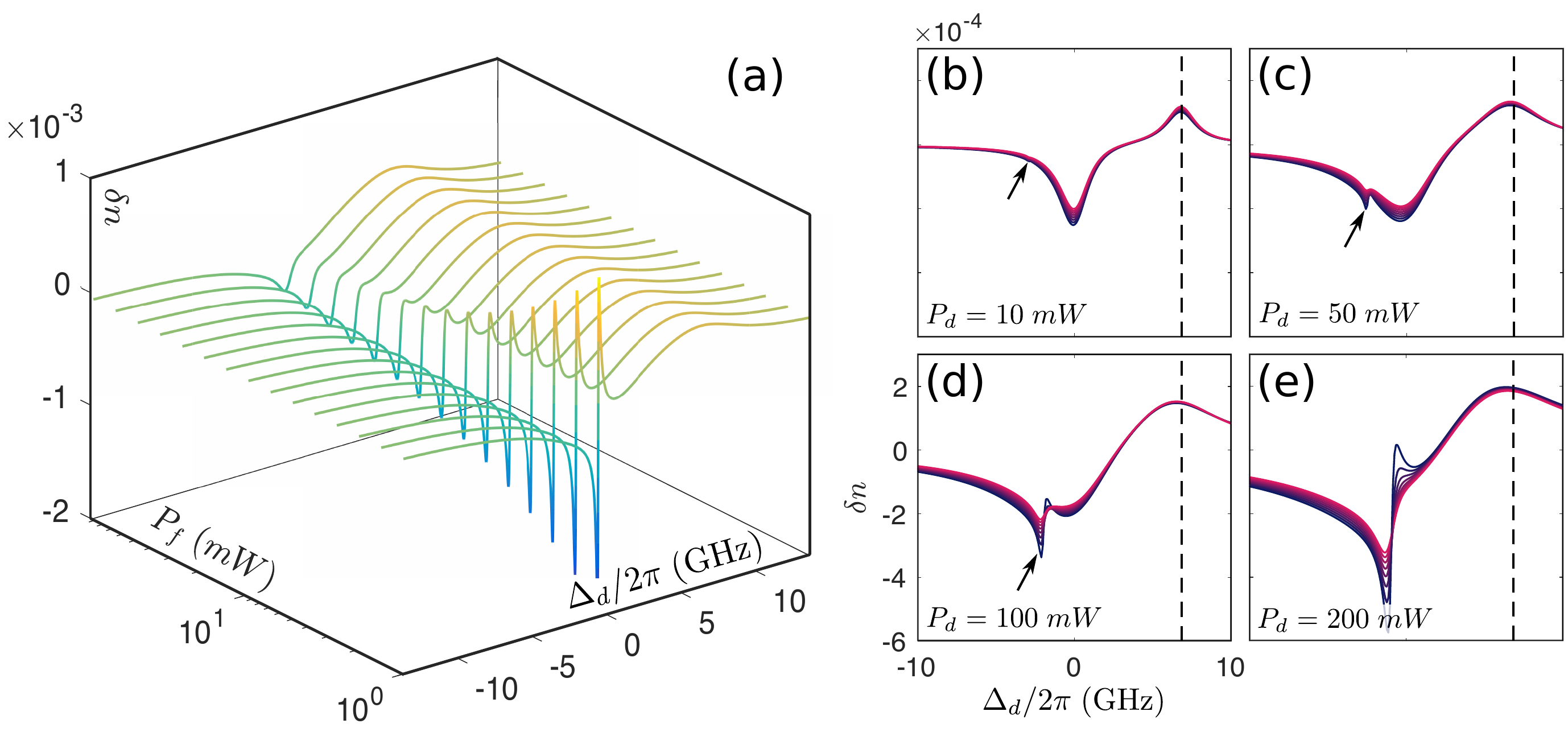} 
\caption{(a) shows the evolution of $\delta n$ as function of the defect detuning $\Delta_{d}$ when the fluid power increases gradually. The saturation by the probe of the $\ket{c,n} \rightarrow \ket{4,n}$ transition makes $\delta n$ strongly depend on $\mathcal{P}_{f}$, close to $\Delta_{d} \simeq - 2\pi \times 1$ GHz. (b)-(e)$\;$show$\;$the same as in (a): $\delta n$ has been plotted as function of $\Delta_{d}$ for different fluid powers, ranging from $25$ (blue) to $200$ mW (magenta). The black arrow indicates the contribution of the $\ket{c, n} \rightarrow \ket{4,n}$ transition line, when it is resonantly driven by the probe field. Parameters: $w_{0,d} = 50$ $\mu$m, $w_{0,f} = 500$ $\mu$m and $T=415$ K. Results obtained for a $^{_{87}}$Rb vapor.}
\label{fig:SaturationMain}. 
\end{figure}

\noindent Both these effects contribute to modify the defect amplitude when the fluid power changes, which is not desirable in photon fluid experiments where the aim is rather to study how a fluid of light flows in a fixed potential landscape $\delta n(\mathrm{r}_{\perp})$ as function of $\Delta n$ in particular. Therefore, the question is: can we change the fluid power while keeping the defect depth (or height, depending on the sign of $\delta n$)$\;$almost$\;$constant? In order to answer this question, we have plotted on figure~\ref{fig:SaturationMain}(a)-(e) the$\,$evolution$\,$of$\,\delta n\,$as$\,$function$\,$of$\,\Delta_{d}\,$and$\,\mathcal{P}_{f}$,$\,$for$\,$different values of $\mathcal{P}_{d}$. In all these plots, the probe detuning is fixed to $\Delta_{f} = -2\pi \times 3$ GHz which is closed to the experimental value. The temperature is $415$ K and the widths of the beams are, as usual, $\omega_{0,d} = 50$ $\mu$m and $\omega_{0,f} = 500$ $\mu$m. Let's first focus on~\ref{fig:SaturationMain}(b). This figure$\;$is obtained for a relatively low defect power, since $\mathcal{P}_{d} = 10$ mW (that is, $\Omega_{d} \simeq 2\pi \times 1$ GHz) in that case. The probe power $\mathcal{P}_{f}$ varies in eight steps from $25$ (blue) to $200$ mW (purple). There are two notable features in this plot.$\,$First,$\,$a dip,$\;$at$\;\Delta_{d} = 0\;$and$\;$second,$\;$a$\;$peak, located$\,$at $\Delta_{d} = \delta_{0}$ (black dashed line). When $\Delta_{d} \simeq0$, atoms are pumped from state$\;\ket{2}\;$to state $\ket{1}$ by the defect beam. The probe beam, red-detuned from the $\ket{2} \rightarrow \ket{4}\,$transition, interacts therefore with a lower density of atoms in state $\ket{2}$ inside than outside the defect cross-section, hence the negative $\delta n$. Reversely, when $\Delta_{d} \simeq \delta_{0}$, atoms are pumped from state $\ket{1}$ to state $\ket{2}$, which induces a positive $\delta n$ this time. Setting the defect detuning to$\;$zero will thus locally generate a repulsive potential in the photon fluid ($\delta n <0$) while setting$\;$it at $\delta_{0}$ will create an attractive (guiding) potential. As you can notice, $\delta n$ does not vary much with the fluid power. The potential height (or depth) will$\;$then$\;$remains$\;$almost constant increasing $P_{f}$, which is precisely what we desire. Figure~\ref{fig:SaturationMain}(c)-(e)$\,$shows$\;$the$\;$same as~\ref{fig:SaturationMain}(b) for higher defect powers (respectively $50$, $100$ and $200$ mW). On~\ref{fig:SaturationMain}(c) and (d), the dip and the peak of~\ref{fig:SaturationMain}(b) are broader but still visible. A new feature, indicated$\;$by$\;$the black arrow, that was barely discernible on~\ref{fig:SaturationMain}(b), starts developing as $P_{d}$ increases, and move from left to right. This local variation of $\delta n$ arises when $\Delta_{d}$ satisfies $\delta_{c}(\Delta_{d}, P_{d}) = \Delta_{f}$. 
When this condition is fulfilled, the defect beam brings the $\ket{c,n} \rightarrow \ket{4,n}$ transition at resonance with the probe field, by shifting the energy of the dressed state $\ket{c,n}$ in such a way that the probe detuning $\Delta_{f}$ is compensated (see figure~\ref{fig:DressedStateSchematic}(a)). When $P_{d}\;$increases, the Autler-Town splitting between the two dressed states $\ket{b,n}$ and $\ket{c,n}$ becomes larger. 

\noindent In other words, the energy shift $\delta_{c}(\Delta_{d}, \mathcal{P}_{d})$, increases with $\mathcal{P}_{d}$. Therefore,$\,$the$\,$point$\,$at$\,$which the$\,$horizontal$\,$line$\,$defined$\,$by$\,\Delta_{f} = - 2\pi \times 3$ GHz$\,$crosses$\,\delta_{c}\,$on$\,$figure~\ref{fig:LinearChangeRefractive}(a) shifts$\,$to$\,$the$\,$right, explaining$\;$why the detuning at which $\delta_{c}(\Delta_{d}, \mathcal{P}_{d}) = \Delta_{f}$ is blue-shifted when $\mathcal{P}_{d}$ increases. Moreover, the contribution of the $\ket{c,n} \rightarrow \ket{4,n}$ transition line to the inside susceptibility $\chi_{_{f,\mathrm{in}}}^{\vphantom{_{(1)}}}$ seems to get larger and larger when the defect$\,$power$\,$steps$\,$up.$\;$When$\,\mathcal{P}_{d} = 200\,$mW$\,$(e), we might be tempted to set the defect detuning where the dip is the deepest.$\;$Nevertheless, at this detuning ($\Delta_{d} \simeq - 2\pi \times 1$ GHz), the index modulation $\delta n$ varies pretty$\;$fast$\;$with$\;$the probe power: from $\mathcal{P}_{f} = 25$ mW (blue) to $\mathcal{P}_{f} = 200$ mW (magenta), $\delta n$ decreases by half, because of the saturation of the $\ket{c,n} \rightarrow \ket{4,n}$ transition. Figure~\ref{fig:SaturationMain}(a) shows this effect more in detail, for the same parameters as in~\ref{fig:SaturationMain}(e).$\;$Let's$\,$summarize$\,$all$\,$that$\,$has$\,$been$\,$said. If we look for generating a repulsive potential in the photon fluid, we can either: 

\newpage

\begin{itemize}
    \item [(1)] drive resonantly the $\ket{2} \rightarrow \ket{3}$ transition with the defect field while keeping its power sufficiently low in order to neglect the contribution of the $\ket{c,n} \rightarrow \ket{4,n}$ transition, as in figures~\ref{fig:SaturationMain}(b) and (c). In that case, the physical mechanism underlying the variation of refractive index at the defect position$\;$is$\;$optical$\;$pumping$\;$from$\,\ket{2}\,$to$\,\ket{1}$. This$\,$configuration$\,$has$\,$the$\,$benefit$\,$of$\,$reducing$\,$the$\,$variations$\;$of$\,\delta n\,$with$\,$the$\,$fluid$\,$power. However, as the $\ket{2} \rightarrow \ket{3}$ transition is not totally saturated in that case, absorption of the defect beam along propagation is expected. 
\end{itemize}
\begin{itemize}
    \item [(2)] completely saturate the $D_{1}$ line and use the large contribution of the $\ket{c,n} \rightarrow \ket{4,n}$ transition when it is resonantly driven by the probe field, that is, when $\Delta_{d}$ satisfies the condition $\delta_{c}(\Delta_{d}, P_{d}) = \Delta_{f}$, as  in figure~\ref{fig:SaturationMain}(e). In that case, variation in the fluid power is not suitable as it will also change the height of the potential. Nevertheless, low absorption and self-effects (focusing, defocusing) on the defect field are expected since the $D_{1}$ line is highly saturated this time.   
\end{itemize}

\subsection{Other routes ?}

\noindent As mentioned in the introduction of this section,$\,$several$\,$other$\,$routes$\,$have$\,$been$\,$investigated to generate local changes of refractive index in rubidium vapors. We should first mention the work of Truscott \textit{et al.}~\cite{6-6Truscott}, in which the guiding$\,$of$\,$a$\,$probe$\,$beam$\,$in$\,$a$\,$donuts-shaped all-optical wave-guide is demonstrated. Truscott and co-workers used the optical pumping between the $D$-lines ground states of rubidium to generate their wave-guide, just$\,$as$\,$we$\,$do.
Their scheme has$\,$been$\,$analysed$\,$in$\,$the$\,$works$\,$of$\,$Kapoor$\,$\textit{et$\,$al.}~\cite{6-12Kapoor}$\,$and$\,$Andersen$\,$\textit{et al.}~\cite{6-13Andersen}, where the optical Bloch equations of the $5$-level system on figure~\ref{fig:4and5Level} are solved$\,$numerically. In this section, we$\,$improve$\,$their$\,$theoretical$\,$description$\,$using$\,$the$\,$dressed-state$\,$formalism. 
\vspace{6pt}
\newline
\noindent More recently, all-optical wave-guiding has been reported$\,$in$\,$warm$\,$atomic$\,$vapors$\,$by$\,$driving a$\,$Raman$\,$transition$\,$off-resonantly~\cite{6-15Vudyasetu}.$\;$This$\,$technique$\,$can$\,$be$\,$used$\,$to$\,$enhance$\,$the$\,$efficiency of nonlinear processes at very low light intensities~\cite{6-14Lukin}. Other schemes taking advantage of the coherences between the atomic levels can be used to induce refractive index changes. Image guiding as for instance be reported using electromagnetically induced transparency (EIT)$\,$in$\,$a$\,$lambda$\,$and$\,$double$\,$lambda$\,$systems~\cite{6-16Firstenberg}.$\;$Following$\,$the$\,$work$\,$of$\,$Sheng$\,$\textit{et$\,$al.}~\cite{6-7Sheng}, Silva and co-workers have investigated the possibility of observing superfluid flows of light in a four-level N-type atomic system~\cite{6-17Silva}. Three$\,$lasers$\,$are$\,$involved$\,$in$\,$this$\,$scheme,$\,$referred to as the probe, the control and$\,$the$\,$switching$\,$fields$\,$in~\cite{6-17Silva}.$\;$The$\,$probe$\,$and$\,$the$\,$control$\,$drive the$\,\ket{1}\rightarrow\ket{3}\,$and$\,\ket{2}\rightarrow\ket{3}\,$transitions $\,$respectively$\,$(EIT$\,$configuration)$\,$while$\,$the switching field adresses the $\ket{2}\rightarrow\ket{4}$ transition. The level schematic is the same as on figure~\ref{fig:4and5Level}(a). The fluid of light is formed by the probe beam here.$\;$The$\,$possibilities$\,$offered$\,$by$\,$this$\,$system are manifold, as it simultaneously allows to control the strength of the$\,$nonlinear$\,$interaction experienced by the probe and to imprint refractive index changes using the switching field. This configuration represents thus a versatile alternative to optical pumping schemes.  

\newpage
    
\section{Optical defect shaping with a Spatial Light Modulator}
    
\noindent In the previous section, we have seen how to induce a potential on the photon fluid forming the probe beam (tuned to the $D_{2}$ line), by driving the $D_{1}$ line with an intense and spatially localized defect field. Up to now, we have put aside the problem of propagation inside the rubidium cell. We only studied the refractive index modulation created over the transverse plane at a given position $z$ on the optical axis. At this position, the defect cross-section as well as the powers of the beams are fixed. Reality is more complex, as we should ideally maintain the potential height (or depth) all along the propagation inside the cell, in order for this potential to be "stationary" with respect$\;$to$\;$the$\;$effective$\;$time$\;\tau = z n_{0}/c$. We$\;$must therefore face issues such as the defect beam absorption $-$ that$\,$will$\,$make$\,$the$\,$local index modulation $\delta n$ vary along $z$ $-$ or its$\;$collimation$\;$over$\;$the$\;$whole$\;$propagation inside the rubidium cell. The question is thus: what are the experimental requirements to fulfill, regarding the probe and the defect beams, in order for the potential to be "stationary"? I will answer this question from a theoretical point of view first, before explaining in a second step how these requirements are implemented in our experiments.   
    
    \subsection{Experimental requirements}
    
\noindent Our primary goal is to locally generate a potential in the photon fluid (that is, an obstacle) in order to study$\,$how$\,$the$\,$fluid$\,$will$\,$flow$\,$around.$\;$In$\,$the$\,$ideal$\,$case,$\,$this$\,$obstacle$\,$should$\,$remain unchanged over the propagation, as mentioned above. This basically suggests two things. First, the defect beam creating the obstacle has to be collimated.$\;$In$\;$other$\;$words,$\,$the$\;$defect cross-section should only slightly changes with the propagation distance.$\;$Second,$\,$the$\,$defect power has to remain (almost) constant along $z$, that is, the absorption of the defect beam should be negligible or somehow compensated over the propagation. Moreover, as we saw in the previous section, increasing the fluid power may also reduce the obstacle amplitude. For the sake of simplicity, I will suppose that the fluid frequency is sufficiently red-detuned from the $\ket{2} \rightarrow \ket{4}$ transition, so that the fluid transmission thought the cell is above$\;70\%$. For$\;$example, let's consider a $1$ mm diameter beam propagating inside a 2.5 cm cell, filled with$\,$a$\,$pure$\,$vapor$\,$of$\,$rubidium$\,$87. In that case, the beam transmission at $415$ K ranges from $83\%$ ($\mathcal{P}_{f} = 1$ $\mu$W) to $89\%$ ($\mathcal{P}_{f} =0.5\;$W), when the laser frequency is $3$ GHz red-detuned from the $\ket{2} \rightarrow \ket{4}$ transition. In such situations,$\,$we$\,$can$\,$neglect$\,$the$\,$dependence$\,$of$\,\delta n\,$on$\,\mathcal{P}_{f}$. Let's therefore focus on the crux of the issue: the defect beam collimation and absorption.

\subsubsection{Defect beam collimation}

\noindent The first thing that comes to mind is to focus a Gaussian beam inside the cell to generate the obstacle. The Rayleigh length $z_{r} = \pi \omega_{0,d}^{2}/\lambda_{d}$ defines, in that case, the typical distance over which the defect beam is collimated. More precisely, it is the distance from the waist at which the beam radius increases by a factor $\sqrt{2}$ (in intensity). If we want to collimate the defect over $2.5$ cm (which is the length of the cell we use in experiments), the minimum width $\omega_{0,\mathrm{lim}}$ we can reach is about $\sqrt{\lambda L/\pi} \simeq 80$ $\mu$m. Using the value of the nonlinear index of refraction $n_{2}$ reported in 2.3.2 iii ($n_{2} \simeq 2.1 \!\times\! 10^{-10}$ m$^{2}$/W),$\,$a$\,$2$\,$mm$\,$diameter$\,$probe$\,$beam of $200$ mW induces an on-axis nonlinear change of refractive index $\Delta n$ of nearly $2.6 \!\times\! 10^{-5}$. The healing length $\xi \sim 1/ k \sqrt{\Delta n}$ is then about 25 $\mu$m, which is less than a third of $\omega_{0,\mathrm{lim}}$.  

\newpage

\noindent As we shall see in chapter 6, probing the superfluidity requires a defect that is comparable in size to $\xi$. Focusing a Gaussian beam inside the cell to generate the obstacle is therefore not suitable for our application. Instead, one can think about using non-diffracting beams such as Airy (parabolic)~\cite{6-19Bandres} or zero-order Bessel beams. The latter have been introduced by Durnin \textit{et al.}~\cite{6-18Durnin}$\,$and result from the interference$\,$of$\,$an$\,$infinite$\,$number$\,$of$\,$plane$\,$waves whose relative wave-vectors constitute the generating lines of the so-called Bessel cone. The radial intensity distribution of zero-order Bessel beams is described by the zero-order Bessel function of the first kind: a high intensity central peak is surrounded by an infinite number of concentric rings of decreasing intensity. Although perfect Bessel beams are only mathematical objects (as they will carry an infinite energy otherwise), spatially limited (or quasi-) Bessel beams can be realized experimentally. Those beams have found various applications$\,-\,$in$\,$optical$\,$trapping~\cite{6-20Civzmar,6-21Garces},$\,$laser$\,$machining~\cite{6-22Courvoisier},$\,$nonlinear$\,$optics~\cite{6-24Porras,6-25Polesana} and imaging~\cite{6-26Dufour, 6-27Zhao} for instance $-$ as their$\,$central$\,$cores$\,$stay$\,$collimated$\,$on$\,$a$\,$distance$\,$that is orders of magnitude longer than the Rayleigh length $z_{r}$. The diffraction-free feature of zeroth-order Bessel beams$\,$make$\,$them$\,$attractive$\,$for$\,$our$\,$application.$\;$They$\,$have$\,$already$\,$been used in~\cite{3-3Michel} to produce non-diffracting obstacle in superfluid light. 

\begin{figure}[h]
\center
\includegraphics[width=\columnwidth]{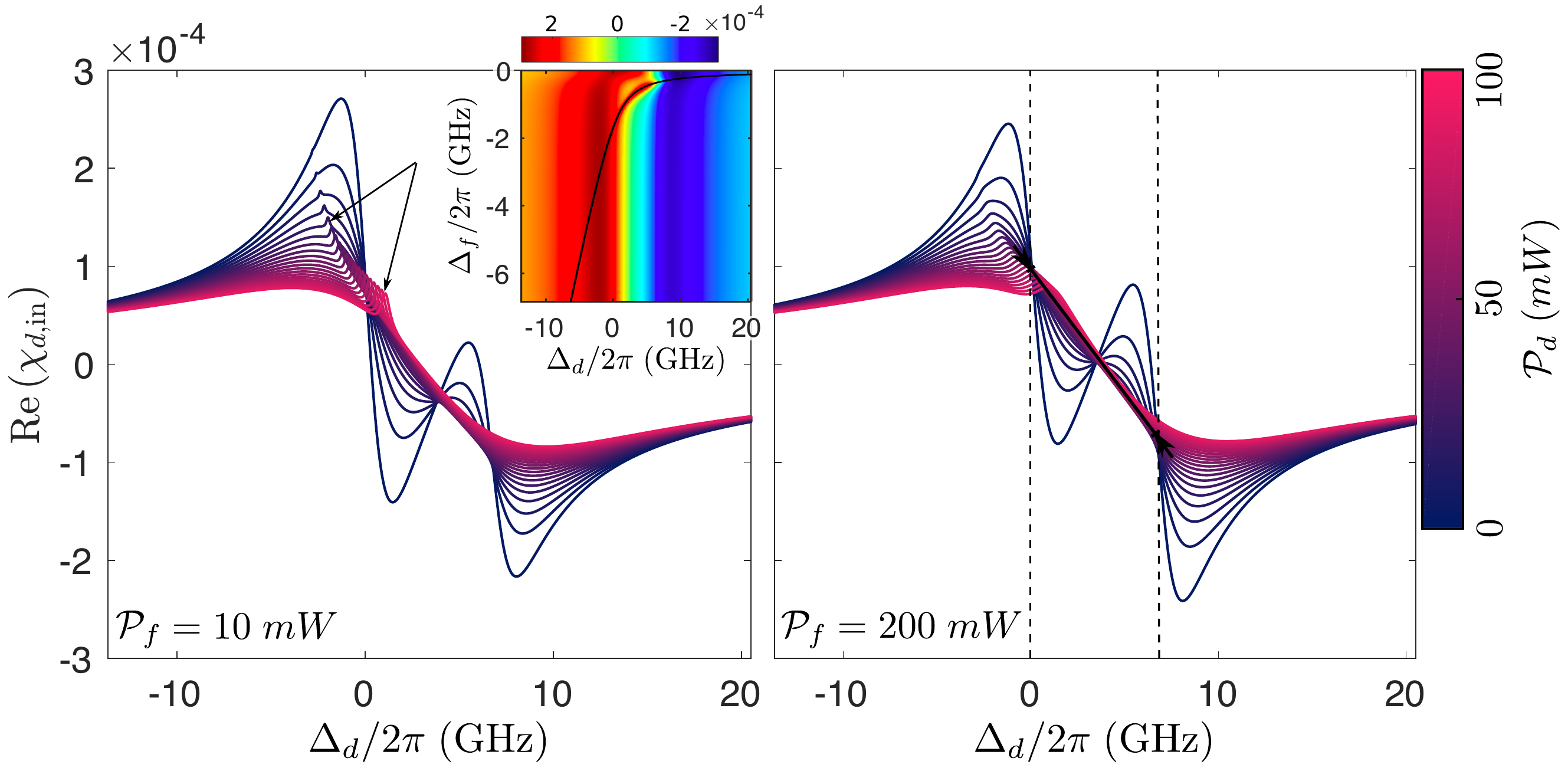} 
\caption{Real part of $\chi_{d, \mathrm{in}}$ as function of $\Delta_{d}$ for different defect powers,$\;$which$\;$varies$\;$in twenty steps from 5 (blue) to 100 mW (magenta). Left: $\mathcal{P}_{f} = 10$ mW.$\;$When$\,$the$\,$probe$\,$is$\,$at resonance with the $\ket{c,n} \rightarrow \ket{4,n}$ transition (that is, when $\Delta_{d}$ satisfies $\delta_{c}(\Delta_{d}, \mathcal{P}_{d}) = \Delta_{f}$), it slightly affects the inside susceptibility $\chi_{d, \mathrm{in}}$, creating peaks in the spectra of $\mathrm{Re} \left[ \chi_{d, \mathrm{in}} \right]$ (black arrows). These peaks move from left to right when $\mathcal{P}_{d}$ increases. The inset shows the variation of $\mathrm{Re} \left[ \chi_{d, \mathrm{in}} \right]$ with $\Delta_{d}$ and $\Delta_{f}$. Right: $\mathcal{P}_{f} = 200$ mW. The peaks are still visible but broader (because of the power broadening of the $\ket{c,n} \rightarrow \ket{4,n}$ transition line). There are points on both graphs at which $\mathrm{Re} \left[ \chi_{d, \mathrm{in}} \right]$ does not depend much on the defect power $\mathcal{P}_{d}$; they are located at $\Delta_{d} = 0$, $\Delta_{d} = \delta_{0}/2$ and $\Delta_{d} = \delta_{0}$. At these detunings, self-effects on the defect beam are expected to be small. The parameters are: $w_{0,d} = 25\;\mu$m, $w_{0,f} = 500\;\mu$m, $T=415$ K and $\Delta_{f} = -2 \pi \!\times\! 3$ GHz. Results obtained for a $^{_{87}}$Rb vapor.} 
\label{fig:SelfDefocusingDefect}.  
\end{figure}

\newpage

\noindent However, Bessel beams $-$ such as Gaussian ones $-$ are sensitive to self-effects. For$\;$instance, a radial broadening (resp. compression) of the Bessel beam central peak is expected along its propagation in self-defocusing (resp. self-focusing) mediums~\cite{6-23Johannisson}. This is problematic as it means that the width $\omega_{0,d}$ of the defect beam will change along $z$. In order to know$\;$the defect detunings at which self-effects are the weakest, the real$\,$part$\,$of$\,$the$\,$susceptibility$\,\chi_{d,in}$ has been plotted on figure~\ref{fig:SelfDefocusingDefect} as function of $\Delta_{d}$ for different defect and probe powers. This quantity characterizes the optical response of the rubidium vapor to the defect field. It also defines the refractive index felt by the defect: $n_{d} = 1+\frac{1}{2} \mathrm{Re} \left[ \chi_{d,in} \right]$. When$\;$the$\;$latter does not depend much on the defect intensity, self-effects are expected$\;$to$\;$be$\;$small.$\;$Indeed, self-focusing or -defocusing arise when the refractive index depends on the beam$\,$intensity. Figure~\ref{fig:SelfDefocusingDefect} shows$\,$the$\,$spectra$\,$of$\,\mathrm{Re} \left[ \chi_{d,in} \right]\,$at$\,$various$\,$defect$\,$powers,$\,$ranging$\,$from$\,\mathcal{P}_{d} = 5\,$mW (blue) to $\mathcal{P}_{d} = 100$ mW (magenta), for $\mathcal{P}_{f} = 10$ mW (left) as well as $\mathcal{P}_{f} = 200\,$mW$\,$(right). The simulation has been performed for $\omega_{0,d} = 25$ $\mu$m (comparable to $\xi$), $\omega_{0,f} = 500$ $\mu$m  and $T=415$ K, as usual. At low defect power, the $\ket{1} \rightarrow \ket{3}$ and $\ket{2} \rightarrow \ket{3}\;$transition$\;$lines are both discernible (as in figure 1.10), but not anymore at high power, where$\;$the$\;D_{1}\;$line$\;$is completely saturated. As you may have noticed, there are values of $\Delta_{d}$ for which $\mathrm{Re} \left[ \chi_{d,in} \right]$ does not evolve much with the defect power. They are located at $\Delta_{d} = 0$, $\Delta_{d} \simeq \delta_{0}/2$ and $\Delta_{d} = \delta_{0}$ respectively. The fact that $\mathrm{Re} \left[ \chi_{d,in} \right]$ (and, by extension, $n_{d}$) does not depend on $\mathcal{P}_{d}$ at $\Delta_{d} = 0$ is particularly interesting, since the best obstacle $-$ at low defect power $-$ is$\,$precisely$\,$obtained$\,$at$\,$this$\,$detuning (see figure~\ref{fig:SaturationMain}(b)$\,$for$\,$example).$\,$Apart$\,$from$\,$the$\,$specific values of $\Delta_{d}$ mentioned above, it$\,$seems hard to$\,$make$\,n_{d}\,$not$\,$dependent$\,$on$\,\mathcal{P}_{d}$.$\;$One$\,$can$\,$think about over-saturating the $D_{1}$ line with the defect, that is, making the on-axis Bessel beam intensity order of magnitude bigger than the saturation intensity. That$\,$way,$\;$the$\,$Bessel$\,$core will roughly speaking see an an uniform refractive index at positions for which the light intensity is higher than the saturation intensity. This is of course a very simplified$\;$picture. As far as I know, there is no reference in the literature studying the impact of saturation on the propagation of Bessel beams. In practice, we do saturate the vapor$\;$and$\;$it$\;$does$\;$not seem to really affect the beam shape (in the cell output plane at least). Moreover, we$\;$have noticed that the best collimation is obtained when the defect is several hundreds of megahertz blue-detuned from the $\ket{2} \rightarrow \ket{4}$ transition. In that case, the variation in the width of the Bessel central peak is less than $10\%$ between$\,$the$\,$input$\,$and$\,$the$\,$output$\,$plane$\,$of$\,$the$\,$cell.

\subsubsection{Defect beam absorption}

\begin{figure}[h]
\center
\includegraphics[width=\columnwidth]{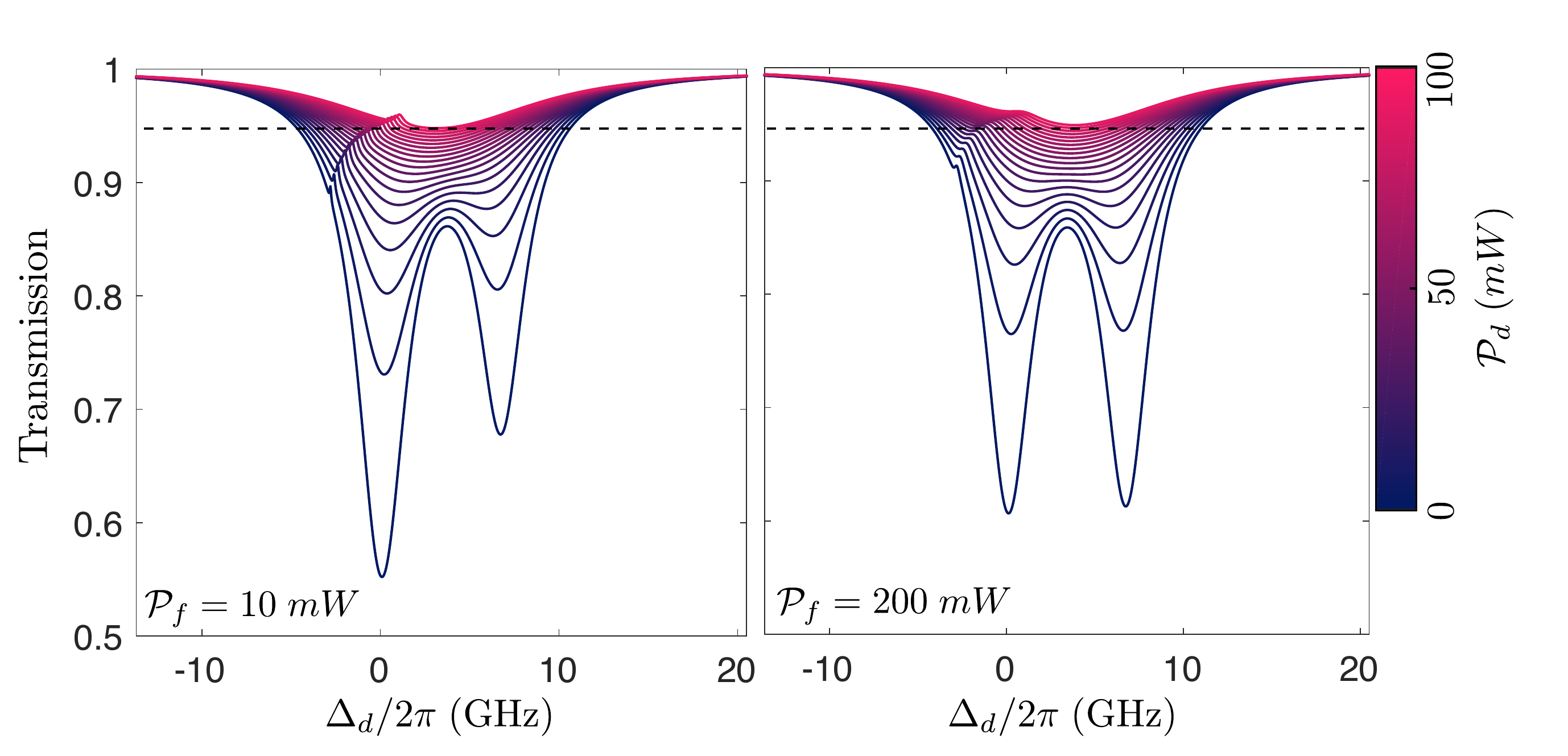} 
\caption{Transmission of the defect beam through a 2.5 cm long cell$\;$as$\;$function$\;$of$\;\Delta_{d}$, for different defect powers. $\mathcal{P}_{d}$ increases$\;$in$\,$twenty$\,$steps$\,$from$\,$5$\,$(blue)$\,$to$\,100\,$mW$\,$(magenta). Left: $\mathcal{P}_{f} = 10$ mW. Right: $\mathcal{P}_{f} = 200$ mW. The same peaks as in figure~\ref{fig:SelfDefocusingDefect} are visible on both graphs. By driving resonantly the $\ket{c,n} \rightarrow \ket{4,n}$ transition when $\delta_{c}(\Delta_{d}, \mathcal{P}_{d}) = \Delta_{f}$, the probe opens a small transparency windows for the defect field. At high defect power, the saturation of the $D_{1}$ line leads to transmissions above $90\%$. In that case, we can safely neglect the absorption of the defect beam. However, at low defect power, absorption is$\;$not negligible anymore and should somehow be compensated (see next subsection for details). The parameters used here are the same as for figure~\ref{fig:SelfDefocusingDefect}. }
\label{fig:SaturationTransmission}. 
\end{figure}

\noindent Another issue to face is the absorption of the defect light inside the vapor cell. Since the defect is tuned almost resonantly to the $D_{1}$ line, strong absorption is expected, at least$\;$at small defect powers. This is indeed what we can see on figure~\ref{fig:SaturationTransmission}, where the transmission $t = \exp(-\alpha_{d}L)\,$has been plotted as function of$\,\Delta_{d}$, for defect powers ranging from$\,5\,$mW (blue) to $100$ mW (magenta). On the left, $\mathcal{P}_{f} = 10$ mW$\;$while$\;$on$\;$the$\;$right, $\mathcal{P}_{f} = 200$ mW. The vapor cell is $2.5$ cm long here. The other parameters are the same as for figure~\ref{fig:SelfDefocusingDefect}. The absorption coefficient $\alpha_{d}$ is defined by: $\alpha_{d} = k_{d} \, \mathrm{Im} \left[ \chi_{d,in} \right]$, where $k_{d}$ is the wave-vector of the defect field. As you can see, absorption is not negligible for $\Delta_{d} = 0$ and $\Delta_{d} = \delta_{0}\;$at low defect powers (and can be even stronger decreasing $\mathcal{P}_{d}$ below $5$ mW). If we choose not to fully saturate the $D_{1}$ line, in order to only take$\;$benefit$\;$of$\;$optical$\;$pumping$\;$between ground states to generate the obstacle (as for example in~\ref{fig:SaturationMain}(b) and (c)), we should then somehow compensated the defect absorption over propagation. A method $-$ based on the on-axis shaping of the Bessel beam intensity profile $-$ is presented to this end in the next subsection. Reversely, if we choose to work in the saturating regime, that is, a high defect powers, absorption is not anymore an issue, since transmission rapidly grows above $90 \%$. I would like to conclude this paragraph explaining the origin of the small amplitude peaks you may have noticed$\,$on$\,$the$\,$transmission$\,$spectra$\,$(on$\,$the$\,$slope$\,$of$\,$the$\,$left$\,$transmission$\,$dip). This small "transparency window" is opened by the probe$\,$field$\,$when$\,$it$\,$resonantly$\,$drives$\,$the $\ket{c,n} \rightarrow \ket{4,n}$ transition (that is, when $\Delta_{d}$ fulfills $\delta_{c}(\Delta_{d}, \mathcal{P}_{d}) = \Delta_{f}$). That is why the peaks position moves to the right when $\mathcal{P}_{d}$ steps up (see~\ref{subsec:SaturationdeltaN}). The width of the peaks increases with the fluid power (because of the power broadening of the $\ket{c,n} \rightarrow \ket{4,n}\,$transition$\;$line). The same features are also visible on the spectra of figure~\ref{fig:SelfDefocusingDefect} (black arrows).

\vspace{-20pt} 
\subsection{Absorption-compensated Bessel Beam}
    \label{subsec:AbsCompBessel} 
    
\noindent As we have seen in the previous subsection, absorption of low power defect beams prevents the obstacle amplitude from being constant along the propagation inside the vapor cell. We therefore started asking ourselves if it was somehow possible to compensate absorption $-$ at least along the optical axis $-$ by shaping the Bessel beam on-axis intensity profile. This questioning goes well beyond the framework of obstacle generation in photon fluids, since absorption (or diffusion) is the main limitation in a wide range of optical applications.

\newpage

\noindent In bio imaging for instance, light-sheet microscopy $-$ which allows selective illumination$\;$of tissues and fast 3D$\,$imaging$\,$of$\,$live$\,$organisms$\,$at$\,$the$\,$cellular$\,$scale~\cite{6-28Huisken}$\,-\,$is$\,$until$\,$now$\,$limited by the field$\,$of$\,$view,$\,$that$\,$is,$\,$the$\,$penetration$\,$depth$\,$of$\,$the$\,$illuminating$\,$beam$\,$inside$\,$the$\,$sample. Tissues$\,$are$\,$highly$\,$diffusive$\,$optical$\,$mediums.$\,$Imaging$\,$through$\,$is$\,$therefore$\,$a$\,$challenging$\,$task. Selective plane illumination microscopy could thus also benefit from the development of attenuation-resistant$\,$Bessel$\,$beams,$\,$that$\,$would$\,$have$\,$a$\,$significantly$\,$better$\,$penetration$\,$depth. In the literature, the generation of such beams with exponential intensity axicons~\cite{6-29Golub} and with computer generated holograms~\cite{6-30Zamboni} has$\,$been$\,$reported. In$\,$both$\,$cases,$\;$the$\,$Bessel$\,$beam on-axis intensity is tailored so as to exponentially increase over the optical axis and exactly counterbalance the exponential decay $\exp(-\alpha z)$ of the Beer-Lambert law.$\;$The$\,$coefficient$\,\alpha$ is the linear attenuation$\,$coefficient,$\,$describing$\,$either$\,$diffusion$\,$or$\,$absorption$\,$losses$\,$inside$\,$the optical medium. In~\cite{6-30Zamboni}, the attenuation-resistant light field (referred$\,$to$\,$as$\,$"frozen$\,$waves") results from the superposition of equal frequency Bessel beams, generated by modulating only the amplitude of an incident plane-wave, using a Spatial Light Modulator (or SLM). In this subsection, I$\,$present$\,$a$\,$more$\,$versatile$\,$method,$\,$based$\,$on$\,$both$\,$the$\,$phase$\,$and$\,$amplitude shaping of an incident Gaussian beam~\cite{6-32Bolduc,6-33Ouadghiri}. It allows compensating attenuation up to $\alpha = 200$~m$^{-1}$ $-$ whatever the loss mechanism $-$ by using real space$\,$shaping$\,$with$\,$a$\,$reflective phase-only SLM. However, it is worth noting that this techniques works for compensating \textbf{linear attenuation} only. We therefore assume $\alpha$ does not depend on the beam$\;$intensity. In the following paragraphs, I will first present the theoretical background on which the method is based, before showing the results we obtained using absorbing (rubidium vapor) and diffusive$\,$(aqueous$\,$milk suspension)$\,$optical$\,$mediums.$\;$The$\,$content$\,$of$\,$this$\,$subsection$\,$has been published in "Attenuation-free non-diffracting Bessel beams", Optics Express, Vol.$\,$27, Issue 21, pp. 30067-30080 (2019)~\cite{6-3Fontaine}.   

\subsubsection{Shaping Bessel beams on-axis intensity}

At a given position $z_{0}$ on the optical axis (from now on, $z_{0} = 0$), the electric field envelope $\mathcal{E}(x,y,z_{0}=0)$ of the laser beam is related to its spatial spectrum $ \mathcal{S}(k_{x},k_{y},z_{0}=0)$ through:
\begin{equation}
\label{ElectricField_0}
    \mathcal{E}(x,y,0) =  \int_{-\infty}^{\infty} \int_{-\infty}^{\infty}
    \mathcal{S}(k_{x},k_{y},0) \, \exp \! \left[-i(k_{x}x+k_{y}y) \right] \mathrm{d} k_{x} \, \mathrm{d}k_{y}.
\end{equation}
\noindent For a radially symmetric laser beam, equation~\ref{ElectricField_0} can be rewritten as follows~\cite{6-33Ouadghiri}:
\begin{equation}
\label{ElectricField}
    \mathcal{E}(r,0) = \frac{1}{2 \pi} \int_{0}^{\infty} \mathcal{S}(k_{\perp},0) \, J_0(r k_{\perp}) \, k_{\perp} \, \mathrm{d} k_{\perp},
\end{equation}
\noindent where $J_0$ is the zero-order Bessel function of the$\,$first$\,$kind, $r\,$the$\,$transverse$\,$radial$\,$coordinate and $k_{\perp} = \sqrt{k_{x}^{_{2}}+k_{y}^{_{2}}}$ the transverse wave-vector.$\,$The$\,$spatial$\,$spectrum$\,\mathcal{S}(k_{\perp},0)\,$is$\,$the$\,$Hankel transform of the electric field envelope $\mathcal{E}(r,0)$. As$\;$mentioned$\;$in~\cite{6-31Civzmar}, equation~\ref{ElectricField} shows that a radially symmetric field can be regarded as a superposition$\,$of$\,$zero-order$\,$Bessel$\,$fields. Each of these Bessel components propagates without diffracting~\cite{6-18Durnin} as $J_0(r k_{\perp}) \, \exp \! \left[ i k_{z} z \right] $, where $k_{z}  = \sqrt{k_{\tiny{0}}^{_{2}}-k_{\tiny{\perp}}^{_{2}}}$ is the$\,$longitudinal$\,$wave-vector$\,$of$\,$a$\,$given$\,$Bessel$\,$mode.$\;$Consequently, the on-axis electric field at position $z$, $\mathcal{E}(r=0,z)$, can be obtained from \eqref{ElectricField} as:
\begin{equation}
  \label{OnAxisField}
    \mathcal{E}(r=0,z) = \frac{1}{\pi} \int_{0}^{\infty} \mathcal{S} \left( \! \sqrt{k_{0}^2 - k_{z}^2},z=0 \! \right) \, \exp \! \left[i k_{z} z\right] \, k_{z} \, \mathrm{d} k_{z}.    
\end{equation}

\newpage

\noindent In Fourier space, the spatial spectrum of an ideal zero-order Bessel beam is a ring$\;$of$\;$radius $k_{\perp} = k_{0} \sin(\theta)$.$\,$Therefore,$\,$the$\,$light$\,$rays$\,$that$\,$describe$\,$the$\,$beam$\,$propagation$\,$are$\,$distributed over a cone of angle $\theta$. This$\,$angle$\,$sets$\,$the$\,$spot$\,$size, that is, the$\,$full$\,$width$\,$at$\,$half-maximum (FWHM) of the central peak in the transverse intensity profile.$\;$It is equal to $2.27/k_{0} \sin (\theta)$ for an ideal zero-order Bessel beam.$\,$Each$\,$of$\,$the$\,$modes$\,$coming$\,$in$\,$the$\,$spectral$\,$decomposition equation \eqref{ElectricField} propagates in free-space with a slightly different longitudinal wave-vector $k_{z} = k_{0} \cos(\theta)$, as can be seen in \eqref{OnAxisField}. They thus merge with different cone angles and at distinct positions along the optical axis. The on-axis electric field results then$\;$from$\;$the interference arising between these individual modes. If one wants to design a Bessel beam with a given on-axis intensity profile $\mathcal{I}(z) = \vert \mathcal{E}(r=0,z) \vert^{2}$ along$\,$the$\,$optical$\,$axis,$\,$the$\,$spatial spectrum $\mathcal{S}$ must be engineered according to the following formula :
\begin{equation}
\label{Spectrum}
     \mathcal{S}(k_{\perp}, z_{0}=0) \! = \! \frac{1}{k_{z}} \int_{0}^{\infty} \sqrt{\mathcal{I}(z)} \, \exp \left[i(k_{z0}-k_{z})z\right] \, \mathrm{d}z.
\end{equation}
\noindent The spectrum $S$ is centered around the longitudinal wave-vector of the target Bessel beam $k_{z0} = k_{0} \cos(\theta_{0})$. This formula gives a physical insight about the engineering process that will be used$\,$to$\,$compensate$\,$attenuation$\,$along$\,$the$\,z$-axis.$\,$The$\,$initial$\,$electric$\,$field$\,\mathcal{E}(r,z_{0}=0)$ that will produce a Bessel beam with a cone angle $\theta_{0}$ and an on-axis intensity profile $I(z)$ can be evaluated using equations \eqref{ElectricField} and \eqref{Spectrum}. In the following paragraph,$\;$we$\;$briefly describe how to generate the target beam, by shaping$\,$in$\,$real-space$\,$the$\,$phase$\,$and$\,$amplitude of a Gaussian beam with a Spatial Light Modulator (SLM). Fourier space shaping may also be considered~\cite{6-31Civzmar}. Nevertheless, in that case, only the incident light distributed over the thin ring forming the intensity distribution of the Bessel beam in Fourier$\,$space$\,$is$\,$used. A large part of the light in the incident Gaussian beam will thus be filtered out$\,$by$\,$the$\,$SLM. Much higher efficiency can then be obtained using real space shaping technique.   
\vspace{10pt}
\newline
\textbf{(a)} \textbf{Phase and amplitude shaping with a phase-only SLM}
\vspace{10pt}
\newline
\noindent We define $z_{0} = 0$ to be the position of the SLM chip  on the optical axis. By discretizing the electric field according to the SLM matrix ($N_x\times N_y$), the target electric field $\mathcal{E}(i,j,z=0^{+})$ (right after the SLM) can be decomposed in amplitude $\mathcal{A}(i,j)$ and phase $\Phi(i,j)$, where $0  \leq  i  \leq  N_{x}$ and $0  \leq  j  \leq  N_y$ are the pixel coordinates. As suggested by Davis \textit{et al.}~\cite{6-34Davis}, locally reducing the phase wrapping contrast allows for a modulation of the light scattered in the first diffraction order, with a single hologram. We apply this method using a phase-only SLM. The expression of the SLM phase mask $\Psi$ can always by written as follows: 
\begin{equation}
\label{PhaseMask}
     \Psi(i,j) = \mathcal{M}(i,j) \, \mathrm{mod} \left[ \mathcal{F}(i,j) + \Phi_{\mathrm{g}}(i,j), \, 2 \pi \right].
\end{equation}
\noindent The function $\mathcal{F}$ contains the phase information of the target electric field and $\Phi_{\mathrm{g}}$ stands$\;$for the grating phase ramp, used to separate the different diffraction orders in Fourier space. The total phase, $\mathcal{F}+\Phi_{g}$, is wrapped by the modulo operation. The diffraction efficiency$\;$is locally tuned by the modulation function $\mathcal{M}$ ($0 \leq \mathcal{M}(i,j) \leq 1$). The complex amplitude$\;$of the field diffracted in the first order can be expressed as follows~\cite{6-32Bolduc, 6-33Ouadghiri}:
\begin{equation}
\label{FirstDiffraction}
    \mathcal{E}_{1}(i,j,z=0^{+})  = \mathcal{A}_{\mathrm{in}}(i,j) \, \mathrm{sinc} \left[\pi \mathcal{M}(i,j) -\pi \right] \ \exp\left[ i \left( \mathcal{F}(i,j) + \pi \mathcal{M}(i,j) \right) \right], 
\end{equation}
\noindent where $\mathcal{A}_{\mathrm{in}}$ is the amplitude of the incident laser beam on the SLM. By identifying $\mathcal{E}_{1}$ with the target electric field, one can obtain the functions $\mathcal{F}$ and $\mathcal{M}$ solving the system:

\newpage

\begin{align}
\label{System}
        \mathcal{M}(i,j)  &= 1 + \frac{1}{\pi} \, \mathrm{sinc}^{-1} \left( \frac{\mathcal{A}(i,j)}{\mathcal{A}_{\mathrm{in}}(i,j)} \right) ,\\ 
        \mathcal{F}(i,j)  &= \Phi(i,j) - \pi \mathcal{M}(i,j) .
\end{align}
\noindent The inverse sinc function ($\mathrm{sinc}^{-1}$) is defined on $[-\pi,0]$ here. Computing this function at each points of the hologram is usually demanding ($N_{x}  \times  N_{y}$ operations). However, if both the incident and the first order diffracted beams are radially symmetric, we only need to determine$\,$the$\,$radial$\,$profile$\,$of$\,$the$\,$modulation$\,$function.$\,$For$\,$beams$\,$centered$\,$on$\,$the$\,$SLM$\,$chip, equation \eqref{System} can be simplified such that: 
\begin{equation}
\label{Eq1System}
m(i) = 1 + \frac{1}{\pi} \, \mathrm{sinc}^{-1} \left( \frac{\mathcal{A}(i,N_y/2)}{\mathcal{A}_{\mathrm{in}}(i,N_y/2)} \right) ,
\end{equation}
\noindent where $i$ is an integer running from 0 to $N_{x}/2$ (we assume $N_{x} \geq N_{y}$ here). The modulation matrix $\mathcal{M}$ can be fully reconstructed from $m$ using a circular interpolation, which requires computing the inverse $\mathrm{sinc}$ function for $N_{x}/2$ points only instead of $N_{x}\times N_{y}$. In practice, we start by cleaning-up the incident laser beam, that is, by filtering out in Fourier space its high transverse wave-vector components thanks to a small pinhole aperture. The incident Gaussian$\,$beam$\,$is$\,$radially$\,$symmetric$\,$in$\,$the$\,$SLM$\,$plane$\,$afterwards$\,$($\omega_{x,y} \simeq 3.3 \pm 0.1\,$mm).
\vspace{10pt}
\newline
\textbf{(b)} \textbf{Target on-axis intensity profile}
\vspace{10pt}
\newline
\noindent In principle, arbitrary on-axis intensity profiles can be generated using the method above. In this paragraph, I introduce the target profile $\mathcal{I}(z)$ we use to maintain the central peak intensity constant along the propagation in a uniform and linear lossy$\,$medium.$\,$Let$\,L\,$and$\,\alpha$ stand respectively for the propagation length and the linear attenuation coefficient of the optical medium. According to the Beer-Lambert's law, the transmittance $t$ of the medium decays exponentially with the propagation distance:$\;t = \exp(-\alpha z)$.$\;$Therefore,$\,$the$\,$on-axis intensity$\,$should$\,$exponentially$\,$increase$\,$along$\,$the$\,z$-axis,$\,$such$\,$that$\,\mathcal{I}(z) \!\sim\! \exp(\alpha z)$,$\,$in$\,$order$\,$to compensate for losses. We ramp the on-axis intensity up (from 0 to $\mathcal{I}(z_{1}) = \mathcal{I}_{0}$), until the entrance plane position $z_{1}$, before making it exponentially increase over the distance$\;L$. We then bring it back to zero smoothly. The full on-axis target profile we designed can finally be described by the following equation:
\begin{align}
\label{SystemProfile}
\mathcal{I}(z) = 
    \begin{cases}
        \mathcal{I}_{0}  \left[ \frac{\sin(C_{1} z/z_{1})}{\sin(C_{1})} \right]^{2} &\mathrm{if} \;\; 0 \leq z \leq z_{1} \\
        \mathcal{I}_{0} \exp \left[ \alpha  (z-z_{1}) \right] &\mathrm{if} \;\; z_{1} \leq z \leq z_{2} \\
        \mathcal{I}_{\max} \sin^{2} \! \left[ C_{2} + (\frac{\pi}{2} - C_{2})  \frac{z-z_{2}}{z_{3}-z_{2}} \right] \; &\mathrm{if} \;\; z_{2} \leq z \leq z_{3} \\
        \mathcal{I}_{\max}\sin^{2} \! \left[\frac{\pi}{2} \left(1- \frac{z-z_{3}}{z_{4}-z_{3}} \right) \right]  &\mathrm{if} \;\; z_{3} \leq z \leq z_{4} .
    \end{cases}
\end{align}
\noindent For all the measurements we performed, we set $z_{1} \, G^2 = 1.5$ cm, $z_{2} \, G^{2} = z_{1} \, G^{2} + L$ and $z_{4} \, G^{2} = 3 \, z_{1} \, G^{2} + L$, where $G = 0.5$ stands for the telescope demagnification factor which optically conjugates the SLM chip and the plane $z=0$. The constants$\;C_{1}\;$and$\;C_{2}\;$as$\;$well$\;$as $z_{3}$ are chosen in order to make the profile continuous and differentiable$\;$(see~\cite{6-3Fontaine} for$\,$details). 

\newpage

\begin{figure}[h]
\centering
\includegraphics[width=0.8\linewidth]{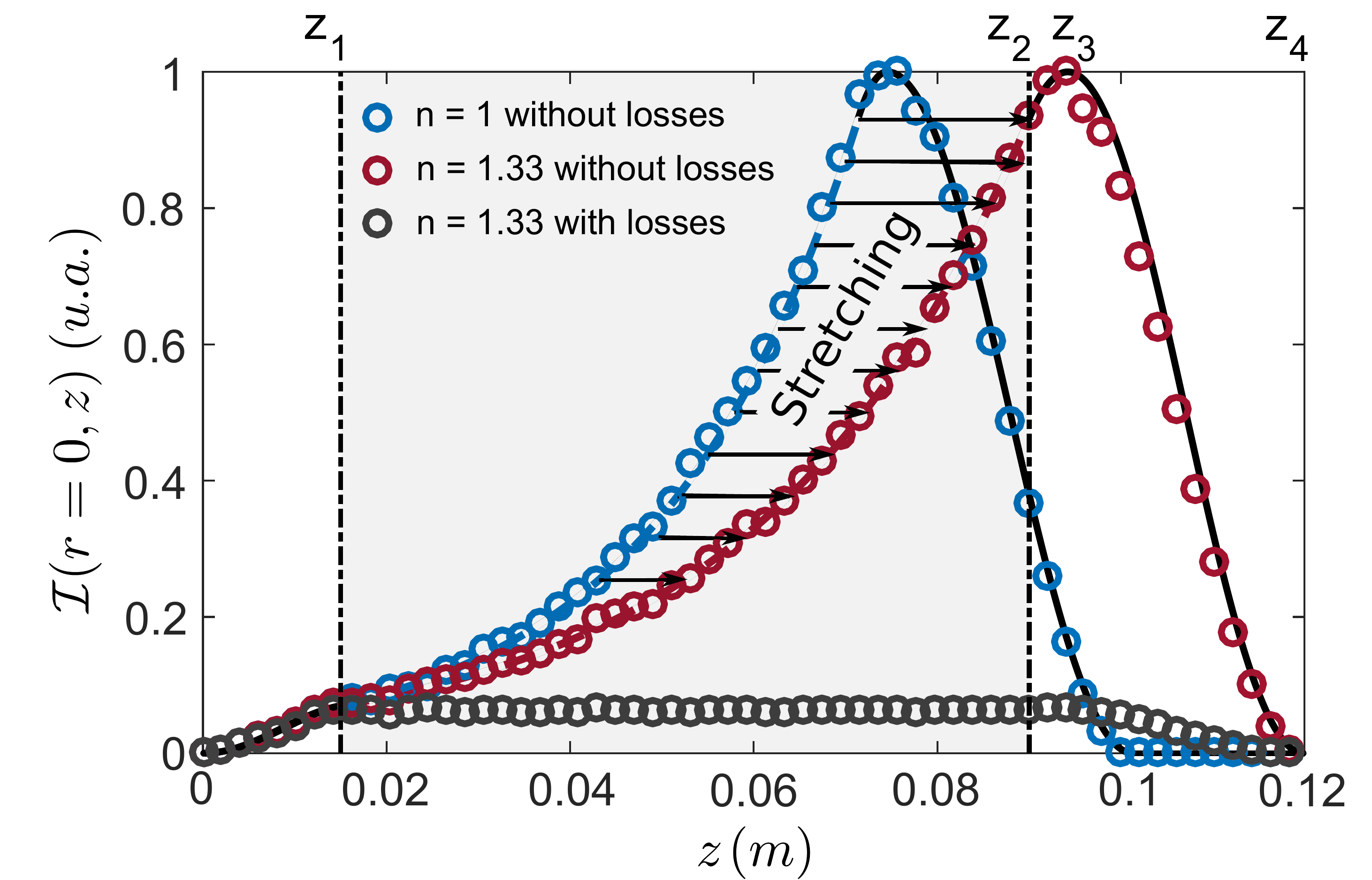}
\caption{Numerical simulation of the longitudinal refractive index stretching of the on-axis target profile. When the medium refractive index $n_{0}$ is not equal to one, the calculated profile must be stretched by $n_{0}$ to compensate for refraction. Blue,$\;$red$\;$and$\;$grey$\;$circles$\;$are obtained by solving numerically the evolution of the Bessel beam in air (blue), in a lossless (red) as well as in a lossy material (grey) of refractive index $n_{0} = 1.33$.$\;$The$\,$simulation$\,$data obtained for $n_{0} \ne 1$ can be deduced from the data obtained for $n_{0} = 1$ by stretching the $z$-axis by a factor $n_{0}$ between $z_{1}$ and $z_{2}$. We adjust the exponentially growing section of the target on-axis profile (blue dotted line) such that the stretched Bessel beam ends up compensating for the good attenuation coefficient $\alpha$. When losses are taken into account, the on-axis intensity remains constant all along the propagation inside the medium.
}
\label{fig:Stretching}
\end{figure}

\noindent In what follows, the target profile has been normalized to 1 dividing $\mathcal{I}(z)$ by the maximum intensity $\mathcal{I}_{\max} = \mathcal{I}_{0} \left[ 1 + \exp(\alpha L) \right]$.$\;$The$\,$spatial$\,$spectrum$\,$associated$\,$to$\,\mathcal{I}(z)\,$has$\,$been$\,$derived analytically in~\cite{6-3Fontaine} using equation~\eqref{Spectrum}. As all the parts composing$\,$the$\,$target$\,$profile$\,$can be$\,$expressed$\,$either$\,$by$\,$an$\,$exponential$\,$rising$\,$function$\,$or$\,$a$\,$sine$\,$square$\,$function,$\,$computing$\,$the spatial spectra associated to the generic functions $\mathcal{I}^{_{i \shortrightarrow j}}_{\,\mathrm{exp}}(z) = \mathcal{I} \exp \left[\alpha (z-z_{i}) \right)]$ as$\;$well$\;$as $\mathcal{I}^{_{i \shortrightarrow j}}_{\,\mathrm{sin}}(z) = \mathcal{I} \, \sin^{2} \left( a \, \frac{z-z_{i}}{z_{j}-z_{i}}+b \right)\,$is$\,$enough. The$\,$derivation$\,$of$\,$the$\,$related$\,$spectra$\;\mathcal{S}^{_{i \shortrightarrow j}}_{\mathrm{sin}}\,$and$\,\mathcal{S}^{_{i \shortrightarrow j}}_{\mathrm{exp}}$ is a bit laborious but straightforward; we only give the final result here: 
\begin{align}
        \mathcal{S}^{_{i \shortrightarrow j}}_{\mathrm{sin}} =& \sqrt{\mathcal{I}} \, \frac{l}{k_{z}}  \left[a \, \frac{\cos(a) - \cos(a+b)}{a^{2}-(\delta k l)^{2}}  
        -i \delta k \frac{\sin(a) \, e^{\,i \delta k z_{i}} - \sin(a+b) \, e^{\,i \delta k z_{j}}}{a^{2}-(\delta k l)^{2}} \right], \\
        \mathcal{S}^{_{i \shortrightarrow j}}_{\mathrm{exp}} =& - \sqrt{\mathcal{I}} \, \frac{2}{k_{z}} \frac{e^{\,i \delta k z_{i}}- \exp \left(\alpha l/2\right) \, e^{\,i \delta k z_{j}}}{\alpha+2i\delta k},
\end{align}
\noindent with $l = z_{j}-z_{i}$ and $\delta k = k_{z0}-k_{z}$.$\;$We obtain$\,$the$\,$spectrum$\,$adding$\,$the$\,$spectral$\,$contributions coming from the different parts of the profile: $\mathcal{S} = \mathcal{S}_{\mathrm{sin}}^{_{0 \shortrightarrow 1}} + \mathcal{S}_{\mathrm{exp}}^{_{1 \shortrightarrow 2}} + \mathcal{S}_{\mathrm{sin}}^{_{2 \shortrightarrow 3}} + \mathcal{S}_{\mathrm{sin}}^{_{3 \shortrightarrow 4}}$.
The target electric$\,$field$\,$is$\,$finally$\,$derived$\,$by$\,$computing$\,$the$\,$inverse$\,$Hankel$\,$transform$\,$of$\,\mathcal{S}\,$with~\eqref{ElectricField}.
\vspace{10pt}
\newline
\textbf{(c)} \textbf{Compensating for the refractive index stretching}
\vspace{10pt}
\newline
\noindent When the linear refractive index $n_{0}$ of the medium is not equal$\,$to$\,$one $-$ as$\,$it$\,$has$\,$implicitly been assumed in the preceding paragraphs $-$ the target$\,$Bessel$\,$beam$\,$will$\,$undergo$\,$refraction at the medium entrance and output planes. From the Snell's$\,$refraction$\,$law,$\,$one$\,$finds$\,$that: $\sin(\theta_{i}) = n_{0} \sin(\theta_{r})$, where $\theta_{i}$ and $\theta_{r}$ stand respectively for the incident and refractive cone angles of a given Bessel mode. By introducing the transverse spatial wave-vector, that is, $k_{\perp} = n_{0} k_{0}  \sin(\theta)$, we easily show that $k_{\perp}^{(i)} = k_{\perp}^{(r)}$. Therefore, according to equation~\eqref{ElectricField}, the transverse shape$\,$of$\,$the$\,$target$\,$Bessel$\,$beam$\,$is$\,$not$\,$modified$\,$by$\,$successive$\,$refractions~\cite{6-35Mugnai}. Nevertheless, the cone angle does change, as$\,$soon$\,$as$\,$the$\,$Bessel$\,$beam$\,$enters$\,$the$\,$medium.
\vspace{6pt}
\newline
\noindent If $n_{0} > 1$, the inner cone angle $\theta_{r}$ is smaller than the external one and the Bessel beam will cover a distance longer than in air. This stretching of the beam inside$\,$the$\,$medium$\,$will necessarily$\,$reduce$\,$the$\,$compensation$\,$coefficient$\,$by$\,$a$\,$factor$\,n_{0}$.$\;$So$\,$as$\,$to$\,$counteract$\,$this$\,$effect, we constrict the exponentially rising part of the target on-axis intensity profile$\,$beforehand by a factor $n_{0}$ (as suggested in~\cite{6-30Zamboni}). In other words, we replace $L$ by$\,L/n_{0}\,$and$\,\alpha\,$by$\,\alpha n_{0}\,$in the second line of equation~\eqref{SystemProfile}. By doing so, the stretching of the beam will compensate exactly the exponential attenuation$\,$in$\,$the$\,$medium,$\,$as$\,$sketched$\,$on$\,$figure~\ref{fig:Stretching}.$\;$Indeed,$\,$using equation\eqref{Spectrum} and the change of variable $z  \rightarrow  \Tilde{z} = n_{0}(z - z_{1})$, one can derive the spectrum $\mathcal{S}_{\mathrm{exp}}^{_{1 \shortrightarrow 2}}$ associated to the exponential rising part of the on-axis profile (between $z_{1}$ and $z_{2}$):

\begin{align}
\label{Spectrum12}
    \mathcal{S}_{\mathrm{exp}}^{_{1 \shortrightarrow 2}} &=  \sqrt{\mathcal{I}_{0}} \, \frac{e^{\, i(k_{z0}-k_{z}) \, z_{1}}}{n_{0} \, k_{z}}  \int_{0}^{L} \exp\left( \alpha \Tilde{z}/2 \right)  \, e^{\, i(k_{z0}-k_{z})\Tilde{z}/n_{0}} \, \mathrm{d}\Tilde{z} \nonumber \\
     & =  - \frac{i}{{n_{0} \, k_{z}}}\frac{\sqrt{\mathcal{I}_{0}} \, e^{\,i(k_{z0}-k_{z}) \, z_{1}}}{\left(k_{z}-k_{z0}\right)\!/n_{0}  +  i\frac{\alpha}{2}}  \left(1-  e^{\,-i\left[\left(k_{z}-k_{z0}\right)/n_{0}+ i \frac{\alpha}{2}\right]L} \right) .
\end{align}
\noindent The on-axis electric field $\mathcal{E}(r=0,z)$ is related to the spectrum $\mathcal{S}$ by the Fourier transform of equation \eqref{OnAxisField}. Using \eqref{Spectrum12} and \eqref{OnAxisField} and the change of variable $\Bar{k}_{z}  =  (k_{z}  -  k_{z0})/n$, we can derive the on-axis electric field $\mathcal{E}_{\mathrm{exp}}^{_{1 \shortrightarrow 2}} (r=0,z)$ associated to $\mathcal{S}_{\mathrm{exp}}^{_{1 \shortrightarrow 2}}$ :

\newpage

$\vphantom{a}$

\begin{equation}
\label{OnAxisExp}
     \mathcal{E}_{\mathrm{exp}}^{_{1 \shortrightarrow 2}} (r=0,\delta z)  =  \sqrt{\mathcal{I}_{0}} \, e^{\,i \, k_{z0} (z_{1}+\delta z/n_{0})} \!\times\! \left[ \frac{-i}{\pi} \int_{0}^{\infty} \frac{1-e^{\,-i\left[\overline{k}_{z}+ i \, \alpha/2 \right]L}}{\overline{k}_{z} +  i \, \alpha/2} e^{\,i \overline{k}_{z} \delta z} \, \mathrm{d}\overline{k}_{z} \right].
\end{equation}
\noindent As $z$ lies in the interval $[z_{1},z_{2}]$ and $z_{2} = z_{1} +L/n_{0}$, $\delta z  =  n_{0}(z-z_{1})$ varies from 0 to$\;L$. The phase $\Phi_{l}  =  k_{z0} \, (z_{1}  +  \delta z/n_{0})$ is the phase accumulated by the Bessel beam along$\;$its propagation from $z_{1}$ to $z_{2}$. Since the medium is supposed to be linear, this$\;$contribution$\;$is the only one we expect. The term inside the brackets in equation~\eqref{OnAxisExp} should$\,$then$\,$be$\,$real. Let's divide the integral in two parts, $I_{1}$ and $I_{2}$, as follows:
\begin{align}
         I_{1}(\delta z)  &=  \frac{-i}{\pi} \int_{0}^{\infty} \frac{\overline{k}_{z}  -  i \, \alpha/2}{\overline{k}_{z}^{2}  +  \left(\alpha/2\right)^{2}} \, e^{\,i \overline{k}_{z} \delta z} \, \mathrm{d}\overline{k}_{z} \label{I1} \\ \label{I2}
         I_{2}(\delta z)  &=  \frac{i}{\pi} \, \exp \left(\alpha L/2 \right) \int_{0}^{\infty} \frac{\overline{k}_{z}  -  i \, \alpha/2}{\overline{k}_{z}^{2}  +  \left(\alpha/2\right)^{2}} \, e^{\,-i \overline{k}_{z} (L-\delta z)} \, \mathrm{d}\overline{k}_{z}  .
\end{align}

\newpage

\noindent From equations~\eqref{I1} and~\eqref{I2}, we can derive the real parts of $I_{1}$ and $I_{2}$: $\mathrm{Re}\left(I_{1}\right) = 0 $ and $\mathrm{Re}\left(I_{2}\right) = \exp \left(\alpha \delta z/2 \right) $. The on-axis electric field $\mathcal{E}_{\mathrm{exp}}^{_{1 \shortrightarrow 2}} (r=0,\delta z)$ is finally given by: 
\begin{align}
    \mathcal{E}_{1,2}\,(r=0,\delta z) ={}& \sqrt{\mathcal{I}_{0}} \, e^{\,i \, k_{z0} (z_{1}+\delta z/n_{0})} \times \left[ \mathrm{Re}\left(I_{1}\right) + \mathrm{Re}\left(I_{2}\right) \right] 
    \nonumber     
\\
\begin{split}
     \phantom{\mathcal{E}_{1,2}\,(r=0,\delta z)} ={}& - \sqrt{\mathcal{I}_{0}} \, e^{\,i \, k_{z0} (z_{1}+\delta z/n_{0})} \, \exp \left( \alpha \delta z / 2 \right).
     \label{eq:3}
\end{split}
\end{align}
\noindent and exponentially increases, as required, over the propagation distance $\delta z$ in the medium. By replacing $L$ with $L/n_{0}$ and $\alpha$ with $\alpha \, n_{0}$ in the expression of the target on-axis intensity profile equation~\eqref{SystemProfile}, we therefore manage to counterbalance the refractive$\;$stretching$\;$of the Bessel beam and compensate for the good attenuation coefficient $\alpha$.

\subsubsection{Experimental setup}

\begin{landscape}

\begin{figure}[h]
\centering
\includegraphics[width=0.8\linewidth]{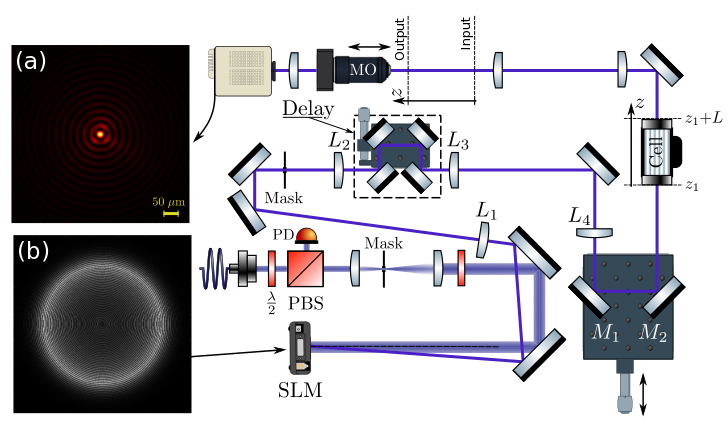}
\caption{Experimental setup. The beam is magnified before reaching the SLM at normal incidence. An example of hologram displayed on the SLM is shown on inset (b) (without grating for clarity). The contrast of the phase wrapping is spatially modulated so as to shape the Bessel beam on-axis intensity. The first diffraction order is selected by filtering out the others in$\;$the$\;$focal$\;$plane$\;$of$\;L_{1}$. The telescopes $\left\{L_{1}, L_{2}\right\}$ and $\left\{L_{3}, L_{4}\right\}$ conjugates the SLM chip with the cell entrance plane. The$\;$delay$\;$line$\;$and$\;$the$\;$translation$\;$stage (over which the mirrors $M_{1}$ and $M_{2}$ are fixed) allows to set precisely $z_{1}$ (starting of the exponentially$\,$rising$\,$section$\,$of$\,$the$\,$target$\,$profile) in the cell entrance plane. The output plane is imaged onto a microscope objective by a 4-f system. Moving the objective along the optical axis enables us to measure the on-axis profile. The transverse profile at $z_{3}$ of the generated Bessel beam is shown$\,$on$\,$inset$\,$(a).
}
\label{fig:BesselExp}
\end{figure}

\end{landscape}

\noindent The experimental setup I built to generate attenuation-resistant Bessel beams has been depicted$\,$on$\,$figure~\ref{fig:BesselExp}. The continuous-wave$\,$laser$\,$beam$\,$produced$\,$by$\,$a$\,$TA$\,$Pro$\,$laser$\,$system is sent onto the optical table through a high power fiber. A polarized beam splitter$\;$(PBS), preceded by a half wave-plate ($\lambda/2$), splits the outgoing beam in two$\,$parts. The$\;$low$\;$power reflection is focused onto a photo-diode (PD) in order to monitor the stability$\;$of the fiber output power. The transmitted part is magnified four times$\,$using$\,$a$\,$telescope$\,$formed$\,$by$\,$two converging lenses of $50$ and $200$ mm focal lengths respectively. A small pinhole$\;$aperture is positioned in the focal plane of the $50$ mm focal lens in order to filter out the$\;$high$\;$transverse wave-vector components of the Gaussian beam.$\;$After$\,$the$\,$telescope,$\,$the$\,$beam$\,$has$\,$a$\,$diameter of $6.6\;$mm and is perfectly radially symmetric. It reflects on the SLM chip, which is strictly perpendicular to the optical axis. The SLM used for the experiment is a liquid crystal on silicon phase-only modulator (LCOS), with an effective area of $1272 \times 1024$ pixels and a pitch of 12.5 $\mu$m. Because of the grating imprinted on the SLM, the first order diffracted beam propagates at a small angle from the incidence axis. A succession of two telescopes $-$ formed respectively by the lenses $L_{1}$ and $L_{2}$ on one side$\,$($f_{1} = 750\,$mm$\,$and$\,f_{2} = 150\,$mm) and by $L_{3}$ and $L_{4}$ on the other ($f_{1} = 100$ mm and $f_{2} = 250$ mm) $-$ conjugates the SLM chip and the $z=0$ plane with a demagnification factor $G=0.5$. In between$\;$the$\;$telescopes, a delay line as been set up in order to roughly position the starting of the exponential$\,$rising section in the entrance plane of the cell. Another translation stage, on which the mirrors $M_{1}$ and $M_{2}$ are mounted, allows the fine tuning of this positioning. The output plane of the medium is imaged by a 4-f arrangement onto a microscope objective which is set up on a computer controlled translation stage. By moving the objective along$\;$the$\;$optical$\;$axis, one can monitor the Bessel beam evolution along $z$. The last$\,$lens$\,$on$\,$the$\,$beam$\,$path$\,$images the plane we look at on the CMOS camera, positioned$\,$in$\,$the$\,$focal$\,$plane.$\;$The$\,$magnification factor $G'\,$of$\,$the$\,$whole$\,$imaging$\,$system$\,$is$\,13.6 \pm 0.1$.$\;$An$\,$example$\,$of$\,$the$\,$Bessel$\,$beam$\,$transverse intensity distribution, captured close to the maximum of the target on axis profile ($z=z_{3}$), is shown on the inset (a) of figure~\ref{fig:BesselExp}. The hologram displayed on the SLM in order$\;$to obtain this image is shown on the inset$\;$(b) (for zero grating). Let's finally mention that the choice of the lenses $L_{1}$, $L_{2}$, $L_{3}$ and $L_{4}$ (and thus, of the demagnification factor $G$) is$\,$conditioned by the length of the lossy medium we deal with. For biological applications, $G$ should at least be divided$\;$by$\;$10, as pointed out in the following paragraph.

\subsubsection{Experimental results}

\begin{figure}[hbt!]
\centering
\includegraphics[width=\columnwidth]{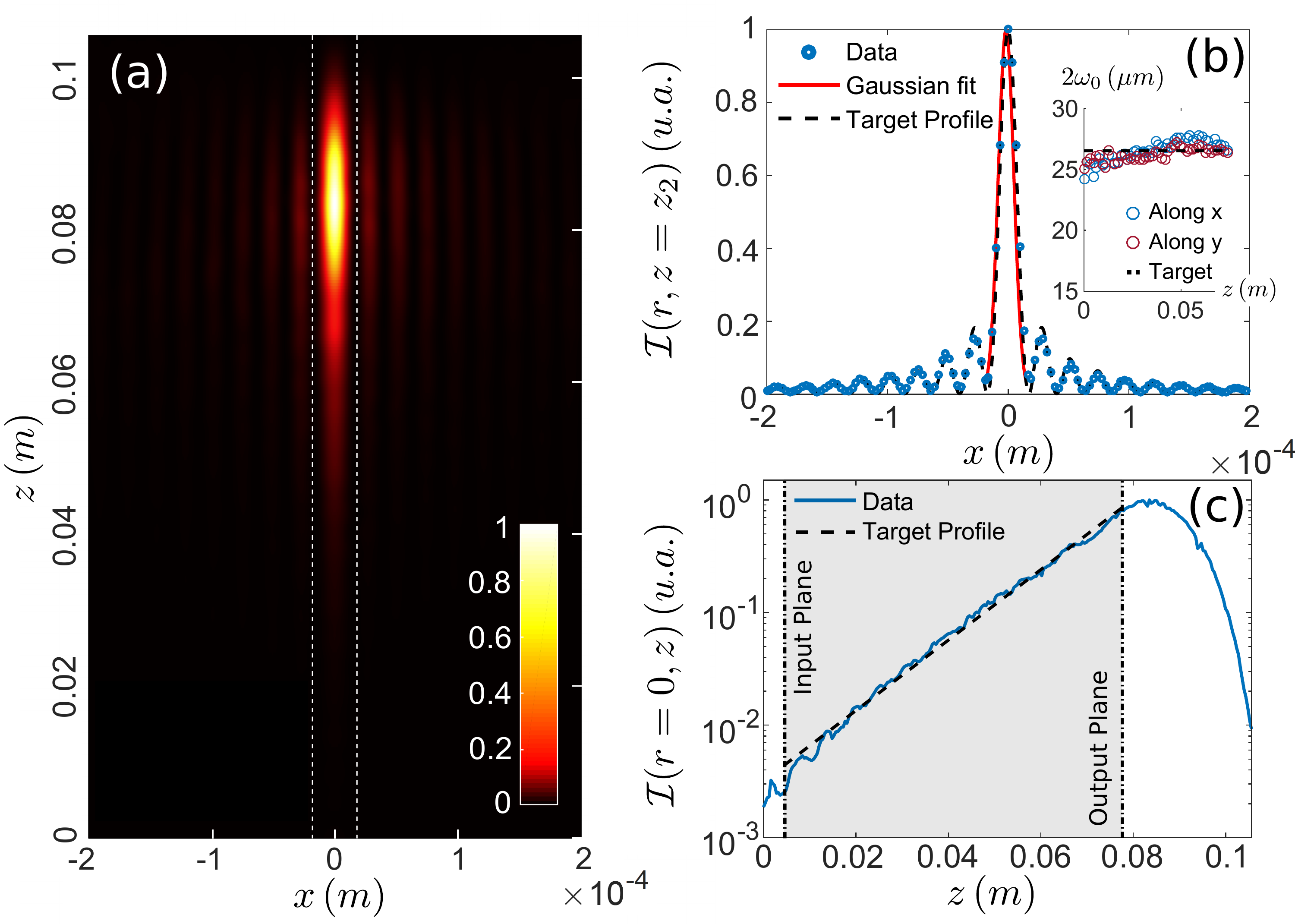}
\caption{Experimental characterization of the reconstructed Bessel beam. The Bessel cone angle $\theta_0$ was set to $(1/G) \times 8.5$ mrad, where $G=0.5$ is the overall demagnification factor of the telescopes $\left\{L_{1}, L_{2}\right\}$ and $\left\{L_{3}, L_{4}\right\}$. The 2D map on figure (a) is obtained by scanning slowly ($v = 2$ mm.s$^{-1}$) the microscope objective along the z axis and capturing a frame every second. The white dotted lines on both sides of the central peak define$\;$the region$\,$where$\,$the$\,$Gaussian$\,$fit$\,$is$\,$performed.$\,$The$\,$transverse$\,$and$\,$longitudinal$\,$intensity$\,$profiles of the reconstructed Bessel beam have been plotted respectively on the figures (b)$\,$and$\,$(c), in the absence of lossy material. (b): Blue dots are experimental data obtained by cutting at $z = 8.5\;$cm the 2D map (a). The dashed line is the target$\,$profile$\,$calculated$\,$numerically and the red solid line a Gaussian fit of the central peak, performed to extract its width. Inset: Dots are the fitted peak diameter $2 \omega_{0}$ as function of $z$ (blue along $x\,$an$\,$red$\,$along$\;y$). The black dashed line is the calculated target peak diameter. Data shows$\,$a$\,$change$\,$of$\,$less than $5\%$ over the length of the medium ($L = 7.5$ cm), confirming$\,$the$\,$non-diffractive$\,$nature of the attenuation-resistant Bessel beam in the transverse plane. (c): The$\,$blue$\,$line$\,$is$\,$a$\,$cut of$\,$the$\,$2D$\,$map$\,$(a)$\,$along$\,$the$\,z$-axis.$\;$It$\,$matches$\,$perfectly$\,$with$\,$the$\,$calculated$\,$on-axis$\,$intensity profile (black dashed line) in the region where it increases exponentially (shaded area).}
\label{fig:ProfileAll}
\end{figure}

\noindent The first step toward$\,$the$\,$on-axis$\,$compensation$\,$of$\,$attenuation$\,$is$\,$to$\,$check$\,$that$\,$the$\,$transverse and longitudinal intensity profiles of the experimentally measured Bessel beam (in air) fit the target ones, obtained by simulations.$\;$We design the target beam$\,$in$\,$order$\,$to$\,$compensate $96\%$ attenuation over a lossy, $7.5$ cm long medium. The 2D maps shown on figure~\ref{fig:ProfileAll}(a) and the profiles of figures ~\ref{fig:ProfileAll}(b) and (c) are obtained by scanning slowly ($v = 2$ mm.s$^{-1}$) the microscope objective along the $z$-axis. Both the transverse (blue circles on figure (b)) and the longitudinal profiles (blue line on figure (c)) of the measured Bessel beam are in excellent agreement with the target profiles (black dashed line).$\;$The$\,$latter$\,$are$\,$obtained$\,$by numerically solving the evolution of the transverse electric field from$\,z=0\,$to$\,L\,$with$\,$the second order split-step method. We take as initial condition a field with$\,$the$\,$SLM$\,$imprinted phase $\Psi$ and the radially symmetric Gaussian envelope of the SLM input beam. In$\,$order$\,$to accurately determine the central peak intensity along $z$ $-$ as presented$\,$in$\,$figure~\ref{fig:ProfileAll}(c)$\,-$ we fit with a Gaussian profile the region delimited by the two white dashed lines on both sides of the central peak (figure~\ref{fig:ProfileAll}(a)), as illustrated on figure~\ref{fig:ProfileAll}(b) (red solid line). The width of the central peak, along the propagation, is found to be constant (at $\pm 5\%$), as$\,$shown on the inset of figure~\ref{fig:ProfileAll}(b). This measurement demonstrates that$\,$we$\,$are$\,$able$\,$to control the longitudinal intensity profile without altering the non-diffracting behavior of the Bessel beam. More importantly, we observe, as required, an exponential$\,$increase$\,$of$\,$the on-axis intensity along $z$.
Nevertheless, small intensity oscillations can be observed at the beginning of the measured on-axis profile on figure~\ref{fig:ProfileAll}(c). They are basically$\,$due$\,$to$\,$high longitudinal frequency truncation~\cite{6-33Ouadghiri}, as $k_{z}$ is upper bounded by$\,$the$\,$laser$\,$wave-vector$\,k_{0}$. The oscillation amplitude can be reduced further  by increasing the Bessel cone angle $\theta_{0}$.
\vspace{6pt}
\newline
\noindent The lossy medium is then positioned on the beam path.$\;$Fitting$\,$the$\,$on-axis$\,$intensity$\,$profile with the function~\eqref{SystemProfile} provides$\,$the$\,$position$\,z_{2}\,$where$\,$the$\,$medium$\,$output$\,$plane$\,$should$\,$set. We then translates this position using$\,$the$\,$translation$\,$stage$\,$sketched$\,$on$\,$figure~\ref{fig:BesselExp}$\,$until$\,z_{2}$ matches the medium output plane position on the optical axis. This plane is imaged on the camera using the imaging system described on figure~\ref{fig:BesselExp}.$\;$The$\,0.5\,$mm$\,$depth$\,$of$\,$field$\,$of$\,$the imaging system and the standard deviation on the fit parameters translate into a $\pm 1$~mm uncertainty on the medium output plane$\,$position.$\;$Three$\,$different$\,$media$\,$(contained$\,$in$\,$three different glass cells) have been used so as to determine$\,$if$\,$attenuation-resistant$\,$Bessel$\,$beams are capable of compensating attenuation along the optical axis. Two cells are filled with isotopically pure rubidium vapors (the first ($7.5$ cm long), with $^{_{87}}$Rb only, and the second ($2.5$ cm long), with $^{_{85}}$Rb only). The third one ($2.5$ cm long) contains a diffusive$\;$aqueous suspension of milk. Rubidium cells are heated up to 140$^{\circ}$C. At$\,$this$\,$temperature,$\,$the$\,$atomic density is large ($n_{a} \simeq 2 \, \mathrm{to} \,5 \times 10^{13}$ atoms/cm$^{3}$). By tuning the laser frequency $\nu_{0}$ over the rubidium $D_{2}$ for instance, we can change the transmission over$\,$several$\,$orders$\,$of$\,$magnitude, without affecting significantly the refractive index of the vapor, which remains close one (with $\pm 1 \%$ fluctuations).$\;$The$\,$transmission$\,$of$\,$the$\,$water-milk$\,$mixture$\,$can$\,$be$\,$tuned$\,$changing the milk concentration. Remaining under highly diluted condition, the medium refractive index stays close to the water one $n_w \simeq 1.33$. As explained above, we should balance in this case the change of refracting index stretching the Bessel beam along the optical axis, replacing beforehand in the target profile $L$ and $\alpha$ with $L/n$ and $\alpha \, n$ respectively. 

\newpage
$\vphantom{a}$
\newline

\newpage

\begin{figure}[hbt!]
\centering
\includegraphics[width=0.8\linewidth]{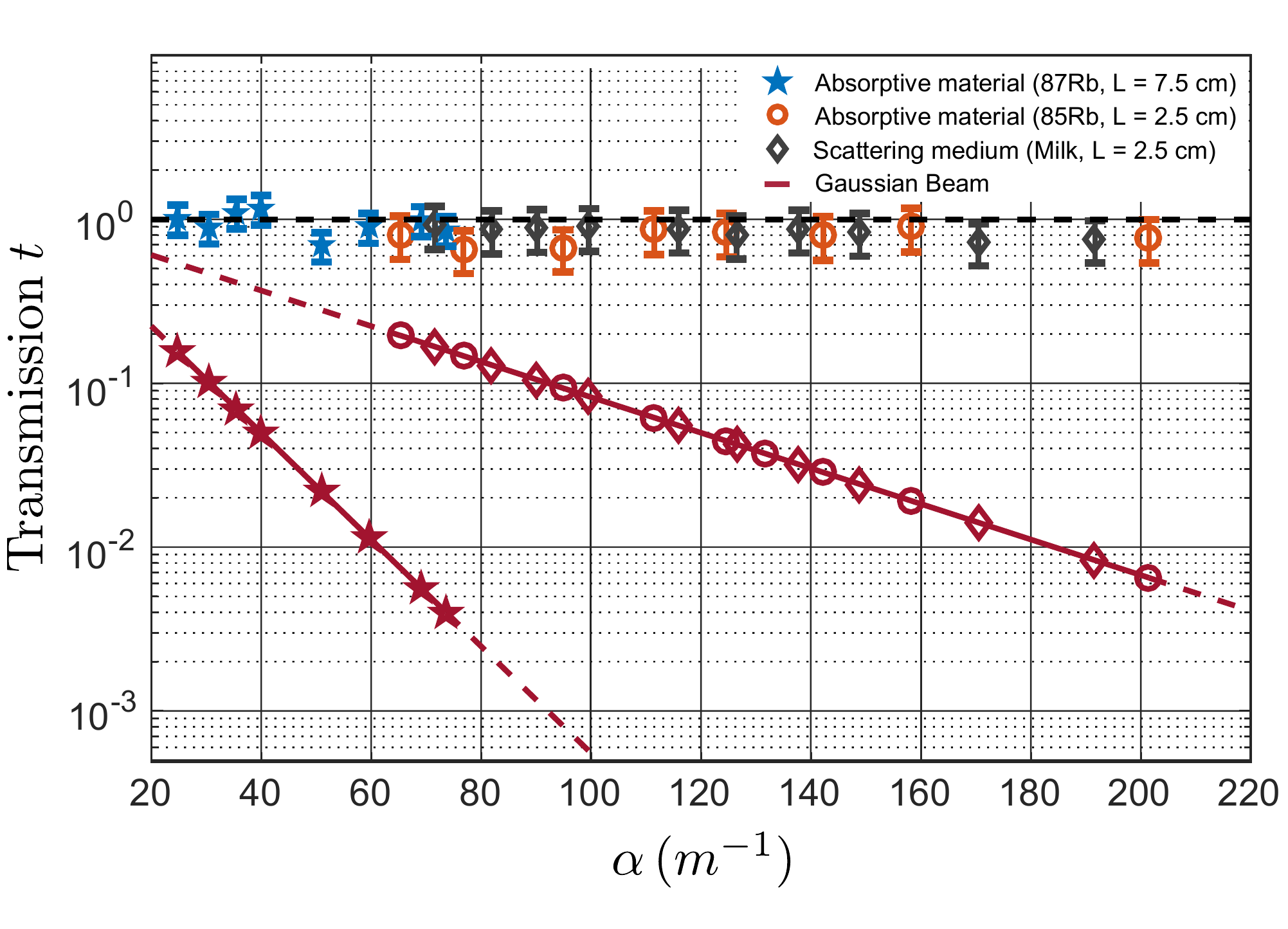}
\caption{Measure of the on-axis transmission $t$ of the Bessel beam as a function of the attenuation$\,$coefficient$\,\alpha$. Data obtained with the rubidium vapors have been plotted in blue stars ($^{_{87}}$Rb) and orange circles ($^{_{85}}$Rb). Data obtained with the water-milk mixture have been plotted in grey diamonds.$\;$The$\,$two$\,$red$\,$lines$\,$show$\,$the$\,$transmission$\,$expected$\,$from the Beer-Lambert law through a 2.5 cm and a$\,$7.5$\,$cm$\,$long$\,$lossy$\,$material}
\label{fig:Transmission}
\end{figure}

\noindent We design the target profile in order to compensate attenuation over 7.5 cm long materials, whatever the length of the cell we use. The overall Bessel beam power is reduced$\;$to$\;$keep the input peak intensity $\mathcal{I}_{0}$ lower than the rubidium on-resonance saturation intensity$\;\mathcal{I}_{s}$ ($\mathcal{I}_{\mathrm{s}} \simeq  2.5$ mW/cm$^{2}$ for linearly polarized light). We finally measure the$\,$peak$\,$intensity$\,$in$\,$the entrance plane (without cell) and in the output plane (with cell) in order to evaluate the on-axis transmission trough the lossy medium. To$\;$do$\;$so,$\,$we$\;$perform$\;$the$\;$fit$\,$on$\,$five$\,$different images of the central Bessel spot with a 2D-Gaussian function. The measured$\,$transmission is$\,$shown$\,$on$\,$figure~\ref{fig:Transmission}.$\;$The$\,$blue$\,$stars$\,$and$\,$the$\,$orange$\,$circles$\,$represent$\,$the$\,$experimental$\,$data obtained$\,$with the $2.5$ cm$\,$and$\,$the$\,7.5\,$cm$\,$long$\,$vapor$\,$cells$\,$respectively.$\;$The$\,$grey$\,$diamonds$\,$are the data we get using the water-milk mixture. The $4 \%$ reflectively of the cell windows has been taken into account. The black dashed line represents a perfect on-axis$\;$compensation (that$\,$is,$\,$an$\,$on-axis$\,$transmission$\,$of$\,$one).$\;$Most$\,$of$\,$the$\,$experimental$\,$points$\,$lie$\,$slightly$\,$under$\,$it. This small$\,$discrepancy$\,$comes$\,$from$\,$the$\,$input$\,$plane$\,$intensity$\,$measurements$\,$rather$\,$than$\,$from the output plane ones. Indeed,$\,$the$\,$on-axis$\,$intensity$\,$is$\,$oscillating$\,$in$\,$the$\,$medium$\,$input$\,$plane (as you can see for instance on figure~\ref{fig:ProfileAll}(c)). Moreover, the relative position of the cell output plane with respect to $z_{2}$ (and$\,$thus$\,$of$\,$the input$\,$plane$\,$with$\,$respect$\,$to$\,z_{1}$)$\,$is$\,$known$\,$with a$\,$precision$\,$of$\,\pm 1$ mm.$\;$This$\,$two$\,$factors$\,$together induce an uncertainty$\,$on$\,$the$\,$input$\,$intensity measurements that is reported on figure~\ref{fig:Transmission} (errorbars). The two red lines represent the transmission of a non-saturating collimated beam (computed$\,$using$\,$the Beer-Lambert$\,$law) through a 2.5 cm and a$\,$7.5$\,$cm$\,$long$\,$lossy$\,$material respectively. 

\newpage

\subsubsection{Applications and limitations}
 
\noindent In previous experiments, compensations of $10 \%$ and $30 \%$ attenuation have been achieved, using attenuation-resistant frozen waves and exicon (that is, exponential intensity axicon) respectively. In comparison, the method we developed allows compensation of attenuation coefficients up to $200$ m$^{-1}$ (which is equivalent to a transmission of$\,5 \!\times\!10^{-3}\,$through$\,$a$\,2.5\,$cm long lossy material). This is a crucial advantage for light-sheet microscopy for instance, since the diffusive coefficients of biological tissues commonly observed with this technique range typically from $50$ to $200$ cm$^{-1}$~\cite{6-36Johns,6-37Nylk}. Indeed, using the same target profile as for the last orange circle on figure~\ref{fig:Transmission} (for which $\alpha \simeq 200$ m$^{-1}$) and reducing the overall demagnification$\,$factor $G$ by a$\,$factor$\,$10$\,$will$\,$constrict$\,$the$\,$length$\,$of$\,$the$\,$generated$\,$Bessel$\,$beam by a factor$\,$100$\,$(as$\,$it$\,$scales$\,$with$\,G^{2}$).$\,$Therefore,$\,$the$\,$latter$\,$will$\,$compensate$\,$for$\,$an$\,$attenuation coefficient 100 times bigger than the$\,$target$\,$one$\,$(that is, for $\alpha = 200$ cm$^{-1}$), over a distance which is$\,$now$\,L/100 = 250\,\mu$m.$\,$For$\,$such$\,$values$\,$of$\,$the$\,$attenuation$\,$coefficient,$\,$the$\,$penetration depth of our attenuation-resistant$\,$Bessel$\,$beam$\,$is$\,$therefore$\,$expected$\,$to$\,$be$\,$more$\,$than$\,$100$\,\mu$m longer than the best value reported in the literature so far~\cite{6-37Nylk}, which would constitute an improvement of almost $170\%$ of the current field of view inside highly diffusive samples. 
\vspace{6pt}
\newline
\noindent Nevertheless, the non-diffracting feature of Bessel beams comes at the$\;$cost$\;$of$\;$the$\;$presence of high-energy side lobes, that are known to degrade the imaging contrast$\;$of$\;$fluorescence light sheet microscopy by inducing photo-bleaching~\cite{6-38Gao}. For ideal zero-order$\;$Bessel$\;$beams, the intensity of the first ring surrounding the central spot is about$\,17\%\,$of the peak$\,$intensity. This can also be an issue in photon fluid experiments. If we use a Bessel beam as a defect, whether or not it compensates for attenuation, the rings around the central peak will also generate a change of refractive index.$\;$This$\,$creates,$\,$at$\,$the$\,$end$\,$of$\,$the$\,$day,$\,$a$\,$complex$\,$potential landscape for the photon fluid, that is not acting anymore as a spatially localized obstacle. We should thus think about a way to cancel the Bessel beam side lobes$\;$while$\;$preserving, at least, its non-diffracting feature. Following the work of G. Di Dominico \textit{et al.} in~\cite{6-39Domenico}, I will present in the next subsection a method based on the generation of "droplet$\;$beams", which are formed by making interfere two Bessel beams with specific cone angles$\;\theta_{1}\;$and$\;\theta_{2}$. For some values of the ratio $\theta_{2}/\theta_{1}$, the intensity of the side lobes is significantly reduced. Obviously, droplet beams do not allow to compensate attenuation along the optical axis. Keeping the amplitude of the obstacle constant requires therefore to saturate the rubidium vapor with the droplet beam all along the optical axis.
    
    \subsection{Droplet beam}
    
    \subsubsection{Optimizing the side lobe cancellation}

\begin{figure}[hbt!]
\centering
\includegraphics[width=\columnwidth]{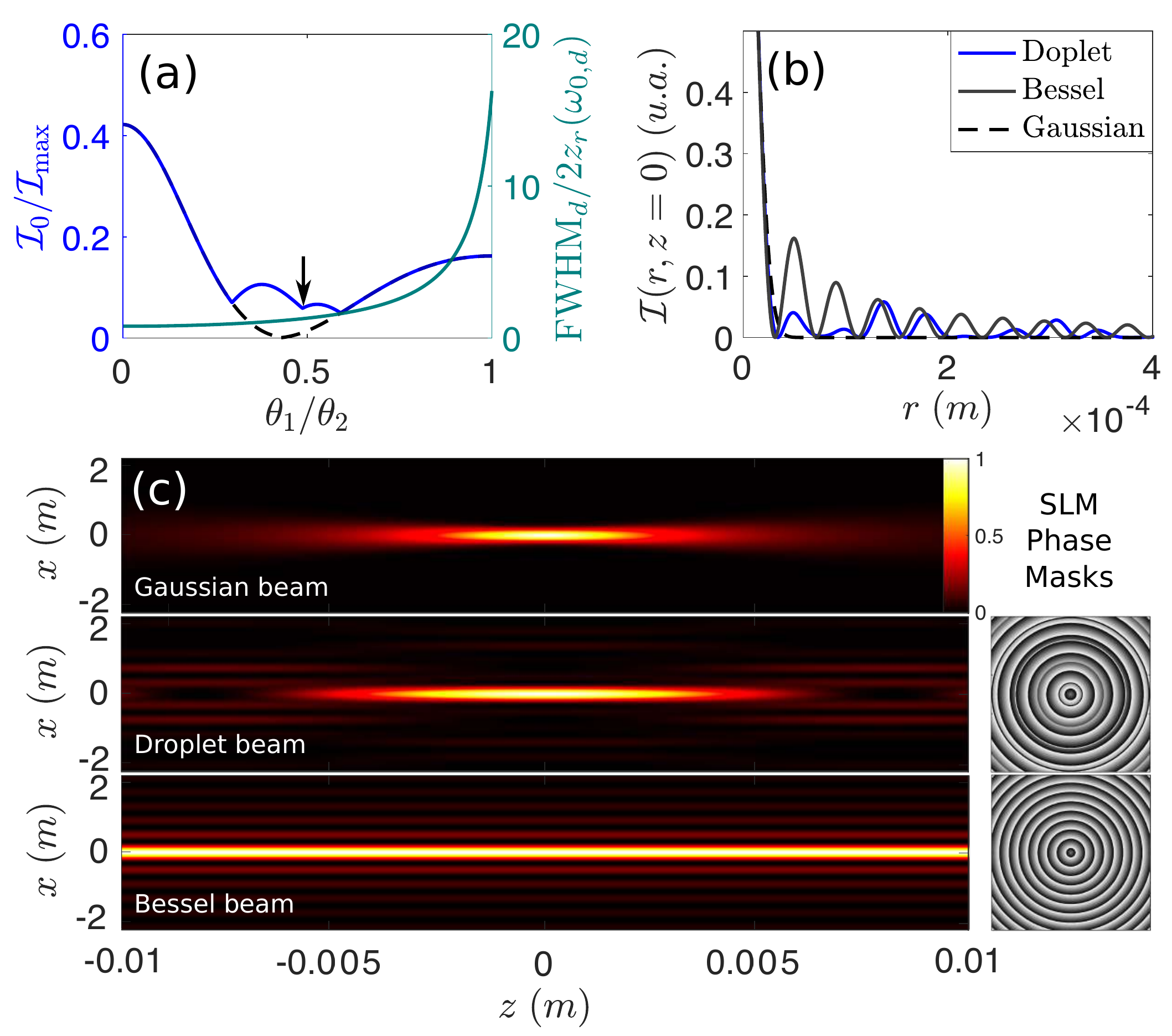}
\caption{(a) Blue curve: plot of the ratio between the droplet peak$\,$intensity$\,\mathcal{I}_{0}\,$and$\,$the intensity of the brightest side-lobe $\mathcal{I}_{\mathrm{max}}$ as function of $\theta_{2}/\theta_{1}$. Black dashed line:$\,$plot$\,$of$\,$the ratio between $\mathcal{I}_{0}$ and the$\,$intensity$\,$of$\,$the$\,$first$\,$side$\,$lobe$\,\mathcal{I}_{1}$.$\,$Both$\,$curves$\,$are$\,$obtained$\,$for$\,z\!=\!0$. The axial FWHM of the droplet beam, $\mathrm{FWHM}_{d}$, is plotted on the same graph (cyan curve) as function of $\theta_{2}/\theta_{1}$.$\,$It$\,$is$\,$normalized$\,$by$\,$the$\,$axial$\,$FWHM$\,$of$\,$a$\,$Gaussian$\,$beam$\,$having$\,$a$\,$waist equal to the droplet radial width $\omega_{0,d}$, \textit{ie}, by twice the Rayleigh length $z_{r}(\omega_{0,d}) = \pi\omega_{0,d}^{2}/\lambda$. The black arrow on figure (a) indicates the value we have chosen for the cone angles ratio in experiments ($\theta_{2}/\theta_{1}\simeq 0.5)$. Figures (b) and (c) show the transverse intensity$\;$profile$\;$and the in-air propagation (top view, $xz$ plane) of a Gaussian, a droplet ($\theta_{2}/\theta_{1} = 0.5$)$\;$and$\;$a Bessel-Gauss beam. They all have the same transverse$\,$width$\,$in$\,$the$\,z=0\,$plane$\,$($\simeq 25 \, \mu$m). The intensities of the droplet first and second$\,$side-lobes$\,$have$\,$been$\,$reduced$\,$by$\,72\%\,$and$\,85\%$ respectively with respect to the Bessel ones. On the right of figure (c), the phase masks displayed on the SLM in order to generate the droplet and the Bessel beams are shown. Parameters: for the droplet and Bessel beams, the width of the Gaussian envelope in~\eqref{BesselGauss} is $G \!\times\! 3.3$ mm and $G \! \times \! \theta_{1} = 0.008$ rad, where $G = 2/3$ is the overall demagnification factor of the telescopes conjugating the$\,$SLM$\,$chip$\,$and$\,$the$\,$cell$\,$input$\,$plane.$\,$Figure$\,$inspired$\,$from~\cite{6-39Domenico}. 
}
\label{fig:DropletMain}
\end{figure}

\noindent As mentioned previously, diffraction-free droplet beams$\,$are$\,$generated$\,$in$\,$practice$\,$by$\,$making interfere a set of plane-waves whose wave-vectors lie on two different co-axial$\,$cones$\,$of$\,$angles $\theta_{1}$ and $\theta_{2}$ respectively. For some specific values of the ratio $\theta_{2}/\theta_{1}$, droplet beams exhibit significantly lower side lobes than simple zero-order Bessel beams, as$\,$a$\,$result$\,$of$\,$the$\,$selective destructive interference between the two Bessel fields forming a droplet. In order$\;$to$\;$find theoretically the ratio $\theta_{2}/\theta_{1}$ providing the minimum amount of power distributed$\;$over$\;$the droplet side-lobes, we can start by considering the generic expression describing$\;$the$\;$in-air electric field $\mathcal{E}(r,z,\theta)$ of a zero-order quasi Bessel beam of cone angle $\theta$~\cite{6-18Durnin}:

\newpage

\begin{equation}
    \begin{split}
        \mathcal{E}(r,z,\theta) ={}& \mathcal{E}_{0} \, \frac{\omega_{0}(0)}{\omega_{0}(z)} \, J_{0} \left( \frac{r k_{r}(\theta)}{1+i z/z_{r}} \right) \exp \left\{ i \left[ \left(k-\frac{1}{2} \frac{k^{2}_{r}(\theta)}{k} \right)z -\Phi_{g}(z) \right] \right\} \\
        & \hspace{2.3cm} \dots \times \exp \left\{ \left[ \frac{i}{2}\frac{k}{R(z)} -\frac{1}{\omega^{2}_{0}(z)} \vphantom{\frac{k^{2}_{r}(\theta)}{k^{2}}} \right]\!\left[ r^{2} + \frac{k^{2}_{r}(\theta)}{k^{2}} z^{2} \right] \right\}.
    \end{split}
    \label{BesselGauss}
\end{equation} 
\noindent In$\,$formula$\,$\eqref{BesselGauss},$\,J_{0}\,$is$\,$the$\,$zero-order$\,$Bessel$\,$function$\,$of$\,$the$\,$first kind, $r\,$the$\,$radial$\,$coordinate and $k_{r}(\theta) = k \sin(\theta)$ the Bessel beam transverse wave-vector. We have also introduced$\;$the width $\omega_{0}(z) = \omega_{0}(0) \sqrt{1+(z/z_{r})^{2}}$ of the Gaussian$\,$envelope$\,$along$\,$the$\,$z-axis$\,$($z=0\,$being$\,$the position of the waist and $z_{r} = \pi \omega_{0}^{2}(0)/\lambda$ the Rayleigh length), the radius of$\;$curvature$\;$of the$\,$beam$\,$wavefront:$\,R(z) = z \left[1+(z_{r}/z)^{_{2}} \right]\,$as$\,$well$\,$as$\,$the$\,$Gouy$\,$phase$\,$shift:$\,\Phi_{g} = \mathrm{atan}(z/z_{r})$. The electric field of the droplet beam is then given by: $\mathcal{E}_{d}(r, z) = \mathcal{E}_{1}(r, z, \theta_{1})+ \mathcal{E}_{2}(r, z, \theta_{2})$ and the total power distributed over its side-lobes, at $z=0$, by:
\begin{equation}
    \mathcal{P}_{d,sl} = 2 \pi \int_{r_{\mathrm{min}}}^{\infty} \left| \mathcal{E}_{1}(r, 0, \theta_{1})+ \mathcal{E}_{2}(r, 0, \theta_{2}) \right|^{2} r \mathrm{d}r,  
\end{equation}
\noindent where $r_{\mathrm{min}}$ is the radial distance associated to the first minimum in the intensity pattern. In~\cite{6-39Domenico}, the authors compare the amount of power $\mathcal{P}_{d,sl}$ in the side-lobes$\,$of$\,$a$\,$droplet$\,$beam with the power $\mathcal{P}_{b,sl}$ in the side-lobes of a Bessel Gauss beam, having the same on-axis intensity. They compute, for different values of $\theta_{2}/\theta_{1}$, the ratio $\mathcal{P}_{d,sl}/\mathcal{P}_{b,sl}$ at the waist position $z=0$.
In our case, we are more interested in the ratio between the droplet peak intensity $\mathcal{I}_{0}$ and the intensity $\mathcal{I}_{1}$ of the first side lobe, as we look for the value of $\theta_{2}/\theta_{1}$ for which it cancels. This ratio is plotted as function of $\theta_{2}/\theta_{1}$ on figure~\ref{fig:DropletMain}(a) (black line), for $z = 0$, where the droplet on-axis intensity is maximal. 
\vspace{6pt}
\newline
\noindent As you can see, it exhibits a minimum close to $\theta_{2}/\theta_{1} = 0.4$.$\;$One$\;$might$\;$then$\;$be$\;$tempted$\;$to choose this value for $\theta_{2}/\theta_{1}$ in experiments. However, the ratio between $\mathcal{I}_{0}$ and the intensity $\mathcal{I}_{\mathrm{max}}$ of the brightest droplet side-lobe$\,$(blue line)$\,$exhibits,$\,$at$\,\theta_{2}/\theta_{1} = 0.4$,$\,$a$\,$local$\,$maximum. We thus compromise,$\,$setting$\,\theta_{2}/\theta_{1}\,$close$\,$to$\,0.5\,$(black$\,$arrow)$\,$in$\,$practice.
\vspace{6pt}
\newline
\noindent On$\,$the$\,$same$\,$graph, the axial FWHM of$\,$the$\,$Droplet$\,$beam,$\,\mathrm{FWHM}_{d}$,$\,$has$\,$also$\,$been$\,$reported. It$\,$is$\,$normalized$\,$by$\,$twice$\,$the$\,$Rayleigh$\,$length$\,z_{r}(\omega_{0,d}) = \pi \omega_{0,d}^{2}/\lambda$,$\,$that$\,$is,$\,$by$\,$the$\,$axial$\,$FWHM of a focused Gaussian beam having a waist equal$\,$to$\,$the$\,$radial$\,$width$\,\omega_{0,d}\,$of$\,$the$\,$droplet$\,$peak. As you can see, $\mathrm{FWHM}_{d}$ increases faster and faster with $\theta_{2}/\theta_{1}$, until it reaches a maximum for  $\theta_{2}/\theta_{1} = 1$ (that is, when the droplet becomes$\,$a$\,$quasi$\,$Bessel$\,$beam$\,$with$\,$a$\,$cone$\,$angle$\,\theta_{1}$). This$\,$behaviour can be intuitively understood as follows. The interference between the two co-axial Bessel fields forming the droplet beam modulates periodically its on-axis$\,$intensity. When $\theta_{2}/\theta_{1}$ decreases, the difference between the longitudinal wave-vectors of these two fields increases and makes therefore the axial extend of the droplet (\textit{ie} $\mathrm{FWHM}_{d}$) smaller. For $\theta_{2}/\theta_{1} = 0.5$, $\mathrm{FWHM}_{d}/\left(2 z_{r}(\omega_{0,d})\right) \simeq 2$. Consequently, even if the axial FWHM of the droplet is smaller than for Bessel beams, it remains twice larger than the Gaussian$\;$one. This$\,$is$\,$of$\,$crucial$\,$importance.$\;$Apart$\,$from$\,$being$\,$perfectly$\,$collimated,$\,$unlike$\,$Gaussian$\,$beams, droplet beams allows to create much smaller defects, for a given cell length $L$, than their Gaussian homologs.$\;$In order to compare more$\,$visually$\,$the$\,$in-air$\,$propagation$\,$of$\,$a$\,$Gaussian, a$\,$droplet and a Bessel beams $-$ having the same transverse width $\omega_{0,d}$ in the plane $z = 0 \, -$ the theoretical intensity distributions of these three fields in the $xz$ plane (top view) have 

\newpage

\newpage

\noindent been plotted on figure~\ref{fig:DropletMain}(c).$\;$The$\,$quasi$\,$Bessel$\,$beam$\,$exhibits$\,$an$\,$invariant$\,$transverse$\,$profile along the propagation axis. Nevertheless, half of the energy it carries is distributed$\,$over$\,$its side lobes~\cite{6-39Domenico}. In contrast, the Droplet maintains an extended (but limited)$\,$depth$\,$of$\,$focus $-\,$which is twice longer than for the Gaussian beam$\,-\,$but provides a$\,$significantly$\,$enhanced energy$\,$confinement$\,$across$\,$the$\,$central$\,$peak.$\,$This$\,$is$\,$confirmed$\,$by$\,$figure~\ref{fig:DropletMain}(b),$\,$on$\,$which$\,$the radial profiles (at $z=0$) of the three beams of figure~\ref{fig:DropletMain}(b) have been plotted.$\,$The$\,$amount of light in the first and second side-lobes  is respectively $72\%$ and $85\%$ lower for the droplet than for the Bessel beam. Moreover, the intensity $\mathcal{I}_{\mathrm{max}}$ of the brightest droplet ring$\,$is$\,$less than $6\%$ of the peak intensity $\mathcal{I}_{0}$, and, within a circle of $120$ $\mu$m radius around the central peak, the intensity does not exceed $4\%$ of $\mathcal{I}_{0}$. All this$\,$suggests$\,$that$\,$droplets$\,$beams$\,$are good candidates to generate localized and collimated defects in photon fluid$\,$experiments. 

    \subsubsection{Experimental results}
 
\begin{figure}[hbt!]
\centering
\includegraphics[width=\columnwidth]{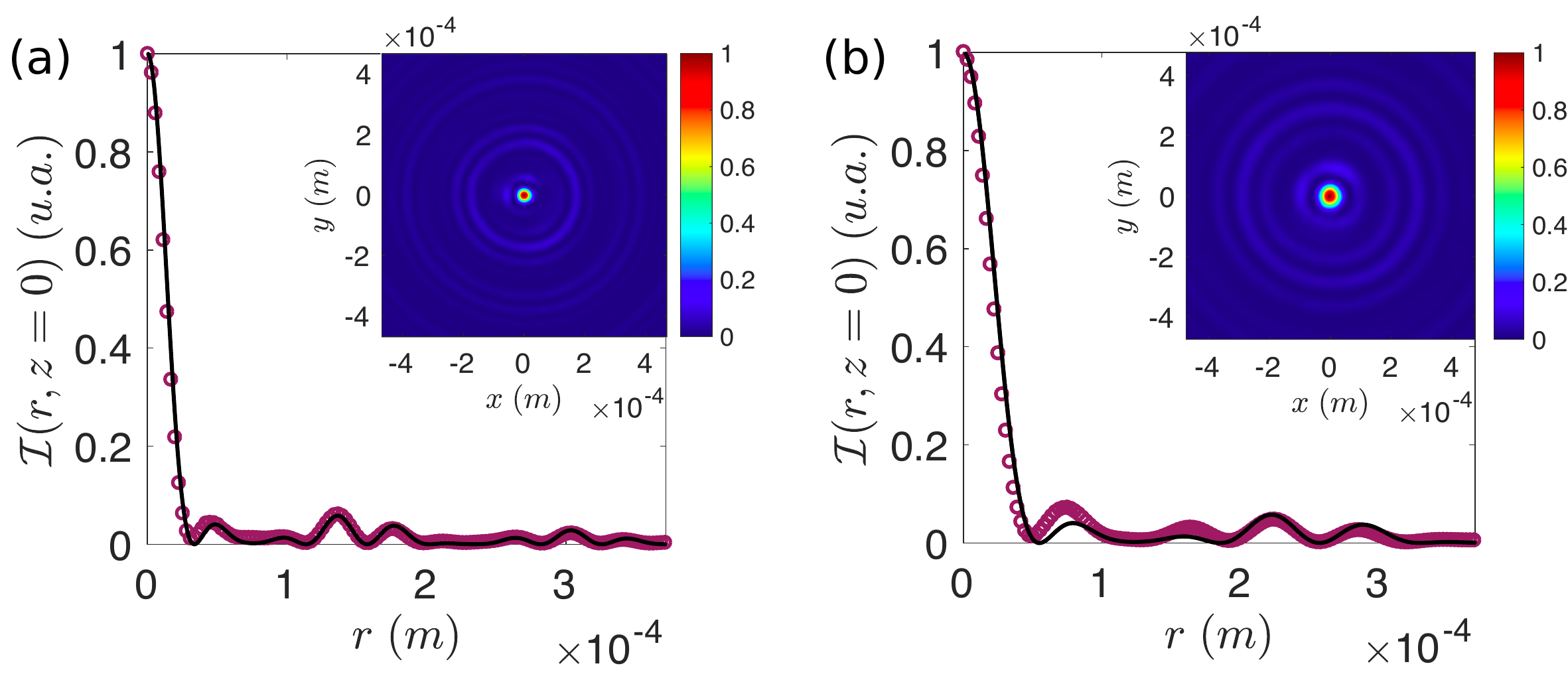}
\caption{Comparison between the droplet target and$\,$measured$\,$radial$\,$intensity$\,$profiles. (a): $G  \times \theta_{1} = 8.0$ mrad and (b): $G \times \theta_{1} = 5.0$ mrad.$\;$In$\,$both$\,$cases,$\,$we$\,$have$\,$set$\,\theta_{2}/\theta_{1} = 0.52$. For $G \! \times \! \theta_{1} = 8.0$ mrad, the radial width of the Droplet peak is $25$ $\mu$m. It is $38$ $\mu$m when $G \! \times \! \theta_{1} = 5.0$ mrad. The data (purple circles) are obtained by radially averaging the 2D intensity$\,$distributions plotted on the top right corner of each graph.$\;$These$\,$two$\,$images have been captured by translating the microscope objective of figure~\ref{fig:BesselExp} along the optical axis until the droplet peak intensity was maximized. The overall magnification factor of the imaging system is equal to $10.8 \pm 0.2$.  
}
\label{fig:ExpDroplet}
\end{figure}

\noindent The experimental setup we use to generate droplet beams is the same as the one sketched on figure~\ref{fig:BesselExp}. Only the focal lengths of the various lenses in~\ref{fig:BesselExp} change. The overall demagnification factor of the telescopes conjugating the SLM chip and the cell entrance plane is now $G = 2/3$, while the magnification factor of the imaging system is $10.8 \pm 0.2$. 
The width of the input Gaussian beam diffracting on the SLM grating is still$\,$about$\,3.3\,$mm. In contrast with what is done in~\cite{6-39Domenico}, the droplet is obtained by shaping the phase of this Gaussian beam in real space directly. An example of grating displayed on the$\,$SLM$\,$to$\,$this end is shown on the right of figure~\ref{fig:DropletMain}(c). It is simply obtained by wrapping the$\,$phase$\,$of: $\exp \left[i \boldsymbol{k_{r}}(\theta_{1}) \!\cdot\! \boldsymbol{r}\right] + \exp \left[i \boldsymbol{k_{r}}(\theta_{2}) \!\cdot\! \boldsymbol{r}\right]$ ($\boldsymbol{k_{r}}(\theta) \!\cdot\! \boldsymbol{r}$ being the phase shift provided by an axicon to create a quasi Bessel beam with a cone angle $\theta$). By moving the microscope objective along the $z$-axis (see figure~\ref{fig:BesselExp}), we can image the plane in which the$\;$droplet$\;$peak$\;$intensity is the highest (that is, the plane $z=0$). Figure~\ref{fig:ExpDroplet} shows the radial$\,$intensity$\,$distribution we measure in that case for different cone angles $\theta_{1}$. The data (purple circles)$\;$are$\;$obtained by radially averaging the 2D map on the top right corner of each graph. The radial profiles on figures~\ref{fig:ExpDroplet}(a) and (b) are obtained for $G \! \times \! \theta_{1} = 8.0$ mrad and $G \! \times \! \theta_{1} = 5.0$ mrad. In$\;$both cases, the ratio $\theta{2}/\theta{1}$ is equal to 0.52. The$\;$target$\;$(theoretical)$\;$profiles$\;$have$\;$been plotted on both graphs in black solid. As you can see, the agreement between the expected and the measured radial intensity distributions is excellent. The side-lobes$\,$are$\,$correctly cancelled as required. The radial width, $\omega_{0,d}$, of the droplet peak when $G \! \times \! \theta_{1} = 8.0\,$mrad is $25$ $\mu$m and $38$ $\mu$m for $G \! \times \! \theta_{1} = 5.0$ mrad. In this last case, the longitudinal extent of the droplet beam is around 22 mm (against 8.4 mm when  $G \! \times \! \theta_{1} = 8.0$ mrad, see figure~\ref{fig:DropletMain}(c)). This fits well with the length of the vapor cell we use in experiments, which$\,$is$\,2.5\,$cm$\,$long. Therefore, we choose the droplet beam beam obtained for $G \! \times \! \theta_{1} = 5.0$ mrad to$\,$generate the obstacle in our photon fluids.$\;$It$\,$results$\,$indeed$\,$from$\,$the$\,$best$\,$balance$\,$we$\,$found$\,$between a small droplet radial width, a good side-lobes cancellation and a long droplet axial$\;$extent. This$\;$last$\;$point$\;$is$\;$actually$\;$what$\,$seriously$\,$restricts$\,$the$\,$choice$\,$of$\,$the$\,$parameters $\theta_{1}\,$and$\,\theta_{2}/\theta_{1}$. Increasing the axial FWHM of the droplet by decreasing $\theta_{1}$ comes at the cost of an increase in the beam radial width. Similarly, increasing it by raising up the ratio $\theta_{2}/\theta_{1}\,$comes at the cost of an increase in the first droplet side-lobe intensity $\mathcal{I}_{1}$, which is not suitable to produce a localized obstacle. 

\newpage

\chapter{Outlook - Probing superfluidity}

\noindent In chapter 4, the measurement of the dispersion relation of density waves travelling onto paraxial photon fluids in warm rubidium vapor has been reported. I have shown that this dispersion relation exhibits a linear trend at low excitation wave-vectors. In other words, density waves travelling onto the photon fluid with those wave-vectors behave as collective phonons in the transverse plane, whose velocity is given by the speed of sound $c_{s} = \sqrt{n_{2} \, \mathcal{I}_{0}}$. According to the Landau criterion for superfluidity, presented in subsection 2.3.3,$\,c_{s}\,$defines a$\,$critical speed $v_{c}$ below which the photon fluid cannot theoretically dissipate energy by emitting sound-like excitations. In other words, a fluid of light moving toward an obstacle much smaller than the healing length $\xi$ at a velocity lower than $c_{s}$ should flow around it without scattering. In$\,$chapter$\,$5,$\,$we$\,$saw$\,$how$\,$to$\,$generate$\,$such$\,$on$\,$obstacle$\,$in$\,$the$\,$photon$\,$fluid by locally changing the refractive index it experiences using non-diffracting droplet beams. Therefore, all the ingredients are there to observe in our system frictionless flows of light around an all-optical obstacle. In this section, I first give a short historical overview about superfluidity. I then briefly review the various theoretical methods proposed to probe$\;$it$\;$in paraxial photon fluids and comment recent results obtained$\,$by$\,$Michel$\,$\textit{et$\,$al.}$\,$in~\cite{3-3Michel}.$\;$I$\,$finally present some preliminary results I obtained using our platform.  

\newpage

\section{Introduction and historical perspective}

\noindent Superfluidity, that is, the$\,$ability$\,$of$\,$a$\,$fluid$\,$to$\,$move$\,$without$\,$any$\,$friction,$\;$is$\,$without$\,$any$\,$doubt one of the most striking phenomenon observed in many-body physics. It was first observed in 1934 in liquid helium-4~\cite{7-1Kapitza} and has given rise since then to a$\,$huge$\,$amount$\,$of$\,$theoretical and experimental studies, as with liquid helium-3~\cite{7-2Osheroff} or ultra-cold atomic vapors~\cite{7-3Bloch}. 
In those systems, superfluidity (as Bose-Einstein condensation) manifests itself when the bosons forming the quantum$\,$fluid$\,$are$\,$cooled$\,$down$\,$below$\,$some$\,$critical$\,$temperature$\,$at$\,$which it$\,$undergoes$\,$a$\,$transition$\,$from$\,$normal$\,$to$\,$superfluid.$\,$At$\,$very$\,$low$\,$temperatures,$\,$the$\,$De$\,$Broglie wavelengths of atoms is of the order of the inter-atomic spacing. This$\,$delocalization$\,$allows to describe the fluid with a single macroscopic wave-function, which plays the role of an order$\,$parameter in the Ginzburg-Landau theory of phase transitions~\cite{3-9Dalfovo}. The dynamics$\;$of the weakly interacting bosons can then be described $-$ in the mean field approximation$\,-$ by the Gross-Pitaevskii equation (named$\,$after$\,$E.$\,$P.$\,$Gross$\,$and$\,$L.$\,$P.$\,$Pitaevskii)$\;$which$\,$drives the space-time evolution of the order parameter. As we saw in chapter 2, this equation is analogous to the nonlinear Schr\"{o}dinger equation. Photon fluids share consequently$\,$strong similarities with atomic Bose Einstein condensates or superfluid helium. The possibility$\;$to observe superfluid motion of light was initially suggested by Y. Pommeau$\,$and S. Rica~\cite{7-4Pomeau} and further investigated by Chiao \textit{et al.}~\cite{9:Chia0} in Kerr$\,$mediums$\,$embedded$\,$in$\,$optical$\,$cavities. Experimental observations of superfluid flows of light were reported few years later in exciton-polariton condensates~\cite{14:Amo,15:Amo}. Recently, the first clear evidence of superfluidity$\,$in cavityless systems has been reported by Michel \textit{et al.} in~\cite{3-3Michel}. A brief review$\,$of$\,$their$\,$work$\,$is presented in subsection 6.2.2.
\vspace{6pt}
\newline
\noindent As mentioned earlier, the most common way of probing superfluidity in photon fluids$\,$is$\,$to introduce a spatially localized$\,$defect$\,$into$\,$the$\,$flow$\,$and$\,$look$\,$at$\,$the$\,$perturbations$\,$it$\,$generates. In exciton-polariton experiments, defects of different sizes and shapes appear naturally in the growth process of microcavity samples. In paraxial fluids$\,$of$\,$light,$\,$the$\,$obstacle$\,$has$\,$to$\,$be created by changing locally the refractive$\,$index$\,$of$\,$the$\,$nonlinear$\,$material$\,$either$\,$by$\,$using$\,$an other light source~\cite{3-3Michel}, or by directly installing a piece of$\,$dielectric$\,$inside$\,$the$\,$medium~\cite{2-23Vocke}.
Depending on the relative value between the flow velocity and the speed of sound,$\,$various nonlinear hydrodynamical phenomena can be observed. At low velocity, the flow$\,$remains practically unaffected by the obstacle~\cite{15:Amo}, which is the hallmark signature of superfluidity. When the flow speed matches the critical velocity, quantized$\,$vortices$\,$start$\,$being$\,$nucleated, as reported in$\,$excitons-polariton~\cite{16:Sanvitto, 7-5Nardin}$\,$and$\,$in$\,$cavityless$\,$systems~\cite{2-23Vocke}.$\;$These$\,$vortices$\,$are topological phase singularities characterized by a quantized circulation. They appear$\;$right after$\,$the$\,$breakdown$\,$of$\,$superfluidity.$\;$At$\,$high$\,$velocity,$\,$the$\,$Cherenkov$\,$emission$\,$of$\,$Bogoliubov waves as well as the generation of dark solitons have been observed~\cite{17:Amo, 7-6Grosso}. Nevertheless, in most of the experimental studies cited above, quantitative measurements probing the fluid/superfluid threshold are missing. It would be interesting, for instance, to measure the effects of the defect diameter and height on the critical speed. This requires however to develop a robust technique to measure the flow velocity $v$ at which the transition from normal to superfluid occurs. In the following section, I thus$\,$briefly$\,$review$\,$the$\,$methods$\,$used so far to characterise this threshold.

\newpage

\section{Flow of light around an all-optical defect}

\noindent As mentioned before, probing superfluidity in paraxial photon fluids requires to generate a local variation of refractive index $-$ acting as an obstacle $-$ all along the Kerr medium and to study$\,$how$\,$the$\,$fluid$\,$flows$\,$around.$\;$The$\,$purpose$\,$of$\,$this$\,$section$\,$is$\,$to$\,$provide$\,$a$\,$short$\,$overview of the various flow regimes we can observe in that case by tuning the parameters, namely, the$\,$flow$\,$velocity,$\,$the$\,$fluid$\,$density$\,$and$\,$the$\,$defect$\,$width$\,$and$\,$height.$\;$A$\,$complete$\,$description$\,$of $\,$the$\,$variety$\,$of phenomena observed past an obstacle in$\,$photon$\,$fluids$\,$can$\,$be$\,$found$\,$in~\cite{3-2Carusotto,18:Carusotto}. Throughout all this chapter, the \textbf{probe} names the beam forming the \textbf{photon fluid}.

\subsection{Theory and simulations}

\begin{figure}[H]

\hspace{-1cm}\includegraphics[scale=0.451]{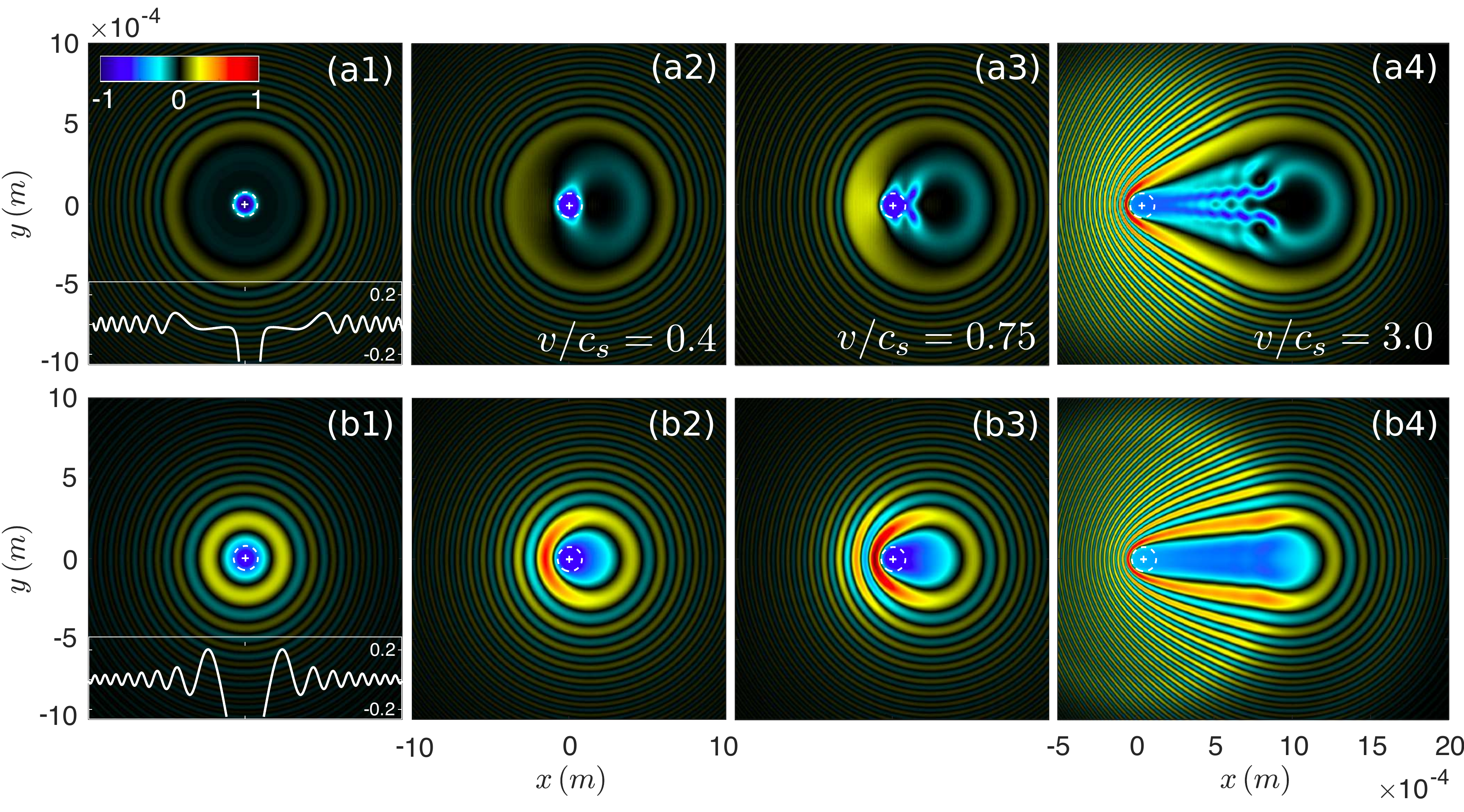}
\caption{Near-field scattering patterns at high (a) and low (b) background$\,$densities$\,$for $v/c_{s} = 0$ (1), $v/c_{s} = 0.4$ (2), $v/c_{s} = 0.75$ (3) and $v/c_{s} = 3.0$ (4).$\;$The$\,$images$\,$(a2)-(a4) show some of the most significant examples of$\,$the$\,$intensity$\,$distribution$\,$in$\,$the$\,$exit$\,$plane.
On all the images (except (a1) and (b1) where the fluid is at rest), the flow goes rightward. (a2): Superfluid regime. The photon fluid moves toward the obstacle (white dotted circle) at a speed$\,$lower$\,$than$\,$the$\,$critical$\,$velocity$\,$and$\,$flows$\,$around$\,$without$\,$any$\,$friction.$\;$This$\,$results in a cancellation of the interference fringes upstream from the obstacle. Those$\;$fringes$\;$are, reversely,$\,$well$\,$contrasted$\,$in$\,$the$\,$linear$\,$regime$\,$(figure$\,$(b2)).$\;$(a2):$\,$Breakdown$\,$of$\,$superfluidity. A pair of quantized vortices is visible downstream from the obstacle. The amount$\,$of$\,$light back-scattered by the defect remains much lower than in the linear case (see figure (b2)). (a3): Highly turbulent$\,$regime. The flow is super-critical (\textit{ie} $v> v_{c}$).$\;$A$\,$Cherenkov$\,$cone$\,$of aperture $\sin (\theta) = c_{s}/v$ forms downstream from the obstacle in that case. On the contrary, in the linear regime (figure (b3)), the fringes resulting from the interference of the incident and scattered light exhibit a standard parabolic shape. Inside the Cherenkov cone,$\;$pairs$\,$of oblique solitons$\,$are$\,$emitted$\,$downstream$\,$from$\,$the$\,$obstacle$\,$and$\,$end$\,$up$\,$breaking$\,$in$\,$quantized vortex/anti-vortex pairs. Parameters: The width of the background beam$\,$is$\,2.0\,$mm$\,$and its on-axis intensity is $\mathcal{I}_{f} = 2.0\!\times\!10^{5}$ W/m$^{2}$. The nonlinear refractive index $n_{2}$ is equal to $1.0\!\times\! 10^{-10}$ m$^{2}/$W, and thus, $\Delta n = 2.0\!\times\! 10^{-5}$.$\;$Consequently,$\,\xi \simeq 30 \, \mu$m$\,$and$\,c_{s} \simeq 4.5\,$mrad. Finally, we set $\omega_{0,d} = \xi$ and $\delta n(0) = 25\!\times\! \Delta n$. The nonlinear medium is 7.5 cm long.
}
\label{fig:SimuSuper}
\end{figure}

\noindent Throughout$\;$this$\;$subsection,$\;$I$\,$consider$\,$that$\,$the$\,$fluid$\,$of$\,$light$\,$is$\,$formed$\,$by$\,$a$\,$wide$\,$and$\,$intense Gaussian probe propagating (in the transverse plane) toward a spatially localized obstacle, located at $\boldsymbol{r}_{\perp} = \boldsymbol{0}$. This obstacle is generated by a Gaussian-shaped modulation $\delta n\,$of$\,$the refractive index of the form:
\begin{equation}
\delta n = \delta n(0) \exp \left[ - \left(r_{\perp}/\omega_{0,d}\right)^{_{2}} \right].    
\end{equation}
\noindent I suppose that the on-axis change of refractive index does not depend on the propagation distance $z$ inside the medium. This condition is fulfilled experimentally$\,$if$\,$the$\,$droplet$\,$beam forming the defect saturates the rubidium vapor across the entire cell. The $z$-evolution of the probe field envelope $\mathcal{E}_{0}$ can be described using the nonlinear Schr\"{o}dinger equation$\,$(2.9). From the hydrodynamical perspective, it is as if the probe beam was flowing toward the defect$\,$at$\,$a$\,$velocity $\boldsymbol{v} = \mathbf{k_{\perp}}/k_{0}\,$(where$\,\mathbf{k_{\perp}} = k_{0} \sin{\theta_{i}} \boldsymbol{e_{x}}\,$is$\,$the$\,$probe$\,$transverse wave-vector). In order to illustrate the various flow$\,$regimes$\,$of$\,$the$\,$photon$\,$fluid$\,$in$\,$this$\,$configuration,$\,$I$\,$solve the NLSE using a second-order split step method. The intensity distribution of the probe beam at the medium output plane is shown on figure~\ref{fig:SimuSuper}, for different sets of parameters.     
The series of images (a1)-(a4) are obtained at high probe power, while the series (b1)-(b4) show the results of the simulation at low probe power, that is, in the quasi-linear regime. On all these figures, the background fluid flows rightward. Its density$\,$has$\,$been$\;$subtracted. 

\noindent The$\,$probe$\,$creates$\,$a$\,$nonlinear$\,$refractive$\,$index$\,$change$\,$of$\,2.0\!\times\! 10^{-5}\,$that$\,$yields$\,$a$\,$healing$\,$length of about $30\, \mu$m and a speed of sound $c_{s}$ of $4.5$ mrad. The obstacle is located at $\boldsymbol{r}_{\perp} = 0$ (white circle).$\;$Its$\,$width$\,\omega_{0,d}\,$is$\,$equal$\,$to$\,$the$\,$healing$\,$length$\,\xi\,$and$\,\delta n(0) = 25 \!\times\! \Delta n$. 
\vspace{6pt}
\newline
\noindent Let's first focus on the series (b1)-(b4). On figure (b1), the photon fluid is at rest ($v=0$). The concentric rings visible on this image are spherical waves emitted from the defect$\,$in$\,$the medium entrance plane. We could observe the formation of a similar pattern by$\,$throwing$\,$a stone in standing water. As you can notice, spherical$\,$waves$\,$are$\,$not$\,$anymore$\,$centered$\,$on$\,$the defect location on figures (b2), (b3) and (b4). The flow velocity is non-zero$\,$on$\,$these$\,$images and steps up$\,$from$\,$left$\,$to$\,$right.$\;$Spherical$\,$waves$\,$are$\,$therefore$\,$dragged$\,$rightward$\,$with$\,$the$\,$flow in that case. At non-zero fluid velocities, we also observe the formation of well contrasted fringes upstream from the obstacle. This is due to interferences between the incoming fluid and the light back-scattered by the defect.
\vspace{4pt}
\newline
\noindent On figure (a1), the photon fluid is at rest but its density is much higher than on$\,$figure$\,$(b1). Spherical waves are also emitted$\,$in$\,$that$\,$case.$\;$However,$\,$they$\,$travel$\,$in$\,$the$\,$transverse$\,$plane$\,$at a speed greater than or equal to the sound velocity $c_{s}$.$\;$This$\,$explains$\,$why$\,$a$\,$region$\,$of$\,$uniform intensity surrounds the defect on figure (a1). This is also clearly visible on the inset profile, 

\newpage

\noindent showing a cut of the intensity distribution along the $x$-axis. On figure (a1), the$\,$flow$\,$velocity is sub-critical ($v<v_{c}$). In that case, the$\,$fluid$\,$flows$\,$around$\,$the$\,$obstacle$\,$without$\,$any$\,$friction. In other words, the incident light$\,$is$\,$not$\,$scattered$\,$by$\,$the$\,$obstacle$\,$in$\,$that$\,$case.$\;$The$\,$absence$\,$of interference fringes upstream from the defect as compared to the linear case (see$\,$figure$\,$(b2)) is a hallmark signature of superfluid flow of light.

\newpage

\noindent On figure (a3), the speed of the flow matches the critical speed. A$\,$pair$\,$of$\,$quantized$\,$vortices (blue$\,$spots)$\,$is$\,$nucleated$\,$downstream$\,$and$\,$starts$\,$being$\,$dragged$\,$by$\,$the$\,$flow.$\;$The$\,$image$\,$(a3)$\,$is therefore obtained right before the breakdown of superfluidity. As you may have noticed, the flow velocity at which this occurs is not equal$\,$to$\,$but$\,$slightly$\,$lower$\,$than$\,c_{s}$ ($v/c_{s} = 0.75$). This$\,$effect$\,$has$\,$been$\,$reported$\,$in~\cite{7-7Frisch}:$\;$in$\,$the$\,$vicinity$\,$of$\,$an$\,$extended$\,$defect$\,$(\textit{ie},$\,$when$\,\xi \lesssim \omega_{0,d}$) the local flow velocity becomes super-critical close to the obstacle, even in the case where the flow is sub-critical (\textit{ie} $v<c_{s}$) far away from it. This is due to the local bending of the streamlines around the extended defect,$\;$which$\,$locally$\,$modifies$\,$the$\,$flow$\,$velocity$\,$of$\,$the$\,$fluid. We consequently expect the critical speed to depend on the fluid velocity as well as on the defect width. This$\,$dependence$\,$is$\,$illustrated$\,$on$\,$figure~\ref{fig:Expectation},$\,$where$\,$the$\,$phase$\,$diagram$\,$obtained by scanning $v$ and $\omega_{0,d}$ has been sketched. The critical speed $v_{c}(v, \omega_{0,d})$, represented by the thick black line, separates the superfluid phase (I) to the regime where$\,$quantized$\,$vortices are generated (II). By increasing continuously the defect size at fixed flow velocity$\,$(red$\,$line, $v/c_{s} = 0.75$), we can thus probe the transition from phase (I) to phase (II).  
\vspace{4pt}
\newline
\noindent Figure$\,$(a4)$\,$illustrates$\,$the$\,$strongly$\,$turbulent$\,$regime$\,$observed$\,$at$\,$high$\,$flow$\,$speeds.$\;$Two$\,$trains of oblique solitons are formed past the obstacle. They end up breaking in pairs$\,$of$\,$quantized vortices$\,$with$\,$opposite$\,$vorticities. We$\,$can$\,$also$\,$notice$\,$the$\,$cone-shaped$\,$structure$\,$of$\,$the$\,$fringes arising from the interference$\,$of$\,$the$\,$incoming$\,$and$\,$scattered$\,$light$\,$downstream$\,$from$\,$the$\,$defect. This effect can be interpreted as a result of Cherenkov radiations of Bogoliubov excitations by the obstacle~\cite{3-2Carusotto} (hence the name of "Cherenkov cone"). The cone angle$\,$is$\,$related$\,$to the sound$\,$velocity$\,$by$\,$the$\,$formula:$\,\sin(\theta) = c_{s}/v$.$\,$On$\,$figure$\,$(b4),$\,$the$\,$intensity$\,$distribution$\,$in$\,$the exit plane obtained$\,$at$\,$low$\,$background$\,$density$\,$is$\,$plotted$\,$for$\,$the$\,$same$\,$flow$\,$velocity$\,$as$\,$on$\,$(a4). In that case, the interference fringes exhibit a standard parabolic shape.   
 
\subsection{Probing the superfluid phase transition}

\begin{figure}[H]
\centering
\includegraphics[scale=0.43]{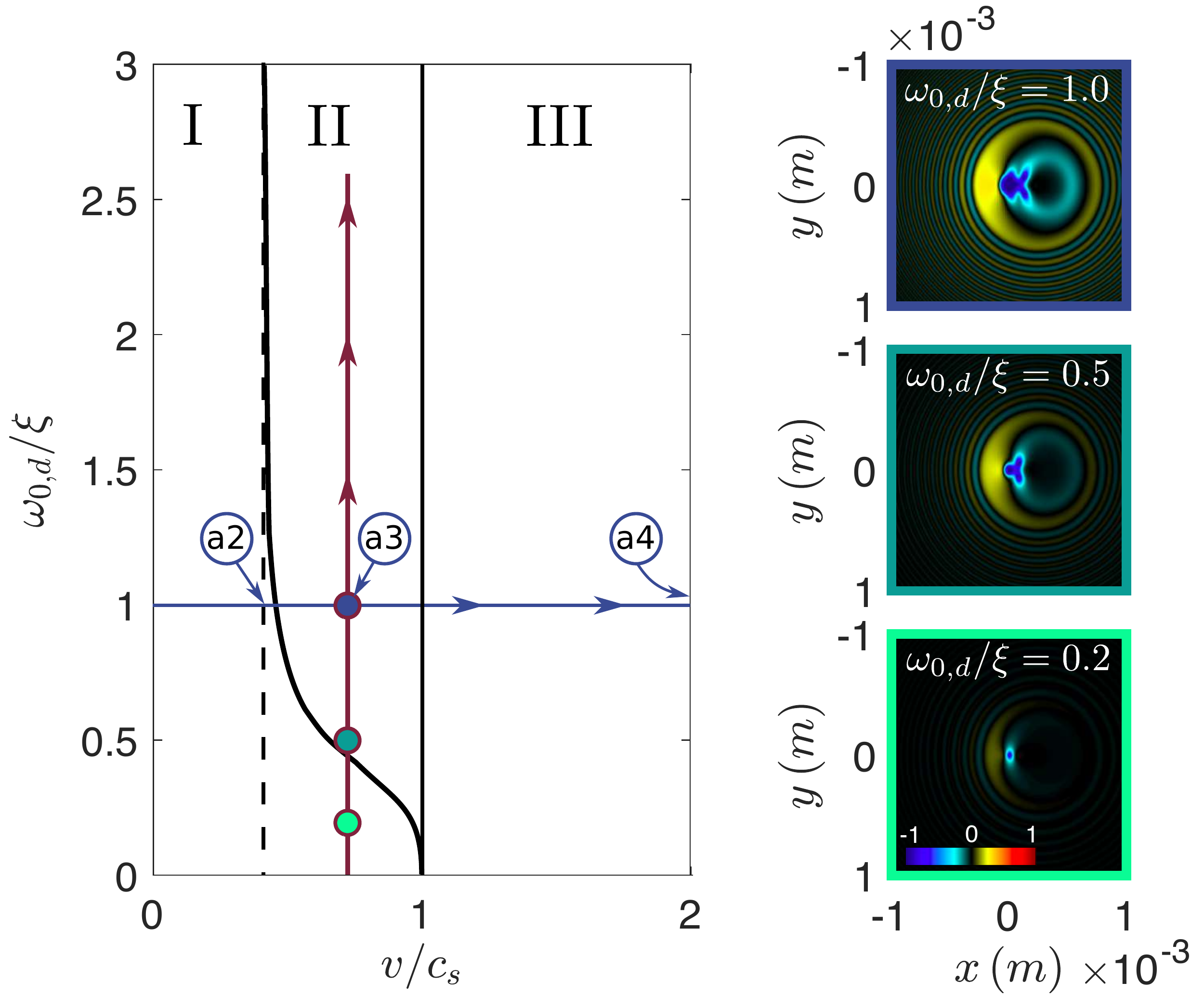}
\caption{Phase diagram obtained by scanning the flow speed$\,v\,$and$\,$the$\,$defect$\,$width$\,\omega_{0,d}$. This diagram splits into three different regions: I,$\,$II$\,$and$\,$III.$\;$The$\,$first$\,$one$\,$(I)$\,$corresponds$\,$to the superfluid phase alone, the second one (II) to the regime where quantized vortices are nucleated$\,$onto$\,$the$\,$superfluid$\,$and$\,$the$\,$third$\,$one$\;$(III)$\;$to$\,$the$\,$supersonic$\,$and$\,$turbulent$\,$regime. By scanning the defect width a fixed background density and flow velocity, we can probe the transition from phase$\,$I$\,$to$\,$II$\,$(red$\,$line).$\;$Indeed,$\,$when$\,\omega_{0,d}/\xi = 0.2\,$(green-framed$\,$image), the fluid flows around the defect without neither scattering nor emitting vortices,$\,$which$\,$is not the case at larger defect width (cyan- and blue-framed$\,$images). Reversely, we$\,$can$\,$fixed the defect size and scan the flow velocity (blue solid line). In that case, both the transitions I/II and II/III are probed. This is what has been done on figures \ref{fig:SimuSuper}(a1)-(a4) for instance. The simulation parameters are the same as for figures~\ref{fig:SimuSuper}.
}
\label{fig:Expectation}
\end{figure}

\noindent The ways of tracking superfluidity in photon fluids are manifold.$\;$In$\,$this$\,$subsection,$\,$I$\,$review some of the experimental methods developed so far for this purpose.
\vspace{-4pt}
\begin{itemize}
    \item [$\bullet$] The detection of quantized vortices in the wake of an obstacle$\,$is$\,$commonly$\,$considered as a hallmark signature of superfluidity. This$\,$is$\,$the$\,$method I$\,$use$\,$to$\,$numerically$\,$probe the$\,$transition$\,$from$\,$phase$\,$(I)$\,$to$\,$phase$\,$(II)$\,$in$\,$figure$\,$\ref{fig:Expectation}.$\;$The$\,$observation$\,$of$\,$such$\,$vortices has been reported in exciton-polariton$\,$systems~\cite{16:Sanvitto, 7-5Nardin}$\,$and$\,$more$\,$recently$\,$in$\,$paraxial photon$\,$fluids~\cite{2-23Vocke}. The dimension of vortices is of the order of the$\,$healing$\,$length~\cite{2-23Vocke}, which is typically few tens of microns in our experiment. Therefore, using standard spatial phase interferometry is enough to locate$\,$vortices in our system, which makes photon$\,$fluids$\,$a$\,$promising$\,$toolbox$\,$for$\,$studying$\,$vortex$\,$nucleation$\,$in$\,$superfluid$\,$light. 
    \vspace{-4pt}
    \item[$\bullet$] As$\,$we$\,$saw$\,$in$\,$the$\,$previous$\,$section,$\,$the$\,$elastic$\,$scattering$\,$on$\,$the$\,$defect$\,$at$\,$low$\,$background densities is no longer expected to occur at high densities (when $v<v_{c}$), because$\,$the light becomes superfluid. Therefore, at the superfluid transition, the amount$\,$of$\,$light scattered by the defect should sharply drop. This can either be measured$\,$in$\,$real$\,$space (see subsection 6.2.3) or in Fourier space, by measuring$\,$the$\,$power$\,$distributed$\,$over$\,$the so-called Rayleigh ring. Amo \textit{et al.} have experimentally demonstrated the collapse$\,$of this scattering ring at the normal/superfluid threshold in~\cite{15:Amo}. In$\,$the$\,$subsection$\,$6.3.2, examples of Rayleigh rings generated by the scattering of the photon fluid on our all-optical defect are shown (see for instance figure~\ref{fig:ScatteringVSPower}(a)). 
\end{itemize}

\newpage

\begin{itemize}
    \item[$\bullet$] In quantum fluids, the normal/superfluid transition is usually observed through the drop of the drag force exerted$\,$by$\,$the$\,$quantum$\,$fluid$\,$on$\,$a$\,$movable$\,$obstacle~\cite{7-8Rayfield,7-9Castelijns,7-10Raman,7-11Miller}. Following theoretical works studying the cancellation of the drag force in superfluid exciton-polariton$\,$condensates~\cite{7-12Wouters,7-13Berceanu,7-14Larre, 7-15Larre},$\,$Larré$\,$\textit{et$\,$al.}$\,$have$\,$recently$\,$proposed$\,$a$\,$concrete way of measuring it in propagating photon fluids~\cite{6-2Larre}. As mentioned in their paper, the propagating geometry offers the possibility of using movable and/or deformable obstacles by directly immersing a piece of dielectric (a rod) in the nonlinear$\,$material. 
    The local variation of the fluid density around the dielectric generates a force$\;$on$\;$the obstacle which, by moving backward, induces$\,$a$\,$spatial$\,$variation$\,$of$\,$the$\,$liquid$\,$pressure. The joint action of both these effects results in a total force per$\,$unit$\,$area$\,$acting$\,$on$\,$the defect that is proportional to the upstream/downstream variation of the$\,$fluid$\,$density. Larré \textit{et al.} predicts that this force linearly increases with the background density in the normal regime before suddenly dropping to zero when$\,$reaching$\,$the$\,$superfluid$\,$one. 
    
    \newpage
    
    However,$\;$installing$\,$a$\,$piece$\,$of$\,$dielectric$\,$inside$\,$a$\,$nonlinear$\,$medium$\,$is$\,$a$\,$challenging$\,$task in practice. In subsection 6.3.3, I briefly present$\,$a$\,$way$\,$of$\,$doing$\,$it$\,$in$\,$rubidium$\,$vapors. An alternative is to generate the defect optically $-$ as we do using droplet beams$\;-$ and$\,$try$\,$to$\,$measure$\,$the$\,$optical$\,$analog$\,$of$\,$the$\,$drag$\,$force$\,$cancellation$\,$by$\,$measuring$\,$either the upstream/downstream variation in the intensity, $\mathcal{I}_{+}-\mathcal{I}_{-}$, or directly the defect displacement in the output plane of the medium. This is what Michel \textit{et al.}$\,$have$\,$done experimentally; they report in~\cite{3-3Michel} the first clear evidence of superfluid flow of light in the paraxial geometry. The following subsection is a review of their work.   
 
\end{itemize}

\subsection{Comments on the results of Michel \textit{et al.}}

\noindent As mentioned previously, Michel \textit{et al.} report in~\cite{3-3Michel} two distinct experimental evidences of the normal/superfluid transition occurring in their propagating fluid of light.$\;$They$\,$use$\,$a $1\,$cm long photo-refractive crystal (SBN:61) as Kerr medium in$\,$their$\,$experiment.$\,$The$\,$defect is generated by saturating the crystal with an intense Bessel beam whose$\,$central$\,$core$\,$radius is about $6 \, \mu$m. The on-axis refractive$\,$index$\,$depletion$\,\delta n(0)\,$it$\,$creates$\,$is$\,$equal$\,$to$\,-2.2 \!\times\! 10^{-4}$. The photon fluid is formed by$\,$a$\,$Gaussian$\,$probe$\,$whose$\,$width$\,$is$\,270 \mu$m.$\;$The$\,$fluid$\,$intensity, $\mathcal{I}_{f}$, can be tuned from $0$ to $350$ mW/cm$^{2}$ and the$\,$flow$\,$velocity,$\,v$,$\,$from$\,0\,$to$\,\pm 1.3 \!\times\! 10^{-2}\,$mrad.

\begin{figure}[H]
\centering
\includegraphics[width=\columnwidth]{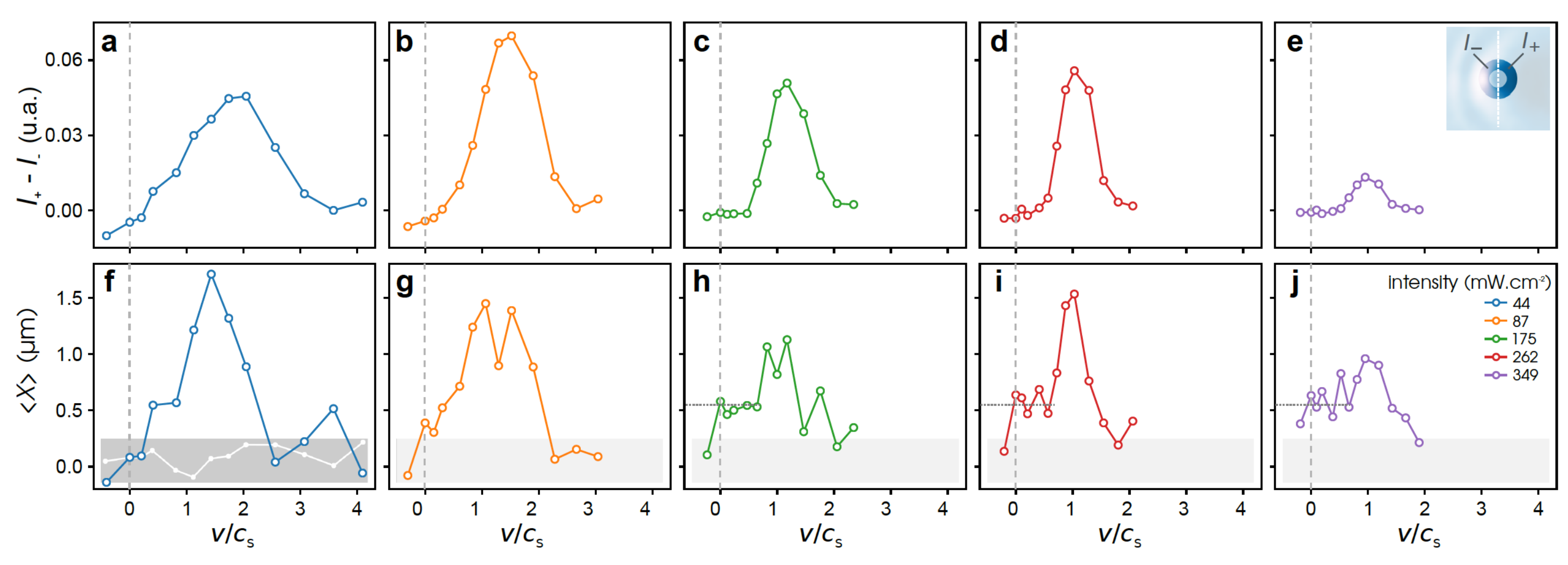}
\caption{Results$\,$obtained$\,$by$\,$Michel$\,$\textit{et$\,$al.}$\,$in~\cite{3-3Michel}.$\;$Figures$\,$(a-e):$\,$Local$\,$intensity$\,$difference $\mathcal{I}_{+}-\mathcal{I}_{-}$ measured at the exit plane of the photo-refractive crystal they use in experiments as function of $v/c_{s}$, for different values of the$\,$fluid$\,$intensity$\,\mathcal{I}_{f}$.$\;$The$\,$basic$\,$measurement$\,$idea is illustrated on the inset of figure (e). The original image is cropped around the obstacle and the intensity is averaged over two$\,$regions$\,$of$\,$space,$\,$downstream$\,$($\mathcal{I}_{-}$)$\,$and$\,$upstream$\,$($\mathcal{I}_{+}$) from the defect. At low background densities (figures$\,$(a)$\,$and$\,$(b)),$\,\mathcal{I}_{+}-\mathcal{I}_{-}\,$starts$\,$increasing from $v/c_{s} = 0$ while at high densities (figures (c) to (e)),$\,$a$\,$plateau$\,$forms$\,$at$\,$low$\,$flow$\,$speeds, which ends when $v$ matches the critical speed $v_{c}$. This is a clear signature of superfluidity. Michel \textit{et al.} go further by probing the transverse displacement $\langle X \rangle\,$of$\,$the$\,$obstacle$\,$induced by the local variations in the intensity of the background fluid$\,$for$\,$the$\,$same$\,$input$\,$conditions as for figures (a) to (g). The results are shown on figures (f) to (j).$\;$While$\,$the$\,$defect$\,$starts being dragged with the flow from $v/c_{s} = 0$ on (f) and (g), its position$\,$remains$\,$unchanged$\,$at low flow velocities on (h) and (i). This constitutes a hallmark signature of the cancellation of the drag force exerted by the fluid on the obstacle at the normal/superfluid transition.}
\label{fig:Bellec}
\end{figure}

\newpage

\noindent When the$\,$fluid$\,$power$\,$is$\,$maximum,$\,\Delta n\,$is$\,$about$\,1\! \times \! 10^{-4}\,$which$\,$yields$ \,\xi \simeq 6.2 \, \mu$m.$\,$Therefore, the$\,$defect$\,$is$\,$always$\,$smaller$\,$than$\,$or$\,$equal$\,$to$\,\xi\,$in$\,$the$\,$experiment$\,$carried out$\,$by$\,$Michel$\,$\textit{et$\,$al.}
\vspace{6pt}
\newline
\noindent Figure~\ref{fig:Bellec}, taken from~\cite{3-3Michel}, shows the results they obtained$\,$using$\,$the$\,$configuration$\,$above. On figures$\,$(a-e),$\,$the$\,$intensity$\,$variation$\,\mathcal{I}_{+}-\mathcal{I}_{-}\,$at$\,$the$\,$crystal$\,$exit$\,$plane$\,$is$\,$plotted$\,$as$\,$function of $v/c_{s}$ for various probe intensities. At low background densities (see figures (a) and (b)), $\mathcal{I}_{+}-\mathcal{I}_{-}$ increases from $v/c_{s} = 0$. In other words, as soon as the flow velocity is non-zero, the$\,$obstacle$\,$starts$\,$being$\,$dragged$\,$by$\,$the$\,$fluid.$\;$On$\,$the$\,$contrary,$\,$at$\,$high$\,$background$\,$densities, 
(see figures (c), (d) and (e)), a plateau at $\mathcal{I}_{+}-\mathcal{I}_{-}=0$ clearly forms at low flow velocities, which indicates that the drag force exerted by the fluid on$\,$the$\,$obstacle$\,$vanishes$\,$in$\,$that$\,$case. This demonstrates that light$\,$is$\,$superfluid$\,$deep$\,$in$\,$the$\,$subsonic$\,$regime.$\;$As$\,$mentioned$\,$in~\cite{3-3Michel}, the$\,$curves$\,$obtained$\,$at$\,$different$\,$intensities$\,$do$\,$not$\,$fall$\,$on$\,$a$\,$single$\,$universal$\,$curve,$\,$although$\,$the flow velocity is normalized on each graph by the related speed of sound. This is expected, as changing the probe intensity affects both the healing length and the relative strength$\,$of the obstacle with respect to the nonlinear change of refractive index, \textit{ie} the ratio $\delta_{n}(0)/\Delta n$. This index matching effect may explain why the maximal amount of scattered light on figure (e) is much lower than on figures (a)-(d). When $\mathcal{P}_{f} = 349$ mW, $\Delta n$ is$\,$of$\,$the$\,$order$\,$of $1.0 \!\times\! 10^{-4}$ which yields $\delta_{n}(0)/\Delta n \simeq 2$. The defect$\,$then$\,$only$\,$induces$\,$a$\,$perturbative$\,$potential. This is why it is barely visible at high fluid power on the image 2(c) of~\cite{3-3Michel}. Nevertheless, index matching can not explain the plateau$\,$observed$\,$on$\,$figures$\,$(a)-(e)$\,$that$\,$really$\,$constitute a clear signature of superfluidity. 
\vspace{6pt}
\newline
\noindent In order to go further, Michel \textit{et al.} have measured the position $\langle X \rangle$ of the obstacle at$\,$the crystal exit plane, as function of $v/c_{s}$ still, for the same probe intensities$\,$as$\,$on$\,$figures$\,$(a-e). The method used to that end is described in details in the supplementary materials of~\cite{3-3Michel}. I thus only comment the experimental results obtained with this second technique,$\,$that$\,$are shown on figures (f-j). As you may have noticed, the position of the defect$\,$at$\,$the$\,$exit$\,$plane starts drifting from $v/c_{s} = 0$ at low background densities, on figures (g) and (h). Reversely, it remains constant at low flow velocities on figures (h) and (i), where$\,$the$\,$density$\,$is$\,$higher. Nevertheless, the defect displacement does not cancel exactly in that case, even$\,$deep$\,$in$\,$the subsonic regime. This surprising effect is not explained$\,$in$\,$text.$\;$However,$\,$the$\,$results$\,$are$\,$still in good agreement with those in figures (a-e). For example, the$\,$critical$\,$speeds$\,$in$\,$(c)$\,$and$\,$(d) correspond fairly well to the ones measured in (h) and (i).$\;$It$\,$would$\,$have$\,$been$\,$interesting$\,$to see if the motion of the defect in real space leads additionally to$\,$a$\,$drift$\,$of$\,$the$\,$ ring-shaped far-field intensity distribution of the Bessel beam forming the obstacle. If it is the case, this could have provided a easier way of probing the normal/superfluid transition.  

\section{Preliminary results}

\noindent In this section, I present the$\,$preliminary$\,$results$\,$we$\,$obtained$\,$recently$\,$using$\,$droplet$\,$beams$\,$to generate an all-optical defect into our photon fluid. I first describe the experimental setup before showing$\,$images$\,$of$\,$the$\,$scattering$\,$patterns$\,$we$\,$observe$\,$both$\,$in$\,$real$\,$and$\,$in$\,$Fourier$\,$space. In this latter case,$\,$we$\,$ measure$\,$the$\,$amount$\,$of$\,$light$\,$scattered$\,$into$\,$the$\,$Rayleigh$\,$ring$\,$at$\,$various flow velocities and show that it$\,$quickly$\,$drops$\,$with$\,$the$\,$fluid$\,$intensity$\,$passing$\,$some$\,$threshold. However, the reasons underlying this phenomenon are still not clear. It could arise from superfluidity as well as from the reduction of the obstacle strength at high fluid intensities because of saturation. The results shown in this section should$\,$thus$\,$be$\,$viewed$\,$with$\,$caution.

\newpage

\subsection{Experimental setup and settings}

\begin{figure}[H]
\centering
\includegraphics[width=\columnwidth]{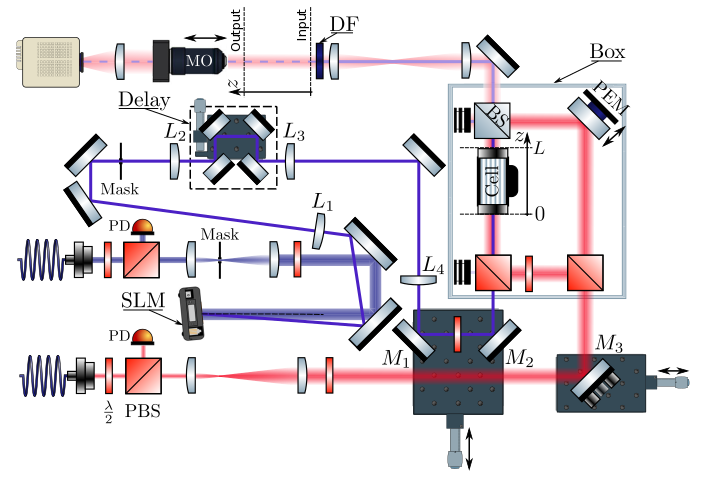}
\caption{Experimental setup. The paths of the defect and the probe beams are sketched in blue and red respectively. The way the$\,$defect$\,$beam$\,$is$\,$generated$\,$by$\,$shaping,$\,$in$\,$real$\,$space, the phase of a wide Gaussian beam with a phase-only SLM is described in subsection 5.2.3. The resulting droplet beam has an axial extent of almost 2.5 cm and a width$\,\omega_{0,d}\,$of$\,40\,\mu$m. Its propagation direction in the cell defines the optical axis. The probe beam is magnified before entering the Mach-Zehnder interferometer. At that point, it splits into a low power (reference) and a high power parts. The low power beam provides the reference we need$\,$to retrieve the phase of the photon fluid at the$\,$medium$\,$output$\,$plane$\,$(using$\,$the$\,$scanning$\,$phase interferometry introduced in paragraph 2.3.3 i). The high power one enters with the defect inside the cell. The angle $\theta_{i}$ between$\,$these$\,$beams$\,-\,$that$\,$is,$\,$the$\,$speed$\,$at$\,$which$\,$the$\,$fluid$\,$flows toward the obstacle $-$ can be finely tuned using$\,$the$\,$piezo-actuated$\,$mount$\,$of$\,$the$\,$mirror$\,M_{3}$.
The diffraction patterns observed on figure~\ref{fig:RealSpace} are obtained by imaging the cell exit plane with the imaging system described in subsection 5.2.3. By removing$\,$the$\,$objective$\,$from$\,$the beam path, we are able to image the k-space and observe the diffraction rings shown on figures~\ref{fig:FourierSpace} and~\ref{fig:ScatteringVSPower}(a). The rings of the droplet beam are barely visible on those$\,$images$\,$as it is filtered out by the bandpass dichroic filter (DF).}
\label{fig:DefectExp}
\end{figure}

\noindent The experimental setup is sketched on figure~\ref{fig:DefectExp}. The way the defect beam$\,$(blue)$\,$is$\,$created is described in subsection 5.2.3. I will therefore$\,$not$\,$comment$\,$the$\,$blue$\,$path$\,$here.$\;$For$\,$details, please refer to figure 5.13. The probe (red) is a continuous-wave laser beam produced by the Ti-Sapphire laser source of subsection 2.1.2. It is sent onto the optical table through$\,$a single-mode polarization-maintaining fiber. The power of the outgoing beam can be tuned turning the half wave-plate facing the fiber output. The$\,$probe$\,$beam$\,$is$\,$magnified$\,$four$\,$times, reflects onto the mirror $M_{3}$ (hold in a piezo-actuated mirror mount) and gets inside the Mach-Zehnder interferometer afterwards. The latter is, as usual,$\,$protected$\,$against$\,$air$\,$flows by$\,$a$\,$box$\,$made$\,$of$\,$Plexiglas$\,$(which$\,$greatly$\,$enhances$\,$the$\,$interferometer$\,$stability).$\,$A$\,$small$\,$part of the probe beam is$\,$transmitted$\,$by$\,$the$\,$PBS$\,$while$\,$the$\,$remaining$\,$high$\,$power$\,$part$\,$is$\,$reflected toward the cell. The transmitted beam (referred to as the reference from now on)$\,$provides the reference we need to retrieve the phase of the probe beam at the medium output plane. It reflects onto a mirror mounted on a piezo-actuated translation stage (PEM). Scanning the high tension applied across the piezo allows to modulate the length$\,$of$\,$the$\,$reference$\,$arm and thus to scan over $2\pi$ the relative phase between probe and reference.$\;$The$\,$probe$\,$and$\,$the defect beams enter together inside the 2.5 cm cell, filled with a pure vapor of rubidium$\,$85. In this configuration, they are co-propagating. The photon fluid generated by the probe flows toward the defect at a velocity $\boldsymbol{v} = \mathbf{k_{\perp}}/k_{0}$, where $\mathbf{k_{\perp}} = k_{0} \sin(\theta_{i}) \boldsymbol{e_{x}}$ stands for the probe transverse wave-vector. In this experiment, the $z$-axis is defined by the propagation direction of the defect beam. The piezo-actuated mount of $M_{3}$ allows to finely tune the angle $\theta_{i}$ between probe and defect. The cell exit plane is imaged onto the CMOS camera using the imaging system described in subsection 5.2.3. The overall magnification factor is $G=10.8 \pm 0.2$. A dichroic bandpass filter (DF) $-$ whose center wavelength$\,$is$\,780\,$nm$\,-$ filters$\,$out$\,$the defect beam,$\,$which$\,$addresses$\,$the$\,$rubidium$\,D_{1}\,$line$\,$($\lambda_{d} \simeq 795\,$nm).$\;$In$\,$this$\,$way, only the scattering pattern created by the obstacle onto the photon fluid is captured.
\vspace{6pt}
\newline
\noindent At$\,$the$\,$time$\,$at$\,$which$\,$preliminary$\,$data$\,$were$\,$acquired,$\,$a$\,$clear$\,$understanding$\,$of$\,$the$\,$mechanism underlying$\,$the$\,$generation$\,$of$\,$the$\,$refractive$\;$index$\,$modulation$\,$in$\,$the$\,$photon$\,$fluid$\,$was$\,$missing. We thus chose the laser detunings in order to fulfill certain basic conditions. We first want the defect to be collimated inside the cell. We then look for a detuning $\Delta_{d}$ at which the variation of the droplet width between the input and output planes does not exceed $10\%$. We also want the height of the repulsive potential $\delta n$ to be as large as$\,$possible$\,$of$\,$course. In$\,$experiments,$\;$the$\,$vapor$\,$temperature$\,$is$\,$about$\,140^{\circ}$C$\,$and$\,$the$\,$probe$\,$is$\,$2.0$\,$GHz$\,$red-detuned from the $F=3 \rightarrow F'$ transition of the $^{_{85}}$Rb $D_{2}$ line. In that case, we$\,$empirically$\,$find$\,$that the best was to blue-detuned the defect beam from the $F_{g} = 3 \rightarrow F_{e} = 3$ transition of the $^{_{85}}$Rb $D_{1}$ line by 400-500 MHz. I$\,$would$\,$like$\,$to$\,$mention$\,$here$\,$that$\,$an$\,$isotopically$\,$pure$\,$vapor$\,$of rubidium 87 is$\,$more$\,$suitable$\,$to$\,$perform$\,$this$\,$experiment. Indeed,$\,$the$\,$vapor$\,$we$\,$used$\,$contains a small but non-zero fraction of $^{_{87}}$Rb atoms, which are almost at resonance with the probe beam when it is 2.0 GHz red detuned from the $F=3 \rightarrow F'$ transition of the$\,^{_{85}}$Rb $D_{2}\,$line. This is clearly visible on figure 2.3(b) for instance.$\;$Even$\,$if$\,$the$\,$fraction$\,$of$\,^{_{87}}$Rb$\,$atoms$\,$is$\,$small, the transmission of the laser beam is greatly reduced when it gets closer to the $F=2 \rightarrow F'$ transition of the $^{_{87}}$Rb $D_{2}$ line. In addition to increasing the absorption of the probe beam, $^{_{87}}$Rb$\,$atoms also affect the refractive index it feels. Using a vapor$\,$of$\,$rubidium$\,$87$\,$would$\,$solve this issue, as red-detuning the probe from the $F=2 \rightarrow F'$ transition of the $^{_{87}}$Rb $D_{2}$ line will not bring it at resonance with any$\,$transition$\,$of$\,^{_{85}}$Rb. 

\newpage

$\vphantom{a}$

\newpage

\subsection{Real-space scattering pattern}

\begin{figure}[H]
\centering
\includegraphics[width=\columnwidth]{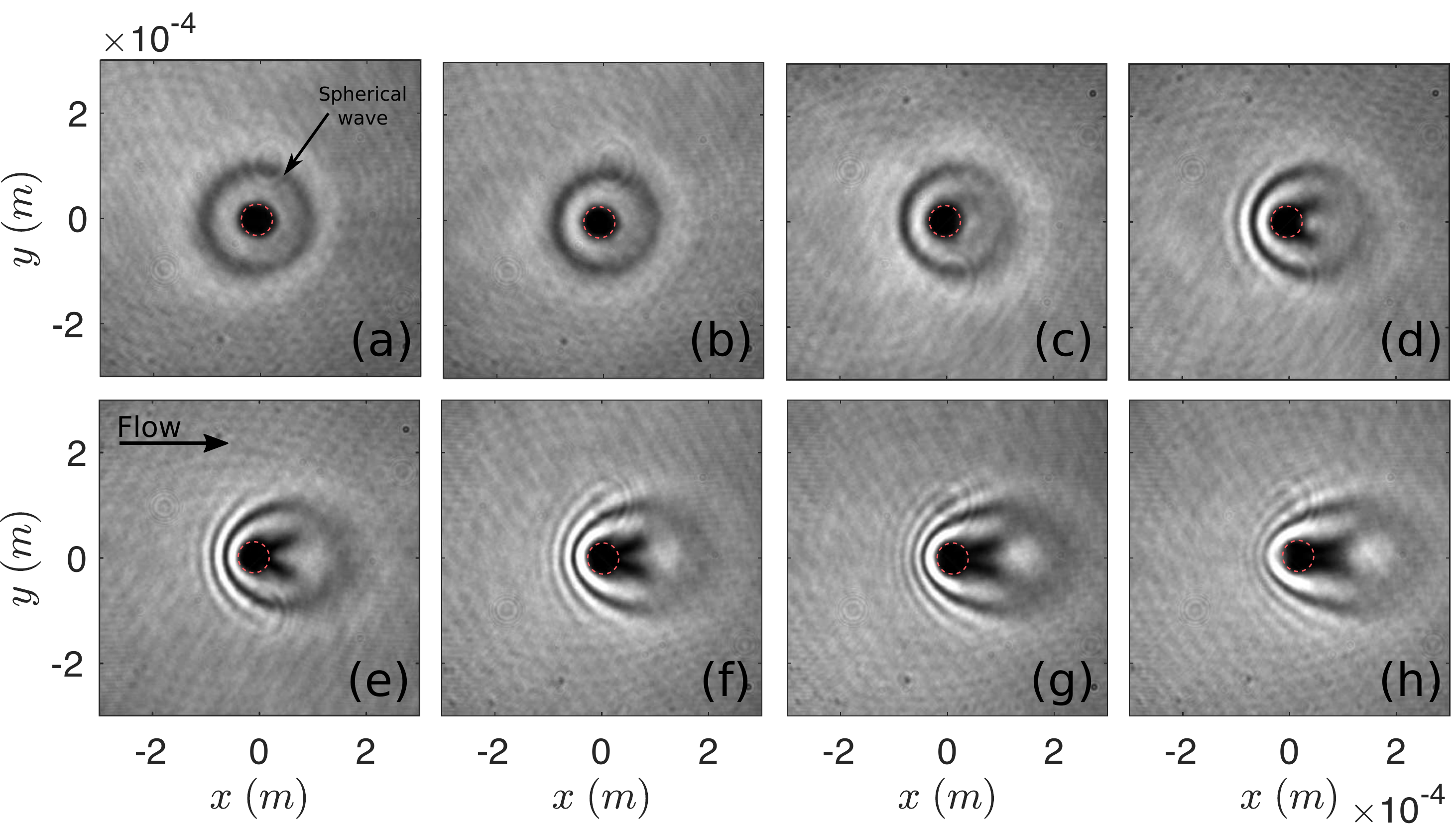}
\caption{Near-field scattering patterns at low background density $\rho_{0}$.$\;$The$\,$droplet$\,$beam generates a negative Gaussian index modulation$\,$($\delta n < 0$),$\,$whose$\,$width$\,\omega_{0,d}\,$is$\,$about$\,40\,\mu$m. The$\,$resulting$\,$potential$\,$is$\,$repulsive$\,$explaining$\,$why$\,$a$\,$dark$\,$spot$\,$is$\,$visible$\,$at$\,$the$\,$position$\,$where the defect lies (red dotted circle). (a): The$\,$photon$\,$fluid$\,$is$\,$at$\,$rest$\,$($v = 0$).$\;$A$\,$spherical$\,$wave is created as soon as probe and defect enter the cell and propagate away from the defect. (b)-(h): The photon fluid is flowing rightward. The flow velocity steps up$\;$from$\;$one$\;$image to the next. On figure (d), interference fringes upstream$\,$from$\,$the$\,$obstacle$\,$start$\,$appearing. Light is scattered backward by the repulsive defect and interferes$\,$with$\,$the$\,$incoming$\,$fluid. On$\,$(g)$\,$and$\,$(h),$\,$the$\,$contrast$\,$of$\,$the$\,$fringes$\,$before$\,$the$\,$obstacle$\,$decreases.$\;$The$\,$kinetic$\,$energy$\,$of the photon fluid is large compared to the height of the potential and light just go through without being back-scattered~\cite{7-16Albert}. We can finally notice that the spherical$\,$wave$\,$visible$\,$on figure (a) is dragged with the flow and drifts rightward.}
\label{fig:RealSpace}
\end{figure}

\noindent All the results I present in this subsection have been obtained by$\,$generating$\,$obstacles$\,$with the droplet beam whose transverse$\,$profile$\,$is$\,$shown$\,$on$\,$figure$\,$5.17(b).$\;$The$\,$resulting$\,$repulsive potential is Gaussian and its width a bit larger than $40 \, \mu$m. The$\,$maximum$\,$on-axis$\,$intensity reachable in experiments using this droplet beam is about $1.3 \!\times\! 10^{7}$ W/m$^{2}$. Only 10$\%\,$of$\,$the beam overall power is distributed$\,$over$\,$the$\,$droplet$\,$central$\,$core.$\;$This$\,$explains$\,$why$\,$we$\,$cannot easily reach higher on-axis intensity experimentally. However,$\,$this$\,$is$\,$enough$\,$to$\,$saturate$\,$the rubidium vapor across the entire cell. The probe beam is much larger than$\,$the$\,$droplet$\,$core ($\omega_{0,f} \simeq 0.9\,$mm)$\,$in$\,$order$\,$for$\,$the$\,$fluid$\,$density$\,$to$\,$be$\,$nearly$\,$uniform$\,$in$\,$the$\,$vicinity$\,$of$\,$the$\,$defect.
Images of the near-field diffraction patterns observed, at low fluid density (linear regime), by imaging the cell exit plane onto the camera are shown on figure~\ref{fig:RealSpace}.

\newpage

\noindent Figure~\ref{fig:RealSpace}(a) shows the image obtained when the photon fluid$\,$is$\,$at$\,$rest$\,$(no$\,$transverse$\,$flow). The obstacle is located at $\boldsymbol{r}_{\perp} = \boldsymbol{0}$ (red dotted circle). As$\,$the$\,$potential$\,$is$\,$repulsive$\,$($\delta n < 0$), the fluid is pushed away$\,$from$\,$the$\,$defect,$\,$thereby$\,$creating$\,$the$\,$dark$\,$central$\,$spot$\,$on$\,$the$\,$image. A spherical wave is emitted in the cell input plane and travels radially onto$\,$the$\,$photon$\,$fluid. On figure (a), the benefit of using droplet beams to generate the obstacle is clearly visible. While standard quasi Bessel-Gauss beams induce a concentric succession of ring-shaped refractive$\,$index modulations in the fluid (as$\,$can$\,$been$\,$seen$\,$on$\,$figure$\,$2(a)$\,$of~\cite{3-3Michel}$\,$for$\,$example), droplet beams only create one main index depletion, since the power distributed over the external rings is drastically reduced, because of interferences. From figure (b) to$\,$figure$\,$(h), the flow velocity of the$\,$photon$\,$fluid$\,$progressively$\,$steps$\,$up$\,$and$\,$interference$\,$fringes$\,$start$\,$thus developing upstream from the obstacle. As in simulations of figure~\ref{fig:SimuSuper}, the spherical wave is dragged rightward by the photon fluid. We can also notice that the$\,$contrast$\,$of$\,$the$\,$fringes in figures (g) and (e) is lower than at smaller flow velocities. At$\,$such$\,$high$\,$speeds,$\,$the$\,$kinetic energy of the fluid is certainly larger than the potential barrier formed$\,$by$\,$the$\,$obstacle$\,$and photons start thus tunneling through. This tunneling effect, theoretically$\,$described$\,$in~\cite{7-16Albert}, have been studied experimentally by Wan \textit{et$\,$al.}~\cite{27:Wan}$\,$in$\,$1D$\,$photon$\,$fluids. 
\vspace{-5pt}
\subsection{Far-field diffraction pattern}

\begin{figure}[H]
\centering
\includegraphics[width=\columnwidth]{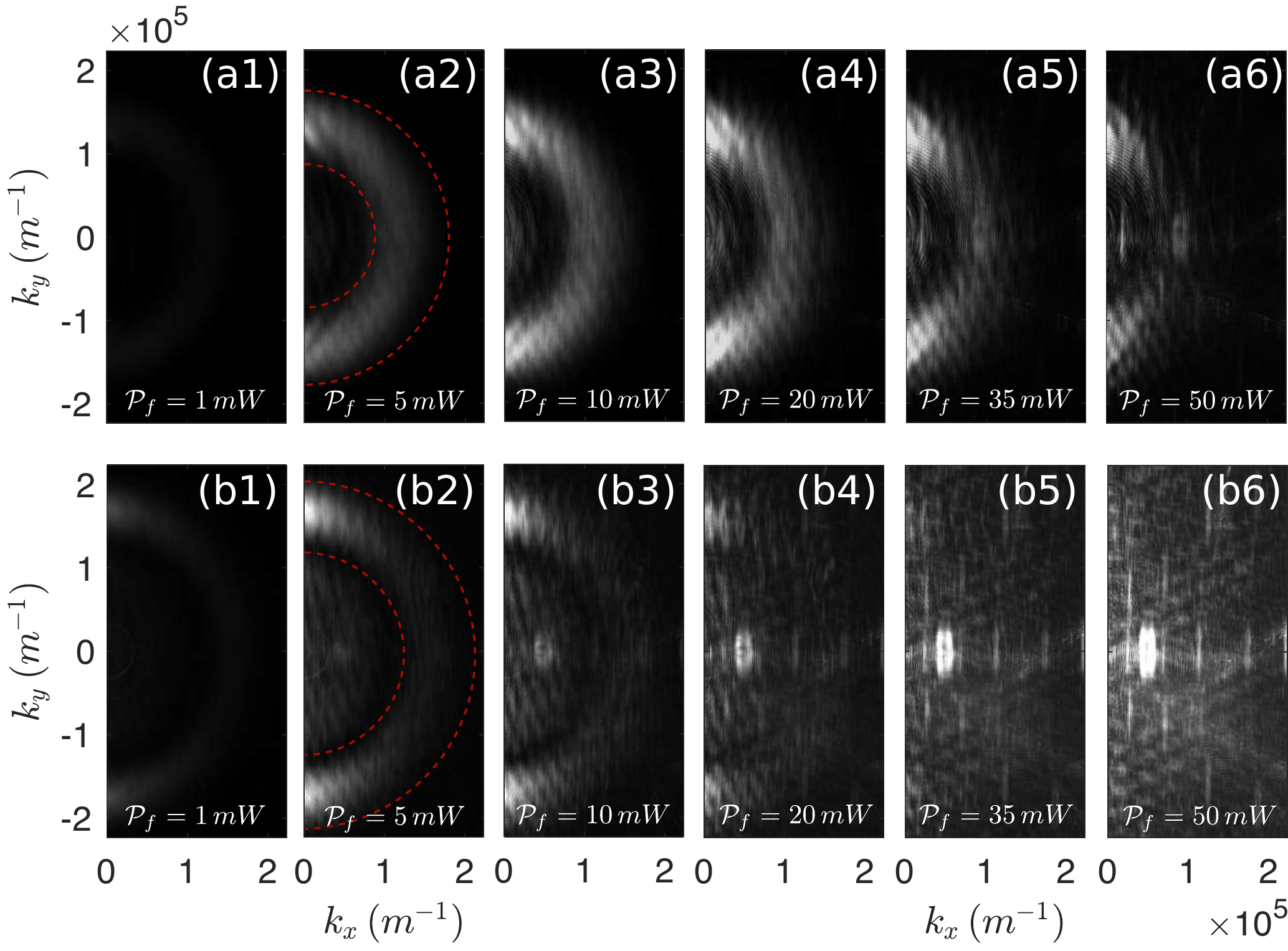}
\caption{Diffraction patterns observed in far-field (k-space) at different$\,$fluid$\,$powers$\,\mathcal{P}_{f}$, for $v = 18$ mrad (a) and $v = 23$ mrad (b). The $k_{x}$/$k_{y}$-axes origin-point lies at the center$\,$of the droplet rings. The probe is located at $k_{x} = - k_{0} v$ and is thus not visible on the images. The exposure is the same for the upper (a) and lower series (b). The$\,$scattering$\,$of$\,$the$\,$fluid on the defect is responsible for the apparition of a ring-shaped structure in far field whose radius is defined by the norm of the probe transverse wave-vector. On series (a) and (b), the power distributed over the scattering ring increases first with $\mathcal{P}_{f}$, reaches a threshold and$\,$then$\,$decreases.$\;$A$\,$more$\,$quantitative$\,$analysis$\,$can$\,$be$\,$found$\,$on$\,$figure$\,$\ref{fig:ScatteringVSPower}(b).$\;$On$\,$series$\,$(a), the$\,$radius$\,$of$\,$the$\,$ring$\,$seems$\,$to$\,$decrease$\,$with$\,\mathcal{P}_{f}$.$\;$A$\,$similar$\,$behaviour$\,$is$\,$reported$\,$in~\cite{3-2Carusotto}.
}
\label{fig:FourierSpace}
\end{figure}

\begin{figure}[H]
\centering
\includegraphics[width=0.99\linewidth]{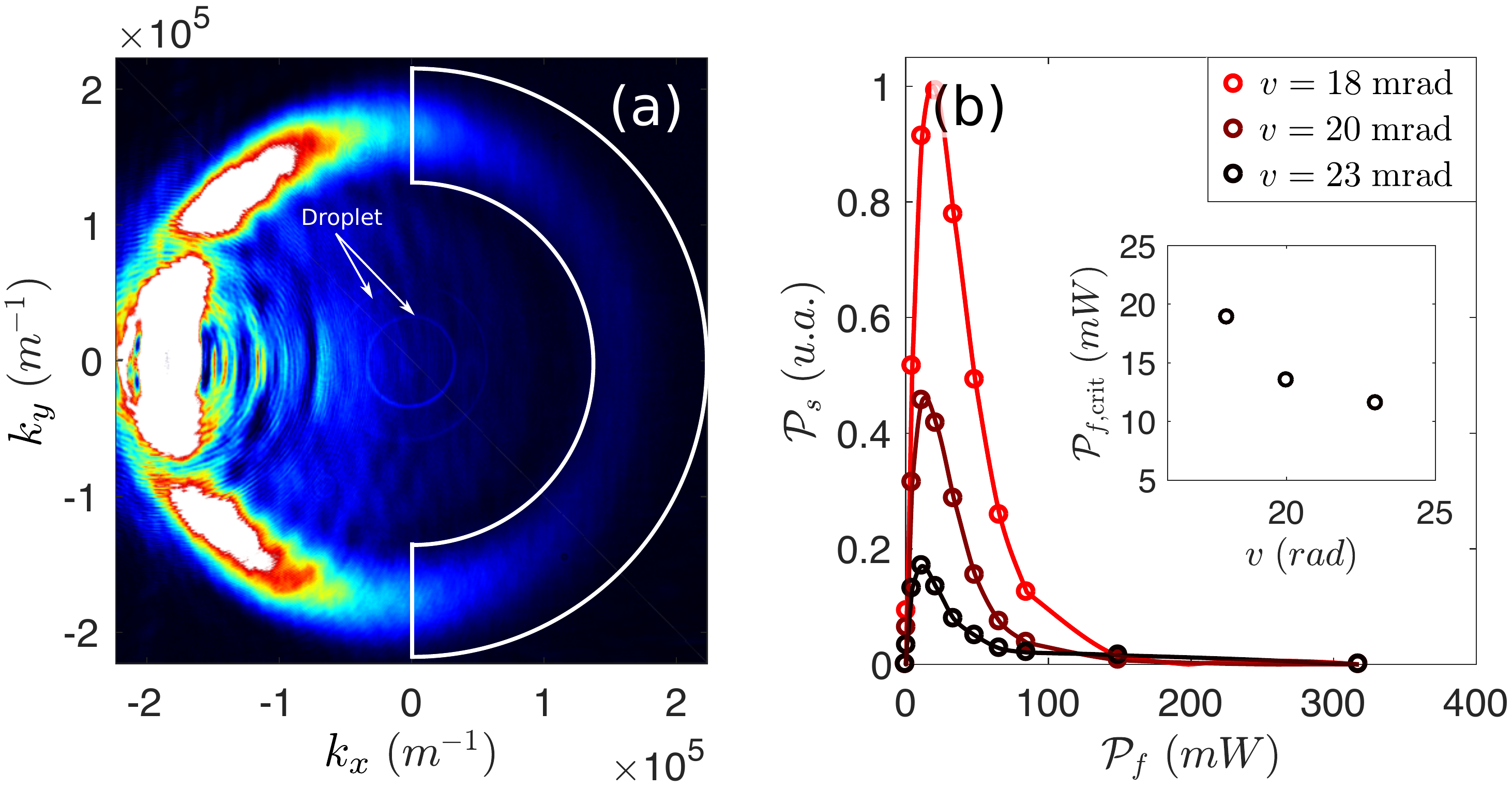}
\caption{(a): Scattering ring observed in far-field for $v=23$ mrad and$\,$for$\,\mathcal{P}_{f} = 1.0\,$mW. The two rings forming the intensity distribution of the droplet beam in k-space are visible. By integrating the intensity within the white frame, we can measure the amount of light back-scattered by the obstacle as function of the fluid power $\mathcal{P}_{f}$. The results are shown$\;$on figure (b) for different flow velocities $v$. When $P_{f}$ increases, the power $\mathcal{P}_{s}$ distributed over the right half of the diffraction ring first increases, reaches a maximum and then$\,$decreases. The critical fluid power $\mathcal{P}_{f, \mathrm{crit}}$ $-$ at which $\mathcal{P}_{s}$ starts dropping $-$ slightly decreases when$\,$the flow velocity steps up. This behaviour is not expected if we suppose that the drop in the back-scattering power is due to superfluidity. 
}
\label{fig:ScatteringVSPower}
\end{figure}

\noindent We also investigate the far-field diffraction pattern formed by the scattering of the fluid on the obstacle. Images of the Rayleigh rings observed in k-space at$\,$different$\,$probe$\,$powers$\,$and flow velocities are shown on figure~\ref{fig:FourierSpace}. On the$\,$upper$\,$series$\,$of$\,$images$\,$(a),$\,v = 18\,$mrad,$\,$while on the lower one, $v = 23$ mrad. The origin of the $k_{x}$- and $k_{y}$-axes lies at the center of the droplet rings (see figure~\ref{fig:ScatteringVSPower}(a)). The full$\,$images$\,$have$\,$been$\,$cropped$\,$in$\,$order$\,$to$\,$show$\,$only$\,$the back scattering part of the Rayleigh ring. Its radius is equal to the transverse wave-vector of$\,$the$\,$probe$\,$beam$\,\mathbf{k_{\perp}} \!= \!k_{0} \sin(\theta_{i}) \boldsymbol{e_{x}}$,$\,$explaining$\,$why$\,$it$\,$is$\,$larger$\,$on$\,$the$\,$series$\,$(a)$\,$than$\,$on$\,$(b).
\vspace{2pt}
\newline
\noindent As you may have seen, the amount of light distributed over the half ring first$\,$increases$\,$with the probe power $\mathcal{P}_{f}$, reaches a maximum and then decreases when further$\,$stepping$\,$up$\,\mathcal{P}_{f}$. We also notice a decrease in the radius of the scattering ring from$\,$figures$\,$(a1)$\,$to$\,$(a6)$\,$that$\,$is observed$\,$in$\,$simulations$\,$too.$\;$Indeed,$\,$when$\,c_{s}\,$rises$\,$toward$\,$the$\,$flow$\,$velocity,$\,$simulations show that the Rayleigh ring is deformed and develops a corner at $\mathbf{k_{\perp}}$. This corner consequently squeezes the ring in the $k_{y}$ direction (see~\cite{3-2Carusotto} for details).
\vspace{2pt}
\newline
\noindent We can finally mentioned the fact that the amount of light back-scattered$\;$by$\;$the$\;$defect$\;$is
lower at$\,$large$\,$flow$\,$velocities$\,$(b)$\,$than$\,$at$\,$small$\,$ones$\,$(a).$\;$This$\,$is$\,$certainly$\,$due$\,$to$\,$the$\,$previously mentioned tunnelling effect. We can go further by quantitatively measuring$\;$the power $\mathcal{P}_{s}$ distributed over the half scattering ring as function of the fluid power $\mathcal{P}_{f}$. The results are presented$\,$on$\,$figure$\,$\ref{fig:ScatteringVSPower}(b),$\,$for$\,$different$\,$flow$\,$velocities.$\;$Figure$\,$\ref{fig:ScatteringVSPower}(a)$\,$shows$\,$the$\,$Rayleight$\,$ring observed in k-space when $v=23$ mrad and $\mathcal{P}_{f} = 1$ mW. The fluid saturates$\,$the$\,$image on the left. The droplet rings, centered at $\mathbf{k}_{\perp} = \boldsymbol{0}$, are also visible. We can check$\,$that$\,$the$\,$ratio between the radius of the droplet outer and inner rings is 0.52 as expected.$\;$The$\,$white$\,$frame enclosing the right-hand part of the scattering ring delineates the$\,$area$\,$where$\,$the$\,$intensity is integrated.$\;$The$\,$points$\,$on$\,$figure$\,$\ref{fig:ScatteringVSPower}(b)$\,$arise$\,$from$\,$this$\,$integration$\,$at$\,$various$\,$probe$\,$powers. The resulting curves are in agreement with$\,$the$\,$qualitative$\,$observations$\,$made on$\,$figure$\,$\ref{fig:FourierSpace}. Indeed, whatever the flow velocity, the power of the back-scattered light increases with$\,\mathcal{P}_{f}$, reaches a maximum value at some critical fluid power $\mathcal{P}_{f,\mathrm{crit}}\,$before$\,$quickly$\,$dropping$\,$to$\,$zero. It is tempting to attribute this collapse of back- scattering$\,$to$\,$superfluidity.$\;$Nevertheless, crucial$\,$information$\,$are$\,$missing,$\,$such$\,$as$\,$the$\,$value$\,$of$\,$the sound velocity $c_{s}$, which prevents us from concluding with certainty. Moreover, we$\,$did$\,$not$\,$capture$\,$the$\,$near-field$\,$images$\,$related to the data points on figure~\ref{fig:ScatteringVSPower}(b). We$\,$are$\,$thus unable to tell if the obstacle is still visible in the cell output plane when $\mathcal{P}_{s}$ is decreasing.

\noindent Otherwise, it would mean that$\,$the$\,$strength$\,$of$\,$the$\,$obstacle$\,\delta n (0)\,$becomes$\,$comparable$\,$to$\,$the nonlinear change of refractive index $\Delta n(\boldsymbol{r}_{\perp})$ surrounding the defect at high probe powers. In that case, the decrease in the back-scattering would be due to the matching$\,$between$\,$the refractive indices inside and outside the obstacle cross-section. This$\,$should$\,$not$\,$be$\,$confused with$\,$superfluidity, which$\,$causes$\,$the$\,$suppression$\,$of$\,$back-scattering$\,$even$\,$when $\delta n(0) \gg \Delta n$, as mentioned previously when discussing the results of Michel \textit{et al.}~\cite{3-3Michel}. On figure~\ref{fig:SimuSuper}(b) for instance, superfluidity is observed despite the$\,$fact$\,$that$\,\delta n(0)/\Delta n = 25$. 

\newpage

\noindent It is also worth mentioning that if the drop of $\mathcal{P}_{s}$ is due to superfluidity, the flow$\,$velocity$\,v$ should then match the critical speed $v_{c}$ when $\mathcal{P}_{f} = \mathcal{P}_{f,\mathrm{crit}}$. As$\,$we$\,$deal$\,$with$\,$extended$\,$defect, we expect $v_{c}$ to be lower than $c_{s}$. Therefore, the sound velocity should, at least,$\,$be$\,$equal$\,$to the flow velocity. This requires $\Delta n >  v^{2} \sim 4 \!\times\! 10^{-4}$. Even if the probe is close to resonance,   
reaching$\,$such$\,$high$\,$values$\,$of$\,\Delta n\,$is$\,$unrealistic$\,$regarding$\,$the$\,$available$\,$power$\,$($\mathcal{P}_{f} < 1\,$W).
\vspace{2pt}
\newline
\noindent Moreover, we can observe that the critical$\,$power$\,\mathcal{P}_{f,\mathrm{crit}}\,$at$\,$which$\,\mathcal{P}_{s}\,$starts$\,$dropping$\,$slightly decreases when$\,$stepping$\,$up$\,$the$\,$flow$\,$velocity.$\;$This$\,$is$\,$not$\,$expected$\,$if$\,$we$\,$attribute$\,$the$\,$collapse of back-scattering to superfluidity. Scanning the probe power as in$\,$figure$\,$\ref{fig:ScatteringVSPower}(b)$\,$affects$\,$both the sound velocity and the healing length. In the phase diagram of figure~\ref{fig:Expectation}, such a scan$\,$of the fluid density $-$ at fixed defect size $\omega_{0,d}$ and flow velocity $v$ $-$ is like following leftward the curve of equation $y(x) = k \omega_{0,d} v/x$ (where $k$ is the probe wave-vector).$\;$For$\,$the$\,$values$\,$of the flow velocity considered here, $k \omega_{0,d} v$ ranges from 5.8 to 7.4. We thus expect$\,$the$\,$curves $x \!\shortrightarrow\! y_{1}(x)$, $x \!\shortrightarrow\! y_{2}(x)\,$and$\,x \!\shortrightarrow\! y_{3}(x)\,$(related$\,$to$\,v_{1} = 18\,$mrad,$\,v_{2} = 20\,$mrad$\,$and$\,v_{3} = 23\,$mrad) to cross the thick black curve$\,$of$\,$the$\,$phase$\,$diagram$\,$almost$\,$at$\,$the$\,$same$\,$abscissa$\,x$.$\;$Therefore, $\frac{v_{1}}{c_{s,1}} \simeq \frac{v_{2}}{c_{s,2}} \simeq \frac{v_{3}}{c_{s,3}}$.$\;$Since $v_{1} < v_{2} < v_{3}$,$\,$the$\,$preceding$\,$equation$\,$requires$\,$that$\,c_{s,1} > c_{s,2} > c_{s,3}$. As the sound velocity scales with$\,$the$\,$square$\,$root$\,$of$\,$the$\,$fluid$\,$density,$\,$we$\,$should$\,$consequently observe that $\mathcal{P}_{f,\mathrm{crit}}$ increases with the flow velocity. This is not what we$\,$measured$\,$however. For all the reasons mentioned above, superfluidity is certainly not the cause of the$\,$collapse of back-scattering on figure~\ref{fig:ScatteringVSPower}(b), which is more$\,$likely$\,$to$\,$arise$\,$from$\,$index$\,$matching effect. 

\newpage

\section{Future works}

\begin{figure}[H]
\centering
\includegraphics[width=\columnwidth]{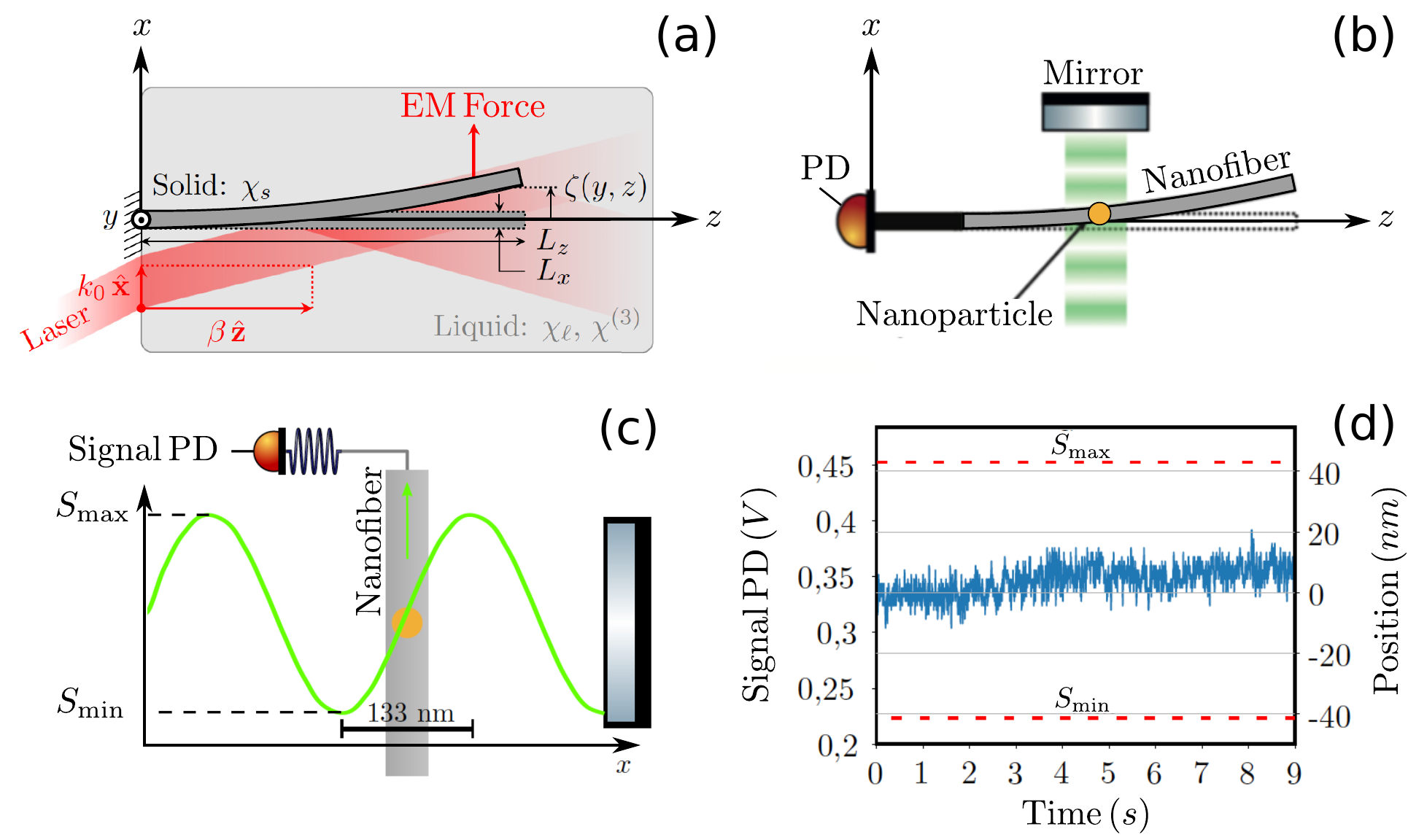}
\caption{Optomechanical$\,$signature$\,$of$\,$superfluidity.$\,$(a):$\,$Sketch$\,$of$\,$the$\,$experimental$\,$setup proposed by Larré \textit{et al.} in~\cite{6-2Larre}. The left end of the obstacle is clamped$\,$to$\,$the$\,$cell$\,$entrance window and its right end is free to move. Initially, the obstacle is aligned along the $z$-axis. When$\,$the$\,$laser$\,$is$\,$switched$\,$on,$\,$the$\,$obstacle$\,$bends$\,$under$\,$the$\,$radiation$\,$pressure.$\;$The$\,$position $\zeta(x,L_{z})$ of the right tip increases with the fluid intensity $\mathcal{I}_{f}$ before dropping to zero when passing through the normal/superfluid transition.$\;$(b)$\,$and$\,$(c):$\,$Displacement$\,$measurement. A gold nanoparticle is dropped onto a nanofiber (obstacle). A standing-wave is$\,$created$\,$by reflecting a laser beam on a mirror. The nanoparticle$\,$diffuses$\,$more$\,$or$\,$less$\,$light$\,$into$\,$the$\,$fiber depending on its position inside this standing-wave. By collecting the scattered light with a photo-diode (PD), we can measure $\zeta(x,L_{z})$ with an accuracy of $\pm 20$ nm (see figure (d)).}
\label{fig:Nanofiber}
\end{figure}

\noindent Rigorously probing the superfluidity of light in rubidium vapors$\,$by$\,$studying$\,$the$\,$scattering of the photon fluid$\,$on$\,$an$\,$all-optical$\,$defect$\,$requires$\,$to$\,$complete$\,$quite$\,$a$\,$few$\,$preliminary$\,$steps. We should first of all characterize the change of refractive index $\delta n$ induced by the droplet beam in the fluid of light (for different sets of parameters) and compare$\,$the$\,$results$\,$with$\,$the theory developed in the preceding chapter. For this purpose, we can for instance$\,$set$\,$to$\,$zero the probe transverse wave-vector and measure the phase accumulated$\,$by$\,$the$\,$photon$\,$fluid$\,$in the vicinity of the obstacle (using the scanning phase interferometry of subsection$\,$2.3.3$\,$i). One of the most crucial point consists in precisely characterizing the effects of$\,$the$\,$probe$\,$on the$\,$defect$\,$strength.$\;$Once$\,$this$\,$is$\,$done,$\,$we$\,$can$\,$start$\,$studying$\,$the$\,$scattering$\,$of$\,$the$\,$fluid$\,$of$\,$light on the obstacle. The first step is to investigate in details the physics of the transient$\,$regime, that is, the emission of spherical waves in the cell input plane. We$\,$can$\,$for$\,$example$\,$measure the$\,$velocity at which those waves propagate in the transverse plane. This could provide$\,$a convenient way of accessing the speed of sound \textit{in situ}. 
\vspace{2pt}
\newline
\noindent
The last but not least challenge$\,$is to probe superfluidity itself by$\,$measuring$\,$simultaneously the scattering patterns in real$\,$and in k-space. To that end, the best is to scan the probe transverse wave-vector while keeping its power constant. By measuring the phase of the photon fluid at the cell output plane, we can also think about observing and studying the spontaneous nucleation of quantized vortices in the wake of$\,$the$\,$defect.$\;$This$\,$would$\,$complete the study of superfluidity in our system.      
\vspace{6pt}
\newline
\noindent As mentioned in subsection 6.2.2, we also currently investigate the possibility of probing optomechanically the normal/superfluid transition in hot rubidium vapors.$\;$Following$\,$the theoretical proposal by Larré \textit{et al.}~\cite{6-2Larre}, we plan to install a nanofiber in a vacuum$\,$chamber filled$\,$with$\,$a$\,$pure vapor$\,$of$\,$rubidium.$\;$A$\,$sketch$\,$of$\,$the$\,$experimental$\,$setup$\,$has$\,$been$\,$depicted$\,$on figure~\ref{fig:Nanofiber}(a)$\,$(top$\,$view).$\,$The$\,$laser$\,$beam$\,$enters$\,$the$\,$Kerr$\,$medium$\,$from$\,$the$\,$left$\,$and$\,$propagates at a small angle with respect to the $z$-axis. The nanofiber is mechanically clamped to the entrance$\,$window.$\,$We$\,$assume$\,$it$\,$is$\,$aligned$\,$along$\,$the$\,z$-axis$\,$when$\,$the$\,$laser$\,$is$\,$off.$\,$Because$\,$of$\,$the radiation pressure, the nanofiber should bend when switching the laser on. We$\,$expect$\,$this bending to increase with the laser intensity until reaching the normal/superfluid$\,$threshold. At that point, Larré \textit{et$\,$al.} predict a fast drop in the fiber deflection $\zeta$, which returns$\,$to$\,$its initial position. In practice, the fiber is hold on both ends. Indeed, it$\,$seems$\,$difficult$\,$if$\,$not impossible to$\,$cut$\,$a$\,$nanofiber$\,$keeping$\,$it$\,$straight$\,$as$\,$it$\,$spontaneously$\,$tends$\,$to$\,$wind$\,$onto$\,$itself. Numerical simulations predicting the deflection $\zeta$ at the center of the$\,$nanofiber$\,$(in$\,$vaccum) has been performed by Maxime Joos in~\cite{7-17Joos} using the Mie scattering theory.$\;$In$\,$its$\,$work,
Maxime Joos considers$\,$a$\,$laser$\,$beam$\,$having$\,$a$\,$normal$\,$incidence$\,$onto$\,$a$\,$10$\,$mm$\,$long$\,$nanofiber. With the available laser resources, we can expect in that$\,$case$\,$a$\,$displacement$\,$ranging$\,$from $0.1$ to $1 \, \mu$m, depending on the laser polarization and on the diameter of the$\,$nanofiber$\,$used in experiments. In fine, however, we would like to make the laser propagates at$\,$a$\,$small$\,$angle with respect to the nanofiber. We expect the displacement $\zeta$ to be much lower$\;$in$\;$that$\;$case. We are thus currently looking for a$\,$way$\,$of$\,$increasing$\,$the$\,$momentum$\,$transferred$\,$by$\,$the$\,$laser to the fiber. A possibility would be to deposit some metallic reflective$\,$coating$\,$on$\,$its$\,$surface. In any event, we must also be able to accurately measure the nanometric deflection of the nanofiber at its$\,$center.$\;$This$\,$is$\,$done$\,$using$\,$an$\,$optical$\,$ruler$\,$whose$\,$principle$\,$is$\,$explained$\,$below.   

\newpage

\noindent A gold nanoparticle is dropped onto the fiber$\,$and$\,$positioned$\,$inside$\,$a$\,$standing-wave$\,$created by reflecting a green laser on a mirror, as shown on figures~\ref{fig:Nanofiber} (b) and (c). The$\,$nanoparticle scatters part of$\,$the$\,$light$\,$coming$\,$from$\,$the$\,$standing-wave$\,$into$\,$the$\,$nanofiber.$\,$As$\,$the$\,$amount$\,$of scattered light depends on the nanoparticle position in the standing-wave, we can retrieve the deflection of the fiber by simply measuring its output$\,$power$\,$using$\,$a$\,$photo-diode$\,$(PD). This method, developed by Maxime Joos~\cite{7-17Joos} and further improved by Chengjie Ding, allows to measure the position of the nanoparticle with an accuracy of $\pm 20$ nm (figure$\,$(d)).

\chapter*{Summary and future works}

\newpage

\section*{General conclusion}


\noindent The primary purpose of this thesis was to study some of the hydrodynamic properties$\;$of$\;$a propagating$\,$photon$\,$fluid$\,$in$\,$hot$\,$rubidium$\,$vapors.$\,$While$\,$photons$\,$are$\,$not$\,$interacting$\,$particles in vacuum, they are inside the rubidium vapor, when the laser is tuned close to an atomic resonance. This interaction between photons, mediates by the atomic ensemble, makes$\,$the laser beam behaveS as a fluid flowing in the plane perpendicular to the optical axis$\,$when$\,$it propagates through the vapor. The dynamics of this fluid of light$\,$is$\,$driven$\,$by$\,$the$\,$nonlinear Schr\"{o}dinger equation, which shares strong similarities with the Gross–Pitaevskii equation. The latter describes the space-time evolution of atomic Bose-Einstein condensates in the mean-field approximation. Those$\,$systems$\,$exhibit$\,$in$\,$particular$\,$the$\,$ability$\,$of$\,$flowing$\,$without experiencing any friction, that is, without dissipating energy. Our desire to observe$\,$such superfluid$\,$flow$\,$of$\,$light$\,$in$\,$our$\,$system$\,$has$\,$driven$\,$the$\,$entire$\,$work$\,$presented$\,$in$\,$this$\,$manuscript.
\vspace{6pt}
\newline
\noindent The strength of the photon-photon interaction in Kerr mediums is characterized by the material$\,$nonlinear$\,$refractive$\,$index$\,n_{2}$.$\;$This$\,$quantity$\,$thus$\,$plays$\,$a$\,$crucial$\,$role$\,$in$\,$experiments. In rubidium vapors, repulsive interactions between photons are obtained by red-detuning some laser beam from one of the $D$-lines. In chapter 1, we first describe the$\,$optical$\,$response of the rubidium vapor under this near-resonance laser excitation with a two-level model. This simplistic description is improved afterwards by taking into account optical pumping between the $D$-line ground states as well as the finite transit time of atoms$\,$across$\,$the$\,$beam. Using this extended model, we derive a general expression for the dielectric susceptibility. In order to further refine our theoretical description, Doppler broadening and nonlocality, arising$\,$from$\,$the$\,$ballistic$\,$transport$\,$of$\,$excited$\,$atoms$\,$in$\,$hot$\,$vapors,$\,$have$\,$also$\,$been$\,$included. 
\vspace{6pt}
\newline
\noindent Chapter 2 begins$\,$with$\,$the$\,$derivation$\,$of$\,$the$\,$nonlinear$\,$Schr\"{o}dinger$\,$equation.$\,$The$\,$connection with the Gross Pitaevskii equation is established and discussed in details.$\;$We$\,$then$\,$focus$\,$on describing the dynamics of small amplitude density waves travelling onto the photon fluid. Using the Bogoliubov transform, we show that those waves obey the so-called Bogoliubov dispersion$\,$relation,$\;$which$\,$exhibits$\,$two$\,$different$\,$regimes.$\;$It$\,$first$\,$starts$\,$by$\,$linearly$\,$increasing at low excitation wave-vectors, where density waves behave as collective phonons$\,$travelling all$\,$at$\,$the$\,$same$\,$speed$\,$(the$\,$sound$\,$velocity),$\,$before$\,$developing$\,$a$\,$quadratic$\,$trend,$\,$characteristic of$\,$a$\,$particle-like$\,$dispersion.$\;$According$\,$to$\,$the$\,$Landau$\,$criterion$\,$for$\,$superfluidity,$\;$showing the existence of the sound velocity would guarantee the observation of superfluid flows of$\;$light in our system. I consequently dedicate an important part of my time to measuring the dispersion relation of density fluctuations. The results are shown in chapter 4.
\vspace{6pt}
\newline
\noindent In chapter 3, the tools and the methods used to generate the photon fluid$\,$on$\,$one$\,$hand$\,$and characterise it on the other are introduced. We start by presenting the rubidium cell and its heating system before describing the laser sources. An important part of this chapter$\,$is dedicated to presenting two techniques we used to access the nonlinear refractive index$\,n_{2}$. Both are based on measuring the self-phase accumulated by a wide Gaussian beam during its propagation inside the vapor cell. The first one consists in counting the number$\,$of$\,$rings appearing$\,$in$\,$the$\,$far-field$\,$intensity$\,$distribution$\,$of$\,$the$\,$beam.$\;$The$\,$correct$\,$counting$\,$procedure, taking into account the Gouy phase shift, is described$\,$in$\,$both$\,$the$\,$1D$\,$and$\,$2D$\,$cases.$\,$To$\,$do$\,$so, we have extended the approach followed by Nicolas Pavloff in an unpublished work of 2018.

\newpage

\noindent The second method consists in measuring the spatial variations of the nonlinear$\,$phase$\,$shift accumulated by the beam using$\,$the$\,$scanning$\,$phase$\,$interferometry. We show additionally, using numerical simulations, that both techniques are accurate as long as the thin medium approximation is fulfilled. This is the case as long as the characteristic propagation length over which self-defocusing starts affecting the beam shape is longer than the cell. 
\vspace{6pt}
\newline
\noindent In chapter 4, the dispersion relation of small amplitude density fluctuations$\,$travelling$\,$onto the fluid of light is measured. Following the work of Vocke$\,$\textit{et al.}~\cite{2-24Vocke},$\,$we$\,$first$\,$try$\,$to$\,$measure the difference in the phase velocity between two plane waves propagating on top of low and high density background fluids. At the medium output plan, this results in$\,$a$\,$shift$\,$between the$\,$crests$\,$of$\,$these$\,$two$\,$waves,$\,$that$\,$depends$\,$on$\,$their$\,$wavelength.$\;$We$\,$show$\,$that$\,$the$\,$theoretical description of Vocke \textit{et al.} is incomplete as they do not account for the propagation$\,$of$\,$the conjugate beam in the cell, which is spontaneously generated (because of four-wave$\,$mixing) in the input plane. Following the work of Larré \textit{et al.}~\cite{3-19Larre}, we derive an exact formula to calculate the shift using the Bogoliubov formalism. This formula$\,$depends$\,$on$\,$the$\,$dispersion relation of density waves but the latter can only be retrieved from the shift$\,$at$\,$the$\,$expense$\,$of a complex numerical inversion. We thus claim that this method is not suitable to measure the dispersion relation of density waves. We still however demonstrate that the$\,$shift$\,$is$\,$not saturating at large modulation wavelength$\,-\,$both experimentally and$\,$numerically$\,-\,$which contradicts the results of Vocke \textit{et al.} In order to access the dispersion, we$\,$propose$\,$a$\,$new experimental$\,$scheme$\,$based$\,$on$\,$measuring$\,$the$\,$group$\,$velocity$\,$of$\,$a$\,$small$\,$Gaussian$\,$wave-packet travelling onto the photon fluid. Beside the fact that this technique overcomes all the main limitations of the shift measurement, it also provides a much deeper understanding about the physics at play. By probing the sound-like regime of the dispersion, the wave-packet splits into a pair of counter-propagating Bogoliubov wave-trains in the cell entrance plane, which both travel at the speed of sound. Reversely, when probing the particle-like regime, the wave-packet behaves as a free particle moving at a velocity that increases linearly with its wave-vector.$\,$This$\,$nonlinear$\,$refraction$\,$law$\,$is$\,$theoretically$\,$described$\,$using$\,$the$\,$Bogoliubov formalism and illustrated by performing numerical simulations. The experimental results are in excellent agreement with theory at low background density, when transport-induced nonlocality is taken into account. The dispersion relation retrieved$\,$using$\,$the$\,$group$\,$velocity measurement exhibits a linear$\,$increase$\,$at$\,$low$\,$excitation$\,$wave-vectors$\,$which$\,$is$\,$characterised by the sound velocity. The way the later depends on the fluid density is investigated and exactly$\,$matches$\,$theory$\,$without$\,$any$\,$fitting$\,$parameter.$\;$At$\,$the$\,$end$\,$of$\,$chapter$\,$4,$\,$we$\,$also$\,$report the observation of quasi-particle interferences occurring$\,$between$\,$the$\,$counter-propagating Bogoliubov wave-packets at the cell exit plane. 
\vspace{6pt}
\newline
\noindent Demonstrating the existence of a sonic regime in the dispersion$\,$relation$\,$is$\,$a$\,$key$\,$requirement for the observation of superfluidity. In order to go one step further, the way the fluid flows around a localized$\,$obstacle$\,$should$\,$be$\,$investigated.$\;$In$\,$chapter$\,$5,$\,$the$\,$method$\,$used$\,$to$\,$generate such an obstacle in our system is described. In photon fluids, any local change$\,$of$\,$refractive index acts as a defect into the flow. In rubidium vapors, localized refractive index changes can be induced by red-detuning the fluid from one of the $D$-line while strongly$\,$driving$\,$the other with an intense defect field. This situation is first theoretically investigated using a open 4-level N-type atomic model. The$\,$optical$\,$Bloch$\,$equations$\,$are$\,$derived$\,$and$\,$solved$\,$using first a perturbative approach and then the dressed-state formalism. This second method helps us getting a deeper insight into the process underlying the generation of the obstacle into the fluid of light. We show$\,$that$\,$the$\,$strength$\,$and$\,$the$\,$sign$\,$of$\,$the$\,$potential$\,$induced$\,$by$\,$the defect field can be tuned by changing$\,$its$\,$power$\,$and/or$\,$its$\,$frequency.$\;$We$\,$also$\,$alert$\,$the$\,$reader to the fact that varying the power of the fluid surrounding$\,$the$\,$defect$\,$also$\,$affects$\,$its$\,$strength. The second section of chapter 5 is dedicated to presenting$\,$the$\,$methods$\,$used$\,$in$\,$experiments to$\,$produce$\,$the$\,$defect$\,$beam.$\,$The$\,$later$\,$should$\,$ideally$\,$fulfill$\,$certain$\,$requirements$\,$such$\,$as$\,$being collimated and keeping the same strength across the full cell. Its diameter should also be comparable to the healing length $-$ which is typically of the order of few tens of$\,$microns.  We chose to use Bessel beams whose diffraction-free features enable to generate$\,$collimated defects of appropriate size all along the cell. Moreover, we show that designing the on-axis intensity profile$\,$of$\,$Bessel$\,$beam$\,$enables$\,$to$\,$compensate$\,$linear$\,$absorption$\,$during$\,$propagation. This technique, that$\,$is$\,$based$\,$on$\,$the$\,$real$\,$space$\,$shaping$\,$of$\,$a$\,$Gaussian$\,$beam$\,$with$\,$a$\,$phase-only spatial light modulator, finds a wide variety of applications, in bio-imaging in particular, where the huge absorption and diffusion coefficients of living tissues make the illumination of such samples challenging. Nevertheless, generating obstacles with attenuation-resistant Bessel beams (or standard quasi-Bessel beams) is not the most suitable option as the rings surrounding the Bessel central core are intense enough$\,$to$\,$also$\,$change$\,$the$\,$refractive$\,$index. In experiments, we prefer instead using droplet beams, which results from the interference between two co-propagating co-axial Bessel beams having different cone angles. They$\,$offer the benefit of being perfectly collimated while reducing the$\,$power$\,$distributed$\,$over$\,$the$\,$rings compared to standard quasi-Bessel beams.  
\vspace{6pt}
\newline
\noindent In$\,$chapter$\,$6,$\,$we$\,$finally$\,$present$\,$some$\,$preliminary$\,$results$\,$obtained$\,$by$\,$bringing$\,$all$\,$the$\,$previous ingredients together. Images of the near-field and far-field$\,$scattering$\,$patterns$\,$generated$\,$by making the fluid flow toward the defect are shown. So far, we$\,$did$\,$not$\,$observe$\,$clear$\,$signature of superfluidity, but effects such as the collapse of the Rayleigh ring in k-space should be further investigated and may provide the experimental evidence we look for.

\newpage

\section*{Future works}

\noindent Propagating photon fluids in nonlinear mediums are$\,$a$\,$versatile$\,$and$\,$highly$\,$tunable$\,$platform to study the rich physics of quantum fluids. However, those systems$\,$are$\,$intrinsically$\,$limited by the length of the nonlinear medium, that fixes, once and for all, the "time" over which they are evolving. Beside the fact that long mediums are often required to make$\,$this$\,$time$\,$of evolution as large as possible, probing the dynamics of the photon$\,$fluid$\,$is$\,$really$\,$challenging, as we can only image in practice the medium exit plane. One possible way of overcoming this issue is to measure the full electric field (intensity plus phase) at the output plane$\;$of$\;$a short cell, by using a shearing interferometric camera~\cite{8-2Hung} for$\,$example.$\;$We$\,$can$\,$then$\,$tailor the input field to make it match the output one, shaping its intensity and/or$\,$phase$\,$using$\,$a digital micro-mirror device and a SLM. This$\,$re-injection$\,$strategy~\cite{8-1Mukherjee} would provide$\,$a$\,$way of extending the evolution  "time"$\,$(by$\,$performing$\,$many$\,$loops)$\,$while$\,$resolving$\,$the$\,$"temporal" dynamics of the photon fluid (using a short cell for instance). However, the role played$\,$by the air/medium interfaces in this time-loop needs to be further investigated. 
\vspace{6pt}
\newline
\noindent The control over the photon fluid temporal evolution can be further improved by adding confinement in the transverse direction. This can be done$\,$by$\,$modifying$\,$the$\,$refractive$\,$index experienced by the fluid as shown in chapter 5.$\;$By$\,$using$\,$ring-shaped$\,$confining$\,$beam,$\,$we$\,$can for instance create a light-induced wave-guide inside the vapor cell$\,$and$\,$trap$\,$the$\,$photon$\,$fluid in$\,$the$\,$resulting$\,$harmonic$\,$potential.$\;$In$\,$these$\,$circumstances,$\,$the$\,$system$\,$is$\,$scale-invariant$\,$and a new type of 2D$\,$breathers,$\,$recently$\,$discovered$\,$in~\cite{3-8Saint},$\,$may$\,$be$\,$observed.$\;$Optically-induced potentials are also promising tools to study the interplay between localization features$\,$and superfluidity. Illuminating the rubidium cell with an amorphous speckle light~\cite{8-3Battista} is a way of creating "stationary" disordered potential landscapes into the photon$\,$fluid,$\,$that$\,$are necessary to investigate the competition between localization and superfluid transport~\cite{6-1Segev}. Using the "light guiding light" strategy also provides a way of implementing evaporative cooling of photon fluids~\cite{8-4Chiocchetta}, which$\,$is$\,$a$\,$key$\,$requirement$\,$for$\,$accelerating$\,$and$\,$thus$\,$observing Bose-Einstein condensation of light. 
\vspace{6pt}
\newline
\noindent An important frontier in fluid of light research is to go beyond$\,$mean$\,$field$\,$and$\,$observe$\,$purely quantum phenomena.$\;$Because$\,$light$\,$propagating$\,$in$\,$atomic$\,$vapors$\,$is$\,$a$\,$well$\,$controlled$\,$system which has already proved to be an excellent source of quantum correlated beams~\cite{2-12Glorieux,8-5Glorieux}, we believe that it could be the first platform to lay the groundwork for studying quantum effects in propagating photon fluids. One step in this direction would be to observe the optical analogue of the dynamical Casimir effect~\cite{3-5Larre}. This can be achieved using the same experimental configuration as in section 4.2, but without stimulating the emission of the Bogoliubov modes this time. In$\,$that$\,$case,$\,$because$\,$of$\,$the$\,$sudden$\,$jump$\,$of$\,$the$\,$photon-photon interaction in the cell entrance plane, pairs of quantum-correlated Bogoliubov excitations are spontaneously emitted in that plane, seeded by vacuum. The main challenge$\,$is$\,$now$\,$to find a technique to measure the correlations between these excitations,$\,$using$\,$certainly$\,$some sophisticated homodyne detection scheme.


\backmatter

\bibliographystyle{unsrt}
\bibliography{sample}

\end{document}